%% file: MekhovDSc.tex
\documentclass[a4paper,12pt]{extreport}

\input{packages}        
\input{styles}          
\input{dataIM}            
\begin{document}
\input{names_eng}          
\input{title_engIM}           
\pagestyle{plain}
\setcounter{page}{3}
\input{contentsIM}            

\input{Introduction}   
\input{newcommandsIM}

\input{Chapter1}        

\input{Chapter2}        

\input{Chapter3}        

\input{Chapter4}        

\input{Chapter5}		  

\input{Conclusions}    
\input{referencesIM}  

\end{document}

%% file: packages.tex
\usepackage{ifxetex}

\usepackage{lscape}                                 
\usepackage{geometry}                               

\usepackage{amsthm,amsfonts,amsmath,amssymb,amscd}  
\usepackage{physics} 
\usepackage{stackengine} 
\usepackage{enumitem} 

\ifxetex
  \usepackage{polyglossia}                          
  \usepackage{fontspec}                             
  \usepackage[utf8]{inputenc}    
  \usepackage{bm}                                   
\else
  \usepackage{cmap}                                 
  \usepackage[T2A]{fontenc}                         
  \usepackage[utf8]{inputenc}                       
  \usepackage[english, russian]{babel}              
  \IfFileExists{pscyr.sty}{\usepackage{pscyr}}{}    

\fi

\usepackage{indentfirst}                            

\usepackage[usenames]{color}
\usepackage{color}
\usepackage{colortbl}

\usepackage{longtable}                              
\usepackage{multirow,makecell,array}                
\usepackage{booktabs}                               

\usepackage[singlelinecheck=off,center]{caption}    
\usepackage{soul}                                   
\usepackage{icomma}                                 
\usepackage{float}                                  

\usepackage{cite}                                   

\usepackage[linktocpage=true,plainpages=false,pdfpagelabels=false]{hyperref}

\usepackage{graphicx}                               

\usepackage{tocloft}

\usepackage{setspace}

\usepackage{fancyhdr}

\usepackage{bm}

\usepackage{comment}

\usepackage{titlesec}

\usepackage[figure,table]{totalcount}

%% file: styles.tex
\ifxetex
  \setmainlanguage{russian}
  \setotherlanguage{english}
  \defaultfontfeatures{Ligatures=TeX,Mapping=tex-text}  
  \newfontfamily\cyrillicfont{Minion Pro}  
  \newfontfamily\cyrillicfontsf{Myriad Pro}  
  \newfontfamily\cyrillicfonttt{Courier New}
  
  \titleformat{\chapter}[display]
  {\normalfont\large\bfseries\sffamily\centering} %
  {\centering\chaptertitlename\ \thechapter}{14pt}{\large}
  \titleformat*{\section}{\normalfont\large\bfseries\sffamily} %
  \titleformat*{\subsection}{\normalfont\bfseries\sffamily} %
\else
  \IfFileExists{pscyr.sty}{}{}
  
  \titleformat{\chapter}[display]
  {\normalfont\large\bfseries\sffamily\centering} %
  {\centering\chaptertitlename\ \thechapter}{14pt}{\large}
  \titleformat*{\section}{\normalfont\large\bfseries\sffamily} %
  \titleformat*{\subsection}{\normalfont\large\bfseries\sffamily} %
  
\fi

\makeatletter
\renewcommand{\@biblabel}[1]{#1.}   
\makeatother

\graphicspath{{../images/}}            

\definecolor{linkcolor}{rgb}{0.9,0,0}
\definecolor{citecolor}{rgb}{0,0.6,0}
\definecolor{urlcolor}{rgb}{0,0,1}
\hypersetup{
    colorlinks, linkcolor={linkcolor},
    citecolor={citecolor}, urlcolor={urlcolor}
}

\newcommand{\todo}[1]{\textcolor{black}{#1}} 

\makeatletter
\let\ps@plain\ps@fancy              
\makeatother
\makeatletter
\providecommand\add@text{}
\newcommand\tagaddtext[1]{%
  \gdef\add@text{#1\gdef\add@text{}}}%
\renewcommand\tagform@[1]{%
  \maketag@@@{\llap{\add@text\quad}(\ignorespaces#1\unskip\@@italiccorr)}%
}
\makeatother

\input{glyphtounicode}

%% file: dataIM.tex
\newcommand{\thesisAuthor}             
{\todo{Мехов Игорь Борисович}}
\newcommand{\thesisUdk}                
{\todo{533.9}}
\newcommand{\thesisTitle}              
{\todo{Квантовая оптика ультрахолодных квантовых газов: открытые системы за рамками диссипации}}
\newcommand{\thesisSpecialtyNumber}    
{\todo{1.3.6.}}
\newcommand{\thesisSpecialtyTitle}     
{\todo{Оптика}}
\newcommand{\thesisDegree}             
{\todo{доктора физико-математических наук}}
\newcommand{\thesisCity}               
{\todo{Санкт-Петербург}}
\newcommand{\thesisYear}               
{\todo{2021}}
\newcommand{\thesisOrganization}       
{\todo{САНКТ-ПЕТЕРБУРГСКИЙ ГОСУДАРСТВЕННЫЙ УНИВЕРСИТЕТ}}

\newcommand{\supervisorFio}            
{\todo{}}
\newcommand{\supervisorRegalia}        
{\todo{}}

\newcommand{\thesisAuthoreng}             
{\todo{Mekhov Igor Borisovich}}
\newcommand{\thesisTitleeng}              
{\todo{Quantum optics of ultracold quantum gases: \\ open systems beyond dissipation}}
\newcommand{\thesisSpecialtyTitleeng}     
{\todo{Optics}}
\newcommand{\thesisDegreeeng}             
{\todo{Doctor of Physical and Mathematical Sciences}}
\newcommand{\thesisCityeng}               
{\todo{Saint Petersburg}}
\newcommand{\thesisOrganizationeng}       
{\todo{SAINT PETERSBURG STATE UNIVERSITY}}
\newcommand{\supervisorFioeng}            
{\todo{}}
\newcommand{\supervisorRegaliaeng}        
{\todo{}}

%% file: names_eng.tex
\renewcommand{\abstractname}{Annotation}
\renewcommand{\alsoname}{см. также}
\renewcommand{\appendixname}{Appendix} 
\renewcommand{\bibname}{References} 
\renewcommand{\ccname}{исх.}
\renewcommand{\chaptername}{Chapter}
\renewcommand{\contentsname}{Table of contents} 
\renewcommand{\enclname}{вкл.}
\renewcommand{\figurename}{Figure} 
\renewcommand{\headtoname}{вх.}
\renewcommand{\indexname}{Предметный указатель}
\renewcommand{\listfigurename}{Список рисунков}
\renewcommand{\listtablename}{Список таблиц}
\renewcommand{\pagename}{Стр.}
\renewcommand{\partname}{Часть}
\renewcommand{\refname}{References} 
\renewcommand{\seename}{см.}
\renewcommand{\tablename}{Table} 

\renewcommand{\cftchappresnum}{Chapter } 
\newlength{\mylen} 
\settowidth{\mylen}{\bfseries\cftchappresnum\cftchapaftersnum} 
\addtolength{\cftchapnumwidth}{\mylen} 

%% file: title_engIM.tex
\thispagestyle{empty}
\pagestyle{empty}
\clearpage

\begin{center}
\par
\end{center}

\begin{center}
{\bf \large Quantum optics of ultracold quantum gases: \\ open systems beyond dissipation. \\
Doctor of Sciences (Habilitation) Thesis.
\par}
\end{center}

\vspace{2mm}
\begin{center}
{\large Igor B. Mekhov}
\end{center}


\vspace{30mm}

\begin{figure}[h]
\centering
\captionsetup{justification=justified}
\includegraphics[trim=20 440 90 130, clip, width=0.7\textwidth]{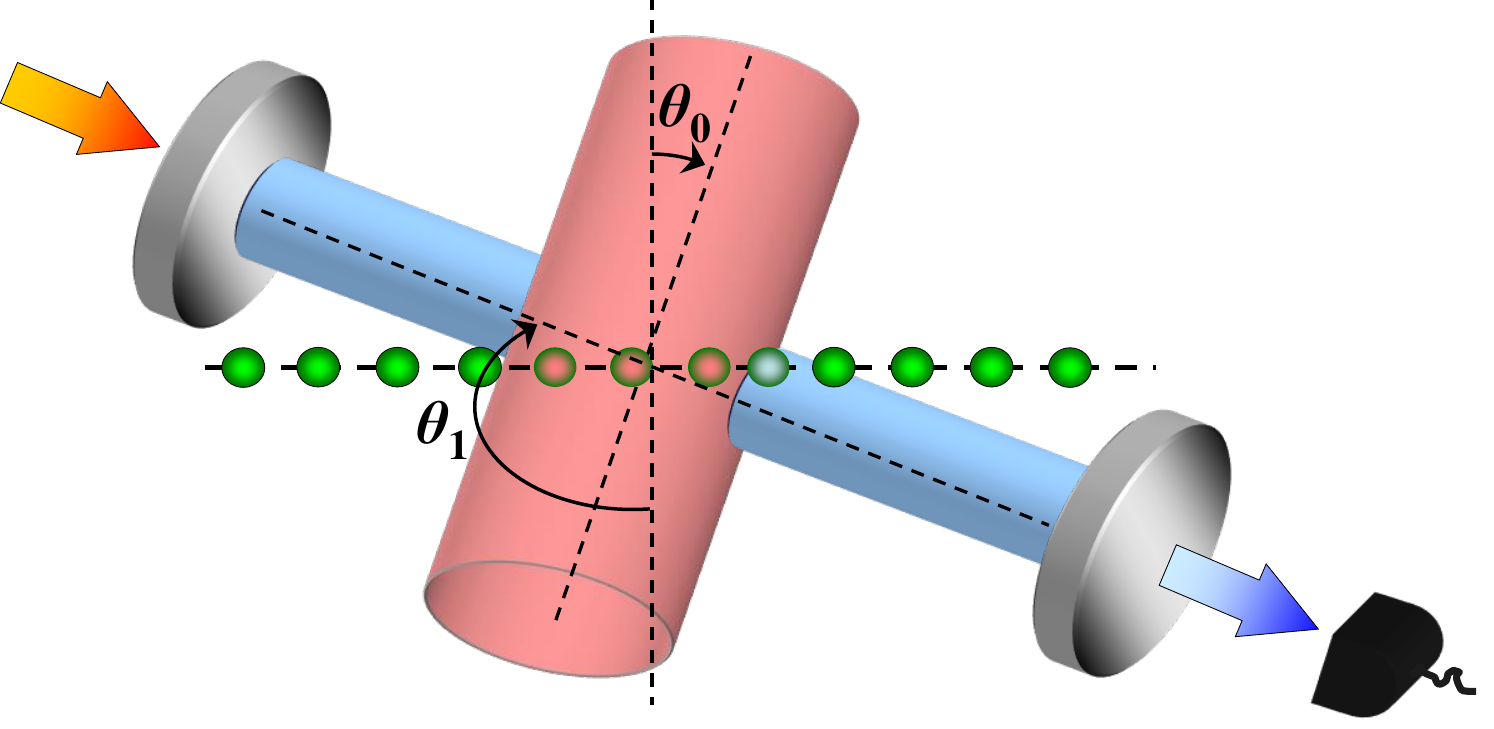}
\end{figure}

\vspace{90mm}

{\it igor.b.mekhov@gmail.com}

\newpage

\begin{center}
{\large    }
\end{center}

\vspace{7mm}

\begin{center}
{\large Abstract}
\end{center}

\vspace{7mm}

\footnote{This work is based on the Doctor of Sciences (Habilitation) Thesis at the St. Petersburg State University. The original papers it is based on (written at Innsbruck, Harvard, Oxford, and St. Petersburg) and all co-authors are acknowledged in the Introduction and main text.}Both quantum optics and physics of ultracold quantum gases are well-established fields of modern quantum science forming the basis for emerging quantum technologies, quantum information, and quantum artificial intelligence. Nevertheless, until recently the interaction between them was practically absent: in all experiments and in the majority of theories of ultracold gases, the quantum nature of light plays no role. In this work, we consider systems and phenomena, where the quantumness of both light and ultracold matter is equally important. Therefore, quantum optics of ultracold quantum gases describes the ultimate quantum level of light--matter interaction predicting phenomena unobtainable in standard problems of ultracold atoms trapped in prescribed optical potentials, e.g., optical lattices.

First, we show that light can serve as a quantum nondemolition (QND) probe of the atomic many-body phases: the phases can be distinguished by measuring their statistical properties from nontrivial correlation functions to full distribution functions of many-body variables (we give examples for bosons, fermions, and dipolar molecules). We demonstrate that light scattering is not only sensitive to the on-site atomic densities, but also to the matter-field interference at its shortest possible distance in an optical lattice. Second, we prove that the backaction of quantum measurements constitutes a novel source of competitions in many-body systems, which is especially pronounced, when the measurements are not QND ones. This leads to a plethora of novel phenomena: macroscopic oscillations of multipartite entangled matter modes, measurement-induced protection and break-up of fermion pairs, generation of antiferromagnetic orders, novel long-range pair correlated tunnelling and entanglement beyond the standard Hubbard models. We prove that the feedback control can induce phase transitions in quantum systems and tune their critical exponents and universality class. Third, the quantization of optical trapping potential (quantum optical lattices) leads to novel many-body phases of ultracold atoms, including both the density orders (e.g. lattice supersolids and density waves) and orders of matter-field coherences (e.g. bond orders such as superfluid and supersolid dimers, trimers, etc.)  

The general results applicable beyond physics of ultracold atoms are the following. We extend the paradigm of feedback control from the quantum state control to the control of phase transitions, including tuning their universality class. We present the backaction of quantum measurements as a novel source of competitions in many-body physics. We merge the paradigms of quantum Zeno dynamics and non-Hermitian physics, and introduce a novel type of quantum Zeno phenomena with Raman-like transitions well beyond the standard concept of Zeno dynamics. We propose a concept of quantum simulators based on the collective light--matter interaction with the interplay of long- and short-range interactions. Our models can be applied to other arrays of quantum particles (qubits). In general, quantum measurements and feedback open a way to obtain novel phenomena untypical to both closed unitary systems and open dissipative ones in the context of many-body physics.



\newpage

%

%% file: contentsIM.tex
\renewcommand{\contentsname}{\large{Table of contents}}

\vspace*{50mm}
\begin{center}
{\it To Nina and Dimitri}
\end{center}

\newpage

\tableofcontents
\clearpage

%% file: Introduction.tex
\chapter*{Introduction}							
\addcontentsline{toc}{chapter}{Introduction}	

Both  quantum optics and physics of ultracold quantum gases represent nowadays two well-established and actively developing fields of modern quantum science forming the basis for emerging quantum technologies, quantum information, and quantum artificial intelligence. Nevertheless, until recently the interaction between these two fields was practically absent. More precisely, the field of ultracold gases is based on manipulation of atoms using optical methods and laser light. However, all the experiments so far and overwhelming majority of theoretical works in this field do not take into account the quantum nature of light at all. The main goal of this work is to merge these two fields and develop a theory of novel phenomena, where the quantum natures of both the atomic matter waves and light waves are equally important. Therefore, the subject of this work -- quantum optics of ultracold quantum gases -- is a direction of research describing the ultimate quantum level of light--matter interaction, which is focused on effects unobtainable in standard setups and theories of ultracold atoms trapped in prescribed optical potentials.

\begin{figure}[h]
\centering
\captionsetup{justification=justified}
\includegraphics[width=0.9\textwidth]{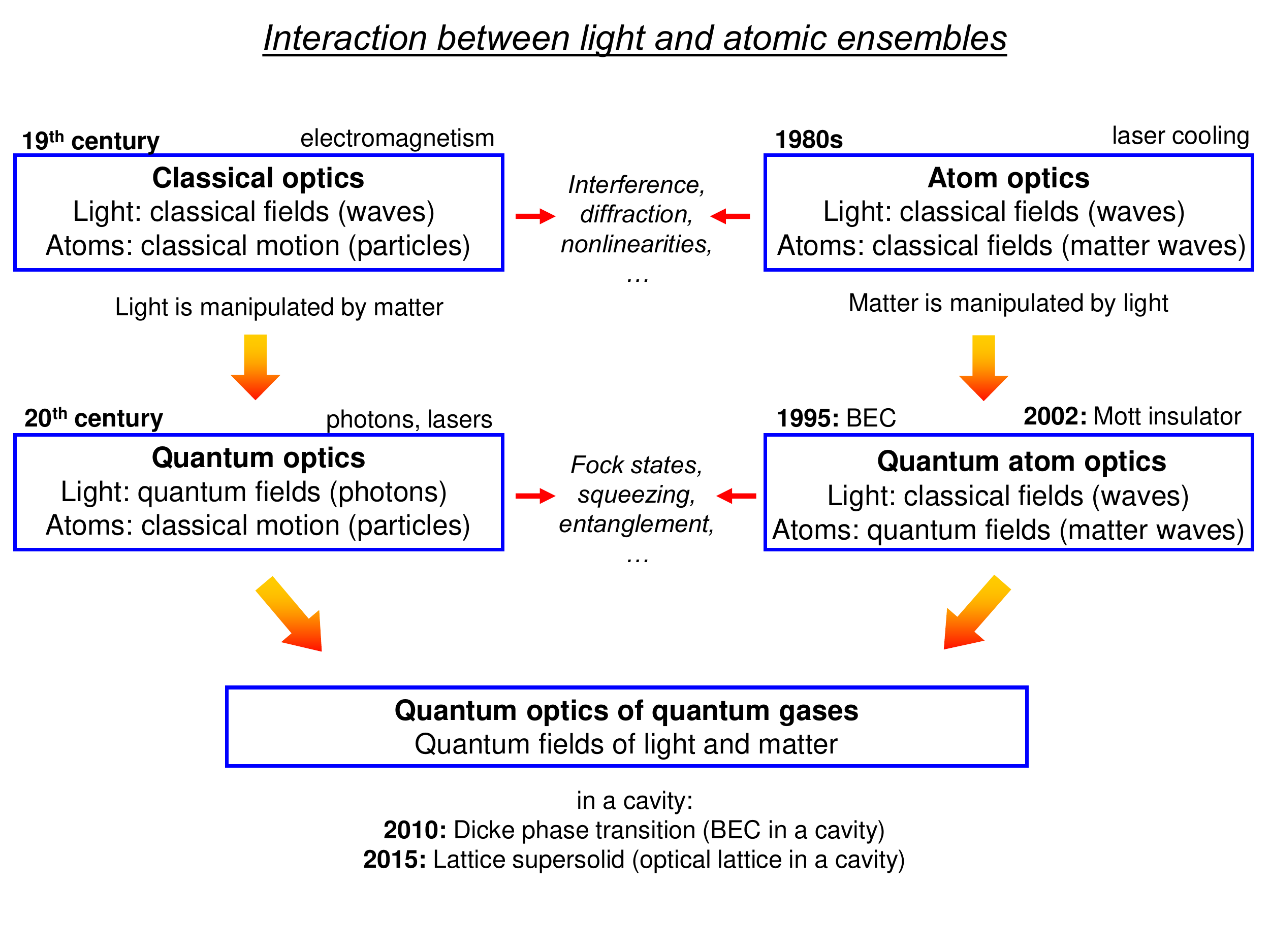}
\caption{\label{0-fig1} The place of quantum optics of quantum gases in the picture of interaction between light and atomic ensembles. In light optics and atom optics, the roles of light and matter are interchanged: many classical and quantum phenomena and states have been obtained for both light and atom waves.  Quantum optics of quantum gases joins the quantization of light as used in quantum optics with the quantization of atomic motion as used in quantum atom optics, thus considering both light and ultracold matter as quantum fields.}
\end{figure}

Historically (cf. Fig.~\ref{0-fig1}), classical optics treating the light as classical electromagnetic waves was created in the 19th century and has become one of the most developed and fruitful fields of physics. It has provided us a lot of technological breakthroughs, e.g., the highest level of measurement precision. A new era in optics started in the 20th century with the creation of quantum theory and invention of laser, when the concept of photons came into existence and became testable experimentally. Currently, quantum optics, which treats the light as quantum field, thus going beyond the classical description of electromagnetic waves, is a well-developed area as well.

The progress in laser cooling techniques in the last decades of the 20th century led to the foundation of a new field of atom physics: atom optics. According to quantum mechanics, at very low temperatures (nanokelvins in modern experiments) the velocity of atoms becomes very small and, consequently, the de Broglie wavelength becomes very large. Thus, massive particles delocalize in space and behave as waves. Therefore, in many senses, they can be treated in a way similar to the light waves are treated in classical optics (thus the term ``atom optics''). Such matter waves can be manipulated using the forces and potentials of laser light beams: The analogues of optical devices such as beam splitters, mirrors, diffraction gratings, and cavities can be now created by laser beams and applied to atomic matter waves. Thus, the roles of light and matter in optics and atom optics are totally reversed: instead of manipulating light by matter, the matter waves are now manipulated by light in a similar manner.

The quantum properties of matter waves beyond the classical matter-field description became accessible after 1995, when the first Bose--Einstein condensate (BEC) and many other fascinating quantum states of bosonic and fermionic ultracold atoms were obtained. An exciting demonstration of a newer filed, ``quantum atom optics,''  was presented in 2002, when the quantum phase transition between two states of atoms with nearly the same mean density, but radically different quantum fluctuations, was demonstrated: superfluid (SF) to Mott insulator (MI) state transition. The analogy between optics and atom optics further extends into their quantum regimes. The famous quantum states introduced by the founders of quantum theory Vladimir Aleksandrovich Fock (Fock states), Roy Glauber (Glauber coherent squeezed states), and Erwin Schr\"odinger (Schr\"odinger cat states), have been realized for both photons and atoms in the 21th century.

Let us underline that, in this work, by the quantization of ultracold matter we mean the quantization of its motion (the external degree of freedom), rather than any internal quantum excitation. Indeed, it is the quantization of atomic motion (the matter wave) that physically characterizes the atoms as being ultracold and, therefore, delocalized in space. 

While the roles of light and matter in quantum optics and atom optics are reversed, in all experiments and most theoretical works on quantum gases so far, the quantum nature of light is not important at all. The role of light is thus reduced to classical auxiliary tools (mimicking beam splitters, mirrors, diffraction gratings, etc.) to prepare intriguing many-body quantum states of atoms. In this context, the periodic micropotentials of light (the famous optical lattices) play a role of cavities in optics enabling one to store and manipulate various atomic quantum states, like the laser light is trapped, for example, between two mirrors of a Fabry--Perot resonator.

Quantum optics of quantum gases (cf. Fig.~\ref{0-fig1}), which is presented and developed in this work, closes the gap between quantum optics and quantum atom optics by addressing phenomena, where the quantum natures of both light and matter play equally important roles. On the one hand, the quantum properties of atoms will be imprinted on the scattered light, which will lead to novel nondestructive methods of probing many-body quantum matter by the light detection. The majority of current methods to probe quantum gases are totally destructive: even for measurement of a single experimental point, the atomic sample is destroyed and needs to be re-created and re-measured enormous number of times to obtain just a simple mean value of some variable. On the other hand, the quantization of light (i.e. the optical potential, where atoms are trapped) will modify atomic many-body dynamics well-known only for classical optical potentials. This will give rise to novel quantum phases and phenomena, which have been never considered in standard problems of quantum gases. Thus the quantum dynamics of light and matter should be found in an essentially self-consistent way, which is the ultimate goal of quantum optics of quantum gases.

Moreover, two quantum systems, light and matter, are entangled. This opens a path for preparing and manipulating nontrivial many-body atomic states using the fundamental notion of quantum mechanics -- the quantum measurement. Indeed, similarly to the one of the most intriguing predictions of quantum theory, the Einstein--Podolsky--Rosen (EPR) paradox, when one of the entangled subsystems (here, the light) is measured, the quantum state of another one (quantum gas) immediately changes as well. In general, we will consider many-body effects in an open quantum system, which nevertheless is not necessarily a dissipative one: the photons which are scattered from atoms are not lost in the environment, but are measured by the detector. Thus, quantum optics of quantum gases provides a realistic system to shift the paradigm of open many-body systems beyond the concept of dissipation and predict novel physical phenomena.

The main system we consider in this work is a quantum gas in an optical lattice, nevertheless, we obtain several results, which can be applicable in a much broader research context. This includes the following findings. We extended the paradigm of feedback control from the state control to the control of quantum phase transitions, including tuning their universality class. We presented the quantum backaction of weak measurements as a novel source of competitions in many-body systems. We merged the paradigms of quantum Zeno dynamics and non-Hermitian physics. We introduced a novel type of quantum Zeno phenomena with Raman-like transitions well beyond the standard concept of Zeno dynamics. We proposed a concept of quantum simulators based on the collective light--matter interaction and, thus, global addressing of quantum particles. Our models for atoms in lattices can be applied to other arrays of quantum particles (qubits) resulting in new methods of quantum measurements and probing, quantum state preparation, generation of genuine multipartite mode entanglement in quantum arrays. In general, quantum measurements and feedback open a way to obtain novel phenomena untypical to both close unitary systems and open dissipative ones in the context of many-body physics.

\section*{The objective and tasks of the work}

The main general objective of this work is to develop a theory of novel phenomena, which joins two broad and intensively developing fields of modern physics: quantum optics and physics of ultracold quantum gases. Such theory treats the light and motion of many-body systems at equally quantum footing. To be as close as possible to the most promising experimental realizations and clear verifications of this theory, we focus on the system of a quantum gas trapped in an optical lattice potential and confined inside an optical cavity. For high-Q (high quality factor) cavities we thus build a theory of cavity quantum electrodynamics (cavity QED) of ultracold quantum gases.

To achieve this objective we formulated the following problems to solve.

1. Develop a model describing the ultracold quantum gas trapped in an optical lattice and interacting with light modes of an optical cavity.

2. Propose new nondestructive schemes for probing properties of quantum gases in an optical lattice by light detection.

3. Investigate the action of truly quantum measurement on dynamics of a many-body atomic system trapped in an optical lattice. Treat the fundamental quantum measurement backaction as an active tool for preparing many-body states of ultracold atoms.

4. Develop a theory of feedback control of strongly correlated atomic states by optical methods.

5. Identify and treat quantum phases of ultracold gases, which can appear in quantum or dynamical trapping potentials.

\section*{Timeliness of the subject and degree of its development} 

Firstly, the experimental progress in this novel direction is very fast. Very recent experiments provide a challenge for theorists to construct models never addressed so far, which can be verified in the near future. Thus, the formulation of novel theories is timely and crucial for this field. Even early theoretical works \cite{JavanainenPRA1995, JavanainenPRL1995, Moore1999, PuPRL2003, YouPRA1995, IdziaszekPRA2000, MustPRA2000, MustPRA2000x2,
CiracPRL1994,CiracPRA1994,SaitoPRA1999, PratavieraPRA2004, Onofrio, Milburn1998} on scattering of quantized light from a spatially homogeneous BEC or a BEC trapped in a double-well potential have not been realized so far. However, it is the setup with a cavity and optical lattice, which we consider in this work, that provides the best interplay between the atom- and light-stimulated quantum effects. In contrast to the single- and double-well systems, this configuration enables study of truly many-body interacting systems with tunable quantum fluctuations (e.g. phase transition from the Mott insulator to superfluid state, etc.) and generation of very nontrivial quantum spatial structures of light and matter with genuine mode entanglement, which are the key points of our work. (Optical lattices were considered in \cite{JavanainenPRL2003}, but still without benefiting from their spatial structure for light scattering). At the time the author started working on this direction in 2005, the very first experiment joining a quantum gas, Bose--Einstein condensate (BEC), and an optical cavity  was performed at the ETH Zurich \cite{Ottl2005} (in parallel with a theory paper \cite{Maschler2005}), showing the feasibility of theories and their timeliness. The mutual stimulation between theory and experiment led to a fast progress in the area, where the author has been one of its initiators. At the time the author has started his theory group at the University of Oxford in 2011, the number of published works in the field reached about fifty. Two first experiments joining a cavity and an optical lattice in a single setup have been reported in 2015 \cite{EsslingerNature2016,Hemmerich2015}. Currently, the number of papers, including articles and PhD Theses, can be estimated as several hundred. Several experimental groups in the world have succeeded to reach this level of light--matter interaction \cite{EsslingerNature2016,Hemmerich2015, Zimmermann2018, Colombe2007, LevPRX2018,Naik2018, Brantut2020}.  The physical origin of this kind of setups can be traced back to the spatial self-ordering (self-organization) of atoms predicted in the work \cite{DomokosPRL2002} and first observed in the work \cite{Vuletic2003} with thermal atoms. Nevertheless, there are many more open questions than the number of already answered ones. At present, several research lines have emerged. In this work, the author's contribution to the two lines is presented: quantum measurements of light and many-atom effects due to the quantumness of trapping potential. The lines developed by other researchers include: various condensed matter effects due to the dynamical cavity mode; phenomena in multimode cavities; numerical simulations of few-atom systems; fermions and spins trapped inside a cavity, self-organization of atoms without a cavity, dissipative many-body effects, etc. (see for a recent review \cite{Review2021}).  So far, experiments demonstrated several new fascinating phenomena due to the cavity-generated dynamical optical lattices, which depend on the many-body atomic state in a self-consistent way. Some of them have been proposed by the author and are described in this work. However, the truly quantum phenomena, which we predict here, still await their demonstration.

Secondly, both quantum optics and physics of quantum gases form a basis of quantum information processing (QIP) and quantum technologies, which right now start entering their industrial phase. This marks the ``second quantum revolution'' in technologies, where the entanglement will be utilized in fabricated devices. Nevertheless, as has been mentioned, these two fields are still rather separated both theoretically and experimentally. Closing this gap in theoretical and experimental approaches, which is the goal of this work, is very timely as it will enable very efficient implementation of new ideas in applied quantum technologies. Moreover, a new generation of experts having a bigger picture of light--matter interaction will be trained and will be ready to apply their knowledge in both fundamental research and applied quantum information and quantum technologies.

\section*{Statements of the work}

1. We derived a model, which describes the interaction of ultracold atoms (bosons and fermions) trapped in an optical lattice with one or several quantized modes of light. This model forms a basis for theoretical study of quantum optics of quantum gases.

2. We prove that under certain conditions the measurement of light represents nondestructive (up to the physically exciting quantum nondemolition, QND, level) probe of many-body variables of an ultracold atomic system. This is in contrast to the absolute majority of modern methods, which are destructive. 

3. We find the relations between the measurable light properties and quantum statistical variables of a quantum gas such as fluctuations and multi-point spatial density correlations. Moreover, we prove that the  distribution functions of various atomic variables can be directly mapped on the transmission spectrum of a high-Q cavity. In general, we prove that light measurements can distinguish between different many-body states of ultracold bosons and fermions, as well as few-body states of molecular complexes.  

4. We show that light scattering is not only sensitive to the on-site atomic densities, but  also to the matter-field interference at its shortest possible distance in an optical lattice (the lattice period), which defines key properties such as tunneling, atom currents, and matter-wave phase gradients.

5. We show that light scattering from atomic arrays constitutes a quantum measurement with a controllable form of the measurement backaction. We thus use the measurement as an active tool to prepare many-body atomic states such as number squeezed and macroscopic superposition states. Moreover, the class of emerging many-body states can be chosen via the optical geometry and light
frequencies.  

6. We prove that the backaction of quantum measurement constitutes a novel source of competitions in many-body atomic systems, in addition to the standard tunneling and short-range atom interaction. As a general physical concept, new competitions can lead to new effects. We demonstrate a plethora of novel phenomena: the generation and macroscopic oscillations of matter modes, long-range correlated tunneling and genuinely multipartite mode entanglement, both protection and break-up of fermion pairs by the measurement, as well as the measurement-induced antiferromagnetic order. 

7. We predict a new unconventional type of quantum Zeno dynamics due to the Raman-like transitions via the virtual states outside the Zeno subspace. We extend the notion of quantum Zeno dynamics into the realm of non-Hermitian quantum mechanics joining the two paradigms. 

8. We extend the concept of feedback control from the quantum state control (as known in quantum metrology) to the control of phase transitions in quantum systems. We show that quantum weak measurements and feedback can induce phase transitions beyond the dissipative ones. Moreover, feedback allows controlling essentially quantum properties of phase transitions such as the critical exponents. Thus, we demonstrate the possibility to actively tune and control the universality class of a phase transition.  

9. We demonstrate that the quantum and dynamical natures of optical trapping potentials lead to new quantum phases of ultracold atoms unobtainable in comparable prescribed classical optical lattices. We demonstrate not only the density orders as lattice supersolid state and density waves, but the orders of the matter-wave amplitudes (bonds) such as superfluid and supersolid dimers. We formulate a concept of quantum simulators based on the collective light--matter interactions.

{\bf The scientific novelty} of this work is determined by the fact that all the results and statements listed in detail above are new theoretical results. 

\section*{Theoretical and practical significance of the work}  

From the theoretical perspective, this work develops a theory of novel phenomena at the crossroad of quantum optics and physics of quantum gases. As has been mentioned, these two fields are still rather separated both theoretically and experimentally. This is probably explained by historically different traditions and approaches of the communities working on them. In quantum optics, one is mainly interested in complicated quantum states, dynamics, and measurement problems, but for a small number of atoms or ensembles of noninteracting particles. In contrast, the field of quantum gases is heavily dominated by theories originating from condensed matter physics, where many-body strongly interacting systems are considered routinely, but delicate effects of quantum measurements and light quantization are ignored because of very strong noise in the original solid state systems and hence theoretical models. Closing this gap in theoretical and experimental approaches is the goal of this work. It will enable the discoveries of new phenomena well beyond the ones based on old condensed matter theories, which are unobtainable neither in quantum optics or condensed matter systems separately. 
 
Going beyond atomic and optical physics, we formulate a new paradigm of controlling quantum phase transitions with tunable universality class. We merged the paradigms of quantum Zeno dynamics and non-Hermitian physics. We introduced a novel type of quantum Zeno phenomena with Raman-like transitions well beyond the standard concept of Zeno dynamics. Moreover, we introduce the fundamental measurement backaction as a new source of competitions in many-body physics. These concepts are of interest in many domains of theoretical and experimental physics and will stimulate their progress further. It will be intriguing to study, how more advanced methods than feedback control can influence quantum systems, for example, applying the digital methods of machine learning and classical or quantum artificial intelligence in real time. 
 
 From the perspective of applications, both quantum optics and physics of ultracold gases form a basis of quantum information processing and quantum technologies, which right now start entering their industrial phase. This marks the ``second quantum revolution'' in technologies, where the entanglement will be utilized in fabricated devices. Developing a new direction at the intersection of both fields will enable very efficient implementation of new ideas in applied quantum technologies. Moreover, a new generation of experts having a bigger picture of light--matter interaction will be trained and will be ready to apply their knowledge in both fundamental and applied research. More precisely, the applications can be found in the following areas. (I) Quantum simulations: a broader range of Hamiltonians can be simulated than it is possible in classical optical lattices (we propose novel simulators of quantum baths and simulators based on the collective light--matter interaction). (II) Quantum metrology and matter interferometry: we expand the concept of feedback control towards the control of phase transitions; we suggest the preparation of macroscopic quantum states of massive particles. (III) Quantum sensing: we suggest nondestructive methods of probing nontrivial many-body systems. (IV) Quantum information and computing: the creation of multipartite entangled states, which we predict, may be useful in this domain. Applications of our models are not limited to atomic systems, but can be extended to other quantum arrays of objects such as superconducting circuits (qubits), ions, semiconductor exciton-polaritons, molecular systems, and hand-made nano-objects.
 
\section*{Theoretical methods}

Having a goal of merging two research directions, we used theoretical approaches from both quantum optics and many-body (originally, condensed matter) physics. The quantum optical methods are Heisenberg--Langevin equations, master equation, stochastic master equation, stochastic Ito and Stratonovich equations, Quantum Monte Carlo Wave Function simulation. The many-body methods include mean-field approaches, multidimensional optimization problems, and Density-Matrix Renormalization Group (DMRG) simulations.

\section*{Reliability of the results and approbation of the work}

The reliability of the results is based on the agreement between available experimental and theoretical results, as well as careful estimation of relevant experimental parameters during the development of theoretical models. 

The results of this work have been presented by the author as invited talks at 14 international conference, more than 20 invited seminars at multiple leading universities at various countries (Russia, Austria, UK, US, Germany, France, etc.), and other presentations at more than 40 international conferences and workshops. My coauthors presented the results at more than 20 conferences. In total, our results have been presented at more than 100 meetings.

Several selected meetings include: 
International Workshop Many-Body Cavity QED (Harvard, USA),
International Program Measurement and Control of Quantum Systems (Paris, France), 
International Workshop Novel paradigms in many-body physics from open quantum systems (Dresden, Germany), 
Meeting of the Quantum Information Division of the Mexican Physical Society,
Workshops Dynamics and Simulation of Ultra-Cold Matter (Windsor, UK),
International Laser Physics Workshops,
International Conferences on Quantum Optics (Obergurgl, Austria),
European Quantum Electronics Conferences CLEO/Europe--EQEC (Munich, Germany),
International Conference on Quantum Information (Rochester, USA),
WE-Heraeus-Seminars (Bad Honnef, Germany), and
International Conferences on Atomic Physics ICAP.

\section*{Publications}

The results on the subject of this work have been published in 29 papers (\cite{MekhovNaturePh2007,MekhovPRL2007,MekhovPRA2007,MekhovEPJD08,MekhovPRL2009, MekhovPRA2009, MekhovLP2009,MekhovLP2010, MekhovLP2011, MekhovPRL2011, MekhovPRA2011,  Mekhov2012,MekhovLP2013,Elliott2015, Caballero2015,Caballero2015a, Kozlowski2015PRA, Atoms, Mazzucchi2016Opt,Mazzucchi2016SciRep,Mazzucchi2016NJP,CaballeroNJP2016,Mazzucchi2016PRA, Kozlowski2016PRAnH, CaballeroPRA2016, PRA2016-4, Kozlowski2017,IvanovSciRep2020,IvanovPRL2020} in the Reference list) in journals indexed in Web of Science and Scopus, including one review article ([12] below and Ref. \cite{Mekhov2012} in the Reference list).
 
\begin{enumerate} 
\item Mekhov I. B., Maschler C., Ritsch H. Probing quantum phases of ultracold atoms in optical lattices by transmission spectra in cavity QED //  \href{http://dx.doi.org/10.1038/nphys571}{Nature Phys.} --- 2007. --- Vol. 3. --- P. 319--323.

\item Mekhov I. B., Maschler C., Ritsch H. Cavity-enhanced light scattering in optical lattices to probe atomic quantum statistics // \href{http://dx.doi.org/10.1103/PhysRevLett.98.100402}{Phys. Rev. Lett.} --- 2007. --- Vol. 98. --- P. 100402.

\item Mekhov I. B., Maschler C., Ritsch H. Light scattering from ultracold atoms in optical lattices as an optical probe of quantum statistics // \href{http://dx.doi.org/10.1103/PhysRevA.76.053618}{Phys. Rev. A}. --- 2007. --- Vol. 76. --- P. 053618.

\item Maschler C., Mekhov I. B., Ritsch H. Ultracold atoms in optical lattices generated by quantized light fields // \href{http://dx.doi.org/10.1140/epjd/e2008-00016-4}{Eur. Phys. J. D}. --- 2008. --- Vol. 46. ---  P. 545--560.

\item Mekhov I. B., Ritsch H. QND measurements and state preparation in quantum gases by light detection // \href{http://dx.doi.org/10.1103/PhysRevLett.102.020403}{Phys. Rev. Lett.} --- 2009. ---
Vol. 102. --- P. 020403. 

\item Mekhov I. B., Ritsch H. Quantum optics with quantum gases: controlled state reduction by designed light scattering // \href{http://dx.doi.org/10.1103/PhysRevA.80.013604}{Phys. Rev. A} ---  2009. --- Vol. 80. --- P. 013604.

\item Mekhov I. B., Ritsch H. Quantum optics with quantum gases // \href{http://dx.doi.org/10.1134/S1054660X09040136}{Laser Phys.} --- 2009. --- Vol. 19. --- P. 610--615.

\item Mekhov I. B., Ritsch H. Quantum optical measurements in ultracold gases: macroscopic Bose--Einstein condensates // \href{http://dx.doi.org/10.1134/S1054660X10050105}{Laser Phys.} ---  2010. --- Vol. 20. --- P. 694--699.

\item Mekhov I. B., Ritsch H. Atom state evolution and collapse in ultracold gases during light scattering into a cavity // \href{http://dx.doi.org/10.1134/S1054660X11150163}{Laser Phys.} --- 2011. --- Vol. 21. --- P. 1486--1490. 

\item Wunsch B., Zinner N. T., Mekhov I. B., Huang S.-J., Wang D.-W., Demler E. Few-body bound states in dipolar gases and their detection // \href{http://dx.doi.org/10.1103/PhysRevLett.107.073201}{Phys. Rev. Lett.} --- 2011. --- Vol. 107. --- P. 073201.

\item Zinner N. T., Wunsch B., Mekhov I. B., Huang S.-J., Wang D.-W., Demler E. Few-body bound complexes in 1D dipolar gases and their non-destructive optical detection // \href{http://dx.doi.org/10.1103/PhysRevA.84.063606}{Phys. Rev. A}. --- 2011. --- Vol. 84. --- P. 063606.

\item Mekhov I. B., Ritsch H. Topical Review. Quantum optics with ultracold quantum gases: towards the full quantum regime of the light--matter interaction // \href{http://dx.doi.org/10.1088/0953-4075/45/10/102001}{Journ. Phys. B}. --- 2012. --- Vol. 45. --- P. 102001.

\item Mekhov I. B. Quantum non-demolition detection of polar molecule complexes: dimers, trimers, tetramers // \href{http://dx.doi.org/10.1088/1054-660X/23/1/015501}{Laser Phys.} --- 2013. --- Vol. 23. --- P. 015501.

\item Elliott T. J., Kozlowski W., Caballero--Benitez S. F., Mekhov I. B.  Multipartite Entangled Spatial Modes of Ultracold Atoms Generated and Controlled by Quantum Measurement // \href{http://dx.doi.org/10.1103/PhysRevLett.114.113604}{Phys. Rev. Lett.} --- 2015. --- Vol. 114. --- P. 113604.

\item Caballero--Benitez S. F., Mekhov I. B. Quantum optical lattices for emergent many-body phases of ultracold atoms // \href{http://dx.doi.org/10.1103/PhysRevLett.115.243604}{Phys. Rev. Lett.} ---  2015. --- Vol. 115. --- P. 243604.

\item Caballero--Benitez S. F., Mekhov I. B. Quantum properties of light scattered from structured many-body phases of ultracold atoms in quantum optical lattices // \href{http://dx.doi.org/10.1088/1367-2630/17/12/123023}{New J. Phys.} --- 2015. --- Vol. 17. --- P. 123023.

\item Kozlowski W., Caballero--Benitez S. F., Mekhov I. B. Probing Matter-Fields and Atom-Number correlations in Optical Lattices by Global Nondestructive Addressing // \href{http://dx.doi.org/10.1103/PhysRevA.92.013613}{Phys. Rev. A}. --- 2015. --- Vol. 92. --- P. 013613.

\item Elliott T. J., Mazzucchi G., Kozlowski W., Caballero--Benitez S. F., Mekhov I. B. Probing and manipulating fermionic and bosonic quantum gases with quantum light. Invited article in the Special Issue Cavity QED with Ultracold Atoms // \href{http://dx.doi.org/10.3390/atoms3030392}{Atoms} --- 2015. --- Vol. 3. --- P. 392--406.

\item Mazzucchi G., Caballero--Benitez S. F., Ivanov D. A., Mekhov I. B. Quantum optical feedback control for creating strong correlations in many-body systems // \href{http://dx.doi.org/10.1364/OPTICA.3.001213}{Optica (OSA)}. --- 2016. --- Vol. 3. --- P. 1213--1219.
 
\item Mazzucchi G., Caballero--Benitez S. F., Mekhov I. B. Quantum measurement-induced antiferromagnetic order and density modulations in ultracold Fermi gases in optical lattices //   \href{http://dx.doi.org/10.1038/srep31196}{Sci. Rep.} --- 2016. --- Vol. 6. --- P. 31196.
 
\item Mazzucchi G., Kozlowski  W., Caballero--Benitez S. F., Mekhov I. B. Collective dynamics of multimode bosonic systems induced by weak quantum measurement // \href{http://dx.doi.org/10.1088/1367-2630/18/7/073017}{New J. Phys.} --- 2016. --- Vol. 18. --- P. 073017. 

\item Caballero--Benitez S. F., Mekhov I. B. Bond-order via light-induced synthetic many-body interactions of ultracold atoms in optical lattices // \href{http://dx.doi.org/10.1088/1367-2630/18/11/113010}{New J. Phys.} --- 2016. --- Vol. 18. --- P. 113010.

\item Mazzucchi G., Kozlowski  W., Caballero--Benitez S. F., Elliott T. J., Mekhov I. B. Quantum Measurement-induced Dynamics of Many-Body Ultracold Bosonic and Fermionic Systems in Optical Lattices // \href{http://dx.doi.org/10.1103/PhysRevA.93.023632}{Phys. Rev. A}. --- 2016. --- Vol. 93. --- P. 023632.

\item Kozlowski  W., Caballero--Benitez S. F.,  Mekhov I. B. Non-Hermitian Dynamics in the Quantum Zeno Limit // \href{http://dx.doi.org/10.1103/PhysRevA.94.012123}{Phys. Rev. A}. --- 2016. --- Vol. 94. --- P. 012123.

\item Caballero--Benitez S. F., Mazzucchi G., Mekhov I. B. Quantum simulators based on the global collective light--matter interaction // \href{http://dx.doi.org/10.1103/PhysRevA.93.063632}{Phys. Rev. A}. --- 2016. --- Vol. 93. --- P. 063632.

\item Elliott T. J., Mekhov I. B., Engineering Many-Body Dynamics with  Quantum light potentials and Measurements // \href{http://dx.doi.org/10.1103/PhysRevA.94.013614}{Phys. Rev. A}. --- 2016. --- Vol. 94. --- P. 013614.

\item Kozlowski  W., Caballero--Benitez S. F.,  Mekhov I. B. Quantum State Reduction by Matter-Phase-Related Measurements in Optical Lattices // \href{http://dx.doi.org/10.1038/srep42597}{Sci. Rep.} --- 2017. --- Vol. 7. --- P. 42597.

\item Ivanov D. A., Ivanova T. Yu., Caballero--Benitez S. F., Mekhov I. B. Cavityless self-organization of ultracold atoms due to the feedback-induced phase transition //  \href{http://dx.doi.org/10.1038/s41598-020-67280-3}{Sci. Rep.} --- 2020. --- Vol. 10. --- P. 10550.

\item Ivanov D. A., Ivanova T. Yu., Caballero--Benitez S. F., Mekhov I. B. Feedback-induced quantum phase transitions using weak measurements // \href{http://dx.doi.org/10.1103/PhysRevLett.124.010603}{Phys. Rev. Lett.} --- 2020. --- Vol. 124. --- P. 010603.
\end{enumerate}

Papers [1-9] were written during my postdoc in the group of Prof. H. Ritsch at the University of Innsbruck: in papers [1-3] and [5-9], I am the the lead author, paper [4] is written jointly. Papers [10,11] were written during my fellowship at Harvard University jointly with the group of Prof. E. Demler: the detection of few-body complexes by light is my contribution presented in this work. Papers [12-27] were written jointly with my team members, during my position as a head of theory group at the University of Oxford. The main strategic ideas of these works were formulated and documented in my successeful proposal for the EPSRC fellowship enabling me to hold this long-term position. Papers [28,29] were written in collaboration with my colleagues in St. Petersburg and Mexico City\footnote{For a more recent work cf. Ivanov D. A., Ivanova T. Yu., Caballero--Benitez S. F., Mekhov I. B. Tuning the universality class of phase transitions by feedback:
Open quantum systems beyond dissipation // \href{http://dx.doi.org/10.1103/PhysRevA.104.033719}{Phys. Rev. A} --- 2021. --- Vol. 104. --- P. 033719.}.

\section*{Structure of the work}

The work consists of the Introduction, five chapters, Conclusions, and Reference list. Each chapter starts with a short introduction explaining the context of research and details its structure, it ends by a short conclusion section.

\begin{figure}[h]
\centering
\captionsetup{justification=justified}
\includegraphics[clip, trim=0cm 3cm 0cm 0cm, width=0.9\textwidth]{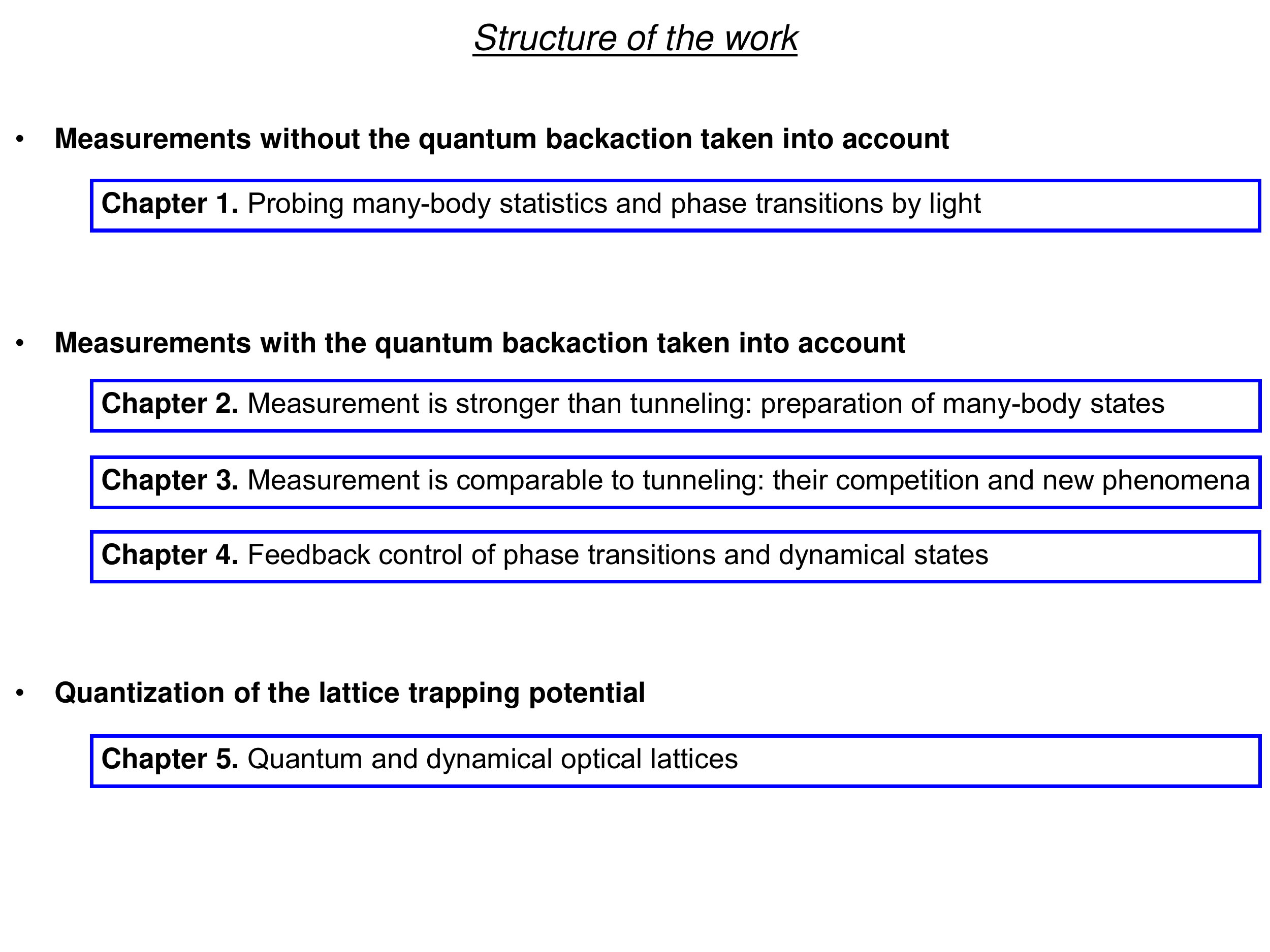}
\caption{\label{0-fig2} Structure of the work. The measurements without quantum backaction are considered in Chapter 1, the measurement backaction is taken into account in Chapters 2, 3, and 4. The quantization of lattice trapping potential is presented in Chapter 5. The interaction of light with both atomic densities and matter-wave interference patterns is considered in all chapters. Atoms are bosons and fermions, molecules are considered in Chapter 1.}
\end{figure}

The structure of the work is determined by the complexity level of physical assumptions (cf. Fig.~\ref{0-fig2}). This enables us to focus in different chapters on various physical aspects of systems and phenomena, where the quantum natures of both quantum gases and light are equally important. Firstly (Chapter 1), we will focus on the standard ``applied'' goal of the measurement: we will establish the relations between the statistical quantities of matter and detected light. Here, the notion of the quantum measurement backaction is not required and the focus is on the quantum mechanical averages. Secondly (Chapters 2, 3, and 4), we will focus on the fundamental quantum measurement backaction during the continuous measurement  at single quantum trajectories without calculating the full statistical averages as before. Thirdly (Chapter 5), we will present quantum optical lattices, where the quantization of even the trapping potential is crucial. In detail, the structure of presentation is the following.

In Chapter 1, we will first develop a rather general theoretical model of the interaction between ultracold atoms trapped in an optical lattice and quantized light modes of a cavity. Both light- and matter-related variables will be written as quantum fields, which in some sense underlines the similarity of both quantum objects and the wave-particle dualism in the many-body context. With various approximations, we will use this model in all chapters of this work. 

We will then focus on the standard ``applied'' goal of the measurement, where the fundamental notion of quantum measurement and its backaction are not required. We will link various measurable statistical quantities of light (the average intensity and quadratures, photon number and quadrature fluctuations) with statistical variables of the ultracold gas (density fluctuations and spatial correlations). Moreover, we will show that full distribution functions of different many-body variables can be mapped on cavity transmission spectra. We show that light is sensitive not only to the on-site density-related variables, but to the inter-site matter-wave coherence as well. We demonstrate that light scattering can distinguish between various many-body phases, for example, Bose glass, Mott insulator, and superfluid. We show that the light detection constitutes the quantum nondemolition (QND) measurement under certain conditions. We apply our results to bosonic and fermionic many-body atomic systems, as well as to the few-body molecular complexes.

As it is usual in quantum mechanics, the determination of statistical quantities requires multiple measurements. Therefore, the repeated preparation of the initial state is necessary, because any quantum measurement (even the QND one) generally affects the quantum state of the system. 

 In Chapter 2, we address the measurement problem from another point of view: instead of repeated measurements and averaging, we consider system dynamics during the continuous light detection (a single quantum trajectory). During the interaction, the light and matter get entangled. According to quantum mechanics (e.g. the paradigmatic Einstein--Podolsky--Rosen (EPR) paradox), due to the entanglement, the measurement of one of the quantum subsystems (light) will affect another quantum subsystem (atoms) as well. This is an example of the fundamental quantum measurement backaction, which we now take into account. We show that light scattering from atomic arrays constitutes a quantum measurement with a controllable form of the measurement backaction. We thus use the measurement as an active tool to prepare many-body atomic states such as the number squeezed and macroscopic superposition states. Moreover, the class of emerging many-body states can be chosen via the optical geometry and light frequencies.

In Chapter 3, we consider the rate of quantum measurement to be on a time scale comparable to the atomic tunneling (which was not the case in the previous chapter, where tunneling was almost neglected). We prove that the backaction of quantum measurement constitutes a novel source of competitions in many-body atomic systems, in addition to standard tunneling and short-range atom interaction. Interestingly, the measurement should not be a QND one to be able to compete with other processes, and we go beyond the framework of QND paradigm. As a general physical concept, new competitions can lead to new effects. We demonstrate a plethora of novel phenomena: the generation and macroscopic oscillations of matter modes with dynamical supersolid properties, long-range correlated tunneling and genuinely multipartite mode entanglement, both protection and break-up of fermion pairs by the measurement, as well as the measurement-induced antiferromagnetic order. We prove that even global, but spatially structured, measurements can compete with short-range processes such as tunneling and atom interactions. We predict a new unconventional type of quantum Zeno dynamics due to the Raman-like transitions via the virtual states outside the Zeno subspace. We extend the notion of quantum Zeno dynamics into the realm of non-Hermitian quantum mechanics joining the two paradigms. 

In Chapter 4, we add the feedback control to the system, which can either amplify or balance the action of the quantum measurement backaction. We extend the concept of feedback control from the quantum state control (as known in quantum metrology) to the control of phase transitions in quantum systems. We show that quantum weak measurements and feedback can induce phase transitions beyond the dissipative ones. Moreover, feedback allows controlling essentially quantum properties of phase transitions such as the critical exponents. Thus, we demonstrate tuning the universality class of a phase transition in a single given setup. We show that our approach will enable creation of novel quantum simulators of quantum baths, simulating effects similar to spin-bath problems, and creating new Floquet time crystals with tunable long-range (long-memory) interactions. 

In Chapter 5, we consider the ultimate regime, where quantization of even the trapping potential is crucial. We demonstrate that the quantum and dynamical natures of optical trapping potentials lead to new quantum phases of ultracold atoms unobtainable in comparable prescribed classical optical lattices. We demonstrate not only density orders as lattice supersolid state and density waves, but the orders of the matter-wave amplitudes (bonds) such as superfluid and supersolid dimers. We show that many-body systems imprint their properties on scattered light, in particular its squeezing. We formulate a concept of quantum simulators based on the collective light--matter interactions with the tunable effective interaction length. This bridges physics of short- and long-range interactions and enables simulating systems with the tunable interaction, which is extremely difficult to achieve in other physical contexts.

Finally, we conclude this work and describe its perspectives in the Conclusions section.

As we already mentioned, during recent years, many works have been performed in the field of quantum optics of quantum gases and, in particular, quantum gases trapped in optical cavities. The reviews of this field can be found, for example, in Refs. \cite{Mekhov2012,ritsch2013,Review2021}. At the same time, various quantum many-body phenomena are still waiting for their demonstrations.

\section*{Acknowledgments}

First of all, I am very grateful to my scientific teachers at the St. Petersburg State University I. A. Chekhonin and V. S. Egorov for helping me to obtain a strong scientific background and develop curiosity and intuition sufficient to start my independent career after obtaining my PhD degree. I would like to thank other colleagues at the St. Petersburg State University for fruitful collaboration and very warm and timely support at various stages: D. A. Ivanov, T. Yu. Ivanova, Yu. M. Golubev, T. Yu. Golubeva, V. A. Averchenko, P. V. Moroshkin, V. N. Lebedev, and A. A. Manshina.  

I would like to thank especially strongly Helmut Ritsch, whose group at the University of Innsbruck I joined as a postdoc and started working on this research direction. I'd like to thank all the members of his team at that time and especially C. Maschler, W. Niedenzu, T. Salzburger, H. Zoubi, C. Genes, A. Vukics, and J. Asboth.  

I would like to thank my theory coauthors during my fellowship at Harvard University E. Demler, B. Wunsch, and N. Zinner, as well as experimental colleagues at MIT, with whom I could discuss the applicability of models, especially, V. Vuleti{\'{c}} and his group and W. Ketterle and his group.

I d'like to thank very strongly the members of my theory group that I led at the University of Oxford, without whom most of the ideas could not be realized in their final and broad forms: S. F. Caballero--Benitez, W. Kozlowski, G. Mazzucchi, T. Elliott, and F. Tennie. I thank other colleagues at Oxford, especially, D. Jaksch and his group, C. Foot and his group, A. Kuhn and his group, J. Nunn, and P. Ewart.

I would like to thank many other colleagues with whom I had very stimulating discussions at various occasions: P. Domokos, T. Esslinger and his team members, T. Donner, A. Hemmerich and his team members, C. Zimmermann, Ph. Courteille, S. Slama, G. Morigi, J. Ruostekoski, E. Polzik, D. Stamper-Kurn, J. Sherson, G. Rempe, S. Ritter, I. Bloch, R. Glauber, D. V. Kupriyanov, I. M. Sokolov, I. Lesanovsky, G. De Chiara,  D. Esteve and his group, P. Zoller and his group, H. C. N\"agerl,  G. Pupillo,  K. Hammerer, F. Piazza, D. Nagy, M. Oberthaler, T. Roscilde, A. Bertoldi, M. Weitz, and many others.

\clearpage

%% file: newcommandsIM.tex
\newcommand{\oop}[1]{\hat{\mathcal{#1}}}	 
\newcommand{\Trx}{\mathrm{Tr}}			       

\newcommand{\ketz}[1]{\mid \! \!  #1 \,\rangle}
\newcommand{\braz}[1]{\langle{#1}|}
\newcommand{\braketz}[2]{\langle{#1}|{#2}\rangle}

\renewcommand{\b}[1]{\mathbf{ #1}}									
\newcommand{\HH}{\mathcal{H}}
\renewcommand{\baselinestretch}{1} 
\newcommand{\md}{\mathrm{d}}
\newcommand{\ketx}[1]{|{#1}\rangle}
\newcommand{\brkt}[2]{\langle{#1}|{#2}\rangle}
\newcommand{\bopk}[3]{\langle{#1}|{#2}|{#3}\rangle}
\newcommand{\leftexp}[2]{{\vphantom{#2}}^{#1}{#2}}
\newcommand{\fv}[1]{\textsf{#1}}
\newcommand{\pd}[2]{\frac{\partial #1}{\partial #2}}
\newcommand{\td}[2]{\frac{\md #1}{\md #2}}
\newcommand{\utilde}[1]{\underset{\widetilde{}}{#1}}
\newcommand{\sv}[1]{\utilde{\bm{#1}}}
\newcommand{\del}{\nabla}
\newcommand{\boxdel}{\square}
\newcommand{\realsum}{\displaystyle\sum}
\newcommand{\realprod}{\displaystyle\prod}
\newcommand{\realtprod}{\displaystyle\bigotimes}
\newcommand{\tens}[1]{\mathbb{#1}}
\newcommand{\figref}[1]{Fig.\ \ref{#1}}
\newcommand{\figsref}[1]{Figs.\ \ref{#1}}

\renewcommand{\d }{\mathrm{d}}										
\newcommand{\der}[2]{\frac{\d #1}{\d #2}}							
\newcommand{\funder}[2]{\frac{\delta #1}{\delta #2}}				
\newcommand{\desude}[1]{\frac{\d }{\d #1}}						
\newcommand{\derpar}[2]{\frac{\partial #1}{\partial #2}}			
\newcommand{\parsupar}[1]{\frac{\partial }{\partial #1}}			

\newcommand{\bk}[2]{\langle \, #1 \! \mid  \! #2 \, \rangle}		
\newcommand{\cop}[1]{#1^{\dagger}\!}								
\newcommand{\aop}[1]{#1}											
\newcommand{\bok}[3]{\langle #1 \! \mid \! #2 \! \mid  \! #3  \rangle}	
\newcommand{\boknoc}[3]{\langle \, #1 \! \mid \! #2 \! \mid  \! #3 \, \rangle}	
\newcommand{\up}{\uparrow}											
\newcommand{\down}{\downarrow}										
\newcommand{\D}{\mathcal{D}}										
\newcommand{\m}[1]{\langle #1 \rangle}								
\newcommand{\rom}[1]{\uppercase\expandafter{\romannumeral #1\relax}} 
\newcommand{\id}{\hat{1}}										
\newcommand{\res}[1]{\mathrm{Res}\left[ #1 \right]}					
\newcommand{\REF}{$\mathrm{\b{(REF!!)}}$}							
\newcommand{\pol}{y}												
\newcommand{\odd}{\mathrm{o}}	
\newcommand{\even}{\mathrm{e}}											
\newcommand{\h}[1]{\hat{#1}}	

\def\a{\hat{a}} 
\def\ad{\hat{a}^\dagger}
\def\bd{b^\dagger} 
\def\btd{\beta^\dagger} 
\def\c{\hat{c}}
\def\cd{\hat{c}^\dagger} 
\def\n{\hat{n}} 
\def\N{\hat{N}}
\def\B{\hat{B}} 
\def\Bd{\hat{B}^\dagger}
\def\D{\hat{D}} 

\def\opo{\hat{o}}

\def\hx{\hat{h}}



\newcommand{\vx}{\mathbf{x}}
\newcommand{\vy}{\mathbf{y}}
\newcommand{\vk}{\mathbf{k}}
\newcommand{\vq}{\mathbf{q}}
\newcommand{\Vol}{\mathcal{V}}
\newcommand{\OO}{\mathcal{O}}
\renewcommand{\Re}{\operatorname{Re}}

%% file: Chapter1.tex
\chapter{Quantum diffraction gratings: quantum nondemolition (QND) measurements in atomic and molecular quantum gases} \label{chapt1}

\section{Introduction and plan of the chapter}

In this Chapter we will first develop a general model describing the interaction between quantized light and atoms of an ultracold quantum gas in Sec. 1.2 \cite{MekhovPRA2007,Mekhov2012}. We will use various aspects of this model in all chapters of this work, and in Chapter 1 we limit ourselves to the first level of complexity of quantum optics of quantum gases described in the Introduction: We will establish the relation between quantum properties of ultracold atoms and the characteristics of scattered light. This will allow the nondestructive (in the sense of quantum nondemolition, QND) probing of a quantum gas by light, which provides a significant step beyond the state-of-the-art, because the absolute majority of experimental methods used so far are totally destructive. We will leave the consideration of the quantum measurement backaction and measurement-based many-body state preparation to Chapters 2--4, and the consideration of quantum optical lattices (quantum trapping potentials) to Chapter 5, which will be based on the model developed in Sec. 1.2 as well.

Note that this model can be generalized for other quantum systems with spatial periodic structures such as superconducting circuits (qubits), ions, Rydberg atoms, polaritons as well as other micro- and nanostructures. 

In Sec. 1.3 \cite{MekhovPRL2007,MekhovPRA2007,Mekhov2012} we will reduce our general model to the case, where the optical lattice is deep and scattered light is determined by the fluctuating atom numbers at multiple lattice sites. Thus, the influence of matter-wave amplitudes (coherences) and tunneling on the scattered light will be neglected. We will introduce the popular quantum nondemolition measurements, but will underline that the non-QND measurement can lead to even more intriguing phenomena considered in Chapters 3 and 4. Here we will give clear physical analogies: in its classical limit, light scattering from atoms in lattices indeed reduces to the well-known diffraction. Therefore, we can consider ultracold atoms in a lattice as a ``quantum diffraction grating''.

In Sec. 1.4 \cite{MekhovPRL2007,MekhovPRA2007} we will derive the relations between the quantum properties of atomic states and angular distributions of variables of scattered light such as the light amplitude, photon number (intensity), quadratures and their variances, and photon number variances. We will show that even simple mean intensity is sensitive to the quantum fluctuations and correlations of atoms, which are radically different for various atomic states. The photon number variance reflects the four-point spatial correlation functions of atoms.

In Sec. 1.5 \cite{MekhovPRL2007,MekhovPRA2007,MekhovLP2009, Mekhov2012} we will demonstrate, how the light scattering can distinguish between various quantum states of ultracold atoms (Mott insulator and superfluid states, as well as the coherent-state approximation to the superfluid state). This will form a basis for nondestructive probing of quantum gases. We will present results for traveling and standing waves for both probe and scattered light modes. The most interesting observation angle turns out to be the diffraction minimum, rather than the maximum as expected from the classical diffraction. In the minimum, the strong classical scattering is suppressed and the light directly reflects the quantum properties of atoms.

In Sec. 1.6 \cite{Kozlowski2015PRA} we present an example of a 3D optical lattice. We will show that, even when the classical Bragg diffraction is completely forbidden, a quantum gas in a lattice scatters light. Moreover, its angular distribution is structured and we derive the generalized Bragg condition beyond the standard one for classical diffraction.

 In Sec. 1.7 \cite{Kozlowski2015PRA} we demonstrate how to map the full phase diagram of Bose glass -- Mott insulator -- superfluid phase transition by light scattering, where all phases are well distinguished from each other.
 
In Sec. 1.8 \cite{Kozlowski2015PRA} we will make an important step further and take into account the influence of matter-wave amplitudes (coherences) on the scattered light. Moreover, we will show how to completely suppress the density contribution to scattering and make the matter-amplitude contribution dominant. We prove that light scattering from an ultracold gas reveals the matter-field interference at its shortest possible distance in an optical lattice (i.e. the lattice period), which defines key properties such as tunneling, atomic currents, and matter-field phase gradients. This signal can be enhanced by concentrating the probe light between lattice sites rather than at density maxima. We demonstrate the nondestructive detection of the matter order parameters, matter-field quadratures and their squeezing.
 
In Sec. 1.9 \cite{Atoms} we extend our model to ultracold fermions and demonstrate the QND probing of strongly interacting fermionic gases.

In Sec. 1.10 \cite{MekhovNaturePh2007} we prove that the transmission spectroscopy can map the full distribution functions of different variables of a quantum gas, while in the previous sections we addressed only integrated variables as means, fluctuations, and correlations. Interestingly, while the famous Voigt lineshape is known in the field of hot gases, it appears to be the lineshape of the ultracold quantum gas as well, though for different physical reasons.

In Sec. 1.11 \cite{MekhovPRL2011, MekhovPRA2011, MekhovLP2013} we extend the QND detection methods for few-body complexes of ultracold polar moleculs. Few-body physics is an interesting filed with phenomena different from the many-body context. We demonstrate that the association and dissociation of nontrivial molecular complexes (dimers, trimers, tetramers, etc.) can be nondestructively detected in a QND way.

We conclude this chapter in Sec. 1.12 and make a link to the next chapter of this work.    

{\it The results of this chapter are based on the papers}  \cite{MekhovNaturePh2007,MekhovPRL2007,MekhovPRA2007,MekhovLP2009, MekhovPRL2011, MekhovPRA2011, Mekhov2012,MekhovLP2013,Kozlowski2015PRA, Atoms}.


\section{General theoretical model of the interaction between quantized light and ultracold atoms}

In this section we will develop a general model describing the interaction between quantized light and atoms of an ultracold quantum gas trapped in an optical lattice \cite{ICAP06,MekhovPRL2007,MekhovPRA2007,MekhovEPJD08}. We will use various aspects of this model in all chapters of this work.

We will take into account the spatial geometry of light in a very general form: the model can describe the interaction inside a cavity or in free space. The consideration of multiple probes and cavities (or cavity modes) strongly expands and corrects the preliminary model \cite{Maschler2005}. The trapping of atoms by fully quantum potential is captured by this model as well. Describing the quantum properties of atoms, we first focus on a particular case: spinless bosons. Later, we will generalize this model for fermions with spins, and specific molecular few-body complexes.

We consider an ensemble of $N$ two-level atoms in an optical lattice with $M$ sites. In quantum optics of quantum gases, the case of optical lattices is very convenient as it allows to precisely describe the many-body atomic state for a broad range of parameters. Moreover, experimentally different setups can be described by the same general model. The typical examples include a multi-site lattice with the low filling factor (e.g. one or two atoms per lattice site as in a typical superfluid to Mott insulator transition setup) and a trapping potential with the high filling factors but small number of sites (e.g. a BEC in a double-well potential setup). Even early theoretical
works on scattering of quantized light from a BEC was not realized
so far \cite{Moore1999,PuPRL2003,YouPRA1995,IdziaszekPRA2000,MustPRA2000,MustPRA2000x2,
JavanainenPRL1995,JavanainenPRA1995,CiracPRL1994,CiracPRA1994,SaitoPRA1999, PratavieraPRA2004,JavanainenPRL2003}. However, it
is the setup with a cavity and optical lattice that will
provide the best interplay between the atom- and light-stimulated
quantum effects, which we consider here.

\begin{figure}[h]
\centering
\captionsetup{justification=justified}
\includegraphics[width=0.7\textwidth]{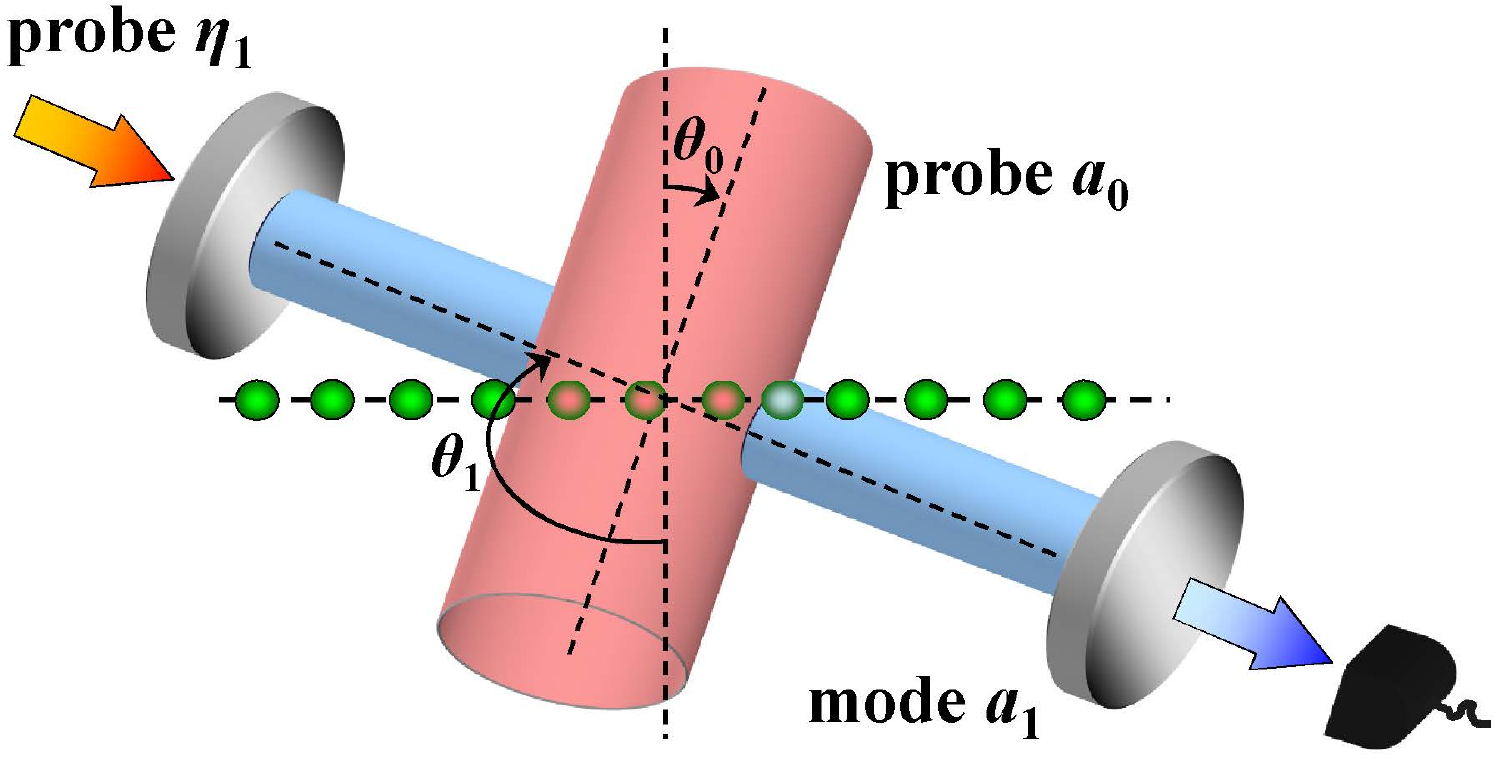}
\caption{\label{1-fig1} Setup. The atoms periodically trapped in a lattice are illuminated
by the transverse probe $a_0$ (the trapping laser beams are not shown). The scattered light mode $a_1$ is collected by a cavity. Another probe $\eta_1$ through a cavity mirror can be present. The photons leaking the cavity are measured by the detector.}
\end{figure}

In general, the atoms are trapped in a lattice potential created by strong lasers as it is usual in the standard cold atom problems. In the presence of such a classical potential,
the atoms are illuminated by light and scatter light at different
directions. As shown in Fig.~\ref{1-fig1}, the atoms in a lattice are illuminated by a probe beam, and the measurements are carried out in the direction of one of scattered light modes. In fact, the probe and scattered light modes can be in free space without the presence of any cavity. In practice, light modes can be selected by traveling- or standing-wave cavities, or even correspond to different modes of the same cavity. One important reason for including a cavity is to enhance light scattering into some particular direction. Another reason, is that the cavity mode can form a fully quantum trapping potential for the atoms (even if no classical trapping potential is present). For definiteness, we will
consider the case, where the light mode functions are determined by cavities,
whose axes directions can be varied with respect to the lattice axis
(the simplest case of two waves, probe beam and cavity mode, at angles
$\theta_0$ and $\theta_1$ is shown in Fig.~\ref{1-fig1}). Instead of
varying the angles, the light wavelengths can be varied with respect
to the wavelength of a trapping beam as well. We also assume, that not all
$M$ lattice sites are necessarily illuminated,
but only a subset of $K\le M$ sites. The selection of $K$ out of the total $M$ sites enriches the picture. In the simplest case, a continuous part of a lattice with $K$ sites can be illuminated. However, the nontrivial selection of the illuminated sites is possible as well: e.g., each second site can be easily illuminated by choosing the light wavelength twice as the wavelength of the trapping beam. Moreover, using several probe beams one can make the optical geometry even more interesting \cite{CaballeroPRA2016}, which we will consider in Chapter 5.

The many-body Hamiltonian of light--matter interaction  $H$ is given by the contributions of two terms: the light-field part of the Hamiltonian $H_f$, and the atom-part $H_a$ written using the formalizm of the second quantization \cite{MekhovPRL2007,MekhovPRA2007,MekhovEPJD08}:
\begin{subequations}\label{1-1}
\begin{gather}
H=H_f +H_a, \\
H_f=\sum_l{\hbar\omega_l a^\dag_l a_l} -i\hbar\sum_l{(\eta^*_l a_l -
\eta_l a^\dag_l)}, \\
H_a=\int{d^3{\bf r}\Psi^\dag({\bf r})H_{a1}\Psi({\bf r})} 
+\frac{2\pi a_s \hbar^2}{m}\int{d^3{\bf r}\Psi^\dag({\bf
r})\Psi^\dag({\bf r})\Psi({\bf r})\Psi({\bf r})}.
\end{gather}
\end{subequations}
In the light-field part of the Hamiltonian $H_f$, $a_l$ are the
annihilation operators of light modes with the frequencies
$\omega_l$, wave vectors ${\bf k}_l$, and mode functions $u_l({\bf
r})$, which can be pumped by coherent fields with amplitudes
$\eta_l$. In the atom part, $H_a$, $\Psi({\bf r})$ is the atomic
matter-field operator, $a_s$ is the $s$-wave scattering length
characterizing the direct interatomic interaction. $H_{a1}$ is
the atomic part of the single-particle Hamiltonian $H_1$, which in
the rotating-wave and dipole approximation has a form
\begin{subequations}\label{1-2}
\begin{gather}
H_1=H_f +H_{a1}, \\
H_{a1}=\frac{{\bf p}^2}{2m_a}+\frac{\hbar\omega_a}{2} \sigma_z -
i\hbar \sum_l{[\sigma^+ g_l a_l u_l({\bf r})-\text{H. c.}}]
\end{gather}
\end{subequations}
Here, ${\bf p}$ and ${\bf r}$ are the momentum and position
operators of an atom of mass $m_a$ and resonance frequency
$\omega_a$, $\sigma^+$, $\sigma^-$, and $\sigma_z$ are the raising,
lowering, and population difference operators, $g_l$ are the
atom--light coupling constants for each mode. 

It is important to underline, that the inclusion of the interaction between an atom and quantum light in the single-particle Hamiltonian is the key step, which is absent in the standard problems of ultracold atoms in classical potentials.

We will consider essentially nonresonant interaction where the
light-atom detunings $\Delta_{la} = \omega_l - \omega_a$ are much
larger than the spontaneous emission rate and Rabi frequencies $g_l
a_l$. Thus, in the Heisenberg equations obtained from the
single-atom Hamiltonian $H_1$ (\ref{1-2}), the atomic population difference $\sigma_z$ can be set to
$-1$ (approximation of linear dipoles, i.e., the dipoles responding linearly to the light amplitude with the negligible population of the excited state). Moreover, the polarization
$\sigma^-$ can be adiabatically eliminated and expressed via the
fields $a_l$. The elimination of $\sigma^-$ and setting $\sigma_z=-1$ is equivalent to the adiabatic elimination of the upper atomic level. The light-atom detunings can be all then replaced by a single value $\Delta_{a} = \omega_p - \omega_a$, where $\omega_p$ is, for example, the frequency of the external probe. An effective single-particle Hamiltonian that gives
the corresponding Heisenberg equation for $a_l$ can be written as
$H_{1\text{eff}}=H_f +H_{a1}$ with
\begin{eqnarray}\label{1-3}
H_{a1}=\frac{{\bf p}^2}{2m_a}+V_{\text {cl}}({\bf r})+\frac{\hbar}{\Delta_{a}}
\sum_{l,m}{u_l^*({\bf r})u_m({\bf r}) g_l g_m a^\dag_l
a_m}.
\end{eqnarray}
Here, we have added a classical trapping potential of the
lattice, $V_{\text {cl}}({\bf r})$, which corresponds to a strong
classical standing wave. This potential can be, of course, derived
from one of the modes $a_l = a_{\text {cl}}$ [in this case $V_{\text
{cl}}({\bf r})=\hbar g^2_{\text {cl}}|a_{\text {cl}} u_{\text {cl}}({\bf
r})|^2/\Delta_{\text {cl}a}$], and it can scatter light into other
modes \cite{NimmrichterNJP2010}. Nevertheless, at this point we will consider $V_{\text
{cl}}({\bf r})$ as an independent potential, which does not affect
light scattering of other modes that will be significantly detuned
from $a_{\text {cl}}$ [i.e. the interference terms between $a_{\text
{cl}}$ and other modes are not considered in the last term of
Eq.~(\ref{1-3})]. A later inclusion of the light scattered by the
trapping wave will not constitute a difficulty, due to the linearity
of dipoles assumed in this model.

If one considers scattering of weak modes from the atoms in a deep
lattice, the fields $a_l$ are much weaker than the field
forming the potential $V_{\text {cl}}({\bf r})$. To derive the
generalized Bose--Hubbard Hamiltonian near zero temperature, we expand the field operator
$\Psi({\bf r})$ in Eq.~(\ref{1-1}), using localized Wannier functions
corresponding to $V_{\text {cl}}({\bf r})$ and keeping only the
lowest vibrational state at each site: $\Psi({\bf
r})=\sum_{i=1}^{M}{b_i w({\bf r}-{\bf r}_i)}$, where $b_i$ is the
annihilation operator of an atom at the site $i$ with the coordinate
${\bf r}_i$. 

Note that, starting from this important point of our derivation, both light and atoms are represented in a very similar way as quantum fileds: 

(I) The light is given by the modes with annihilation operators  $a_l$ and spatial mode functions $u_l({\bf r})$. 

(II) The atoms are given by the modes with annihilation operators $b_i$ and spatial mode functions $w({\bf r}-{\bf r}_i)$.
While the light mode functions $u_l({\bf r})$ are delocalized in space (e.g. they are given by travelling-wave exponents), the atomic mode functions $w({\bf r}-{\bf r}_i)$ are very tightly localized in space near the lattice sites $i$.

Substituting this expansion for $\Psi({\bf r})$ in Eq.~(\ref{1-1}) with
$H_{a1}$ (\ref{1-3}), we get
\begin{gather}
H=H_f+\sum_{i,j=1}^M{J_{i,j}^{\text {cl}}b_i^\dag b_j} 
+ \frac{\hbar}{\Delta_{a}}
\sum_{l,m}{g_l g_m a^\dag_l
a_m}\left(\sum_{i,j=1}^K{J_{i,j}^{lm}b_i^\dag
b_j}\right)  +\frac{U}{2}\sum_{i=1}^M{b_i^\dag b_i(b_i^\dag b_i-1)},\label{1-4}
\end{gather}
where the coefficients $J_{ij}^{\text {cl}}$ correspond to the
quantum motion of atoms in the classical potential and are typical
for the Bose--Hubbard Hamiltonian \cite{JakschPRL1998}:

\begin{equation}\label{1-5}
J_{i,j}^{\text {cl}}=\int{d{\bf r}}w({\bf r}-{\bf
r}_i)\left(-\frac{\hbar^2\nabla^2}{2m}+V_{\text {cl}}({\bf
r})\right)w({\bf r}-{\bf r}_j).
\end{equation}
However, in contrast to the usual Bose--Hubbard model, one has new
terms depending on the coefficients $J_{ij}^{lm}$, which describe an
additional contribution arising from the presence of light modes:
\begin{equation}\label{1-6}
J_{i,j}^{lm}=\int{d{\bf r}}w({\bf r}-{\bf r}_i) u_l^*({\bf
r})u_m({\bf r})  w({\bf r}-{\bf r}_j).
\end{equation}
In the last term of Eq.~(\ref{1-4}), only the on-site interaction was
taken into account and $U=4\pi a_s\hbar^2/m_a \int{d{\bf r}|w({\bf
r})|^4}$.

Note, that if the contribution of the quantized light is not much weaker than the contribution of the classical potential, or if the classical potential is not present at all, the Wannier functions should be determined in a self-consistent way taking into account the mean depth of the quantum potential, generated by the quantum light modes. This indeed significantly complicates the theoretical treatment \cite{MekhovEPJD08}. Although the form of the above equations still holds, the coefficients (\ref{1-5}) and (\ref{1-6}) will depend on the quantum state of light as well.

As a usual approximation, we restrict atom tunneling to the nearest neighbor sites. Thus, coefficients (\ref{1-5}) do
not depend on the site indices ($J_{i,i}^{\text {cl}}=J_0^{\text
{cl}}$ and $J_{i,i\pm 1}^{\text {cl}}=J^{\text {cl}}$), while
coefficients (\ref{1-6}) are still index-dependent. The Hamiltonian
(\ref{1-4}) then reads
\begin{gather}
H=H_f+J_0^{\text {cl}}\hat{N}+J^{\text {cl}}\hat{B} 
+\frac{\hbar}{\Delta_{a}}
\sum_{l,m}{g_l g_m a^\dag_l
a_m}\left(\sum_{i=1}^K{J_{i,i}^{lm}\hat{n}_i}\right)  \nonumber \\
+\frac{\hbar}{\Delta_{a}} \sum_{l,m}{g_l g_m a^\dag_l
a_m}\left(\sum_{<i,j>}^K{J_{i,j}^{lm}b_i^\dag
b_j}\right)
+\frac{U}{2}\sum_{i=1}^M{\hat{n}_i(\hat{n}_i-1)},\label{1-7}
\end{gather}
where $<i,j>$ denotes the sum over neighboring pairs,
$\hat{n}_i=b_i^\dag b_i$ is the atom number operator at the $i$-th
site, and $\hat{B}=\sum_{i=1}^M{b^\dag_i b_{i+1}}+{\text {H.c.}}$
While the total atom number determined by
$\hat{N}=\sum_{i=1}^M{\hat{n}_i}$ is conserved, the atom number at
the illuminated sites, determined by
$\hat{N}_K=\sum_{i=1}^K{\hat{n}_i}$, is not necessarily a conserved
quantity.

The Heisenberg equations for $a_l$ and $b_i$ can be obtained from
the Hamiltonian (\ref{1-7}) as
\begin{subequations}\label{1-8}
\begin{gather}
\dot{a}_l= -i\left( \omega_l
+\frac{g_l^2}{\Delta_{a}}\sum_{i=1}^K{J_{i,i}^{ll}\hat{n}_i}+
\frac{g_l^2}{\Delta_{a}}\sum_{<i,j>}^K{J_{i,j}^{ll}b_i^\dag
b_j}\right) a_l    \nonumber \\
-i\frac{g_l}{\Delta_{a}}\sum_{m \ne
l}g_m a_m\left(\sum_{i=1}^K{J_{i,i}^{lm}\hat{n}_i}\right)
-i\frac{g_l}{\Delta_{a}}\sum_{m \ne
l}g_m a_m\left(\sum_{<i,j>}^K{J_{i,j}^{lm}b_i^\dag
b_j}\right)+ \eta_l -\kappa_l a_l, \label{1-8a}    \\
\dot{b}_i=-\frac{i}{\hbar}\left( J_0^{\text {cl}}+\frac{\hbar}{\Delta_{a}}
\sum_{l,m}{g_l g_m a^\dag_l a_m J_{i,i}^{lm}}+U\hat{n}_i\right) b_i  \nonumber \\
-\frac{i}{\hbar}\left( J^{\text {cl}}+\frac{\hbar}{\Delta_{a}}\sum_{l,m}{g_l g_m a^\dag_l a_m J_{i,i+1}^{lm}}\right)b_{i+1} 
-\frac{i}{\hbar}\left( J^{\text {cl}}+\frac{\hbar}{\Delta_{a}}\sum_{l,m}{g_l g_m a^\dag_l a_m J_{i,i-1}^{lm}}\right)b_{i-1}, \label{1-8b}
\end{gather}
\end{subequations}
where we phenomenologically included the decay rate $\kappa_l$ of the mode $a_l$. Here, we do not add the corresponding Langevin noise term, since we will be interested in the normal-ordered quantities only (when it will be necessary later, we will add the Langevin noise term). The decay of the atoms can be included in a similar way as well, but it is usually much smaller than the cavity relaxation.

In Eq.~(\ref{1-8a}) for the electromagnetic fields $a_l$, the two last
terms in the first parentheses correspond to the phase shift of the light
mode due to nonresonant dispersion (the second term) and due to
tunneling to neighboring sites (the third one). The second term in
Eq.~(\ref{1-8a}) describes scattering of all modes into $a_l$, while
the third term takes into account corrections to such scattering
associated with tunneling due to the presence of additional light
fields. In Eq.~(\ref{1-8b}) for the matter field operators $b_i$, the
first term gives the phase of the matter-field at the site $i$, the
second and third terms describe the coupling to neighboring sites.

It is important to underline that except for the direct coupling
between neighboring sites, as it is usual for the standard
Bose--Hubbard model, Eqs.~(\ref{1-8}) also take into account
the long-range interaction between the sites, which does not decrease with
the distance and is provided by the common light modes $a_l$, which in turn
are determined by the whole set of matter-field operators $b_i$.
Such a cavity-mediated long-range interaction and nonlocal correlations between the operators $b_i$, which are introduced by the general Eqs.~(\ref{1-8}), can give rise to new many-body effects beyond the standard Bose--Hubbard model.

In this chapter we will use the developed model to establish the link between the light- and matter-related variables to focus on the nondestructive probing of ultracold matter by light. In Chapters 2--4 we will see, how the atomic dynamics (tunneling) and interaction presented here will compete with the backaction of quantum measurement. In Chapter 5 we will analyze novel many-body phases appearing due to the quantumness of the optical trapping potential.


\section{QND measurements in deep lattices and classical optical analogies}

As the results of the previous section show, the light--matter interaction leads to the joint evolution of the light and atomic variables. The light and matter get correlated. As this is a fully quantum problem, in general, the light and matter get entangled. Thus, observing the light, one can obtain information about the quantum states of the atoms. The goal of this section is to establish relations between the characteristics of scattered light and those of the many-body atomic system.

Considering measurements in quantum mechanics, the first goal is to obtain the expectation values of desired quantities or their distribution functions. Thus, such a type of probing is intrinsically associated with multiple measurements of the same quantity and averaging over the measurement outcomes. The type of the quantum measurement considered in this section corresponds to the quantum nondemolition (QND) \cite{BrunePRA1992} observation of different atomic variables by detecting light. In contrast to the completely destructive schemes usually used (e.g. time of flight measurements, absorption images), the QND measurement by light scattering only weakly perturbs the atomic sample and many consecutive measurements can be carried out with the same atoms without preparing a new sample for each measurement. This is in sharp contrast, for example, to the method of time of flight measurements: In this method, the lattice is completely turned off, the atoms fall down freely, and the interference of atomic matter waves is measured. (The interference of atoms falling down from a two-dimensional optical lattice is very similar to the diffraction pattern of light scattered from a two-dimensional diffraction grating.) Accordingly, to measure an average value at just one point, it is necessary to repeatedly turn off the trap and destroy the entire atomic system many times.

However, an important statement of quantum mechanics is, that any measurement, even a QND one, affects the quantum state of the system. Therefore, to measure the expectation values or statistics of some variable, one should prepare the system in the same quantum state before each measurement (or wait until the initial state will restor due to the system evolution). In the next chapter, we focus on the essentially quantum properties of the measurement process itself going beyond the simple goal of measuring only the expectation values. We will analyze, how the many-body atomic state changes during the measurement process, i.e., we will consider the measurement at a single quantum trajectory (single run of a measurement) without a requirement of the statistical averaging over the many runs of an optical experiment.

To focus on the question how to establish the relation between the light and atom observables, we simplify the model. We consider only two light modes: the probe $a_0$ and the scattered light $a_1$. We assume that the tunneling of atoms between the neighboring sites is much slower than light scattering, and tunneling can be neglected during light scattering. Physically this means that the quantum properties of the atomic state are determined by tunneling and interaction, but are frozen during the measurement time. In practice, the tunneling and scattering times can be different in orders of magnitude. For a deep lattice the coefficients $J_{i,i}^{lm}$ (\ref{1-6}) reduce to
$J_{i,i}^{lm}=u_l^*({\bf r}_i)u_m({\bf r}_i)$ neglecting atom spreading determined by Wannier functions (i. e. the Wannier functions can be approximated by the delta-functions); for $i\ne j$, $J_{i,j}^{lm}=0$ as well as the classical tunneling coefficient $J^{\text {cl}}=0$. The influence of the atom spreading within each site on the light signal can be studied even by classical scattering \cite{KetterlePRL2011}.

After those simplifications, the Hamiltonian Eqs.~(\ref{1-7}) takes
the form:
\begin{gather}
H=\hbar(\omega_1 + U_{11} \hat{D}_{11}) a^\dag_1 a_1+
\hbar U_{10}(\hat{D}^*_{10}a^*_0a_1 + \hat{D}_{10}a_0a^\dag_1) 
-i\hbar(\eta_1^* a_1 - \eta_1 a^\dag_1),\label{1-1PRA09}
\end{gather}
where $a_1$ is the cavity-mode annihilation operator. The term $J_0^{\text {cl}}\hat{N}$ gives a constant energy shift and was omitted in the Hamiltonian. The external probe is assumed to be in a coherent state, thus its amplitude is given by
a c-number $a_0$. $U_{lm}=g_lg_m/\Delta_a$ ($l,m=0,1$), $\eta_1$
is the amplitude of the additional probing through a mirror at the
frequency $\omega_p$ (the probe-cavity detuning is
$\Delta_{p}=\omega_{p}-\omega_1$). The operators
$\hat{D}_{lm}= \sum_{j=1}^K{u_l^*({\bf r}_j)u_m({\bf
r}_j)\hat{n}_j}$ sum contributions from all illuminated sites with
the atom-number operators $\hat{n}_j$ at the position ${\bf r}_j$.

The first term in the Hamiltonian describes the atom-induced shift of
the cavity resonance. The second one reflects scattering
(diffraction) of the probe $a_0$ into a cavity mode $a_1$. The key feature of a quantum gas is that the frequency shift and probe-cavity coupling coefficient are operators, which leads to different light scattering amplitudes for various atomic quantum states.

The Hamiltonian (\ref{1-1PRA09}) describes QND measurements of the
variables associated with the operators $\hat{D}_{lm}$ by detecting the photon number
$a^\dag_1a_1$ or light amplitude $a_1$ related quantities (e.g. quadratures). In order for a measurement to be a QND one, several conditions should be fulfilled for the ``signal observable'' of interest $A_S$ ($\hat{D}_{lm}$ in this case), ``probe observable'' $A_P$, which is actually detected, (here, $a^\dag_1a_1$ or $a_1$) and the coupling between them through the interaction Hamiltonian \cite{BrunePRA1992}. According, for example, to Ref. \cite{BrunePRA1992}, the interaction Hamiltonian should be a function of $A_S$; the interaction between the signal and probe should affect the dynamics of $A_P$ ($[A_P,H]\ne 0$), whereas the signal observable should not be affected by the coupling to the probe ($[A_S,H] = 0$). In addition, measuring $A_S$, its conjugate variable is affected in an uncontrollable way, therefore the evolution should not depend on that uncontrolled variable at all. In this case, the signal observables $\hat{D}_{lm}$ depend only on the atom numbers, and their conjugate variables related to the atomic phase are not present in the Hamiltonian. One can see that all those conditions of the QND measurement are fulfilled for the Hamiltonian (\ref{1-1PRA09}).

QND measurements (considered in this chapter) have became a popular subject. Physically, they are called ``nondemolition,'' because they destroy (perturb, introduce noise to) the conjugate variable, which is not in the Hamiltonian. Thus, the noisy (destroyed) variable does not affect the evolution in an uncontrollable way. This constitutes an advantage for, e.g., quantum metrology. Nevertheless, we would like to underline that non-QND measurements can lead to much more interesting quantum evolution. In Chapter 3 we will show how the measurement-induced perturbation of the conjugate variable leads to very interesting dynamical effects. This will demonstrate that the measurement backaction constitutes a novel source of competitions in many-body systems (in addition to standard tunneling and atom interaction in classical optical lattices). In this chapter we can not discuss such competition as the tunneling has been simply neglected.

Note that one has a QND access to various
many-body variables, as $\hat{D}_{lm}$ strongly depends on the
lattice and light geometry via $u_{0,1}({\bf r})$. This is an
advantage of the lattice comparing to single- or double-well setups \cite{Moore1999,PuPRL2003,YouPRA1995,IdziaszekPRA2000,MustPRA2000,MustPRA2000x2,
JavanainenPRL1995,JavanainenPRA1995,CiracPRL1994,CiracPRA1994,SaitoPRA1999, PratavieraPRA2004, JavanainenPRL2003}.

For example, one can consider a 1D lattice of the period $d$ with
atoms trapped at $x_j=jd$ ($j=1,2,..,M$). In this case, the
geometric mode functions can be expressed as follows: $u_{0,1}({\bf
r}_j)=\exp (ijk_{0,1x}d+\phi_{0,1j})$ for traveling waves, and
$u_{0,1}({\bf r}_j)=\cos (jk_{0,1x}d +\phi_{0,1j})$ for standing
waves, where $k_{0,1x}=|{\bf k}_{0,1}|\sin\theta_{0,1}$,
$\theta_{0,1}$ are the angles between mode wave vectors ${\bf
k}_{0,1}$ and a vector normal to the lattice axis. In the plane-wave
approximation, additional phases $\phi_{0,1j}$ are $j$-independent.

For some geometries, $\hat{D}_{11}$ simply reduces to the operator
$\hat{N}_K=\sum_{j=1}^K\hat{n}_j$ of the atom number at $K$ sites
 (if $a_1$ is a traveling wave at an arbitrary
angle to the lattice, or the standing wave with atoms trapped at the
antinodes). If the probe and cavity modes are coupled at a
diffraction maximum (Bragg angle), i.e., all atoms scatter light in
phase, $u_1^*({\bf r}_j)u_0({\bf r}_j)=1$, the probe-cavity coupling
is maximized, $\hat{D}_{10}=\hat{N}_K$. If they are coupled at a
diffraction minimum, i.e., the neighboring atoms scatter with opposite phases $0$ and $\pi$, $\hat{D}_{10}=\sum_{j=1}^K (-1)^{j+1}\hat{n}_j=\hat{N}_\text{odd}-\hat{N}_\text{even}$ is the
operator of number difference between odd and even sites. Thus, the
atom number as well as number difference can be nondestructively
measured. Note, that those are just two of many examples of how a
QND-variable can be chosen by the geometry in a many-body system (for more cases see e.g. \cite{Elliott2015} and Sec. 2.9).

From the Hamiltonian, the Heisenberg equation for the scattered light can be obtained as follows:
\begin{eqnarray}\label{1-2NatPhys}
\dot{a}_1= -i\left(\omega_1 +U_{11}\hat{D}_{11}\right)a_1
-iU_{10}\hat{D}_{10}a_0 -\kappa a_1+\eta_1,
\end{eqnarray}
where $\kappa$ is the cavity decay rate. In classical physics, such an equation is directly analogous to the Maxwell's equation for the light amplitude of the cavity mode, and the classical meaning of the $\hat{D}_{lm}$ operators (i.e. the frequency dispersion shift and coupling coefficient between two modes) is obvious. Here, in the fully quantum problem, both light- and matter-related quantities are treated as operators indeed.

The stationary solution for the operator of the light amplitude oscillating at the probe frequency takes the form
\begin{eqnarray}\label{1-11}
a_1=\frac{\eta_1-iU_{10} a_0\hat{D}_{10}}{i(U_{11}
\hat{D}_{11}-\Delta_p)+\kappa},
\end{eqnarray}
which gives us a direct relation between the light operator and various atom number-related operators.
It is clear that if a light-related observable is a linear function of the atom number operators $\hat{n}_j$, then the measurement of that observable will depend only on the mean atom numbers (i.e., their expectation values $\langle\hat{n}_j\rangle$). Such a measurement will carry information only about the mean atom density, which can be similar for various quantum states, and thus is not of interest for the scope of this work. Therefore, the question is to find light observables, which depend nonlinearly on the atom number operators. In this case, the measurement will reveal the higher moments of the atom number operator, which carry information about the quantum state of ultracold atoms.

This suggests us to consider the following optical configurations, where the light observables are sensitive to various atomic states.

(I) Transverse probing. Here we neglect the dispersive frequency shift. In this case, the light amplitude $\langle a_1 \rangle$ is a linear function of the atom numbers and is not of great interest. However, the light intensity (given by the mean photon number $n_\Phi= \langle a_1^\dag a_1 \rangle$) already depends on the atom density-density correlations $\langle\hat{n}_i\hat{n}_j\rangle$, which differ for various atom states. Moreover, the photon number variance carries the information about the four-point correlation function $\langle\hat{n}_i\hat{n}_j\hat{n}_k\hat{n}_l\rangle$, which is even a more exciting result. We will address this case in the following sections.

(II) Probing through a mirror, where the dispersive frequency shift plays a key role ~\cite{MekhovNaturePh2007,MekhovLP2009}. In this case, even the light amplitude nonlinearly depends on the atom numbers (see the denominator in Eq.~(\ref{1-11})). Here, the measurement of light can directly reveal the full atom number distribution function. The measurement of frequency shifts in a cavity configuration is similar to the measurements of the phase shifts in the free-space geometry. We will consider this case in Sec. 1.10.

Let us consider the configuration (I), where only the transverse probe $a_0$ is present (no probing through a mirror $\eta_1=0$) and the dispersive frequency shift is small (the term $U_{11}\hat{D}_{11}$ can be neglected) in Eq.~(\ref{1-11}). In this case the probe-cavity detuning is $\Delta_p=\omega_0-\omega_1$.

Thus, a stationary solution (\ref{1-11}) has a form
\begin{eqnarray}\label{1-10PRA}
a_1=C\hat{D}_{10}, 
\end{eqnarray}
where we replaced the operators $a_{0,1}(t)$ by their slowly varying
envelopes $\tilde{a}_{0,1}(t)$
[$a_{0,1}(t)=\tilde{a}_{0,1}\exp(-i\omega_0 t)$] skipping in the
following notations all tilde signs. Therefore, the light amplitude operator is linear in the atom number operators, $a_1=C\hat{D}_{10}=\sum_{j=1}^K A_j\hat{n}_j$, where $C=iU_{10}a_0/(i\Delta_p-\kappa)$ and we introduce the coefficients $A_j=u_1^*({\bf r}_j)u_0({\bf r}_j)$. The expectation value of the light amplitude operator $\langle a_1 \rangle$ measures the mean atom numbers $\langle\hat{n}_j\rangle$ only. However, the number of photons scattered into a cavity depends quadratically on the atom operators:
\begin{eqnarray}\label{1-3PRL09-2}
n_\Phi= \langle a_1^\dag a_1 \rangle=|C|^2\langle\hat{D}_{10}^*\hat{D}_{10}\rangle=|C|^2\sum_{i,j=1}^K{A_i^* A_j \langle\hat{n}_i\hat{n}_j\rangle} .
\end{eqnarray}
Importantly, already a simple quantity as the mean light intensity (mean photon number $n_\Phi$) depends on the atom density-density correlations, which are second moments of the atom operators and are different for various atomic quantum states.

Expressing the light operators in terms of the atomic ones in
Eq.~(\ref{1-10PRA}) is an important result, which we will use to study
the properties of the scattered field. The dependence of the light
Heisenberg operators on the atomic operators reflects the
entanglement between light and matter during the light--matter
interaction. 

In the following, we will consider a 1D lattice (3D case will be considered later in Sec. 1.6)  of the period $d$
with atoms trapped at $x_m=md$ ($m=1,2,..,M$). The result for the
field operator $a_1$ (\ref{1-10PRA}) has an analogy in classical diffraction. For scattering
of a traveling wave $a_0$ in the direction of a traveling wave $a_1$
from a lattice with $\langle \hat{n}_i\rangle =n$ at each site, the
expectation value of the field is given by
\begin{gather}
\langle a_1\rangle =C\langle \hat{D}_{10}\rangle
=C\sum_{m=1}^K{e^{im\delta k_x d}\langle \hat{n}_m\rangle}
=Cn e^{i(K+1)\alpha_-/2}
\frac{\sin{(K\alpha_-/2})}{\sin{(\alpha_-/2})},\label{1-11PRA}
\end{gather}
where $\alpha_-=\delta k_x d$, and $\delta k_x =({\bf k}_0-{\bf
k}_1)_x=k(\sin\theta_0 - \sin\theta_1)$ is the projection of the
difference between two wave vectors on the lattice direction,
$\theta_{0,1}$ are the angles between wave vectors and a vector
normal to the lattice direction (cf. Fig.~\ref{1-fig1}), $k=\omega/c$
for $\omega_0=\omega_1=\omega$.

Equation (\ref{1-11PRA}) simply describes classical diffraction of the
traveling wave $a_0$ on a diffraction grating formed by equally
spaced atoms with positions of diffraction maxima and minima (i.e.
scattering angles $\theta_1$) determined by the parameter $\alpha_-$
depending on the geometry of incident and scattered waves and
diffraction grating through $\theta_{0}$, $|{\bf k}_{0,1}|$, and
$d$. A more general form of the operator $\hat{D}_{10}$ describes also diffraction of a standing wave $a_0$ into another mode $a_1$, which can be formed, for example, by a
standing--wave or ring optical cavity.

Equation (\ref{1-11PRA}) shows that the expectation value of the
scattered field is sensitive only to the mean number of atoms per
site $n$ and reflects a direct analogy of light scattering from a
classical diffraction grating. Nevertheless, the photon number
(intensity) and photon statistics of the field $a_1$ are sensitive
to higher moments of the number operators $\hat{n}_i$ as well as to
the quantum correlations between different lattice sites, which
determines quantum statistical properties of ultracold atoms in an
optical lattice and will be considered in the next sections.


\section{Relation between quantum statistics of atoms and characteristics
of scattered light}

In this section we will derive the relations between quantum properties of atomic states and angular distributions of variables of scattered light such as the light amplitude, photon number (intensity), quadratures and their variances, and photon number variances. We will show that even simple mean intensity is sensitive to the quantum fluctuations and correlations of atoms, which are radically different for various atomic states. The photon number variance will reflect the four-point spatial correlation functions of atoms. In this section, we will derive general formulas. We will illustrate them in the next section using particular examples.

\subsection{Probing quantum statistics by intensity measurements}

According to Eq.~(\ref{1-10PRA}), the expectation value of the photon
number $a_1^\dag a_1$ is proportional to the expectation value of
the operator $\hat{D}_{10}^*\hat{D}_{10}$. We have already introduced coefficients
$A_i(\theta_0,\theta_1)$ responsible for the geometry of the
problem:
\begin{gather}
\hat{D}_{10}=\sum_{i=1}^K{A_i \hat{n}_i}, \quad
A_i(\theta_0,\theta_1)\equiv u_1^*({\bf r}_i)u_0({\bf r}_i),
\quad
A(\theta_0,\theta_1)\equiv \sum_{i=1}^K{A_i(\theta_0,\theta_1)},\label{1-12}
\end{gather}
where  $u_{0,1}({\bf r}_m)=\exp (imk_{0,1x}d+\phi_{0,1m})$ for
traveling waves, and $u_{0,1}({\bf r}_m)=\cos (mk_{0,1x}d
+\phi_{0,1m})$ for standing waves ($m=1,2,...M$), $k_{0,1x}=|{\bf
k}_{0,1}|\sin\theta_{0,1}$, $\theta_{0,1}$ are the angles between
mode wave vectors and a vector normal to the lattice axis; in the
plane-wave approximation, additional phases $\phi_{0,1m}$ are
$m$-independent.

The expectation values of $\hat{D}_{10}$ and $\hat{D}_{10}^*\hat{D}_{10}$ then read
\begin{subequations}\label{1-13}
\begin{gather}
\langle \hat{D}_{10} \rangle =\sum_{i=1}^K{A_i\langle\hat{n}_i}\rangle= nA, \label{1-13a}\\
\langle \hat{D}_{10}^*\hat{D}_{10} \rangle = \sum_{i,j=1}^K{A_i^* A_j
\langle\hat{n}_i\hat{n}_j\rangle}   \label{1-13b}\\
=\langle \hat{n}_a\hat{n}_b\rangle |A|^2+(\langle\hat{n}^2\rangle -
\langle \hat{n}_a\hat{n}_b\rangle)\sum_{i=1}^K{|A_i|^2}, \label{1-13c}\\
R(\theta_0, \theta_1)\equiv \langle\hat{D}_{10}^*\hat{D}_{10} \rangle - |\langle \hat{D}_{10} \rangle|^2 =  
(\langle \hat{n}_a\hat{n}_b\rangle - n^2)
|A|^2+(\langle\hat{n}^2\rangle - \langle
\hat{n}_a\hat{n}_b\rangle)\sum_{i=1}^K{|A_i|^2}  \label{1-13d}\\
=\langle \delta\hat{n}_a\delta\hat{n}_b\rangle
|A|^2+(\langle\delta\hat{n}^2\rangle - \langle
\delta\hat{n}_a\delta\hat{n}_b\rangle)\sum_{i=1}^K{|A_i|^2}.\label{1-13e}
\end{gather}
\end{subequations}
In Eqs.~(\ref{1-13}) we have used the following assumptions about the
atomic quantum state $|\Psi\rangle$: (i) the expectation values of
the atom number at all sites are the same, $\langle\hat{n}_i\rangle
= n$ (thus, the expectation value of atom number at $K$ sites is
$\langle\hat{N}_K\rangle=N_K\equiv nK$), (ii) the nonlocal pair
correlations between atom numbers at different sites
$\langle\hat{n}_i\hat{n}_j\rangle$ are equal to each other for any
$i\ne j$ and will be denoted as $\langle\hat{n}_a\hat{n}_b\rangle$
(with $a\ne b$). Although the approximation that pair fluctuations are equal for all sites is not general, it holds for several interesting examples such as Mott insulator state (MI), superfluid state (SF) and the coherent state approximation to the SF state. Importantly, all those states have the same mean atom numbers $n$ at each site, but very different atom number fluctuations. We will show that such differences are readily captured by the light scattering. We will go beyond such an approximation of equal correlations in Sec. 1.7. In addition, we introduced the fluctuation operators
$\delta\hat{n}_i=\hat{n}_i - n$, which gives
$\langle\delta\hat{n}^2\rangle$ equal to the variance $(\Delta
n_i)^2=\langle\hat{n}_i^2\rangle-n^2$.

Equation (\ref{1-13a}) reflects the fact that the expectation value of
the field amplitude (\ref{1-10PRA}) is sensitive only to the mean atom
numbers and displays the angle dependence of classical diffraction
given by the factor $A(\theta_0,\theta_1)$, which depends on the
mode angles and displays pronounced diffraction maxima and minima.
Equation (\ref{1-13b}) shows that the number of scattered photons
(intensity) at some angle is determined by the density--density
correlations. In the simplest case of two traveling waves, the
prefactors $A_i^* A_j= \exp[i\delta k_x(x_j-x_i)]$ with $\delta k_x=k_{0x}-k_{1x}$. In this case, Eq.~(\ref{1-13b}) gives the so-called
structure factor (function), which was considered in the works on
light scattering from homogeneous BEC \cite{JavanainenPRL1995,JavanainenPRA1995}. Here we
essentially focus on optical lattices. Moreover, it will be shown,
that the more general Eq.~(\ref{1-13b}), which includes scattering of
standing waves, contains new measurable features different from
those of a usual structure factor.

Equation (\ref{1-13c}) shows, that the angle dependence of the
scattered intensity consists of two contributions. The first term
has an angle dependence $|A(\theta_0,\theta_1)|^2$ identical to that
of the expectation value of the field amplitude squared (\ref{1-13a}).
The second term is proportional to the quantity
$\langle\hat{n}^2\rangle - \langle \hat{n}_a\hat{n}_b\rangle$ giving
quantum fluctuations and has a completely different angle dependence
$\sum_{i=1}^K{|A_i|^2}$. The expression (\ref{1-13c}) has a form
similar to the one considered in papers
\cite{YouPRA1995,MustPRA2000,MustPRA2000x2} on light scattering from a
homogeneous BEC, where the scattered intensity consisted of two
parts: ``coherent'' (i.e. depending on the average density) and
``incoherent'' one (i.e. depending on the density fluctuations).
Nevertheless, in the present case of a periodic lattice, this
similarity would be exact only in a particular case where there are
no nonlocal pair correlations $\langle \hat{n}_a\hat{n}_b\rangle =
n_a n_b = n^2$ ($\langle \delta\hat{n}_a\delta\hat{n}_b\rangle =0$),
which in general is not true and leads to observable difference
between states with and without pair correlations.

Further insight into a physical role of nonlocal pair correlations
can be obtained from Eqs.~(\ref{1-13d}) and (\ref{1-13e}) for the
``noise quantity'' (or ``quantum addition'') $R(\theta_0, \theta_1)\equiv
\langle\hat{D}_{10}^*\hat{D}_{10} \rangle - |\langle \hat{D}_{10} \rangle|^2$,
where we have subtracted the classical (averaged) contribution
$|\langle \hat{D}_{10} \rangle|^2$ to the intensity
$\langle\hat{D}_{10}^*\hat{D}_{10} \rangle$. Equation (\ref{1-13e}) shows that,
in the noise quantity, a term with the classical angular
distribution $|A(\theta_0,\theta_1)|^2$ appears only if the pair
correlations are nonzero. The physical meaning of this result is
that, in an optical lattice, it is not only the density distribution
that displays spatial periodic structure leading to diffraction
scattering, but also the distribution of number fluctuations
themselves. In the framework of our assumption about equal pair
correlations, the spatial distribution of fluctuations $\langle
\delta\hat{n}_a\delta\hat{n}_b\rangle$ can be either the same as the
density distribution (with a lattice period $d$) or zero. In the
former case, pair correlations contribute to the first term in
Eqs.~(\ref{1-13d}) and (\ref{1-13e}) with classical distribution
$|A(\theta_0,\theta_1)|^2$, in the latter case, $\langle
\delta\hat{n}_a\delta\hat{n}_b\rangle = 0$, and the only signal in
the noise quantity is due to on-site fluctuations
$\langle\delta\hat{n}^2\rangle$ with a different angle dependence
$\sum_{i=1}^K{|A_i|^2}$. Note that, in general, the spatial
distribution of fluctuations can be different from that of the
average density and can have a period proportional to the lattice
period $d$. This will lead to additional peaks in the angular
distribution of the noise quantity (\ref{1-13d}),~(\ref{1-13e}). The
generalization of those formulas is straightforward.

Even with spatially incoherent probe $a_0$, the intensity of the
scattered mode $a_1^\dag a_1$ is sensitive to the on-site atom
statistics. To model this situation, the quantum expectation value
$\langle\hat{D}_{10}^*\hat{D}_{10}\rangle$ (\ref{1-13b}) should be additionally
averaged over random phases $\phi_{0,1m}$ appearing in the
definition of mode functions in Eq.~(\ref{1-12}). In Eq.~(\ref{1-13b}),
only terms with $i=j$ will then survive and the final result reads
\begin{eqnarray}\label{1-14}
\langle \hat{D}^*\hat{D} \rangle_\text{inc} =p_0 K \langle
\hat{n}^2\rangle ,
\end{eqnarray}
where $p_0$ is equal to 1 for two traveling waves, 1/2 for a
configuration with one standing wave, and 1/4, when both modes
$a_{0,1}$ are standing waves.

\subsection{Quadrature measurements}

The photon number $a_1^\dag a_1$ is determined by the expectation
value $\langle\hat{D}_{10}^*\hat{D}_{10} \rangle$, whereas $\langle \hat{D}_{10}
\rangle$ gives the field $\langle a_1 \rangle$ (\ref{1-10PRA}). While
photon numbers can be directly measured, a field $\langle a_1
\rangle$ measurement requires a homodyne scheme. Such a measurement
then makes $\langle \hat{D}_{10} \rangle$ experimentally accessible.
Actually for a quantum field only the expectation values of
quadratures of $a_1$ that are Hermitian operators and can be
measured. Using Eq.~(\ref{1-10PRA}) and the commutation relation
$[a_1,a_1^\dag]=1$, the quadrature operator $X_\phi$ and its
variance $(\Delta X_\phi)^2$ can be written as
\begin{subequations}\label{1-15}
\begin{gather}
X_\phi \equiv \frac{1}{2}\left(a_1 e^{-i\phi}+a_1^\dag
e^{i\phi}\right)=|C|\hat{X}^D_{\phi-\phi_C},  \label{1-15a}\\
X_\phi^2 =\frac{1}{4}+|C|^2 (\hat{X}^D_{\phi-\phi_C})^2, \label{1-15b}\\
(\Delta X_\phi)^2 \equiv \langle X_\phi^2\rangle - \langle
X_\phi\rangle^2 =\frac{1}{4}+|C|^2 (\Delta X^D_{\phi-\phi_C})^2,
\label{1-15c}
\end{gather}
\end{subequations}
where $C=|C|\exp({i\phi_C})$ and the quadratures of $\hat{D}_{10}$ are
\begin{subequations}\label{1-16}
\begin{gather}
\hat{X}_\beta^D \equiv \frac{1}{2}\left(\hat{D}_{10}
e^{-i\beta}+\hat{D}_{10}^*e^{i\beta}\right), \label{1-16a}\\ (\Delta
X_\beta^D)^2 \equiv \langle (\hat{X}_\beta^D)^2\rangle - \langle
\hat{X}_\beta^D\rangle^2.\label{1-16b}
\end{gather}
\end{subequations}
In Eqs.~(\ref{1-15}), the phase $\phi$ is related to the homodyne
reference phase, while $\phi_C$ is determined by the phase of the
probe $a_0$ and parameters of the field--matter system [cf.
Eq.~(\ref{1-10PRA})]. Hence, the phase $\beta = \phi-\phi_C$ entering
Eqs.~(\ref{1-15}) can be controlled by varying the phase difference
between the probe and homodyne fields.

Using Eq.~(\ref{1-12}), the quadrature operator $\hat{X}_\beta^D$
reads
\begin{gather}
\hat{X}_\beta^D =\sum_{i=1}^K{A_i^\beta \hat{n}_i}, \quad
A_i^\beta(\theta_0,\theta_1)\equiv |A_i|\cos{(\phi_{A_i}-\beta)},
\nonumber\\
A^\beta (\theta_0,\theta_1)\equiv
\sum_{i=1}^K{A_i^\beta(\theta_0,\theta_1)},\label{1-17}
\end{gather}
where $A_i=|A_i|\exp (i\phi_{A_i})$, and we defined new quantities
$A_i^\beta(\theta_0,\theta_1)$ and $A^\beta(\theta_0,\theta_1)$.

Since Eq.~(\ref{1-17}) for $\hat{X}_\beta^D$ and Eq.~(\ref{1-12}) for
$\hat{D}_{10}$ have a similar structure, the Eqs.~(\ref{1-13}) for the
quantities $\langle \hat{D}_{10}\rangle$,
$\langle\hat{D}_{10}^*\hat{D}_{10}\rangle$, and $R$ can be rewritten for the
quantities $\langle\hat{X}_\beta^D\rangle$, $\langle
(\hat{X}_\beta^D)^2\rangle$, and $(\Delta X_\beta^D)^2$,
respectively, with the change of parameters $A_i(\theta_0,\theta_1)$
and $A(\theta_0,\theta_1)$ to $A_i^\beta(\theta_0,\theta_1)$ and
$A^\beta (\theta_0,\theta_1)$. Thus, the above discussion of
Eqs.~(\ref{1-13}) can be repeated in terms of the quadrature operators
with the only difference that coefficients
$A_i^\beta(\theta_0,\theta_1)$ and $A^\beta (\theta_0,\theta_1)$ now
depend also on the homodyne phase. An advantage of this
reformulation is that the expectation value of the non-Hermitian
operator $a_1$, which determines $\langle \hat{D}_{10}\rangle$, is now
replaced by the expectation value of the Hermitian operator
$X_\phi$, which is consistent with a procedure of measuring
quadratures of the quantum field $a_1$. The well-known relations
between intracavity and outcoupled fields can be found, e.g., in
Ref.~\cite{walls2008quantum} for linear systems, which is the case in our work.

\subsection{Photon number fluctuations}

While the intensity of the scattered light is sensitive to the
second moments of the number operators $\hat{n}_i$, quantum
statistics of the field reflects the higher-order moments. The
variance $(\Delta n_\text{ph})^2$ of the photon number
$n_\Phi=a_1^\dag a_1$ depends on the atomic fourth-order moments and is given by
\begin{gather}
(\Delta n_\Phi)^2=\langle n_\Phi^2\rangle - \langle
n_\Phi\rangle^2
=:(\Delta n_\Phi^2): +\langle n_\Phi\rangle \nonumber\\
= |C|^4(\langle \hat{D}_{10}^{*2}\hat{D}_{10}^2\rangle -\langle
\hat{D}_{10}^*\hat{D}_{10}\rangle^2)+|C|^2\langle \hat{D}_{10}^*\hat{D}_{10}\rangle,\label{1-18}
\end{gather}
where $:(\Delta n_\Phi^2): = \langle a_1^{\dag 2}a_1^2 \rangle
- \langle a_1^{\dag}a_1\rangle^2=|C|^4(\langle
\hat{D}_{10}^{*2}\hat{D}_{10}^2\rangle -\langle \hat{D}_{10}^*\hat{D}_{10}\rangle^2)$ is
a normal ordered photon-number variance. Thus, the problem is
reduced to measurements of the photon number $|C|^2\langle
\hat{D}_{10}^*\hat{D}_{10} \rangle$ and quantity $|C|^4\langle
\hat{D}_{10}^{*2}\hat{D}_{10}^2\rangle$, which after straightforward
calculations is given by
\begin{gather}
\langle \hat{D}_{10}^{*2}\hat{D}_{10}^2\rangle =\left|\sum_{i=1}^K
A_i\right|^4\langle n_an_bn_cn_d\rangle  \nonumber\\
+2\left[\left(\sum_{i=1}^K
|A_i|^2A_i\right)\sum_{i=1}^K A_i^*+ \text{c.c.}\right](2\langle
n_an_bn_cn_d\rangle-3\langle n_a^2n_bn_c\rangle+\langle
n_a^3n_b\rangle) \nonumber\\
+\left[\left(\sum_{i=1}^K A_i^2\right)\left(\sum_{i=1}^K
A_i^*\right)^2+ \text{c.c.}\right](-\langle
n_an_bn_cn_d\rangle+\langle n_a^2n_bn_c\rangle) \nonumber
\end{gather}
\begin{gather}
+2\left(\sum_{i=1}^K|A_i|^2\right)^2(\langle
n_an_bn_cn_d\rangle-2\langle n_a^2n_bn_c\rangle+\langle
n_a^2n_b^2\rangle) \nonumber\\
+\left|\sum_{i=1}^K A_i^2\right|^2(\langle
n_an_bn_cn_d\rangle-2\langle n_a^2n_bn_c\rangle+\langle
n_a^2n_b^2\rangle) \nonumber\\
+4\left|\sum_{i=1}^K A_i\right|^2\sum_{i=1}^K
|A_i|^2(-\langle n_an_bn_cn_d\rangle+\langle n_a^2n_bn_c\rangle) \nonumber\\
+\sum_{i=1}^K |A_i|^4(-6\langle n_an_bn_cn_d\rangle+12\langle
n_a^2n_bn_c\rangle-4\langle n_a^3n_b\rangle-3\langle
n_a^2n_b^2\rangle+\langle n^4\rangle),\quad
\label{1-19}
\end{gather}
where we assumed again that correlations do not depend on site
indices, and sites with the indices $a$, $b$, $c$, and $d$ are
different. In Eq.~(\ref{1-19}), each prefactor containing geometrical
coefficients $A_i$ (\ref{1-12}) determines different angle dependences
of a corresponding term.

Thus, varying the geometry of a problem (e.g. angles of two modes,
wavelengths of the modes or that of trapping potential determining
the lattice period), one has access to different statistical
quantities characterizing the quantum state of ultracold atoms.

\subsection{Quantum statistical properties of typical atomic distributions}

Let us briefly summarize some key statistical properties of typical
states of $N$ atoms at $M$ lattice sites, i.e: the Mott insulator
state (MI), superfluid state (SF), and a multisite coherent-state
approximation to the SF state (cf. Table~\ref{1-table1}).

\begin{table}[h]
\centering
\captionsetup{justification=justified}
\begin{tabular}{|l|l|l|l|}\hline
& {\bf MI} & {\bf SF} & {\bf Coherent}\\ \hline \hline
$|\Psi\rangle$ & $\displaystyle\prod_{i=1}^M |n_i\rangle_i$ &
$\displaystyle\frac{1}{\sqrt{M^N N!}}(\sum_{i=1}^M
b_i^\dag)^N|0\rangle$ & $\displaystyle
e^{-\frac{N}{2}}\prod_{i=1}^M{e^{\sqrt{\frac{N}{M}}b_i^\dag}}
|0\rangle_i$
\\\hline $\langle\hat{n}_i^2\rangle$ &  $n^2$   & $n^2(1-1/N)+n$ &
$n^2+n$
\\\hline
$\left(\Delta n_i\right)^2$ &  0 & $n(1-1/M)$ & $n$
\\\hline
$\langle \hat{N}_K^2\rangle$ & $N_K^2$ & $N_K^2(1-1/N)+N_K$ &
$N_K^2+N_K$ \\\hline $(\Delta N_K)^2$ &  0 & $N_K(1-K/M)$ & $N_K$
\\\hline
$\langle\hat{n}_a\hat{n}_b\rangle$ & $n^2$ & $n^2(1-1/N)$ & $n^2$
\\\hline
$\langle\delta \hat{n}_a\delta \hat{n}_b\rangle$ & 0 & $-N/M^2$ & 0
\\\hline
\end{tabular}
\caption{\label{1-table1}Statistical quantities of typical atomic
states: Mott insulator (MI), superfluid (SF), and coherent states.}
\end{table}

The MI state represents a simple product of local Fock states at
each site with precisely $n_i$ atoms at a site $i$. As a
consequence, atom numbers at each site $\hat{n}_i$ (as well as the
number of atoms at $K$ sites $\hat{N}_K$) do not fluctuate, and
there is no quantum correlations between sites.

Similarly to the pair correlations, all two-, three-, and four-site
quantities in Eq.~(\ref{1-19}) factorize. From the light--scattering
point of view, this is the most classical atomic state, which
corresponds to periodically ordered pointlike atoms. We will further
consider the commensurate filling with $n_i=N/M$ atoms at each site,
neglecting possible random vacancies. This can be made if one has
some additional information that quantum fluctuations dominate over
other, thermal or technical, sources of noise.

The SF state corresponds to a BEC where each atom is in the zero
quasi-momentum Bloch--state of the lowest band and is equally
delocalized over all sites. Hence, the atom numbers at a given site
(and the number of atoms at $K<M$ sites) fluctuate. As a consequence
of the total atom number conservation, the numbers of particles at
two different sites $a\ne b$ are anticorrelated. All two-, three-,
and four-site quantities in Eq.~(\ref{1-19}) also do not factorize.

The expectation values in the SF state can be calculated using
normal ordering and the following relations:
\begin{subequations}
\begin{gather}
b_i|\Psi_\text{SF}(N,M)\rangle=\sqrt{\frac{N}{M}}|\Psi_\text{SF}(N-1,M)\rangle,
\nonumber\\
\langle\Psi_\text{SF}|b_i^{\dag m}b_i^m|\Psi_\text{SF}\rangle =
\frac{N(N-1)...(N-m+1)}{M^m},\nonumber
\end{gather}
\end{subequations}
where the first equation relates SFs with $N$ and $N-1$ atoms.

We will introduce another, coherent, quantum state, which is often
considered as an approximation to the SF state, and represents a
product of local coherent states at each site. In this approximate
state, the numbers of particles at a given site and at any $K\le M$
sites fluctuate. Moreover, the total number of particles at $M$
sites is also a fluctuating quantity, which is a disadvantage of
this approximation. Similarly to the MI state, correlations between
several different sites are absent. In the coherent state, one has
\begin{subequations}
\begin{gather}
b_i|\Psi_\text{Coh}(N,M)\rangle=\sqrt{\frac{N}{M}}|\Psi_\text{Coh}(N,M)\rangle,
\nonumber\\
\langle\Psi_\text{Coh}|b_i^{\dag m}b_i^m|\Psi_\text{Coh}\rangle =
\frac{N^m}{M^m}.\nonumber
\end{gather}
\end{subequations}

Comparing properties of the SF and coherent states in
Table~\ref{1-table1}, we can state that under the approximation
$N,M\rightarrow \infty$, but finite $N/M$, the coherent state is a
good approximation for local one-site quantities and correlations
between different sites. Moreover, if $K\ll M$, the SF expectation
values related to the nonlocal $\hat{N}_K$ operator are also well
approximated by corresponding quantities in the coherent state.
Nevertheless, if the number of sites $K$ is of the order of $M$, the
coherent-state approximation fails for those quantities.

One can prove even a more general statement for the functions
$\langle \hat{D}_{10}^*\hat{D}_{10} \rangle$ (\ref{1-13}) and $\langle
\hat{D}_{10}^{*2}\hat{D}_{10}^2\rangle$ (\ref{1-19}), which determine the
intensity and statistics of light and are the most important
quantities in this chapter. If the number of sites illuminated by
light, $K$, is much smaller than the total number of lattice sites
$M$, the coherent-state is a good approximation for calculating
characteristics of scattered light in the limit $N,M\rightarrow
\infty$, but finite $N/M$. If, in opposite, the number of sites
interacting with light is of the order of the total number of sites
in the lattice, this approximation, in general, gives wrong results.
As will be shown, it fails for light scattering in the directions of
diffraction maxima. The proof of the statement is based on the
consideration of the orders of sums in Eqs.~(\ref{1-13}), (\ref{1-19}),
which contain geometrical coefficients $A_i$ and are proportional to
the powers of $K$, whereas factors containing atom fluctuations have
powers of $M$ in denominators.

Thus, light scattering from the region of a SF optical lattice with
$K\ll M$ sites is equivalent to the light scattering from the atoms
in the coherent state (in absolute values both $K$ and $M$ can be
very large). Moreover, in the directions outside diffraction maxima,
the coherent-state approximation works well even in the case where
any number of sites is illuminated.

In the following, discussing all states, we will use the notations
$n=N/M$ for the atomic ``density'' (expectation value of the
particle number at each site) and $N_K=KN/M=nK$ for the expectation
value of the particle number at $K$ sites. These two parameters
fully characterize light scattering in the MI and coherent states,
while all three parameters $N$, $M$, and $K$ are necessary to
characterize scattering in the SF phase. For definitiveness, we will
discuss a case with large values of $N$, $M$, and $K$ where
difference between odd and even number of lattice sites vanishes.
Nevertheless, note that physical problems including BECs with large
atom number loaded into lattices with small site numbers are also of
great importance \cite{CenniniPRA2005,AlbiezPRL2005,MorschRMP2006}. Results for this
case, can be obtained from expressions of this section and
Eqs.~(\ref{1-13}) and (\ref{1-19}).

We will now consider specific applications of the expressions derived in this section.


\section{Angular distributions of scattered light}

In this section we will demonstrate, how light scattering can distinguish between various quantum states of ultracold atoms (Mott insulator and superfluid states, as well as the coherent-state approximation to the superfluid state). The most interesting observation angle turns out to be the diffraction minimum, rather than the maximum as expected from the classical diffraction. In the minimum, the strong classical scattering is suppressed and the light directly reflects the quantum properties of atoms.

\subsection{Example: 1D optical lattice in a transversely probed cavity}

Before considering a general angular distribution of scattered
light, we would like to present the most striking prediction of our
model describing the difference between atomic quantum states,
observable by light scattering. Let us consider a configuration of
Fig.~\ref{1-fig1} where the probe (traveling or standing wave) is
orthogonal to the lattice ($\theta_0=0$), and the scattered light is
collected along the lattice axis ($\theta_1=\pi/2$) by a standing-
or traveling-wave cavity. This geometry coincides with the one
considered in the context of cavity cooling
\cite{DomokosPRL2004,BlackJPB2005,AsbothPRA2005}. Atoms are assumed to be trapped at each
lattice site ($d=\lambda/2$) at the field antinodes.

In this case, the operator $\hat{D}_{10}$ (\ref{1-12}) is reduced to
$\sum_{k=1}^K(-1)^{k+1}\hat{n}_k$, which, independently on an atomic
state, gives zero for the expectation value of the field amplitude
proportional to $\langle\hat{D}_{10}\rangle$ (here we assume even $K$).
This corresponds to the classical destructive interference between
atoms separated by $\lambda/2$. In contrast, the photon number in a
cavity $a_1^\dag a_1$ is proportional to
$\langle\hat{D}_{10}^*\hat{D}_{10}\rangle =(\langle\hat{n}^2\rangle - \langle
\hat{n}_a\hat{n}_b\rangle)K$ [cf. Eq.~(\ref{1-13c})], which is
determined by statistics of a particular state, and is equal to zero
for the MI state and to $N_K$ for the SF state.

Thus, atoms in a MI state scatter no photons into a cavity, while a
SF scatters number of photons proportional to the atom number:
\begin{eqnarray}
\quad \langle a_1\rangle_\text{MI}&=&\langle a_1\rangle_\text{SF}=0,
\quad \text{but} \nonumber\\
\langle a_1^\dag a_1\rangle_\text{MI}&=&0, \quad \langle a_1^\dag
a_1\rangle_\text{SF}=|C|^2N_K . \nonumber
\end{eqnarray}

Hence, already the mean photon number provides information about a
quantum state of ultracold atoms.

The photon number fluctuations $(\Delta n_\Phi)^2$ (\ref{1-18})
are also different for various states. In the MI state, the variance
$(\Delta |D_{10}|^2)^2=\langle \hat{D}_{10}^{*2}\hat{D}_{10}^2\rangle -\langle
\hat{D}_{10}^*\hat{D}_{10}\rangle^2$ is zero, $(\Delta |D_{10}|^2)^2_\text{MI}=0$,
whereas in the SF state, Eq.~(\ref{1-19}) gives a very strong noise
$(\Delta |D_{10}|^2)^2_\text{SF}=2N_K^2$ (in highest order of $N_K$).

Nonlinear light--matter dynamics in a cavity can lead to a new
self-organized phase \cite{DomokosPRL2002,BlackJPB2005} where all
atoms occupy only each second site leading to doubling of the
lattice period, $d=\lambda$. The operator $\hat{D}_{10}$ (\ref{1-12}) is
then reduced to $\sum_{k=1}^K\hat{n}_k=\hat{N}_K$. Thus, if the
final self-organized state is a MI with $d=\lambda$, the photon
number in a cavity is $\langle a_1^\dag
a_1\rangle_\text{Self-org}=|C|^2 N_K^2$, which is proportional to
the atom number squared and has a superradiant character. This
result coincides with the theory of self-organization with classical
center-of-mass motion \cite{DomokosPRL2002}.

In the following we will compare light scattering from atoms in the
following states: MI, SF with all sites illuminated ($K=M$ using the
notation SF$_M$), and partially illuminated SF under the
approximation $N,M\rightarrow \infty$, finite $n=N/M$, $K\ll M$,
which will be denoted as the ``coherent''. The results for the SF$_K$ state with
any $K$ can be obtained from the general Eqs.~(\ref{1-13}) and
(\ref{1-19}). We will restrict ourselves to the case of plane waves.
Distinguishing between atomic states using light modes with more
complicated spatial profiles can be analyzed by the general expressions.

\subsection{Two traveling waves and discussion of essential physics}

For two traveling waves, which can be free-space modes or fixed by
ring cavities, the geometrical coefficients (\ref{1-12}) are
$A_m=\exp(im \alpha_-)$ ($\alpha_-=k_{0x}d\sin\theta_0 -
k_{1x}d\sin\theta_1$), and Eq.~(\ref{1-13e}) for the noise quantity is
reduced to
\begin{eqnarray}\label{1-25}
R=\langle \delta\hat{n}_a\delta\hat{n}_b\rangle
\frac{\sin^2{(K\alpha_-/2})}{\sin^2{(\alpha_-/2})}
+(\langle\delta\hat{n}^2\rangle - \langle
\delta\hat{n}_a\delta\hat{n}_b\rangle)K,
\end{eqnarray}
where the first term has the angle dependence of classical
diffraction (\ref{1-11PRA}), and the angle dependence in the second term
in Eq.~(\ref{1-13e}) is reduced to a constant (isotropic) one, $K$. In
the MI and coherent states, where pair correlations $\langle
\delta\hat{n}_a\delta\hat{n}_b\rangle$ are absent, the first term is
zero. In the MI state, on-site density fluctuations
$\langle\delta\hat{n}^2\rangle$ are also zero giving the zero value
of the noise quantity (\ref{1-25}), while in the coherent state, it is
the on-site fluctuations $\langle\delta\hat{n}^2\rangle=n$ that give
isotropic contribution to $R$. Thus, we have
\begin{subequations}\label{1-26}
\begin{gather}
R_\text{MI}=0, \label{1-26a}\\
R_\text{Coh}=nK=N_K, \label{1-26b}\\
R_{\text{SF}_K}= -\frac{N}{M^2}
\frac{\sin^2{(K\alpha_-/2})}{\sin^2{(\alpha_-/2})}
+\frac{N}{M}K.\label{1-26c}
\end{gather}
\end{subequations}

It is important to note, that in the SF state (\ref{1-26c}), even in a
large optical lattice with $N,M\rightarrow \infty$, very small pair
correlations $\langle \delta\hat{n}_a\delta\hat{n}_b\rangle=-N/M^2$
can give a significant angle-dependent contribution to the noise
quantity, which occurs near a diffraction maximum ($\alpha_-=2\pi l,
l=0,1,..$), where the geometrical factor is equal to $K^2$, and if
the number of the illuminated sites $K$ is of the order of $M$. This
demonstrates the importance of nonlocal correlations and invalidity
of the coherent-state approximation under those conditions. Outside
the diffraction maximum, where the geometrical factor is small, pair
correlations do not play any role and the coherent-state
approximation works well even for all sites illuminated.

\begin{figure}[h]
\centering
\captionsetup{justification=justified}
\includegraphics[width=0.7\textwidth]{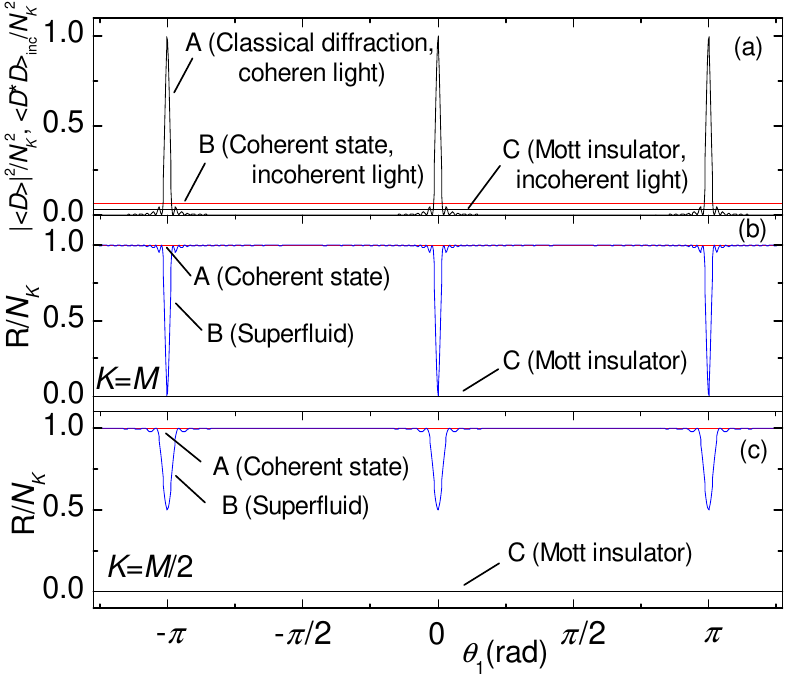}
\caption{\label{1-fig2PRA} Intensity angular distributions
for two traveling waves, the probe is transverse to the lattice
($\theta_0=0$). For a more illustrative representations of several curves see Fig. \ref{1-FigA}. (a) Intensity of classical diffraction of coherent
light (curve A), isotropic intensity of incoherent light scattering,
Eq.~(\ref{1-14}), for coherent atomic state (line B) and MI state
(line C); (b) noise quantity, Eq.~(\ref{1-26}), for coherent atomic
state (constant value 1, line A), SF with all sites illuminated
$K=M$ (curve B), and MI (constant value 0, line C); (c) the same as
in (b) but for partially illuminated SF with $K=M/2$. $N=M=30$.}
\end{figure}

\begin{figure}[h]
\centering
\captionsetup{justification=justified}
\includegraphics[clip, trim=2cm 11cm 1cm 2cm, width=0.7\textwidth]{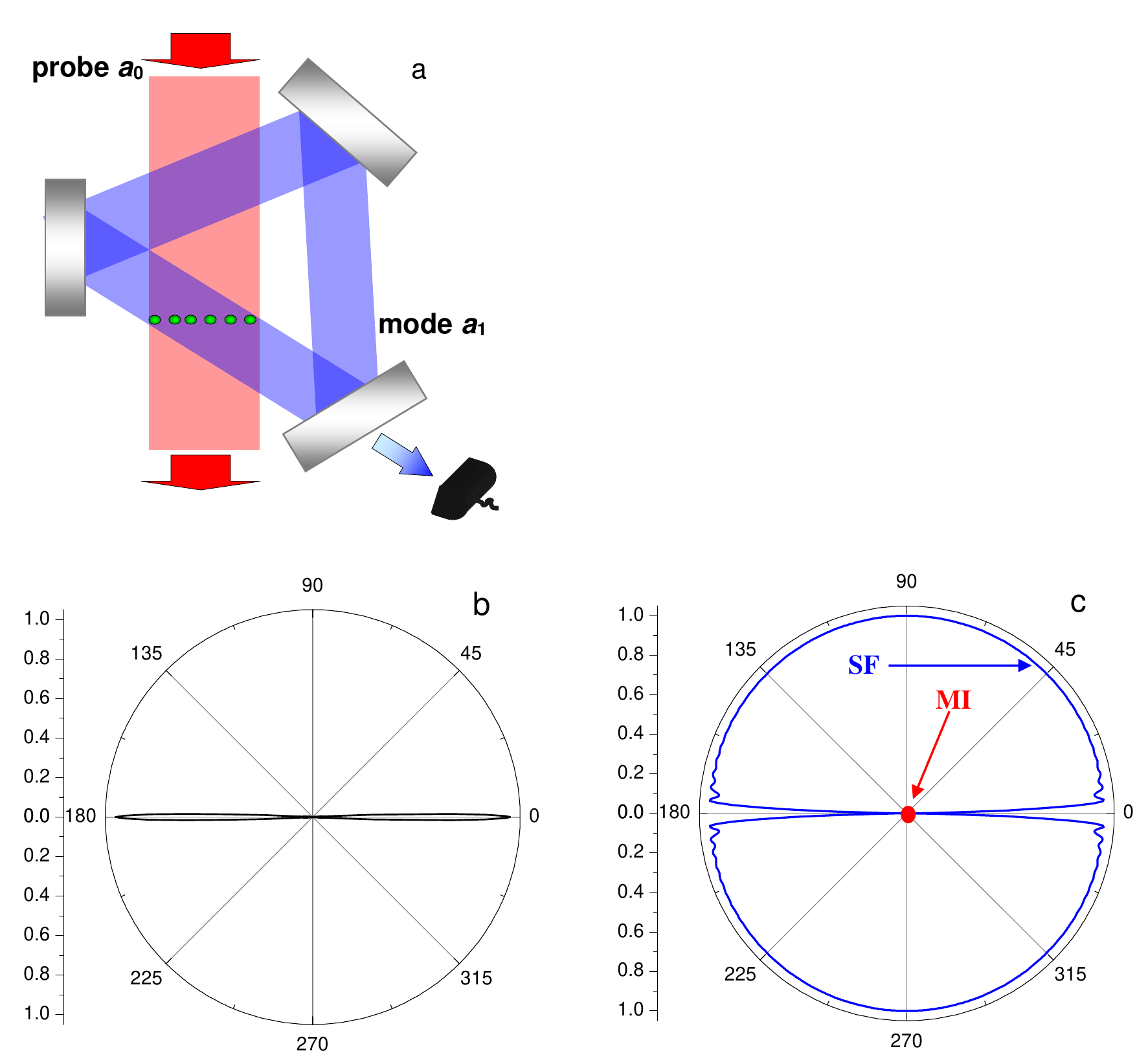}
\caption{\label{1-FigA} Angular distribution of scattered light. A more illustrative representation of some curves from Fig. \ref{1-fig2PRA}. (a) Two traveling waves are used for probing and measurement. (b) Intensity of classical scattering shows usual diffraction peaks. (c) Quantum addition to classical scattering $R(\theta_1)$. While the Mott insulator (MI) state shows no (zero) addition, the quantum addition for scattering from the superfluid (SF) state shows the anisotropic signal proportional to the atom number $N$. $N=M=K=30$, lattice period is $d=\lambda/2$, the probe angle is $\theta_0=0$.} 
\end{figure}

Figure~\ref{1-fig2PRA} shows several angle dependences of the scattered
light in the case of two traveling waves. For illustrative purposes we repeat some of the curves in Fig.~\ref{1-FigA} presented in the polar plot. As an example, in all
figures, we will consider atoms at each lattice sites providing
$d=\lambda_{0,1}/2$. In Fig.~\ref{1-fig2PRA}(a), the angular distribution
of classical diffraction $|\langle D\rangle|^2$ (curve A) is shown.
In the case of $d=\lambda_{0,1}/2$ and the probe being orthogonal to
the lattice ($\theta_0=0$), only the zero-order diffraction maxima
at $\theta_1=0, \pi$ are possible in the classical picture.
Corresponding noise quantities $R$ for the coherent (constant lines
A) and SF$_K$ (curves B) states are displayed in Figs.~\ref{1-fig2PRA}(b)
and \ref{1-fig2PRA}(c) (in MI, the noise is zero, which is displayed by
lines C). According to Eq.~(\ref{1-26}), the intensity fluctuations
are isotropic for the coherent atomic state, while there is
suppression of intensity noise under scattering from the SF. The
suppression occurs in the regions of diffraction maxima. For all
sites illuminated, $K=M$ [cf. Fig.~\ref{1-fig2PRA}(b)], the suppression
is total, while for $K=M/2$ it is only partial [cf.
Fig.~\ref{1-fig2PRA}(c)]. Outside the maxima, the dependence for SF$_K$
is well approximated by that for the coherent state for any $K$.

It is important to underline, that in a broad range of angles, the
number of scattered photons from the SF (or coherent) state is
nonzero, even if the expectation value of the electromagnetic field
vanishes, which manifests the appearance of nonclassical
entanglement between the light and manybody atomic system. Moreover,
in contrast to MI state, atoms in SF state scatter photons at
angles, where the classical diffraction does not exist.

For example, in a simple configuration considered before, where
the probe is orthogonal to the lattices ($\theta_0=0$), and the
scattered light is collected by a cavity along the lattice axis
($\theta_1=\pi/2$), the atoms in the MI state scatter no photons as
in classical diffraction minimum. In contrast, atoms in the SF$_K$
state scatter the number of photons $a_1^\dag a_1=|C|^2\langle
\hat{D}_{10}^*\hat{D}_{10}\rangle=|C|^2 N_K$, proportional to the number of
the atoms illuminated [cf. Eq.~(\ref{1-26}) and Fig.~\ref{1-fig2PRA}(b) at
the angle $\theta_1=\pi/2$].

For two traveling waves, the expression for $\hat{D}_{10}$ (\ref{1-12}), in
a diffraction maximum where all atoms radiate in phase with each
other and $\alpha_-=2\pi l$, is reduced to the operator $\hat{N}_K$.
Thus, the quantity $\langle \hat{D}_{10}\rangle=N_K=nK$ is the
expectation value of the atom number at $K$ sites and proportional
to the average atom number at a single site. The intensity of the
light scattered into a diffraction maximum is determined by
$\langle\hat{D}_{10}^*\hat{D}_{10} \rangle=\langle N_K^2\rangle$, while noise
$R = (\Delta N_K)^2$ gives the atom number variance at $K$ sites.
The latter statement corresponds to Figs.~\ref{1-fig2PRA}(b) and
\ref{1-fig2PRA}(c) displaying the total noise suppression in SF$_M$
state, where the total atom number at all sites $K=M$ does not
fluctuate, while for $K<M$, $N_K$ is a fluctuating quantity and the
noise suppression is only partial.

At the angle of a classical diffraction ``minimum'' (for $K\gg 1$
this is approximately valid for any angle outside narrow regions of
maxima), the expectation value of the field amplitude is zero, as
well as the first terms in Eqs.~(\ref{1-13c}), (\ref{1-13d}),
(\ref{1-13e}), and both the intensity $\langle\hat{D}_{10}^*\hat{D}_{10}
\rangle$ and noise $R$ are proportional to the quantity
$\langle\hat{n}^2\rangle - \langle \hat{n}_a\hat{n}_b\rangle$ giving
the difference between local and nonlocal fluctuations. For two
traveling waves, the coefficient of proportionality is isotropic and
equal to $K$ [cf. Eq.~(\ref{1-25})].

For scattering of incoherent light (\ref{1-14}), the intensity is
proportional to the local quantity $\langle \hat{n}^2\rangle$ and is
shown in Fig.~\ref{1-fig2PRA}(a) for MI (curve C) and coherent, almost
the same as in SF, (curve B) states. This quantity can be also
obtained under coherent scattering of two traveling waves, if one
tunes the angles such that the geometrical factor of the first term
in Eq.~(\ref{1-25}) is equal to $K$. Practically, this variant is easy
to achieve only for a diffraction pattern with diffraction maxima,
which are not too narrow.

Hence, in an optical experiment, both global statistical quantities
related to $K\le M$ sites, local quantities reflecting statistics at
a single site, and pair correlations can be obtained. It is
important, that local statistics can be determined by global
measurements, i.e., an optical access to a single site is not
necessary.

Therefore, light scattering gives a possibility to distinguish
different quantum states of ultracold atoms. As demonstrated by
Eq.~(\ref{1-26}) and Fig.~\ref{1-fig2PRA}, MI and SF$_M$ states are
distinguishable in diffraction ``minima'' and in incoherent light,
while they are indistinguishable (for traveling waves) in maxima,
because the total atom number contributing to the maximum does not
fluctuate. The SF$_M$ and coherent states can be distinguished in
diffraction maxima only. The MI and coherent states can be
distinguished in any angle of the scattering pattern.

Measurements of the noise quantity discussed or, alternatively,
related quantities for quadratures (\ref{1-16}) or photon number
variance (\ref{1-18}), give the values, which are different in orders
of the emitter number $N_K$ for different quantum states.
Nevertheless, for large $N_K$, there could be practical problems in
the subtraction of large values in a diffraction maximum to get the
noise contribution. In some papers, a similar problem even led to a
conclusion about state indistinguishability by intensity
measurements in BEC \cite{IdziaszekPRA2000,CiracPRL1994,CiracPRA1994} and, hence,
to a necessity to measure photon statistics. A rather involved
method to suppress the strong classical part of scattering using a
dark-state resonance in BEC was proposed in Ref.~\cite{MustPRA2000x2}.
In contrast to homogeneous ensembles, here, for optical lattices, this
problem has a natural solution: measurements outside diffraction
maxima are free of the strong classical-like part and thus directly
reflect density fluctuations.

\subsection{Physical interpretation and role of the entanglement
between light and matter}

The classical analogy of the difference in light scattering from
different atomic states consists in various density fluctuations in
different states. In particular, classical density fluctuations
would also lead to impossibility of obtaining a perfect diffraction
minimum, where contributions from all sites should precisely cancel
each other.

Scattering at diffraction maxima can be treated as superradiant one,
since the intensity of the scattered light is proportional to the
number of phase-synchronized emitters squared $N_K^2$. In
diffraction minima, destructive interference leads to the total
(subradiant) suppression of coherent radiation for MI state; whereas
for SF$_K$ state, the intensity is nonzero and proportional to the
number of emitters $N_K$, which is analogous to the emission of
independent (non-phase-synchronized) atoms.

Nevertheless, the quantum treatment gives a deeper insight into the
problem.

The expression for the SF state in Table~\ref{1-table1} can be
rewritten in the following from:
\begin{eqnarray}
|\Psi_\text{SF}\rangle=\frac{1}{(\sqrt{M})^N}\sum_{q_i}
\sqrt{\frac{N!}{q_1!q_2!...q_M!}}|q_1,q_2,..q_M\rangle,\nonumber
\end{eqnarray}
where the sum is taken over all $q_i$ such that $\sum_i^M q_i=N$. It
shows that the SF state is a quantum superposition of all possible
multisite Fock states corresponding to all possible distributions of
$N$ atoms at $M$ lattice sites. Under the light--matter interaction,
the Fock states corresponding to different atomic distributions
become entangled to scattered light of different phases and
amplitudes.

For example, in a simple case of two atoms at two sites,
$|\Psi_\text{SF}\rangle =
1/2|2,0\rangle+1/\sqrt{2}|1,1\rangle+1/2|0,2\rangle$. In the simple example
configuration, where the orthogonal probe illuminates
lattice sites separated by $\lambda/2$ (diffraction minimum), the
wave function of the whole light--matter system reads
\begin{gather}
|\Psi_\text{matter-light}\rangle=1/2|2,0\rangle|\gamma\rangle+
1/\sqrt{2}|1,1\rangle|0\rangle 
+1/2|0,2\rangle|-\gamma\rangle.\nonumber
\end{gather}

Here, if we assume that the distribution $|2,0\rangle$ is entangled
to the coherent state of light $|\gamma\rangle$, the distribution
$|0,2\rangle$ will be entangled to the similar light state with the
opposite phase $|-\gamma\rangle$, and the distribution $|1,1\rangle$
will be entangled to the vacuum field $|0\rangle$, because the
fields emitted by two atoms cancel each other.

In contrast to the classical case, light fields entangled to various
atomic distributions do not interfere with each other, which is due
to the orthogonality of the Fock states, providing a sort of
which-path information. This leads to a difference from the
classical (or MI with the only Fock state $|1,1\rangle$) case and
nonzero expectation value of the photon number even in the
diffraction minimum ($|\gamma|^2/2$ in the above example). The
absence of interference gives also an insight into the similarity of
scattering from the SF state to the scattering from independent
(non-phase-synchronized) atoms, where interference is also absent.

\subsection{Standing waves}

If at least one of the modes is a standing wave, the angle
dependence of the noise becomes richer. In an experiment, this
configuration corresponds to a case where the scattered light is
collected by a standing-wave cavity, whose axis can by tuned with
respect to the lattice axis. Except for the
appearance of new classical diffraction maxima represented by the
first terms in Eqs.~(\ref{1-13c}), (\ref{1-13d}), (\ref{1-13e}), which
depend on the phase parameters $\alpha_{\pm}=k_{0x}d\sin\theta_0 \pm
k_{1x}d\sin\theta_1$, the angle dependence of the second term is
also not an isotropic one, as it was for two traveling waves. This
second, ``noise,'' term includes a sum of the geometrical
coefficients squared, which is equivalent to the effective doubling
of the lattice period (or doubling of the light frequency) and leads
to the appearance of new spatial harmonics in the light angular
distribution. Such period doubling leads to the appearance of the
peaks in the noise distribution at the angles, where classical
diffraction does not exists.

\begin{figure}[h!]
\centering
\captionsetup{justification=justified}
\includegraphics[width=0.4\textwidth]{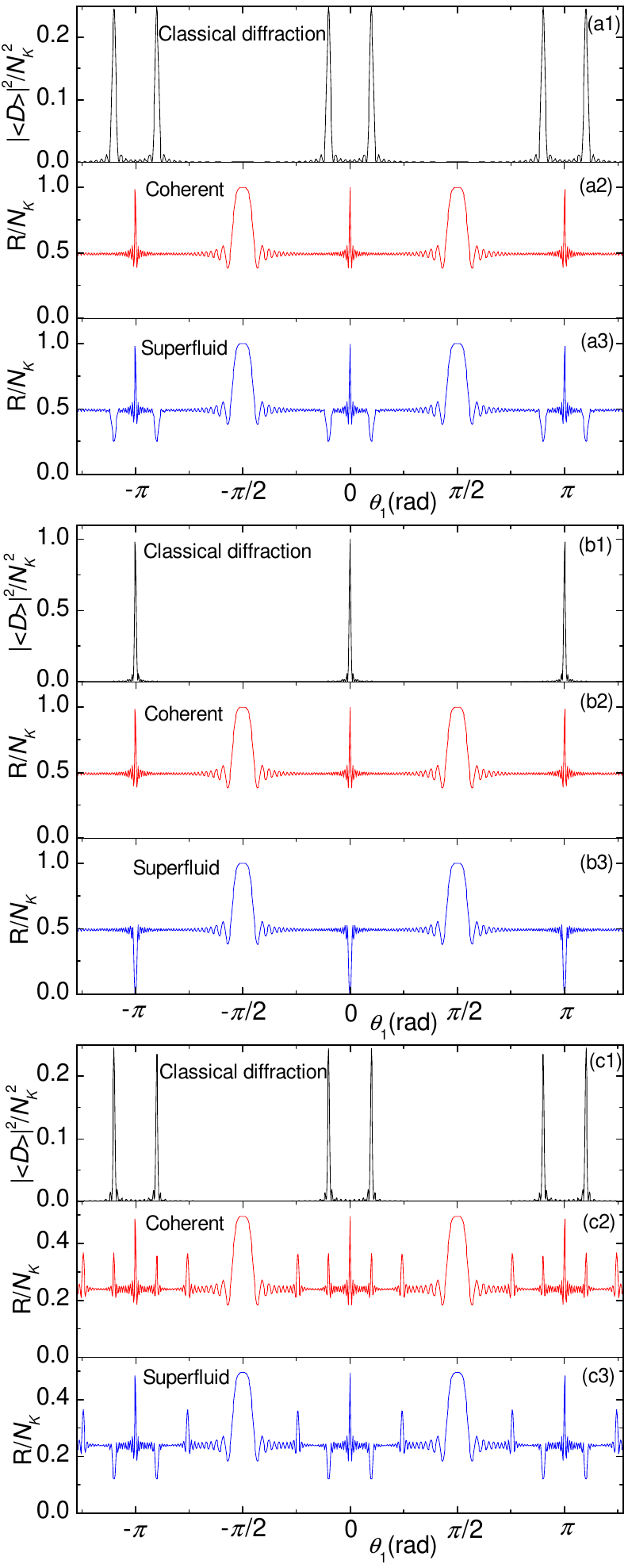}
\caption{\label{1-fig3PRA}Intensity angular distributions
for scattering into a standing-wave cavity. (a) Traveling-wave probe
at $\theta_0=0.1\pi$. For a more illustrative representation of this case as a polar plot see Fig.~\ref{1-FigB}. (b) Traveling or standing-wave probe at
$\theta_0=0$. (c) Standing-wave probe at $\theta_0=0.1\pi$. For a more illustrative representation of this case as a polar plot see Fig.~\ref{1-FigC}. Intensities of classical diffraction are shown in Figs. (a1), (b1),
and (c1); noise quantities for coherent state are shown in Figs.
(a2), (b2), and (c2) and for SF in Figs. (a3), (b3), and (c3).
$N=M=K=30$.}
\end{figure}

\begin{figure}[h!]
\centering
\captionsetup{justification=justified}
\includegraphics[clip, trim=2cm 11cm 1cm 2cm, width=0.7\textwidth]{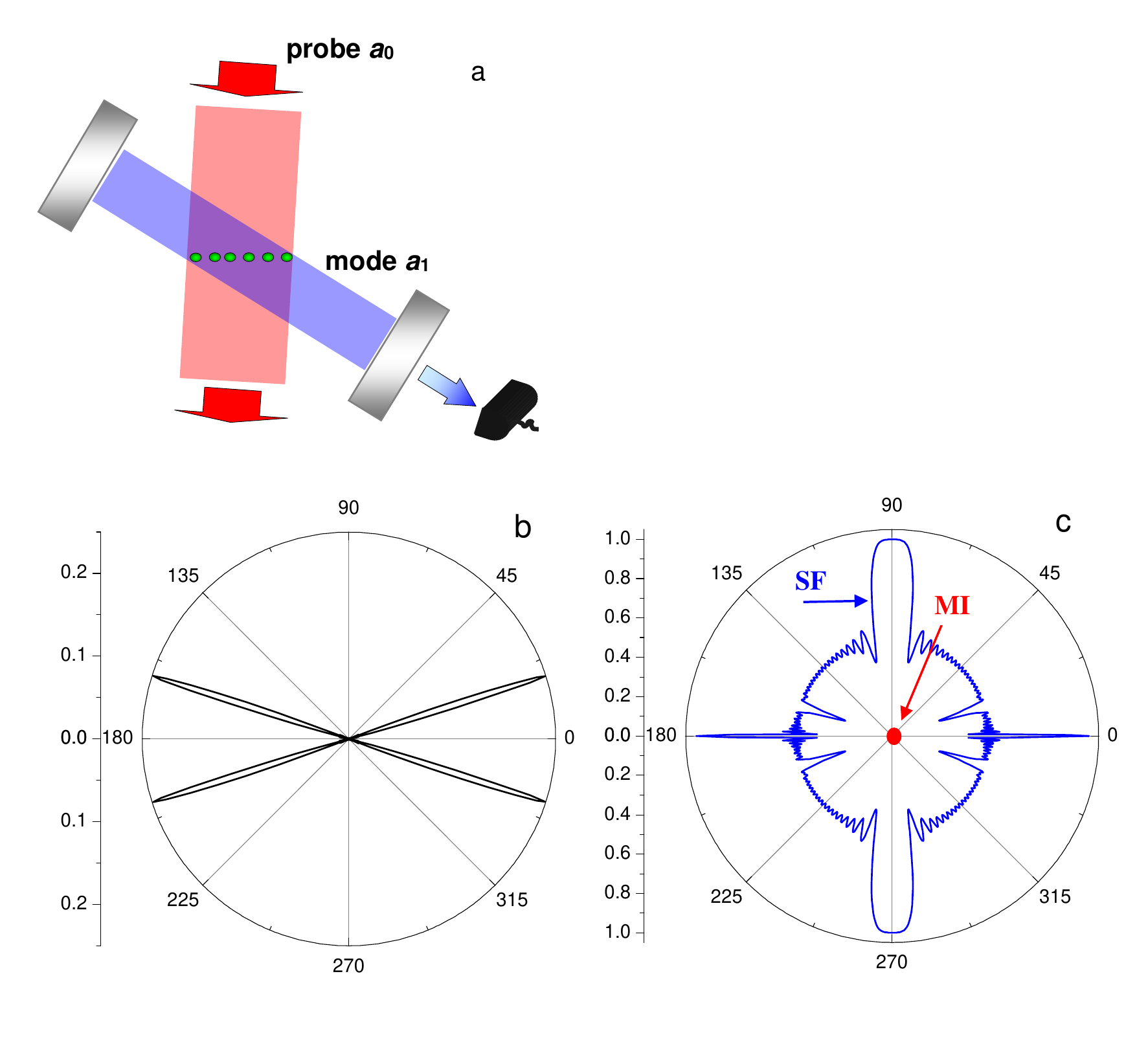}
\caption{\label{1-FigB} Angular distribution of scattered light. A more illustrative representation of some curves from Fig. \ref{1-fig3PRA}(a). (a) Travelling-wave probe scatters light into a standing-wave cavity. (b) Intensity of classical scattering shows usual diffraction peaks. (c) Quantum addition to classical scattering $R(\theta_1)$. While the Mott insulator (MI) state shows no (zero) addition, the quantum addition for scattering from the superfluid (SF) state shows the anisotropic signal proportional to the atom number $N$. In contrast to the traveling wave configuration (Fig. \ref{1-FigA}), the angular distribution of quantum scattering is much richer: the features appear even at the angles, where the classical diffraction does not exist. $N=M=K=30$, lattice period is $d=\lambda/2$, probe angle is $\theta_0=0.1\pi$.}
\end{figure}

In Fig.~\ref{1-fig3PRA}(a), angular distributions of the scattered light
are shown for a traveling-wave probe, which is almost orthogonal to a
lattice ($\theta_0=0.1\pi$), while the scattering is in a standing wave. For illustrative purposes we repeat some of the curves in Fig.~\ref{1-FigB} presented in the polar plot.
Classical diffraction pattern [cf. Fig.~\ref{1-fig3PRA}(a1)] is
determined by $|A|^2$ through the parameters $\alpha_\pm$ and shows
zero-order diffraction maxima in transmission ($\theta_1=\theta_0$
and its counterpart due to the presence of the standing-wave cavity
at $\theta_1=\pi+\theta_0$) and reflection ($\theta_1=\pi-\theta_0$
and the counterpart at $-\theta_0$). The intensity noise for atoms
in the coherent state [cf. Fig.~\ref{1-fig3PRA}(a2)] is determined by
$\sum_{i=1}^K{|A_i|^2}$ through another parameter
$2\alpha_1=2k_{1x}d\sin\theta_1$ and has different characteristic
features at $\theta_1=0,\pi$, and $\pm \pi/2$. It is the latter
feature that corresponds to the effective frequency doubling and
appears at an angle, where classical diffraction has a minimum. In
the case of SF$_M$ state [cf. Fig.~\ref{1-fig3PRA}(a3)], pair
correlations in Eqs.~(\ref{1-13d}) and (\ref{1-13e}) are nonzero, hence,
both geometrical factors contribute to the noise distribution, which
has the features at angles characteristic to both classical
scattering and the light noise of the coherent-state case. Outside
the characteristic features, the noise distribution is isotropic and
takes a nonzero value similar to the case of two traveling waves
[cf. Fig.~\ref{1-fig2PRA}]. Figure~\ref{1-fig3PRA}(b) shows a simpler
situation, where the probe is precisely orthogonal to the lattices
($\theta_0=0$).

\begin{figure}[h!]
\centering
\captionsetup{justification=justified}
\includegraphics[clip, trim=2cm 11cm 1cm 2cm, width=0.7\textwidth]{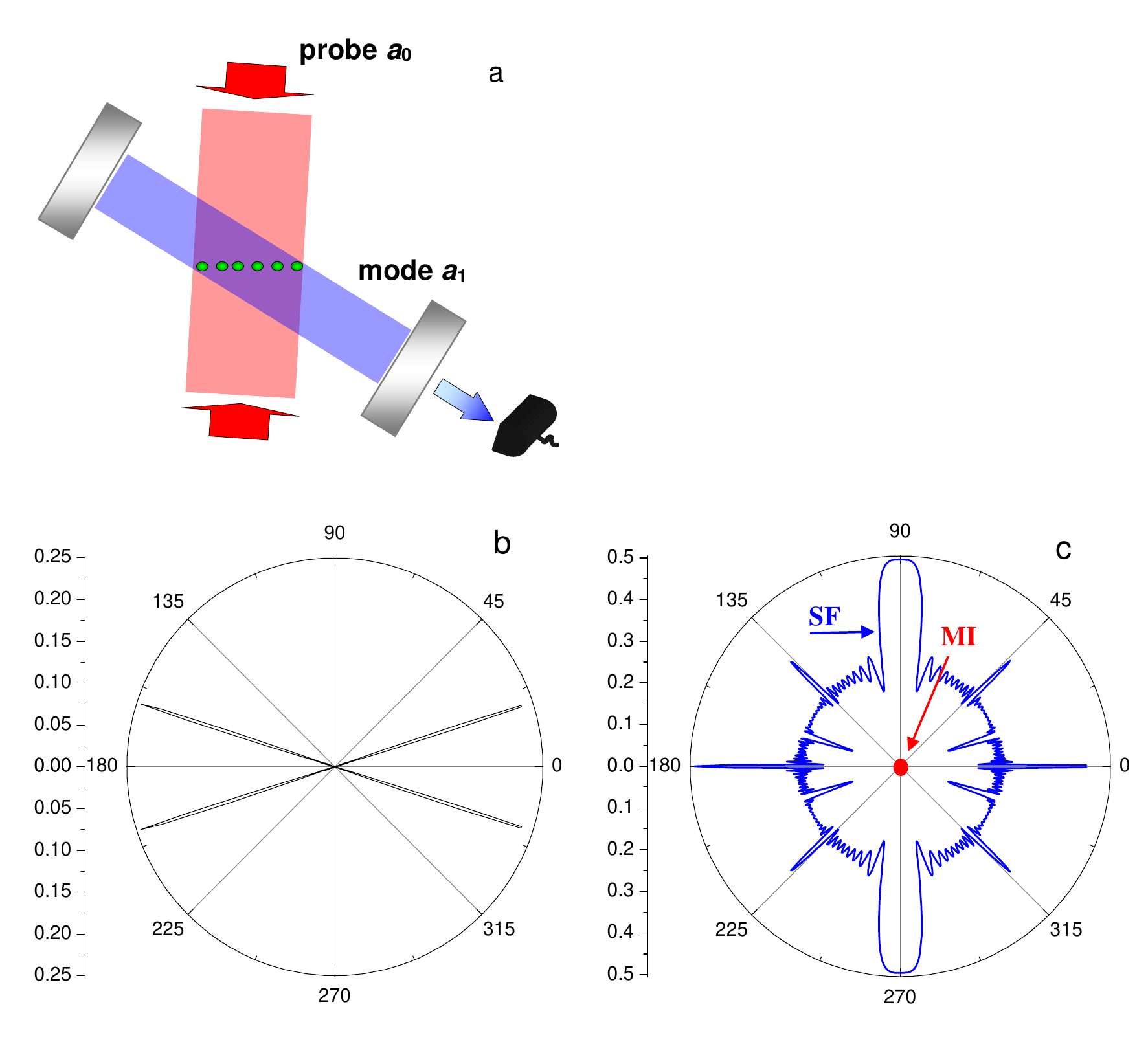}
\caption{\label{1-FigC} Angular distribution of scattered light. A more illustrative representation of some curves from Fig. \ref{1-fig3PRA}(c). (a) Standing-wave probe scatters light into a standing-wave cavity. (b) Intensity of classical scattering shows usual diffraction peaks. (c) Quantum addition to classical scattering $R(\theta_1)$. While the Mott insulator (MI) state shows no (zero) addition, the quantum addition for scattering from the superfluid (SF) state shows the anisotropic signal proportional to the atom number $N$. In contrast to the traveling wave configuration (Fig. \ref{1-FigA}), the angular distribution of quantum scattering is much richer: the features appear even at the angles, where the classical diffraction does not exist. $N=M=K=30$, lattice period is $d=\lambda/2$, probe angle is $\theta_0=0.1\pi$.}
\end{figure}

In Fig.~\ref{1-fig3PRA}(c), a situation similar to Fig.~\ref{1-fig3PRA}(a) is
shown for the case where both the probe and scattered light are standing waves. For illustrative purposes we repeat some of the curves in Fig.~\ref{1-FigC} presented in the polar plot.
While classical diffraction still depends on the parameters
$\alpha_\pm$, the factor $\sum_{i=1}^K{|A_i|^2}$ determining the
intensity noise depends on four parameters
$2\alpha_{0,1}=2k_{0,1x}d\sin\theta_{0,1}$ and $2\alpha_{\pm}$.
Thus, in the light noise from a lattice in the coherent and SF
states, the features are placed at the positions of classical
zero-order diffraction maxima and the angles, which would correspond
to the classical scattering from a lattice with a doubled period
$d=\lambda$, where the appearance of first-order diffraction maxima
is possible. Similar to Fig.~\ref{1-fig3PRA}(a), features at
$\theta_1=0,\pi$, and $\pi/2$ also exist. In the case $\theta_0=0$,
the angular distribution for two standing waves is identical to that
of one standing wave shown in Fig.~\ref{1-fig3PRA}(b).

In the SF$_M$ state, there are two types of diffraction maxima. In
the first one, the noise can be completely suppressed due to the
total atom number conservation, similarly to the case of traveling
waves. This occurs, if the condition of the maximum is fulfilled for
both of two traveling waves forming a single standing wave [cf.
Fig.~\ref{1-fig3PRA}(b)]. In the second type, even for $K=M$, only
partial noise suppression is possible, since only one of the
traveling waves is in a maximum, while another one, being in a
minimum, produces the noise [cf. Figs.~\ref{1-fig3PRA}(a) and
\ref{1-fig3PRA}(c)]. In contrast to two traveling modes, in the second
type of maxima, one can distinguish between SF$_M$ and MI states,
since MI produces no noise in any direction.

\subsection{Quadratures and photon statistics}

\begin{figure}[h!]
\centering
\captionsetup{justification=justified}
\includegraphics[width=0.4\textwidth]{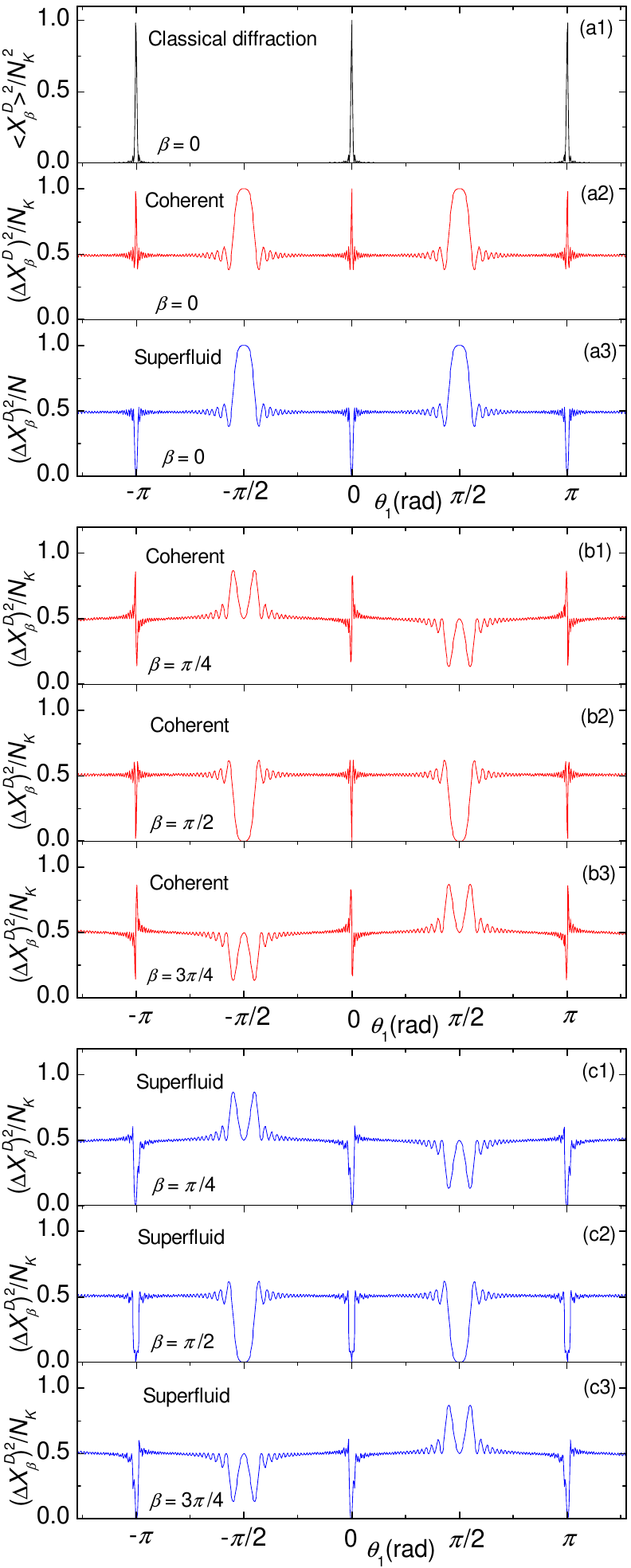}
\caption{\label{1-fig4PRA}Quadrature angular distributions
for two traveling waves. (a) Quadrature for classical diffraction
(a1), quadrature variance for coherent (a2) and SF (a3) states,
probe-homodyne phase difference $\beta=0$; (b) quadrature variance
for coherent state for $\beta=\pi/4$ (b1), $\beta=\pi/2$ (b2),
$\beta=3\pi/4$ (b3); (c) quadrature variance for SF for
$\beta=\pi/4$ (c1), $\beta=\pi/2$ (c2), $\beta=3\pi/4$ (c3).
$\theta_0=0$, $N=M=K=30$.}
\end{figure}

An analysis of the angular distribution of the quadrature variance
$(\Delta X_\beta^D)^2$ (\ref{1-16b}) shows, that even for two
traveling waves, new peaks due to effective period doubling appear
[see Fig.~\ref{1-fig4PRA}(a)]. Additionally, the amplitude of noise
features can be varied by the phase difference between the probe and
homodyne beams $\beta$, which is shown in Figs.~\ref{1-fig4PRA}(b) for
the coherent and in Fig.~\ref{1-fig4PRA}(c) SF$_M$ states. In the
coherent state, all peaks are very sensitive to $\beta$. In the
SF$_M$ state, the noise suppression at diffraction maxima is
insensitive to variations of $\beta$, whereas other peaks are
$\beta$-dependent. The relation of $(\Delta X_\beta^D)^2$ to the
quadrature variance of the light field is given by Eq.~(\ref{1-15c}).

\begin{figure}[h!]
\centering
\captionsetup{justification=justified}
\includegraphics[width=0.7\textwidth]{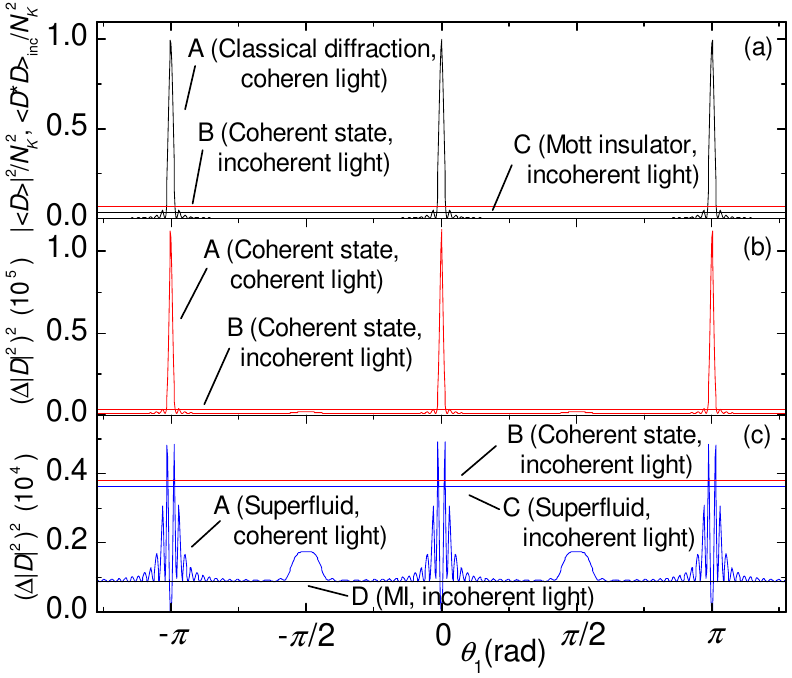}
\caption{\label{1-fig5PRA} Angular distributions of
photon-number variances for two traveling waves. (a) Intensity of
classical diffraction (curve A), isotropic intensity of incoherent
light scattering, Eq.~(\ref{1-14}), for coherent atomic state (line B)
and MI state (line C); (b) normal ordered photon-number variance for
coherent atomic state under scattering of coherent, Eq.~(\ref{1-27})
(curve A), and incoherent (line B) light; (c) normal ordered
photon-number variance for SF state under scattering of coherent,
Eq.~(\ref{1-27}) (curve A), and incoherent (line C) light, variance
for coherent (line B) and MI (curve D) states under scattering of
incoherent light. Normal ordered photon-number variance for MI state
under scattering of coherent light is zero for all angles.
$\theta_0=0$, $N=M=K=30$.}
\end{figure}

The angle dependence of the variance $(\Delta |D_{10}|^2)^2=\langle
\hat{D}_{10}^{*2}\hat{D}_{10}^2\rangle -\langle \hat{D}_{10}^*\hat{D}_{10}\rangle^2$,
which is proportional to the normal ordered photon-number variance
and determines the light statistics (\ref{1-18}), also shows
anisotropic features due to frequency doubling even for two
traveling waves (Fig.~\ref{1-fig5PRA}). In this case, Eq.~(\ref{1-19}) is
reduced to
\begin{gather}
\langle \hat{D}_{10}^{*2}\hat{D}_{10}^2\rangle =
\left(\frac{\sin(K\alpha_-/2)}{\sin(\alpha_-/2)}\right)^4 \langle
n_an_bn_cn_d\rangle \nonumber\\
+2\left(\frac{\sin(K\alpha_-/2)}{\sin(\alpha_-/2)}\right)^3
\frac{\cos(K\alpha_-/2)}{\cos(\alpha_-/2)}(\langle
n_a^2n_bn_c\rangle-\langle n_an_bn_cn_d\rangle) \nonumber
\end{gather}
\begin{gather}
-4\left(\frac{\sin(K\alpha_-/2)}{\sin(\alpha_-/2)}\right)^2
[(K-2)\langle n_an_bn_cn_d\rangle-(K-3)\langle n_a^2n_bn_c\rangle -
\langle n_a^3n_b\rangle] \nonumber\\
+ \left(\frac{\sin K\alpha_-}{\sin\alpha_-}\right)^2 (\langle
n_an_bn_cn_d\rangle-2\langle n_a^2n_bn_c\rangle + \langle
n_a^2n_b^2\rangle)+ 2K^2(\langle n_an_bn_cn_d\rangle \nonumber\\
-2\langle n_a^2n_bn_c\rangle+\langle n_a^2n_b^2\rangle) \nonumber\\
+K(-6\langle n_an_bn_cn_d\rangle+12\langle n_a^2n_bn_c\rangle-
4\langle n_a^3n_b\rangle-3\langle n_a^2n_b^2\rangle+\langle
n^4\rangle),\label{1-27}
\end{gather}
where the first four terms has features at angles typical to
classical diffraction, the fourth term is also responsible for the
doubled-frequency feature, and the last two terms contribute to the
isotropic component.

For the coherent state, the light scattered into a diffraction
maximum displays a very strong noise (equal to $4N_K^3+6N_K^2+N_k$
because $\langle
\hat{D}_{10}^{*2}\hat{D}_{10}^2\rangle=N_K^4+6N_K^3+7N_K^2+N_k$ and $\langle
\hat{D}_{10}^*\hat{D}_{10}\rangle=N_K^2+N_K$), which is much stronger than the
isotropic component ($N_K^2$ in highest order of $N_K$) and the
features at $\theta_1=\pm \pi/2$ ($2N_K^2$ in highest order of
$N_K$) [Fig.~\ref{1-fig5PRA}(b)]. In SF$_M$ state, the noise at maxima
can be suppressed, while at other angles, in highest order of $N_K$,
it is equal to that of the coherent state [Fig.\ref{1-fig5PRA}(c)]. In MI
state, the variance $(\Delta |D|^2)^2$ is zero for all angles.
Conclusions about state distinguishing by measuring light statistics
are very similar to those drown from the intensity and amplitude
measurements, which we have discussed before, including
scattering of incoherent light (see Fig.~\ref{1-fig5PRA} and the
discussion of Fig.~\ref{1-fig2PRA}).

In experiments, the nontrivial angle dependence of the noise can
help in the separation of the light noise reflecting atom statistics
from technical imperfections.


\section{Generalized Bragg condition for scattering from a 3D lattice}

In this section we will demonstrate the angular distributions of scattered light in the case of a 3D lattice. As it is well-known, in 3D gratings the diffraction is possible at Bragg angles only. For quantum gases, we will show that the scattering is possible even, when the Bragg diffraction is forbidden, and derive the new generalizd  Bragg conditions for the peaks in the diffraction pattern. Moreover, we give the parameter estimation for the scattering in free space (without any cavity).

Let us summarize the expressions derived previously. The Hamiltonian (\ref{1-7}) can be rewritten in the following form:
	\begin{multline}
	\label{1-eq:H}
		H = H_\text{BH} + \sum_l \hbar \omega_l a^\dag_l a_l + \hbar\sum_{l,m} U_{lm} a^\dag_l a_m \hat{D}_{lm}, \\
		H_\text{BH} = - J^\text{cl} \sum^M_{\langle i, j \rangle}  b^\dag_i b_j + \frac{U}{2} \sum^M_i \hat{n}_i (\hat{n}_i - 1)  - \mu 				\sum^M_i \hat{n}_i.  \nonumber
	\end{multline}
Here $H_\text{BH}$ is the standard Bose--Hubbard Hamiltonian, where we added the term with the chemical potential $\mu$. We have neglected the dispersive cavity shift due to $\hat{D}_{ll}$ and the influence of the atom tunneling on light scattering, as we have done in all calculations so far.   

The stationary light amplitude $a_1$ is given by $\hat{D}_{10}$ (\ref{1-11}) (when the dispersive cavity shift is neglected). In a cavity with the decay rate $\kappa$ and probe-cavity detuning $\Delta_p$, 
\begin{equation}
		a_1 = \frac{U_{10} a_0} {\Delta_p + i \kappa} \hat{D}_{10} = C \hat{D}_{10}.
\end{equation} 
In free space, the electric field operator in the far-field point $r$ is given by a similar expression
\begin{equation}\label{freespace}
		\hat{E}_1= \frac{\omega^2_ad_A^2E_0} {8\pi\hbar\epsilon_0c^2\Delta_a r} \hat{D}_{10} = C_E\hat{D}_{10},
\end{equation}
where $d_A$ is the dipole moment and $E_0$ is probe electric field \cite{ScullyBook}.

The light quadrature operators Eq.~(\ref{1-15}) $\hat{X}_\phi = (a_1 e^{-i \phi} + a^\dag_1 e^{i \phi})/2$  can be expressed via $\hat{D}_{10}$ quadratures, $\hat{X}^D_\beta$, 
\begin{equation}
\hat{X}_\phi = |C| \hat{X}^D_\beta = |C|(\hat{D}_{10}e^{-i\beta} + \hat{D}_{10}^\dagger e^{i \beta})/2, 
\end{equation} 
where $\beta = \phi - \phi_C$, $C = |C|\exp(i\phi_C)$, and $\phi$ is the local oscillator phase. The means of amplitude and
quadrature, $\langle a_1\rangle$ and $\langle\hat{X}_\phi\rangle$, only depend on atomic mean values. In contrast, the means of light
intensity $\langle a_1^\dag a_1\rangle=|C|^2\langle\hat{D}_{10}^\dag\hat{D}_{10}\rangle$ and quadrature variance 
	\begin{equation}
		(\Delta X_ \phi)^2 = \langle \hat{X}_\phi^2 \rangle - \langle \hat{X}_\phi \rangle^2 = 1/4 + |C|^2 (\Delta X^D_\beta)^2
	\end{equation}
reflect atomic correlations and fluctuations, which is our main focus. Alternatively, one can measure the light intensity, where the ``quantum addition'' to light due atom quantum fluctuations (classical diffraction signal is subtracted), $R=\langle a^\dagger a \rangle - |\langle a \rangle|^2$, behaves similarly to $(\Delta X^D_\beta)^2$.

In a deep lattice, Eq.~(\ref{1-12}), 
	\begin{equation}
		\hat{D}_{10}=\sum_i^K u_1^*({\bf r}_i) u_0({\bf r}_i) \hat{n}_i,
	\end{equation}
which for travelling [$u_l({\bf r})=\exp(i{\bf k}_l{\bf r}+i\varphi_l)$] or standing [$u_l({\bf r})=\cos({\bf k}_l{\bf r}+\varphi_l)$] waves is just a density Fourier transform at one or several wave vectors $\pm({\bf k}_1\pm {\bf k}_0)$. The quadrature for two travelling waves is reduced to 
	\begin{equation}
		\hat{X}^D_\beta=\sum_i^K \hat{n}_i\cos[({\bf k}_1-{\bf k}_0) \cdot {\bf r_i}-\beta].
	\end{equation}
Note that different light quadratures are differently coupled to the atom distribution, hence varying local oscillator phase and detection angle, one scans the coupling from maximal to zero. An identical expression exists for $\hat{D}_{10}$ for a standing wave, where $\beta$ is replaced by $\varphi_l$, and scanning is achieved by varying the position of the wave with respect to atoms. Thus, the variance $(\Delta X^D_\beta)^2$ and quantum addition $R$ have a non-trivial angular dependence, showing more peaks than classical diffraction and the peaks can be tuned by the light-atom coupling.

\begin{figure}[h!]
\centering
\captionsetup{justification=justified}
	\begin{center}
		\includegraphics[width=0.7\linewidth]{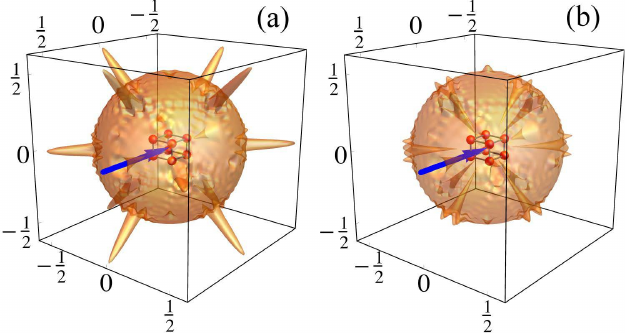}
	\end{center}
	\caption{Light intensity scattered into a standing wave mode from a SF in a 3D lattice (units of $R/N_K$). Arrows denote incoming travelling wave probes. The Bragg condition, $\Delta {\bf k} = {\bf G}$, is not fulfilled, so there is no classical diffraction, but intensity still shows multiple peaks, whose heights are tunable by simple phase shifts of the optical beams: (a) $\varphi_1=0$; (b) $\varphi_1=\pi/2$. Interestingly, there is also a significant uniform background level of scattering which does not occur in its classical counterpart. }
	\label{1-fig:Scattering}
\end{figure}

Fig. \ref{1-fig:Scattering} shows the angular dependence of $R$ for standing and travelling waves in a 3D optical lattice. The isotropic background
gives the density fluctuations [$R=K( \langle\hat{n}^2\rangle-\langle\hat{n}\rangle^2)/2$ with inter-site correlations
neglected]. The radius of the sphere changes from zero, when it is a MI with suppressed fluctuations, to half the atom number at $K$ sites, $N_K/2$, in the deep SF. There exist peaks at angles different than the classical Bragg ones and thus, can be observed without being masked by classical diffraction. Interestingly, even if 3D diffraction \cite{KetterlePRL2011} is forbidden (Fig. \ref{1-fig:Scattering}), the peaks
are still present. As $(\Delta X^D_\beta)^2$ and $R$ are squared variables, the generalized Bragg conditions for the peaks are $2
\Delta {\bf k} = {\bf G}$ for quadratures of travelling waves, where $\Delta {\bf k}={\bf k}_0 - {\bf k}_1$ and ${\bf G}$ is the reciprocal lattice vector, and $2 {\bf k}_1 = {\bf G}$ for standing wave $a_1$ and travelling $a_0$, which is clearly different from the classical
Bragg condition $\Delta {\bf k} = {\bf G}$. The peak height is tunable by the local oscillator phase or standing wave shift as seen in Fig.
\ref{1-fig:Scattering}(b).

Using Eq. (\ref{freespace}), we estimate \cite{ScullyBook} the mean photon number per second integrated over the solid angle for two experiments on light diffraction from truly ultracold bosons where the measurement object was light, and not the atoms, as it is the case in most experiments using the time-of-flight measurements: 
	\begin{equation}
		n_{\Phi}= \left(\frac{\Omega_0}{\Delta_a}\right)^2 \frac{\Gamma K}{8} (\langle\hat{n}^2\rangle-\langle\hat{n}\rangle^2),
	\end{equation}
where $\Omega_0=d_A E_0/\hbar$ and $\Gamma$ is the atomic relaxation rate.  The background signal should reach $n_\Phi \approx 10^6$ s$^{-1}$ in Ref. \cite{Weitenberg2011} (150 atoms in 2D), and $n_\Phi \approx 10^{11}$ s$^{-1}$ in Ref. \cite{KetterlePRL2011} ($10^5$ atoms in 3D). These are the examples of two setups, where the atom numbers are very different. In both cases, the predicted photon number can be detected.


\section{Mapping the quantum phase diagram: Bose glass, Mott insulator, superfluid}

In this section we will show, that light scattering can distinguish all phases in the Mott insulator -- superfluid -- Bose glass phase transition and thus map the full phase diagram of this scenario.

\begin{figure}[h!]
\centering
\captionsetup{justification=justified}
	\centering
		\includegraphics[width=0.7\linewidth]{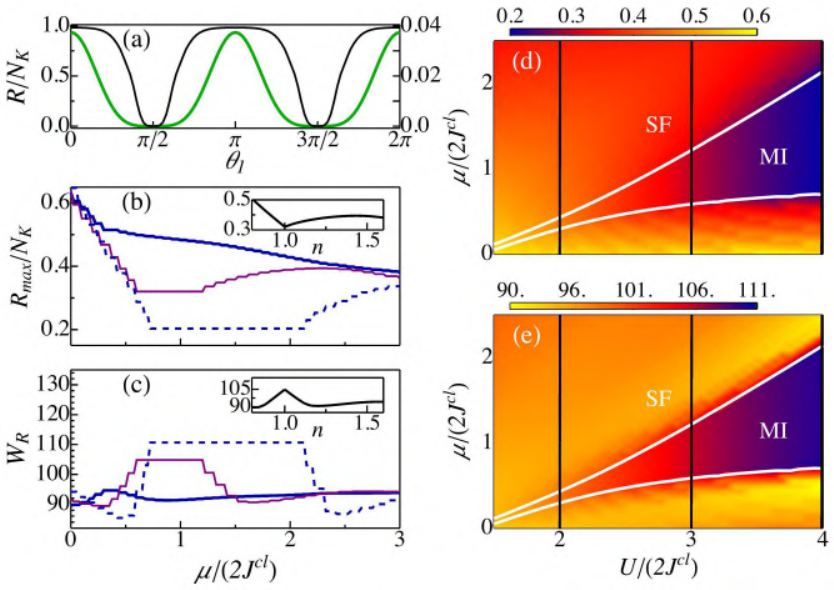}
	\caption{(a) The angular dependence of scattered light $R$ for SF (thin black, left scale, $U/2J^\text{cl} = 0$) and MI (thick green, right scale, $U/2J^\text{cl} =10$). The two phases differ in both their value of $R_\text{max}$ as well as $W_R$ showing that density correlations in the two phases differ in magnitude as well as extent. Light scattering maximum $R_\text{max}$ is shown in (b, d) and the width $W_R$ in (c, e).  It is very clear that varying chemical potential $\mu$ or density $\langle n\rangle$ sharply identifies the SF-MI transition in both quantities. (b) and (c) are cross-sections of the phase diagrams (d) and (e) at $U/2J^\text{cl}=2$ (thick blue), 3 (thin purple), and 4 (dashed blue). Insets show density dependencies for the $U/(2 J^\text{cl}) = 3$ line. $K=M=N=25$.}
	\label{1-fig:SFMI}
\end{figure}

Detecting the boundaries of quantum phases is especially important in low dimensions, where the mean-field (MF) approaches do not work. A prominent example is the Bose--Hubbard model in 1D \cite{Cazalilla2011,Ejima2011, Kuhner2000, Pino2012, Pino2013}. Observing the transition in 1D by light at fixed density was considered to be difficult \cite{DeChiaraPRA2014} or even impossible \cite{BurnettPRA2003}. By contrast, here we propose to vary the density or chemical potential, which sharply identifies the transition. We perform these calculations numerically by calculating the ground state using the Density Matrix Renormalization Group (DMRG) methods from which we can compute all the necessary atomic observables. Experiments typically use an additional harmonic confining potential on top of the OL to keep the atoms in place which  means that the chemical potential will vary in space. With careful consideration of the full ($\mu/2J^\text{cl}$, $U/2J^\text{cl}$) phase diagrams in Fig. \ref{1-fig:SFMI}(d,e) our analysis can still be applied to the system \cite{Batrouni2002}.

The 1D phase transition is best understood in terms of two-point correlations \cite{Giamarchi}. In the MI phase, the two-point correlations $\langle b_i^\dagger b_j \rangle$ and $\langle \delta \hat{n}_i \delta \hat{n}_j \rangle$ ($\delta \hat{n}_i =\hat{n}_i-\langle \hat{n}_i\rangle$) decay exponentially with $|i-j|$. On the other hand the SF will exhibit long-range order which in dimensions higher than one, manifests itself with an infinite correlation length. However, in 1D only pseudo long-range order happens and both the matter-field and density fluctuation correlations decay algebraically \cite{Giamarchi}.

The method we propose gives us direct access to the structure factor, which is a function of the two-point correlation $\langle \delta \hat{n}_i \delta \hat{n}_j \rangle$, by measuring the light intensity. For two travelling waves maximally coupled to the density, the quantum addition is given by
\begin{equation}
	R =\sum_{i, j} \exp[i (\mathbf{k}_1 - \mathbf{k}_0) (\mathbf{r}_i - \mathbf{r}_j)] \langle \delta \hat{n}_i \delta \hat{n}_j \rangle.
\end{equation}

The angular dependence of $R$ for a MI and a SF is shown in Fig. \ref{1-fig:SFMI}(a), and there are two variables distinguishing the states. Firstly, maximal $R$, $R_\text{max} \propto \sum_i \langle \delta \hat{n}_i^2 \rangle$, probes the fluctuations and compressibility $\kappa'$ ($\langle \delta \hat{n}^2_i \rangle \propto \kappa' \langle \hat{n}_i \rangle$).  The MI is incompressible and thus will have very small on-site fluctuations and it will scatter little light leading to a small $R_\text{max}$. The deeper the system is in the MI phase (i.e. that larger the $U/2J^\text{cl}$ ratio is), the smaller these values will be until ultimately it will scatter no light at all in the $U \rightarrow \infty$ limit. In Fig. \ref{1-fig:SFMI}(a) this can be seen in the value of the peak in $R$. The value $R_\text{max}$ in the SF phase ($U/2J^\text{cl} = 0$) is larger than its value in the MI phase ($U/2J^\text{cl} = 10$) by a factor of $\sim$25. Figs. \ref{1-fig:SFMI}(b,d) show how the value of $R_\text{max}$ changes across the phase transition. We see that the transition shows up very sharply as $\mu$ is varied.

Secondly, being a Fourier transform, the width $W_R$ of the dip in $R$ is a direct measure of the correlation length $l$, $W_R \propto 1/l$. The MI being an insulating phase is characterised by exponentially decaying correlations and as such it will have a very large $W_R$. However, the SF in 1D exhibits pseudo long-range order which manifests itself in algebraically decaying two-point correlations \cite{Giamarchi} which significantly narrows the dip in the $R$. This can be seen in Fig. \ref{1-fig:SFMI}(a) and we can also see that this identifies the phase transition very sharply as $\mu$ is varied in Figs. \ref{1-fig:SFMI}(c,e). One possible concern with experimentally measuring $W_R$ is that it might be obstructed by the classical diffraction maxima which appear at angles corresponding to the minima in $R$. However, the width of the diffraction maximum peak is much smaller as its width is proportional to $1/M$.

It is also possible to analyse the phase transition quantitatively using our method. Unlike in higher dimensions where an order parameter can be easily defined within the mean-field approximation there is no such quantity in 1D. However, a valid description of the relevant 1D low energy physics is provided by Luttinger liquid theory \cite{Giamarchi}. In this model correlations in the SF phase as well as the SF density itself are characterised by the Tomonaga--Luttinger parameter, $K_b$. This parameter also identifies the phase transition in the thermodynamic limit at $K_b = 1/2$. This quantity can be extracted from various correlation functions and in our case it can be extracted directly from $R$ \cite{Ejima2011}. By extracting this parameter from $R$ for various lattice lengths from numerical DMRG calculations it was even possible to give a theoretical estimate of the critical point for commensurate filling, $N = M$, in the thermodynamic limit to occur at $U/2J^\text{cl} \approx 1.64$ \cite{Ejima2011}. Our proposal provides a method to directly measure $R$ in a lab which can then be used to experimentally determine the location of the critical point in 1D.

So far both variables we considered, $R_\text{max}$ and $W_R$, provide similar information. Next, we present a case where it is very different. Bose glass (BG) is a localized insulating phase with exponentially decaying correlations but large compressibility and on-site fluctuations in a disordered OL. Therefore, measuring both $R_\text{max}$ and $W_R$ will distinguish all the phases. In a BG we have finite compressibility, but exponentially decaying correlations. This gives a large $R_\text{max}$ and a large $W_R$. A MI will also have exponentially decaying correlations since it is an insulator, but it will be incompressible. Thus, it will scatter light with a small $R_\text{max}$ and large $W_R$. Finally, a SF will have long range correlations and large compressibility which results in a large $R_\text{max}$ and a small $W_R$.

\begin{figure}[h!]
\centering
\captionsetup{justification=justified}
	\centering
		\includegraphics[width=0.7\linewidth]{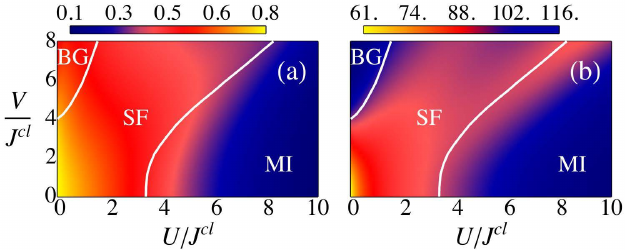}
	\caption{The MI-SF-BG phase diagrams for light scattering maximum $R_\text{max}/N_K$ (a) and width $W_R$ (b). Measurement of both quantities distinguish all three phases. Transition lines are shifted due to finite size effects \cite{Roux2008}, but it is possible to apply well known numerical methods to extract these transition lines from such experimental data extracted from $R$ \cite{Ejima2011}. $K=M=N=35$.}
	\label{1-fig:BG}
\end{figure}

We confirm this in Fig. \ref{1-fig:BG} for simulations with the ratio of superlattice- to trapping lattice-period $r\approx 0.77$ for various disorder strengths $V$ \cite{Roux2008}. Here, we only consider calculations for a fixed density, because the usual interpretation of the phase diagram  in the ($\mu/2J^\text{cl}$, $U/2J^\text{cl}$) plane for a fixed ratio $V/U$ becomes complicated due to the presence of multiple compressible and incompressible phases between successive MI lobes \cite{Roux2008}. This way, we have limited our parameter space to the three phases we are interested in: SF, MI, and BG. From Fig. \ref{1-fig:BG} we see that all three phases can indeed be distinguished. In the 1D BHM there is no sharp MI--SF phase transition in 1D at a fixed density \cite{Cazalilla2011,Ejima2011, Kuhner2000, Pino2012, Pino2013} just like in Figs. \ref{1-fig:SFMI}(d,e) if we follow the transition through the tip of the lobe which corresponds to a line of unit density. However, despite the lack of an easily distinguishable critical point it is possible to quantitatively extract the location of the transition lines by extracting the Tomonaga--Luttinger parameter from the scattered light, $R$, in the same way it was done for an unperturbed Bose--Hubbard model \cite{Ejima2011}.

Only recently \cite{Derrico2014} BG was studied by combined measurements of coherence, transport, and excitation spectra, all of which are destructive techniques. Our method is simpler as it only requires measurement of the quantity $R$ and additionally, it is nondestructive.


\section{Beyond atom densities: measurements of the matter fields in optical lattices}

In this section we make a very important step forward, beyond all the examples we have considered so far.  After we derived the general Hamiltonian (\ref{1-7}) and Heisenberg equations (\ref{1-8}), we have totally neglected the influence of atomic tunneling on light scattering. In other words, we have considered the influence of the density operators $\hat{n}_i$ on scattering, while the influence of the matter-wave amplitude operators $b_i$ was completely neglected. In this section, the measurement of matter-field operators and correlations will be our central point.

We will now show that light scattering from an ultracold gas reveals not only density correlations, but also matter-field interference at its 	shortest possible distance in an optical lattice (i.e. the lattice period), which defines key properties such as tunneling, atomic currents, and matter-field phase gradients. This signal can be enhanced by concentrating probe light between lattice sites rather than at density maxima. As addressing between two single sites is challenging, we focus on global nondestructive scattering, allowing probing order parameters, matter-field quadratures and their squeezing.

The general Hamiltonian (\ref{1-7}) can be rewritten in the following form:
	\begin{multline}
	\label{1-eq:H}
		H = H_\text{BH} + \sum_l \hbar \omega_l a^\dag_l a_l + \hbar\sum_{l,m} U_{lm} a^\dag_l a_m \hat{F}_{lm}, \\
		H_\text{BH} = - J^\text{cl} \sum^M_{\langle i, j \rangle}  b^\dag_i b_j + \frac{U}{2} \sum^M_i \hat{n}_i (\hat{n}_i - 1)  - \mu 				\sum^M_i \hat{n}_i. 
	\end{multline}
Here $H_\text{BH}$ is the standard Bose--Hubbard Hamiltonian, where we added the term with the chemical potential $\mu$. We have neglected the dispersive cavity shift due to $\hat{D}_{ll}$. In contrast to all previous examples in this work, we will now keep the contribution of the atom tunneling via the operator $\hat{F}_{lm} = \hat{D}_{lm} + \hat{B}_{lm}$:
	\begin{equation}
	\label{1-eqDB}
  		\hat{D}_{lm} = \sum_{i=1}^K J^{lm}_{i,i} \hat{n}_i,
		\,\,\,\,
  		\hat{B}_{lm} = \sum_{\langle i,j \rangle}^K J^{lm}_{i,j} b^\dag_i b_j,
	\end{equation}
which comes from overlaps of light mode functions $u_l({\bf r})$ and density operator $\hat{n}({\bf r})=\hat{\Psi}^\dag({\bf r})\hat{\Psi}({\bf r})$, after the matter-field operator is expressed via Wannier functions: $\hat{\Psi}({\bf r})=\sum_i b_i w({\bf r}-{\bf r}_i)$. $\hat{D}_{lm}$ sums the density contributions $\hat{n}_i$, while $\hat{B}_{lm}$ sums the matter-field interference terms. Let us remind that $J^{lm}_{i,j}$ are the convolutions of Wannier and light mode functions and are given by (\ref{1-6})
	\begin{equation}
	\label{1-Jcoeff}
		J^{lm}_{i,j} = \int w({\bf r} - {\bf r}_i) u^*_l({\bf r}) u_m({\bf r}) w({\bf r} - {\bf r}_j)\mathrm{d}{\bf r}. 
	\end{equation}

Typically, the dominant term in $\hat{F}_{10}$ is the density-term $\hat{D}$, rather than inter-site matter-field interference $\hat{B}$
\cite{MekhovPRA2007,MorigiPRA2010Scat,TrippenbachPRA2009, RuostekoskiPRL2009, MekhovLP2009}, because the Wannier functions' overlap is small (in this section, we drop the subscripts in $\hat{F}$, $\hat{D}$, and $\hat{B}$). Our aim is to enhance the $\hat{B}$-term in light scattering by suppressing the density signal.

\begin{figure}[h!]
\centering
\captionsetup{justification=justified}
	\begin{center}
		\includegraphics[width=0.7\linewidth]{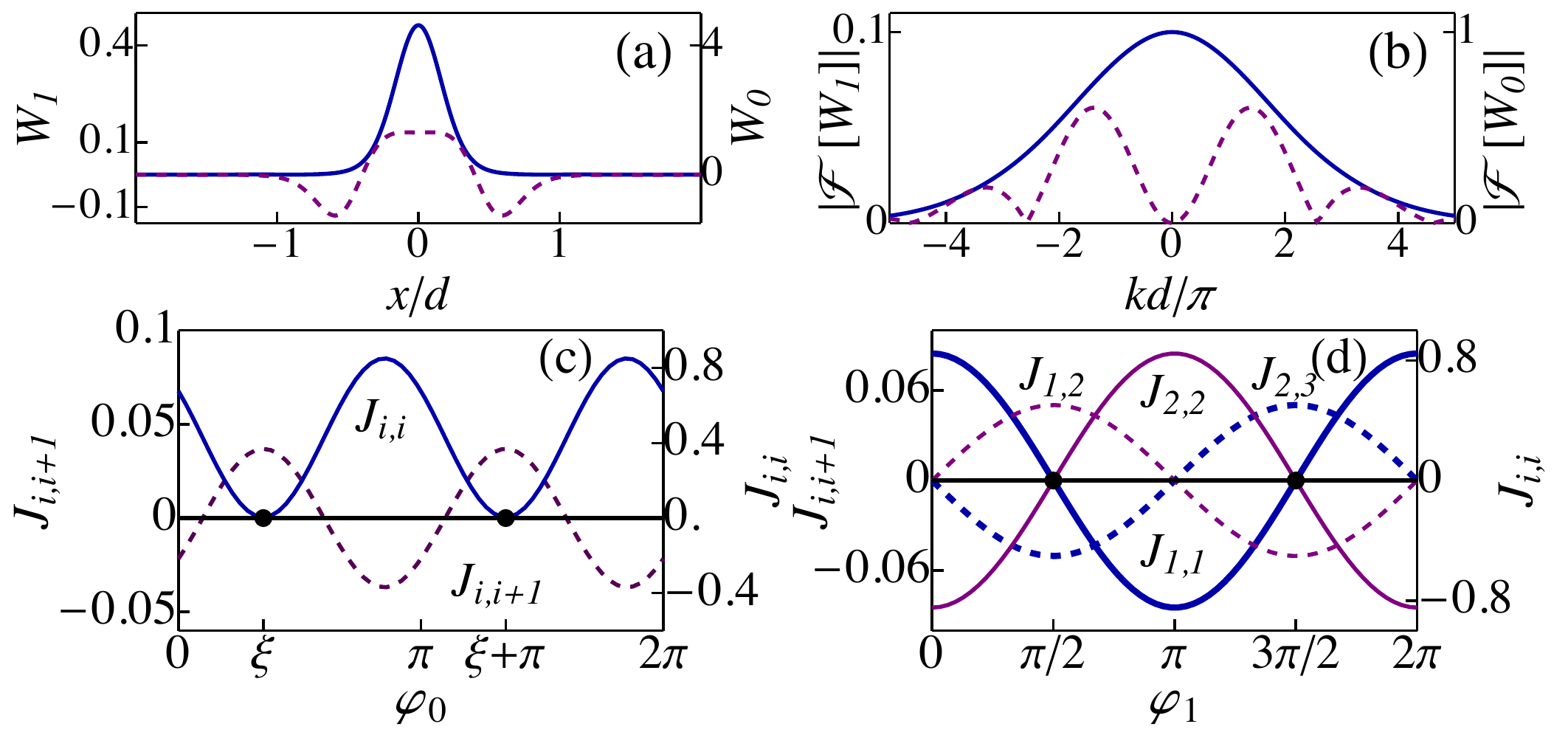}
	\end{center}
	\caption{The Wannier function products: (a) $W_0(x)$ (solid line, right axis), $W_1(x)$ (dashed line,  left axis) and their (b) Fourier transforms $\mathcal{F}[W_{0,1}]$. The Density $J_{i,i}$ and matter-interference $J_{i,i+1}$ coefficients  (\ref{1-eqDB}) in diffraction maximum (c) and minimum (d) as are shown as functions  of standing wave shifts $\varphi$ or, if one were to measure the quadrature variance $(\Delta X^F_\beta)^2$, the local oscillator phase  $\beta$. The black points indicate the positions, where light measures matter interference $\hat{B} \ne 0$, and the density-term is suppressed, $\hat{D} = 0$. The trapping potential depth is approximately 5 recoil energies.}
	\label{1-FJ}
\end{figure}

We now focus on enhancing the interference term $\hat{B}$ in the operator $\hat{F}$. For clarity we will consider a 1D lattice, but the results can be applied and generalised to higher dimensions. Central to engineering the $\hat{F}$ operator are the coefficients $J_{i,j}$ given by Eq. (\ref{1-Jcoeff}). The operators $\hat{B}$ and $\hat{D}$ depend on the values of $J_{i,i+1}$ and $J_{i,i}$ respectively. These coefficients are determined by the convolution of the light mode product, $u_1^*({\bf r})u_0({\bf r})$, with the relevant Wannier function overlap shown in Fig. \ref{1-FJ}(a). For the $\hat{B}$ operator we calculate the convolution with the nearest neighbour overlap, $W_1({\bf r}) \equiv w({\bf r} - {\bf d}/2) w({\bf r}+{\bf d}/2)$, and for the $\hat{D}$ operator we calculate the convolution with the square of the Wannier function at a single site, $W_0({\bf r}) \equiv w^2({\bf r})$. Therefore, in order to enhance the $\hat{B}$ term we need to maximise the overlap between the light modes and the nearest neighbour Wannier overlap, $W_1({\bf r})$. This can be achieved by concentrating the light between the sites rather than at atom positions. Ideally, one could measure between two sites similarly to single-site addressing \cite{GreinerNature2009,BlochNature2011}, which would measure a single term $\langle b^\dag_i b_{i+1}+b_i b^\dag_{i+1}\rangle$, e.g., by superposing a deeper optical lattice shifted by $d/2$ with respect to the original one, catching and measuring the atoms in the new lattice sites. A single-shot success rate of atom detection will be small. As single-site addressing is challenging, we proceed with the global scattering.

In order to calculate the $J_{i,j}$ coefficients we perform numerical calculations using realistic Wannier functions \cite{Walters2013}. However, it is possible to gain some analytic insight into the behaviour of these values by looking at the Fourier transforms of the Wannier function overlaps, $\mathcal{F}[W_{0,1}]({\bf k})$, shown in Fig \ref{1-FJ}(b). This is because the light mode product, $u_1^*({\bf r})u_0({\bf r})$, can be in general decomposed into a sum of oscillating exponentials of the form $e^{i {\bf k} \cdot {\bf r}}$ making the integral in Eq. (\ref{1-Jcoeff}) a sum of Fourier transforms of $W_{0,1}({\bf r})$. We consider both the detected and probe beam to be standing waves which gives the following expressions for the $\hat{D}$ and $\hat{B}$ operators
\begin{gather}
\label{1-FTs}
	\hat{D} = \frac{1}{2}[\mathcal{F}[W_0](k_-)\sum_m\hat{n}_m\cos(k_- x_m +\varphi_-)  
	+\mathcal{F}[W_0](k_+)\sum_m\hat{n}_m\cos(k_+ x_m +\varphi_+)], \nonumber\\
	\hat{B} = \frac{1}{2}[\mathcal{F}[W_1](k_-)\sum_m\hat{B}_m\cos(k_- x_m +\frac{k_-d}{2}+\varphi_-)  \nonumber\\
	+\mathcal{F}[W_1](k_+)\sum_m\hat{B}_m\cos(k_+ x_m +\frac{k_+d}{2}+\varphi_+)],
\end{gather}
where $k_\pm = k_{0x} \pm k_{1x}$, $k_{(0,1)x} = k_{0,1} \sin(\theta_{0,1})$, $\hat{B}_m=b^\dag_mb_{m+1}+b_mb^\dag_{m+1}$, and $\varphi_\pm=\varphi_0 \pm \varphi_1$. The key result is that the $\hat{B}$ operator is phase shifted by $k_\pm d/2$ with respect to the $\hat{D}$ operator since it depends on the amplitude of light in between the lattice sites and not at the positions of the atoms, allowing to decouple them at specific angles.

Firstly, we will use this result to show how one can probe $\langle \hat{B} \rangle$ which in the mean-field (MF) approximation gives information about the matter-field amplitude, $\Phi = \langle b \rangle$. The simplest case is to find a diffraction maximum where $J_{i,i+1} = J_B$. This can be achieved by crossing the light modes such that $\theta_0 = -\theta_1$ and $k_{0x} = k_{1x} = \pi/d$ and choosing the light mode phases such that $\varphi_+ = 0$. Fig. \ref{1-FJ}(c) shows the value of the $J_{i,j}$ coefficients under these circumstances. In order to make the $\hat{B}$ contribution to light scattering dominant we need to set $\hat{D} = 0$ which from Eq. (\ref{1-FTs}) we see is possible if $\varphi_0 = -\varphi_1 = \arccos[-\mathcal{F}[W_0](2\pi/d)/\mathcal{F}[W_0](0)]/2$. This arrangement of light modes maximizes the interference signal, $\hat{B}$, by suppressing the density signal, $\hat{D}$, via interference compensating for the spreading of the Wannier functions. Hence, by measuring the light quadrature we probe the kinetic energy and, in MF, the matter-field amplitude (order parameter) $\Phi$: $\langle \hat{X}^F_{\beta=0} \rangle = | \Phi |^2 \mathcal{F}[W_1](2\pi/d) (K-1)$.

\begin{figure}[h!]
\centering
\captionsetup{justification=justified}
	\begin{center}
  		\includegraphics[width=0.7\linewidth]{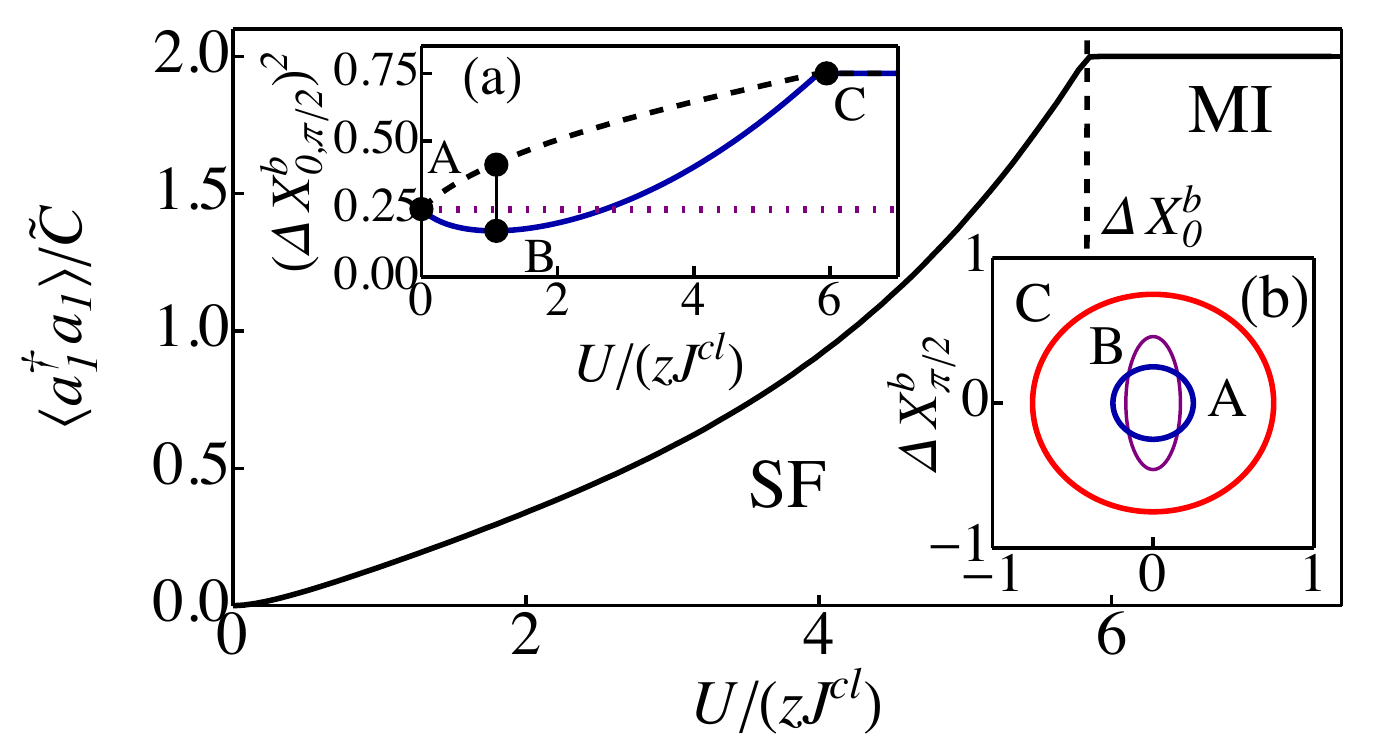}
 	\end{center}
	\caption{Photon number scattered in a diffraction minimum, given by Eq. (\ref{1-intensity}), where $\tilde{C} = 2 |C|^2 (K-1) \mathcal{F}^2 [W_1](\pi/d)$. 	More light is scattered from a MI than a SF due to the large uncertainty in phase in the insulator. (a) The variances of quadratures $\Delta X^b_0$ (solid) and $\Delta X^b_{\pi/2}$ (dashed) of the matter field across the phase transition. Level 1/4 is the minimal (Heisenberg) uncertainty. There are three important points along the phase transition: the coherent state (SF) at A, the amplitude-squeezed state at B, and the Fock state (MI) at C. (b) The uncertainties plotted in phase space.}
	\label{1-Quads}
\end{figure}

Secondly, we show that it is also possible to access the fluctuations of matter-field quadratures $\hat{X}^b_\alpha = (b e^{-i\alpha} + b^\dag e^{i\alpha})/2$, which in MF can be probed by measuring the variance of $\hat{B}$. Across the phase transition, the matter field changes its state from Fock (in MI) to coherent (deep SF) through an amplitude-squeezed state as shown in Fig. \ref{1-Quads}(a,b). We consider an arrangement where the beams are arranged such that $k_{0x} = 0$ and $k_{1x} = \pi/d$ which gives the following expressions for the density and interference terms
\begin{eqnarray}
\label{1-DMin}
	\hat{D} = \mathcal{F}[W_0](\pi/d) \sum_m (-1)^m \hat{n}_m \cos(\varphi_0) \cos(\varphi_1), \nonumber \\
	\hat{B} = -\mathcal{F}[W_1](\pi/d) \sum_m (-1)^m \hat{B}_m \cos(\varphi_0) \sin(\varphi_1).
\end{eqnarray}
The corresponding $J_{i,j}$ coefficients are shown in Fig. \ref{1-FJ}(d) for $\varphi_0=0$. It is clear that for $\varphi_1 = \pm \pi/2$, $\hat{D} = 0$, which is intuitive as this places the lattice sites at the nodes of the mode $u_1({\bf r})$. This is a diffraction minimum as the light amplitude is also zero, $\langle \hat{B} \rangle = 0$, because contributions from alternating inter-site regions interfere destructively. However, the intensity $\langle a^\dag_1 a \rangle = |C|^2 \langle \hat{B}^2 \rangle$ is proportional to the variance of $\hat{B}$ and is non-zero. Assuming $\Phi$ is real in MF:
\begin{gather}
\label{1-intensity}
	\langle a_1^\dag a_1\rangle = 2 |C|^2(K-1)\mathcal{F}^2[W_1](\frac{\pi}{d}) 
	 [ ( \langle b^2 \rangle - \Phi^2 )^2 + ( n - \Phi^2 ) ( 1 +n - \Phi^2 ) ],
\end{gather}
and it is shown as a function of $U/(zJ^\text{cl})$ in Fig. \ref{1-Quads}. Thus, since measurement in the diffraction maximum yields $\Phi^2$ we can deduce $\langle b^2 \rangle - \Phi^2$ from the intensity. This quantity is of great interest as it gives us access to the quadrature variances of the matter-field 
\begin{equation}
	(\Delta X^b_{0,\pi/2})^2 = 1/4 + [(n - \Phi^2) \pm (\langle b^2 \rangle - \Phi^2)]/2,
\end{equation}
where $n=\langle\hat{n}\rangle$ is the mean on-site atomic density.

Alternatively, one can use the arrangement for a diffraction minimum described above, but use travelling instead of standing waves for the probe and detected beams and measure the light quadrature variance. In this case $\hat{X}^F_\beta = \hat{D} \cos(\beta) + \hat{B} \sin(\beta)$, and by varying the local oscillator phase, one can choose which conjugate operator to measure. For $\beta=\pi/2$, $(\Delta X^F_{\pi/2})^2$ looks identical to Eq. (\ref{1-intensity}).

Probing $\hat{B}^2$ gives us access to kinetic energy fluctuations with 4-point correlations ($b^\dag_i b_j$ combined in pairs). Measuring the photon number variance, which is standard in quantum optics, will lead up to 8-point correlations similar to 4-point density correlations in Sec. 1.4.3. These are of significant interest, because it has been shown that there are quantum entangled states that manifest themselves only in high-order correlations.

Surprisingly, inter-site terms scatter more light from a MI than a SF Eq. (\ref{1-intensity}), as shown in Fig. (\ref{1-Quads}), although the mean inter-site density $\langle \hat{n}(\bf{r})\rangle $ is tiny in a MI. This reflects a fundamental effect of the boson interference in Fock states. It indeed happens between two sites, but as the phase is uncertain, it results in the large variance of $\hat{n}(\bf{r})$ captured by light as shown in Eq. (\ref{1-intensity}). The interference between two macroscopic BECs has been observed and studied theoretically \cite{Horak1999}. When two BECs in Fock states interfere a phase difference is established between them and an interference pattern is observed which disappears when the results are averaged over a large number of experimental realizations. This reflects the large shot-to-shot phase fluctuations corresponding to a large inter-site variance of $\hat{n}(\mathbf{r})$. By contrast, our method enables the observation of such phase uncertainty in a Fock state directly between lattice sites on the microscopic scale in-situ.

Thus, we proposed how to measure the matter-field interference by concentrating light between the sites. This corresponds to interference at the shortest possible distance in an optical lattice. By
contrast, standard destructive time-of-flight measurements deal with far-field interference and a relatively near-field one was used in
Ref. \cite{KetterlePRL2011}. This defines most processes in optical lattices. E.g. matter-field phase changes may happen not only due to external
gradients, but also due to intriguing effects such quantum jumps leading to phase flips at neighbouring sites and sudden cancellation of
tunneling \cite{VukicsNJP2007}, which should be accessible by our method. In mean-field, one can measure the matter-field amplitude (order
parameter), quadratures and squeezing. This can link atom optics to areas where quantum optics has already made progress, e.g., quantum
imaging \cite{GolubevPRA2010,KolobovRMP1999}, using an optical lattice as an array of multimode nonclassical matter-field sources with a high degree of entanglement for quantum information processing.


\section{Probing ultracold fermions}

In this section we show, how to extend the model developed before for bosons for the case of ultracold fermions.

Let us remind again that for bosons the general Hamiltonian (\ref{1-7}) can be rewritten in the following form:
	\begin{multline}
	\label{1-eq:H1}
		H = H_\text{BH} + \sum_l \hbar \omega_l a^\dag_l a_l + \hbar\sum_{l,m} U_{lm} a^\dag_l a_m \hat{F}_{lm}, \\
		H_\text{BH} = - J^\text{cl} \sum^M_{\langle i, j \rangle}  b^\dag_i b_j + \frac{U}{2} \sum^M_i \hat{n}_i (\hat{n}_i - 1). 
	\end{multline}
Now, for fermions, instead of using the Bose--Hubbard Hamiltonian $H_\text{BH}$, we will use the Hubbard Hamiltonian 

\begin{equation}
H_\text{H}= - J^\text{cl}\sum_{\sigma\in\{\uparrow,\downarrow\}}\sum_{\langle i,j\rangle}(\hat f^\dagger_{i\sigma}\hat f^{\phantom{\dagger}}_{j\sigma}+h.c.)+U\sum_i\hat n_{i\uparrow}\hat n_{i\downarrow},
\end{equation}
where $\hat b_i$ and $\hat f_{i,\sigma}$ are the bosonic and fermionic annihilation operators at site $i$ respectively,  $J^\text{cl}$ is the nearest-neighbour hopping amplitude, $U$ is the on-site atom-atom interaction energy, the bosonic (fermionic) number operator $\hat{n}_i=\hat{b}_i^\dagger \hat{b}_i$ ($\hat{n}_{i\sigma}=\hat{f}_{i\sigma}^\dagger \hat{f}_{i\sigma}$), and the sum $\langle i,j\rangle$ is taken over neighbouring pairs of sites. 

For bosons, we introduced the operator $\hat F_{lm}=\hat D_{lm}+\hat B_{lm}$ with
 
\begin{align}
\hat D_{lm}=\sum_{i=1}^M J_{ii}^{lm} \hat n_i, \qquad \hat B_{lm}=\sum_{\langle i,j\rangle }^M J_{ij}^{lm} \hat b_i^\dagger \hat b_j.
\end{align}
The operator $\hat D_{lm}$ describes the scattering from on-site density, while  $\hat B_{lm}$ that from small intersite densities. In this section, we will consider well-localized atoms and neglect the term $\hat B_{lm}$.

In the fermionic case, we introduce the light polarisation as an additional degree of freedom for the light field, which allows for probing different spin species. In particular, we use linearly polarised light $a_{1x}$ and $a_{1y}$ which couple differently to the two spin densities $\hat{n}_{i\uparrow}$ and $\hat{n}_{i\downarrow}$ because of selection rules that constrain the allowed transitions between different hyperfine states of the atoms \cite{GreinerPRL2005,LeePRA2012,Meineke2012,Sanner2010,Sanner2011,Sanner2012}. In contrast to the (spinless) bosonic case, this allows us to probe different linear combinations via the operators $\hat{D}_{lm,x} = \sum_i J_{ii}^{lm} \hat{\rho}_i$ or $\hat{D}_{lm,y} = \sum_i J_{ii}^{lm} \hat{m}_i$ where $\hat{\rho}_i = \hat{n}_{i\uparrow}+\hat{n}_{i\downarrow}$ (density) and $\hat{m}_i = \hat{n}_{i\uparrow}-\hat{n}_{i\downarrow}$ (magnetisation). In the fermionic case, we also consider attractive particle-particle interactions.

In order to emphasise the difference between classical and quantum scattering contributions we focus on two-point correlation functions and define as before the quantum addition

\begin{align}\label{1-Rdef}
R= \langle \hat{D}^\dagger \hat{D} \rangle - |\langle \hat{D}\rangle|^2 = \sum_{i,j} J_{ii}^* J_{jj} \left(\langle \hat{n}_i \hat{n}_j  \rangle - \langle \hat{n}_i \rangle \langle \hat{n}_j \rangle \right).
\end{align}

This quantity depends on the scattering angle via the coefficients $J_{ii}$ and is directly related to the fluctuations of the atom number. Therefore, quantum light scattering distinguishes between different quantum states such as the superfluid and Mott insulator states \cite{Mekhov2012}. Here we extend the bosonic model \cite{Mekhov2012,MekhovLP2009, MekhovLP2013} to describe spin-1/2 fermionic atoms and define the quantum additions $R_x$ and $R_y$, related to the fluctuations in density and magnetisation respectively.
If the mode functions of the probe and scattered light are travelling waves with wave-vectors $\b{k_\mathrm{in}}$ and $\b{k_\mathrm{out}}$ respectively, the coefficients  $J_{jj}$ are proportional to $\exp\left[ i (\b{k_\mathrm{in}}-\b{k_\mathrm{out}}) \b{r}_j \right]$, and the quantum addition $R$ is proportional to a structure factor, i.~e.~ the Fourier transform of the density-density correlations. 

\begin{figure}[h!]
\centering
\captionsetup{justification=justified}
\includegraphics[width=0.8\textwidth]{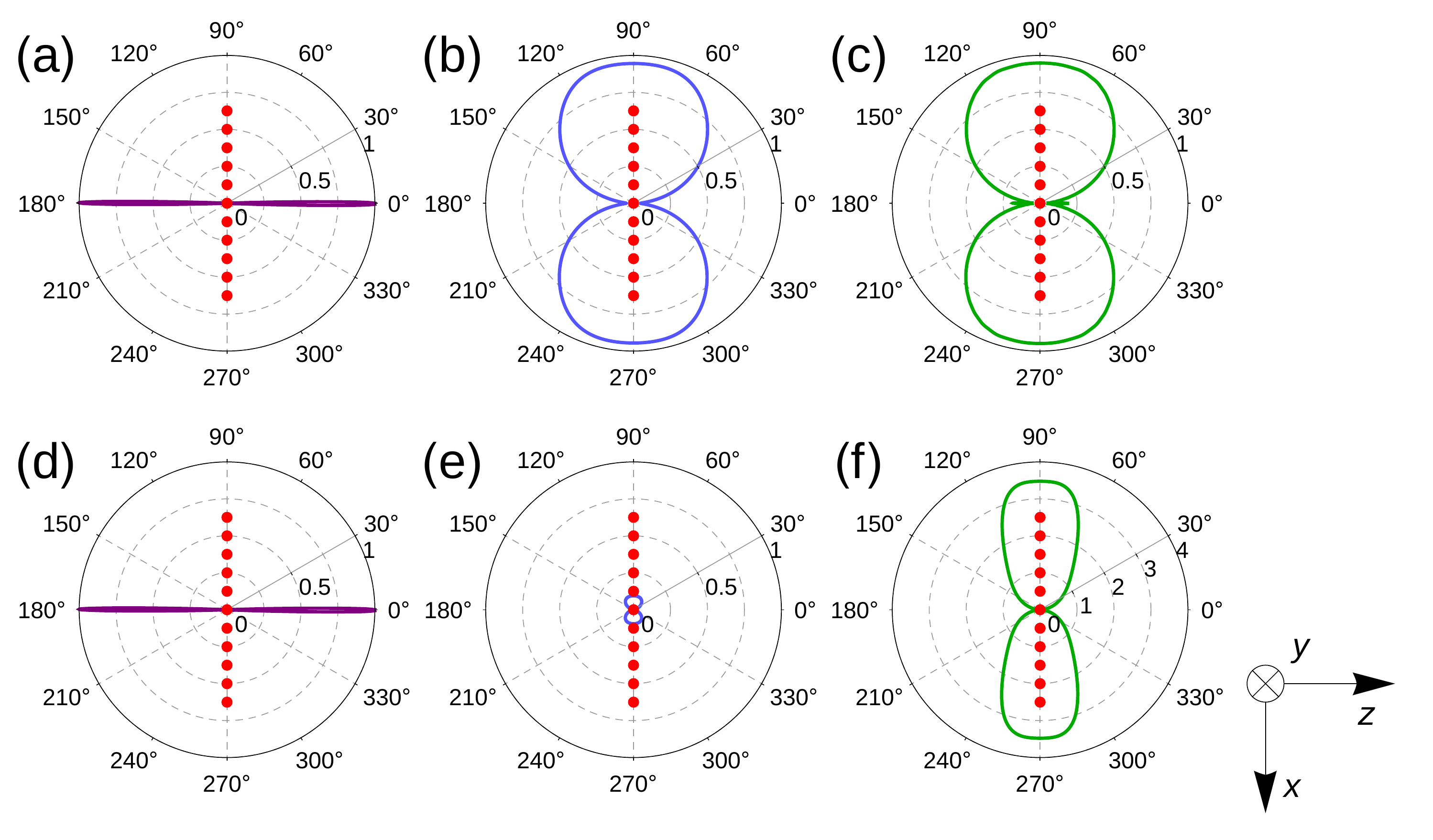}
\caption{Classical diffraction patterns (normalised to $N^2$) and quantum additions (normalised to $N$) for $N$ fermions in a one-dimensional optical lattice  (45 sites, half filling) in the non-interacting regime (first row) and for $U/J^\text{cl}=10$ (second row). The lattice extends into the vertical direction and the probe is along the horizontal, i.e.~the angle marked with $0^{\circ}$ indicates the light intensity in the forward direction. The classical diffraction patterns (a), (d) do not depend on the interaction while the quantum additions $R_y$ (b), (e) and $R_x$ (c), (f) distinguish between different ground states. Note that the scale in (f) is different from (a-e): the fluctuations in the atom number increase as the attraction between the atoms favours doubly occupied sites. Red circles indicate the orientation of the lattice.}
\label{1-fig:addition}
\end{figure} 

Figure \ref{1-fig:addition} compares the classical diffraction pattern and quantum additions $R_x$ and $R_y$ for fermions in a one-dimensional optical lattice at half filling. The scattering patterns were calculated for the ground state, obtained via imaginary time evolution using the TNT library. In particular, we describe the system using the Hubbard model and focus on the ground state of the system in two different regimes: non-interacting  ($U/J^\text{cl}=0$); and strongly attractive interactions ($U/J^\text{cl}=10$). Note that in both cases the local magnetisation of the system is zero ($\langle \hat{m}_i \rangle=0$), and as a consequence the classical diffraction pattern for linear-$y$ polarised light vanishes, i.e.~$ \langle \hat  a_{1y} \rangle=0$. Nevertheless, the quantum addition $R_y$ is non-zero and depends on the quantum state of the system. In particular, we find that $R_y$ decreases with increasing values of $U/J^\text{cl}$, as the attraction between the atoms favours doubly occupied sites, i.e.~the  formation of pairs of fermions with opposite spins, which suppress the fluctuations in the magnetisation. Furthermore, the classical diffraction pattern due to the atomic density does not depend on $U/J^\text{cl}$ as the local density is independent of the interaction for a translational invariant  optical lattice ($\langle \hat{\rho}_i \rangle=1$). However, the presence of doubly occupied sites increases the fluctuations in the atom number, leading to a stronger signal for $R_x$.


\section{Mapping the full distribution function of a quantum gas}

In this section we will relax another assumption used in this work so far: we will take into account the dispersive shift of the cavity due to the presence of ultracold atoms. It was taken into account in the general model (\ref{1-7}), (\ref{1-8}). It was taken into account also in Eqs. (\ref{1-1PRA09})--(\ref{1-11}), where we neglected the influence of the tunneling on light as we will do in this section as well. 

In this section we will prove that atomic quantum statistics can be mapped on transmission spectra of high-Q cavities, where atoms create a quantum refractive index (exactly due to the cavity dispersion shift). More precisely, the dispersion shift of a cavity mode depends on the atom
number. If the atom number in some lattice region fluctuates from
realization to realization, the modes get a fluctuating frequency
shift. Thus, in the cavity transmission-spectrum, resonances appear
at different frequencies directly reflecting the atom number
distribution function. Until this section, we demonstrated the cases, where light provides information about various statistical variables (such as variances), but not the full distribution function. 

Different phases of a degenerate gas possess similar mean-field
densities but different quantum amplitudes. This will lead to a
superposition of different transmission spectra, which e.g. for a
superfluid state (SF) will consist of numerous peaks reflecting the
discreteness of the matter-field. Analogous discrete spectra
reversing the role of atoms and light, thus reflecting the photon
structure of electromagnetic fields, were obtained in cavity QED
with Rydberg atoms~\cite{BrunePRL19961800} and solid-state superconducting
circuits~\cite{GambettaPRA2006}. We will show that a quantum phase transition towards a
Mott insulator state (MI) is characterized by a reduction of the
number of peaks towards a single resonance, because atom number
fluctuations are significantly suppressed.

In this section we use Eq. (\ref{1-2NatPhys}), where the dispersive frequency shift is taken into account. It describes dynamics of the cavity mode $a_1$ coupled to the mode $a_0$, which corresponds to our main setup in Fig. \ref{1-fig1}. First, we will proceed further with a single dynamical mode $a_1$, but in the second part of this section we will assume that the probe mode $a_0$ is dynamical as well (it can be formed by the same or another cavity). Therefore, we rewrite Eq. (\ref{1-2NatPhys}) for two modes straight away ($l=0,1$):

\begin{gather}
\dot{a}_l= -i\left(\omega_l +U_{ll}\hat{D}_{ll}\right)a_l
-iU_{lm}\hat{D}_{lm}a_m -\kappa a_l+\eta_l(t), \label{1-2NP}\\
\text{with} \quad \hat{D}_{lm}\equiv \sum_{i=1}^K{u_l^*({\bf
r}_i)u_m({\bf r}_i)\hat{n}_i},\nonumber
\end{gather}
where $l\ne m$, $\eta_l(t)=\eta_l e^{-i\omega_{lp}t}$ gives the external probe.

In a classical limit, Eq.~(\ref{1-2NP}) corresponds to Maxwell's
equations with the dispersion-induced frequency shifts of cavity
modes $U_{ll}\hat{D}_{ll}$ and the coupling coefficient between
them $U_{10}\hat{D}_{10}$. For a quantum gas those quantities are
operators, which will lead to striking results: atom number
fluctuations will be directly reflected in such measurable
frequency-dependent observables. Thus, cavity transmission-spectra
will reflect atomic statistics.

Eq.~(\ref{1-2NP}) allows to express the light operators $a_l$ as a
function $f(\hat{n}_1,...,\hat{n}_M)$ of atomic occupation number
operators and calculate their expectation values for prescribed
atomic states $|\Psi\rangle$. We start with the well known examples
of MI and SF states and generalize to any $|\Psi\rangle$ later.

From the viewpoint of light scattering, the MI state behaves almost
classically as, for negligible tunneling, precisely
$\langle\hat{n}_i\rangle_\text{MI}=q_i$ atoms are well localized at
the $i$th site with no number fluctuations. It is represented by a
product of Fock states, i.e. $|\Psi\rangle_\text{MI}=\prod_{i=1}^M
|q_i\rangle_i\equiv |q_1,...,q_M\rangle$, with expectation values
\begin{eqnarray}\label{1-3NP}
\langle f(\hat{n}_1,...,\hat{n}_M)\rangle_\text{MI}=f(q_1,...,q_M),
\end{eqnarray}
since $\hat{n}_i|q_1,...,q_M\rangle=q_i|q_1,...,q_M\rangle$. For
simplicity we consider equal average densities
$\langle\hat{n}_i\rangle_\text{MI}=N/M\equiv n$
($\langle\hat{N}_K\rangle_\text{MI}=nK\equiv N_K$).

In our second example, SF state, each atom is delocalized over all
sites leading to local number fluctuations at a lattice region with
$K<M$ sites. Mathematically it is a superposition of Fock states
corresponding to all possible distributions of $N$ atoms at $M$
sites: $|\Psi\rangle_\text{SF}
=\sum_{q_1,...,q_M}\sqrt{N!/M^N}/\sqrt{q_1!...q_M!}
|q_1,...,q_M\rangle$. Although its average density
$\langle\hat{n}_i\rangle_\text{SF}=N/M$ is identical to a MI, it
creates different light transmission spectra. Expectation values of
light operators can be calculated from
\begin{eqnarray}\label{1-4NP}
\langle f(\hat{n}_1,...,\hat{n}_M)\rangle_\text{SF}=\frac{1}{M^N}
\sum_{q_1,...,q_M}\frac{N!} {q_1!...q_M!}f(q_1,...,q_M),
\end{eqnarray}
representing a sum of all possible ``classical'' terms. Thus, all
these distributions contribute to scattering from a SF, which is
obviously different from $\langle
f(\hat{n}_1,...,\hat{n}_M)\rangle_\text{MI}$ (\ref{1-3NP}) with only a
single contributing term.

\begin{figure}[h!]
\centering
\captionsetup{justification=justified}
\includegraphics[width=0.7\textwidth]{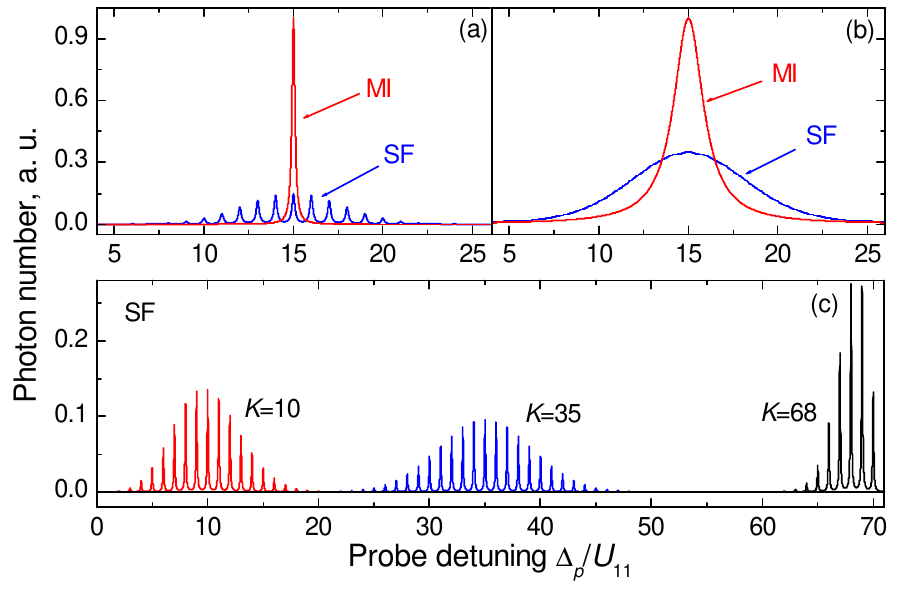}
\caption{\label{1-fig2NP} Transmission spectra of a cavity. The spectra directly map out the full atom number distribution functions of an ultracold gas. (a) Single Lorentzian for MI reflects the non-fluctuating atom number.
Many Lorentzians for SF reflect the atom number fluctuations,
which are imprinted on the positions of narrow resonances in the spectrum. Here, the cavity is good and all satellites are resolved (the cavity decay rate $\kappa$ is smaller than the satellite separation $U_{11}$, $\kappa=0.1U_{11}$). $N=M=30$, $K=15$. (b) The same as in (a) but the cavity is worse ($\kappa=U_{11}$), which gives smooth broadened contour for SF. Although the satellites are not resolved, the spectra for SF and MI states are very different. (c) Spectra for SF with $N=M=70$ and different number of sites illuminated $K=10,35,68$. The transmission spectra have different forms, since different atom distribution functions correspond to different $K$. $\kappa=0.05U_{11}$.}
\end{figure}

In the simple case of only one mode $a_1$ ($a_0\equiv 0$), the
stationary solution of Eq.~(\ref{1-2NP}) for the photon number reads
\begin{eqnarray}\label{1-5NP}
a^\dag_1a_1=f(\hat{n}_1,...,\hat{n}_M)=
\frac{|\eta_1|^2}{(\Delta_p-U_{11}\hat{D}_{11})^2 +\kappa^2},
\end{eqnarray}
where $\Delta_p$ is the probe-cavity detuning.
We present transmission spectra in Fig.~\ref{1-fig2NP} for the case,
where $|u_{1}({\bf r}_i)|^2=1$, and $\hat{D}_{11}=\sum_{i=1}^K
\hat{n}_i$ reduces to $\hat{N}_K$. For a 1D lattice (see
Fig.~\ref{1-fig1}), this occurs for a traveling wave at any angle, and
standing wave transverse or parallel to the lattice with atoms trapped at field maxima.

For MI,  averaging of Eq.~(\ref{1-5NP}) according to Eq.~(\ref{1-3NP})
gives the photon number $\langle a^\dag_1a_1\rangle_\text{MI}$, as a
function of the detuning, as a single Lorentzian described by
Eq.~(\ref{1-5NP}) with the width $\kappa$ and the frequency shift given by
$U_{11}\langle\hat{D}_{11}\rangle_\text{MI}$ (equal to
$U_{11}N_K$ in Fig.~\ref{1-fig2NP}). Thus, for MI, the spectrum
reproduces a simple classical result of a Lorentzian shifted due to the
dispersion.

In contrast, for a SF, the averaging procedure of Eq.~(\ref{1-4NP})
gives a sum of Lorentzians with different dispersion shifts
corresponding to all atomic distributions $|q_1,...,q_K\rangle$. So,
if each Lorentzian is resolved, one can measure a comb-like
structure by scanning the detuning $\Delta_p$. In Figs.~\ref{1-fig2NP}(a)
and \ref{1-fig2NP}(c), different shifts of the Lorentzians correspond to
different possible atom numbers at $K$ sites (which due to the atom
number fluctuations in SF, can take all values 0,1,2,...,$N$). The
Lorentzians are separated by $U_{11}$. Thus, we see that atom
number fluctuations lead to the fluctuating mode shift, and hence to
multiple resonances in the spectrum. For larger $\kappa$ the
spectrum becomes continuous (Fig.~\ref{1-fig2NP}(b)), but broader than
that for MI.

Scattering of weak fields does not change the atom number
distribution. However, as the SF is a superposition of different
atom numbers in a region with $K$ sites, a measurement projects the
state into a subspace with fixed $N_K$ in this region (we will show this in Chapter 2), and a
subsequent measurement on a time scale short to tunneling between
sites will yield the same result. One recovers the full spectrum of
Fig.~\ref{1-fig2NP} by repeating the experiment or with sufficient delay
to allow for redistribution via tunneling. Such measurements will
allow a time dependent study of tunneling and buildup of long-range
order. Alternatively, one can continue measurements on the reduced
subspace after changing a lattice region or light geometry.

\begin{figure}[h!]
\centering
\captionsetup{justification=justified}
\includegraphics[clip, trim=1cm 16cm 1cm 1cm, width=0.7\textwidth]{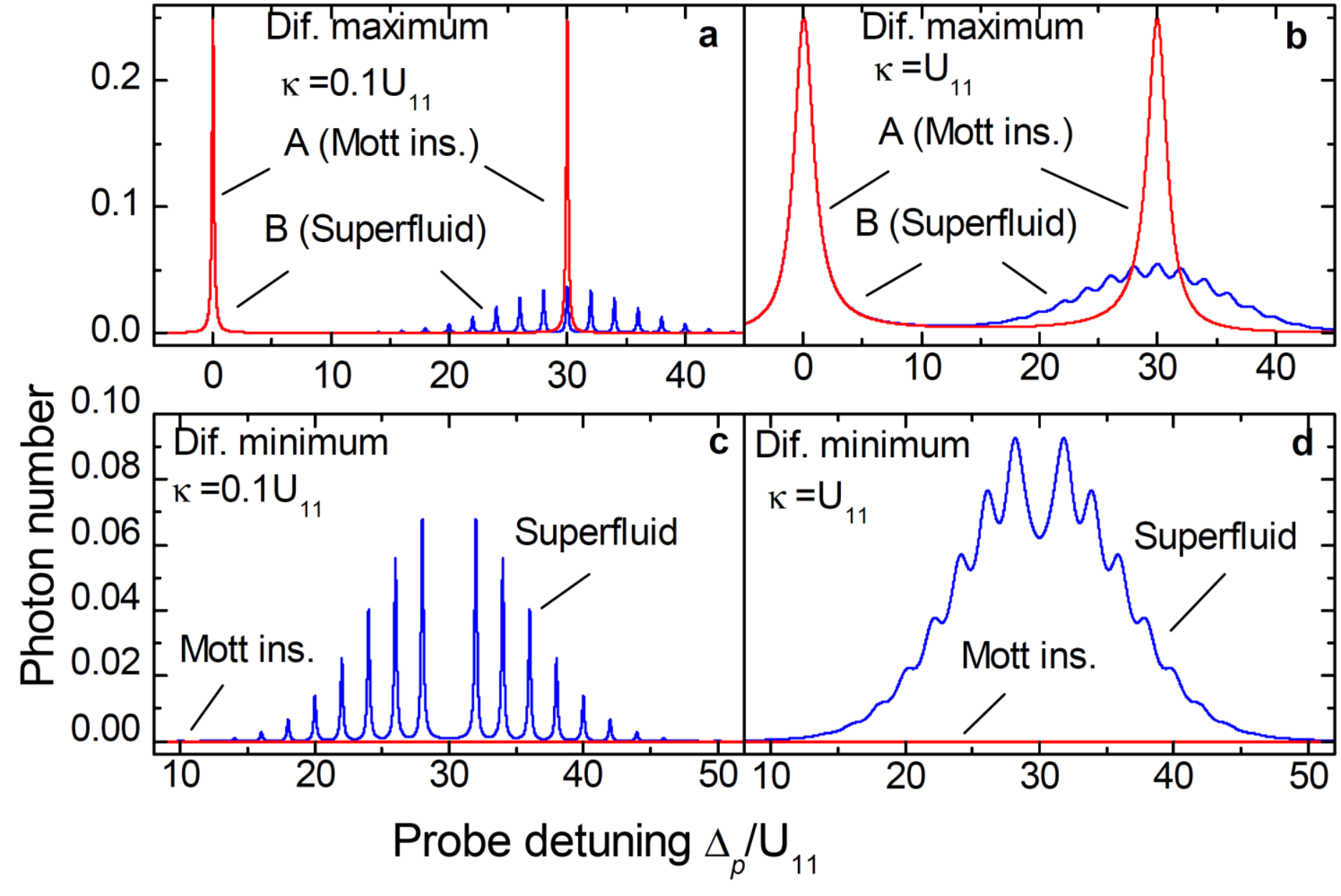}
\caption{\label{1-fig3NP}
Photon number in one of two strongly coupled modes. (a)
Diffraction maximum, doublet for MI (curve A) and spectrum with
structured right satellite for SF (curve B). Structure in the
satellite reflects atom number fluctuations in SF, while narrow
spectrum for MI demonstrates vanishing fluctuations. Here $\kappa$
is smaller than satellite separation $2U_{11}$
($\kappa=0.1U_{11}$), $K=15$. (b) The same as in (a) but
$\kappa=\delta_0$ gives broadened satellite for SF. (c)
Diffraction minimum, zero field for MI and structured spectrum for
SF. Nonzero structured spectrum for SF reflects fluctuating
difference between atom numbers at odd and even sites, which exists
even for the whole lattice illuminated, $K=M$. Here $\kappa$ is
smaller than satellite separation $2U_{11}$
($\kappa=0.1U_{11}$), $K=30$. (d) The same as in (c) but
$\kappa=\delta_0$ gives broadened contour for SF. $N=M=30$ in all
figures.}
\end{figure}

We now consider two modes with $\omega_0=\omega_1$, the probe $\eta_0$
injected only into $a_0$ and the mentioned
geometries where $\hat{D}_{00}=\hat{D}_{11}=\hat{N}_K$ (see
Fig.~\ref{1-fig3NP}). From Eq.~(\ref{1-2NP}), the stationary photon number
$a^\dag_1a_1=f(\hat{n}_1,...,\hat{n}_M)$ is
\begin{eqnarray}\label{1-6NP}
a^\dag_1a_1=\frac{U_{10}^2\hat{D}^\dag_{10}\hat{D}_{10}|\eta_0|^2}
{[\hat{\Delta}'^2_p-
U_{10}^2\hat{D}^\dag_{10}\hat{D}_{10}-\kappa^2]^2+4\kappa^2\hat{\Delta}'^2_p},
\end{eqnarray}
where $\hat{\Delta}'_p=\Delta_p-U_{11}\hat{D}_{11}$.

In a classical (and MI) case, Eq.~(\ref{1-3NP}) gives a two-satellite
contour (\ref{1-6NP}) reflecting normal-mode splitting of two
oscillators $\langle a_{0,1}\rangle$ coupled through atoms. Here, two strongly coupled oscillators are the two light modes, which is different from the usual cavity QED normal-mode splitting (vacuum Rabi splitting), where the two coupled oscillators are light and near-resonant atoms \cite{OS1,OS2,PRA2003,PRA2004,LP2005,Conf,OS2018,Bagayev2018,Bagayev2019,Bagayev2020}. This was observed~\cite{KlinnerPRL2006} for collective strong coupling,
i.e., the splitting $U_{10}\langle\hat{D}_{10}\rangle$ exceeding
$\kappa$. The splitting depends on the geometry (see Eq.~(\ref{1-2NP}))
representing diffraction of one mode into another. Thus, our results
can be treated as scattering from a ``quantum diffraction grating''
generalizing Bragg scattering, well-known in different disciplines.
In diffraction maxima (i.e. $u_1^*({\bf r}_i)u_0({\bf r}_i)=1$) one
finds $\hat{D}_{10}=\hat{N}_K$ providing the maximal classical
splitting. 

In diffraction minima, one finds
$\hat{D}_{10}=\sum_{i=1}^K(-1)^{i+1}\hat{n}_i$, which is the difference between the number of atoms at odd and even sites,  providing both the
classical splitting and photon number are almost zero. Note, that for SF, this difference changes with a step two (as the total atom number is fixed). In contrast, for the approximation of the coherent atomic state with unfixed atom number, this difference can change with the step one. Thus the frequency shifts (the distance between the comb components) will be different for the SF state and the coherent state approximation, which can be captured by our method. 

In SF, Eq.~(\ref{1-4NP}) shows that $\langle
a^\dag_1a_1\rangle_\text{SF}$ is given by a sum of all classical
terms with all possible normal mode splittings. In a diffraction
maximum (Figs.~\ref{1-fig3NP}(a,b)), the right satellite is split into
components corresponding to all possible $N_K$ or extremely
broadened. In a minimum (Figs.~\ref{1-fig3NP}(c,d)), the splittings are
determined by all differences between atom numbers at odd and even
sites $\sum_{i=1}^K(-1)^{i+1}q_i$. Note that there is no classical
description of the spectra in a minimum, since here the classical
field (and $\langle a^\dag_1a_1\rangle_\text{MI}$) are simply zero
for any $\Delta_p$. Thus, for two cavities coupled at diffraction
minimum, the difference between the SF and MI states is even more
striking: one has a structured spectrum instead of zero signal.
Moreover, the difference between atom numbers at odd and even sites
fluctuates even for the whole lattice illuminated, giving nontrivial
spectra even for $K=M$.

In each of the examples in Figs.~\ref{1-fig2NP} and \ref{1-fig3NP}, the
photon number depends only on one statistical quantity, now called
$q$, $f(q_1,...,q_M)=f(q)$. For the single mode and two modes in a
maximum, $q$ is the atom number at $K$ sites. For two modes in a
minimum, $q$ is the atom number at odd (or even) sites. Therefore,
expectation values for some state $|\Psi\rangle$ can be reduced to
$\langle f\rangle_\Psi= \sum_{q=0}^Nf(q)p_\Psi (q)$, where $p_\Psi
(q)$ is the distribution function of $q$ in this state.

In high-Q cavities ($\kappa \ll \delta_0=g^2/\Delta_{0a}$), $f(q)$
is given by a narrow Lorentzian of width $\kappa$ peaked at some
frequency proportional to $q$ ($q=0,1,...,N$). The Lorentzian hight
is $q$-independent. Thus, $\langle f\rangle_\Psi$ as a function of
$\Delta_p$ represents a comb of Lorentzians with the amplitudes
simply proportional to $p_\Psi (q)$.

This is our central result of this section. It states that the transmission spectrum
of a high-Q cavity $\langle a^\dag a (\Delta_p)\rangle_\Psi$
directly maps the distribution function of ultracold atoms $p_\Psi
(q)$, e.g., distribution function of atom number at $K$ sites.
Various atomic statistical quantities characterizing a particular
state can be then calculated: mean value (given by the spectrum
center), variance (determined by the spectral width) and higher
moments. Furthermore, transitions between different states will be
reflected in spectral changes. Deviations from idealized MI and SF
states are also measurable.

For SF, using $p_\text{SF}(q)$, we can write the
envelopes of the comb of Lorentzians shown in Figs.~\ref{1-fig2NP}(a,c)
and \ref{1-fig3NP}(a,c). As has been mentioned, in all examples
presented in Figs.~\ref{1-fig2NP} and \ref{1-fig3NP}, the photon number
depends only on a single statistical quantity, which we denote as
$q$. Using this fact, the multinomial distribution in Eq.~(\ref{1-4NP})
reduces to a binomial, which can be directly derived from
Eq.~(\ref{1-4NP}): $\langle f\rangle_\text{SF}=
\sum_{q=0}^Nf(q)p_\text{SF}(q)$ with
$p_\text{SF}(q)=N!/[q!(N-q)!](Q/M)^q(1-Q/M)^{N-q}$ and a single sum
instead of $M$ ones. Here $Q$ is the number of specified sites: $Q$
is equal to $K$ for one mode and two modes in a maximum; $Q$ is the
number of odd (or even) sites for two modes in a minimum ($Q=M/2$
for even $M$). This approach can be used for other geometries, e.g.,
for two modes in a minimum and $K<M$, where Eq.~(\ref{1-4NP}) can be
reduced to a trinomial distribution.

As a next approximation we consider $N,M\gg1$, but finite $N/M$,
leading to the Gaussian distribution
$p_\text{SF}(q)=1/(\sqrt{2\pi}\sigma_q)\exp[-(q-\tilde{q})^2/2\sigma_q^2]$
with central value $\tilde{q}=NQ/M$ and width
$\sigma_q=\sqrt{N(Q/M)(1-Q/M)}$.

In high-Q cavities ($\kappa \ll \delta_0=g^2/\Delta_{0a}$), $f(q)$
is a narrow Lorentzian of width $\kappa$ peaked at some
$q$-dependent frequency, now called $\Delta_p^q$. Since the
Lorentzian hight is $q$-independent, $\langle f\rangle_\text{SF}$ as
a function of $\Delta_p$ is a comb of Lorentzians with the
amplitudes proportional to $p_\text{SF}(q)$.

Using the Gaussian distribution $p_\text{SF}(q)$, we can write the
envelope of such a comb. For a single mode [Fig.~\ref{1-fig2NP}(a,c),
Eq.~(\ref{1-5NP})], we find $\Delta_p^q\approx U_{11}q$ with the
envelope
\begin{eqnarray}
\langle a_1^\dag a_1 (\Delta_p^q)\rangle_\text{SF}= \frac{\alpha
U_{11}}{\sqrt{2\pi}\sigma_\omega}
e^{-(\Delta_p^q-\tilde{\Delta}_p)^2/2\sigma_\omega^2},\nonumber
\end{eqnarray}
where the central frequency $\tilde{\Delta}_p=U_{11}N_K$, spectral
width $\sigma_\omega=U_{11}\sqrt{N_K(1-K/M)}$, and
$\alpha=|\eta_1|^2/\kappa^2$. So, the spectrum envelopes in
Fig.~\ref{1-fig2NP}(a,c) are well described by Gaussians of widths
strongly depending on $K$.

For $K\rightarrow 0$ and $K\rightarrow M$, the binomial distribution
$p_\text{SF}(q)$ is well approximated by a Poissonian distribution,
which is demonstrated in Fig.~\ref{1-fig2NP}(c) for $K=10$ and $K=68$. For
$K=M$ the spectrum shrinks to a single Lorenzian, since the total
atom number at $M$ sites does not fluctuate.

In other examples (Figs.~\ref{1-fig3NP}(a) and \ref{1-fig3NP}(c)), the above
expression is also valid, although with other parameters. For two
modes in a diffraction maximum (Fig.~\ref{1-fig3NP}(a)), the central
frequency, separation between Lorentzians and width are doubled:
$\tilde{\Delta}_p=2U_{11}N_K$, $\Delta_p^q\approx 2U_{11}q$ and
$\sigma_\omega=2U_{11}\sqrt{N_K(1-K/M)}$;
$\alpha=|\eta_0|^2/(2\kappa^2)$. The left satellite at $\Delta_p=0$
has a classical amplitude $|\eta_0|^2/(4\kappa^2)$.

The nonclassical spectrum for two waves in a diffraction minimum
(Fig.~\ref{1-fig3NP}(c)) is centered at $\tilde{\Delta}_p=U_{11}N$, with
components at $\Delta_p^q\approx 2U_{11}q$, and is very broad,
$\sigma_\omega=U_{11}\sqrt{N}$; $\alpha=|\eta_0|^2/\kappa^2$.

For bad cavities ($\kappa\gg U_{11}$), the sums can be replaced by
integrals with the same parameters $\tilde{\Delta}_p$ and
$\sigma_\omega$ as for $\kappa < U_{11}$. For a single mode,
Fig.~\ref{1-fig2NP}(b) represents a Voigt contour, well-know in spectroscopy of hot gases (here, the
``inhomogeneous broadening'' is a striking contribution of quantum
statistics):
\begin{eqnarray}
\langle a_1^\dag a_1 (\Delta_p)\rangle_\text{SF}=
\frac{|\eta_1|^2}{\sqrt{2\pi}\sigma_\omega}\int_0^\infty
\frac{e^{-(\omega-\tilde{\Delta}_p)^2/2\sigma_\omega^2}d\omega}{(\Delta_p-\omega)^2
+\kappa^2}.\nonumber
\end{eqnarray}
For two modes in a diffraction minimum the photon number
(Fig.~\ref{1-fig3NP}(d)) is
\begin{eqnarray}
\langle a_1^\dag a_1\rangle_\text{SF}=
\frac{|\eta_0|^2}{\sqrt{2\pi}\sigma_\omega}\int_{-\infty}^\infty
\frac{\omega^2e^{-\omega^2/2\sigma_\omega^2}d\omega}
{(\Delta'^2_p-\omega^2-\kappa^2)^2 +4\kappa^2\Delta'^2_p},\nonumber
\end{eqnarray}
where $\Delta'_p=\Delta_p-\tilde{\Delta}_p$, while in a maximum
(Fig.~\ref{1-fig3NP}(b))
\begin{eqnarray}
\langle a_1^\dag a_1\rangle_\text{SF}=
\frac{|\eta_0|^2}{4\sqrt{2\pi}\sigma_\omega}\int_0^\infty
\frac{\omega^2e^{-(\omega-\tilde{\Delta}_p)^2/2\sigma_\omega^2}d\omega}
{[\Delta_p(\Delta_p-\omega)+\kappa^2]^2 +\kappa^2\omega^2}.\nonumber
\end{eqnarray}

The condition $\kappa < U_{11}$ is already met in
present experiments. In the work~\cite{BourdelPRA2006}, where
setups of cavity QED and ultracold gases were joined to probe
quantum statistics of an atom laser with $^{87}$Rb atoms, the
parameters are $(g,\Delta_{0a},\kappa)=2\pi\times(10.4,30,1.4)$ MHz.
The setups of cavity cooling~\cite{RempeNature2004,HoodScience2000} are also very
promising.

In summary of this section, we exhibited that transmission spectra of cavities
around a degenerate gas in an optical lattice are distinct for
different quantum phases of even equal densities. Similar
information is also contained even in the amplitudes of the transmitted field $\langle
a_{0,1}\rangle$ (and not only in the photon numbers as it was the case in the previous sections of this work). This
reflects (i) the orthogonality of Fock states corresponding to
different atom distributions and (ii) the different frequency shifts
of light fields entangled to those states. In general also other
optical phenomena and quantities depending nonlinearly on atom
number operators should similarly reflect the underlying quantum
statistics \cite{KUPRIYANOV2017}.


\section{QND detection of few-body polar molecule complexes: dimers, trimers, tetramers}

In this section we extend the QND detection methods presented above to the systems of ultracold molecules, which is valid for both bosonic and fermionic few-body complexes.

We present an optical nondestructive method for in situ detection of the bound states of ultracold polar molecules. It promises a minimally destructive measurement scheme up to a physically exciting quantum nondemolition (QND) level. The detection of molecular complexes beyond simple pairs of quantum particles (dimers, known, e.g., as Cooper pairs from the BEC-BCS theory of superconductivity) is suggested, including three-body (trimers) and four-body (tertramers) complexes trapped by one-dimensional tubes. The intensity of scattered light is sensitive to the molecule number fluctuations beyond the mean-density approximation. Such fluctuations are very different for various complexes, which leads to radically different light scattering. This type of research extends quantum optics of quantum gases to the field of ultracold molecules.

The study of ultracold polar molecules attracts significant attention because of their long-range anisotropic interaction, which can lead to the creation of exotic quantum phases of ultracold particles. The phase diagram is expected to be much richer than that for atomic gases with the short-range interaction. The existence of several few-body bound states of polar molecules for a low dimensional geometry was proved in the joint works \cite{MekhovPRL2011,MekhovPRA2011}. Being important in the context of few-body physics, those results can help to get insight into the many-body problems as well, where the elementary few-body building blocks can play a crucial role. For example, going beyond the two-body complexes and predicting the existence of bound states consisting of more than two particles (as trimers and tetramers), those results can modify the standard description of the BCS-BEC crossover to superconductivity in certain systems, which is usually based on the picture of Cooper pairs (i.e., the dimers) only. In contrast to extensively studied Efimov-type states with short-range contact interaction, the states appearing due to the anisotropic long-range dipole-dipole interaction are less investigated.

The use of optical methods to detect the states of polar molecules promises the development of nondestructive in situ measurement schemes, which can be used to probe the system dynamics in real time. Moreover, as we first suggested in Refs. \cite{MekhovPRL2011,MekhovPRA2011}, the optical non-destructive detection of ultracold molecules can be developed up to the physically exciting quantum nondemolition (QND) level. Such an ultimately quantum measurement scheme affects the quantum state in a minimally destructive way and triggers the intriguing fundamental questions about the quantum measurement backaction and the entanglement between the light and ultracold molecules. Other probing methods such as time-of-flight measurements or lattice shaking are usually destructive. Focusing on a simple physical picture of the light--matter interaction, we show how the main characteristics of the light scattering can be estimated analytically, using a simple statistical approach. Moreover, those results should be valid even in the many-body systems with a large number of ultracold molecules, at least, in the low-density regime. 

Few-body physics is a very interesting field rather different from many-body one. Few-body bound states play a crucial role in determining properties of many physical systems. 
In quantum chromodynamics (QCD) and nuclear physics, 
quarks bind into nucleons and nucleons into nuclei. In chemistry and biology, 
chemical reactions are determined by 
properties of complexes of atoms and molecules. In soft condensed matter physics, 
self-organization of elementary 
objects into chains determines properties 
of rheological electro- and magnetofluids. 
In semiconducting nanostructures, like quantum wells, dots or nanotubes, 
few-body states like charged excitons and biexcitons
affect optical properties. 

A special feature of cold atom ensembles is the possibility to tune the 
two-particle interaction strength, which controls the properties of few-particle complexes.  
While most of the earlier work focused on Efimov states in systems with short-range contact interactions, recent 
experimental progress with polar molecules and Rydberg atoms open interesting possibilities for studying few-particle complexes in systems 
with long-range interactions. 
These systems can provide insights into  many-body systems
with long-range forces in the intriguing but poorly
understood regime of intermediate interaction strengths.
Multiparticle bound states require strong enough interactions
to form composite objects, but not too strong to avoid locking
molecules into a Wigner crystal. Studying
dynamics  of formation of the mutiparticle composites can
help to understand open questions of chemical reactions in
reduced dimensions. 

The stability of few-body states of ultracold polar molecules with long-range dipole interactions 
in a low-dimensional setup consisting of two one-dimensional tubes was shown in Refs. \cite{MekhovPRL2011,MekhovPRA2011}. This geometry can be produced by optical lattices or atomic chip traps.

\subsection{Model for light scattering from molecules}

\begin{figure}[h!]
\centering
\captionsetup{justification=justified}
\includegraphics[width=0.4\textwidth]{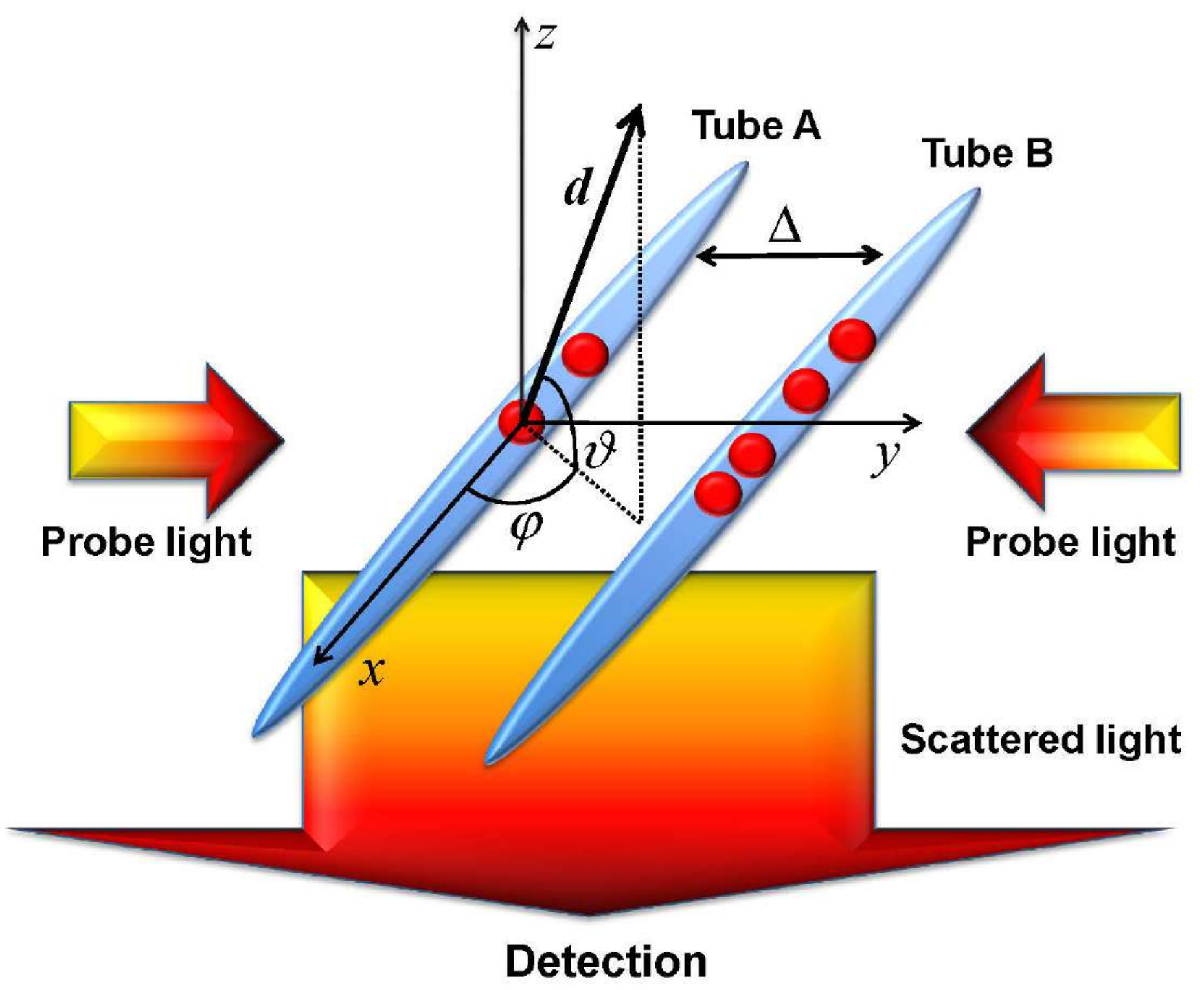}
\caption{\label{1-fig1LP13}Setup. The molecules with dipole moment ${\bf d}$ are trapped in the potential of two 1D tubes. The probe and detection are in the plane perpendicular to the tubes.}
\end{figure}

We consider ultracold dipolar molecules trapped in the potential of two one-dimensional (1D) tubes (cf. Fig. \ref{1-fig1LP13}). As described in details in Refs. \cite{MekhovPRL2011,MekhovPRA2011}, even for the repulsive interaction between the molecules within each tube, they can form bound complexes due to the attractive dipole-dipole interaction with the molecules in a different tube. Thus, several repulsing molecules in one tube can be bound by the presence of a molecule in another tube, with which they interact attractively. The association of molecules into various stable complexes was proved: dimers ``1-1'' (with one molecule in each tube), trimers ``1-2'' (with one molecule in one tube and two molecules in the other tube) and tetramers ``1-3'' (with one molecule in one tube and three molecules in the other tube) and ``2-2'' (with two molecules in each tube).

We will show that few-body complexes can be detected using light scattering. Our general method presented in the previous sections is very relevant to the present system, as it explicitly uses the sensitivity of light scattering to the relative position of the particles forming a complex. This is due to the constructive or destructive interference of the light waves scattered from the different particles. This method can be directly applied for extended periodic structures (many equidistantly spaced tubes or layers) and many-body systems, which makes the experimental realization promising. In contrast to Refs. \cite{deVegaPRA2008,BruunPRL2009,Roscilde2009} and experiments \cite{Sanner2011,Sanner2012} with spin ensembles, our proposal does not rely on any state-selective (e.g., spin-selective) light scattering, but is sensitive to the particle position.

We consider the scattering of the probe light with the amplitude given by the Rabi frequency $\Omega_p=d_0E_p/\hbar$ ($E_p$ is the probe-light electric field amplitude and $d_0$ is the induced dipole moment), cf. Fig.~ \ref{1-fig1LP13}. To increase the signal, the scattered light can be collected by a cavity, and the photons leaking from the cavity are then measured. Alternatively, the measurement of photons scattered can be made in a far-field region without the use of a cavity.

Using the approach of the second quantization for the molecule-field operator, the amplitude of the scattered light (i.e., the annihilation operator of the scattered photon) is given by (compare with the adiabatically eliminated light Eq. (\ref{1-10PRA}) and its derivation)

\begin{eqnarray}\label{1-1.A}
a_s=C\int d{\bf r} \hat{\Psi}^\dag({\bf r})u_s^*({\bf r})u_p({\bf r})\hat{\Psi}({\bf r}),
\end{eqnarray}
where $\hat{\Psi}({\bf r})$ is the matter-field operator at the point ${\bf r}$. For the free space scattering, the value of $C$ corresponds to the Rayleigh scattering \cite{ScullyBook}. Adding a cavity to the setup the scattering is increased and $C=-ig_s\Omega_p/(\Delta_a\kappa)$ with $\kappa$ being the cavity decay rate, $g_s$ is the molecule-light coupling constant, and $\Delta_a$ is the light detuning from the resonance, cf. Refs.~\cite{MekhovPRL2007,MekhovPRA2007}. In Eq.~(\ref{1-1.A}), $u_{p,s}({\bf r})$ are the mode functions of probe and scattered light, which contain the information about the propagation directions of probe and scattered light waves with respect to the tube direction. For the simplest case of two traveling light waves, the product of two mode functions takes the well-known form from classical light scattering theory: $u^*_s({\bf r})u_p({\bf r})=\exp[i({\bf k}_p-{\bf k}_s){\bf r}]$, where ${\bf k}_{p,s}$ are the probe and scattered light wave vectors.

One can express the matter-field operator in the basis of the functions corresponding to the transverse distribution of molecules within two tubes A and B:
\begin{eqnarray}\label{1-1.B}
\hat{\Psi}({\bf r})=\hat{\Psi}_A(x)w(\rho-\rho_A)+\hat{\Psi}_B(x)w(\rho-\rho_B),
\end{eqnarray}
where $\hat{\Psi}_{A,B}(x)$ are the matter-field operators within each tube with the coordinate $x$ alone the tube, where the molecules can move (cf. Fig.  \ref{1-fig1LP13}); $w(\rho)$ gives the distribution of a molecule in the transverse direction ($\rho$ is the transverse coordinate). In all previous sections about atoms, we used the Wannier functions instead. Substituting this expression in Eq.~(\ref{1-1.A}), we can describe the light scattering taking into account the possible overlap of the molecules between two tubes (overlapping $w(\rho-\rho_A)$ and $w(\rho-\rho_B)$) and the nontrivial overlap between the molecule distribution $w(\rho)$ and the light modes $u_{p,s}({\bf r})$. However, following Refs. \cite{MekhovPRL2011,MekhovPRA2011}, we assume that two tubes do not overlap at all, and they are well localized with respect to the light wave.

Thus, after several assumptions (the small tube radius, far off-resonant light scattering, detection in the far field zone), the light scattering has a simple physical interpretation. The scattered light amplitude is given by the sum of the light amplitudes, scattered from each molecule (cf. Fig.~ \ref{1-fig1LP13}). Each term has a phase and amplitude coefficient depending on the position of the molecule as well as on the  direction and amplitude of the incoming and outgoing light waves:
\begin{eqnarray}\label{1-1.5}
a_s=C\sum_{i=A,B}\int dx \hat{n}_i(x) u^*_s(x,\rho_i)u_p(x,\rho_i),
\end{eqnarray}
where the sum is over two tubes A and B, $\hat{n}_i(x)=\hat{\Psi}^\dag_i(x)\hat{\Psi}_i(x)$ is the operator of particle linear density. In Eq.~(\ref{1-1.5}), $u_{p,s}(x,\rho_i)$ are the mode functions of probe and scattered light at the tube positions $\rho_{A,B}$. Earlier, for atoms trapped in all three dimensions, we had a similar expression (\ref{1-10PRA}) with $\hat{D}_{10}$ operators summing all $\hat{n}_i$. Here, we keep the integration over the tube length, because the molecules are not trapped in this dimension.

Equation~(\ref{1-1.5}) is valid for any optical geometry and can describe the angular distribution of the scattered light. However, our important conclusion is that some information about the many-body state can be obtained even by a simple measurement of the photon number scattered at a single particular angle, which is fully enough for our purpose. Moreover, as it was shown, the particularly convenient angle of measurement corresponds to the direction of a diffraction minimum, rather than Bragg angle (diffraction maximum). At the directions of diffraction minimum any classical (possibly very strong) scattering is suppressed, and the light signal exclusively reflects the quantum fluctuations of the particles.

We now fix the optical geometry as follows (cf. Fig.  \ref{1-fig1LP13}). The incoming probe light is a traveling or standing wave propagating at the direction perpendicular to the tubes, which gives $u_{p}({\bf r})=R(x)\exp(ik_py)$ (for the traveling wave) or $u_{p}({\bf r})=R(x)\cos(k_py)$ (for the standing wave) and includes the transverse probe profile $R(x)$ of an effective width $W$. To perform the measurements at the direction of a diffraction minimum, the scattered light is measured along $z$ direction. For the free space detection, or the traveling-wave cavity, this gives $u_{s}({\bf r})=\exp(ik_sz)$, while for the case of a standing wave cavity, $u_{s}({\bf r})=\cos(k_sz)$. Without loss of generality, we can assume $u_{s}({\bf r})=1$ at the tube position $z=0$. The absolute values of the wave vectors are equal to their vacuum quantities $k_{p,s}=2\pi/\lambda_\text{light}$.

An important property of such a configuration (illumination and detection at the directions perpendicular to the tubes), is that all molecules within one tube scatter light with the same phase independently of their longitudinal position $x$ within the tube. Thus, the light scattered from the molecules within one tube interferes fully constructively. As a consequence, all molecules within two different tubes scatter light with a fixed phase difference with respect to each other. Due to this fact, the averaging over the probabilistic position of the complex does not involve the light phase and all complexes of the same type scatter light identically. Moreover, averaging over the probabilistic relative positions within each complex does not involve the dependence on the light phase as well. At other directions, both those kinds of phase averaging are important and would decrease the optical signal and the distinguishability of the complex types. The simple scattering picture also allows the generalization of the model for an array of several tubes.

The operator of the light amplitude reduces to
\begin{eqnarray}\label{1-1.6}
a_s=C\left(u_p(y_A)\hat{N}_A(W)+u_p(y_B)\hat{N}_B(W)\right),
\end{eqnarray}
where $\hat{N}_{A,B}(W)$ are the operators of the effective particle numbers in the tubes A and B within the region illuminated by the laser beam,
\begin{eqnarray}\label{1-1.7a}
\hat{N}_{A,B}(W)=\int_{-\infty}^{\infty}\hat{n}_{A,B}(x)R(x)dx.
\end{eqnarray}
If the laser profile can be approximated by a constant in the interval $(-W/2, W/2)$, the operators $\hat{N}_{A,B}(W)$ exactly correspond to the atom number operators in two tubes within the laser beam. (Comparing to the previous sections, this is similar to $\hat{N}_K$ operator for the number of atoms at illuminated sites.)

The classical condition of the diffraction minimum is fulfilled, when the expectation value of the light-amplitude operator (\ref{1-1.6}) is zero due to the perfect cancelation of the expectation values of two terms in Eq.~(\ref{1-1.6}) (i.e. the total destructive interference between the scatterers in two tubes). This is achieved for $u_p(y_B)/u_p(y_A)=-\langle\hat{N}_A\rangle/\langle\hat{N}_B\rangle$. We introduce the atom number ratio $\alpha=\langle\hat{N}_A\rangle/\langle\hat{N}_B\rangle$. For the equal mean atom numbers (the few-body complexes 1-1 and 2-2), the optical geometry should be chosen such that $u_s(y_B)/u_s(y_A)=-1$, which can be achieved if, e.g., the tube spacing is the half of the light wavelength, $\Delta=\lambda_\text{light}/2$. For the few-body complex 1-2, $\alpha=1/2$, and the diffraction minimum is achieved if the light wavelength and tube spacing satisfy the condition $\cos(k_py_B)/\cos(k_py_A)=-1/2$. This can be achieved, e.g., if the position of the tube A corresponds to the antinode of the standing wave $\cos(k_py_A)=1$, while that of tube B corresponds to $k_py_B=2\pi/3$ or $4\pi/3$, leading to the ratios between the tube spacing and light wavelength as $\Delta=\lambda_\text{light}/3$ or $2\lambda_\text{light}/3$. Similarly, for the 1-3 complex, that ratio can be $\Delta\approx 0.3\lambda_\text{light}$ or $0.7\lambda_\text{light}$. All those example ratios can be indeed larger, taking into account the periodicity of the light wave.

The expectation value of number of photons scattered at the direction of diffraction minimum $n_{\Phi}$ is then given by
\begin{gather}
n_{\Phi}=\langle a^\dag_s a_s\rangle = 
\left|C\right|^2 \left|u_p(y_A)\right|^2
\left\langle\left(\hat{N}_A(W)-\alpha\hat{N}_B(W)\right)^2\right\rangle,\label{1-1.7}
\end{gather}
where $u_p(y_A)$ can be easily chosen as 1. This expression manifests that the number of photons scattered in the diffraction minimum is proportional to the second moment of the ``rated'' particle number difference between two tubes in the laser-illuminated region. The mean light amplitude is sensitive to the mean values of the particle number and is precisely zero at the diffraction minimum: $\langle a_s\rangle \sim \left\langle\left(\hat{N}_A(W)-\alpha\hat{N}_B(W)\right)\right\rangle=0$. However, in general, the photon number (\ref{1-1.7}) is non-zero. It directly reflects the particle number fluctuations and correlations between the tubes. Thus, the number of photons reflects the quantum state of ultracold molecules.

\subsection{Applications for dimers, trimers, and tetramers}

In the joint works \cite{MekhovPRL2011,MekhovPRA2011}, we presented the results of numerical simulations for light scattering from few-body complexes for particular parameters. We have shown that, while the photon number in the diffraction minimum is zero for a bound state, it immediately increases, when the complex dissociates into a smaller complex and a free molecule. Although after such a dissociation, the mean particle number stays the same (and the light amplitude would not change), the fluctuations of the particle number inside the laser beam change strongly after the dissociation: instead of one bound complex, one gets another complex and a free particle, whose positions are uncorrelated. The particle fluctuations increase the intensity of the scattered light.

In this section, we demonstrate that the values of light intensity for stable complexes and free molecules can be estimated analytically using the statistical calculations. Such estimations are valid for many molecules in each tube (at least in the low-density regime) and agree well with the numerical simulations made for real systems, but the tiny number of molecules per tube \cite{MekhovPRL2011,MekhovPRA2011}. The approach developed in this papers also gives a possibility to get a deeper physical insight into the problem. Although the development of modern trapping techniques targets the manipulation of ultracold atoms at a single-particle level, the many-particle realization is still more realistic.

Expression (\ref{1-1.7}) can be written in the form
\begin{gather}
n_{\Phi}/\left|C\right|^2 = \langle(\hat{N}_A -\alpha\hat{N}_B)^2\rangle = 
 \langle\hat{N}^2_A\rangle +\langle\hat{N}^2_B\rangle - 2\alpha \langle\hat{N}_A\hat{N}_B\rangle,\label{1-1.N1}
\end{gather}
which underlines the correlations between the molecule numbers in two different tubes.

Let us start with the example of dimers ``1-1'' and consider the equal number of molecules in two tubes ($\langle\hat{N}_A\rangle=\langle\hat{N}_B\rangle=N$, $\alpha=1$). When all molecules are strongly bound into dimers, they appear within the laser beam only in pairs, or do not appear there at all. Thus, the fluctuations of the molecule number difference is zero (one can think about the two number operators as identical ones, $\hat{N}_A=\hat{N}_B$, i.e., all their moments coincide) and so does the light intensity: $n_{\Phi}/\left|C\right|^2 = \langle(\hat{N}_A -\alpha\hat{N}_B)^2\rangle = \langle(\hat{N}_A -\alpha\hat{N}_B)\rangle=0$. On the other hand, when a dimer dissociates into two independent free molecules, the two operators are different, and the term with the intertube correlation function in Eq. (\ref{1-1.N1}) decorrelates into a product: $\langle\hat{N}_A\hat{N}_B\rangle=\langle\hat{N}_A\rangle\langle\hat{N}_B\rangle=N^2$. One can assume that the number fluctuations of the independent free molecules are Poissonian, $\langle\hat{N}^2_{A,B}\rangle=\langle\hat{N}_{A,B}\rangle^2+\langle\hat{N}_{A,B}\rangle=N^2+N$. Then, the number of scattered photons Eq. (\ref{1-1.N1}) gets $n_{\Phi}/\left|C\right|^2 = 2N$.

\begin{figure}[h!]
\centering
\captionsetup{justification=justified}
\includegraphics[width=0.4\textwidth]{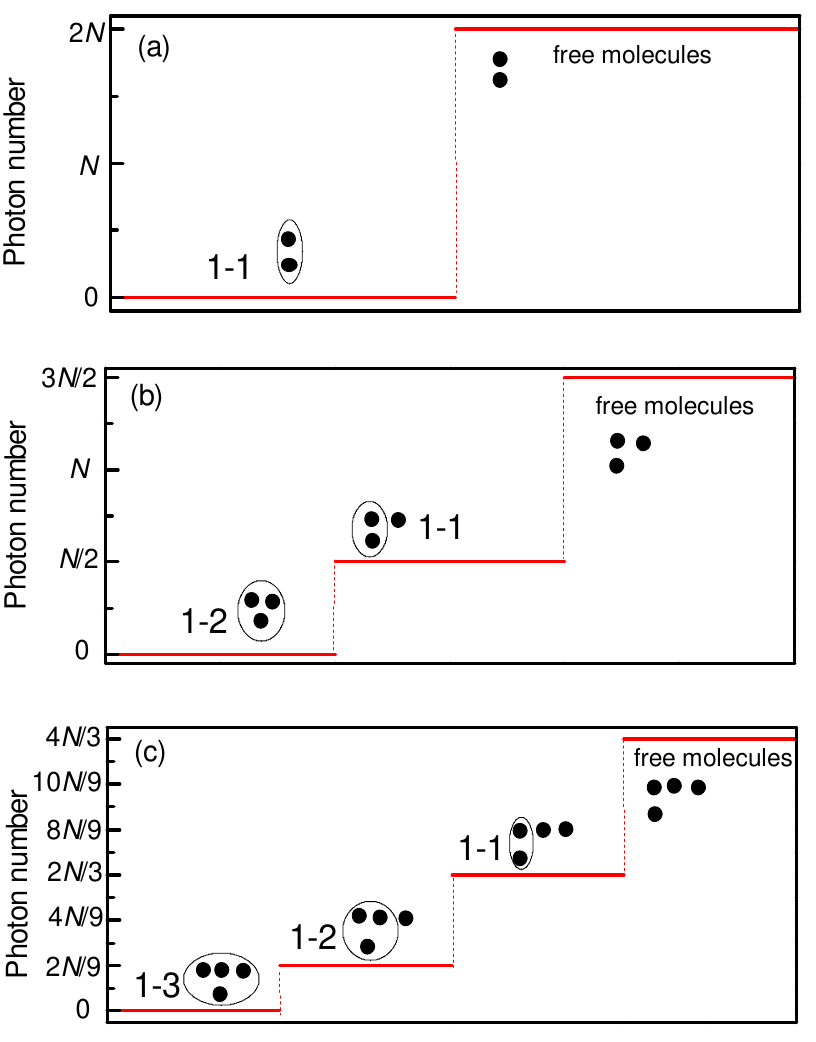}
\caption{\label{1-fig2LP13}QND measurement of ultracold polar molecule complexes. Intensity of scattered light (i.e. the relative photon number $n_{\Phi}/\left|C\right|^2$) depending on the existence of various few-body complexes. The variable on the horizontal axis is schematic. It can correspond to several parameters, which allow to scan the system through the regimes, were different complexes exist (e.g., the dipole orientation angle or dipole-dipole interaction strength as shown in Refs. \cite{MekhovPRL2011,MekhovPRA2011}). The results of numerical simulations for continuous dissociation are shown in the next figure. (a) Dissociation of dimers ``1-1'' into free molecules corresponds to the change of light intensity from $n_{\Phi}/\left|C\right|^2=0$ to $n_{\Phi}/\left|C\right|^2=2N$. (b) Dissociation of trimers ``1-2'' into dimers ``1-1'' and free molecules, and then into all free molecules corresponds to the intensity jumps as $n_{\Phi}/\left|C\right|^2=0$, $n_{\Phi}/\left|C\right|^2=N/2$, and $n_{\Phi}/\left|C\right|^2=3N/2$. (c) Dissociation of tetramers ``1-3'' into trimers ``1-2'' and free molecules, then into dimers ``1-1'' and free molecules, and finally into all free molecules correponds to the intensity values $n_{\Phi}/\left|C\right|^2=0$, $n_{\Phi}/\left|C\right|^2=2N/9$, $n_{\Phi}/\left|C\right|^2=6N/9$, and $n_{\Phi}/\left|C\right|^2=12N/9$. Inversely, the association of those complexes will correspond to the suppression of light scattered into the diffraction minimum.}
\end{figure}

Therefore, we see that the light intensity jumps from zero to $n_{\Phi}/\left|C\right|^2 = 2N$, when the dimers dissociate into free molecules. Such a change of light intensity for two different phases of ultracold molecules is schematically demonstrated in Fig. \ref{1-fig2LP13}(a). Physically, the strongly bound complex does not scatter light, because the geometry corresponds to the diffraction minimum. Thus, the fluctuation of the complex number within the laser beam does not change the light intensity (it is zero if both the complex is within the beam, and, obviously, outside the beam). However, when the complex dissociate into two independent species (two free molecules in this example), the species can be within or outside the beam independently from each other. Thus, the condition of the total diffraction minimum is not satisfied anymore, because, probabilistically, the numbers of molecules within the beam can be nonequal in two tubes (even though they are always equal in average) and the complete destructive interference of light is not possible anymore. 

\begin{figure}[h!]
\centering
\captionsetup{justification=justified}
\includegraphics[clip, trim=1cm 22cm 1cm 1cm, width=0.9\textwidth]{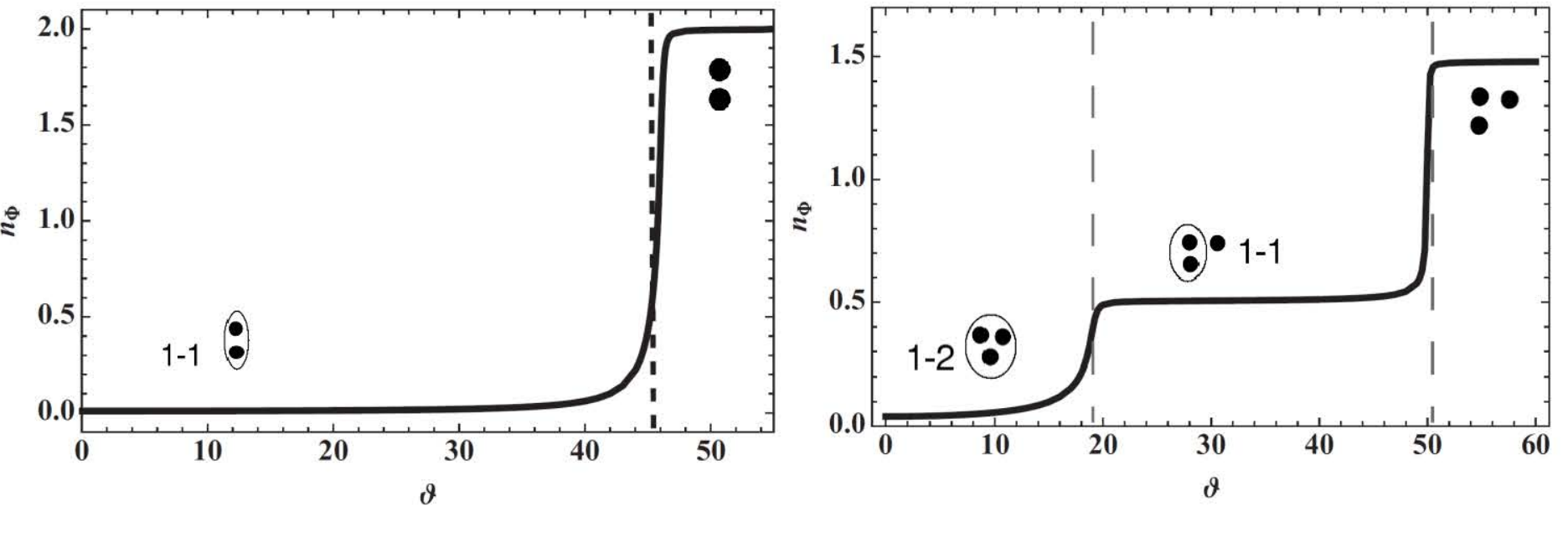}
\caption{\label{1-fig3LP13} QND measurement of ultracold polar molecule complexes. Intensity of scattered light depending on the existence of various few-body complexes. The results of numerical simulations from the joint works \cite{MekhovPRL2011,MekhovPRA2011} corresponding to the analytical plots in Fig. \ref{1-fig2LP13}(a,b).} 
\end{figure}

Note, that this result agrees very well with the numerical calculations presented in joint works \cite{MekhovPRL2011,MekhovPRA2011} carried out for two molecules in two tubes. Those numerical results indeed show not only the constant values of the light intensity, but also describe the continuous transition between them, when the dimer dissociates. As shown in Refs.  \cite{MekhovPRL2011,MekhovPRA2011}, to go through all the dissociation stages and see the intensity jumps one can change the orientation of the dipoles (e.g., the angle $\vartheta$ in Fig. \ref{1-fig1LP13}), or the strength of the dipole-dipole interaction between the molecules. Fig. \ref{1-fig3LP13} shows the continuous dissociation of a dimer, when the angle $\vartheta$ is tuned.

Let us now consider the case of trimers ``1-2'', when the populations of two tubes are imbalanced: $\langle\hat{N}_A\rangle=N$, $\langle\hat{N}_B\rangle=2N$, $\alpha=1/2$. When the molecules are strongly bound into a trimer, they appear in the laser beam only all three together, or do not appear at all. (Here we indeed neglect the small effects when the trimer is large and can overlap with the laser beam only partially. This however could be captured by the numerical simulations in Ref. \cite{MekhovPRL2011,MekhovPRA2011}, and was shown to introduce only small corrections to the result.) Therefore, the fluctuations of the operator $(\hat{N}_A -\alpha\hat{N}_B)^2$ are zero and the number of scattered photons is zero as well: $n_{\Phi}/\left|C\right|^2 = \langle(\hat{N}_A -1/2\hat{N}_B)^2\rangle = \langle(\hat{N}_A -1/2\hat{N}_B)\rangle=0$.

The trimer can dissociate into a dimer ``1-1'' and a free particle, which are independent from each other. The operator of the number of particles in the tube B can be split into two parts: $\hat{N}_B=\hat{N}^D_B+\hat{N}^F_B$, where the operator $\hat{N}_B=\hat{N}^D_B$ corresponds to the molecules, which form a dimer with another molecule in the tube A, and $\hat{N}_B=\hat{N}^F_B$ corresponds to the free molecules. To calculate the expectation value for the photon number, we can group the molecule number operators in Eq. (\ref{1-1.N1}) such that they would correspond to the same species (dimers or free molecules). Then, $\hat{N}_A-1/2\hat{N}_B=\hat{N}_A-1/2\hat{N}^D_B-1/2\hat{N}^F_B=1/2(\hat{N}^D- \hat{N}^F)$, where we introduced the operators for the number of dimers, $\hat{N}^D=\hat{N}^D_B=\hat{N}_A$, and number of free molecules, $\hat{N}^F=\hat{N}^F_B$.

After introducing the operators for different independent species (dimers and free molecules), we can calculate the expectation value in Eq. (\ref{1-1.N1}), assuming that the species are uncorrelated ($\langle\hat{N}^D\hat{N}^F\rangle=\langle\hat{N}^D\rangle\langle\hat{N}^F\rangle=N^2$) and each of them displays the Poissonian fluctuations ($\langle(\hat{N}^{D,F})^2\rangle=\langle\hat{N}^{D,F}\rangle^2+\langle\hat{N}^{D,F}\rangle= N^2+N$). The result reads: $n_{\Phi}/\left|C\right|^2 = N/2$. So, we see, how the light intensity jumps from zero to this non-zero value, when the trimer dissociates into a dimer and a free molecule.

Those dimers and free molecules can dissociate further into three independent molecules. Taking into account the mean values of the free molecules in two tubes, $\langle\hat{N}^{F}_A\rangle=N$ and  $\langle\hat{N}^{F}_B\rangle=2N$, the expectation value of the light intensity reads $n_{\Phi}/\left|C\right|^2 = 3N/2$. That is, it jumps further upwards.

The consecutive dissociation of the trimers is schematically shown in Fig. \ref{1-fig2LP13}(b). All three phases can be distinguished by the light intensity: it is zero for bound trimers, proportional to $N/2$ for dimers and free particles, and to $3N/2$ for all free particles. This result agrees with the numerical simulations \cite{MekhovPRL2011,MekhovPRA2011}. Fig. \ref{1-fig3LP13} shows the continuous dissociation of a trimer, when the angle $\vartheta$ is tuned. 

Let us now expand the consideration for the the case of tetramers ``1-3'', when the populations of two tubes are imbalanced: $\langle\hat{N}_A\rangle=N$, $\langle\hat{N}_B\rangle=3N$, $\alpha=1/3$. The numerical simulations for that situation were not reported in Refs. \cite{MekhovPRL2011,MekhovPRA2011}. As before, when the complex is strongly bound, it does not scatter light into the diffraction minimum and $n_{\Phi}/\left|C\right|^2 = \langle(\hat{N}_A -1/3\hat{N}_B)^2\rangle = 0$. The following steps of a tetramer dissociation are possible: 1) a trimer ``1-2'' and a free molecule, 2) a dimer ``1-1'' and two free molecules and 3) three free molecules.

The tetramers first dissociate into the trimers ``1-2'' and free molecules. Proceeding as before, the number operator in the tube B can be split into two statistically independent operators: $\hat{N}_B=\hat{N}^T_B+\hat{N}^F_B$, where $\hat{N}^T_B$ corresponds to the molecules in the tube B, which form trimers with molecules in A, and $\hat{N}^F_B$ corresponds to free molecules. As before, we introduces the operator of the trimer number $\hat{N}^T$. All molecules in the tube A participate in the trimer creation: $\hat{N}_A=\hat{N}^T$, while the number of molecules forming the trimer in the tube B is two times larger: $\hat{N}^T_B=2\hat{N}^T$. Proceeding as before, assuming that the trimers and free molecules are not correlated ($\langle\hat{N}^T\hat{N}^F\rangle=\langle\hat{N}^T\rangle\langle\hat{N}^F\rangle=N^2$), and obey the Poissonian fluctuations, we arrive to the photon number as $n_{\Phi}/\left|C\right|^2 = \langle(\hat{N}_A -1/3\hat{N}_B)^2\rangle = 2N/9$.

After that, this four-body state can dissociate further into dimers ``1-1'' and two free molecules. All molecules in the tube A will form the dimer, $\hat{N}_A=\hat{N}^D$, and the number of molecules from the tube B forming the dimers will be the same, $\hat{N}^D_B=\hat{N}^D$. In this case, the mean values are: $\langle\hat{N}^D\rangle=N$, $\langle\hat{N}^F\rangle=2N$. The photon number jumps upwards: $n_{\Phi}/\left|C\right|^2 = \langle(\hat{N}_A -1/3\hat{N}_B)^2\rangle = 2N/3$.

Similarly, the last step of dissociation leading to all free molecules increases the intensity of scattered light further due to even stronger fluctuations of the molecule number within the beam: $n_{\Phi}/\left|C\right|^2 = 4N/3$. (To derive this expression, note that $\langle\hat{N}^F_A\rangle=N$, while $\langle\hat{N}^F_B\rangle=3N$).

The dependence of the light intensity on the molecule state is schematically shown in Fig. \ref{1-fig2LP13}(c). The plateaus with four different values are expected: $n_{\Phi}/\left|C\right|^2=0$ for the tetramers ``1-3'', $n_{\Phi}/\left|C\right|^2=2N/9$ for trimer ``1-2'' and free molecules, $n_{\Phi}/\left|C\right|^2=6N/9$ for dimers ``1-1'' and free molecules, and $n_{\Phi}/\left|C\right|^2=12N/9$ for the totally dissociated system.

We have seen that the dissociation of a complex increases the particle number fluctuations, which leads to the jump of the light intensity. Inversely, the observation of the consecutive association would correspond to the stepwise suppression of the light intensity, which reflects the decrease of the number fluctuations. As shown in Ref. \cite{MekhovPRL2011,MekhovPRA2011}, to go through all those stages one can change the orientation of the dipoles (e.g., the angle $\vartheta$ in Fig. \ref{1-fig1LP13}), or the strength of the dipole-dipole interaction between the molecules. Interestingly, in contrast to the light intensity in the diffraction minimum, the mean light amplitude would not change at all and would stay zero for all states considered above. This is an example of a quantum optical problem, where one has a zero light amplitude $\langle a_s\rangle=0$, but non-zero photon number $\langle a^\dag_s a_s\rangle \ne |\langle a_s\rangle|^2$ due to the matter-induced photon fluctuations.

In summary of this section, we presented the optical nondestructive scheme for probing bound states of ultracold polar molecules. Based on the off-resonant light scattering it promises the in situ measurement of the molecular dynamics in real time up to a physically exciting QND level. The detection of association and dissociation of molecular pairs (dimers), three-body states (trimers) and four-body states (tetramers) has been demonstrated. In contrast to other QND schemes \cite{deVegaPRA2008,BruunPRL2009,Roscilde2009,Sanner2011,Sanner2012} requiring the state-selective (e.g. spin-selective) light scattering, this method is originally based on our proposal presented in the previous sections  and is not sensitive to the internal-level structure, which is its advantage. The light scattering directly reflects the relative spatial positions of the complex parts and measures the quantum fluctuations of the molecule numbers beyond the mean-density approximation. Development of such QND techniques opens the field of quantum optics of quantum gases  for ultracold molecular gases and raises intriguing questions about the quantum measurement backaction and preparation of the exotic many-body phases using the entanglement between the light and many-body molecular states.


\section{Concluding remarks of Chapter 1}

In this chapter, we formulated a rather general model for interaction between the quantized light and ultracold atoms trapped in an optical lattices. We will continue using this model in next chapters as well. 

We showed that scattering from different atomic quantum states creates different quantum states of scattered light, which can be distinguished by measurements of the spatial intensity distribution, quadrature variances, photon statistics, or spectral measurements. In particular, angle-resolved
intensity measurements reflect global statistics of atoms (total number of radiating atoms) as well as local statistical quantities (single-site statistics even without an optical access to a single site) and pair correlations between different sites. As a striking example we considered scattering from transversally illuminated atoms into an optical cavity mode. For the Mott insulator state, similar
to classical diffraction, the number of photons scattered into a cavity is zero due to destructive interference, while for the superfluid state it is nonzero and proportional to the number of
atoms. Moreover, we showed that all three phases in the Bose glass -- Mott insulator -- superfluid phase transition can be distinguished by light scattering. We demonstrated that light scattering can have a nontrivial angle dependence, including the appearance of narrow features at angles, where classical diffraction predicts zero. We derived a generalized Bragg condition for such cases. We proved that the transmission spectroscopy can map the full distribution functions of different variables of a quantum gas.  

We showed that light scattering reveals not only density correlations, but also the matter-field interference at its shortest possible distance in an optical lattice, which defines key properties such as tunnelling and matter-field phase gradients. We demonstrated probing the order parameters, matter-field quadratures, and their squeezing. We further extended our model for fermionic atoms and few-body complexes of polar molecules, demonstrating their nondestructive detection. 

The measurement procedure presented here corresponds to the quantum nondemolition (QND) measurement of various atomic variables observing light. Nevertheless, as quantum mechanics states, any measurement (even a QND one) affects the system, which we neglected in this chapter. In the next chapter we will take into account such a measurement backaction and will use the measurement as an active tool to prepare nontrival many-body atomic states. 

\clearpage

%% file: Chapter2.tex
\chapter{Weak quantum measurements as a tool for preparing many-body atomic states} \label{chapt2}

\section{Introduction and plan of the chapter}

In all problems considered above, the physical mechanism of light scattering establishes a relation between the light and matter observables. As we have shown, the light provides information about various statistical quantities of the quantum states of atoms: different correlation functions (given by the expectation values of some operators) and distribution functions of different variables can be measured. As it is usual in quantum mechanics, the determination of such statistical quantities requires multiple measurements. Therefore, the repeated preparation of the initial state is necessary, because any quantum measurement (even a QND one) generally affects the quantum state of the system. Indeed, such nondestructive methods are already advantageous in comparison to the standard destructive methods, because the preparation of the initial state can be done with the same atomic sample. (The destructive methods destroy the atomic system completely and require new atoms.) This can be important, as the atoms can carry additional information in other degrees of freedom (e.g. electronic or spin excitation), which will be preserved, when the initial motional state is prepared with the same atoms. In the case of destructive detection, such an additional information will be lost with the atomic sample. 

In this chapter, we will address the problem from another point of view going beyond the standard goal of measuring the expectation values and distribution functions. Here, we focus on system dynamics during a single run of the optical measurement (i.e. the continuous detection of scattered photons), without taking the average (expectation values) over many realizations as we were doing so far. Indeed, the result of a single-run measurement is important, as it is the first result one obtains in an experiment before the averaging procedure.

During the interaction, the light and matter get entangled. According to quantum mechanics (e.g. the paradigmatic Einstein--Podolsky--Rosen (EPR) paradox), due to the entanglement, the measurement of one of the quantum subsystems (light) will affect the other quantum subsystem (atoms) as well. This is an example of the quantum measurement backaction. While the measurement backaction often is an unwanted perturbation, we will use it to change the atomic state in a desired way. We will consider the light measurement as an active method to prepare particular many-body quantum states of ultracold atoms.

In Sec. 2.2 \cite{MekhovPRA2009, Mekhov2012} we will develop a general model of the quantum measurement backaction on atoms in a lattice, when the light is detected. We will show, how the atomic state changes at a single experimental run (quantum trajectory), which consists of quantum jumps and the non-Hermitian evolution between them.

In Sec. 2.3 \cite{MekhovPRL2009, MekhovPRA2009, MekhovLP2011, MekhovLP2011} we will show, that choosing the optical geometry enables selecting the class of emerging quantum many-body atomic states. Light detection along the
angle of a diffraction maximum (Bragg angle) creates an atom-number
squeezed state, while light detection at diffraction minima leads to
the macroscopic superposition states (Schr{\"o}dinger cat states) of
different atom numbers.

In Sec. 2.4 \cite{MekhovPRA2009} we will show how various types of photon statistics (both conditional and unconditional ones) change during the measurement and thus during the process of quantum state reduction. 

In Sec. 2.5 \cite{MekhovPRA2009} we consider the measurement of cavity transmission. We show that it can lead to both atom-number squeezed and macroscopic superposition states depending on its outcome. We find that the transmission measurement yields more robust
and controllable superposition states than the ones obtained by transverse 
scattering at a diffraction minimum.

In Sec. 2.6 \cite{MekhovPRA2009, MekhovLP2010} we analyze the robustness of the macroscopic superposition states with respect to the limited efficiency of a photodetector and thus photon losses. We demonstrate that even if several photons are lost, the preparation of the superposition states is still of interest. 

In Sec. 2.7 \cite{Atoms} we will generalize our model for ultracold fermions and show that the preparation of nontrivial states of spin magnetization is possible by quantum measurements. 

In Sec. 2.8 \cite{Atoms} we will describe the quantum measurements of light quadratures rather than direct detection of photons considered so far. We will show the preparation of cat states, where the state robustness can be changed by selecting appropriate light quadratures. 

In Sec. 2.9 \cite{Elliott2015} we will show how to generate multiple many-body spatial modes of ultracold atoms. The multipartite mode entanglement properties and their nontrivial spatial overlap can be varied by tuning the optical geometry in a single setup. This can be used to engineer quantum states and dynamics of matter fields. We will provide examples of multimode generalizations of parametric down-conversion, Dicke, and other states, investigate the entanglement properties of such states, and show how they can be transformed into a class of atomic generalized squeezed states, which can be hardly obtained in quantum optics of light.

We will conclude this chapter in Sec. 2.10 and make a link to the next chapter of this work. 

{\it The results of this chapter are based on the papers}  \cite{MekhovPRL2009, MekhovPRA2009, MekhovLP2010, MekhovLP2011, Mekhov2012,Elliott2015, Atoms}.


\section{Model of the quantum measurement backaction on the many-body atomic system in a lattice}

In this section we will derive a model for the system shown in Fig. 1.1. We will show, how the quantum man-body state of atoms will change during the continous measurement of photons leaking from the cavity. Even early theoretical
works on scattering of quantized light from a BEC was not realized
so far \cite{Moore1999,PuPRL2003,YouPRA1995,IdziaszekPRA2000,MustPRA2000,MustPRA2000x2,
JavanainenPRL1995,JavanainenPRA1995,CiracPRL1994,CiracPRA1994,SaitoPRA1999, PratavieraPRA2004,JavanainenPRL2003}. However, it
is the setup with a cavity and optical lattice that will
provide the best interplay between the atom- and light-stimulated
quantum effects.

We start with the Hamiltonian Eqs.~(1.9), where the influence of atomic tunneling on light scattering has been neglected
\begin{gather}
H=\hbar(\omega_1 + U_{11} \hat{D}_{11}) a^\dag_1 a_1+
\hbar U_{10}(\hat{D}^*_{10}a^*_0a_1 + \hat{D}_{10}a_0a^\dag_1) 
-i\hbar(\eta_1^* a_1 - \eta_1 a^\dag_1).\label{2-2-1PRA09}
\end{gather}

The tunneling of atoms in the lattice
potential plays indeed an important role in establishing a
particular quantum atomic state. However, after the state is
established, one can assume that the scattering of the probe occurs
on the time-scale faster than slow tunneling. Thus, from the light
scattering point of view, the atomic distribution can be considered
frozen and the tunneling is not important. It is especially
reasonable to neglect the tunneling dynamics in this section, because
we will show that even after that rather obvious dynamics is
neglected, there is still nontrivial dynamics (quantum jumps and non-Hermitian evolution)
exclusively associated with the quantum measurement process, which
is the main subject of this section. Moreover, in Chapter 3 we will show that the measurement backaction constitutes a novel source of competitions in many-body physics (in addition to the standard tunneling and atom-atom interaction).

The Hamiltonian (\ref{2-2-1PRA09}) describes QND measurements of the
variables related to $\hat{D}_{lm}$ measuring the photon number
$a^\dag_1a_1$. Note, that one has a QND access to various
many-body variables, as $\hat{D}_{lm}$ strongly depend on the
lattice and light geometry via $u_{0,1}({\bf r})$. This is an
advantage of the lattice comparing to single- or double-well setups,
where the photon measurement backaction was considered
\cite{RuostekoskiPRA1998,Milburn1998,Onofrio,HuangPRA2006}. Moreover, such a geometrical
approach can be extended to other quantum arrays, e.g., ion strings, arrays of superconducting qubits, Rydberg atoms, etc.

We will present the solution for the quantum state of the
coupled light--matter system including the measurement process.
Importantly, it is possible to obtain an analytical solution with a
very transparent physical meaning thanks to the approximations used
(slow tunneling and coherent state of the external probes).

The initial motional state of the ultracold atoms trapped in the
periodic lattice potential at the time moment $t=0$ can be
represented as

\begin{eqnarray}\label{2-2}
|\Psi^a(0)\rangle =\sum_{q}c_q^0 |q_1,..,q_M\rangle,
\end{eqnarray}
which is a quantum superposition of the Fock states corresponding to
all possible classical configurations $q=\{q_1,..,q_M\}$ of $N$
atoms at $M$ sites, where $q_j$ is the atom number at the site $j$.
For each classical configuration $q$, the total atom number is
conserved: $\sum_j^M q_j=N$. This superposition displays the
uncertainty principle, stating that at ultralow temperatures even a
single atom can be delocalized in space, i.e., there is a
probability to find an atom at any lattice site. We will show, how
this atomic uncertainty is influenced by the light detection.

For example, for a limiting case of the MI state, where the atom
numbers at each lattice site are precisely known, only one Fock
state will exist in Eq.~(\ref{2-2}): $|\Psi_\text{MI}\rangle
=|1,1,..,1\rangle$ for the MI with one atom at each site. On the
other hand, the SF state is given by the superposition of all
possible classical configurations with multinomial coefficients:

\begin{eqnarray}\label{2-3}
|\Psi^a_\text{SF}\rangle=\frac{1}{(\sqrt{M})^N}\sum_{q}
\sqrt{\frac{N!}{q_1!q_2!...q_M!}}|q_1,q_2,..q_M\rangle.
\end{eqnarray}
Thus, the atom number at a single site as well as the atom number at
$K<M$ sites are uncertain in the SF state.

As an initial condition, we assume that at the time moment $t=0$ the
light and matter are disentangled, and the initial state of light is
a coherent state with the amplitude $\alpha_0$. Thus, the initial
quantum state of the system is given by the product state
$|\Psi(0)\rangle=|\Psi^a(0)\rangle |\alpha_0\rangle$. In particular,
initially, the light can be in the vacuum state $|0\rangle$.

We use the open system approach \cite{Carmichael} to describe the
continuous counting of  photons leaking out the cavity of the
cavity decay rate $\kappa$. According to this approach, when a
photon is detected at the moment $t_i$, the quantum jump occurs, and
the state instantaneously changes to a new one obtained by applying
the cavity photon annihilation operator $|\Psi_c(t_i)\rangle
\rightarrow a_1|\Psi_c(t_i)\rangle$ and renormalization (the
subscript $c$ underlines that we deal with the state conditioned on
the photocount event). Between the photocounts, the system evolves
with a non-Hermitian Hamiltonian $H-i\hbar\kappa a^\dag_1a_1$. Such
an evolution gives a quantum trajectory for $|\Psi_c(t)\rangle$
conditioned on the detection of photons at times $t_1,t_2,...$. The
probability of the photon escape within the time interval $t$ is
$2\kappa t\langle a^\dag_1a_1\rangle_c$, where $\langle
a^\dag_1a_1\rangle_c$ is the conditional photon number in the
cavity, i.e., the photon number calculated for the conditional
quantum state $|\Psi_c(t)\rangle$.

The state $|\Psi_c(t)\rangle$ should be found by solving the
Schr{\"o}dinger equation with the non-Hermitian Hamiltonian for
no-count intervals and applying the jump operator $a_1$ at the
moments of photocounts. Thanks to the slow tunneling approximation,
the Hamiltonian (\ref{2-2-1PRA09}) does not mix the Fock states in the
expression (\ref{2-2}). So, the problem is significantly reduced to
separate finding solutions for each classical atomic configuration
$q=\{q_1,..,q_M\}$, after that the full solution will be given by
the superposition of those solutions.

The next simplification appears thanks to the use of the external
probes in the coherent state. It is
known~\cite{DenisOC2007,GardinerZoller} that, if a coherent probe
illuminates a classical atomic configuration $q$ in a cavity, the
light remains in a coherent state, however, with some pre-factor:
$\exp[\Phi_q(t)]|\alpha_q(t)\rangle$. The pre-factor is indeed not
important for a single classical configuration as it disappears
after the renormalization, but it will play a role, when the
superposition of classical solutions with different pre-factors will
be considered. Moreover, the light amplitude $\alpha_q(t)$ is simply
given by the solution of a classical Maxwell's equation:
\begin{gather}
\alpha_q(t)=\frac{\tilde{\eta_1}-iU_{10}\tilde{a}_0D^q_{10}}{i(U_{11}
D^q_{11}-\Delta_p)+\kappa}e^{-i\omega_pt}+ 
\left(\alpha_0 - \frac{\tilde{\eta_1}-iU_{10}
\tilde{a}_0D^q_{10}}{i(U_{11}
D^q_{11}-\Delta_p)+\kappa}\right)e^{-i(\omega_1+U_{11}D^q_{11})t-\kappa
t},\label{2-4}
\end{gather}
where we introduced the constant probe amplitudes $\tilde{a}_0$ and
$\tilde{\eta_1}$ as $a_0=\tilde{a}_0\exp(-i\omega_pt)$ and
$\eta_1=\tilde{\eta_1}\exp(-i\omega_pt)$; $D^q_{lm}=
\sum_{j=1}^K{u_l^*({\bf r}_j)u_m({\bf r}_j)q_j}$ is a realization of
the operator $\hat{D}_{lm}$ at the configuration $q=\{q_1,..q_M\}$.
As in classical optics, the first term in Eq.~(\ref{2-4}) gives the
oscillations at the probe frequency, while the second term gives the
transient process with the oscillations at the cavity frequency
shifted by the dispersion, $\omega_1+U_{11}D^q_{11}$, which decays
with the rate $\kappa$. In the following, we introduce the slowly
varying light amplitude $\tilde{\alpha}_q(t)$ as
$\alpha_q(t)=\tilde{\alpha}_q(t)\exp(-i\omega_pt)$ and, for the
notation simplicity, will drop the tilde sign in all amplitudes
$\tilde{a}_0$, $\tilde{\eta_1}$, and $\tilde{\alpha}_q(t)$.

The function $\Phi_q(t)$ in the pre-factor is a complex one and is
given by
\begin{eqnarray}\label{2-5}
\Phi_q(t)=\int_0^t\left[\frac{1}{2}(\eta_1\alpha^*_q-iU_{10}
a_0D^q_{10}\alpha^*_q-\text{c.c.})-\kappa|\alpha_q|^2\right]dt,
\end{eqnarray}
where the light amplitude $\alpha_q(t)$ is given by Eq.~(\ref{2-4}).

First, let us consider the solution for the atomic state initially
containing a single Fock state $|q_1,..,q_M\rangle$. The solution
for light is given by the solution for the classical configuration
$q=\{q_1,..,q_M\}$. So, the evolution is given by the product state
$|q_1,..,q_M\rangle|\alpha_q(t)\rangle$. An important property of
this solution is that a quantum jump does not change the state,
since applying the jump operator $a_1$ simply leads to the
pre-factor $\alpha_q(t)$, which disappears after the
renormalization. Therefore, even in the presence of photocounts
(i.e. quantum jumps), the time evolution is continuous and is given
by Eq.~(\ref{2-4}): after a transient process for $t<1/\kappa$, the
steady state for $\alpha_q(t)$ is achieved. Note, that in contrast
to many problems in quantum optics \cite{Carmichael}, where the
steady state is a result of averaging over many quantum trajectories,
here, the steady state appears even at a single quantum trajectory
for $t>1/\kappa$. This is a particular property of the coherent
quantum state. The continuity of evolution of the state
$|q_1,..,q_M\rangle|\alpha_q(t)\rangle$, and, hence, more generally,
the unnormalized state including the pre-factor,
$\exp[\Phi_q(t)]|q_1,..,q_M\rangle|\alpha_q(t)\rangle$,
independently of the presence of  quantum jumps provides us a
significant mathematical simplification. Moreover, we will use the
result that after the time $t>1/\kappa$, all light amplitudes
$\alpha_q$ are constant for all Fock states $|q_1,..,q_M\rangle$.

Let us now consider the full initial state given by the
superposition (\ref{2-2}). As stated, the evolution of each term is
independent and contains the continuous part
$\exp[\Phi_q(t)]|q_1,..,q_M\rangle|\alpha_q(t)\rangle$. Thus,
applying the jump operators at the times of the photodetections
$t_1,t_2,..t_m$ leads to the following analytical solution for the
conditional quantum state at the time $t$ after $m$ photocounts:
\begin{gather}
|\Psi_c(m,t)\rangle
=\frac{1}{F(t)}\sum_{q}\alpha_q(t_1)\alpha_q(t_2)...\alpha_q(t_m)
  e^{\Phi_q(t)}c_q^0|q_1,...,q_M\rangle|\alpha_q(t)\rangle,\label{2-6}
\end{gather}
where
\begin{eqnarray}
F(t)=
\sqrt{\sum_{q}|\alpha_q(t_1)|^2|\alpha_q(t_2)|^2...|\alpha_q(t_m)|^2
|e^{\Phi_q(t)}|^2|c_q^0|^2} \nonumber
\end{eqnarray}
is the normalization coefficient.

In contrast to a single atomic Fock state, the solution (\ref{2-6}),
in general, is not factorizable into a product of the atomic and
light states. Thus, in general, the light and matter are entangled.
Moreover, in contrast to a single Fock state, the quantum jump
(applying $a_1$) changes the state, and the evolution of the full
$|\Psi_c(m,t)\rangle$ is not continuous.

The general solution (\ref{2-6}) valid for all times simplifies
significantly for $t>1/\kappa$, when all $\alpha_q$ are constants,
and if the first photocount occurred at $t_1>1/\kappa$, when all
$\alpha_q$ has already become constants. The latter assumption is
especially probable for the small cavity photon number, since the
probability of the photon escape within the time interval $t$ is
$2\kappa t\langle a^\dag_1a_1\rangle_c$. This solution takes the
form
\begin{gather}
|\Psi_c(m,t)\rangle =\frac{1}{F(t)}\sum_{q}\alpha_q^m e^{\Phi_q(t)}
c_q^0 |q_1,...,q_M\rangle|\alpha_q\rangle, \label{2-7}\\
\alpha_q=\frac{\eta_1-iU_{10} a_0D^q_{10}}{i(U_{11}
D^q_{11}-\Delta_p)+\kappa}, \label{2-8}\\
\Phi_q(t)=-|\alpha_q|^2\kappa t+(\eta_1\alpha^*_q-iU_{10}
a_0D^q_{10}\alpha^*_q-\text{c.c.})t/2, \label{2-9}
\end{gather}
with the normalization coefficient
\begin{eqnarray}
F(t)= \sqrt{\sum_{q}|\alpha_q|^{2m}e^{-2|\alpha_q|^2\kappa
t}|c_q^0|^2}. \nonumber
\end{eqnarray}

The solution (\ref{2-7}) does not depend on the photocount times
$t_1,t_2,..t_m$ any more. Note however, that even for $t>1/\kappa$,
when all $\alpha_q$ reached their steady states and are constants,
the solution (\ref{2-7}) is still time-dependent. Thus, the time
$t=1/\kappa$ is not a characteristic time scale for the steady state
of the full solution (\ref{2-6}) and (\ref{2-7}). The stationary light
amplitudes $\alpha_q$ in (\ref{2-8}) are given by the Lorentz
functions in the absolute correspondence with classical optics. The
function $\Phi_q(t)$ has also simplified and contains the first real
term responsible for the amplitudes of the coefficients in the
quantum superposition, and the second imaginary term responsible for
their phases.

The solutions (\ref{2-6}) and (\ref{2-7}) show, how the probability to
find the atomic Fock state $|q_1,..,q_M\rangle$ (corresponding to
the classical configuration $q$) changes in time. Such a change in
the atomic quantum state appears essentially due to the measurement
of photons and is a direct consequence of the light--matter
entanglement: according to quantum mechanics, by measuring one of
the entangled subsystem (light) one also affects the state of
another subsystem (atoms). Now we can focus on that purely
measurement-base dynamics, since other obvious sources of the
time-evolution (e.g. tunneling) were neglected in our model. (We will include the tunneling in Chapter 3). The
initial probability to find the Fock state $|q_1,..,q_M\rangle$ is
$p_q(0)=|c^0_q|^2$. From Eq.~(\ref{2-6}) the time evolution of this
probability is given by
\begin{gather}
p_q(m,t)= |\alpha_q(t_1)|^2|\alpha_q(t_2)|^2...|\alpha_q(t_m)|^2
|e^{\Phi_q(t)}|^2p_q(0)/F^2(t).\label{2-10}
\end{gather}
For $t,t_1>1/\kappa$ [cf. Eq.~(\ref{2-7})] it reduces to
\begin{eqnarray}\label{2-11}
p_q(m,t)= |\alpha_q|^{2m}e^{-2|\alpha_q|^2\kappa t}p_q(0)/F^2(t).
\end{eqnarray}

In the following, we will demonstrate the applications of the
general solutions (\ref{2-6}) and (\ref{2-7}) in the examples, where,
for simplicity, only a single statistical quantity is important,
instead of the whole set of all possible configurations $q$. As
particular examples, we will consider the cases, where that
statistical quantity (let us now call it $z$) is the atom number at
$K$ lattice sites or the atom number difference between odd and even
sites. Thus, instead of the huge number of detailed probabilities
$p_q(m,t)$, we will be interested in the probability $p(z,m,t)$ to
find a particular value of $z$. For the initial state (\ref{2-2}) at
$t=0$, $p_0(z)=\sum_{q'} |c_{q'}^0|^2$ with the summation over all
configurations $q'$ having the same $z$. As under our assumptions
all light amplitudes $\alpha_q(t)$ depend only on $z$, we change
their subscript to $z$ and write the probability to find a
particular value of $z$ at time $t$ after $m$ photocounts:
\begin{gather}
p(z,m,t)= |\alpha_z(t_1)|^2|\alpha_z(t_2)|^2...|\alpha_z(t_m)|^2
e^{2\text{Re}\Phi_z(t)}p_0(z)/F^2(t),\label{2-12}
\end{gather}
For $t,t_1>1/\kappa$ it reduces to
\begin{eqnarray}\label{2-13}
p(z,m,t)= |\alpha_z|^{2m}e^{-2|\alpha_z|^2\kappa
t}p_0(z)/F^2(t), \\
F(t)= \sqrt{\sum_{z}|\alpha_z|^{2m}e^{-2|\alpha_z|^2\kappa
t}p_0(z)}. \nonumber
\end{eqnarray}

In the following we will consider the solution (\ref{2-7}). When the
time progresses, both $m$ and $t$ increase with an essentially
probabilistic relation between them. The Quantum Monte Carlo method
\cite{Carmichael} establishes such a relation, thus giving a quantum
trajectory. Note, that thanks to the simple analytical solution
(\ref{2-7}), the method gets extremely simple. The evolution is split
into small time intervals $\delta t_i$. In each time step, the
conditional photon number is calculated in the state Eq.~(\ref{2-7}),
and the probability of the photocount within this time interval
$2\kappa \langle a^\dag_1a_1\rangle_c \delta t_i$ is compared with a
random number $0<\epsilon_i<1$ generated in advance, thus, deciding
whether the detection (if $2\kappa \langle a^\dag_1a_1\rangle_c
\delta t_i>\epsilon_i$) or no-count process (otherwise) has
happened.


\section{Preparation of the atom-number squeezed and Schr{\"o}dinger cat states by transverse probing}

In this section we will show, that choosing the optical geometry enables selecting the class of emerging quantum many-body atomic states. Light detection along the
angle of a diffraction maximum (Bragg angle) creates an atom-number
squeezed state, while light detection at diffraction minima leads to
the macroscopic superposition states (Schr{\"o}dinger cat states) of
different atom numbers in the cavity mode.

We will consider a case, where only the transverse
probe $a_0$ is present, while the probe through the mirror does not
exist, $\eta_1=0$. We neglect the dispersive frequency shift assuming
that $U_{11} D^q_{11}\ll\kappa$ or $\Delta_p$. Thus, the light
amplitudes $\alpha_q$ will only depend on the quantity $D^q_{10}$.
In Sec. 2.5, in contrast, we will focus on the case, where the
dispersive mode shift is very important.

\subsection{Atom-number squeezing at the diffraction maximum}

The condition of a diffraction maximum for the scattering of light
from the probe wave $a_0$ into the cavity mode $a_1$ is the
following: the atoms at all lattice sites scatter the light in phase
with each other. For the plain standing or traveling waves, this
condition means that in the expression for the operator
$\hat{D}_{10}= \sum_{j=1}^K{u_1^*({\bf r}_j)u_0({\bf
r}_j)\hat{n}_j}$, $u_1^*({\bf r}_j)u_0({\bf r}_j)=1$ for all sites
$j$. Thus, $\hat{D}_{10}=\hat{N}_K$ is reduced to the operator of
the atom number at $K$ illuminated sites. In Eq.~(\ref{2-7}), after
neglecting the dispersive frequency shift, the only statistical
quantity is $D^q_{10}$ giving the atom number at $K$ sites for the
configuration $q$. We will call this single statistical quantity as
$z$: $D^q_{10}=z$, which varies between 0 and $N$ reflecting all
possible realizations of the atom number at $K$ sites.

From Eq.~(\ref{2-8}), the light amplitudes in the diffraction maximum
are proportional to the atom number $z$:
\begin{eqnarray}\label{2-14}
\alpha_z=Cz, \text {with   }
C=\frac{iU_{10}a_0}{(i\Delta_p-\kappa)}.
\end{eqnarray}
Thus, the probability to find the atom number $z$ is given from
Eq.~(\ref{2-13}) by
\begin{eqnarray}\label{2-15}
p(z,m,t)=z^{2m}e^{-z^2\tau}p_0(z)/\tilde{F}^2,
\end{eqnarray}
where we introduced the dimensionless time $\tau=2|C|^2\kappa t$ and
new normalization coefficient $\tilde{F}$ such that $\sum_{z=0}^N
p(z,m,t)=1$. When time progresses, both $m$ and $\tau$ increase with a
probabilistic relation between them.

If the initial atom number $z$ at $K$ sites is uncertain, $p_0(z)$
is broad. For the SF state the probability to find the atom number
$z$ at the lattice region of $K$ sites is given by the binomial
distribution
\begin{eqnarray}\label{2-16}
p_\text{SF}(z)=\frac{N!}{z!(N-z)!}\left(\frac{K}{M}\right)^z
\left(1-\frac{K}{M}\right)^{N-z}.
\end{eqnarray}
For a lattice with the large atom and site numbers $N,M\gg1$, but
finite $N/M$, it can be approximated as a Gaussian distribution
\begin{eqnarray}\label{2-17}
p_\text{SF}(z)=\frac{1}{\sqrt{2\pi}\sigma_z}e^{-\frac{(z-z_0)^2}{2\sigma_z^2}}
\end{eqnarray}
with the mean atom number $\langle \hat{N}_K\rangle = z_0=NK/M$ and
$\sigma_z=\sqrt{N(K/M)(1-K/M)}$ giving the full width at a half
maximum (FWHM) $2\sigma_z\sqrt{2\ln 2}$. The atom number variance in
the SF state is $(\Delta N_K)^2=\langle \hat{N}_K^2\rangle - \langle
\hat{N}_K\rangle^2=\sigma_z^2$.

\begin{figure}[h!]
\centering
\captionsetup{justification=justified}
\includegraphics[width=0.6\textwidth]{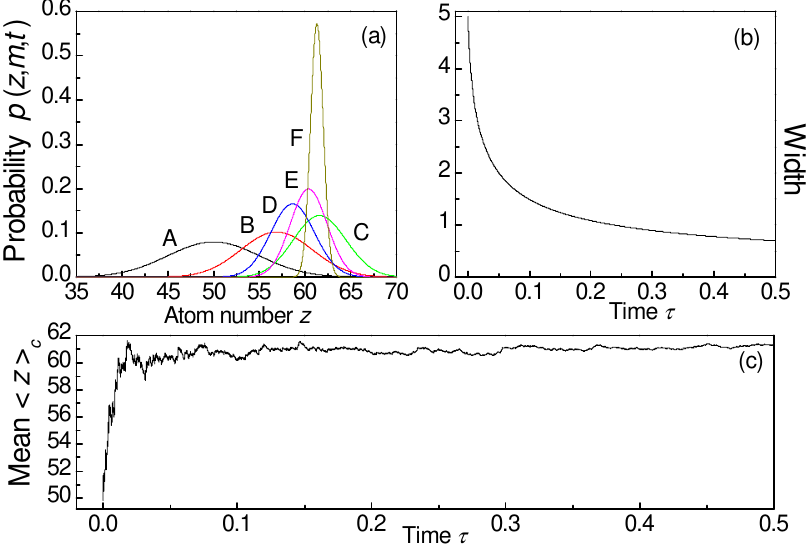}
\caption{\label{2-fig2PRL09}Photodetections at diffraction
maximum. (a) Shrinking atom number distribution at different times
$\tau=$ 0, 0.005, 0.018, 0.03, 0.05, 0.5 (A-F); (b) decreasing width
$\delta z$; (c) stabilizing mean atom number $\langle z\rangle_c$.
Initial state: SF, $N=100$ atoms, $K=M/2=50$ illuminated sites.}
\end{figure}

Eq.~(\ref{2-15}) shows how the initial distribution $p_0(z)$ changes
in time. The function $z^{2m}\exp(-z^2\tau)$ has its maximum at
$z_1=\sqrt{m/\tau}$ and the FWHM $\delta z \approx
\sqrt{2\ln2/\tau}$ (for $\delta z \ll z_1$). Thus, multiplying
$p_0(z)$ by this function will shrink the distribution $p(z,m,t)$ to
a narrow peak at $z_1$ with the width decreasing in time (Fig.~\ref{2-fig2PRL09}).

Physically, this describes the projection of the atomic quantum
state to a final state with the squeezed atom number at $K$ sites (a
Fock state $|z_1,N-z_1\rangle$ with the precisely known atom number at
$K$ sites $z_1$ and $N-z_1$ atoms at $M-K$ sites).  In contrast to the results in spin squeezing \cite{Eckert2007},
which can be also obtained for thermal atoms
\cite{PolzikHot,Holland}, in our work, the quantum nature of ultracold
atoms is crucial, as we deal with the atom number fluctuations
appearing due to the delocalization of ultracold atoms in space.

After the distribution shrinks to a single $z_1$, the light
collapses to a single coherent state $|\alpha_{z_1}\rangle$, and the
atoms and light get disentangled with a factorized state
\begin{eqnarray}\label{2-18}
|\Psi_c\rangle=|z_1,N-z_1\rangle|\alpha_{z_1}\rangle.
\end{eqnarray}
 {\it A priori} $z_1$ is unpredictable. However,
measuring the photon number $m$ and time $t$, one can determine
$z_1$ of the quantum trajectory.

As the final state contains only one coherent state of light, the light statistics evolves from super-Poissonian to Poissonian.
The conditioned (i.e., at a single trajectory) cavity photon number
$\langle a^\dag_1a_1\rangle_c(t)=|C|^2 \sum_{z=0}^Nz^2p(z,m,t)$ is
given by the second moment of $p(z,m,t)$. Its dynamics [very similar
to $\langle z\rangle_c$ in Fig.~\ref{2-fig2PRL09}(c)] has jumps, even
though all $\alpha_{z}(t)$ are continuous. In the no-count process,
$\langle a^\dag_1a_1\rangle_c$ decreases, while at one-count it
jumps upwards, which is a signature of super-Poissonian statistics.
Finally, it reduces to $\langle a^\dag_1a_1\rangle_c=|C|^2z_1^2$,
reflecting a direct correspondence between the final atom number and
cavity photon number, which is useful for experiments.

Even the final Fock state (\ref{2-18}) can contain the significant
atom-atom entanglement, as this is still a many-body state. In
general, the Fock state $|z_1,N-z_1\rangle$ cannot be factorized
into the product of two states for $K$ illuminated and $M-K$
unilluminated lattice sites. Thus, the entanglement can survive even
between two lattice regions, which depends on the value of $z_1$
realized at a particular quantum trajectory. For some cases, the
factorization is possible. For
example, the SF state can be represented as $|SF\rangle_{N,M}=\sum_z
\sqrt{B_z} |SF\rangle_{z,K}|SF\rangle_{N-z,M-K}$ ($B_z$ are binomial
coefficients). After the measurement, it ends up in
$|SF\rangle_{z_1,K}|SF\rangle_{N-z_1,M-K}$, i.e., the product of two
uncorrelated superfluids.

Our measurement scheme determines (by squeezing) the atom number at
a particular lattice region and projects the initial atomic state to
some subspace. However, the atom number at different regions keeps
quantum uncertainty. So, the quantum structure of the final state
can be revealed in a further optical or matter-wave experiment.
Thanks to the lattice geometry, one can change the illuminated
region, and further study measurement-induced collapse of the state
in the remaining subspace.

Note, that our model does not specify how $K$ sites were selected,
which is determined by the lattice and light geometry. The simplest
case is to illuminate a continuous region. However, one can also
illuminate each second site by choosing the probe wavelength twice
as lattice period and get an atom number squeezing at odd and even
sites. In this way, one gets a measurement-prepared product of two
SFs ``loaded'' at sites one by one (e.g. atoms at odd sites belong
to one SF, while at even sites to another). While the initial SF, as
usual, shows the long-range coherence $\langle b^\dag_i b_j\rangle$
with the lattice period, the measurement-prepared state will
demonstrate the doubled period in $\langle b^\dag_i b_j\rangle$
($b_j$ is the atom annihilation operator such that
$b^\dag_jb_j=\hat{n}_j$). Thus, even though our model does not
include the matter filed operators $b_j$, but only the atom number
operators $\hat{n}_j$, the matter coherence can be still affected
and modified by our QND measurement scheme in a nontrivial way.

This example shows, that the phase of the matter field can be manipulated by this type of QND measurements, even though we measure only the occupation number-related operators. This is a direct consequence of the many-body nature of the atomic state. As usual in the QND measurements, by measuring the number-related variables, one typically destroys the conjugate variable (in this case, the phase). However, here we have a very rich choice of the number-related variables for the measurement. Measuring some of the variables, other variables can be left untouched. Therefore, one can carefully destroy only particular phase information in our system, while other phase-related variables stay unaffected by the measurement. This is demonstrated in the last example, where the phase coherence survives after the measurement, but changes its period. In this example, the coherence between even sites was preserved (as well as between the odd sites), but the coherence between even and odd sites was destroyed, since they belong to different superfluid states.

\subsection{Schr{\"o}dinger cat states at the diffraction minimum}

The condition of a diffraction minimum for
the scattering of light from the probe wave $a_0$ into the cavity
mode $a_1$ is the following: the atoms at the neighboring lattice
sites scatter the light with the phase difference $\pi$. For the
plain standing or traveling waves, this condition means that in the
expression for the operator $\hat{D}_{10}= \sum_{j=1}^K{u_1^*({\bf
r}_j)u_0({\bf r}_j)\hat{n}_j}$, $u_1^*({\bf r}_j)u_0({\bf
r}_j)=(-1)^{j+1}$. Thus, $\hat{D}_{10}=\sum_{j=1}^M
(-1)^{j+1}\hat{n}_j$ is the operator of atom number difference
between odd and even sites (in this subsection, we consider all
sites illuminated, $K=M$). Similarly to the diffraction maximum case, in
Eq.~(\ref{2-7}), after neglecting the dispersive frequency shift, the
only statistical quantity is $D^q_{10}$ giving the atom number
difference for the configuration $q$. We will call this single
statistical quantity as $z$: $D^q_{10}=z$, which varies between $-N$
and $N$ with a step 2 reflecting all possible realizations of the
atom number difference.

Equations (\ref{2-14}) and (\ref{2-15}) keep their form for the
diffraction minimum as well, however, with a different meaning of
the statistical variable $z$, which is now a realization of the atom
number difference, and $p(z,m,t)$ is its probability.

For the SF state the probability to find the atom number at odd (or
even) sites $\tilde{z}$ [$\tilde{z}=(z+N)/2$ because the atom number
difference is $z$ and the total atom number is $N$] is given by the
binomial distribution
\begin{eqnarray}\label{2-19}
p_\text{SF}(\tilde{z})=\frac{N!}{\tilde{z}!(N-\tilde{z})!}
\left(\frac{Q}{M}\right)^{\tilde{z}}
 \left(1-\frac{Q}{M}\right)^{N-\tilde{z}},
\end{eqnarray}
where $Q$ is the number of odd (or even) sites. For even $M$,
$Q=M/2$ and Eq.~(\ref{2-19}) simplifies. For a lattice with the large
atom and site numbers $N,M\gg1$, but finite $N/M$, this binomial
distribution, similarly to the previous subsection, can be
approximated by a Gaussian function. Changing the variable as
$z=2\tilde{z}-N$ we obtain the Gaussian function for the probability
to find the atom number difference $z$:
\begin{eqnarray}\label{2-20}
p_\text{SF}(z)=\frac{1}{\sqrt{2\pi}\sigma_z}e^{-\frac{z^2}{2\sigma_z^2}}
\end{eqnarray}
with the zero mean $z$ and $\sigma_z=\sqrt{N}$ giving the FWHM of 
$2\sigma_z\sqrt{2\ln 2}$. The variance of the atom number difference
in the SF state is $\sigma_z^2=N$.

The striking difference from the diffraction maximum is that our
measurement and the probability (\ref{2-15}) are not sensitive to the
sign of $z$, while the amplitudes $\alpha_z=Cz$ are. So, the final
state obtained from Eq.~(\ref{2-7}) is a superposition of two Fock
states with $z_{1,2}=\pm \sqrt{m/\tau}$ and different light
amplitudes: $\alpha_{z_2}=-\alpha_{z_1}$,
\begin{eqnarray}\label{2-21}
|\Psi_c\rangle=\frac{1}{\sqrt{2}}(|z_1\rangle|\alpha_{z_1}\rangle+
(-1)^m|-z_1\rangle|-\alpha_{z_1}\rangle).
\end{eqnarray}

\begin{figure}[h!]
\centering
\captionsetup{justification=justified}
\includegraphics[width=0.6\textwidth]{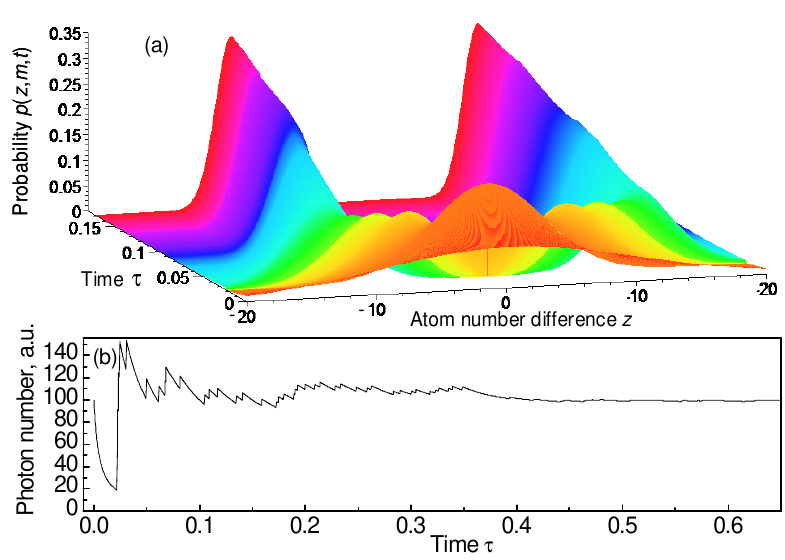}
\caption{\label{2-JPBcat} Photodetection at diffraction minimum leading to the generation of a Schr{\"o}dinger cat state. Results for a single quantum trajectory (Quantum Monte Carlo simulations). (a) Shrinking  distribution of the atom-number difference between odd and even sites for various times. The doublet corresponds to Schr{\"o}dinger cat state. (b) Conditional photon number with quantum jumps. Initial state: SF, $N=100$ atoms, $K=M=100$ sites.}
\end{figure}

Figure \ref{2-JPBcat} shows the collapse to a doublet probability $p(\pm
z_{1},m,t)$ and the photon-number trajectory, where upward jumps and
no-count decreases can be seen.

In contrast to a maximum, even in the final state, the light and
matter are not disentangled. In principle, one can disentangle light
and matter by switching off the probe and counting all leaking
photons. Then both $|\alpha_{z_1}\rangle$ and
$|-\alpha_{z_1}\rangle$ will go to the vacuum $|0\rangle$. Thus one
can prepare a quantum superposition of two macroscopic atomic states
$(|z_1\rangle+(-1)^m|-z_1\rangle)/\sqrt{2}$, which is a
Schr{\"o}dinger cat state that, in the notation of odd and even
sites, reads
\begin{eqnarray}
\frac{1}{\sqrt{2}}\left(|\frac{N+z_1}{2},\frac{N-z_1}{2}\rangle +
(-1)^m|\frac{N-z_1}{2},\frac{N+z_1}{2}\rangle \right). \nonumber
\end{eqnarray}

We have shown that the detection of photons at the direction of a
diffraction minimum leads to the preparation of the Schr{\"o}dinger
cat state (\ref{2-21}). The physical reason for this is that the
quantum measurement of photons determines the absolute value of the
atom number difference $|z|$. However, as the photon number is not
sensitive to the sign of $z$, one ends up in the superposition of
states with the positive and negative values of $z$.

Unfortunately, Eq.~(\ref{2-21}) demonstrates a very strong
disadvantage of such a method, which makes it difficult to realize
experimentally. Each photodetection flips the sign between two
components of the quantum superposition in Eq.~(\ref{2-21}). This
means, that if one loses even a single photocount, which is very
probable for a realistic photodetector, one ends up in the mixture
of two state (\ref{2-21}) with plus and minus signs. Such a mixed
state does not contain any atomic entanglement any more, in contrast
to the pure state (\ref{2-21}), which is a highly entangled one.

Formally, the appearance of the sign flip $(-1)^m$ in Eq.~(\ref{2-21})
originates from Eq.~(\ref{2-7}), which contains the coefficient
$\alpha_z^m$ in the pre-factor of each Fock state. Since, in the
final state, two components with opposite signs of light amplitudes
survive ($\alpha_{z_2}=-\alpha_{z_1}$), the term $\alpha_z^m$
produces the coefficient $(-1)^m$. Therefore, even one
photodetection changes the phase between two components in
Eq.~(\ref{2-21}) by $\Delta \varphi_1=\pi$, which is the maximal
possible phase difference. The idea to make the preparation scheme
more stable with respect to the photon losses is based on the
possibility to make this phase jump $\Delta \varphi_1$ less than
$\pi$. In this case, loosing one photon will also lead to the
mixture of two cat states. However, if those cat states are not very
different (i.e. the phase difference $\Delta \varphi_1$ is small),
the mixed state will still contain significant atomic entanglement.

In Sec. 2.5, we will present a scheme to prepare the Schr{\"o}dinger cat
state with a phase difference between two components less than
$\pi$. Thus, such a scheme is more practical and robust with respect
to the photon losses and, hence, decoherence.

The atom number squeezed states (\ref{2-18}) prepared by observing
light at a diffraction maximum are indeed more robust than the
Schr{\"o}dinger cat state (\ref{2-21}) obtained at a diffraction
minimum, as the former do not have any phase jump. However, the
convenient property of the measurement at a minimum is that during
the same time interval (e.g., during the shrinking time which is the
same for a maximum and minimum, $\delta z \approx
\sqrt{2\ln2/\tau}$) the number of photons scattered at a diffraction
minimum ($\langle a_1^\dag a_1\rangle =|C|^2N$) is much smaller than
the one scattered at a maximum ($\langle a_1^\dag a_1\rangle
=|C|^2N_K^2$) \cite{MekhovPRL2007,MekhovPRA2007,MekhovPRL2009}. Thus, the cat state is not as rapidly destroyed by the decoherence as one may naively expect from the coherent Bragg scattering picture.

Considering the projection as a tool for state preparation, one can prepare various partitions of the systems. In the examples above, one prepared the bipartite systems: atoms in the region of $K$ sites and outside that region, atoms in odd and even sites. More complicated partitions are also possible. For example, the light scattering from a limited lattice region and detection at a diffraction minimum will prepare a tripartite system (if the total atom number is fixed): the atoms can be in odd or even sites, within or outside of the lattice region. The initial state of such a system can be expressed via the trinomial coefficients. Therefore, using light scattering, one can simplify the initial problem of numerous atoms and sites to the simpler problems of just several subsystems. Importantly, the manipulation of particular variables can leave other variables untouched by the light scattering, allowing the consecutive operations on the same sample in the standard QND sense. We will further develop this idea in Sec. 2.9.

\subsection{Macroscopic quantum gases: Quantum Monte Carlo made analytically}

In this section, we will emphasize how the usually complicated numerical procedure of Quantum Monte Carlo simulations can be reduced to
an analytical solution after some assumptions and approximations
valid for macroscopic Bose--Einstein condensates (BEC) with large
atom numbers. We will consider a case, where the initial atomic state
is a macroscopic superfluid (SF) with the atom number $N\gg1$. Note,
that the total number of lattice sites $M$ and the number of sites
illuminated $K$ can be any. Thus, the theory presented below can be
applied for lattices with both low filling factors (e.g. one atom
per lattice site in average) and very high filling factors (e.g. a
BEC in a double-well potential).

We will start with the case of a diffraction minimum. We have shown (\ref{2-20}) that  for a lattice with the large atom number $N\gg1$, the binomial distribution to find the atom number at odd (or even) sites can be approximated by a Gaussian function:

\begin{eqnarray}\label{2-20x}
p_\text{SF}(z)=\frac{1}{\sqrt{2\pi}\sigma_z}e^{-\frac{z^2}{2\sigma_z^2}}.
\end{eqnarray}

As we already discussed, when the time progresses, both the number of photodetections $m$ and time $t$ increase with an
essentially probabilistic relation between them. The Quantum Monte
Carlo method, which establishes such a relation thus giving a
quantum trajectory, consists in the following. The evolution is
split into small time intervals $\delta t_i$. In each time step, the
conditional photon number $\langle a^\dag_1a_1\rangle_c(t)$ is calculated using the probability
distribution (\ref{2-15}):
\begin{gather}
p(z,m,t)=z^{2m}e^{-z^2\tau}p_0(z)/\tilde{F}^2, \label{2-5x}\\
\tilde{F}^2=\sum_z z^{2m}e^{-z^2\tau}p_0(z), \nonumber
\end{gather}
\begin{gather}
\langle a^\dag_1a_1\rangle_c(t)=|C|^2 \sum_z z^2p(z,m,t),\label{2-6x}
\end{gather}
which is proportional to the second moment of $p(z,m,t)$. The
probability of the next, $(m+1)$th, photocount within this time
interval $P_{m+1}=2\kappa \langle a^\dag_1a_1\rangle_c \delta t_i$
is then compared with a random number $0<\epsilon_i<1$ generated in
advance, thus, deciding whether the detection (if $2\kappa \langle
a^\dag_1a_1\rangle_c \delta t_i>\epsilon_i$) or no-count process
(otherwise) has happened.

For the large atom number, the summations in Eqs.~(\ref{2-5x}) and
(\ref{2-6x}) can be replaced by the integrals over all $-\infty
<z<\infty$, which gives the following probability of the next
photocount:
\begin{eqnarray}\label{2-9x}
P_{m+1}= \frac{\int_{-\infty}^\infty{z^{2m+2}e^{-z^2\tau}
e^{-\frac{z^2}{2\sigma^2}}dz}}{\int_{-\infty}^\infty{z^{2m}e^{-z^2\tau}
e^{-\frac{z^2}{2\sigma^2}}dz}}\delta\tau_i,
\end{eqnarray}
where $\delta\tau_i=2|C|^2\kappa\delta t_i$. Taking into account the
following relation \cite{Gradstein}:
\begin{eqnarray}\label{2-10x}
\int_{0}^\infty{x^{2n}e^{-px^2}dx}=\frac{(2n-1)!!}{2(2p)^n}\sqrt{\frac{\pi}{p}},
\end{eqnarray}
the integrals can be calculated and the probability of the next
photocount reads
\begin{eqnarray}\label{2-11x}
P_{m+1}=\frac{m+1/2}{\tau+1/\sigma^2}\delta\tau_i.
\end{eqnarray}

Thus, we see that the Quantum Monte Carlo method, which is usually
expected to be a basis for hard numerical simulations, has reduced
to an extremely simple form. After splitting our time axis into
intervals $\delta\tau_i$ and generating the random numbers
$0<\epsilon_i<1$, one has simply to substitute the current time
$\tau$ and the photocount number $m$ in the trivial algebraic
expression (\ref{2-11x}) and realize if the next photocount happened or
not. Proceeding this way one establishes the relation between the
photocount number and time $m(\tau)$ at the quantum trajectory
corresponding to the generated set of random numbers $\epsilon_i$.
Knowing the relation between $m$ and $\tau$, one can calculate
various conditional expectation values using Eq.~(\ref{2-10x}) and the
complementary expression
\begin{eqnarray}\label{2-12x}
\int_{0}^\infty{x^{2n+1}e^{-px^2}dx}=\frac{n!}{2p^{n+1}}.
\end{eqnarray}

We now switch to the case of the light detection at the direction of
a diffraction maximum. In this
case one measures the atom number $z$ in the lattice region of $K$
illuminated sites.

For the initial SF state the probability (\ref{2-17}) to find the atom number $z$
at the lattice region of $K$ sites can be
approximated as a Gaussian distribution
\begin{eqnarray}\label{2-14x}
p_\text{SF}(z)=\frac{1}{\sqrt{2\pi}\sigma}e^{-\frac{(z-z_0)^2}{2\sigma^2}}.
\end{eqnarray}

Similarly to the case of a diffraction minimum, for the large atom
number, the summations in Eqs.~(\ref{2-5x}) and (\ref{2-6x}) can be
replaced by the integrals over $-\infty<z<\infty$, which gives the
following probability of the next photocount:
\begin{eqnarray}\label{2-15x}
P_{m+1}= \frac{\int_{0}^\infty{z^{2m+2}e^{-z^2\tau}
e^{-\frac{(z-z_0)^2}{2\sigma^2}}dz}}{\int_{0}^\infty{z^{2m}e^{-z^2\tau}
e^{-\frac{(z-z_0)^2}{2\sigma^2}}dz}}\delta\tau_i.
\end{eqnarray}
Taking into account the following relation \cite{Gradstein}:
\begin{gather}
\int_{-\infty}^\infty{x^{n}e^{-px^2+2qx}dx}=\frac{1}{2^{n-1}p}\sqrt{\frac{\pi}{p}}
\frac{d^{n-1}}{dq^{n-1}}\left(qe^\frac{q^2}{p}\right) \\\nonumber
=n!e^\frac{q^2}{p}\sqrt{\frac{\pi}{p}}\left(\frac{q}{p}\right)^n\sum_{k=0}^{E(n/2)}
\frac{1}{(n-2k)!k!}\left(\frac{p}{4q^2}\right)^k,\label{2-16x}
\end{gather}
where $E(n/2)$ is the integer part of $n/2$, the integrals can be
calculated and the probability of the next photocount reads
\begin{eqnarray}\label{2-17x}
P_{m+1}=(2m+1)(2m+2)a^2\sum_{k=0}^{m+1}
\frac{b^k}{(2m+2-2k)!k!}/\sum_{k=0}^{m} \frac{b^k}{(2m-2k)!k!},
\end{eqnarray}
where the parameters are
\begin{eqnarray}
a=\frac{z_0}{2\sigma^2(\tau+1/(2\sigma^2))}=\frac{1}{2(1-K/M)(\tau+1/(2\sigma^2))},\nonumber\\
b=\frac{\tau+1/(2\sigma^2)}{z_0^2}\sigma^4=(\tau+1/(2\sigma^2))\left(1-\frac{K}{M}\right)^2.\nonumber
\end{eqnarray}
Expression (\ref{2-17x}) is more complicated than Eq.~(\ref{2-11x}) for
the diffraction minimum. However, it is also very simple as it
includes only summation over the photocount number. Thus, we were
able to replace the summation over the atom number, which can rich
the values of $N=10^6$, and even the numerical integration over the
atom number. For the far off-resonant interaction considered here,
the number of photocounts $m$ will be many orders of magnitude less
than the atom number. So, the sum in Eq.~(\ref{2-17x}) will contain
only a small number of terms.


\subsection{The discreteness of matter field: exponentially fast state collapse for small atom number}

In this section we will show that the final stage of the state collapse happens extremely quickly: the slow square root dependence changes to the exponential one. This is a direct consequence of the discrete nature of the atomic matter field.

We have shown that any initially broad atom-number distribution $p_0(z)$ shrinks and approaches some very narrow distribution, which corresponds to the atom number squeezing. We have shown as well that the cental value of the final distribution $z_0$ is given by $z_{0}=\sqrt{m/\tau}$, while the full width at half maximum
(FWHM) of the distribution can be estimated as $\delta z \approx \sqrt{2\ln2/\tau}$, which shows the shrinking of the distribution in time. This type of shrinking (as $\sqrt{1/\tau}$) is rather typical for such photocounting schemes \cite{Onofrio}. This approximate formula works well if one assumes that (i) the distribution is already rather narrow, $\delta z \ll z_0$, but still (ii) the atom-number probabilities in Eq.~(\ref{2-15}) can be replaced by the continuous functions of $z$, which is a good approximation for very large atom numbers.

Nevertheless, it is clear that while the distribution function continues to shrink, its width can reach the values of $\delta z \sim 1$, where the approximation of continuous functions obviously fails. In this case, one should explicitly take into account the discrete nature of the atomic ensemble. If one starts with a macroscopic quantum gas, it is probably not very practical to expect that the distribution $p(z,m,t)$ can really shrink to the widths of $\delta z \sim 1$, because various destructive mechanisms will prevent the final state collapse to the many-body Fock state $|z_0, N-z_0\rangle$ with the precisely known atom number $z_0$ at the $K$ illuminated lattice sites and the rest $N-z_0$ atoms at the rest of $M-K$ sites. Instead, a distribution with some rather small $\delta z$ will establish. Thus the approximation $\delta z \approx \sqrt{2\ln2/\tau}$ can work well for the large atom numbers even till the last stage of the conditional time evolution. However, if the atom number and number of illuminated lattice sites are small, the situation $\delta z \sim 1$ is reasonable and practical experimentally \cite{GreinerNature2009}.

\begin{figure}[h!]
\centering
\captionsetup{justification=justified}
\includegraphics[width=0.7\textwidth]{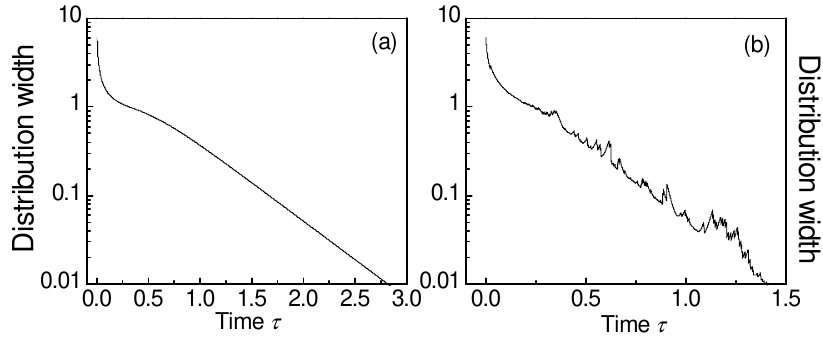}
\caption{\label{2-fig1LP2011}Width of the atom number distribution function during the photodetection. Decreasing width corresponds to the atom number squeezing. At the initial stage, the shrinking is as $1/\sqrt{\tau}$, while at the final stage the shrinking is exponential. (a) Quantum trajectory without quantum jumps; (b) quantum trajectory with quantum jumps. Photodetection is at the diffraction minimum. Total number of atoms $N=100$, $K=M=100$ sites.}
\end{figure}

The numerical results at a single quantum trajectory are presented in Fig. \ref{2-fig1LP2011}. They clearly show that the time evolution of the distribution function width has two different parts. First, the approximation of the square root decrease works very well. However, after that, the decrease and thus the state collapse, becomes much faster. In the logarithmic scale figure, it is clear that the shrinking of the distribution becomes even exponential.

Such a behavior of the distribution function can be explained using the fact that the discrete functions of $z$ should be taken into account. Let us assume that the atom number $z$ changes discretely with a step $Z$. For example, for the atom number at $K$ sites, which can be measured at the directions of diffraction maxima, the step is $Z=1$ atom. In contrast, for the atom number difference between odd and even sites, which can be measured at the directions of diffraction minima, the step is $Z=2$ atoms, which reflects the total atom number conservation. Note that, when the width of the distribution function is of the order of 1, introducing the FWHM becomes meaningless [e.g., the values of $p(z_0\pm Z)$ can be already less than $p(z_0)/2$]. Thus, the distribution width should be characterized directly by the square root of the atom number variance:
\begin{eqnarray}\label{2-6xx}
\langle z^2\rangle - \langle z\rangle^2 = \langle (z-\langle z\rangle)^2\rangle=\sum_z(z-\langle z\rangle)^2p(z).
\end{eqnarray}
The insight in the exponential shrinking can be made as follows. If one assumes that at the final stage of the state collapse only the atom numbers of $z_0$ and $z_0\pm Z$ are non-negligible, Eq.~(\ref{2-6xx}) reduces to
\begin{eqnarray}\label{2-7xx}
\langle z^2\rangle - \langle z\rangle^2 \approx Z^2p(z_0-Z)+Z^2p(z_0+Z).
\end{eqnarray}
Using Eq.~(\ref{2-15}) (expanding the expressions in Taylor series in $Z/z_0$, taking into account that the normalization factor $\tilde{F}$ is a function of $m$, $\tau$, and $z_0$, and substituting $m$ as $m=z^2_0\tau$), one can get an estimation $\langle z^2\rangle - \langle z\rangle^2 \sim \exp(-Z^2\tau)$. This expression supports the exponential shrinking of the atom number distribution width at the final stage of the quantum state collapse, which is demonstrated in Fig. \ref{2-fig1LP2011}.

As a result, the exponential shrinking of distribution function demonstrated here is a direct consequence of the discrete nature of the atomic matter field.


\subsection{The light-induced nonlinearity in the evolution of matter-field phase}

We will now show that the light--matter coupling leads to the nonlinear (in the atom number) evolution of the matter-wave phase. Importantly, in standard prescribed optical lattices (without any light dynamics) such a nonlinearity is only possible due to the direct atom-atom interaction, which is not necessary here at all, and has been completely neglected.   

We address a question about the unitary (coherent) evolution of the light--matter state in the short-time limit, where the photon escapes are not important ($t\ll 1/\kappa$). We will be interested in the scattering into the direction of the diffraction maximum, where $D^q_{10}=N_K$ is the fluctuating atom number at $K$ lattice sites. We will consider only the transverse probe $a_0$ (while probing through the mirror is not present, $\eta_1=0$). The dispersion frequency shift $U_{11}D^q_{11}$ will be neglected in this configuration. Using the general solution (\ref{2-7}-\ref{2-9}) with those assumptions, and setting $\kappa=0$, $m=0$, one gets

\begin{eqnarray}\label{2-8xx}
|\Psi(t)\rangle
=\frac{1}{F(t)}\sum_{q}e^{\Phi_q(t)}c_q^0|q_1,...,q_M\rangle|\alpha_q(t)\rangle,
\end{eqnarray}
where
\begin{eqnarray}\label{2-9xx}
\alpha_q(t)= CD^q_{10}\left( 1-e^{i\Delta_pt}\right)e^{-i\omega_pt},
\end{eqnarray}
\begin{eqnarray}\label{2-10xx}
\Phi_q(t)=-i\Delta_p|CD^q_{10}|^2t+i|CD^q_{10}|^2 \sin\Delta_p t,
\end{eqnarray}
and $C=U_{10}a_0/\Delta_p$. For times such that $\Delta_pt \gg 1$, the phase gets even a simpler form: $\Phi_q(t)=-i\Delta_p|CD^q_{10}|^2t$. Thus, the evolution of the state for the interaction at the diffraction maximum reads:

\begin{eqnarray}\label{2-11xx}
|\Psi(t)\rangle
=\frac{1}{F(t)}\sum_{q}e^{-i\Delta_pC^2N_K^2t}c_q^0|q_1,...,q_M\rangle|\alpha_q(t)\rangle.
\end{eqnarray}

An important property of this solution is that the state components with various atom numbers at $K$ lattice sites have different phase evolutions, which quadratically depends on the atom number at $K$ sites $N_K$. For the superfluid state with the large atom number, if the number of illuminated sites is much smaller than the total site number, $K\ll M$, the statistics of $N_K$ is nearly Poissonian. Thus, the situation resembles the problems of a macroscopic BEC, where the collapse and revival of the matter field can be observed \cite{GreinerCollapse}. In both cases, the nontrivial dynamics is a consequence of the particular (i.e. quadratic in the atom number) modification of the phase evolution of the atom number components, initially constituting the Poissonian atom number distribution. Nevertheless, the physical reasons of the phase modifications in two cases are completely different. In the case of Ref. \cite{GreinerCollapse}, it is the atom-atom interaction that brings the nonlinearity in the problem in the form of the phase prefactor $\exp[{-iUN(N-1)t/2\hbar}]$, where $U$ is the atom-atom interaction energy. In our case, the atom-atom interaction has been completely neglected, and it is the atom-light interaction which leads to the phase dependence quadratic in $N_K$. As can be traced from the full solution (\ref{2-9}), the phase depends on the product of probe-mode coupling coefficient proportional to $D^q_{10}$ (which is $N_K$ in this particular case), and the cavity mode amplitude $\alpha_q$. As we consider the light amplitudes as dynamical variables (and not as prescribed quantities as it is made in many problems of atoms in strong optical lattices), the light amplitudes are proportional to the atom number $N_K$. Thus, the product of the coupling coefficient and light amplitude gives us the term quadratic in the atom number $N_K$. In other words, one can say that the dynamical nature of the light mode leads to the effective atom-atom interaction.

It is probable, that similarly to Ref. \cite{GreinerCollapse}, one can obtain the collapse and revival of the matter field in our system, if some matter wave interference measurement will be carried out. Since we use the basis with the fixed total atom number $N$ (in contrast to Ref. \cite{GreinerCollapse}, where the coherent atomic state was assumed for the BEC), the calculation of the matter interference will be nontrivial. To calculate the interference pattern, one needs to include the atom counting procedure. This procedure is well-developed (see, e.g., Refs. \cite{Dalibard,Horak1999}) and leads to the results rather similar to what is expected from the coherent-state approximation for the initial atomic state. In addition, the methods to disentangle light and matter can be used to observe the matter wave interference.

As a result, for observing some intriguing phenomena, the light--matter interaction can effectively replace the direct atom-atom interaction, which might be small and difficult to control (as it can be the case of standard classical optical lattices with no light dynamics taken into account).


\section{Photon statistics}

In this section, we consider three kinds of photon statistics: (i)
statistics $p_\Phi(n,m,t)$ of the photon number $n$ in a cavity
after $m$ photons were detected outside the cavity, (ii) statistics
$P(m,t)$ of the photocount number $m$, and (iii) statistics
$\tilde{P}_T(m,t)$ of the photocount number $m$ if, after the time
measurement $T$, $m_T$ photons were detected.

First, let us consider the statistics of the number of photons in a
cavity after the measurement time $t$ and $m$ photodetections. The
joint probability to find a number of photons in a cavity $n$
together with finding the atomic state in the Fock state
$|q\rangle=|q_1,..,q_M\rangle$ is obtained by projecting the general
solution (\ref{2-6}) on the state $|q\rangle|n\rangle$ and is given by
\begin{eqnarray}
W_q(n,m,t)=\frac{|\alpha_q(t)|^{2n}}{n!}e^{-|\alpha_q(t)|^2}p_q(m,t),\nonumber
\end{eqnarray}
where the probability to find the atomic Fock state $p_q(m,t)$ is
given by Eq. (\ref{2-10}). However, the probability to find $n$
photons in a cavity independently of the atomic state is obtained by
projecting the solution (\ref{2-6}) on the light Fock state
$|n\rangle$ and taking the trace over the atomic states. Thus, the
cavity photon number distribution function is given by
\begin{eqnarray}
p_{\Phi}(n,m,t)=\sum_q\frac{|\alpha_q(t)|^{2n}}{n!}e^{-|\alpha_q(t)|^2}p_q(m,t),
\nonumber
\end{eqnarray}
where the sum is taken over all possible configurations $q$.

If, as in the previous section, the only atomic statistical quantity
is $z$, the sum is simplified and the probability $p_{\Phi}(n,m,t)$
to find $n$ photons in a cavity after the measurement time $t$ and
$m$ photodetections reads as
\begin{eqnarray}\label{2-1x}
p_{\Phi}(n,m,t)=\sum_z\frac{|\alpha_z(t)|^{2n}}{n!}e^{-|\alpha_z(t)|^2}p(z,m,t),
\end{eqnarray}
where the atom number distribution $p(z,m,t)$ is given by
Eq.~(\ref{2-12}).

In general, $p_{\Phi}(n,m,t)$ is a super-Poissonian distribution.
During the measurement, the atomic distribution $p(z,m,t)$ shrinks
to one or two symmetric peaks corresponding to the atom-number
squeezed or macroscopic superposition states. Thus, after some time,
only a single term (or two equal terms) survives in the sum, and the
cavity photon statistics $p_{\Phi}(n,m,t)$ evolves from
super-Poissonian to Poissonian one. This fact can be checked
experimentally.

During the measurement, the mean conditional photon number $\langle
a_1^\dag a_1\rangle_c$ approaches the value $|\alpha_{z_1}|^2$,
which enables one to determine the final atom number $z_1$ by
measuring the photon number in a cavity.

Let us now consider the probability $P(m,t)$ to detect $m$ photons
within the time $t$, if the initial atom number distribution is
$p_0(z)$. As shown, for example, in Ref.~\cite{UedaPRA1992}, this
probability can be obtained from the state (\ref{2-6}) using the
integration over all detection moments $t_1,t_2,..,t_m$ from 0 to
$t$ (because we are not interested in the time moments, but only in
the total number of the photocounts $m$) and taking the trace. The
simple result can be obtained for the case $t,t_1>1/k$, since the
solution (\ref{2-7}) does not depend on the detection times. The
probability reads
\begin{eqnarray}\label{2-2x}
P(m,t)=\sum_z\frac{(2\kappa|\alpha_z|^2t)^m}{m!}e^{-2\kappa|\alpha_z|^2t}p_0(z),
\end{eqnarray}
where the powers of $m$ appear due to the $m$ time-integrations.

In contrast to the probability $p_{\Phi}(n,m,t)$, Eq.~(\ref{2-1x}),
which characterizes the conditional distribution of the cavity
photons at a particular trajectory, the probability $P(m,t)$ depends
on the initial atom number distribution $p_0(z)$ and is not a
characteristic of a particular quantum trajectory, but rather of an
ensemble average. In general, this is a super-Poissonian
distribution, which does not approach any Poissonian one. From
Eq.~(\ref{2-2x}), the increase of the mean photocount number with time
is given by
\begin{eqnarray}
\langle m\rangle_0 = 2\kappa t \langle a_1^\dag
a_1\rangle_0,\nonumber
\end{eqnarray}
where $\langle a_1^\dag a_1\rangle_0$ is not a conditional photon
number, but the one calculated for the initial atomic state. In
average, the photocount number linearly increases in time. However,
as the distribution is not Poissonian, the fluctuations of the
photocount rate $m/t$ do not decrease to zero.

One can also introduce another statistical distribution: the
distribution $\tilde{P}_T(m,t)$ of the photocount number $m$ if,
after the measurement time $T$, $m_T$ photons have been already
detected. Similar approach as in Eq.~(\ref{2-2x}), leads to the
following result:
\begin{gather}
\tilde{P}_T(m,t)=\sum_z\frac{[2\kappa|\alpha_z|^2(t-T)]^m}{m!}e^{-2\kappa|\alpha_z|^2(t-T)}
 p(z,m_T,T),\label{2-3x}
\end{gather}
which, in contrast to Eq.~(\ref{2-2x}), depends not on the initial
atomic distribution, but on that at time $T$. Thus, this probability
combines the quantum trajectory evolution up to the time $T$, and
the ensemble average after that. As we know, the atomic distribution
$p(z,m_T,T)$ approaches the single peak for the Fock state with
increasing $T$. Therefore, this photocount probability approaches
the Poissonian distribution with increasing $T$, the fluctuations of
$m$ grow in time as $\sqrt{m}\sim \sqrt{t}$, and the fluctuations
of the photocount rate $m/t$ vanish with increasing time.


\section{State preparation by cavity transmission measurement}

In this section we will switch to a different probing scheme, corresponding to the transmission spectroscopy considered in Sec. 1.10, where we have shown that the transmission spectrum has a comb-like structure and maps out the full atom number distribution function. Such a spectrum is demonstrated in Fig. 1.15. As discussed before, this spectrum reflects the result of multiple measurements. The question now is what one can get in a single run of the optical measurement. Naively, one can expect that in each run one gets one of the peaks in Fig. 1.15, which would correspond to the collapse to the single-peak distribution function. Only after averaging over many runs one would recover the full comb-like spectrum in Fig. 1.15. Nevertheless, the situation is much more interesting in detail. Depending on the realization (i.e. for different quantum trajectories), the distribution function after a single-run of the transmission spectrum measurement can consist either of one peak or of two peaks. A one-peak distribution function corresponds to the atom number squeezing, while a two-peak function signals the generation of the Schr{\"o}dinger cat state. Thus, probing through a mirror represents an interesting example, where either a number squeezed state or a Schr{\"o}dinger cat state is prepared depending on the measurement outcome. Knowing the properties of the system after the measurement, one can understand, which kind of the states was prepared. Moreover, the cat state obtained using this configuration has a more general form than that in the case of the detection at the diffraction minimum: the amplitudes of two cat components can be non-equal and the phase difference between them can take various values (not restricted to $\pi$ as in the diffraction minimum). The latter property makes the preparation of the cat states more robust to the decoherence and photon loss.

We consider the probe through
the mirror with the amplitude $\eta_1$ (Fig. 1.1), and assume no
transverse probe, $a_0=0$. From Eq.~(\ref{2-8}) we see that the light
amplitudes depend only on the single statistical quantity $D^q_{11}=
\sum_{j=1}^K{|u_1({\bf r}_j)|^2q_j}$, which reduces to the atom
number at $K$ sites for the traveling wave $a_1$ at any angle to the
lattice, or for the standing wave $a_1$ with atoms trapped at the
antinodes. Thus, in this case, the statistical quantity is
$z=D^q_{11}$, which changes between 0 and $N$. The term
$U_{11}D^q_{11}$ in Eq.~(\ref{2-8}) has a meaning of the dispersive
frequency shift of the cavity mode, due to the presence of atoms in
a cavity.

The light amplitude $\alpha_z$ from Eq.~(\ref{2-8}) can be rewritten
as a function of the atom number at $K$ sites $z$:
\begin{eqnarray}\label{2-22}
\alpha_z=C'\frac{\kappa/U_{11}}{i(z-z_p)+\kappa/U_{11}} \text{ with
} C'=\frac{\eta_1}{\kappa},
\end{eqnarray}
and the parameter $z_p=\Delta_p/U_{11}$ is fixed by the probe-cavity
detuning $\Delta_p$. The probability to find the atom number $z$
(\ref{2-13}) takes the form
\begin{eqnarray}\label{2-23}
p(z,m,t)=\frac{e^{-\tau'(\kappa/U_{11})^2/[(z-z_p)^2+(\kappa/U_{11})^2]}}
{[(z-z_p)^2+(\kappa/U_{11})^2]^m} p_0(z)/F'^2,
\end{eqnarray}
where the dimensionless time is $\tau'=2|C'|^2\kappa t$.

The pre-factor function in front of $p_0(z)$ in Eq.~(\ref{2-23}) has a
form more complicated than the one we had for the transverse probing
in Eq.~(\ref{2-15}). However, it provides us a richer physical
picture. In contrast to the transverse probing, this function allows
us a collapse to both the singlet and doublet distribution. Thus,
the measurement of the cavity transmission contains both cases of
transverse probing at a diffraction maximum and minimum considered before.

If for a particular quantum trajectory the number of photocounts is
large such that $m/\tau'\ge 1$, the distribution $p(z,m,t)$
collapses to a single-peak function at $z_p$. This singlet shrinks
(FWHM) as $\delta z\approx 2(\kappa/U_{11})\sqrt[4]{2\ln{2}/\tau'}$,
which can be very fast for a high-Q cavity with the small $\kappa$.
In the estimation of $\delta z$, we used the assumption that the
peak has already become narrow, $\delta z \ll \kappa/U_{11}$, and
that $m\approx 2\kappa\langle a_1^\dag a_1\rangle_c t$. Similar to
the transverse probing, in the final state, the light and atoms are
disentangled: $|z_p,N-z_p\rangle|\alpha_{z_p}\rangle$, which
corresponds to the atom-number squeezed (Fock) state and coherent
light state.

If, in contrast, the number of photocounts is small, $m/\tau'< 1$, the
distribution $p(z,m,t)$ collapses to a doublet centered at $z_p$
with two satellites at $z_{1,2}=z_p\pm\Delta z$ with
\begin{eqnarray}\label{2-24}
\Delta z = \frac{\kappa}{U_{11}}\sqrt{\frac{\tau'}{m}-1}.
\end{eqnarray}
When the doublet has become well-separated (i.e. $\delta z \ll
\Delta z$), each of its component shrinks in time as
\begin{eqnarray}
\delta z \approx \Delta z \left(1+\frac{\kappa^2}{U^2_{11}\Delta
z^2}\right) \sqrt{\frac{2\ln{2}}{\tau'}\left(1+\frac{U^2_{11}\Delta
z^2}{\kappa^2}\right)}. \nonumber
\end{eqnarray}

Physically, tuning the probe at $\Delta_p$, we may expect scattering
from the atom number $z_p$ providing such a frequency shift
$\Delta_p$. If the photocount number is large ($m/\tau'\ge1$),
indeed, the atom number is around $z_p$ and it collapses to this
value. However, if $m$ is small, we gain knowledge that the atom
number $z$ is inconsistent with this choice of $\Delta_p$, but two
possibilities $z<z_p$ or $z>z_p$ are indistinguishable. This
collapses the state to a superposition of two Fock states with
$z_{1,2}$, symmetrically placed around $z_p$.

Thus, the transmission measurement scheme allows one to prepare both
the atom number squeezed state and the Schr{\"o}dinger cat state.
The appearance of the singlet for squeezed state or the doublet for
cat state can be determined by measuring $m$ and $t$, or the final
photon number $\langle a^\dag_1a_1\rangle_c=|\alpha_{z_1}|^2$. From
Eq.~(\ref{2-22}), we see that there is a direct correspondence between
the final cavity photon number and the parameter of the doublet
$\Delta z$:
\begin{eqnarray}\label{2-xx}
\langle a^\dag_1a_1\rangle_c = \frac{\eta_1^2}{U^2_{11}\Delta
z^2+\kappa^2},
\end{eqnarray}
which makes the experimental determination of the doublet position
possible. The parameter $\Delta z$ can be also determined by
measuring the photocount number $m$ and time $t$ using
Eq.~(\ref{2-24}).

\begin{figure}[h!]
\centering
\captionsetup{justification=justified}
\includegraphics[width=0.8\textwidth]{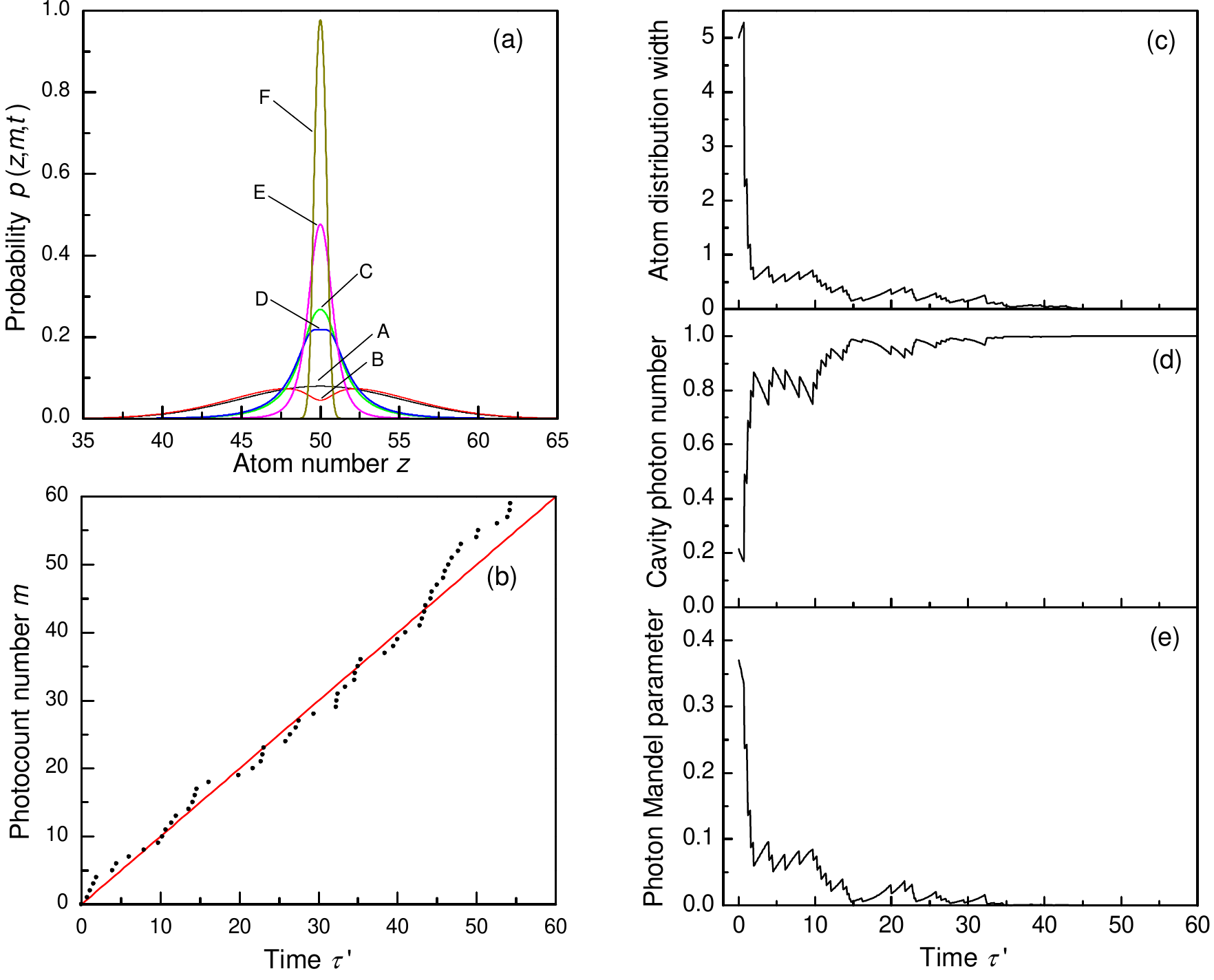}
\caption{\label{2-fig2}Photodetection trajectory
leading to a single-peak distribution (atom number squeezing). The
probe-cavity detuning is chosen such that $z_p=50$ coincides with
the initial distribution center. (a) Shrinking atom number
distribution at different times $\tau'=$ 0 (number of photocounts is
$m=0$), 0.7 (just before the first count, $m=0$), 0.7 (just after
the first count, $m=1$), 1.1 ($m=1$), 1.1 ($m=2$), 14.6 ($m=17$)
(A-F); (b) time dependence of the photocount number $m$, the red
line is for $m=\tau'$; (c) decreasing width of the atom number
distribution; (d) reduced conditioned photon number $\langle
a^\dag_1 a_1\rangle_c/|C'|^2$ with quantum jumps; (e) photon Mandel
parameter. Initial state: SF, $N=100$ atoms, $M=100$ lattice sites,
$K=M/2=50$ illuminated sites.}
\end{figure}

\begin{figure}[h!]
\centering
\captionsetup{justification=justified}
\includegraphics[width=0.8\textwidth]{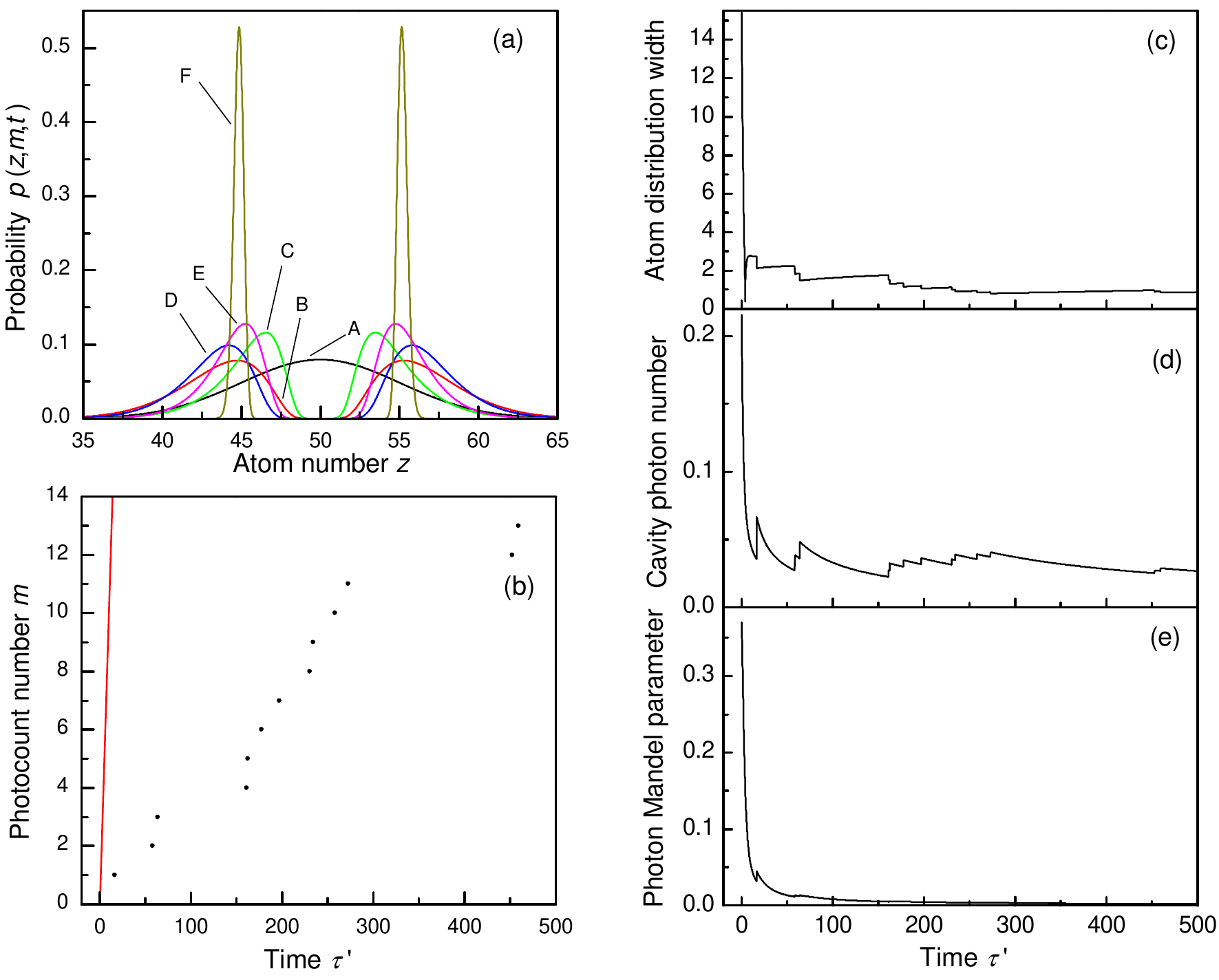}
\caption{\label{2-fig3}Photodetection trajectory
leading to a doublet distribution (Schr{\"o}dinger cat state). The
probe-cavity detuning is chosen such that $z_p=50$ coincides with
the initial distribution center. (a) Shrinking atom number
distribution at different times $\tau'=$ 0 (number of photocounts is
$m=0$), 16.4 (just before the first count, $m=0$), 16.4 (just after
the first count, $m=1$), 58.2 ($m=1$), 58.2 ($m=2$), 2017.6 ($m=73$)
(A-F); (b) time dependence of the photocount number $m$, the red
line is for $m=\tau'$; (c) decreasing width of one of the peaks in
the atom number distribution; (d) reduced conditioned photon number
$\langle a^\dag_1 a_1\rangle_c/|C'|^2$ with quantum jumps; (e)
photon Mandel parameter. Initial state: SF, $N=100$ atoms, $M=100$
lattice sites, $K=M/2=50$ illuminated sites.}
\end{figure}

\begin{figure}[h!]
\centering
\captionsetup{justification=justified}
\includegraphics[width=0.4\textwidth]{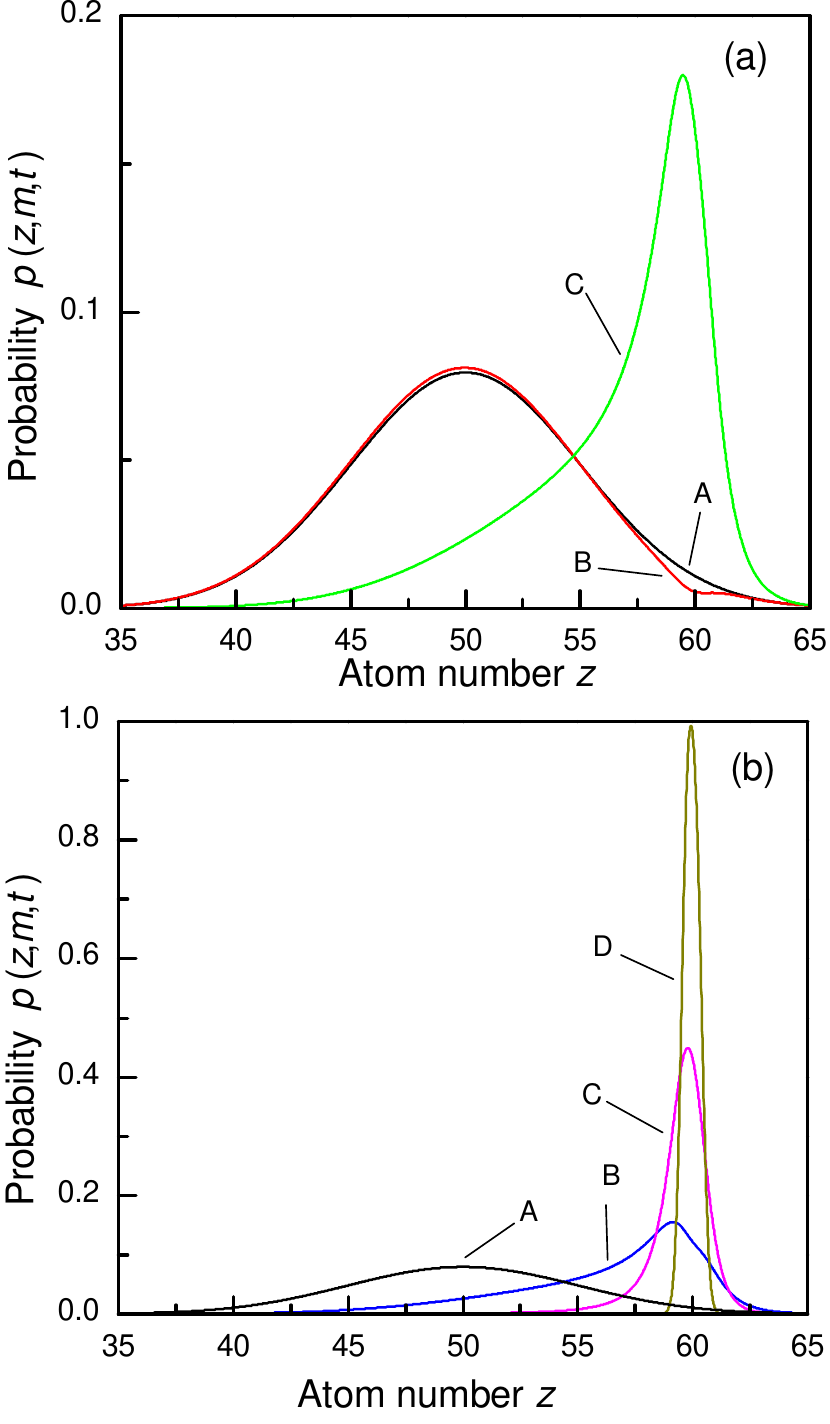}
\caption{\label{2-fig4}Photodetection trajectory
leading to a single-peak distribution (atom number squeezing). The
probe-cavity detuning is chosen such that $z_p=60$ is at the wing of
the initial distribution. Shrinking atom number distribution at
different times, (a) $\tau'=$ 0 (number of photocounts is $m=0$),
0.7 (just before the first count, $m=0$), 0.7 (just after the first
count, $m=1$) (A-C); (b) $\tau'=$ 0 ($m=0$), 1.1 ($m=1$), 1.1
($m=2$), 14.6 ($m=17$) (A-D). Initial state: SF, $N=100$ atoms,
$M=100$ lattice sites, $K=M/2=50$ illuminated sites.}
\end{figure}

\begin{figure}[h!]
\centering
\captionsetup{justification=justified}
\includegraphics[width=0.8\textwidth]{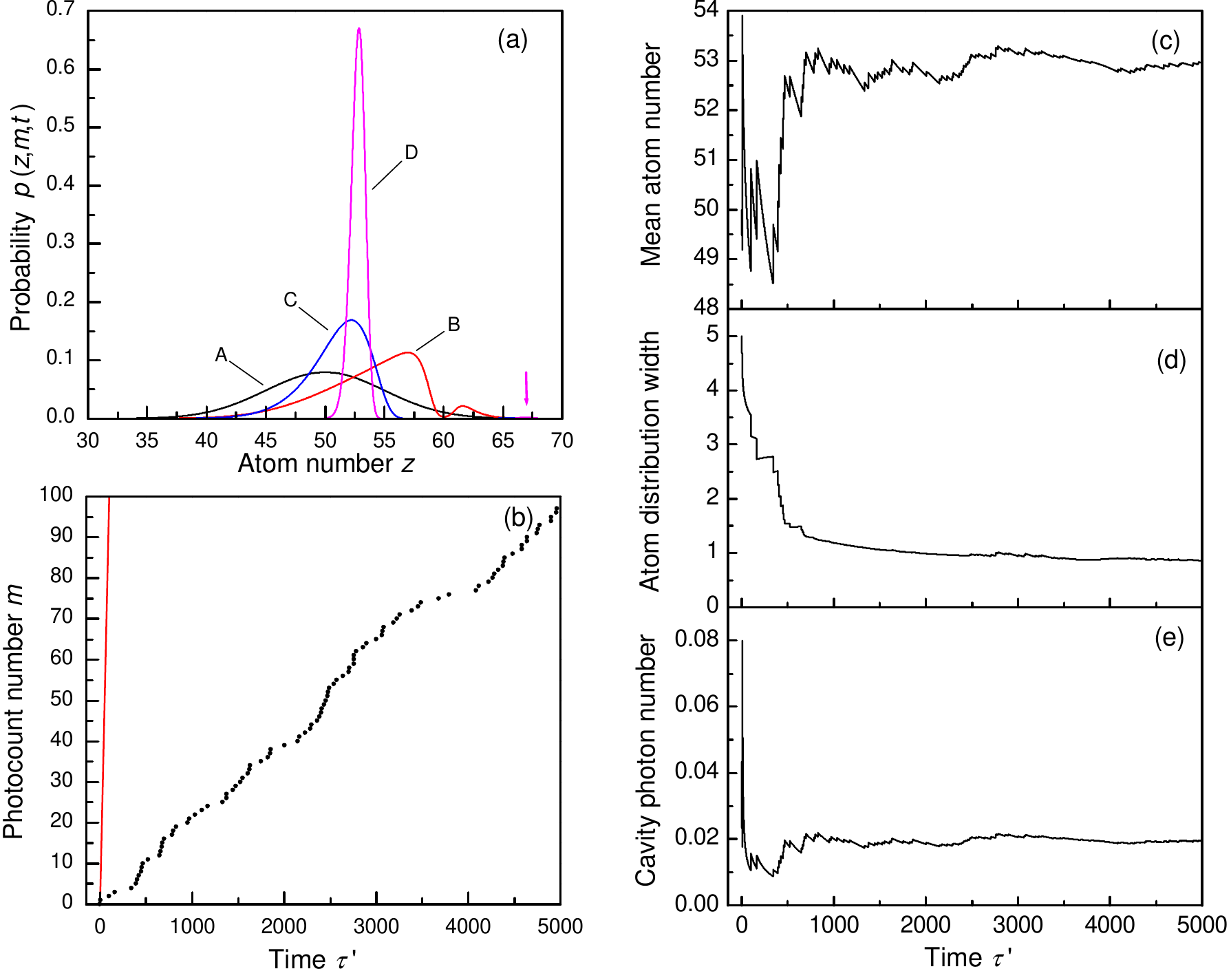}
\caption{\label{2-fig5}Photodetection trajectory
leading to a doublet distribution (Schr{\"o}dinger cat state). The
probe-cavity detuning is chosen such that $z_p=60$ is at the wing of
the initial distribution. (a) Shrinking atom number distribution at
different times $\tau'=$ 0 (number of photocounts is $m=0$), 5.4
($m=1$), 163.5 ($m=3$), 2006.9 ($m=39$) (A-D); the arrow shows the
position of the very small doublet component, which is almost
invisible; (b) time dependence of the photocount number $m$, the red
line is for $m=\tau'$; (c) conditioned mean atom number; (d)
decreasing width of the atom number distribution; (e) reduced
conditioned photon number $\langle a^\dag_1 a_1\rangle_c/|C'|^2$
with quantum jumps. Initial state: SF, $N=100$ atoms, $M=100$
lattice sites, $K=M/2=50$ illuminated sites.}
\end{figure}

In Figs. \ref{2-fig2}-\ref{2-fig5}, we present the results, where the quantum trajectories
of qualitatively different kinds were realized. In all figures, the
initial state is the superfluid with the atom number $N=100$ at
$M=100$ lattice sites, the half of all lattice sites $K=50$ are
illuminated by the cavity mode. The initial distribution of the atom
number at $K$ sites is given by Eq.~(\ref{2-16}) and can be well
approximated by the Gaussian distribution (\ref{2-17}) with the mean
value $\langle \hat{N}_K\rangle = z_0=NK/M=50$ and
$\sigma_z=\sqrt{N(K/M)(1-K/M)}=5$.

Figure \ref{2-fig2} presents the results for a quantum trajectory, which leads
to the collapse to a single-peak distribution. In this case, the
probe is detuned in such way, that the probe-cavity detuning
corresponds to the center of the atom number distribution:
$z_p=\Delta_p/U_{11}=50$. In Fig. \ref{2-fig2}(a), the evolution of the atom
number probability $p(z,m,t)$, Eq.~(\ref{2-23}), is shown. The curve A
is the initial atom number distribution having the Gaussian shape.
Curve B shows the non-Hermitian evolution of the probability just
before the first jump ($m=0$). As, at this time interval, the number
of photocounts is small ($m/\tau'<1$), the distribution tends to a
doublet. This is assured by the exponential factor in Eq.~(\ref{2-23})
leading to the suppression of the component at $z=z_p$ as one does
not record the photocounts at the expected detuning. However, just
after the photocount ($m=1$) this distribution instantly changes to
the curve C, which has already a single peak as for this trajectory
it turns out that $m/\tau'>1$. The switch to a single peak is
assured by the Lorentzian factor in Eq.~(\ref{2-23}). After that,
before the second jump, the probability decreases at $z=z_p$ and
broadens (curve D), but jumps upwards again and narrows, when the
second jump occurs (curve E). Finally, after many jumps, the
probability distribution becomes narrow and has a single peak at
$z=z_p$ (curve F).

Figure \ref{2-fig2}(b) shows the number of photocounts $m$ growing in time. It
is clear that $m$ stays always near the line $m=\tau'$. The
appearance of the singlet is assured by the fact, that at the
initial stage $m/\tau'>1$.

Figure \ref{2-fig2}(c) shows the evolution of the width of the atom number
distribution, $\sqrt{(\Delta N_K)^2}$, which decreases to zero
reflecting the shrinking distribution. Figure \ref{2-fig2}(d) shows the reduced
conditioned photon number in the cavity $\langle a_1^\dag
a_1\rangle_c/|C'|^2$. One sees that for the initial atom
distribution it starts from a relatively small value. However, as
the atomic state goes to the Fock state with atom number $z_p$
exactly corresponding to the detuning $\Delta_p$, it approaches the
maximal possible value $\langle a_1^\dag a_1\rangle_c/|C'|^2=1$.
Moreover, one can easily see the quantum jumps in the initial stage,
when the light field consists of several coherent-state components.
Finally, the jumps disappear as the light state approaches a single
coherent state $|\alpha_{z_p}\rangle$, when the atomic state
approaches a Fock state. It is interesting to note, that the photon
escape from the cavity (photocount) leads to the increase of the
conditional photon number in the cavity, while the no-count process
leads to its decrease. This is a counter-intuitive characteristic
feature of the super-Poissonian photon statistics and is determined
by the conditional nature of the probabilities considered
\cite{UedaPRA1992}. Figure \ref{2-fig2}(e) shows the reduced Mandel parameter
$Q/|C'|^2$ characterizing the photon number variance: $Q=(\langle
n_{\Phi}^2\rangle-\langle n_{\Phi}\rangle^2)/\langle
n_{\Phi}\rangle-1$, where $n_{\Phi}=\langle a_1^\dag a_1\rangle_c$
is the conditioned photon number. One can see how it decreases to
zero corresponding to the coherent state of light. The parameters
shown in Figs. \ref{2-fig2}(b)-\ref{2-fig2}(e) can be measured experimentally thus
presenting the verification of our theory.

Figure \ref{2-fig3} presents the results for the case, where the detuning also
corresponds to the atom number distribution center $z_p=50$, but the
quantum trajectory leads to the doublet distribution (Sch{\"o}dinger
cat state). Figure \ref{2-fig3}(a) shows the evolution of the atom number
distribution $p(z,m,t)$. Curve A is the initial Gaussian
distribution. Before the first jump (curve B) the distribution
evolves to a doublet-like, similar to the curve B in Fig. \ref{2-fig2}(a).
However, in contrast to Fig. \ref{2-fig2}(a), the first photocount does not
return the distribution back to the single peak, and it stays
doublet-like (curve C). This is so, because the first jump occurs
rather late such that $m/\tau'<1$. Next jumps are rather late as
well, so the distribution evolves to a doublet (curves D, E, and F).

Figure \ref{2-fig3}(b) shows the number of photocounts $m$ growing in time. It
is clear that, in contrast to Fig. \ref{2-fig2}(b), $m$ is always much smaller
than the line $m=\tau'$ (red line), which gives the experimental
possibility to claim the appearance of the doublet.

Figures \ref{2-fig3}(c), \ref{2-fig3}(d), and \ref{2-fig3}(e), similarly to Fig. \ref{2-fig2}, show the decrease
of the width of one of two peaks of the atom number distribution,
conditioned cavity photon number with disappearing jumps, and the
Mandel parameter approaching zero. Note, that in contrast to Fig.
\ref{2-fig2}(d), the conditioned photon number $\langle a_1^\dag
a_1\rangle_c/|C'|^2$ does not approach the maximal value 1, but
rather decreases to a smaller value given by the doublet splitting
$\Delta z$, Eq.~(\ref{2-xx}). Thus, the appearance of the doublet can
be characterized by experimentally measuring $m$ and $t$ [Fig. \ref{2-fig3}(b)]
or the cavity photon number [Fig. \ref{2-fig3}(d)].

Figures \ref{2-fig4} and \ref{2-fig5} present another situation, where the probe-cavity
detuning is chosen such that $z_p=60$, which corresponds to a wing
of the atomic distribution function.

Figure \ref{2-fig4} shows the evolution of the atom number distribution in
case, where it collapses to a singlet at $z_p$. Thus, here even the
conditioned mean atom number changes from 50 to $z_p=60$. Other
characteristics of this process look very similar to the ones
presented in Figs. \ref{2-fig2}(b)-(e).

Figure \ref{2-fig5} shows the collapse to a doublet distribution around
$z_p=60$ with $\Delta z=7$. Thus, two satellites in Fig. \ref{2-fig5}(a) are
placed at $z_1=67$ and $z_2=53$. However, while the satellite at
$z_2=53$ is near the maximum of the initial distribution and is well
seen, the second satellite at $z_1=67$ falls on the far wing and is
practically invisible. As a result, the final distribution looks as
a singlet at $z_2=53$, while the second satellite is very small. The
fact that one has indeed a doublet can be verified by measuring the
photocount number [Fig. \ref{2-fig5}(b)], which is obviously less than $\tau'$,
or by measuring the cavity photon number [Fig. \ref{2-fig5}(e)], which is less
than the maximal value 1 and depends on $\Delta z$. The measurement
of the mean atom number [Fig. \ref{2-fig5}(c)] or width of atomic distribution
[Fig. \ref{2-fig5}(d)] would not distinguish between the singlet and doublet.

This suggests us a method to prepare the macroscopic superposition
state with unequal amplitudes of two components. Choosing the
detuning $\Delta_p$ such that $z_p$ is not at the center of the
initial atom number distribution, one can expect that one of the
satellites will be very probable near the distribution center, while
another one will fall on its wing.

The preparation of the particular state is a probabilistic process. However, importantly, varying the detuning $\Delta_p$ between the probe frequency and the cavity resonance, one can change the distribution of possible outcomes and increase or decrease the probability of a certain state to appear. In such a way, one can, e.g., increase the probability of the cat state with a given imbalance of the amplitudes of two components. One can always chose the detuning in such a way that the amplitudes will be equal to each other independently of the measurement outcome.

The most important difference of the Schr{\"o}dinger cat state
prepared by the transmission measurement from the one prepared by
transverse probing (\ref{2-21}) is that the phase difference between
two components of the quantum superposition is not limited by the
values of $0$ and $\pi$ as in Eq.~(\ref{2-21}). Although the light
amplitudes corresponding to $z_1$ and $z_2$ has equal absolute
values, their phases are opposite:
$\alpha_{z_1}=|\alpha_{z_1}|\exp(i\varphi)$ and
$\alpha_{z_2}=|\alpha_{z_1}|\exp(-i\varphi)$, where from
Eq.~(\ref{2-22}) the light phase $\varphi$ is
\begin{eqnarray}\label{2-25}
\varphi=-\arctan{\frac{U_{11}\Delta z}{\kappa}}.
\end{eqnarray}
Moreover, the imaginary parts of the functions $\Phi_q$ in
Eqs.~(\ref{2-7}) and (\ref{2-9}) are also different for $z_1$ and $z_2$
and have opposite signs: $\Phi(t)=\text{Im}\Phi_{z_1}(t)=
-\text{Im}\Phi_{z_2}(t)=\text{Im}(\eta_1 \alpha_{z_1}^*)$. Using
Eq.~(\ref{2-22}),
\begin{eqnarray}\label{2-26}
\Phi(t)=|\alpha_{z_1}|^2 U_{11}\Delta z t.
\end{eqnarray}
Thus, using Eq.~(\ref{2-7}), the cat state is given by
\begin{gather}
|\Psi_c\rangle=\frac{1}{F'}[e^{im\varphi +
i\Phi(t)}|z_1\rangle|\alpha_{z_1}\rangle \sqrt{p_0(z_1)} 
+ e^{-im\varphi - i\Phi(t)}|z_2\rangle|\alpha_{z_2}\rangle
\sqrt{p_0(z_2)}],\label{2-27}
\end{gather}
which is a macroscopic superposition of two Fock states with the
atom numbers $z_1$ and $z_2$ at $K$ sites.

Using the probe-cavity detuning $\Delta_p$ one can chose the center
of the doublet $z_p=\Delta_p/U_{11}$. Moreover, Eq.~(\ref{2-27}) shows
that using this detuning one can influence, at least,
probabilistically, the ratio between two components in the
superposition (\ref{2-27}). In particular, if we tune the probe
frequency such that $z_p$ coincides with the center of the initial
atom number distribution $p_0(z)$, the probabilities of the
symmetric doublet components will be equal, $p_0(z_1)=p_0(z_2)$,
providing the equal amplitudes of the cat components in
Eq.~(\ref{2-27}).

The phases $\Phi(t)$ and $m\varphi$ have opposite evolution in time.
The pase term $\Phi(t)$, Eq.~(\ref{2-26}), grows in time linearly and
deterministically, while the term $m\varphi$ grows in time
stochastically according to the growth of the photodetection number
$m$. In average, $\langle m\rangle = 2\kappa |\alpha_{z_1}|^2t$.
Thus, in general, the phase difference between two cat components
grows in time linearly. However, for some parameters, the growth of
the average $\langle m\rangle\varphi$ and $\Phi(t)$ can be
compensated. From Eqs.~(\ref{2-25}) and  (\ref{2-26}), $\langle
m\rangle\varphi + \Phi(t)=0$ for $U_{11}\Delta z/\kappa\approx
2.33$. So, for the particular cat components, the phase difference
does not grow in average. However, as $m$ is a stochastic quantity,
its uncertainty grows in time as $\sqrt{t}$. Thus, the
phase difference will still grow in time as $\sqrt{t}$, which is
nevertheless much slower than the linear growth.

The problem of photon losses addressed in the previous sections is
related to the stochastic quantity $m\varphi$. A single photocount
changes the phase difference between two components by $\Delta
\varphi_1=2\varphi$. In contrast to the state (\ref{2-21}), where the
phase jump is always maximal $\Delta \varphi_1=\pi$, here, this jump
can be rather small, if the condition $U_{11}\Delta z/\kappa <1$ is
fulfilled. This means, that to provide the robustness of the state
with respect to the photon losses, the doublet should not be split
too strongly. In the next section we quantitatively analyze the
robustness of the cat state.


\section{Robustness of the Schr{\"o}dinger cat states}

In this section we will show, that even if several photons are lost, the preparation of the cat states is still interesting.

The macroscopic superposition state (\ref{2-27}) is a pure state. In
principle, if the measurement is perfect and all photons $m$ leaking
the cavity are counted by a photodetector, it will evolve according
to Eq.~(\ref{2-27}) staying pure. However, if one loses one or more
photons, the state becomes a mixture of several states corresponding
to several lost counts $l$. Thus, if $L$ photons are lost, the state
is a mixture of $L+1$ states of the following form for $0<l<L$:
\begin{eqnarray}\label{2-28}
|\Psi_c\rangle_l=\frac{1}{\sqrt{2}}[e^{il\varphi +
i\gamma}|z_1\rangle|\alpha_{z_1}\rangle + e^{-il\varphi -
i\gamma}|z_2\rangle|\alpha_{z_2}\rangle],
\end{eqnarray}
where, for simplicity, we assumed the symmetric superposition with
$p_0(z_1)=p_0(z_2)$ and included all known phases [for $m$ measured
photons and deterministic $\Phi(t)$] in to the term $\gamma=m\varphi
+ \Phi(t)$. The density matrix of the state (\ref{2-28}) is
\begin{gather}
\rho_l=\frac{1}{2}(|z_1\rangle|\alpha_{z_1}\rangle \langle
z_1|\langle\alpha_{z_1}|+|z_2\rangle|\alpha_{z_2}\rangle \langle
z_2|\langle\alpha_{z_2}| \nonumber \\
+ e^{i2l\varphi + i2\gamma}|z_1\rangle|\alpha_{z_1}\rangle \langle
z_2|\langle\alpha_{z_2}| 
+ e^{-i2l\varphi - i2\gamma}|z_2\rangle|\alpha_{z_2}\rangle \langle
z_1|\langle\alpha_{z_1}|).\label{2-29}
\end{gather}
The density matrix of the mixture state describing $L$ lost photons
is given by a sum of the density matrices (\ref{2-29}):
\begin{gather}
\rho^{(L)}=\frac{1}{2}(|z_1\rangle|\alpha_{z_1}\rangle \langle
z_1|\langle\alpha_{z_1}|+|z_2\rangle|\alpha_{z_2}\rangle \langle
z_2|\langle\alpha_{z_2}| \nonumber \\
+ \frac{e^{i2\gamma}}{L+1}\sum_{l=0}^Le^{i2l\varphi}
|z_1\rangle|\alpha_{z_1}\rangle \langle
z_2|\langle\alpha_{z_2}| 
+ \frac{e^{-i2\gamma}}{L+1}\sum_{l=0}^Le^{-i2l\varphi}
|z_2\rangle|\alpha_{z_2}\rangle \langle z_1|\langle\alpha_{z_1}|).\label{2-30}
\end{gather}

The quantity characterizing how close is a mixture state to a pure
state is the so-called purity: $P=\text{Tr}(\rho^2)$. For a pure
state it is maximal and equal to 1, while for a maximally mixed
state it is minimal and equal to $1/2$ (in our case of the
two-component states). The purity of the state (\ref{2-30}) is given
by
\begin{eqnarray}\label{2-31}
P_L=\frac{1}{2}\left[1+\frac{1}{(L+1)^2}\left|\sum_{l=0}^L
e^{i2l\varphi}\right|^2\right],
\end{eqnarray}
where the sum can be calculated leading to the following result for
the purity of the mixed state corresponding to $L$ lost photons:
\begin{eqnarray}\label{2-32}
P_L=\frac{1}{2}\left[1+\frac{1}{(L+1)^2}\frac{\sin^2(L+1)\varphi}{\sin^2\varphi}\right].
\end{eqnarray}

For example, in the simplest case, where one photon is lost, the
density matrix of the mixed state is given by the sum of two terms
(\ref{2-29}) with $l=0$ and $1$:
\begin{gather}
\rho^{(1)}=\frac{1}{2}(|z_1\rangle|\alpha_{z_1}\rangle \langle
z_1|\langle\alpha_{z_1}|+|z_2\rangle|\alpha_{z_2}\rangle \langle
z_2|\langle\alpha_{z_2}| \nonumber \\
+ \frac{1}{2}e^{i2\gamma}(1+e^{i2\varphi})
|z_1\rangle|\alpha_{z_1}\rangle \langle
z_2|\langle\alpha_{z_2}|
+ \frac{1}{2}e^{-i2\gamma}(1+e^{-i2\varphi})
|z_2\rangle|\alpha_{z_2}\rangle \langle z_1|\langle\alpha_{z_1}|),\label{2-33}
\end{gather}
which has a purity
\begin{eqnarray}\label{2-34}
P_1=\frac{1}{2}(1+\cos^2\varphi)=\frac{1}{2}
\left(1+\frac{1}{1+(U_{11}\Delta z/\kappa)^2}\right).
\end{eqnarray}

Equations (\ref{2-33}) and (\ref{2-34}) show that if the phase jump
associated with the one-photon lost $\Delta\varphi_1=2\varphi$ is
maximal, $\Delta\varphi_1=\pi$, the state (\ref{2-33}) is maximally
mixed, because all non-diagonal terms responsible for the quantum
coherence between the states $|z_1\rangle|\alpha_{z_1}\rangle$ and
$|z_2\rangle|\alpha_{z_2}\rangle$ are zero. Its purity (\ref{2-34}) is
$1/2$, thus no entanglement survived after the single photon lost.
This is a situation of the transverse probing in the diffraction
minimum, Eq.~(\ref{2-21}), where the phase jump is $\pi$, which makes
the preparation scheme practically difficult.

However, if $\varphi$ is small, the purity (\ref{2-34}) and the
non-diagonal coefficients in Eq.~(\ref{2-33}) can be rather large and
close to 1. Thus, after the photon has been lost, one gets a mixed
state, but that of the high purity. More generally, for $L$ photon
losses, Eqs.~(\ref{2-30}) and (\ref{2-32}), the total phase jump
$\Delta\varphi_L=2L\varphi$ should be small. Using the expression
for $\varphi$ (\ref{2-25}), one can estimate the condition for the
high purity of the mixed state as $\varphi<\pi/(2L)$, which for a
small $\varphi$ is approximately the condition $U_{11}\Delta
z/\kappa<\pi/(2L)$. As a result, for the doublet, which is split not
too strongly, the high purity can be preserved even with photon
losses.

\begin{figure}[h!]
\centering
\captionsetup{justification=justified}
\includegraphics[width=0.4\textwidth]{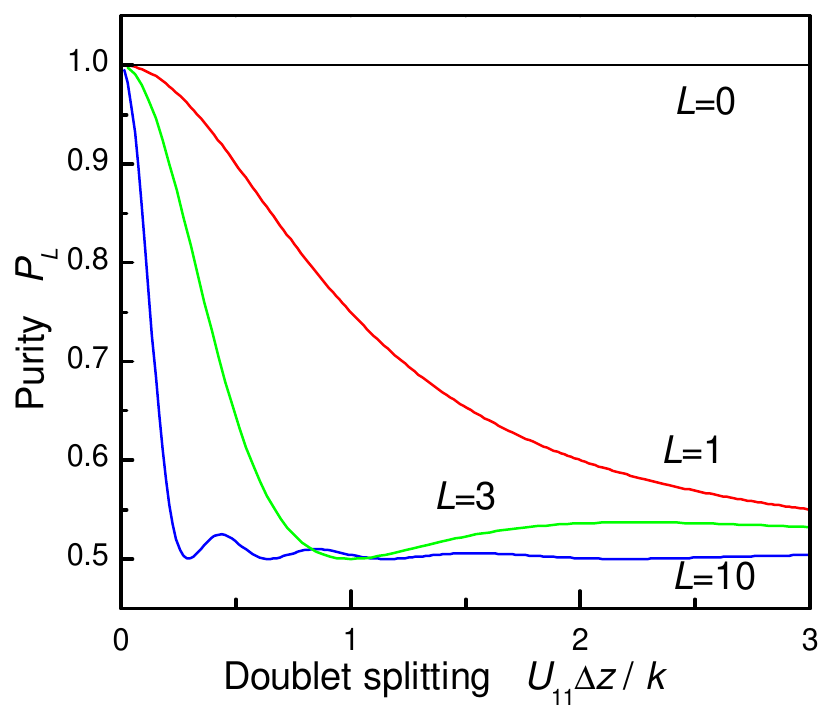}
\caption{\label{2-fig6}Purity of the mixed state as a
function of the doublet splitting $\Delta z$ for different numbers
$L$ of the photocounts lost: $L=0$, 1, 3, and 10.}
\end{figure}

The purity (\ref{2-32}) as a function of the doublet splitting $\Delta
z$ for $L=0$, 1, 3, and 10 photon losses is shown in Fig. \ref{2-fig6}, where
$\varphi$ is given by Eq.~(\ref{2-25}). One can see the decrease of
the purity with increasing doublet splitting and number of photons
lost. Note the non-monotonous character of its decrease, which means
that the larger splitting does not automatically leads to a smaller
purity.

Let us now estimate the purity of the atomic part of the final state (\ref{2-27}) before any photon has leaked. As before, we write 
\begin{eqnarray}\label{2-18x}
|\Psi_c\rangle=\frac{1}{\sqrt{2}}[e^{i\gamma}|z_1\rangle|\alpha_{z_1}\rangle
+ e^{- i\gamma}|z_2\rangle|\alpha_{z_2}\rangle],
\end{eqnarray}
where $\gamma$ is some phase, $|z_{1,2}\rangle$ are the atomic Fock
states with precisely known atom numbers $z_1$ and $z_2$.
$|\alpha_{z_{1,2}}\rangle$ are the corresponding coherent states of
light such that the light amplitudes have equal absolute values, but
opposite phases: $\alpha_{z_1}=|\alpha_{z_1}|\exp(i\varphi)$ and
$\alpha_{z_2}=|\alpha_{z_1}|\exp(-i\varphi)$.

Here, we analyze the state obtained by tracing out the light field
directly in Eq. (\ref{2-18x}), without waiting for photon leakage. This
corresponds to a scheme, where the measurement of light is not
performed.

The trace over the light variables can be calculated using the
photon Fock basis, in which the coherent light states read
\begin{eqnarray}\label{2-19x}
|\alpha_{z_{1,2}}\rangle=e^{-|\alpha_{z_1}|^2/2}\sum_{n=0}^\infty\frac{|\alpha_{z_{1}}|^ne^{\pm
n\varphi}}{\sqrt{n!}}|n\rangle.
\end{eqnarray}
The density matrix of the state after tracing out the light is
$\rho=\sum_{n=0}^\infty\langle n|\Psi_c\rangle \langle
\Psi_c|n\rangle$. Using the expression
\begin{eqnarray}\label{2-20xx}
\sum_{n=0}^\infty\frac{|\alpha_{z_1}|^{2n}}{n!}e^{-|\alpha_{z_1}|^2}e^{-2in\varphi}=
e^{|\alpha_{z_1}|^2(e^{-2i\varphi}-1)},
\end{eqnarray}
the density matrix takes the form
\begin{gather}
\rho=\frac{1}{2}(|z_1\rangle\langle z_1| + |z_2\rangle\langle
z_2|+  \nonumber \\ 
e^{|\alpha_{z_1}|^2(e^{2i\varphi}-1)}e^{2i\gamma}|z_1\rangle\langle
z_2|+e^{|\alpha_{z_1}|^2(e^{-2i\varphi}-1)}e^{-2i\gamma}|z_2\rangle\langle
z_1|).\label{2-21x}
\end{gather}

The purity of state (\ref{2-21x}) is given by
\begin{eqnarray}\label{2-22x}
P=\frac{1}{2}\left(1+e^{-4|\alpha_{z_1}|^2\sin^2\varphi}\right).
\end{eqnarray}

The purity depends on the amplitude of the coherent light states in
Eq. (\ref{2-18x}) and the phase difference between them. In a trivial
case, where two coherent states are indistinguishable ($\varphi=0$),
the purity is maximal, $p=1$, and the state is pure (however, in
this case, $z_1=z_2$ and the state is not a macroscopically
entangled one). In nontrivial cases, where the coherent states
differ by the phase $2\varphi$, the purity decreases with increase
of the light amplitude and phase difference between them. One can
estimate the minimal possible purity as follows. In a coherent
state, the uncertainty of the $X$-quadrature is $1/2$. Thus, two
coherent states can be well distinguished, if
$|\alpha_{z_1}|\sin\varphi>1/4$. Substituting the minimal value 1/4
in Eq. (\ref{2-22x}), one sees that the maximal purity can reach 0.89,
which is a rather high value.


\section{Measurement-based preparation of fermionic states}

In this section we will generalize our model for the case of ultracold fermions. We have already considered fermionic atoms in the previous chapter. Now we will show that the preparation of nontrivial states of magnetization is possible due to the quantum measurement backaction. 

Let us consider fermions in a one-dimensional optical lattice, inside an optical cavity with decay $\kappa_y$ such that only photons linearly polarised along the $y$ axis are allowed to escape it, and thus the measurement scheme probes the magnetisation of the atomic sample (cf. Sec. 1.9). We illuminate $K$ lattice sites at equal intesity with a coherent beam and we detect the $y$-polarised scattered light in the diffraction maximum, effectively probing the magnetisation of the illumined sites as $a_{1y}=C(\hat{N}_{K\uparrow}-\hat{N}_{K\downarrow})=C \hat{M}_K$, where $N_K$ and $M_K$ are the total atomic occupation number and magnetisation of all illuminated sites. If the $i$th photon is detected at time $t_i$, the state of the system changes instantaneously: the cavity annihilation operator $ {a}_{1y}$ is applied to it as $\ketz{\Psi_c(t_i)}\rightarrow  {a}_{1y}\ketz{\Psi_c(t_i)}$ (and the state is subsequently normalised). Moreover, the evolution of the system between two consecutive photocounts is determined by the non-Hermitian Hamiltonian 

\begin{align}
H_\text{eff}=H-i \kappa_y  a_{1y}^\dagger {a}_{1y}.
\end{align}

As we discussed before for the case of bosons, the photocount events and the evolution due to $H_\text{eff}$ have opposite contributions to the expectation value of $\hat M_K$ as the former tends to increase the magnetisation of the illuminated area while the latter tends to decrease it. The full dynamics of the system is therefore conditioned by the photodetections occurring at times $t_1, t_2$,.., which are determined stochastically. The initial wavefunction of the light--matter system can be written as

\begin{align}
\ketz{\Psi(0)} = \sum_{\b{k},\b{q}} c_{\b{q k}}^{(0)}\ketz{\b{q} \uparrow} \ketz{\b{k} \downarrow} \ketz{\alpha_{\b{q k}R}(0)}\ketz{\alpha_{\b{q k}L}(0)} ,
\end{align}
where $\ketz{\alpha_{\b{q k}p}(0)}$ are the coherent states of the light field with polarisation $p$ (in the circular polarisation basis) and $\ketz{\b{q} \sigma}$ are the Fock states for fermions of species $\sigma$. Since we are neglecting atomic tunnelling, the Hamiltonian $H_\text{eff}$ does not mix the different atomic Fock states, and it is possible to write an analytical expression for the conditional evolution of $\ketz{\Psi_c(t)}$. Specifically, the coherent states $\ketz{\alpha_{\b{q k}p}(0)}$ acquire a phase factor $\exp [\Phi_{\b{qk}}(t)]$ which depends only on the atomic configuration  $\b{qk}$: this phase reaches a steady state value after $t\gg1/\kappa_y$ for each quantum trajectory, allowing us to write the conditional wavefunction of the system after $m$ photodetection events as

\begin{align}\label{2-cond}
\ketz{\Psi_c(m,t)} \propto \sum_{\b{k},\b{q}} c_{\b{q k}}^{(0)} \, (\alpha_{\b{kq}R} - \alpha_{\b{qk}L})^{m} \, \mathrm{e}^{\Phi_\b{qk}(t)} \ketz{\b{q} \uparrow} \ketz{\b{k} \downarrow}\ketz{\alpha_{\b{qk}R}(t)}\ketz{\alpha_{\b{qk}L}(t)}.
\end{align}

Due to the steady state in all the light amplitudes $\alpha_{\b{qk}p}$, this expression does not depend on the specific detection times $t_i$. The state of the system is not factorisable into a product of matter and light states as the two remain entangled. Moreover,  the cavity introduces an effective interaction between the two spin species since different atomic configurations with different spins are coupled as a result of the photodetection. Focussing on the properties of the atomic system, we compute the probability distribution of having $z_\sigma$ fermions with spin $\sigma$ in the illuminated area of the optical lattice at time $t$. This can be obtained summing the absolute values of the coefficients of all the configurations $\b{q}^\prime\b{k}^\prime$ with the same $z_\sigma$ and introducing the initial probability distribution $p_0 (z_\sigma)=\sum_{\b{q}^\prime,\b{k}^\prime} |c_{\b{q}^\prime \b{k}^\prime}^{(0)}|^2$.
Hence, the conditional probability distribution is 

\begin{align}
p(t,z_\uparrow,z_\downarrow)=\frac{1}{\mathcal{N}} |z_\uparrow -z_\downarrow|^{2m} \exp\left( -\tau| z_\uparrow -  z_\downarrow|^2 \right) \, p_0(z_\uparrow) p_0(z_\downarrow).
\end{align}
where $\mathcal{N}$ is a normalisation constant and $\tau= 2 |C|^2 \kappa_y t$. Rewriting this expression in terms of the magnetisation of the system we have 

\begin{align}
p(t,M_K)=\frac{1}{\mathcal{N}} |M_K|^{2m}\exp\left( -\tau M_K^2 \right) \, p_0(M_K).
\end{align}

\begin{figure}[h!]
\captionsetup{justification=justified}
\centering
\includegraphics[width=0.8\textwidth]{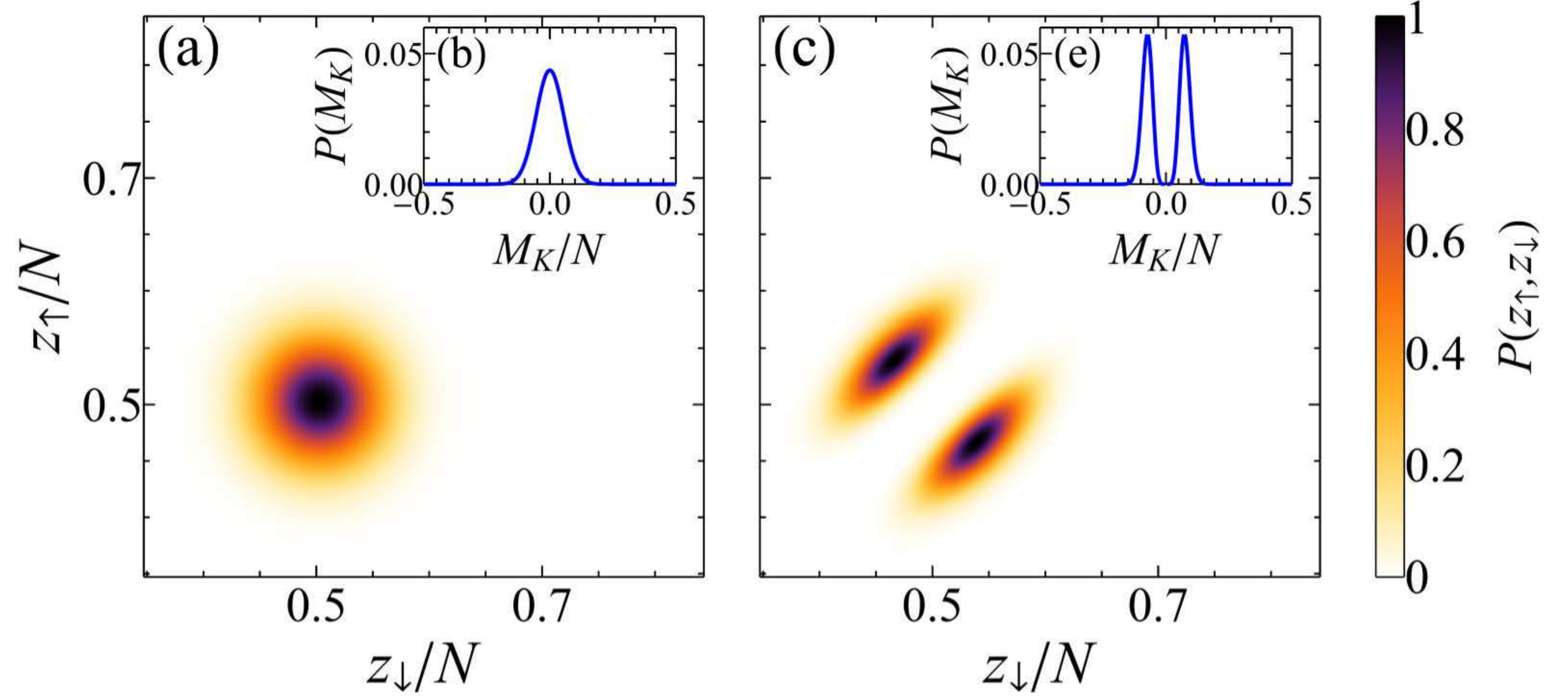}
\caption{Probability distribution $p(z_\uparrow,z_\downarrow)$ without measurement (a) and with measurement (c) for a fermionic system. The insets show the probability distribution of the magnetization $p(M_K)$ for the initial state (b) and after photons are detected (e). Measurement backaction creates a Schr\"odinger cat state.}
\label{2-fig:cat}
\end{figure}

The relation between $m$ and $t$ follows a stochastic process which defines a single quantum trajectory. The measurement process strongly modifies $p_0(M_K)$: when the first photon is detected this probability distribution becomes bimodal and all the configurations with $M_K=0$ are forbidden (Figure \ref{2-fig:cat}). Moreover, as time progresses $p(t,M_K)$ shrinks to two narrow peaks around the values $M_{1,2}=\pm \sqrt{m/\tau}$ with decreasing width (note, the stochastic nature of this expression arises through $m$). Thus, the final state of the system has a well-defined absolute magnetisation as the measurement process projects the atomic state to a superposition of two states with magnetisations $M_1$ and $-M_1$: a Schr\"odinger cat state.

In summary, in this section we showed how to extend the measurement-based quantum manipulation of the atom density to the manipulation of fermion magnetization.


\section{Quantum state preparation by the homodyne detection}

In this section we will briefly describe the homodyne detection scheme rather than the direct photocount scheme considered so far. 

\begin{figure}[h!]
\centering
\captionsetup{justification=justified}
 \includegraphics[width=0.5\linewidth]{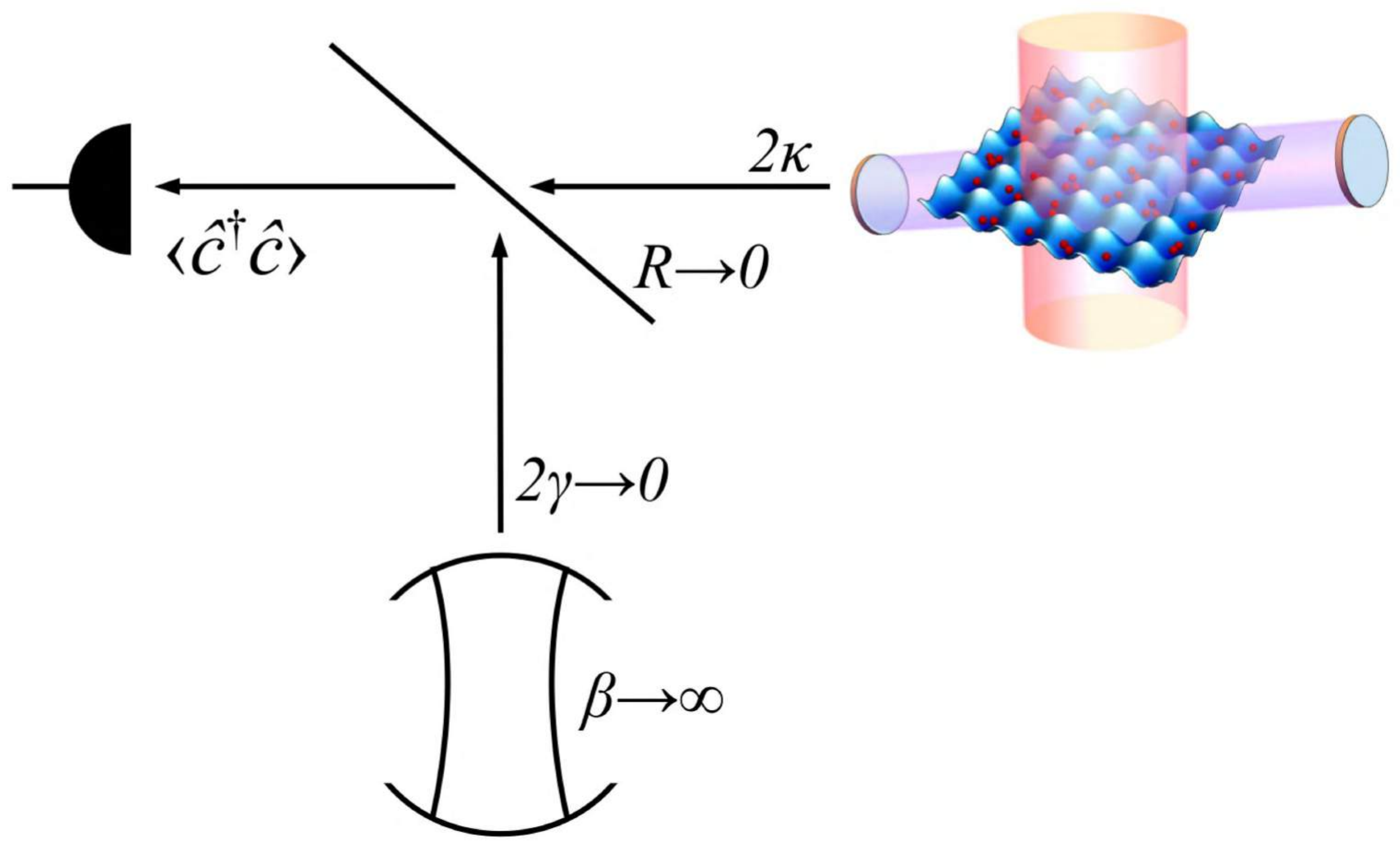}
  \caption{The homodyne detection scheme. The probed cavity contains an atomic gas in an optical lattice potential. The second cavity radiates a coherent local oscillator field $|\beta \rangle$.}
  \label{2-fig:Homodyne}
\end{figure}

The measurement scheme introduced before is very flexible and can be easily extended. For example, we consider a setup which includes a local oscillator cavity, and
another cavity containing the atomic system as illustrated in Fig.~\ref{2-fig:Homodyne}, where $\kappa$ and $\gamma$ are cavity decay constants for the atomic system and local oscillator respectively and
$\beta$ is the local oscillator coherent amplitude. We combine the two light modes with a beam splitter with reflectivity $R$, and in order for all the atomic cavity photons to reach the detector in our system we let $R \rightarrow 0$. However, this also requires that the local oscillator amplitude, $\beta \rightarrow \infty$ and $\gamma\rightarrow 0$ such that the local oscillator flux at the detector, $\mathcal{F}= R^2\gamma|\beta|^2$, remains constant \cite{Carmichael}. Assuming that the atomic dynamics is much slower than the light scattering, it can be shown that the evolution of the light--matter system is described by the non-Hermitian Hamiltonian

\begin{equation}
H_\text{eff} = H - i \hbar \kappa \hat a_1 ^\dagger \hat a_1 - i \hbar e^{-i \theta} e^{i\omega_p t} \sqrt{2 \kappa \mathcal{F}} \hat a_1,
\end{equation}
where $\theta$ is the local oscillator phase at the detector. In this case, the jump operator that is applied to the quantum state at each photodetection is \cite{Carmichael}

\begin{equation}
  \hat c=\sqrt{\mathcal{F}} e^{i \theta} e^{-i\omega_p t} +\sqrt{2 \kappa} \hat a_1.
\end{equation}

The state of the system at time $t$ after $m$ photodetections is

\begin{equation}
  | \Psi_c(m,t) \rangle \propto c^0_{z_+}\left[1 + e^{i \delta \phi}   z_+ \right]^m e^{i\varphi t} |z_+ \rangle | \alpha_{z_+} \rangle  + c^0_{z_-}\left[1 + e^{i \delta \phi} z_- \right]^m  e^{-i\varphi t} |z_- \rangle | \alpha_{z_-} \rangle,
\end{equation}
where $\varphi$ is a constant which can be derived from the Hamiltonian,

\begin{align}
z_\pm &= \sqrt{\frac{\mathcal{F}}{2 \kappa |C|^2}} \zeta_\pm, \\
\zeta_\pm &= \pm \sqrt{ \frac{m/t}{\mathcal{F}} - \sin^2(\delta \phi)} - \cos(\delta \phi),
\end{align}
where $\delta \phi = \phi_C - \theta$ is the phase difference between $\alpha_0$ and the local oscillator, $C = |C|e^{i \phi_C}$ and $z$ are the unique eigenvalues of the $\hat{D}$ operator and as such they are a suitable label of the measurement eigenstates.

The homodyne measurement scheme allows us the freedom to choose the phase difference between the local oscillator and the photons emitted from the atomic cavity system, $\delta \phi = \phi_C - \theta$, by adjusting the local oscillator phase. There are two interesting special cases depending on the value of this phase difference. Firstly, we can set the difference to an integer multiple of $2\pi$, $\delta \phi = 2 \pi n$. In this case the final state is given by

\begin{gather}
  | \Psi_c(m,t) \rangle \propto c^0_{z_+} e^{i \varphi t} |z_+ \rangle  | \alpha_{z_+} \rangle + c^0_{z_-} (-1)^m e^{-i\varphi t} |z_- \rangle | \alpha_{z_-} \rangle,\label{2-fragile}
\end{gather}
with $z_\pm = \sqrt{1 / 2 \kappa |C|^2} ( \pm \sqrt{ m/t } - \sqrt{\mathcal{F}})$, which is a cat state, albeit still entangled with the light. The light and matter can then be disentangled though, by switching off the probe, and detected all leaked photons, leaving the cavity in the vacuum mode. However, each photocount flips the phase difference between the two components by $\pi$, making this state particularly fragile with respect to decoherence: with each missed photocount the state loses its purity and becomes a mixed state, losing any atomic entanglement it possessed. Another notable case is if the two phases are offset by $\pi/2$, $\delta \phi = \pi/2 + 2 \pi n$. In this case \mbox{$z \equiv z_+ = -z_- = \sqrt{1 / 2 \kappa |C|^2} \sqrt{ m/t - \mathcal{F}}$} and the state evolution conditioned to the measurement leads to

\begin{gather}
  | \Psi_c(m,t) \rangle \propto c^0_{z}\left[\sqrt{\mathcal{F}} + i \sqrt{ m/t -\mathcal{F}}\right]^m e^{i \varphi t} |z \rangle | \alpha_{z} \rangle  \nonumber \\
  + c^0_{-z}\left[\sqrt{\mathcal{F}} - i \sqrt{ m/t - \mathcal{F}}\right]^m e^{-i \varphi t} |- z \rangle |- \alpha_{z} \rangle.\label{2-robust}
\end{gather}

Importantly,  $m/t - \mathcal{F}$ is small and each photocount imparts only a small change in phase between the two components.  The state (\ref{2-robust}) is therefore more robust than (\ref{2-fragile}) as it does not suffer much from the decoherence problem associated with a missed photocount and this scheme may prove a more useful method to prepare cat states.

As a result, the detection of light quadratures can be indeed used instead of the direct photocounting for the measurement-based state preparation. We will demonstrate the striking physically new phenomena due to the homodyne detection in Chapter 4 devoted to the feedback control of quantum phase transitions and quantum simulations of quantum baths.


\section{Generation of multipartite entangled spatial modes of ultracold atoms and their applications}

In several sections of this work we considered two main configurations: measurements of the atom number at a certain lattice region and measurements of the atom number difference at odd and even sites. Indeed, these are just two convenient examples of a bigger picture. In this section we will present a more general case, where a lattice region and atoms at odd or even sites are the examples of more general spatial atomic modes.

In this section, we will show that the effect of measurement backaction results in the generation of multiple many-body spatial modes of ultracold atoms trapped in an optical lattice, when scattered light is detected. The multipartite mode entanglement properties and their nontrivial spatial overlap can be varied by tuning the optical geometry in a single setup. This can be used to engineer quantum states and dynamics of matter fields. We provide examples of multimode generalizations of parametric down-conversion, Dicke, and other states, investigate the entanglement properties of such states, and show how they can be transformed into a class of generalized squeezed states, which is practically impossible to obtain in optics of light.

Measurement backaction was exploited in the breakthrough cavity QED experiments \cite{HarocheBook}, where atoms were used as probes of quantum states of light. Intriguing Fock and Schr{\"o}dinger cat states were prepared in a single cavity using quantum nondemolition (QND) methods. However, scaling to a large number of cavities provides an extreme challenge in this type of experiments. 

In contrast, we consider a case where the roles of light and matter are reversed: ultracold atoms are trapped in an optical lattice, and light is used as a global QND probe. Thus, the lattice sites represent the storage of multiple quantum states of matter fields, and the number of illuminated sites can be tuned from few to thousands, enabling scaling. We show how the quantum nature of light manifest in the measurement backaction can be used to establish a rich mode structure of the matter fields, with nontrivial delocalization over many sites and entanglement properties. These modes can be used for quantum state engineering, including multimode generalizations of parametric down-conversion (PDC) and Dicke states. We focus on the mode entanglement properties of these states, which exhibit genuine multipartite mode entanglement \cite{ghirardi2004,amico2008}, and contrary to the entanglement inherent to the symmetrization of indistinguishable particles, may be extracted for use in quantum information processing (QIP). In contrast to setups with atomic ensembles, we consider optical lattices, which enables the modes to have a significant amount of spatial overlap, and the light allows us to introduce effective long-range interactions, allowing for new schemes to be realized. 

\subsection{Multiple spatial atomic modes}

Let us summarize the results of the general model of measurement backaction presented in Sec. 2.2. We will consider only transverse probing ($\eta_1=0$), neglect the dispersive  frequency shift $U_{11}\hat{D}_{11}$, but analyze detection at any angle (without limiting ourselves to the measurements at diffraction maximum or diffraction minimum at $90^{\circ}$). As ultracold particles are delocalized, the atomic state is a superposition of Fock state configurations $\ketx{\bf{q}}=\ketx{q_1,..,q_M}$ corresponding to different occupations $q_j$ at $M$ sites. If the light probe is in a coherent state and the atoms in a Fock state, the scattered light will also be coherent with an amplitude $\alpha_{\bf{q}}$  dependent on the particular matter configuration: $\alpha_{\bf{q}}=C\braz{\bf{q}} \hat{D}_{10}\ketx{\bf{n}}$, where $\hat{D}_{10}=\sum_j u^*_1({\bf r}_j)u_0({\bf r}_j)\hat{n}_j$ sums density-dependent contributions from illuminated sites, $u_{1,0}(\bf{r})$ are the mode functions of probe and scattered light, and $C$ is the Rayleigh scattering coefficient into a cavity or free space \cite{Kozlowski2015PRA}. Due to the linearity of quantum mechanics, since the general atomic state is in a superposition of Fock states, the light and matter become entangled, with joint state $\ketx{\Psi}= \sum_{\{\bf{q}\}} c^0_{\bf{q}}\ketx{\bf{q}}\ketx{\alpha_{\bf{q}}}$.

When light is scattered into a cavity with a decay rate $\kappa$, detection of the escaped photons alters the probability amplitudes $c_{\bf{q}}$:  $c_{\bf{q}}(\tilde{m},t) =\alpha_{\bf{q}}^{\tilde{m}} e^{-|\alpha_{\bf{q}}|^2\kappa t}c^0_{\bf{q}}/F$ ($F$ is the normalization constant), the first factor reflecting $\tilde{m}$ quantum jumps (photons detected) and the second the non-Hermitian evolution during a time $t$. The measurement of lihgt hence changes the state of matter, this being the measurement backaction. While light amplitude and phase is measured, the distribution of $\alpha_{\bf{q}}$ is narrowed and the state is gradually projected towards terms with only one $\alpha$ (squeezing below the standard limit is not required). With continuous measurement light is pinned by the quantum Zeno effect, and the atoms will undergo Zeno dynamics \cite{Facchi2008, schafer2014, Vengalattore}, which is constrained such that they may only evolve within the region of Hilbert space with configurations $\ketx{\bf{q}}$ corresponding to the measured $\alpha$. Thus, the result of coherent collective scattering strikingly contrasts the outcome of incoherent light scattering, where atoms are localized to a mixed state and coherence is destroyed \cite{PichlerDaley2010}.

If both light modes are travelling waves, $\hat{D}_{10}=\sum_m e^{im\delta}\hat{n}_m$, where $\delta=(\bf{k}_{\mathrm{out}}-\bf{k}_{\mathrm{in}})\bf{d}$ (for two wave vectors $\bf{k}$ and lattice vector $\bf{d}$). This is a consequence of diffraction: depending on the angle, the light will show diffraction maxima ($\delta=0, 2\pi,..$) and minima in between. Thus, the choice of $\delta$ (via angles or light frequencies) determines the states to which light can be projected, and hence the corresponding atomic configurations. This choice thus controls the region of Hilbert space to which atomic dynamics is restricted during continuous light measurement. 

\begin{figure}[h!]
\captionsetup{justification=justified}
\centering
\includegraphics[clip, trim=1cm 21cm 6cm 1cm, width=0.5\textwidth]{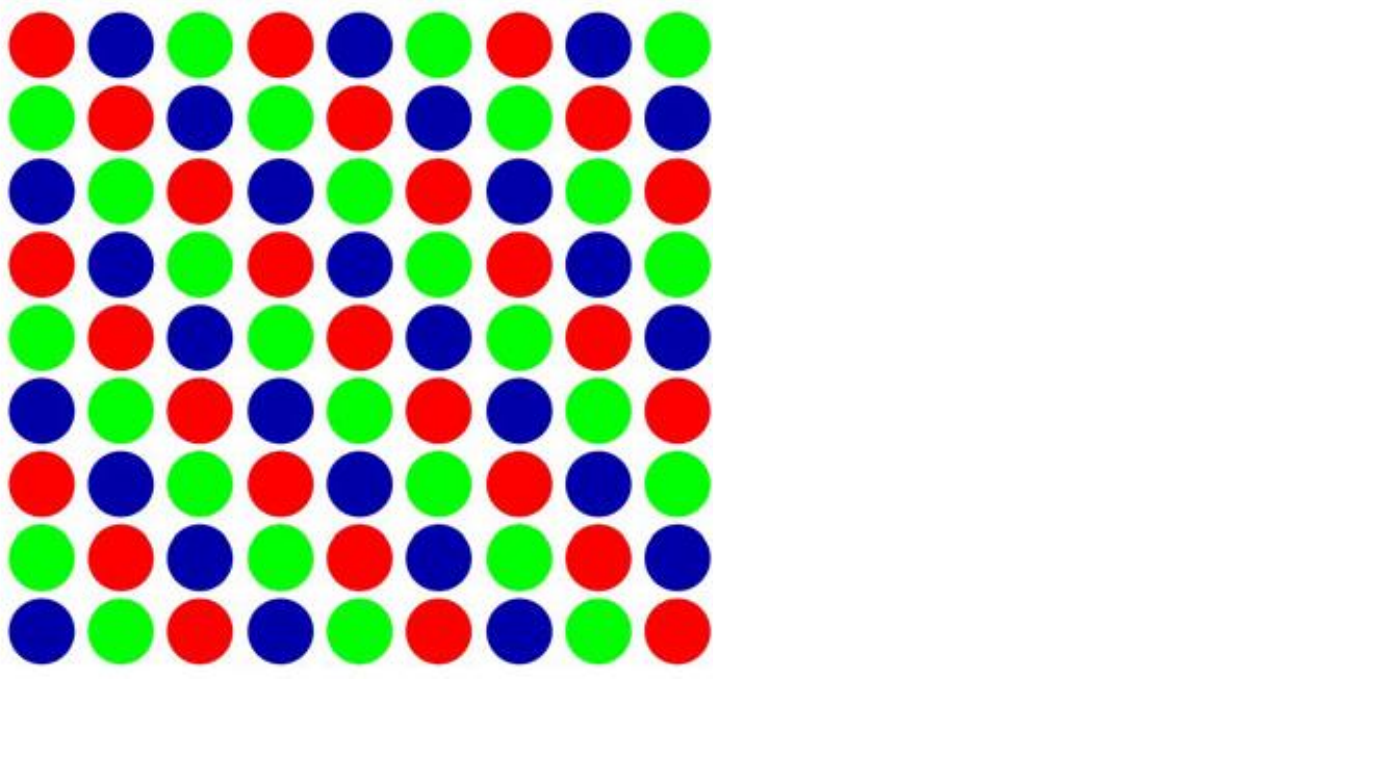}
\caption{Generated spatial structure of three matter-field modes in 2D. Sites of the same colour are indistinguishable to light scattering and thus belong to the same mode.}
\label{2-figmodel}
\end{figure}

Importantly, when $\delta=2\pi/R$ ($R$ is an integer), the atoms at sites $j+mR$ scatter light with the same phase and amplitude, and are therefore indistinguishable to light scattering. This crucially gives rise to $R$ spatial atomic modes: the atoms at indistinguishable sites belong to the same mode, while different atomic modes scatter light with different phases. Physically, $\delta=2\pi/R$ corresponds to the angles of multiple diffraction minima and small number of maxima. For example, for $\delta=2\pi$ (diffraction maximum) one mode ($R=1$) is formed as all atoms scatter light in phase, $\hat{D}_{10}=\hat{N}_K$ is the atom number operator for all $K$ illuminated sites (the remaining $M-K$ non-illuminated sites can be considered as an additional mode).  For $\delta=\pi$ (diffraction minimum for orthogonal light waves), two modes are generated ($R=2$) as the atoms at neighboring sites scatter light with opposite phases:  $\hat{D}_{10}=\sum_m (-1)^m\hat{n}_m= \hat{N}_\mathrm{even}-\hat{N}_\mathrm{odd}$ gives the number difference between even and odd sites. These are two examples, which we have already considered several times in this work, and which are now explained in the terms of the atomic matter modes. Importantly, for other diffraction minima, more spatially overlapping modes are generated as shown in \figsref{2-figmodel} for $R=3$ modes. 

\subsection{Fixing the atom numbers at all sites by sequential global measurements}

Multiple operators $\hat{D}_{10}$ can be measured. It is possible to fully characterize $R$ modes by measuring all of $\delta=2\pi m/R,$ with integer $m\in\left[0,(R-1)/2\right]$, with the $r$th mode composed of sites $j$ satisfying $j$mod$R=r$. The measured operators can be written in the form of an invertible Vandermonde matrix \cite{macon1958}. The inverse then reveals the occupation numbers for each mode (one has a system of linear equations to determine all atom numbers). Ultimately, for $R=M$, this leads to the determination of atom numbers at all lattice sites $n_j$ without the requirement of single-site resolution \cite{GreinerNature2009,BlochNature2011,Ashida2015}. In this case, the quantum measurement will project the state to a single multisite Fock state with well defined atom number at all sites $\ketx{n_1,..n_M}$.  This is a direct multimode analogy of preparation of a photon Fock state in a single cavity \cite{HarocheBook}. After this state is achieved, the quantum dynamics under continuous measurement is usually finished.

Although such effective single-site access is useful, here we focus on essentially many-body states. Importantly, even after the atom number in the modes is defined, the modes are still given by superpositions of Fock states, and thus quantum dynamics is not extinct, in contrast to the single-cavity QED case \cite{HarocheBook}. The Zeno effect prevents the mode atom numbers from changing, and thus prevents interactions between modes, but not within modes. This results in multiple ``virtual'' lattices on a single physical lattice (cf. Fig. \ref{2-figmodel}). By forgoing measurements with certain $\delta$, the restriction on the interaction between modes is partially lifted. This control over the interaction between modes allows for engineering of desired dynamics.

\subsection{Atomic analogies of optical states: beating the limits of quantum optics effectively increasing the nonlinearity and entanglement}

First, we show how an atomic state akin to photonic parametric down conversion (PDC) state \cite{gerry2005} ($\ketx{\Psi_\mathrm{PDC}}= \sum_n c_n\ketx{n}\ketx{n}$) can be realized by measurement, and readily generalized to a multimode case. The initial state is a superfluid (SF) delocalized over all sites. For a large lattice, this can be approximated by the Gutzwiller (mean field) ansatz \cite{Lewenstein} with a product over all sites, where each site is in a coherent state: $\ketx{\Psi}=\prod_{i=1}^M \sum_n e^{-\nu/2}\nu^{n/2}/\sqrt{n!}\ketx{n}_i$, where $\nu$ is the lattice filling factor. This state can be prepared with an external phase reference \cite{bartlett2007}. Measuring amplitude and phase for $\delta=\pi$ and either post-selecting or using feedback (modifying the trapping potential) \cite{Ivanov2014} to get $\langle\hat{D}\rangle_{\delta=\pi}=0$, we project to the state
\begin{equation}
\label{2-eqpdcpostselect}
\ketx{\Psi}=\frac{1}{\mathcal{N}}\realsum_n\frac{e^{-\lambda}}{n!}\lambda^n \ketx{n}\ketx{n},
\end{equation}
where $\lambda=\nu K/R$ is the average initial occupation number of each mode (here $R=2$). The two modes are defined as odd and even sites. Note that while post-selection or feedback is needed to get equal mode occupation, the measurement will deterministically project to a state with a fixed difference in occupation $\Delta N$:
\begin{equation}
\label{2-eqpdcdeterministic}
\ketx{\Psi}=\frac{1}{\mathcal{N}}\realsum_n\frac{e^{-\lambda}}{\sqrt{n!}\sqrt{n+\Delta N!}}\lambda^{n+\Delta N/2} \ketx{n}\ketx{n+\Delta N}.
\end{equation}
For large $N$, $\Delta N$ will become vanishingly small compared to the average number ($\Delta N\leq\mathcal{O}(\langle N\rangle^{1/2})$ and thus $\Delta N\ll\lambda$), hence the states (\ref{2-eqpdcpostselect}) and (\ref{2-eqpdcdeterministic}) exhibit many similar properties, such as their entanglement, and post-selection and feedback are not strictly necessary. 

We now generalize the procedure to several modes. By measuring all $\hat{D}_{10}$ operators required to characterize $R$ modes, excluding $\delta=0$ (which measures the total atom number), and again post-selecting equal numbers in all modes, we obtain the state
\begin{equation}
\label{2-eqmpdc}
\ketx{\Psi}=\frac{1}{\mathcal{N}}\realsum_n\left(\frac{e^{-\lambda}\lambda^n}{n!}\right)^{R/2}\ketx{n}^{\otimes R}.
\end{equation}
It is truly multimode, and has genuine multipartite entanglement (that is, any possible bipartitioning of modes shows nonzero entanglement \cite{horodecki2009}). This entanglement can be expressed by the entanglement entropy  \cite{amico2008}, the von Neumann entropy of the reduced density matrix of one of the subsystems (identical for the choice of either subsystem) $E(\ketx{\Psi}_{AB})=S(\rho_A)=-\mathrm{Tr}(\rho_A\log_2\rho_A).$ For all bipartitionings, if  $\lambda\gg1$, $E=(1/2)\log_2\left(2\pi e {\lambda}/{R}\right).$ As with the two mode case, even without post-selection, the state will still deterministically be projected to one with fixed number differences between each mode, and share similar properties to the ``ideal'' case (\ref{2-eqmpdc}).

In quantum optics, simple multimode PDC or four-wave mixing produce multipartite entanglement that is not genuine (entanglement exists between mode pairs, but not between all of them) \cite{Boyer2015}, as photons are produced in pairs, while higher nonlinearities are challenging to achieve.  In contrast, our system produces a kind of genuinely entangled multimode squeezed state which is generally non-Gaussian. In optics, similar continuous variable (CV) Gaussian-like states are obtained using multiple beam splitters, which complicates the scaling to many modes \cite{BraunsteinRMP}, or frequency combs \cite{Averchenko2014}. The atom-optics system we suggest here may provide advantages using the mode entanglement of quantum matter fields.

This method can also be used to create states similar to generalized squeezed states \cite{braunstein1987}, which in optics are expected to be formed from the highly nonlinear process described by the Hamiltonian $H=ga^k+g^*(a^\dagger)^k$ (which is extremely difficult experimentally). By taking the multimode case above (with $R$ modes taking the place of the $k$-photon process), and then lowering the lattice potential between the modes (but leaving a global trap), the atoms will behave as a single mode, with the state $\ketx{\Psi}=\sum_n c_n \ketx{N_0+nR}$, where $N_0$ and $c_n$ depend on the measured light. This generalizes squeezed vacuums containing even numbers of photons (pairs) to triplets, quadruplets, etc. for increasing $R$.

\subsection{Genuine multipartite mode entanglement}

Measurement at different angles results in other states. A case $\delta=0$ (diffraction maximum) reveals the total number of atoms illuminated, and thus projects to the ``fixed atom-number SF" or multimode generalization of ``spin coherent state". Using several such measurements, or in combination with measurements for other $\delta$ [e.g., 
measuring the total atom number in (\ref{2-eqpdcpostselect}) - (\ref{2-eqmpdc})], one prepares a product state of several SFs with $N_i$ atoms: $\ketx{\Psi}=\prod_{i=1}^R\ketx{N_i}_i$ in the mode basis. This corresponds to the multimode generalization of Dicke states if written in the symmetrized particle basis. Note that those SF modes may be noncontinuous in space, e.g. one SF can occupy each third site (\figref{2-figmodel}), which in time-of-flight measurements would be revealed as the period change in the matter-wave interference.

Another interesting case is when for $\delta=\pi$ the amplitude of $\alpha$ is measured, but not the phase. For two modes, this gives the magnitude of the number difference between the two modes $|\Delta N|$, but not its sign. This results in a superposition of cat states
$\sum_n c_n(\ketx{n,n+\Delta N}\ketx{\alpha_{\Delta N}} \pm \ketx{n+\Delta N,n}\ketx{-\alpha_{\Delta N}})/\sqrt{2}.$ If the light is turned off, this leads to the atomic cat state, which could be maintained by freezing matter dynamics by other means (e.g. by ramping up the lattice depth). 

\begin{figure}[h!]
\captionsetup{justification=justified}
\centering
\includegraphics[width=0.8\linewidth]{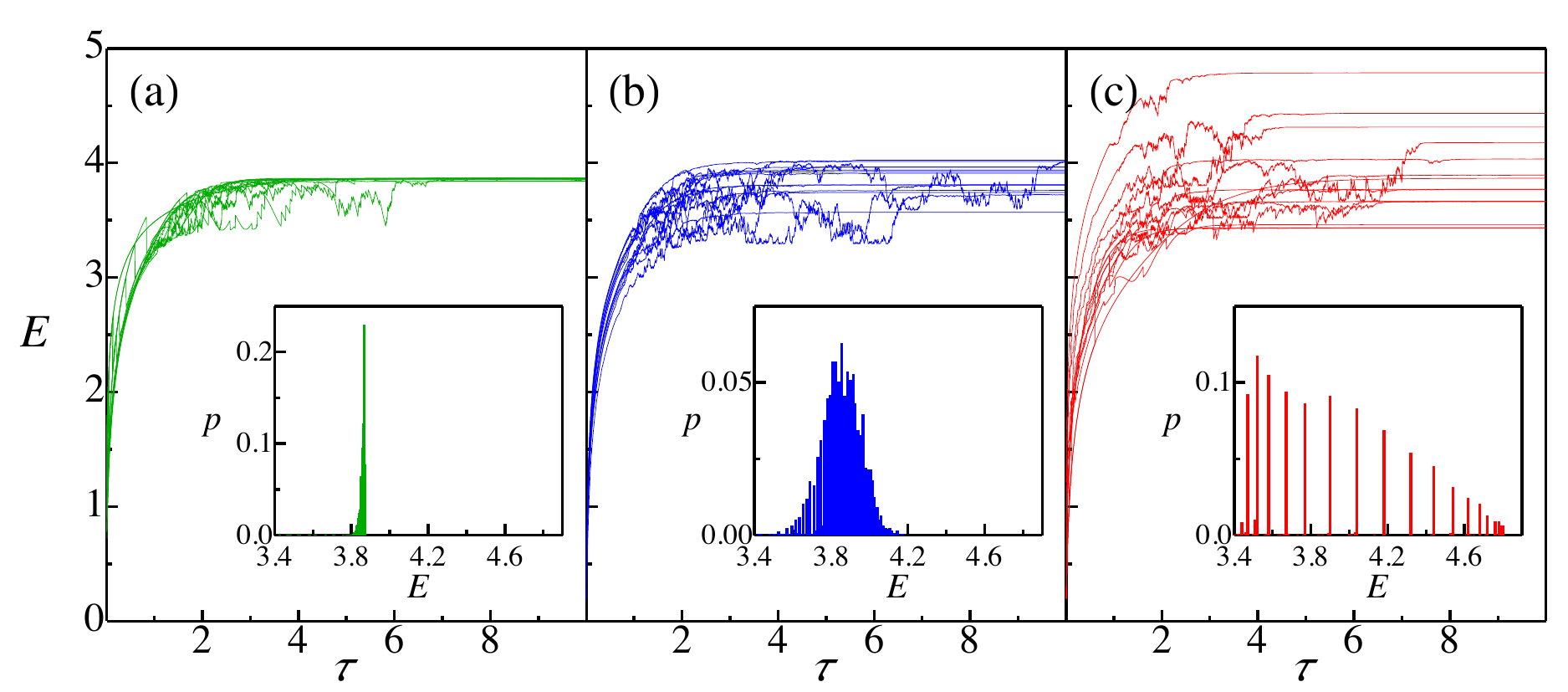}
\caption{Quantum trajectories for the growth of entanglement between two modes where the detected variables are (a) atom number difference in diffraction minimum; (b) total atom number in diffraction maximum; (c) absolute value of atom number difference in minimum. Insets show the final entanglement distribution functions. $\langle N\rangle$=50, $\tau=2|C|^2\kappa t$.}
\label{2-figtrajectories}
\end{figure}

In \figref{2-figtrajectories}, we present quantum trajectories of the evolution of entanglement entropy (using Quantum Monte Carlo Wave Function simulation \cite{Carmichael}) for these three cases: (a) phase-sensitive measurement at the diffraction minimum ($\delta=\pi$); (b) diffraction maximum ($\delta=0$); and (c) phase insensitive measurement at the minimum ($\delta=\pi$) (all for $R=2$ modes). We see that during the measurement, when light--matter entanglement obviously degrades, matter-matter entanglement between initially separable modes is established and grows significantly. While the final average entanglement is similar for all cases, the distribution widths are clearly different (see insets). In \figref{2-figtrajectories}(a), the distribution is very narrow, as amplitudes of Fock state in Eq. (\ref{2-eqpdcdeterministic}) only have a weak dependence on $\Delta N$ for the typical $\Delta N \ll \langle N\rangle$. In \figref{2-figtrajectories}(b), the distribution is broad, as the total atom number measurement projects to SFs with different $N$, and because of the scaling as $\log_2 N^{1/2}$ \cite{simon2002}, the final $E$ depends on the measured atom number. In \figref{2-figtrajectories}(c), the distribution is even broader as the state is a superposition of cat states. It has the highest probability to be projected to a state with large entanglement even though the average for all scenarios is the same. Importantly, for large $N$, all three widths vanish due to the logarithmic scaling. Thus, the entanglement in effect evolves deterministically, and simulation of a single quantum trajectory will be enough to describe the entanglement evolution, providing a significant numerical simplification.

Next, we show how the multimode structure of matter fields enables entanglement to be established between several spatially separated many-body systems, even if they are initially in Fock states without any phase coherence between them. Importantly, our method does not require any particle subtraction or light-induced particle exchange between the systems, in contrast to previous proposals \cite{Horak1999,RuostekoskiPRA1997}. The idea is to introduce additional sub-modes in the systems, which is possible in our setup. We start with two SFs with fixed atom numbers $N_A$ and $N_B$ (they can be prepared by measuring light at $\delta=0$). We define two new sub-modes within each SF, composed of the odd and even sites, and write the state of each subsystem in the basis $\ketx{N_{\mathrm{odd}},N_{\mathrm{even}}}$. Measurement for $\delta=\pi$ across the two subsytems then projects to a state with fixed atom number difference between odd and even sites across the two subsytems: $\ketx{\Psi}=\sum_k{N_A \choose k}^{1/2}{N_B \choose l(k)}^{1/2}\ketx{k,N_A-k}\ketx{l(k),N_B-l(k)}/\mathcal{N}$, where $l(k)=(N_A+N_B-\Delta N)/2 -k$, and $\Delta N$ is the measured number difference. The state of subsystem $A$ is then 
$\rho_A=\sum_k {N_A \choose k} {N_B \choose l(k)} \ketx{k, N_A-k} \braz{k, N_A-k}/\mathcal{N}^2$. There is thus entanglement between $A$ and $B$ if this reduced state has a non-zero entropy, i.e. ${N_A \choose k} {N_B \choose l(k)}\neq0$ for at least two $k$. This occurs whenever $|\Delta N|\neq (N_A+N_B)$.  Such a scheme for entanglement generation will readily work for multiple initially separable systems.

So far, we have used light detection defined by the on-site ($\hat{n}_i$) density-dependent  operators $\hat{D}_{10}$. Note that it is also possible to measure the combinations of conjugate operators $b^\dag_ib_{i+1}$ ($\hat{n}_i=b^\dag_ib_i$) by concentrating the probe light between sites \cite{Kozlowski2015PRA} as we presented in the previous chapter in Sec. 1.8. Detecting combinations of these operators may enable generation of cluster-like and other states used for quantum information processing (QIP) \cite{BraunsteinRMP,Sanpera2009}. We will consider the QND measurement backaction, when the matter-amplitude variables are detected, in Sec. 3.7.

In summary, in this section, we have shown how to use the quantum measurement of light to construct a multimode structure for ultracold atoms. We have demonstrated how this may be controlled to engineer the states and dynamics of the matter, and provided examples of multimode generalizations of down-conversion, Dicke, SF, and other states. We can produce states, which are nearly impossible to obtain in quantum optics of light due to the absence of extremely strong optical nonlinearities required. In our case, it is the measurement backaction that produces the effective nonlinearity, while no real atomic nonlinearity is taken into account. We have also shown how the nontrivial spatial overlap between matter-field modes can be exploited to produce  genuine multipartite mode entanglement.


\section{Concluding remarks of Chapter 2}

In this Chapter we demonstrated that cavity-enhanced light scattering off an ultracold gas in an optical
lattice constitutes a quantum measurement with a controllable form
of the measurement backaction. Time-resolved counting of scattered
photons alters the state of the atoms without particle loss
implementing a quantum nondemolition (QND) measurement, which we considered for both bosonic and fermionic atoms. The conditional dynamics is given by the interplay between
photodetection events (quantum jumps) and no-count processes. The
class of emerging atomic many-body states can be chosen via the
optical geometry and light frequencies. Light detection along the
angle of a diffraction maximum (Bragg angle) creates an atom-number
squeezed state, while light detection at diffraction minima leads to
the macroscopic superposition states (Schr{\"o}dinger cat states) of
different atom numbers in the cavity mode. A measurement of the
cavity transmission intensity can lead to atom-number squeezed or
macroscopic superposition states depending on its outcome. We
analyzed the robustness of the superposition with respect to missed
counts and find that the transmission measurement yields more robust
and controllable superposition states than the ones obtained by
scattering at a diffraction minimum.

We showed that the measurement backaction results in the generation of multiple many-body spatial modes of ultracold atoms. The multipartite mode entanglement properties and their nontrivial spatial overlap can be varied by tuning the optical geometry in a single setup. This can be used to engineer quantum states and dynamics of matter fields. We provided examples of multimode generalizations of parametric down-conversion, Dicke, and other states, investigated the entanglement properties of such states, and showed how they can be transformed into a class of generalized atomic squeezed states, well beyond the abilities of standard quantum optics of light.

An important point of this section was to show that measurement-induced dynamics exists even if other sources of the evolution (tunneling in our case) has been completely neglected. In the next chapter we will include tunneling and show that the quantum measurement backaction constitutes a novel source of competitions in many-body systems. Moreover, such a new type of competitions will lead to a novel type of phase transitions in quantum systems, which we will address in Chapter~4.

\clearpage

%% file: Chapter3.tex
\chapter{Quantum measurements as a novel source of competition in many-body systems} \label{chapt3}

\section{Introduction and plan of the chapter}

In this Chapter we will significantly extend our study of the systems subject to quantum measurement backaction by including the atomic tunneling and direct atom-atom interaction, which we neglected in the previous chapter. 

We show that the quantum measurement constitutes a novel source of competitions in many-body systems in addition to the usual competition between the tunneling and atom interaction. More precisely, we prove that the backaction of global measurement is able to efficiently compete with intrinsic short-range dynamics of an atomic system. The competition becomes possible due to the ability to change the spatial profile of a global measurement at a microscopic scale comparable to the lattice period without any need of single site addressing. In coherence with a general physical concept, where new competitions typically lead to new phenomena, we demonstrate a plethora of novel nontrivial dynamical effects such as large-scale multimode oscillations, long-range entanglement and correlated tunneling, as well as selective suppression and enhancement of dynamical processes beyond the projective limit of the quantum Zeno effect. 

For Fermi gases we demonstrate both the break-up and protection of strongly interacting fermion pairs by measurement. In addition, we present the measurement-induced generation of the antiferromagnetic order and density waves even without the requirement of strong atom interaction. 
 
 The measurement process generates spatial modes of matter fields that can be considered as designed systems and reservoirs, opening the possibility of controlling dissipations in ultracold atomic systems, without resorting to atom losses and collisions which are difficult to manipulate. The continuous measurement of the light field introduces a controllable decoherence channel into the many-body dynamics. Global light scattering from multiple lattice sites creates nontrival, spatially nonlocal coupling to the environment, which is impossible to obtain with local interactions~\cite{Daley2014,Diehl2008,RempeScience2008,Ashida2015}.  Such a quantum optical approach can broaden the field even further, allowing quantum simulation of models unobtainable using classical light, and the design of novel systems beyond condensed matter analogues. For example, both designed systems and reservoirs are represented by many-body strongly correlated systems with internal long-range entanglement. This will raise novel type of questions and stimulate further research on dynamics and states in physics of open systems and quantum engineering.

Note, that novel competition is only possible if the measurement is not a QND one. It is exactly the weak noise due to demolition of quantum variables that pushes the system to intriguing dynamical states. In the framework of QND paradigm, such effects are unattainable. Our quantum optical approach introduces into many-body physics novel processes, objects, and methods of quantum engineering, including the design of many-body entangled baths for open systems.

In Sec. 3.2 \cite{Mazzucchi2016PRA} we present a model of spatially structured global measurement. We explain, how the global long-range measurement covering the whole lattice is able to compete with extremely short-range (one lattice period) tunneling. This happens due to the short-range spatial structure imposed on the profile of the global measurement.

In Sec. 3.3 \cite{Mazzucchi2016PRA} we focus on the weak measurements, which nevertheless quickly push the system to very strong oscillations of macroscopic variables. Essentially, the measurement is not a QND one, and it is its destructive action that introduces weak noise and pushes the system to oscillate. When the atom-number-related variables are detected, the measurement can induce dynamical macroscopic superpositions (multimode Schr{\"o}dinger cat states). The resulting states have both oscillating density-density correlations with nontrivial spatial periods and long-range coherence, thus having properties of the supersolid state~\cite{EsslingerNat2010}, but in the essentially dynamical version. In addition we go beyond the atom-number-related measurements and introduce the matter-phase-related measurements. Using this, we demonstrate the wave-particle dualism of quantum mechanics in the many-body context.

In Sec. 3.4 \cite{Mazzucchi2016PRA} we present our results on the full competition between the measurement, tunneling, and atom interaction. For bosons, we show how the measurement affects the Mott insulator -- superfluid phase transition.  For strongly interacting fermions, we demonstrate how to destroy and protect fermion pairs, clearly showing the interplay between many-body dynamics of fermions and the quantum measurement backaction. 

In Sec. 3.5 \cite{Mazzucchi2016PRA} we switch to very strong measurement, which nevertheless is not fully projective. We demonstrate new phenomena beyond the paradigm of quantum Zeno dynamics and quantum Zeno effect. Even in the strong measurement regime, where the backaction dominates the system evolution, we find that long-range correlated tunneling emerges, leading to dynamical generation of long-range entanglement between distant sites (which can exist even in one-dimensional systems). We propose methods to engineer many-body baths using the measurements.

In Sec. 3.6 \cite{Mazzucchi2016NJP} we give details of analytical calculation of large-scale macroscopic oscillations presented in Sec. 3.3 (including stochastic Ito and Stratonovich equations). In addition, here we confront the quantum measurements and dissipation, by analyzing the role of detector efficiency. We also outline the analogies with purely photonic systems used, e.g., for boson sampling and other multi-path interference experiments of modern quantum technologies.

In Sec. 3.7 \cite{Kozlowski2017} we analyze in details the quantum measurements of matter-phase-related variables, in contrast to atom-number-related observables considered so far. We present both QND and non-QND scenarios and find the locking of the system in nontrivial subspaces, different from the results of quantum Zeno and dissipative state engineering.

In Sec. 3.8 \cite{Kozlowski2016PRAnH} we give more details about rather strong measurements presented in Sec. 3.5. We go beyond conventional
quantum Zeno dynamics. When the measurement is near,
but not in the projective limit, the system is still confined to a
Zeno subspace, nevertheless intermediate transitions are allowed via virtual
Raman-like processes. We show that this can be approximated by a
non-Hermitian Hamiltonian without quantum jumps thus extending the notion of quantum Zeno
dynamics into the realm of non-Hermitian quantum mechanics joining the
two paradigms.

In Sec. 3.9 \cite{Mazzucchi2016SciRep} we demonstrate the measurement-induced antiferromagnetic order and density modulations in Fermi gases, which can appear even without strong interatomic interactions.

We conclude this chapter in Sec. 3.10 and make a link to the next chapter of this work.    

{\it The results of this chapter are based on the papers} \cite{Mazzucchi2016SciRep,Mazzucchi2016NJP,Mazzucchi2016PRA, Kozlowski2016PRAnH, PRA2016-4, Kozlowski2017}.


\section{Model of spatially structured global measurement}

We use the general model developed in Sec. 1.1, and reduce it using the approximations already explained and used in the previous sections. The main difference from the previous chapter is that in the part of the Hamiltonian, which is responsible for unitary dynamics (without measurements), we will now take into account the tunneling and atom interaction. 

For the transverse probing (neglecting the dispersive frequency shift), similarly to classical optics, the light amplitude is given by a sum of scatterings from all atoms with coefficients dependent on their positions: $a = C (\hat{D} + \hat{B})$, where $a$ is the photon annihilation operator, $C$ is the Rayleigh scattering coefficient (compare with expressions in Sec. 1.3 and 1.8), and  
\begin{align}\label{3-op}
\hat{D}=\sum_{j=1}^L J_{jj} \hat{n}_j, \quad \hat{B}=\sum_{\langle i,j \rangle}^L J_{ij} b_i^{\dagger} b_j, 
\end{align}
where $b_j$ and $\hat{n}_j=b_j^{\dagger} b_j$ are the atomic annihilation and number operators at site $j$ (${\langle i,j \rangle}$ sums over neighboring sites) and $J_{ij}$ are given by
\begin{align}
J_{ij}=\int \! w (\b{r} -\b{r}_i) u_{\mathrm{out}}^*(\b{r})u_{\mathrm{in}}(\b{r}) w (\b{r} -\b{r}_j) \, \mathrm{d}  \b{r},
\end{align}
where $w (\b{r})$ are the localized Wannier functions and $u_{\mathrm{in,out}}(\b{r})$ are the mode functions of incoming (probe) and outgoing (scattered) light respectively (e.g., $u_l(\b{r})=\exp({i\b{k}_l\b{\cdot} \b{r}})$ for traveling and $u_l(\b{r})=\cos({\b{k}_l\b{\cdot} \b{r}})$ for standing waves with wave vectors $\b{k}_l$). To simplify the notations of the previous chapters, here we omitted the subscripts of operators $a_1$ and $D_{10}$, as well as the superscripts of the operators $J_{ij}^{lm}$. The number of sites is denoted as $L$ to avoid confusion with the magnetisation. 

Let's remind that in Eq.~(\ref{3-op}) $\hat{D}$ describes scattering from the on-site densities, while $\hat{B}$ that from the inter-site coherence terms. For well-localized atoms, the second term is usually neglected, and $a=C \hat{D}$ with $J_{jj}=u_{\mathrm{out}}^*(\b{r}_j) u_{\mathrm{in}}(\b{r}_j)$ (as in Sec. 1.3). For spin-$\frac{1}{2}$ fermions we use two light polarizations $a_{x,y}$ that couple differently to two spin densities $\hat{n}_{\uparrow j}$, $\hat{n}_{\downarrow j}$ allowing measurement of their linear combinations, e.g.,  $a_x=C \hat{D}_x=C\sum_{j=1}^L J_{jj} \hat{\rho}_{j}$ and $a_y=C \hat{D}_y=C\sum_{j=1}^L J_{jj} \hat{m}_j$, where $\hat{\rho}_{j}=\hat{n}_{\uparrow j}+\hat{n}_{\downarrow j} $ and $ \hat{m}_j=\hat{n}_{\uparrow j}-\hat{n}_{\downarrow j}$ are the mean density and magnetisation. This property has been used to investigate spin-spin correlations in Fermi gases \cite{Meineke2012, Sanner2012}.

As in the previous chapter, we focus on a single run of a continuous measurement experiment using the quantum trajectories technique. The evolution is determined by a stochastic process described by quantum jumps (the jump operator $c=\sqrt{2 \kappa} a$  is applied to the state when a photodetection occurs, where for the normalization purpose we introduce a prefactor $\sqrt{2 \kappa}$) and non-Hermitian evolution with the Hamiltonian 
\begin{equation}\label{3-Heff}
\hat{H}_{\mathrm{eff}}=\hat{H}_0- i \hbar c^\dagger c /2
\end{equation}
between jumps, where $H_0$ is the usual Bose--Hubbard Hamiltonian for bosons and Hubbard Hamiltonian for fermions. For the bosonic case,
\begin{align}\label{3-hamiltonianB}
\hat{H}_0=-\hbar J \sum_{\langle i,j\rangle}b_j^\dagger b_i +  \frac{\hbar U}{2} \sum_i \hat{n}_i \left(\hat{n}_i-1\right),
\end{align}
while for the fermionic case
\begin{align}\label{3-hamiltonian}
\hat{H}_0=-\hbar J \sum_{\sigma=\uparrow,\downarrow} \sum_{\langle i,j\rangle} f_{j,\sigma}^\dagger f_{i,\sigma} -\hbar U \sum_i \hat{n}_{i,\uparrow}\hat{n}_{i,\downarrow},
\end{align}
where $b$ and $f_\sigma$ are respectively the bosonic and fermionic annihilation operators, and $U$ and $J$ are the on-site interaction and tunneling coefficients.

Importantly, the measurement introduces a new energy and time scale $\gamma=|C|^2 \kappa$, which competes with the two other standard scales responsible for unitary dynamics of closed systems (tunneling $J$ and on-site interaction $U$). If different atoms scatter light independently, independent jump operators $c_j$ would be applied to each site, projecting the atomic system to a state, where the long-range coherence degrades~\cite{PichlerDaley2010}. This is a typical scenario for spontaneous emission~\cite{PichlerDaley2010,Sarkar2014}, or rather analogous local~\cite{Daley2014,Bernier2014,Hofstetter2014,Hartmann2012,RempeScience2008,Sherson2015} and fixed-range~\cite{LesanovskyPRL2012,LesanovskyPRB2014} addressing and interactions. Additionally, if the light is scattered without a cavity, i.e.~in uncontrolled directions, we lose the ability to choose the measurement operator and the jump operator applied would depend on the direction of the detected photon. In contrast, here we consider global coherent scattering, where the single global jump operator $c$ is given by the sum over all sites, and the local coefficients $J_{jj}$ (\ref{3-op}) responsible for the atom-environment coupling (via the light mode $a$) can be engineered by optical geometry. Thus, atoms are coupled to the environment globally, and atoms that scatter light with the same phase are indistinguishable to light scattering (i.e. there is no ``which-path information''). As a striking consequence, the long-range quantum superpositions are strongly preserved in the final projected states, and the system splits into several spatial modes, where all atoms belonging to the same mode are indistinguishable, while being distinguishable from atoms belonging to different modes, which we explained in details in Sec. 2.9.

We engineer the atom-environment coupling coefficients $J_{jj}$ using standing or traveling waves at different angles to the lattice. A key mechanism, which will allow us to construct the effective competition between global measurement and local processes, is the ability to modify these couplings at very short microscopic distances. This is in striking contrast to typical scenarios of Dicke and Lipkin-Meshkov-Glick models~\cite{ParkinsPRL2008}, where the coupling is global, but rather homogeneous in space. If both probe and scattered light are standing waves crossed at such angles to the lattice that projection $\b{k}_\mathrm{in} \b{\cdot} \b{r}$ is equal to $\b{k}_\mathrm{out}\b{\cdot} \b{r}$ and shifted such that all even sites are positioned at the nodes (do not scatter light), one gets $J_{jj}=1$ for odd and $J_{jj}=0$ for even sites. Thus we measure the number of atoms at odd sites only (the jump operator is proportional to $\hat{N}_\text{odd}$), introducing two modes, which scatter light differently: odd and even sites. The coefficients $J_{jj}=(-1)^j$ are designed by crossing light waves at $90^\circ$ such that atoms at neighboring sites scatter light with $\pi$ phase difference, giving $\hat{D}=\hat{N}_{\mathrm{even}}-\hat{N}_{\mathrm{odd}}$, introducing the same modes, but with different coherence between them. Moreover, using travelling waves crossed at the angle such that each $R$-th site is indistinguishable ($(\b{k}_\mathrm{in} - \b{k}_\mathrm{out})\b{\cdot} \b{r}_j=2\pi j/R$), introduces $R$ modes with macroscopic atom numbers $\hat{N}_l$: $\hat{D}=\sum_{l=1}^R \hat{N}_l e^{i2\pi l/R}$. Here two (odd- and even-site modes) appear for $R=2$. Therefore, we reduce the jump (measurement) operator from being a sum of numerous microscopic contributions from individual sites to the sum of smaller number of macroscopically occupied modes with a very nontrivial spatial overlap between them. 

In the following, we will show how globally designed measurement backaction introduces spatially long-range interactions of the modes, and demonstrate novel effects resulting from the competition of mode dynamics with standard local processes in a many-body system.


\section{Large-scale dynamics due to weak measurement}

In this section we show, how the competition between the measurement backaction and tunneling leads to the macroscopic oscillations and exchange of atoms between different atomic modes. Here we will focus on the physical side of the phenomena and present the results of numerical simulations. More technical aspects of these effects and analytical approaches will be presented in Sec. 3.6. for the measurements of density-related operators and in Sec. 3.7 for the matter-field-related operators.

\subsection{Measurements of the atom-number-related variables}

\begin{figure}[h!]
\captionsetup{justification=justified}
\centering
\includegraphics[width=0.6\textwidth]{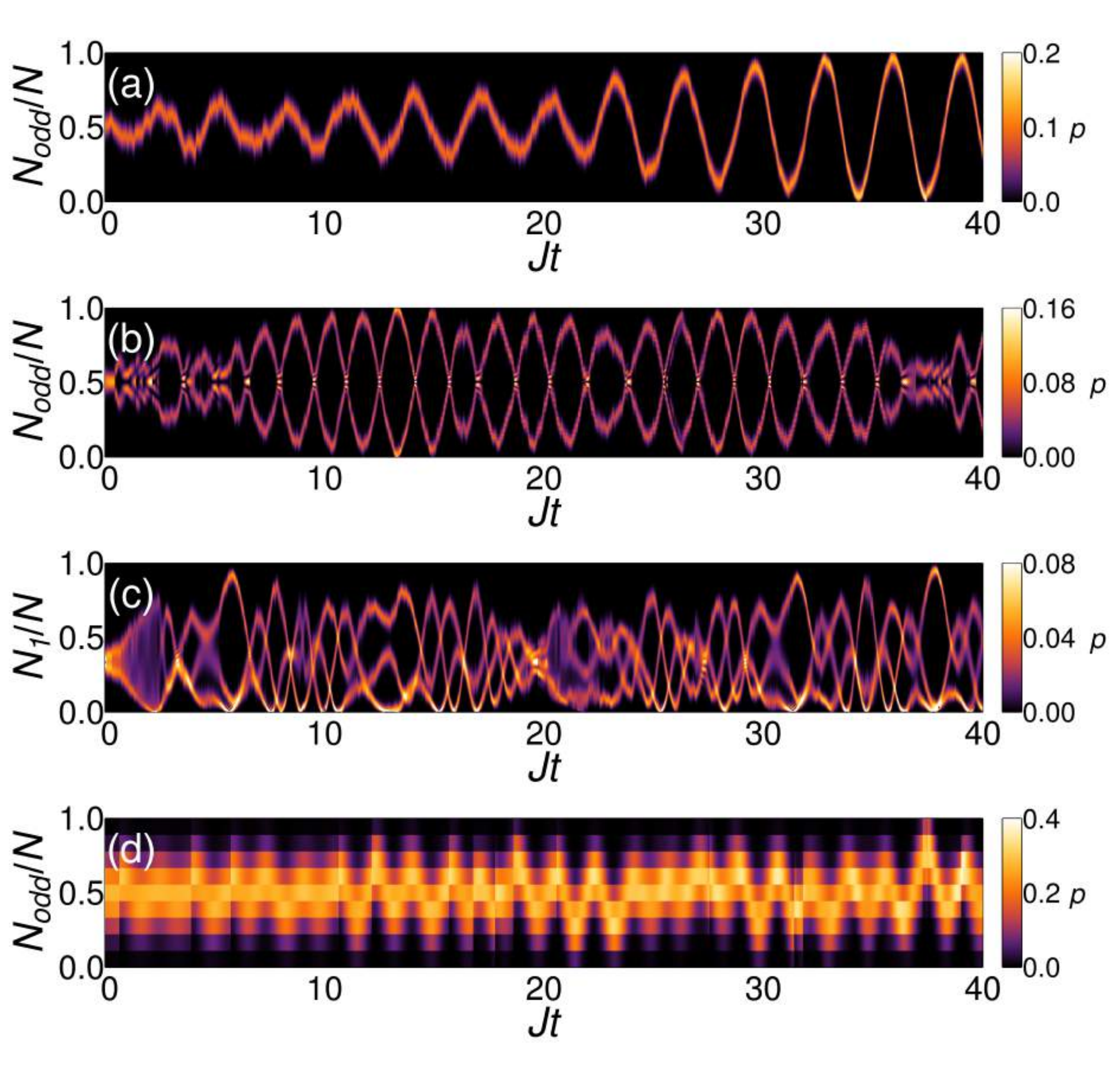}
\caption{Large oscillations between the measurement-induced spatial modes resulting from the competition between tunneling and weak-measurement backaction. The plots show single quantum trajectories. (a)-(d) Atom number distributions $p(N_l)$ (the curves have changing widths) in one of the modes, which show various number of well-squeezed components, reflecting the creation of macroscopic superposition states depending on the measurement configuration ($U/J=0$, $\gamma/J=0.01$, $L=N$, initial states: superfluid for bosons, Fermi sea for fermions). (a) Measurement of the atom number at odd sites $\hat{N}_\text{odd}$ creates one strongly oscillating component in $p(N_\text{odd})$ ($N=100$ bosons, $J_{jj} = 1$ if $j$ is odd and 0 otherwise). (b) Measurement of $(\hat{N}_\text{odd}-\hat{N}_\text{even})^2$ introduces $R=2$ modes and preserves the superposition of positive and negative atom number differences in $p(N_\text{odd})$ ($N=100$ bosons, $J_{jj}=(-1)^{j+1}$). (c) Measurement for $R=3$ modes (see text) preserves three components in $p(N_1)$ ($N=108$ bosons, $J_{jj}=e^{i j 2 \pi /3}$).  (d) Measurement of $\hat{N}_\text{odd}$ for fermions leads to oscillations in $p(N_\text{odd})$, though not as clearly defined as for bosons because of Pauli blocking ($L=8$ sites, $N_{\uparrow}=N_{\downarrow}=4$ fermions, $J_{jj} = 1$ if $j$ is odd and 0 otherwise). 1D lattice.} \label{3-panel1}
\end{figure}

We start with non-interacting bosons ($U/J=0$), and demonstrate that the competition between tunneling and global measurement strongly affects the dynamics of the atomic system. The weak measurement ($\gamma \ll J$) is unable to freeze the atom numbers projecting the atomic state and quantum Zeno dynamics~\cite{Facchi2008, Raimond2010, Raimond2012, schafer2014, Signoles2014} cannot be established. In contrast, the measurement leads to giant oscillations of particle number between the modes. Figures~\ref{3-panel1}(a)-(c) illustrate the atom number distributions (note that the curves have changing widths) in one of the modes for $R=2$ ($N_\text{odd}$) and $R=3$ ($N_1$). Without continuous monitoring, these distributions would spread out significantly and oscillate with an amplitude proportional to the the initial imbalance, i.e.~tiny oscillations for a tiny initial imbalance. In contrast, here we observe (i) full exchange of atoms between the modes independent of the initial state and even in the absence of initial imbalance, (ii) the distributions consist of a small number of well-defined components, and (iii) these components are squeezed even by weak measurement.
Depending on the quantities addressed by the measurement, the state of the system has a multi-component structure which is a consequence of the photon number (intensity) $a^\dag a$ not being sensitive to the light phase. In other words, the measurement does not distinguish between all permutations of mode occupations that scatter light with the same intensity. The number of components of the atomic state, i.e.~the degeneracy of $a^\dag a$, can be computed from the eigenvalues of  $\hat{D}=\sum_{l=1}^R \hat{N}_l e^{i2\pi l/R}$ noting that they can be represented as the sum of vectors on the complex plane with phases that are integer multiples of $2\pi /R$: $N_1e^{i2\pi/R},N_2e^{i4\pi/R}, \dots  N_R $. Since the sum of these vectors is invariant under rotations by $2\pi l /R$, and reflection in the real axis, the state of the system is two-fold degenerate for $R=2$ and $2R$-fold degenerate for $R>2$.
Figure~\ref{3-panel1}(b) shows the superposition of two states with positive and negative $N_\text{odd}-N_\text{even}$, while Fig.~\ref{3-panel1}(c) illustrates the superposition in the three-mode case. 

One can get a physical insight into the origin of oscillations by constructing the following model (full details of analytical calculations are given in Sec. 3.6). For macroscopically occupied modes of non-interacting atoms initially in the superfluid state, the $R$-modes problem can be treated analytically reducing it to $R$ effective sites. In particular, for $R=2$ (effective double-well \cite{Milburn1998,Julia-Diaz2012}), we can write the atomic state as 
\begin{align}\label{3-wf}
|\psi\rangle =\sum_{l=0}^{N} q_l |l,N-l\rangle,
\end{align}
where the ket $|l,N-l\rangle$ represents a superfluid with $l$  atoms in the first and $N-l$ atoms in the second spatial mode. In the limit $N \gg 1$  we can describe the evolution of the system using continuous variables~\cite{Julia-Diaz2012} and define the wave function $\psi(x=l/N)=\sqrt{N} q_l$. Introducing the relative population imbalance between the two wells $z=2 x -1$ and starting from the superfluid state, the solution of the Schr\"odinger equation with Hamiltonian $\hat{H}_{\mathrm{eff}}$ (\ref{3-Heff}) is
\begin{align}
\psi(z,t)\propto \exp \left[{i a(t)+\frac{i z c(t)}{2 b^2(t)} +\frac{i z^2 \phi(t)}{2 b^2(t)}-\frac{(z-z_0(t))^2}{2  b^2(t)}}\right],
\end{align}
where $b(t)$ is the width of atomic distribution, $z_0(t)$ is the average population imbalance and $c(t)$ and $\phi(t)$ are real phases while $a(t)$ is complex and takes into account the decay of norm of the wave function due to the non-Hermitian term present in $\hat{H}_{\mathrm{eff}}$. Expanding the Hamiltonian in powers of $1/N$ up to second order we obtain a system of coupled differential equations for $a(t), \,b(t), \,c(t), \,z_0(t)$, and $\phi(t)$ which, in the weak measurement limit, can be linearized, leading to 
\begin{align}\label{3-population}
 z_0(t)=\frac{1}{2} \mathrm{e}^{-\frac{N \gamma}{2}t } \left[c(0) \sin (2 J t)+2 z_0(0) \cos ( 2 J t)\right].
\end{align}
This expression describes the evolution of the population imbalance between two quantum jumps: The atoms oscillate between the two spatial modes with decreasing amplitude and tend to restore a balanced distribution. The quantum jumps strongly affect the dynamics of the system via measurement backaction, driving the oscillations to their possible maximum amplitude, i.e.,~all the atoms oscillate between the two spatial modes. We explain this effect noting that (i) the average time between two jumps ($\sim2 / N^2 \gamma$) is much smaller than the damping time in (\ref{3-population}), (ii) the probability of a jump depends on the imbalance itself $p_\mathrm{jump}\propto (1+z_0)^2$ and (iii) each jump tends to increase the population imbalance. In order to calculate how the photodetection events affect the evolution of $z_0$, we compute the effect of average jump rate as $\delta z_0 / \delta t= p_{\mathrm{jump}}\, \Delta z_0 / \delta t $, where $\Delta z_0$ is the change in $z_0$ due to the jump itself. Solving this equation leads to
\begin{align}\label{3-jumps}
 z_{0,\mathrm{jumps}}(t)=-1+(1+z_0(0))\mathrm{e}^{N \gamma t}.
\end{align}
Therefore, the amplitude of oscillations of $z_0$ increase since the difference between the exponents in (\ref{3-population}) and (\ref{3-jumps}) is positive and the system leaves the stable point $z_0=\dot{z_0}=0$. Note that the dynamics described here is a feature of single (but all!) trajectories only. This is in contrast to averaging over many runs, which corresponds to the master equation solution,  which masks these effects. This happens because the oscillation phase changes from realization to realization as is known from works involving single and multiple measurements~\cite{RuostekoskiPRA1997}.

The competition between measurement and atomic dynamics allows realization of multicomponent macroscopic superpositions (Schr{\"o}dinger cat or NOON states), which are useful in quantum metrology and information. Such multimode superpositions are a purely quantum effect. The multimode dynamics may recover the semiclassical character~\cite{Ruostekoski2014}, when the number of modes is reduced to one. The method we propose does not require external control \cite{Pedersen2014,Ivanov2014, Ivanov2016} for preparing these states: By continuously monitoring the light intensity it is possible to determine when the splitting in the components reaches its maximum value (corresponding to the maximal macroscopicity \cite{Frowis2012}) and further oscillatory dynamics can be stopped by ramping up the lattice depth. Note that for $R>1$ spatial modes each photocount changes the phase difference between the various components of the atomic state, making it fragile to photon losses. However, the measurement setup can be modified to make these states more robust as we have already seen \cite{MekhovPRA2009}. In addition, the example in Fig.~\ref{3-panel1}(a) consists of only one component (here $N_\text{odd}$ is measured directly) and is not therefore very sensitive to decoherence due to photon losses.

It is interesting to note that the oscillating state we have described shows spatial periodicity in the density-density correlation function depending on the measurement configuration. For $R$ modes the spatial period is $Rd$ ($d$ is the lattice period). Moreover, the state also has long-range coherence between distant sites. Therefore, it has properties of the supersolid state~\cite{EsslingerNat2010} but with a nontrivial period, and it exists in the essentially dynamical version.

In contrast to bosons, dynamics for two modes of non-interacting fermions does not show well-defined oscillations (Fig.~\ref{3-panel1}(d)) due to Pauli exclusion. However, while the initial ground state is a product of $\uparrow$ and $\downarrow$ wave functions (Slater determinants), the measurement introduces an effective interaction between two spin components and the state becomes entangled by measurement. The dynamics of a one-dimensional fermionic system strongly depends on the spatial profile of the measurement operator. For example, if only the central part of the lattice is illuminated (diffraction maximum) the atom transfer between the modes induced by the measurement is suppressed as the presence of an atom at the edges of the illuminated area completely forbids the atomic tunneling between the two modes, greatly decreasing the fluctuations of the measurement operator.

\subsection{Measurements of the matter-wave-related variables}

Carefully choosing geometry (as we explained in Sec. 1.8), one can suppress the on-site contribution to light scattering and effectively concentrate light between the sites, thus \emph{in-situ} measuring the matter-field interference $b^\dag_ib_{i+1}$. In this case, $a=C\hat{B}$, and the coefficients $J_{ij}$ for $i\ne j$ from Eq. (\ref{3-op}) can be engineered (we will give all analytical results in Sec. 3.7). 

Thus, here we will not measure the number-related variable $\hat{D}$ as just before, but its conjugate matter-amplitude or matter-phase related variable $\hat{B}$. In the density-related measurement, the phase was destroyed and became noisy, which pushed the system to compete with tunneling resulting in the macroscopic density oscillations. Here, in contrast, the measurement of phase-related variable will introduce noise to its conjugate number variable, which will lead to the spread of the atomic density. These effects are direct consequences of the wave-particle dualism in quantum mechanics. In this work, we demonstrate them in the context of many-body physics. Depending on the variable chosen for measurement (the particle-like atom number or wave-like matter phase) we can get either particle-like density oscillations as just before, or wave-like spread of atoms across the lattice, as we will show now.  

\begin{figure}[h!]
\captionsetup{justification=justified}
\centering
\includegraphics[width=0.6\textwidth]{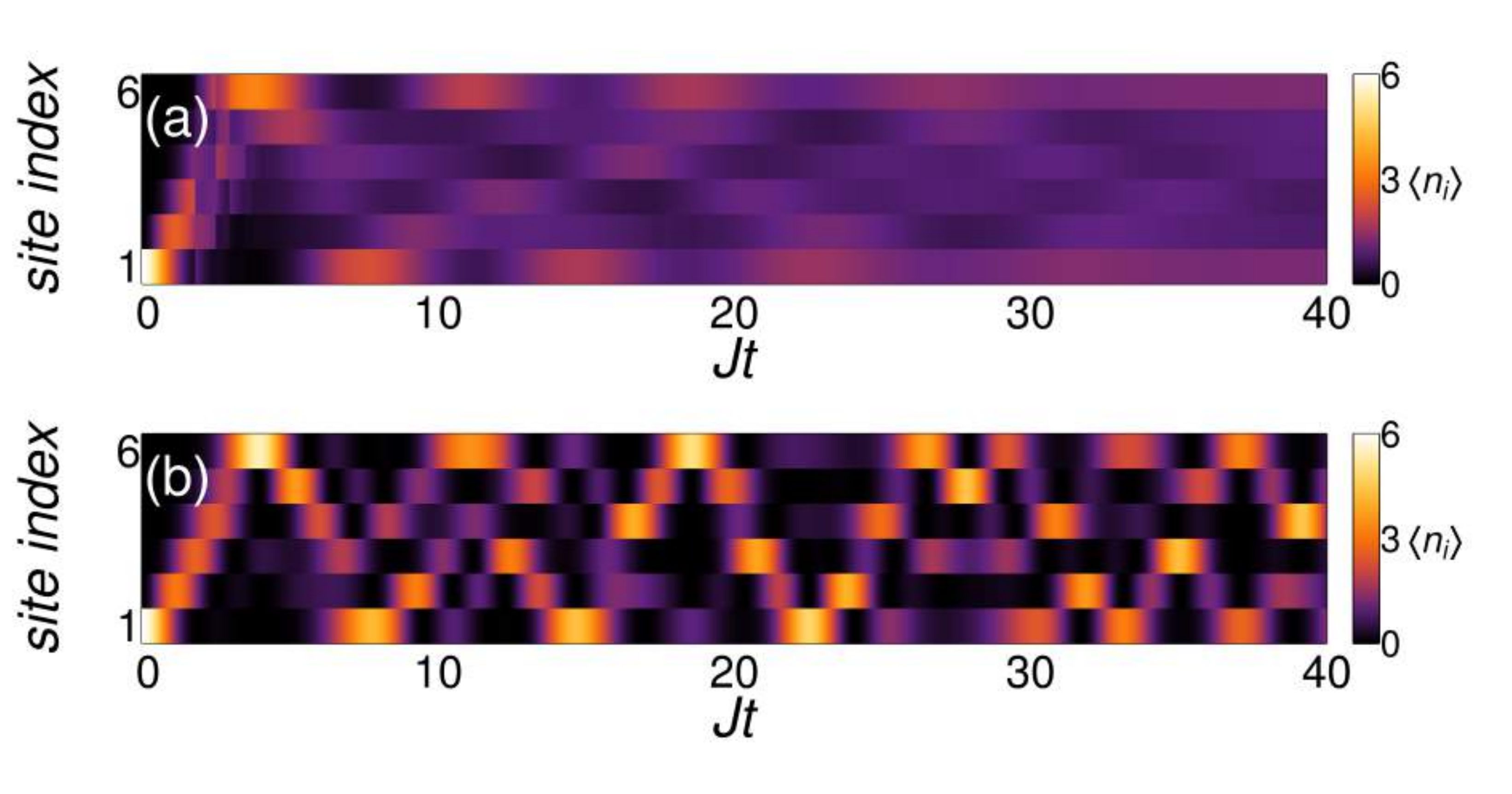}
\caption{Evolution of the on-site atomic density, while measuring the matter-field coherence between the sites $\hat{B}$. The plots show single quantum trajectories.  (a) Atoms, all initially at the edge site, are quickly spread across the whole lattice leading to the large uncertainty in the atom number, while the matter-phase related variable is defined (projected) by the measurement (bosons, $N=L=6$, $U/J=0$, $\gamma/J=0.1$, $J_{jj}=1$). (b) Atomic density spread and revival due to coherent tunnelling in the absence of measurement (bosons, $N=L=6$, $U/J=0$, $\gamma/J=0$, $J_{jj}=1$). Simulations for 1D lattice.} \label{3-panel1bis}
\end{figure}

For $J_{ij}=1$, the jump operator is proportional to the kinetic energy $\hat{E}_K=-2 \hbar J \sum_{k} b^\dagger _k b_k \cos(ka)$ and tends to freeze the system in eigenstates of the non-interacting Hamiltonian. The measurement projects to a superposition of two states with different kinetic energies: a superposition of matter waves propagating with different momenta. The measurement does not distinguish between these two states, because in a lattice, the two momentum states $|k\rangle$ and $|\pi/a-k\rangle$ interfere in the same way, but with opposite phase in between the lattice sites. The measurement freezes dynamics for any $\gamma/J$, since the jump operator and $\hat{H}_{\mathrm{eff}}$ have the same eigenstates. As a result of the detection, the atoms quickly spread across the lattice, and the density distribution becomes uncertain (Fig.~\ref{3-panel1bis}(a)), clearly illustrating the quantum uncertainty relation between the number- and phase-related variables ($\hat{n}_i$ and $b^\dag_ib_{i+1}$). Note that, in the  absence of measurement, such a distribution presents a periodic spread and revival due to coherent tunneling and the system does not reach a steady state (Fig.~\ref{3-panel1bis}(b)). Therefore, engineering $J_{ij}$ can lead to the measurement-based preparation of peculiar multicomponent momentum (or Bloch) states. Further details are given in Sec. 3.7.

\section{Competition between interaction, tunneling, and measurement}

In this section we will show, how the measurement competes with both tunneling and on-site interaction at the same time. In particular, we will show, how the measurement can either protect or break-up pairs of strongly interacting fermions.

\subsection{Bosons: Mott insulator -- superfluid phase transition under the quantum measurement backaction}

As we turn on the inter-atomic interactions, $U/J\neq 0$, the atomic dynamics changes.  One approach is to study the ground or steady state of the system in order to map a quantum or dissipative phase diagram. This is beyond the scope of this chapter, because here, we adopt a quantum optical approach with the focus on the conditional dynamics of a quantum trajectory corresponding to a single experimental realization.
The resulting evolution does not necessarily reach a steady state and can occur far from the ground state of the system. 
Again, each quantum trajectory evolves differently as the detection process is determined stochastically and even states with similar expectation values of $\hat{D}$ can have minimal overlap. However, even though each trajectory is different they all have one feature in common: The uncertainty in the measured operator, $\hat{D}$, is only a function of the Hamiltonian parameters, $\gamma$, $J$, and $U$. Therefore, we average its variance over many realizations ($\langle \sigma^2_D \rangle_{\mathrm{traj}}$) as this quantity effectively describes the squeezing of the atomic distribution due to measurement. Importantly, it is not possible to access this quantity using the master equation solution: The uncertainty in the final state is very large and it completely hides any information on the spread of a single trajectory. In other words, the master equation addresses the variance of the average value of $\hat{D}$ over the trajectories ensemble ($\langle \hat{D}^2 \rangle_{\mathrm{traj}}-\langle \hat{D} \rangle_{\mathrm{traj}}^2$) and not the squeezing of a single trajectory conditioned to the measurement outcome. This again highlights the fact that interesting physics happens only at the single trajectory level, but not in the average over many trajectories, which would correspond to the standard dissipation. In this section we show results for the measurement of  $\hat{D}=\hat{N}_{\mathrm{odd}}$, which is robust to photon losses. Specifically, we compute the width of the atomic distribution ($\langle \sigma^2_D \rangle_{\mathrm{traj}}$) in the limit $t\rightarrow \infty$, when its value does not change significantly in time even if the atomic imbalance is not constant. Moreover, we use the ground state of the system as initial state since this is a realistic starting point and a reference for explaining the measurement-induced dynamics.

\begin{figure}[h!]
\captionsetup{justification=justified}
\centering
\includegraphics[width=0.8\textwidth]{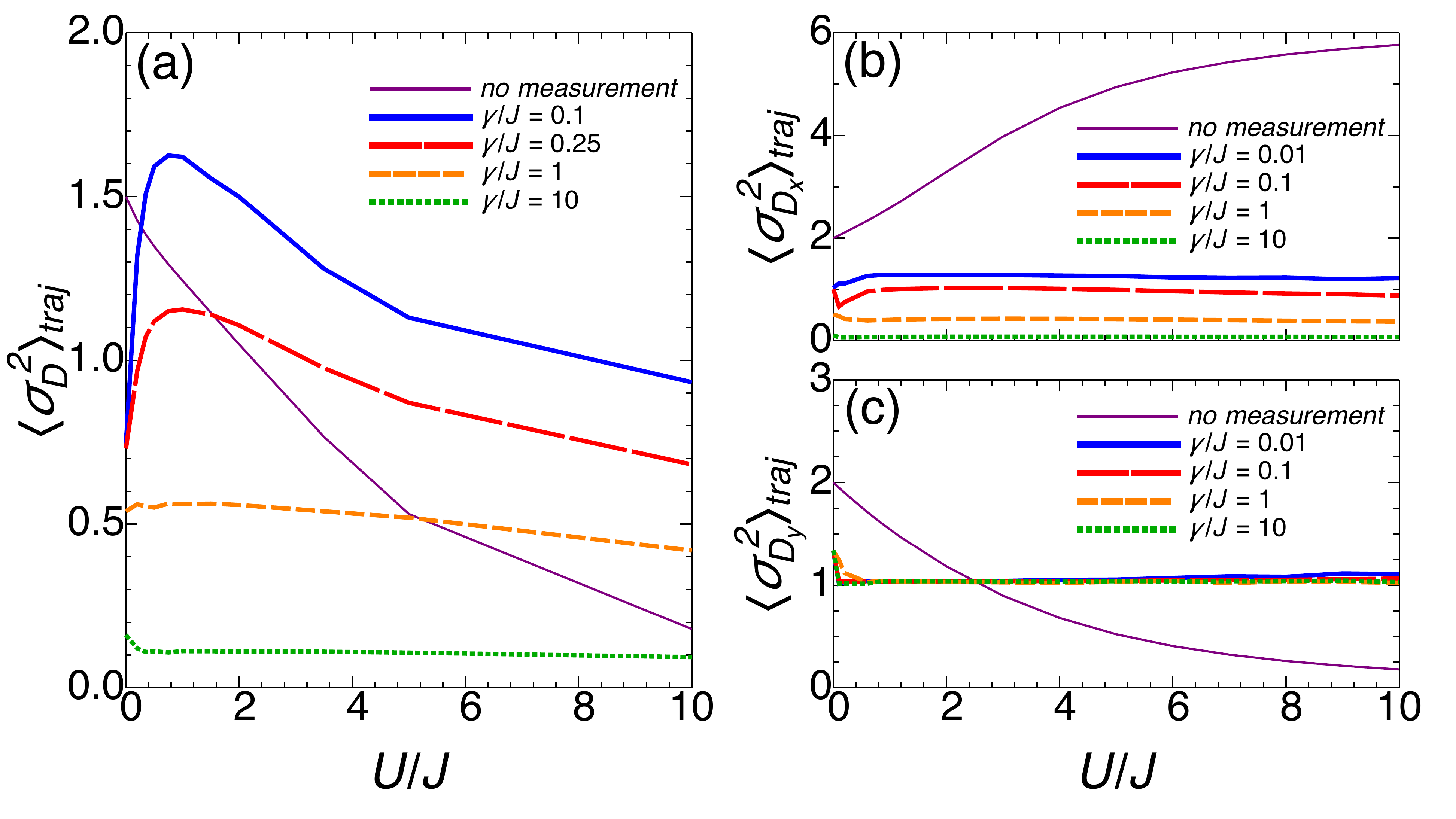}
\caption{\label{3-panelfluc} Atom number fluctuations demonstrating the competition of global measurement with local interaction and tunneling. Atom number variances at single trajectories are averaged over many trajectories (which is impossible to obtain from the master equation, see text).   (a) Bose--Hubbard model with repulsive interaction. The fluctuations of the atom number at odd sites $\hat{N}_\text{odd}$ in the ground state without a measurement (thin solid line) decrease as $U/J$ increases, reflecting the transition between the superfluid and Mott insulator phases. For the weak measurement, $\langle \sigma^2_D \rangle_{\mathrm{traj}}$ is squeezed below the ground state value, but then increases and reaches its maximum as the atom repulsion prevents oscillations and makes the squeezing less effective. In the strong interacting limit, the Mott insulator state is destroyed and the fluctuations are larger than in the ground state. (100 trajectories, $N=L=6$, $J_{jj} = 1$ if $j$ is odd and 0 otherwise.) (b), (c) Fermionic Hubbard model with attractive interaction; fluctuations of the total atom number at odd sites $\hat{D}_x=\hat{N}_{\uparrow \text{odd}}+ \hat{N}_{\downarrow \text{odd}}$ (b) and of the magnetization at odd sites $\hat{D}_y=\hat{M}_\mathrm{odd}=\hat{N}_{\uparrow \text{odd}}-\hat{N}_{\downarrow \text{odd}}$ (c). Without measurement, the interaction favors formation of doubly occupied sites so the density fluctuations in the ground state are increasing while the magnetization ones are decreasing. The measurement creates singly occupied sites decreasing the density fluctuations and increasing the magnetization ones, which manifests the break-up of fermion pairs by measurement. The measurement-based protection of fermion pairs is shown in the next figure. (100 trajectories, $L=8$, $N_{\uparrow}=N_{\downarrow}=4$, $J_{jj} = 1$ if $j$ is odd and 0 otherwise.) Simulations for 1D lattice.} 
\end{figure}

\begin{figure}[h!]
\captionsetup{justification=justified}
\centering
\includegraphics[width=0.5\textwidth]{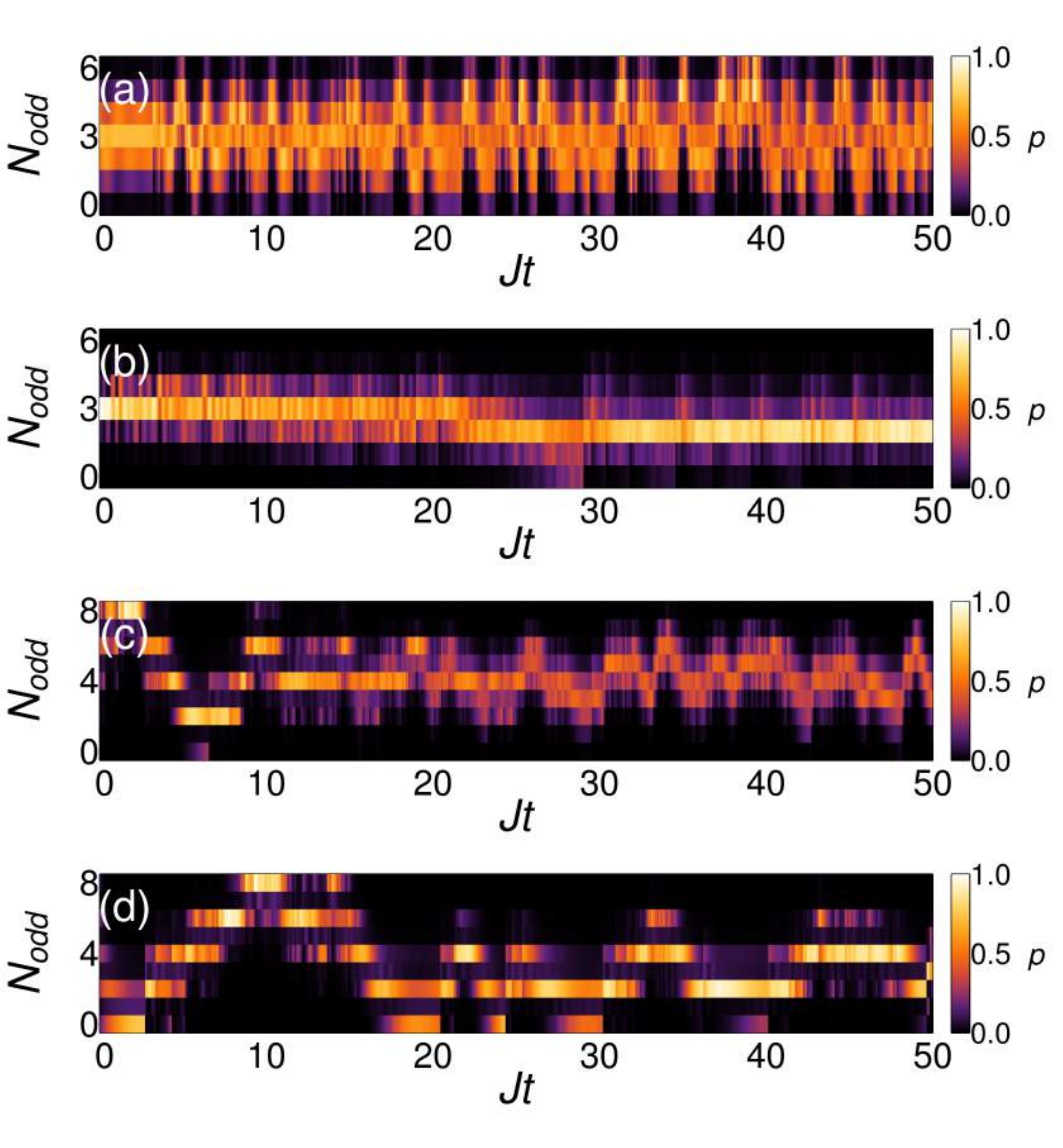}
\caption{\label{3-panel2} Conditional dynamics of the atom-number distributions at odd sites illustrating competition of the global measurement with local interaction and tunneling (single quantum trajectories, initial states are the ground states). (a) Weakly interacting bosons: the on-site repulsion prevents the formation of well-defined oscillations in the population of the mode. As states with different imbalance evolve with different frequencies, the squeezing due to the measurement is not as efficient as one observed in the non-interacting case ($N=L=6$, $U/J=1$, $\gamma/J=0.1$, $J_{jj} = 1$ if $j$ is odd and 0 otherwise). (b) Strongly interacting bosons: oscillations are completely suppressed and the number of atoms in the mode is rather well-defined, although less squeezed than in the Mott insulator. ($N=L=6$, $U/J=10$, $\gamma/J=0.1$, $J_{jj} = 1$ if $j$ is odd and 0 otherwise). (c) Attractive fermionic Hubbard model in the strong interaction limit. Measuring only the total population at odd sites $\hat{D}_x=\hat{N}_{\uparrow \text{odd}}+ \hat{N}_{\downarrow \text{odd}}$ quickly creates singly occupied sites, demonstrating measurement-induced break-up of fermion pairs ($L=8$, $N_{\uparrow}=N_{\downarrow}=4$, $U/J=10$, $\gamma/J=0.1$, $J_{jj} = 1$ if $j$ is odd and 0 otherwise). (d) The same as in (c), but with added measurement of the magnetization at odd sites $\hat{D}_y=\hat{M}_\mathrm{odd}=\hat{N}_{\uparrow \text{odd}}-\hat{N}_{\downarrow \text{odd}}$. This protects the doubly occupied sites, thus, demonstrating protection of fermion pairs by measurement. The distribution of $N_{\mathrm{odd}}$ vanishes for odd numbers, implying that the fermions tunnel only in pairs. Simulations for 1D lattice.}
\end{figure}

For bosons (Fig.~\ref{3-panelfluc}(a)), the number fluctuations $\sigma^2_D$ calculated in the ground state decrease monotonically for increasing $U$, reflecting the superfluid to Mott insulator quantum phase transition. The measured state on the other hand behaves very differently and $\langle \sigma^2_D \rangle_{\mathrm{traj}}$ varies non-monotonically. For weak interaction, the fluctuations are strongly squeezed below those of the ground state; then they quickly increase, reach their maximum and subsequently decrease as the interaction becomes stronger. We explain this effect looking at the dynamics of single trajectories (cf. Fig.~\ref{3-panel2}). For small values of $U/J$, the population imbalance between odd and even sites oscillates and its uncertainty is squeezed by the measurement as described in Sec. 3.3. However, when such oscillations reach maximal amplitude the local atomic repulsion spreads the atomic distribution and prevents the formation of states with large atom number in one of the two modes (Fig.~\ref{3-panel2}(a)). Since the interaction is a nonlinear term in the atomic dynamics, states with different imbalances oscillate  with different frequencies and the measurement is not able to squeeze the fluctuations of $\hat{N}_\text{odd}$ as efficiently as in the non-interacting case.
This behavior in the weak measurement and weak interaction limit is in contrast with what the effective two-sites model introduced in the previous section predicts (in a double well fluctuations are monotonically decreasing with $U$). However, the solution to this model is only valid in the $N \gg 1$ limit which suggests that this difference is due to the fact that in a double well with a large occupancy the population transfer is sequential, i.e.~atoms transfer one by one from one well to the other, increasing the total interaction energy, which goes up as $\langle\hat{n}_i^2\rangle,$ very steadily. On the other hand, in a lattice a collective excitation of a single atom from each site in one mode to another site in the other mode will increase the energy by $KU$, where $K$ is the number sites in one mode. This is an increase by a factor of $K$ compared to a single particle-hole excitation pair.  Therefore, a lattice and a global, long-range measurement scheme are necessary to observe such a collective transfer of atoms.

For weak measurement, but in the strongly interacting limit, we note that the measurement leads to a significant increase in fluctuations compared to the ground state. Both, the local interaction and measurement squeeze fluctuations, but as the measurement destroys the Mott insulator, the fluctuations are larger than in the ground state. Using first-order perturbation theory the ground state of the system is
\begin{align}
| \Psi_{J/U} \rangle = \left[ 1 + \frac{J}{U} \sum_{\langle i, j \rangle} b^\dagger_i b_j \right] | \Psi_{0} \rangle,
\end{align}
where $| \Psi_{0}\rangle$ is the Mott insulator state and the second term represents a uniform distribution of particle-hole excitation pairs across the lattice. The action of a single photocount will amplify the present excitations increasing the fluctuations in the system. In fact, consecutive detections lead to an exponential growth of these excitations as for $K\gg1$ and unit filling, and the atomic state after $m$ quantum jumps becomes $\hat{c}^m| \Psi_{J/U} \rangle \propto | \Psi_{J/U} \rangle  +  | \Phi_m \rangle$ where 
\begin{align}
\label{3-eq:exc}
 | \Phi_m \rangle = \frac{2^mJ}{KU} \sum_{i \,\text{odd}} \left ( b^\dagger_i b_{i-1}  - b^\dagger_{i-1} b_i - b^\dagger_{i+1} b_i + b^\dagger_i b_{i+1} \right)| \Psi_0 \rangle.
\end{align}
In the weak measurement regime the effect of the non-Hermitian decay is negligible compared to the local atomic dynamics combined with the quantum jumps and so there is minimal dissipation occurring. Therefore, because of the exponential growth of the excitations, even a small number of photons arriving in succession can destroy the Mott insulator state very quickly. This will always happen given sufficient time, and provided there are finite fluctuations present in the initial state and so this will happen at any value of $U/J$, except for $J=0$ when the ground state becomes a single Fock state with no particle-hole excitations.

In the strong measurement regime ($\gamma \gg J$) the measurement becomes more significant than the local dynamics and the system will freeze the state in the measurement operator eigenstates. In this case, the squeezing will always be better than in the ground state, because measurement and on-site interaction cooperate in suppressing fluctuations. For low interaction strengths this should be obvious from the fact that in a superfluid ground state the atoms are spread out over the entire lattice and thus the uncertainty in atom number is large whereas measurement eigenstates have a well-defined occupation number. However, the strongly interacting regime is much less evident, especially since we have demonstrated how sensitive the Mott insulating state is to the quantum jumps when the measurement is weak.

To understand the strongly interacting case we will again use the first-order perturbation theory and consider a postselected $\langle \hat{D}^\dagger \hat{D} \rangle = 0$ trajectory. This corresponds to a state that scatters no photons and so the non-Hermitian correction to the Hamiltonian is sufficient to understand the measurement. Since squeezing depends on the measurement strength and is common to all possible trajectories we can gain some insight by considering this specific case. However, we will now consider $\hat{D} = \Delta \hat{N} = \hat{N}_\text{odd} - \hat{N}_\text{even}$ as this measurement also has only two modes, $R = 2$, but its $\langle \hat{D}^\dagger \hat{D} \rangle = 0$ trajectory corresponds to the Mott insulating ground state. According to perturbation theory the modified ground state is now
\begin{align}
\label{3-perturbation}
| \Psi_{(J,U,\gamma)} \rangle = \left[ 1 + \frac{J}{U - i4\gamma} \sum_{\langle i, j \rangle} b^\dagger_i b_j \right] | \Psi_{0} \rangle.
\end{align}
The variance of the measurement operator for this state is given by
\begin{align}
\sigma^2_{\Delta N} = \frac{8 J^2 L}{U^2 + 16 \gamma^2} \nu(\nu+1),
\end{align}
where $\nu$ is the filling factor. From the form of the denominator we immediately see that both interaction and measurement squeeze with the same quadratic dependence and that the squeezing is always better than in the ground state (corresponding to $\gamma = 0$) regardless of the value of $U/J$. Also, depending on the ratio of $\gamma/U$ the squeezing can be dominated by measurement ($\gamma/U \gg 1$) or by interaction ($\gamma/U \ll 1$) or both processes can contribute equally ($\gamma/U \approx 1$). The $\hat{D} = \hat{N}_\text{odd}$ measurement should have a similar dependence on $\gamma$ and $U$ and be proportional to $(U^2 + \gamma^2)^{-1}$ since the $\gamma$ coefficient in the perturbative expansion depends on the value of $(J_{i,i} - J_{i-1,i-1})^2$. We can see the system transitioning into the strong measurement regime in Fig. \ref{3-panelfluc}(a) as the $U$-dependence flattens out with increasing measurement strength. In typical many-body systems in the absence of measurement, strong correlations are a product of large interactions and the non-interacting limit reduces essentially to a single particle theory. In contrast, here, the measurement is another mechanism generating entanglement and strong correlations. This is more evident in weak interacting limit, where the measurement-induced interaction takes over the standard local interaction (as presented in the case of giant oscillations). We will also see in the next section how strong measurement leads to long-range correlated tunneling events.

\subsection{Fermions: protection and break-up of fermion pairs by measurement}

For fermions (Figs.~\ref{3-panelfluc}(b,c)), the ground state of the attractive Hubbard model in the strong interacting regime contains mainly doubly occupied sites (pairs) and empty sites. Therefore, in the  absence of measurement, the fluctuations in the atom population $\hat{D}_x=\hat{N}_{\uparrow \text{odd}}+ \hat{N}_{\downarrow \text{odd}}$ ($\sigma^2_{D_{x}}$) increase with $U/J$ while the ones in the magnetization 
$\hat{D}_y=\hat{M}_\mathrm{odd}=\hat{N}_{\uparrow \text{odd}}-\hat{N}_{\downarrow \text{odd}}$  ($\sigma^2_{D_{y}}$) decrease because singly occupied sites become more improbable. The measurement induces two different kinds of dynamics using the same mode functions since, depending on the light polarization, we can address either the total population, or both total population and magnetization. 

In the first case (Figs.~\ref{3-panelfluc}(b,c)), the weak measurement quickly squeezes $\sigma^2_{D_{x}}$, but destroys the pairs as it does not distinguish between singly or doubly occupied sites. Figure~\ref{3-panel2}(c) illustrates such a measurement-induced break-up: the initial state contains mainly even values of $\hat{N}_\text{odd}$, corresponding to a superposition of empty and doubly occupied lattice sites. As time progresses and photons are scattered from the atoms, the variance $\sigma^2_{D_{x}}$ is squeezed and unpaired fermions are free to tunnel across the lattice allowing odd values for  $\hat{N}_\text{odd}$. 

In contrast, probing both density and magnetization reduces both their fluctuations and increase the lifetime of doubly occupied sites. The resulting measurement-induced dynamics is illustrated in Figure~\ref{3-panel2}(d): Atoms tunnel only in pairs with opposite spin as the probability distribution of $\hat{N}_\text{odd}$ contains only even values, hence demonstrating the measurement-based protection of fermion pairs. In both cases, the dynamics of the system is not a result of the projective quantum Zeno effect (here the measurement is weak) but is a manifestation of the squeezing of the atomic population and magnetization of the two macroscopically occupied modes.


\section{Emergent long-range correlated tunneling}

In the previous sections we showed that even very weak measurement can lead to very strong dynamical phenomena. The measurement weakness was important to assure that it competes with tunneling at the same time and energy scales. In this section we will switch to strong measurements, but not strong enough to fully freeze the atomic evolution (as in the Zeno effect) or totally confine it to the quantum Zeno subspace (as in the Zeno dynamics). Thus, here we consider new phenomena, which arise beyond the standard quantum Zeno paradigm. We will show the appearance of long-range tunneling and long-range correlations and entanglement, which would be extinct by the strong measurement in the Zeno regime. We will demonstrate the many-body bath engineering by quantum measurements.

When $\gamma \gg J$, the photodetections freeze the modes' atom number, and decorrelate the populations of different modes. In the quantum Zeno limit of projective measurement ($\gamma \rightarrow \infty$) tunneling through the boundaries between modes would be fully suppressed. However, by considering a finite $\gamma/J$ we observe additional dynamics while the usual atomic tunneling is still strongly Zeno-suppressed. In this regime, we observe the following effects. First, the evolution between nearest neighbors within each mode is basically unperturbed by the measurement process and it is determined by the usual tunneling. Importantly, by engineering the light mode functions, it is possible to forbid part of this first-order dynamics and select which processes participate in the quantum Zeno dynamics. Therefore, one can design the Zeno subspace of the Hilbert space where the system evolves. Second, tunneling between different spatial modes is possible only via higher-order, long-range correlated tunneling events that preserve the eigenvalue of $\hat{D}$. In other words, atoms tunnel across distant sites of the lattice via a virtual state and they behave as delocalized correlated pairs. 

We can, once more, gain insight into this process from the non-Hermitian Hamiltonian (cf. details and extensions in Sec. 3.8). By looking at a second order expansion \cite{Auerbach,Kozlowski2016PRAnH} of the Hamiltonian confined to a Zeno subspace of a two-mode measurement, $R=2$, for $U/J = 0$ we obtain the following effective Hamiltonian
\begin{align}
\label{3-eq:hz}
\hat{H}_Z = \hat{P}_0 \left[ -J \sum_{\langle i, j \rangle} b^\dagger_i b_j - i \frac{J^2} {A \gamma} \sum_{\varphi} 
\sum_{\substack{\langle i \in \varphi, j \in \varphi^\prime \rangle \\ \langle k \in \varphi^\prime, l \in \varphi \rangle}} b^\dagger_i b_j b^\dagger_k b_l \right] \hat{P}_0,
\end{align}
where $\hat{P}_0$ is the projector into the Zeno subspace, $A = (J_\varphi - J_{\varphi^\prime})^2$ is a constant that depends on the measurement scheme, $\varphi$ denotes a set of sites belonging to a single mode and $\varphi^\prime$ is the set's complement (e.g. odd and even sites).

We can infer a lot of information from this simple expression. First, we see that first order tunneling will only survive between neighboring sites that belong to the same mode, because $\hat{P}_0 b^\dagger_i b_j \hat{P}_0 = 0$ otherwise. Second, the second-order term exists between different modes and occurs at a rate $\sim J^2/\gamma$. The imaginary prefactor means that this tunneling behaves like an exponential decay (overdamped oscillations). This picture is consistent with the projective limit ($\gamma \rightarrow \infty$), where the atomic state is constrained to an eigenspace of the measurement operator.

Crucially, what sets this effect apart from usual many-body dynamics with short-range interactions is that first order processes are selectively suppressed by the global conservation of the measured observable and not by the prohibitive energy costs of doubly-occupied sites, as is the case in the $t$-$J$ model \cite{Auerbach}. This has profound consequences as this is the physical origin of the long-range correlated tunneling events represented in \eqref{3-eq:hz} by the fact that sites $j$ and $k$ can be very distant. This is because the projection $\hat{P}_0$ is not sensitive to individual site occupancies, but instead enforces a fixed value of the observable, i.e.~a single Zeno subspace.

\begin{figure}[h!]
\captionsetup{justification=justified}
\centering
\includegraphics[width=0.99\textwidth]{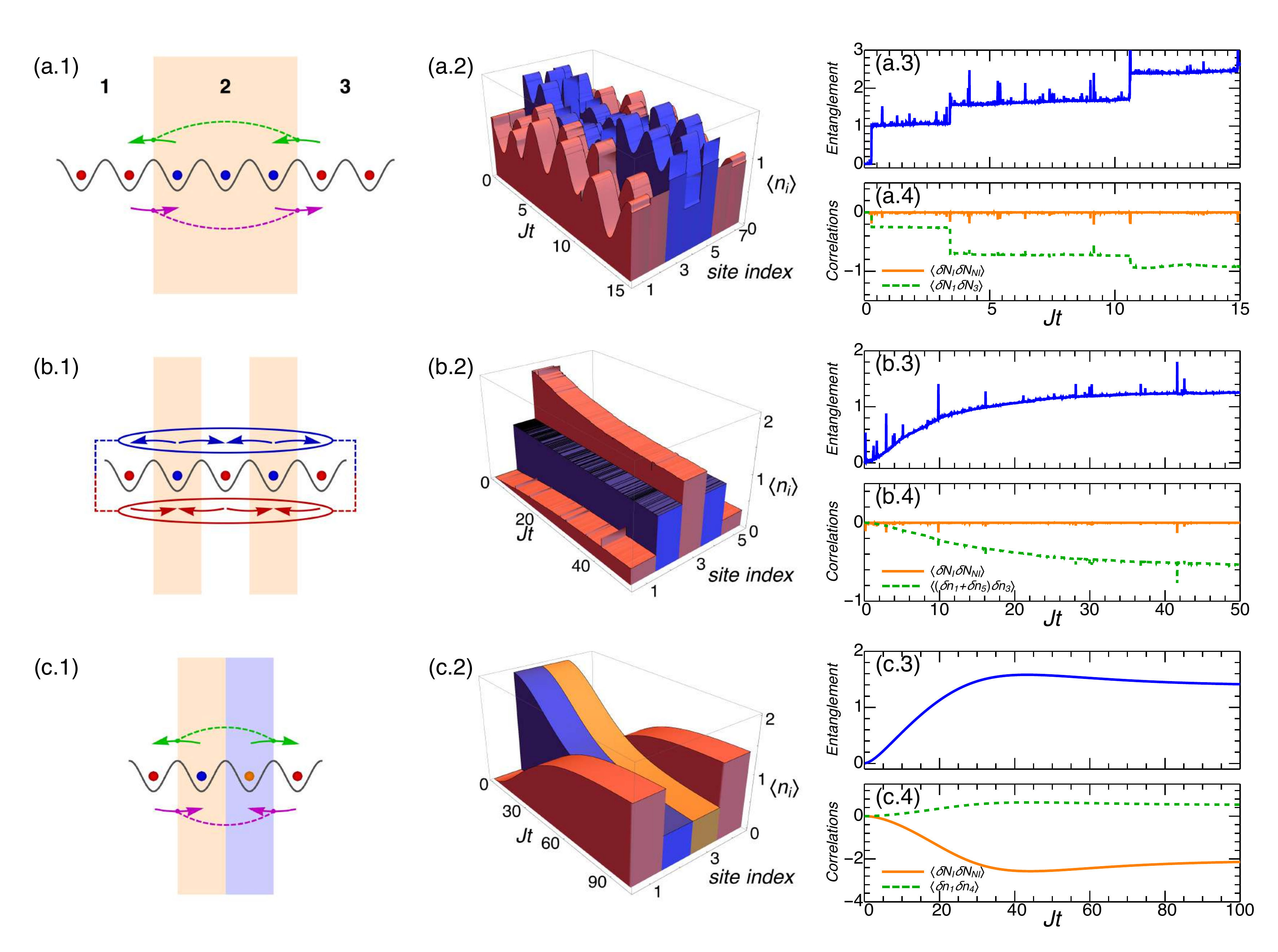}
\caption{\label{3-panelfreezing} Long-range correlated tunneling and entanglement, induced by strong global measurement in a single trajectory. Panels (a), (b), and (c) show different measurement geometries, implying different constraints (superselection rules). Panels (1): Schematic representation of long-range tunneling, when standard tunneling between different zones is Zeno-suppressed. Panels (2): Evolution of on-site densities; atoms effectively tunnel between disconnected regions. Panels (3): Entanglement entropy growth between illuminated and non-illuminated regions. Panels (4): Correlations between different modes (solid orange line) and within the same mode (dashed green line); atom number $N_I$ ($N_{NI}$) in illuminated (non-illuminated) mode. Please, see the detailed description of this large figure directly in the text.}
\end{figure}

We demonstrate several examples of the long-range correlated tunneling in Fig. \ref{3-panelfreezing}. Before explaining the physical processes, we give the detailed description of this large figure here in the text. Panels (a), (b), and (c) show different measurement geometries, implying different constraints (superselection rules). Panels (1): Schematic representation of long-range tunneling, when standard tunneling between different zones is Zeno-suppressed. Panels (2): Evolution of on-site densities; atoms effectively tunnel between disconnected regions. Panels (3): Entanglement entropy growth between illuminated and non-illuminated regions. Panels (4): Correlations between different modes (solid orange line) and within the same mode (dashed green line); atom number $N_I$ ($N_{NI}$) in illuminated (non-illuminated) mode. 
(a) Atom number in the central region is frozen: The system is divided into three regions and correlated tunneling occurs between non-illuminated zones (a.1). Standard dynamics happens within each region, but not between them (a.2). Entanglement build up (a.3). Negative correlations between non-illuminated regions (dashed green line) and zero correlations between the $N_I$ and $N_{NI}$ modes (solid orange line) (a.4). Initial state: $|1,1,1,1,1,1,1 \rangle$, $\gamma/J=100$, $J_{jj}=[0,0,1,1,1,0,0]$.
(b) Even sites are illuminated, freezing $N_\text{even}$ and $N_\text{odd}$. Long-range tunneling is represented by any pair of one blue and one red arrow (b.1). Correlated tunneling occurs between non-neighboring sites without changing mode populations (b.2). Entanglement build up (b.3). Negative correlations  between edge sites (dashed green line) and zero correlations between the modes defined by $N_\text{even}$ and $N_\text{odd}$ (solid orange line) (b.4). Initial state: $|0,1,2,1,0 \rangle$,  $\gamma/J=100$, $J_{jj}=[0,1,0,1,0]$.
(c) Atom number difference between two central sites is frozen. Correlated tunneling leads to exchange of long-range atom pairs between illuminated and non-illuminated regions (c.1,2). Entanglement build up (c.3). In contrast to previous examples, sites in the same zones (illuminated/ non-illuminated) are positively correlated (dashed green line), while atoms in different zones are negatively correlated (solid orange line) (c.4). Initial state: $|0,2,2,0 \rangle$, $\gamma/J=100$, $J_{jj}=[0,-1,1,0]$. 1D lattice, $U/J=0$.

We now turn to the explanation of physical processes. Illuminating only the central region of the optical lattice and detecting light in the diffraction maximum, we freeze the atom number $\hat{N}_\text{illum}$~\cite{MekhovPRL2009,MekhovPRA2009} (Fig.~\ref{3-panelfreezing}(a)). The measurement scheme defines two different spatial modes: the non-illuminated zones $1$ and $3$ and the illuminated one~$2$. Figure~\ref{3-panelfreezing}(a.2) illustrates the evolution of the mean density at each lattice site: typical dynamics occurs within each region but the standard tunneling between different zones is suppressed. Importantly, processes that do not change $N_\text{illum}$ are still possible since an atom from $1$ can tunnel to $2$, if simultaneously one atom tunnels from $2$ to $3$. Thus, effective long-range tunneling between two spatially disconnected zones $1$ and $3$ happens due to two-step processes $1\rightarrow2\rightarrow3$ or $3\rightarrow2\rightarrow1$. These transitions are responsible for the negative (anti-)correlations $\langle \delta N_1 \delta N_3\rangle = \langle N_1 N_3\rangle-\langle N_1 \rangle\langle N_3\rangle$ showing that an atom disappearing in 1 appears in 3, while there are no number correlations between illuminated and non-illuminated regions,  $\langle( \delta N_1 +\delta N_3)\delta N_2\rangle =0$ (Fig.~\ref{3-panelfreezing}(a.4)). In contrast to fully-projective measurement, the intermediate (virtual) step in the correlated tunneling process builds long-range entanglement between illuminated and non-illuminated regions (Fig.~\ref{3-panelfreezing}(a.3)). This resembles two-photon (Raman) processes known in optics (cf. Sec. 3.8 for details).

To make correlated tunneling visible even in the mean atom number, we suppress the standard Bose--Hubbard dynamics by illuminating only the even sites of the lattice~(Fig.~\ref{3-panelfreezing}(b)). Even if this measurement scheme freezes both $N_\text{even}$ and $N_\text{odd}$, atoms can slowly tunnel between the odd sites of the lattice, despite them being spatially disconnected. This atom exchange spreads correlations between non-neighboring lattice sites on a time scale $\sim \gamma/J^2$. The schematic explanation of long-range correlated tunneling is presented in Fig.~\ref{3-panelfreezing}(b.1): the atoms can tunnel only in pairs to assure the globally conserved values of $N_\text{even}$ and $N_\text{odd}$, such that one correlated tunneling event is represented by a pair of one red and one blue arrow. Importantly, this scheme is fully applicable for a lattice with large atom and site numbers, well beyond the numerical example in Fig.~\ref{3-panelfreezing}(b.1), because  as we can see in \eqref{3-eq:hz} it is the geometry of quantum measurement that assures this mode structure (in this example, two modes at odd and even sites) and therefore underlying pairwise global tunneling. Such long-range correlations and long-range entanglement (even in a 1D system) can develop essentially because of the coherent global addressing. In contrast, the local uncorrelated probing of individual sites would decreases the probability of individual tunneling events being correlated. 

This global pair tunneling may play a role of a building block for more complicated many-body effects. For example, a pair tunneling between neighbouring sites has been recently shown to play important role in the formation of new quantum phases, e.g., pair superfluid \cite{LewensteinPSFNJP} and lead to formulation of extended Bose--Hubbard models \cite{LewensteinExtBHM}. The search for novel mechanisms providing long-range interactions is crucial in many-body physics. One of the standard candidates is the dipole-dipole interaction in, e.g., dipolar molecules, where the mentioned pair tunneling between even neighboring sites is already considered to be long-range \cite{LewensteinPSFNJP,LewensteinExtBHM}. In this context, our work suggests a fundamentally different mechanism originating from quantum optics: the backaction of global and spatially structured measurement, which as we prove can successfully compete with other short-range processes in many-body systems. This opens promising opportunities for future research.

The scheme in Fig.~\ref{3-panelfreezing}(b.1) can help to design a nonlocal reservoir for the tunneling (or ``decay'') of atoms from one region to another. For example, if the atoms are placed only at odd sites,  according to \eqref{3-eq:hz} their tunnelling is suppressed since the multi-tunneling event must be successive, i.e.~an atom tunnelling into a different mode, $\varphi^\prime$, must then also tunnel back into its original mode, $\varphi$. If, however, one adds some atoms to even sites (even if they are far from the initial atoms), the correlated tunneling events become allowed and their rate can be tuned by the number of added atoms. This resembles the repulsively bound pairs created by local interactions \cite{Winkler2006,BlochNature2007}. In contrast, here the atom pairs are long-range correlated due to the global measurement. Additionally, these long-range correlations are a consequence of the dynamics being constrained to a Zeno subspace: the virtual processes allowed by the measurement entangle the spatial modes nonlocally. Since the measurement only reveals the total number of atoms in the illuminated sites, but not their exact distribution, these multi-tunelling events cause the build-up of long range entanglement. This is in striking contrast to the entanglement caused by local processes which can be very confined, especially in 1D where it is typically short range. This makes numerical calculations of our system for large atom numbers really difficult, since well-known methods such as Density Matrix Renormalization Group and Matrix Product States \cite{Schollwock} (which are successful for short-range interactions) rely on the limited extent of entanglement.

These types of configurations open intriguing opportunities for quantum engineering of system -- bath interactions: here both the system and reservoir, represented by different modes, are many-body systems with internal long-range entanglement.

The negative number correlations are typical for systems with constraints (superselection rules) such as fixed atom number. The effective dynamics due to our global, but spatially structured, measurement introduces more general constraints to the evolution of the system. For example, in Fig.~\ref{3-panelfreezing}(c) we show the generation of positive number correlations (shown in Fig.~\ref{3-panelfreezing}(c.4)) by freezing the atom number difference between the sites ($N_\text{odd}-N_\text{even}$, by measuring at the diffraction minimum). Thus, atoms can only enter or leave this region in pairs, which again is possible due to correlated tunneling (Figs.~\ref{3-panelfreezing}(c1-2)) and manifests positive correlations. Since this corresponds to a no photon trajectory, we can solve exactly for a small system and for two atoms, these correlations grow as $\langle \delta \hat{n}_1 \delta \hat{n}_4 \rangle \approx [1 - \text{sech}^2(4J^2t/\gamma)]/4$. As in the previous example, two edge modes in Fig.~\ref{3-panelfreezing}(c) can be considered as a nonlocal reservoir for two central sites, where a constraint (superselection rule) is applied. Note that, using more modes, the design of higher-order multi-tunneling events is possible.


\section{Details of the weak-measurement-induced macroscopic dynamics of atomic modes}

In this section we provide more analytical results about the large-scale oscillations induced by weak quantum measurements presented in Sec. 3.3. We show that, in contrast to the fully projective limit of strong quantum measurement, where the evolution is locked to a small subspace (quantum Zeno dynamics), or even frozen completely (quantum Zeno effect), the weak non-projective measurement can effectively compete with standard unitary dynamics leading to nontrivial effects. Even if the measurement strength remains constant, the quantum measurement backaction acts on the atomic ensemble quasi-periodically and induces collective oscillatory dynamics of all the atoms. We introduce an effective model for the evolution of the spatial modes and present an analytical solution showing that the quantum jumps drive the system away from its stable point. We confirm our finding describing the atomic observables in terms of stochastic differential equations. In addition, we comment on the relation between the quantum measurements and dissipation in this system.


\subsection{Effective dynamics of the macroscopic spatial modes}
We start by considering the evolution of a quantum gas with $N$ atoms initially in the superfluid state
\begin{eqnarray}\label{3-eq: superfluid}
\ketz{\Phi(N)}=\frac{1}{\sqrt{M^NN!}}\left( \sum_{i=1}^N b_i^\dagger \right)^N \! \! \ketz{0},
\end{eqnarray}
where $\ketz{0}$ is the vacuum state for the operators $b_i$. We continuously monitor this system using  traveling waves so that the measurement scheme defines $R$ macroscopically occupied spatial modes and the jump operator is  $\h{c} \propto a=C \sum_{j=1}^R e^{i 2 \pi j /R} \h{N}_j$, where $\h{N}_j$ is the occupation of the mode $j$. Making use of the multinomial expansion for the sum in equation (\ref{3-eq: superfluid}) and assuming that each mode has the same number of lattice sites, we can group the creation operators that operates on the same mode so that $\ketz{\Phi(N)}$ can be rewritten as 
\begin{eqnarray}
\ketz{\Phi(N)}= \sqrt{\frac{N!}{R^N}}  \sum_{\sum_i N_i=N} \sqrt{\frac{1}{N_1! N_2! \dots N_R!}}  \prod_{i=1}^R \ketz{\Phi_i(N_i)},
\end{eqnarray}
where $\ketz{\Phi_i(N_i)}$ is a superfluid in the spatial mode $i$ with $N_i$ atoms. In other words, we decompose a superfluid state in a linear combination of ``smaller'' superfluids that are defined in each spatial mode. This choice is particularly convenient because the states   $\prod_{i=1}^R \ketz{\Phi_i(N_i)}$ are eigenvectors of the jump operator
\begin{eqnarray}
\h{c} \prod_{i=1}^R \ketz{\Phi_i(N_i)}= \sqrt{2 \kappa} C\left(\sum_{j=1}^R e^{i 2 \pi j /R} N_j \right) \prod_{i=1}^R \ketz{\Phi_i(N_i)}.
\end{eqnarray}
Therefore, defining $\mathcal{S}_R$  to be the subspace of the Hilbert space that is spanned by the vectors $\left\{ \prod_{i=1}^R \ketz{\Phi_i(N_i)} \right\}_{N_i}$, the dynamics due to the quantum jumps is internal to  $\mathcal{S}_R$. This property enable us to formulate an effective description of the atomic evolution, greatly reducing the computational cost of each quantum trajectory and allowing us to formulate an analytically solvable model. 

Carefully engineering the coefficients $J_{jj}$, we can partition the optical lattice in two spatial modes depending on the parity of the lattice sites. This can be achieved using traveling waves as mode functions for the probe and the cavity (i.~e. $u_l(\b{r})=e^{i \b{k}_l \cdot \b{r}}$) where the wave vectors $\b{k}_0$ and $\b{k}_1$ are orthogonal, corresponding to the detection of the photons scattered in the diffraction minimum and the operator $a=C(\h{N}_\even- \h{N}_\odd)$. Alternatively, one can obtain the same spatial mode structure considering standing waves (i.~e. $u_l(\b{r})=\cos(\b{k}_l \cdot \b{r})$) crossed in such a way that $\b{k}_0$ and $\b{k}_1$ have the same projections on the lattice direction and are shifted in such a way that the even sites of the optical lattice are positioned at the nodes, so that the scattered light operator is $a=C\h{N}_\odd$. 
In this case, we can decompose the initial state of the system (superfluid) as 
\begin{eqnarray}\label{3-eq: superfluid2}
\ketz{\Phi(N)}= \sum_{j=1}^N \sqrt{\frac{N!}{2^N j!(N-j)!}} \ketz{\Phi_\odd(j),\Phi_\even(N-j)}.
\end{eqnarray}
If the interaction between the atoms can be neglected ($U=0$), the dynamics resulting from the competition between the measurement process and the usual nearest-neighbors tunneling preserves the mode structure and it is possible to describe the system in term of collective variables. To clarify this, we rewrite the tunneling term as 
\begin{eqnarray}\label{3-eq: partition}
  \sum_{\langle i,j\rangle}b_j^\dagger b_i =  \sum_{i \in \odd} \sum_{j:i} \cop{b}_{j}\aop{b}_i+ \sum_{i \in \even} \sum_{j:i} \cop{b}_{j}\aop{b}_i ,
\end{eqnarray}
where $j:i$ indicates that the sites $j$ and $i$ are nearest neighbors. The  dynamics generated by this expression and the non-Hermitian term is internal to the space $\mathcal{S}_2$ if the initial state of the system belongs to $\mathcal{S}_2$. In fact,  by applying $\h{H}_\mathrm{eff}$ (\ref{3-Heff}) to the product of two superfluid states one has
\begin{gather}
  \h{H}_\mathrm{eff}  \ketz{\Phi_{\odd}(l),\Phi_{\even}(m)}= 
  - \hbar J \sqrt{l(m+1)}\ketz{\Phi_{\odd}(l-1),\Phi_{\even}(m+1)}  \nonumber \\
  - \hbar J \sqrt{m(l+1)}\ketz{\Phi_{\odd}(l+1),\Phi_{\even}(m-1)}
   -i \hbar \frac{\gamma}{2} |\beta_1 l +\beta_2 m|^2 \ketz{\Phi_{\odd}(l),\Phi_{\even}(m)},\label{3-eq:Heff2}
\end{gather}
where $\beta_1$ and $\beta_2$ depend on the measurement scheme ($\beta_1=-\beta_2=1$ for probing in the diffraction minimum while $\beta_1=1$ and $\beta_2=0$ if only the odd sites are addressed). 
Therefore, the quantum state of the atoms can be expressed as $\ketz{\psi}= \sum_{j=0}^N \alpha_j \ketz{\Phi_\odd(j),\Phi_\even(N-j)}$. This allows us  to reformulate the conditional dynamics of the system in term of an effective double-well problem where the occupation of the two wells corresponds to the population of the spatial modes. However, this approach does not allow us to compute any spatial correlations for the system considered: only ``collective'' properties can be calculated.

The generalization of (\ref{3-eq:Heff2}) to the case of $R$ modes is straightforward: from the spatial structure of the jump operator one can compute the amplitude of the tunneling processes between different spatial modes and build an effective $R-$well problem. Moreover, the coupling between the effective wells can be tuned changing the spatial structure of the measurement operator, allowing us to go beyond simple double-well systems \cite{Milburn1998,Ruostekoski2001}. For example, considering the case of $R=3$ spatial modes generated by a measurement scheme where the light mode functions are traveling waves and the jump operator is $\h{c}\propto \sum_{j=1}^3 e^{i 2 \pi j/3} \h{N}_j$, one has that the modes alternates across the lattice as $RGBRGB\dots$. Therefore, each lattice site belonging to the mode $R$ is connected to one site of mode $G$ and one of mode $B$ (indexes can be cycled) so that tunneling processes are allowed between all the modes. This reduces the system to a three-well problem where a specific linear combination of the atomic populations $\h{N}_j$ is monitored while the atoms are free to hop between the three different wells.  Importantly, this is not the only possible way to divide the optical lattice in three modes: if probe and scattered light are standing waves such that $\b{k}_{1,0} \cdot  \b{r} = \pi/4$, the resulting $J_{jj}$ coefficients are $[0,1/2,1,1/2,0,\dots]$ and the modes alternates as $RGBGRGBGR\dots$. In this case, tunneling is not allowed between mode $R$ and $B$ and the coupling between the effective three-wells describing the conditional dynamics must take this into account by forbidding atomic transfer between two of the effective wells.

\begin{figure}[h]
\centering
\captionsetup{justification=justified}
\includegraphics[width=0.8\textwidth]{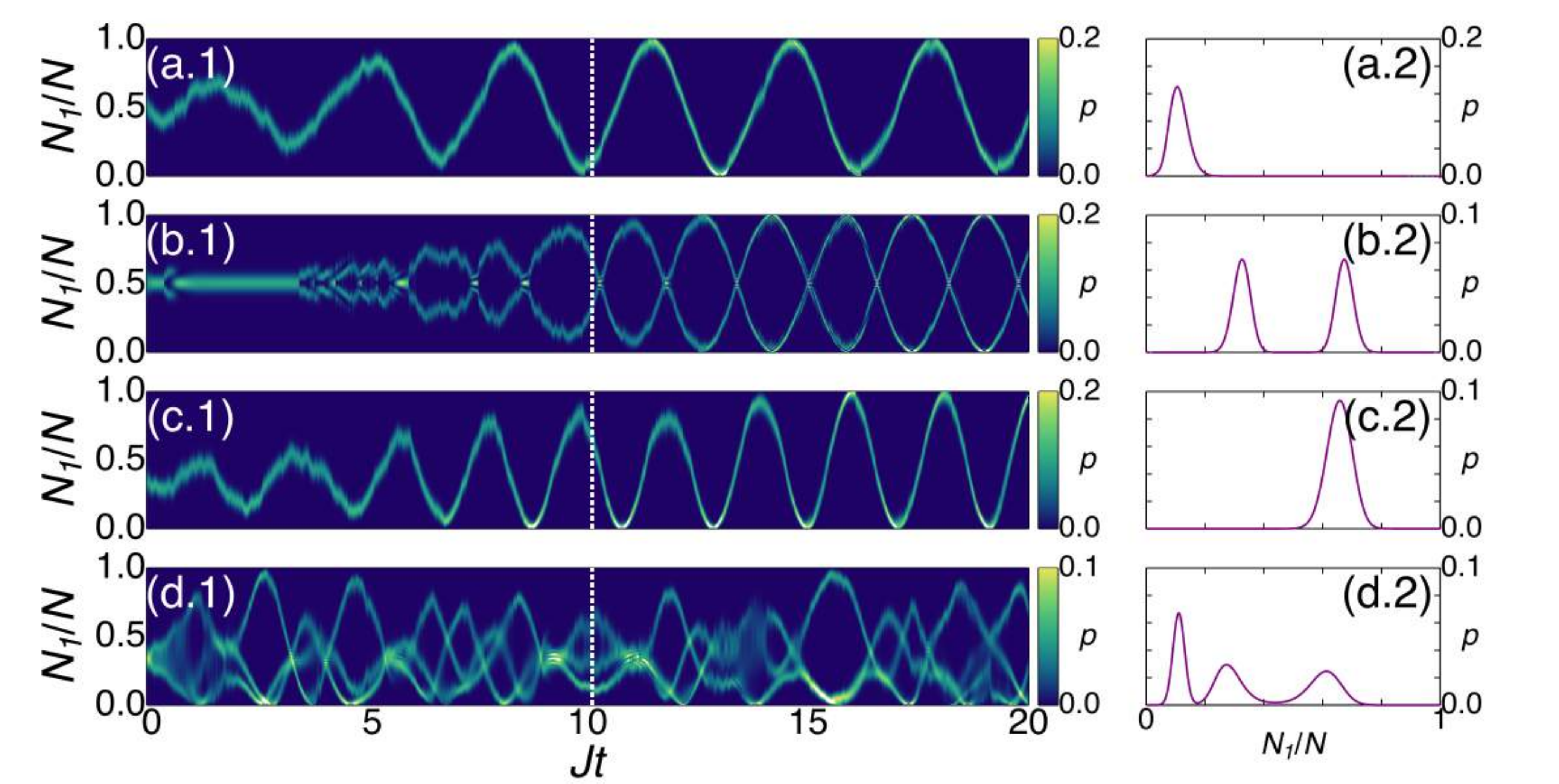}
\caption{ Oscillatory dynamics induced by the measurement process. Probability distribution (the curves have changing widths) for the occupation of one of the modes if the optical lattice is partitioned in two (a)-(b) or three spatial modes (c)-(d). Panels (1) illustrate the the full probability distribution for all times while panels (2) focuses on the time indicated by the dashed vertical line.  Depending on the spatial structure of the jump operator, the detection process can lead to multimode macroscopic superpositions. (a) $\gamma/J=0.02, \, N=100, \, J_{jj}=[1,0,1,0,1\dots]$, (b) $\gamma/J=0.02, \, N=100, \, J_{jj}=(-1)^j$, (c) $\gamma/J=0.02, \, N=99, \, J_{jj}=[0,1/2,1,1/2,0\dots]$, (d) $\gamma/J=0.02, \, N=99, \, J_{jj}=e^{i 2 \pi j/3}$ }\label{3-fig:osc}
\end{figure}

The measurement scheme we consider addresses global atomic observables and therefore preserves the long-range coherence. Moreover, the spatial structure of the jump operator determines the conditional dynamics: eventual degeneracy of the light intensity $\m{\h{a}_1^\dagger \h{a}_1}$ are imprinted on the state of the atoms. For example, if one detects the photons scattered in the diffraction minimum, so that the measurement addresses the difference in the population of the two spatial modes defined by the odd and even lattice sites, i. e. $\h{a}_1=C (\h{N}_\even -\h{N}_\odd) $, the photon number operator is not sensitive to the sign of such difference. This is because states with opposite $\m{\h{N}_\even -\h{N}_\odd}$ scatter light with different phase but with the same intensity. As a consequence of the measurement backaction, the atomic state becomes a superposition of two macroscopically occupied components: a Schr\"odinger cat state. If the monitoring scheme defines more than two degenerate modes, the conditional evolution leads to an atomic wavefunction that can be expressed as a superposition of multiple macroscopically occupied components. Importantly, this property is a consequence of the spatial structure of the jump operator and does not only depend on the number of modes defined by the measurement. For example, considering the case of three modes arranged as $RGBRGB\dots$ ($J_{jj=}e^{i 2 \pi j/3}$), the photon number $\m{\h{a}_1^\dagger \h{a}_1}$ is invariant under the exchange of any two light modes and, as a consequence, the probability distribution of the population of each mode presents three oscillating peaks, indicating that the atomic state conditioned to the measurement is a multimode Schr\"odinger cat state. However, if the spatial modes have a different structure such as $RGBGRGBGR\dots$ ($J_{jj}=[0,1/2,1,1/2,0,\dots]$) this is not the case: only the modes $B$ and $G$ can be exchanged without affecting the intensity of the detected light. Therefore, in this case the probability distribution of the occupation of the mode $R$ has a single peak while the ones for modes $G$ and $B$ are bimodal. This is illustrated in Figure~\ref{3-fig:osc} where, depending on the measurement scheme, the photodetection induces different dynamics.

The local interaction term tends to localize the atoms on each lattice site and cannot be included exactly in the approximation we presented. This is because the dynamics described by the interaction operator $ \sum_i \h{n}_i ( \h{n}_i-1)$ is not internal to the subspace $\mathcal{S}_R$. For weak interactions and assuming that the measurement scheme partitions the optical lattice in $R$ spatial modes, we find that the interaction energy $U$ is rescaled by the number of lattice sites belonging to each mode ($M_j$) so that
\begin{eqnarray}
\frac{\hbar U}{2}  \sum_{i=1}^M \h{n}_j ( \h{n}_j-1) \approx \frac{\hbar}{2}  \sum_{j=1}^R \frac{U}{M_j} \h{N}_j ( \h{N}_j-1).
\end{eqnarray}
Note that this expression is approximate and does not allow to describe the strong interacting limit where the atoms form a Mott insulator state. Despite of this limitation, the model we presented allows to describe interacting systems where synthetic interactions mediated by the light field couple different spatial modes \cite{CaballeroPRA2016}.

\subsection{Probing the odd sites of the optical lattice}\label{3-sec:mf}

If the measurement operator probes the population of the odd sites of the optical lattice, i.~e. $a=C\h{N}_\odd$, we can formulate an analytical description of the atomic dynamics~\cite{Julia-Diaz2012}. In between the quantum jumps, the evolution of the system is deterministic and it is determined by the matrix elements of the non-Hermitian Hamiltonian~(\ref{3-eq:Heff2}):
\begin{gather}
\bok{\Phi_\odd(s),\Phi_\even(N-s)}{\h{H}_\mathrm{eff}}{\psi}=- J \hbar \left( \alpha_{s-1} b_{s-1} + \alpha_{s+1} b_{s}\right)  \nonumber\\ 
+\frac{2 \hbar U}{M} \left[ s( s-1) + (N-s)(N-s-1) \right] \alpha_s- i \hbar\frac{ \gamma}{2} s^2 \alpha_s,\label{3-eq:mat_el}
\end{gather}
where $b_s= \sqrt{(s+1)(N-s)}$. Note that this expression does not contain a chemical potential since we will solve the Schr\"odinger equation for a fixed number of particles, as emphasized by the left hand side of equation~(\ref{3-eq:mat_el}). In the limit $N\gg1$ we can replace the index $s$ with the continuous variable $x=s/N$ which represents the relative occupation of the odd sites of the optical lattice. Moreover, we define the  wavefunction $\psi(x=s/N)=\sqrt{N}\alpha_s$ where the $\sqrt{N}$ prefactor ensures that $\psi(x)$  is normalized, i.~e.
\begin{eqnarray}
\sum_{s=0}^N |\alpha_s|^2=\int_{0}^{1} \left| \psi(x)\right|^2 \d  x= 1.
\end{eqnarray}
Introducing $b(x)=\sqrt{(x+h)(1-x)}$, $h=1/N$, $\Lambda=U N /M$, and $\Gamma = N \gamma/2$, and neglecting constant shifts we can rewrite (\ref{3-eq:mat_el}) as 
\begin{eqnarray}
\bok{x}{\h{H}_\mathrm{eff}}{\psi}=&- \sqrt{N} J \hbar \left[ b(x-h) \psi(x-h)+\psi(x+h)b(x) \right] \nonumber \\
& +2 \sqrt{N}  \hbar \Lambda  x (x-1) \psi(x) - i \sqrt{N} \hbar \Gamma x^2 \psi(x).
\end{eqnarray}
Expanding this expression up to the second order in $h$ and defining the normalized atom imbalance $z=(N_\odd-N_\even)/N=2x-1$, we obtain an effective non-Hermitian Hamiltonian that describes the dynamics of the two macroscopically occupied spatial modes. Specifically, we describe the evolution of the system between two quantum jumps with the effective Schr\"odinger equation 
\begin{eqnarray}\label{3-Schroedinger}
i h \desude{t} \psi(z,t)=H(z) \psi(z,t),
\end{eqnarray}
where the Hamiltonian $H(z)$ is 
\begin{gather}
H(z)\psi(z)= -2 J h^2 \desude{z}\left(\sqrt{1-z^2} \desude{z} \psi(z) \right) + \frac{1}{2} \Lambda z^2
 + V(z) \psi(z)   - i \Gamma \frac{(z+1)^2}{4} \psi(z)\label{3-hamiltonianX}
\end{gather}
and the effective potential $V(z)$ is given by
\begin{eqnarray}
V(z)=-J \sqrt{1-z^2} \psi(z) \left[ 1+ \frac{h}{1-z^2}-\frac{h^2(1+z^2)}{(1-z^2)^2}\right].
\end{eqnarray}
The dynamics of the spatial modes is therefore equivalent to the motion of a particle with the effective mass $\sqrt{1-z^2}$ in the real potential $V(z)$ and imaginary potential  $- i \Gamma (z+1)^2 /4$. Using the same approximations, we find that the initial state (\ref{3-eq: superfluid2}) in the limit $N\gg1$ reduces to the Gaussian function
\begin{eqnarray}\label{3-eq:initial}
\psi(z,0)=\left( \frac{1}{\pi b_0^2} \right)^{1/4} e^{-z^2/(2 b_0^2)},
\end{eqnarray}
describing a perfectly balanced population between the two spatial modes (i. e. $\m{\h{N}_\odd-\h{N}_\even}=0$) with the variance $b_0^2=2h$. In order to give an analytical expression of $\psi(z,t)$, we take the limit of small population unbalance so the mass term becomes $\sqrt{1-z^2}\approx 1$ and we expand the potential $V(z)$ up to the second order in $z$:
\begin{eqnarray}
V(z)\approx-1-h + \frac{1}{8}\omega^2 z^2, \qquad \omega=2 \sqrt{1+\Lambda-h}. 
\end{eqnarray}
Therefore, the dynamics of the atomic system is mapped to the evolution of a Gaussian wave packet in an harmonic potential and subjected to dissipation via the non-Hermitian term due to the measurement. Within these assumptions, the wavefunction of the system remains Gaussian at all times and it can be expressed as 
\begin{eqnarray}\label{3-eq:ansatz}
\psi(z,t)=\left( \frac{1}{\pi b^2(t)} \right)^{1/4} \exp \left[i a(t)+\frac{i z c(t)+ i z^2 \phi(t)-(z-z_0(t))^2}{2 b^2(t)} \right].
\end{eqnarray}
The functions $b^2(t)$, $z_0(t)$, $c(t)$, $\phi(t)$, and $a(t)$ describe the collective dynamics of the system. Specifically, $b^2(t)$ is proportional to the width of the atomic distribution, $z_0(t)$ is the mean value of the unbalance (i. e. $\m{\h{N}_\odd-\h{N}_\even}$), while $c(t)$ and $\phi(t)$ are phase differences between superfluid states with different populations. Moreover, $\operatorname{Re}\left[a(t) \right]$ describes the global phase of the wavefunction and $\operatorname{Im}\left[a(t) \right]$ its norm. Importantly, all these functions are real with the exception of $a(t)$ which is complex. Finally, from the Schr\"odinger equation~(\ref{3-Schroedinger}) we obtain the differential equations that dictate the evolution of  $b^2(t)$, $z_0(t)$, $c(t)$, $\phi(t)$, and $a(t)$. Specifically, one can prove that
\begin{eqnarray}
	&\dot{(b^2)}=8 h J \phi -  \frac{\Gamma}{2 h} b^4 \label{3-eq:b1}, \\
	&\dot{\phi}=-\frac{J \omega^2}{4h} b^2 - \frac{\Gamma}{ 2 h} b^2 \phi +  \frac{4 h J}{b^2}(1 + \phi^2)\label{3-eq:phi1}, \\
	&\dot{z_0}=- \frac{\Gamma}{2h} b^2 (1 + z_0) + \frac{2 h J}{b^2} (2 z_0 \phi + c)\label{3-eq:z01}, \\
	&\dot{c}=- \frac{\Gamma}{2h} b^2 c +  \frac{4hJ}{b^2}(\phi c - 2 z_0)\label{3-eq:c1}.
\end{eqnarray}
Because of the dissipation, the norm of the wavefunction is not conserved and it is decreasing according to $\exp \left(-2 \operatorname{Im}\left[a(t) \right] \right)$ where 
\begin{eqnarray}
    &\operatorname{Im}\left(\dot{a}\right)=\frac{\Gamma}{4 h} \left[ \left( 1+z_0 \right)^2 + \frac{b^2}{2} \right],
\end{eqnarray}
which determines when a photon escapes the cavity and a quantum jump occurs.

Equations (\ref{3-eq:b1})--(\ref{3-eq:c1}) can be solved analytically introducing the auxiliary variables  $p=(1- i \phi)/b^2$ and $q=(z_0+i c/2 )/b^2$. Substituting in (\ref{3-eq:b1})--(\ref{3-eq:c1}), we find that the four equations describing the dynamics of the atomic state reduce to two differential equations:
\begin{eqnarray}
 -2 J h^2 p^2 + \left( \frac{J  \omega^2}{8} - \frac{i \Gamma}{4}\right) + \frac{i h}{2} \frac{\d  p }{\d  t} = 0 \label{3-eq:p}, \\
 4 J h^2 p q - \frac{i \Gamma}{2} - i h\frac{\d  q}{\d  t} =0. \label{3-eq:q}
\end{eqnarray}
Defining, $\zeta^2=1-i 2 \Gamma / (J\omega^2)$ and making use of standard integrals, one can prove that the solution of the first equation is 
\begin{eqnarray}
p(t)=\frac{\zeta \omega }{4 h} \frac{\left( \zeta \omega + 4 h p(0) \right) e^{i 2 \zeta \omega t} -  \left( \zeta \omega - 4 h p(0) \right)}{\left( \zeta \omega + 4 h p(0) \right) e^{i 2 \zeta \omega t} + \left( \zeta \omega - 4 h p(0) \right)}. \label{3-eq:p_sol}
\end{eqnarray}
Furthermore, the equation that determines the evolution of $q(t)$ can be solved noting that (\ref{3-eq:q}) can be rewritten as 
\begin{eqnarray}
\frac{\d }{\d  t} \left( I q \right) = - \frac{\Gamma}{2 h} I,
\end{eqnarray}
where the integrating factor $I$ is given by
\begin{gather}
I=\exp \left[ i 4 h \int p(t) \d  t \right] \\
=\left( \zeta \omega + 4 h p(0) \right) e^{i  \zeta \omega t} + \left( \zeta \omega - 4 h p(0) \right) e^{-i  \zeta \omega t},
\end{gather}
so that $q(t)$ is given by
\begin{gather}
q(t)= \frac{1}{2 h \zeta \omega} \frac{A}{ \left( \zeta \omega + 4 h p(0) \right) e ^{i \zeta \omega t} +  \left( \zeta \omega - 4 h p(0) \right) e ^{-i \zeta \omega t}}\label{3-eq:q_sol}, \\
A=  i \Gamma \left[ \left( \zeta \omega + 4 h p(0) \right) e ^{i \zeta \omega t} -  \left( \zeta \omega - 4 h p(0) \right) e ^{-i \zeta \omega t}\right] 
 +4 h \zeta^2 \omega^2 q(0) -  i 8 h \Gamma p(0). 
\end{gather}
Finally, from these solutions we can extract the physical observables of equations (\ref{3-eq:b1})--(\ref{3-eq:c1}) as $ b^2(t)=1/\operatorname{Re}\left[ p(t) \right]$,  $ \phi(t)=-\operatorname{Im}\left[ p(t) \right] /\operatorname{Re}\left[ p(t) \right]$, $z_0(t)=\operatorname{Re}\left[ q(t) \right] /\operatorname{Re}\left[ p(t) \right]$, and 
$c(t)=2 \operatorname{Im}\left[ q(t) \right] /\operatorname{Re}\left[ p(t) \right] $.

Instead of focusing on the full solution, here we give a qualitative description of the dynamics generated by these equations. Specifically, we compute the eigenvalues of the Jacobian matrix of the system (\ref{3-eq:b1})--(\ref{3-eq:c1}) in its stationary points.  Studying the stationary point of the dynamical equations  (\ref{3-eq:b1})--(\ref{3-eq:c1}) or   (\ref{3-eq:p})--(\ref{3-eq:q}), we find that there is only one physical critical point. Defining the parameter 
\begin{eqnarray}
\alpha=\sqrt{- \frac{1}{2} +\frac{1}{2} \sqrt{\frac{4 \Gamma ^2}{J^2 \omega ^4}+1}},
\end{eqnarray}
one can prove that 
\begin{eqnarray}
&b^2(\infty)= \frac{4 h J \omega \alpha}{\Gamma} \label{3-eq:critb},\\
&\phi(\infty)=\frac{J \omega^2 \alpha^2}{\Gamma}\label{3-eq:critphi},\\
&z_0(\infty)= -1 + \frac{1}{2 \alpha^2+1}\label{3-eq:critz0},\\
&c(\infty)= \frac{4 \Gamma}{J\omega^2 (2 \alpha^2+1)}.\label{3-eq:critc}
\end{eqnarray}
Note that these expressions can be obtained also by taking the limit $t\rightarrow \infty$ of the exact solutions (\ref{3-eq:p_sol}) and (\ref{3-eq:q_sol}). The eigenvalues of the Jacobian matrix computed in the critical point are
\begin{eqnarray}
\lambda_{1,2}=\pm \frac{i \Gamma}{\omega \alpha}-J  \omega \alpha   \qquad \mbox{and} \qquad \lambda_{3,4}=\pm \frac{2 i \Gamma}{\omega \alpha}-2 J \omega \alpha.
\end{eqnarray}
Since all of them have a non-positive real part, the point $(b^2,\phi,z_0,c)$ defined by (\ref{3-eq:critb})--(\ref{3-eq:critc}) is stable. Therefore, the evolution of the system in the long time limit will tend to damped oscillations around the stationary point with frequency $\Omega=2\Gamma/(\omega \alpha)$ and decay time $\Delta t_d=1/(2 J \omega \alpha)$.  The ratio between the measurement strength  $\Gamma$ and the tunneling amplitude $J$ determines whether the oscillatory behavior is under- or over-damped. The predictions of this analytical model agree quantitatively with the dynamics described by (\ref{3-eq:mat_el}) only if the population imbalance between the two spatial modes is small. Despite of this, we find that this simple formulation captures the qualitative behavior of $b^2$ and $z_0$ and help explaining the emergence of the collective oscillations between odd and even sites.

\begin{figure}[h]
\captionsetup{justification=justified}
  \centering
  \includegraphics[width=0.7\linewidth]{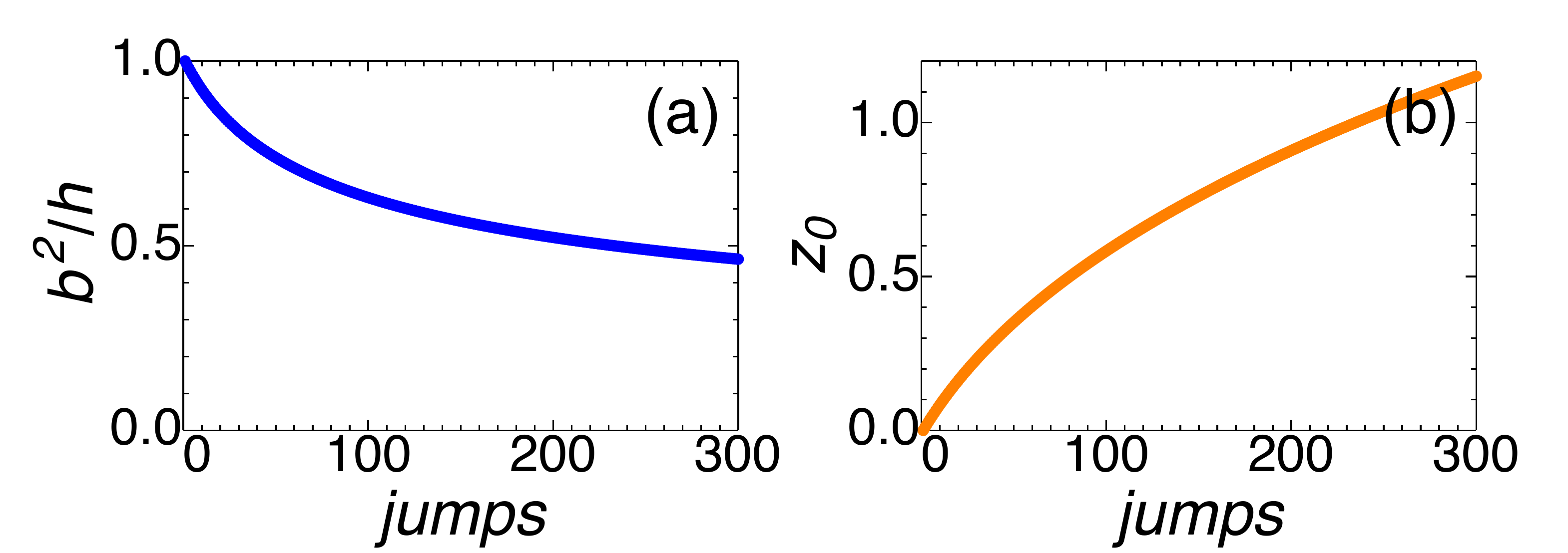}
  \caption{Effect of the quantum jumps on the atomic observables neglecting the effective non-Hermitian dynamics. The uncertainty associated with the number of atoms in each spatial mode decreases (a) while the system prefers configurations with larger imbalance between odd and even sites (b).}
  \label{3-fig:jumps}
\end{figure}

The quantum jumps substantially contribute to the evolution of the atomic state and drastically alter the dynamics described by (\ref{3-eq:b1})--(\ref{3-eq:c1}). Their effect can be included in the model by expanding the jump operator $a=C(z+1)/2$ around the peak of the Gaussian wavefunction (\ref{3-eq:ansatz}) as
\begin{eqnarray}
\frac{(1+z)}{2} \approx  \frac{1}{2} \exp\left[ \ln(1+z_0) + \frac{z-z_0}{(1+z_0)} - \frac{(z-z_0)^2}{2(1+z_0)^2}\right].
\end{eqnarray}
Using this expression, we compute the effect of the jumps on the functions $b^2(t)$, $z_0(t)$, $c(t)$, and $\phi(t)$, and we obtain a set of equations that determines the change in initial condition for (\ref{3-eq:b1})--(\ref{3-eq:c1}) due to the detection of one photon:
\begin{eqnarray}
	&b^2 \rightarrow \frac{b^2 (1+z_0)^2}{(1+z_0)^2+b^2} \label{3-eq:bjump1}, \\
	&\phi \rightarrow \frac{\phi(1+z_0)^2}{(1+z_0)^2+b^2}\label{3-eq:phijump1}, \\
	&z_0 \rightarrow z_0 + \frac{b^2(1+z_0)}{(1+z_0)^2+b^2}\label{3-eq:z0jump1}, \\
	&c  \rightarrow  \frac{c(1+z_0)^2}{(1+z_0)^2+b^2}\label{3-eq:cjump1}.
\end{eqnarray}
Neglecting the non-Hermitian dynamics these equations imply that each quantum jump tends to squeeze the width of the atomic distribution while it increases the atom imbalance between odd and even sites (see~Figure~\ref{3-fig:jumps}). As a consequence, the measurement process decreases the uncertainty in the population of the spatial modes and the atomic state becomes a product of two superfluid states with well-defined atom number.  In the next paragraphs, we will discuss how the quantum jumps compete with the effective non-unitary dynamics, leading to the creation of states  where the atomic population collectively oscillates between odd and even sites.

\subsubsection{Case $J=0$}

We first start by considering the case, when the atomic tunneling is much slower than the measurement, and $J$ can be neglected. Within this assumption, the evolution of the system between two quantum jumps is solely determined by the non-Hermitian dynamics which, together with the quantum jumps, decreases the variance of the population imbalance between odd and even sites. Therefore, the final state of the system is a product of two superfluids with a well-defined number of atoms for each quantum trajectory and the behavior of $b(t)$ is almost deterministic \cite{Onofrio}. However, the imbalance between odd and even sites is not the same for each quantum trajectory since it is determined by the specific sequence of quantum jumps. In fact, the photodetections or the non-Hermitian decay dominates the dynamics of $z_0$ in opposite regimes: the first effect is predominant if the occupation in the odd sites is large while the second one is favored by a large occupation of the even sites.  In each quantum trajectory, these two phenomena balance each other and, in the long time limit, $z_0$ reaches a stationary value which follows the Gaussian probability distribution defined by the initial state (\ref{3-eq:initial}), favoring states with small population difference between odd and even sites.

\subsubsection{Case $J\neq0$}

\begin{figure}[h]
\captionsetup{justification=justified}
  \centering
  \includegraphics[width=0.7\linewidth]{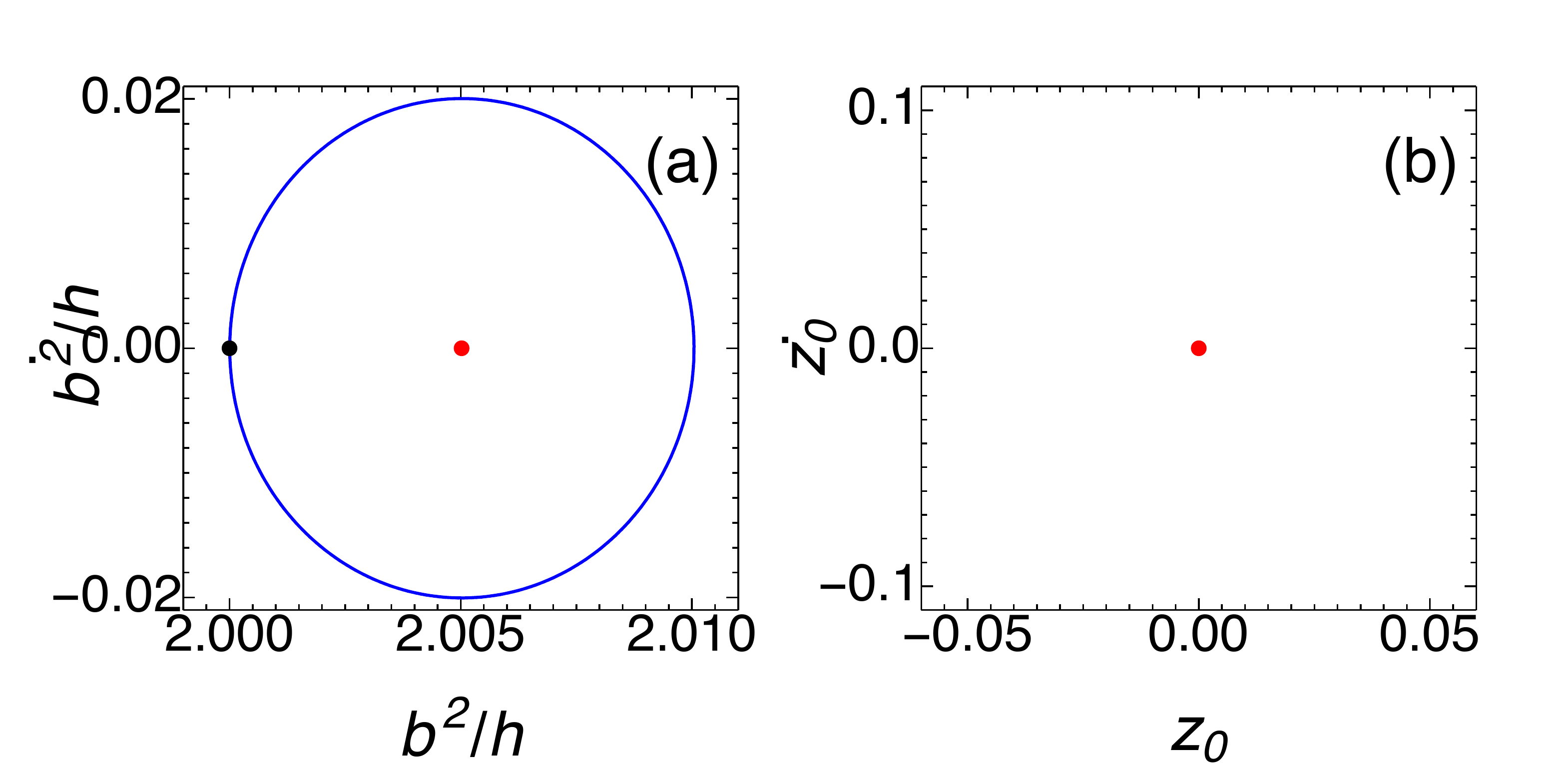}
  \caption{Close orbits around the stable point for the width of the atomic distribution $(b^2,\dot{b^2})$ (a) and the population imbalance $(z_0,\dot{z_0})$ (b) in a single trajectory setting $\Gamma=0$ and without jumps. The black point marks the initial state $(b^2=2 h, \, z_0=0)$ while the red one marks the stationary point. Panel (b) does not show any dynamics since the initial state and the stationary point coincide.}
  \label{3-fig:hermitianNoJumps}
\end{figure}

If the tunneling amplitude $J$ cannot be neglected, the detection process competes with the usual atomic dynamics. 
We first consider the weak measurement limit $\Gamma \ll J$ so that we can describe the evolution of the system between two quantum jumps setting $\Gamma \approx 0$ in (\ref{3-eq:b1})--(\ref{3-eq:c1}).  From the stability analysis we find that the stable point of the system is $b(\infty)\approx 4h/\omega$ and $z_0(\infty)\approx 0$ while the eigenvalues of the Jacobian matrix are purely imaginary, i.~e. $\lambda_{1,2}=~\pm i J \omega$ and $\lambda_{3,4}=\pm 2 i J \omega$. Therefore, in the absence of  quantum jumps, the solutions of (\ref{3-eq:b1})--(\ref{3-eq:c1}) are oscillating around the stable point without damping (see~Figure~\ref{3-fig:hermitianNoJumps}). The photodetections perturb this regular oscillations and drive the system quasi-periodically, leading to giant oscillations in the population of the spatial modes. Specifically, the quantum jumps tend to increase the value of $z_0$ according to (\ref{3-eq:bjump1})--(\ref{3-eq:cjump1}) and consequently, the radius of the oscillations in the $(z_0,\dot{z_0})$ plane is increasing if a jump happens when $z_0>0$ while it is decreasing when $z_0<0$.  Importantly, these two processes do not happen with the same rate because the probability for the emission of a photon in the time interval $\delta t$ depends on the atomic state and it is given by
\begin{eqnarray}\label{3-prob}
p_{\mathrm{jump}}=\frac{\Gamma}{2 h} \left[ \left( 1+z_0 \right)^2 + \frac{b^2}{2} \right] \delta t.
\end{eqnarray}

\begin{figure}[h]
\captionsetup{justification=justified}
  \centering
  \includegraphics[width=0.7\linewidth]{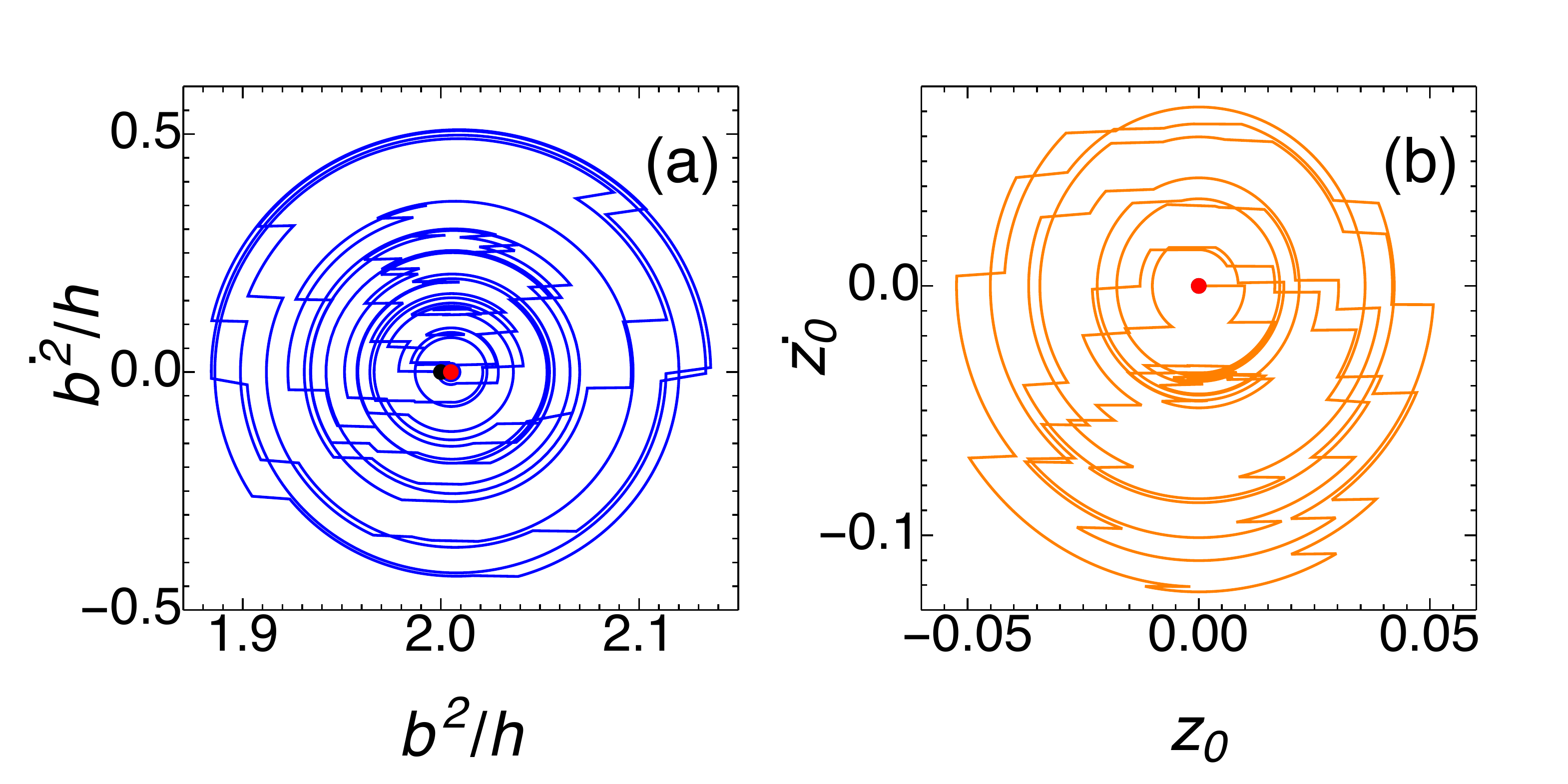}
  \caption{Oscillations of $(b^2,\dot{b^2})$ (a) and $(z_0,\dot{z_0})$ (b) in a single quantum trajectory setting $\Gamma=0$ and applying jumps according to the exact diagonalization solution (\ref{3-eq:Heff2}). The black point represents the initial state $(b^2=2 h, \, z_0=0)$ while the red one marks the stationary point. The solutions rotate around the stable point with increasing amplitude. Note that the jumps are from right to left for $b^2$ while they are from left to right for $z_0$.}
  \label{3-fig:hermitianWithJumps}
\end{figure}

\begin{figure}[h]
\captionsetup{justification=justified}
  \centering
  \includegraphics[width=.7\linewidth]{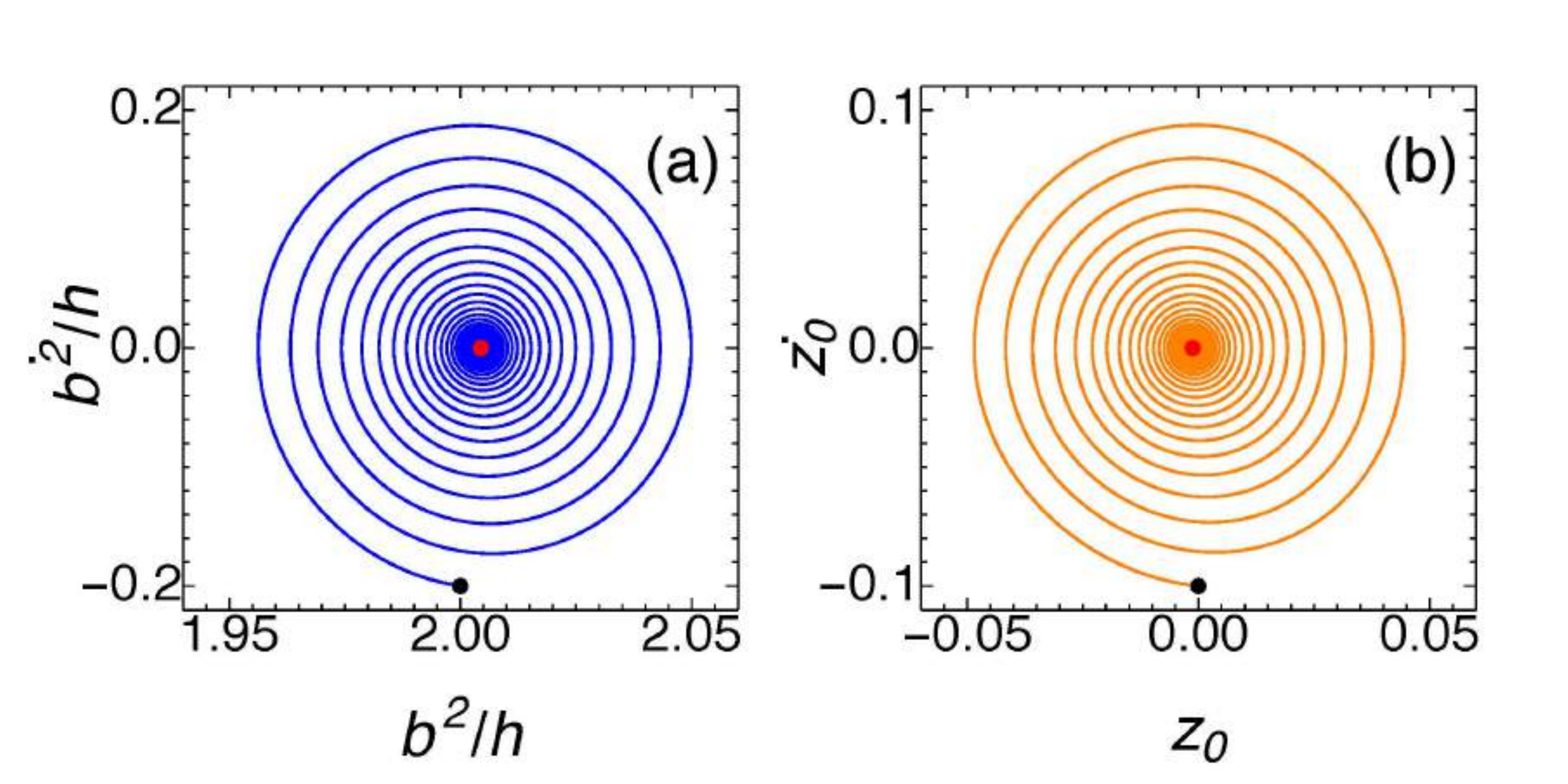}
  \caption{Under-damped oscillations of $(b^2,\dot{b^2})$ (a) and $(z_0,\dot{z_0})$ (b) in a single quantum trajectory in the weak measurement regime ($\Gamma=0.001J$) without quantum jumps. The black point represents the initial state $(b^2=2 h, \, z_0=0)$ while the red one marks the stationary point.}
  \label{3-fig:weakNoJumps}
\end{figure}

Therefore, jumps that increase the radius of the oscillations happen more often and increase the amplitude of the oscillations of $z_0(t)$ (see~Figure~\ref{3-fig:hermitianWithJumps}).
In order to confirm this prediction, we now turn to the full measurement problem. Taking into account the non-Hermitian dynamics in the differential equations for $b(t)$ and $z(t)$, the radius of the orbits shown in Figure~\ref{3-fig:hermitianNoJumps} decreases exponentially. Therefore, we can identify three different time scales in the evolution of the system: (i) the oscillation frequency $\Omega=2\Gamma/(\omega \alpha)$, (ii) the damping time $\Delta t_d=1/(2 J \omega \alpha)$ and (iii) the average time interval between two quantum jumps $\Delta t_j=2h/\Gamma$. The ratios between these quantities determine which process is dominating the physics of the system. 
Considering again the weak measurement regime ($\Gamma \ll J$) but taking into account the terms depending on $\Gamma$ in equations (\ref{3-eq:b1})--(\ref{3-eq:c1}), we find that both $b(t)$ and $z_0(t)$ oscillate around the stationary point with decreasing radius. In this limit, one has $\Omega \Delta t_d \approx J \omega^2/\Gamma \gg 1$, indicating that the system behaves like an under-damped oscillator (Figure~\ref{3-fig:weakNoJumps}). Importantly, there  are many photocounts during each oscillation ($\Omega \Delta t_j \approx \Gamma h /(J \omega^3) \ll 1$) and the quantum jumps can counteract the damping, driving the atomic system towards states with high population imbalance. In order to prove this, we describe the \emph{average} effect of a quantum jump on the width of the atomic distribution and the relative imbalance as 
\begin{eqnarray}
\delta b^2= \Delta b^2 \, p_{\mathrm{jump}} \qquad \mbox{and} \qquad \delta z_0= \Delta z_0 \, p_{\mathrm{jump}},
\end{eqnarray}
where $\Delta b^2$ and  $\Delta z_0$ are the effect of a single jump on $b^2$ and $z_0$ computed using (\ref{3-eq:bjump1}) and (\ref{3-eq:z0jump1}). From these expressions we find that the average photocurrent affects $b^2$ and $z_0$ as
\begin{eqnarray}
&\frac{\delta b^2}{\delta t}= - \frac{\Gamma}{2 h} b^4(t) \left[1- \frac{1}{2}\frac{b^2(t)}{(z_0(t)+1)^2+b^2(t)}\right] \label{3-eq:mean1}, \\
&\frac{\delta z_0}{\delta t}= \frac{\Gamma}{2 h} b^2(t) (z_0(t)+1) \left[1- \frac{1}{2}\frac{b^2(t)}{(z_0(t)+1)^2+b^2(t)}\right] \label{3-eq:mean2}.
\end{eqnarray}

\begin{figure}[h]
\captionsetup{justification=justified}
  \centering
  \includegraphics[width=.7\linewidth]{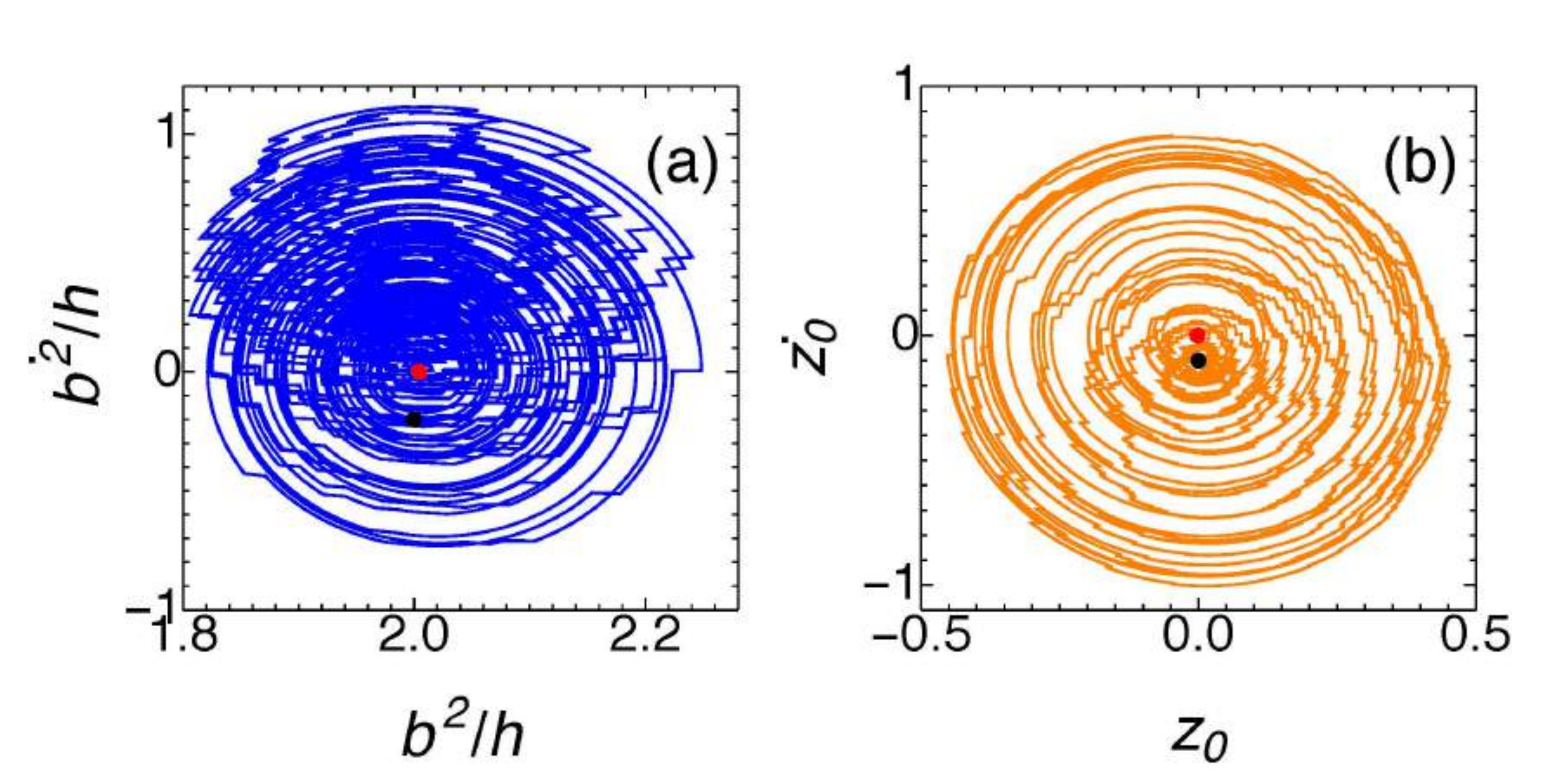}
  \caption{Full conditional dynamics of $(b^2,\dot{b^2})$ (a) and $(z_0,\dot{z_0})$ (b) of a single trajectory in the weak measurement regime ($\Gamma=0.001J$). The black point represents the starting point $(b^2=4 h, \, z_0=0)$ while the red one marks the stationary point.}
  \label{3-fig:weakWithJumps}
\end{figure}

Note that these equations are consistent with the case $J=0$: the measurement process decreases the width of the atomic distribution and, once $b^2$ reaches its stationary value ($b^2=0$), the unbalance between odd and even sites becomes a constant. 
We compare the exponential damping towards the stable point to the effect of the jumps described by (\ref{3-eq:mean1}) and (\ref{3-eq:mean2}). Specifically, solving these equations at first order in $b^2$ we find
\begin{eqnarray}\label{3-eq:jumps_exp1}
z_{0,\mathrm{jumps}}(t)=-1+(1+z_0(0))\mathrm{e}^\frac{b^2 \Gamma t}{2h}.
\end{eqnarray}
The exponent in this expression should be compared with the one describing the exponential decay of $z_0(t)$. In the weak measurement regime, the evolution between two quantum jumps follows 
\begin{eqnarray}\label{3-eq:diff_exp1}
z_0(t)=\frac{1}{2} \mathrm{e}^{-\frac{\Gamma}{\omega}t } \left[c(0) \sin (J \omega t)+2 z_0(0) \cos (J \omega t)\right].
\end{eqnarray}
Therefore, the difference between the exponents in (\ref{3-eq:jumps_exp1}) and (\ref{3-eq:diff_exp1}) is
\begin{eqnarray}
\Gamma \left(\frac{b^2}{2h}-\frac{1}{\omega}\right),
\end{eqnarray}
which, since $\omega=2 \sqrt{(1-h)}$ and $b^2\sim 2 h$, is positive. This confirms that the jumps increase the amplitude of the oscillations driving the system away from the stable point (Figure~\ref{3-fig:weakWithJumps}).

In order to estimate the behavior of the imbalance in the large time limit taking into account both the effective dynamics and the quantum jumps, we compute analogous equations to (\ref{3-eq:mean1}) and (\ref{3-eq:mean2}) for the phases $\phi$ and $c$, and we incorporate them in the system (\ref{3-eq:b1})--(\ref{3-eq:c1}). Expanding the resulting expressions at first order in $h$ we find 
\begin{eqnarray}\label{3-syst}
	&\dot{(b^2)}=8 h J \phi -  \frac{\Gamma}{h} b^4, \\
	&\dot{\phi}=-\frac{J \omega^2}{4h} b^2 - \frac{\Gamma}{ h} b^2 \phi +  \frac{4 h J}{b^2}(1 + \phi^2), \\
	&\dot{z_0}= \frac{2 h J}{b^2} (2 z_0 \phi + c), \\
	&\dot{c}=- \frac{\Gamma}{h} b^2 c +  \frac{4hJ}{b^2}(\phi c - 2 z_0).
\end{eqnarray}
In the large time limit, the width of the atomic distribution becomes constant since the squeezing due to the measurement and the spreading due to the tunneling balance each other so that $\dot{(b^2)}=0$ and $\dot{\phi}=0$. Rearranging the equation for $z_0(t)$ in this limit we find
\begin{eqnarray}
\ddot{z_0}=-\omega^2z_0,
\end{eqnarray}
i. e. the population oscillates between the spatial modes without decaying (Figure~\ref{3-fig:weakWithJumps}).  

\begin{figure}[h]
\captionsetup{justification=justified}
  \centering
  \includegraphics[width=.7\linewidth]{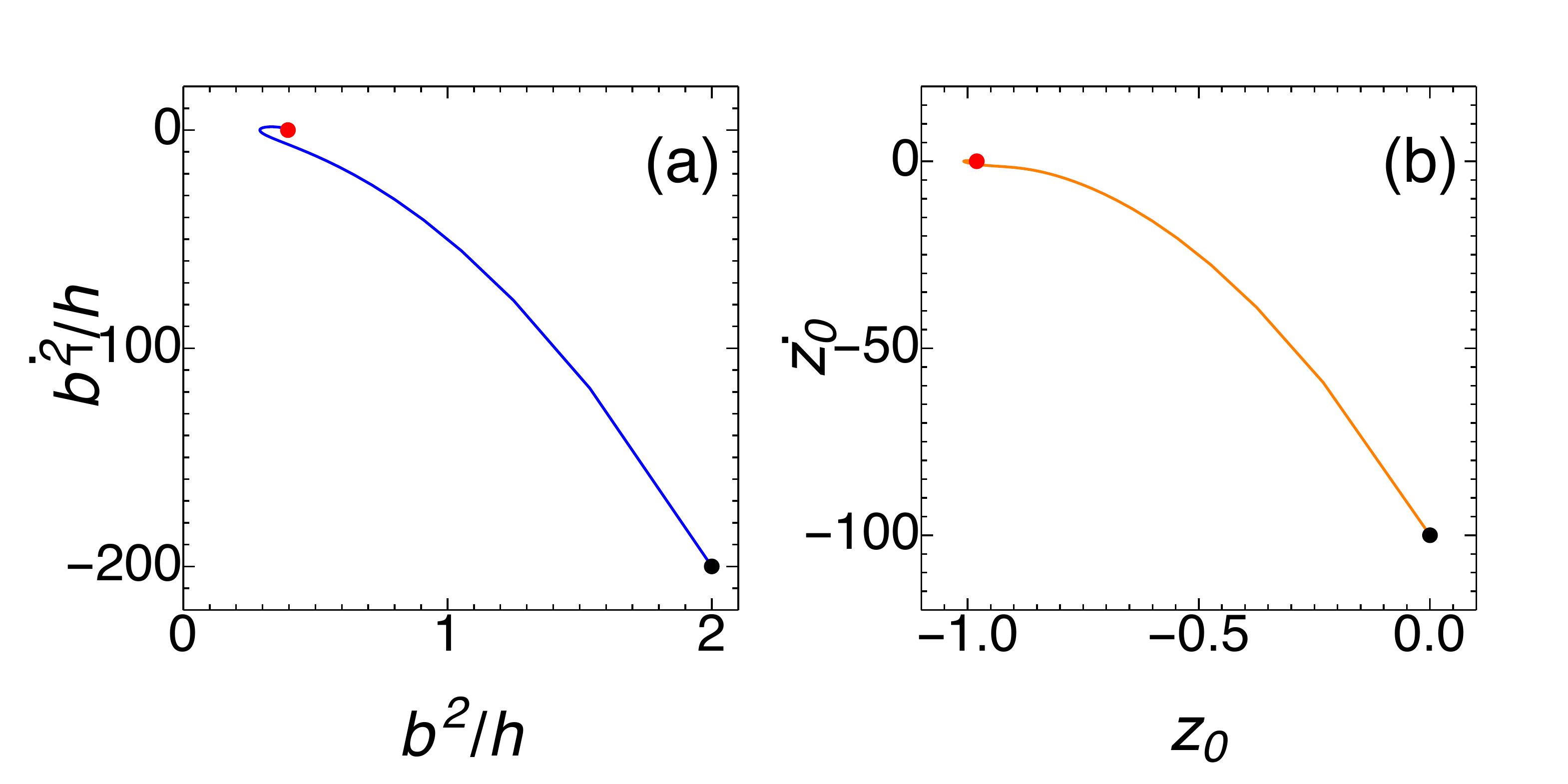}
  \caption{Over-damped oscillations of $(b^2,\dot{b^2})$ (a) and $(z_0,\dot{z_0})$ (b) of a single trajectory in the strong measurement regime ($\Gamma=100J$). The black point represents the starting point $(b^2=4 h, \, z_0=0)$ while the red one marks the stationary point.}
  \label{3-fig:strongNoJumps}
\end{figure}

\begin{figure}[h]
\captionsetup{justification=justified}
  \centering
  \includegraphics[clip, trim=0.5cm 19cm 1cm 1cm, width=0.7\linewidth]{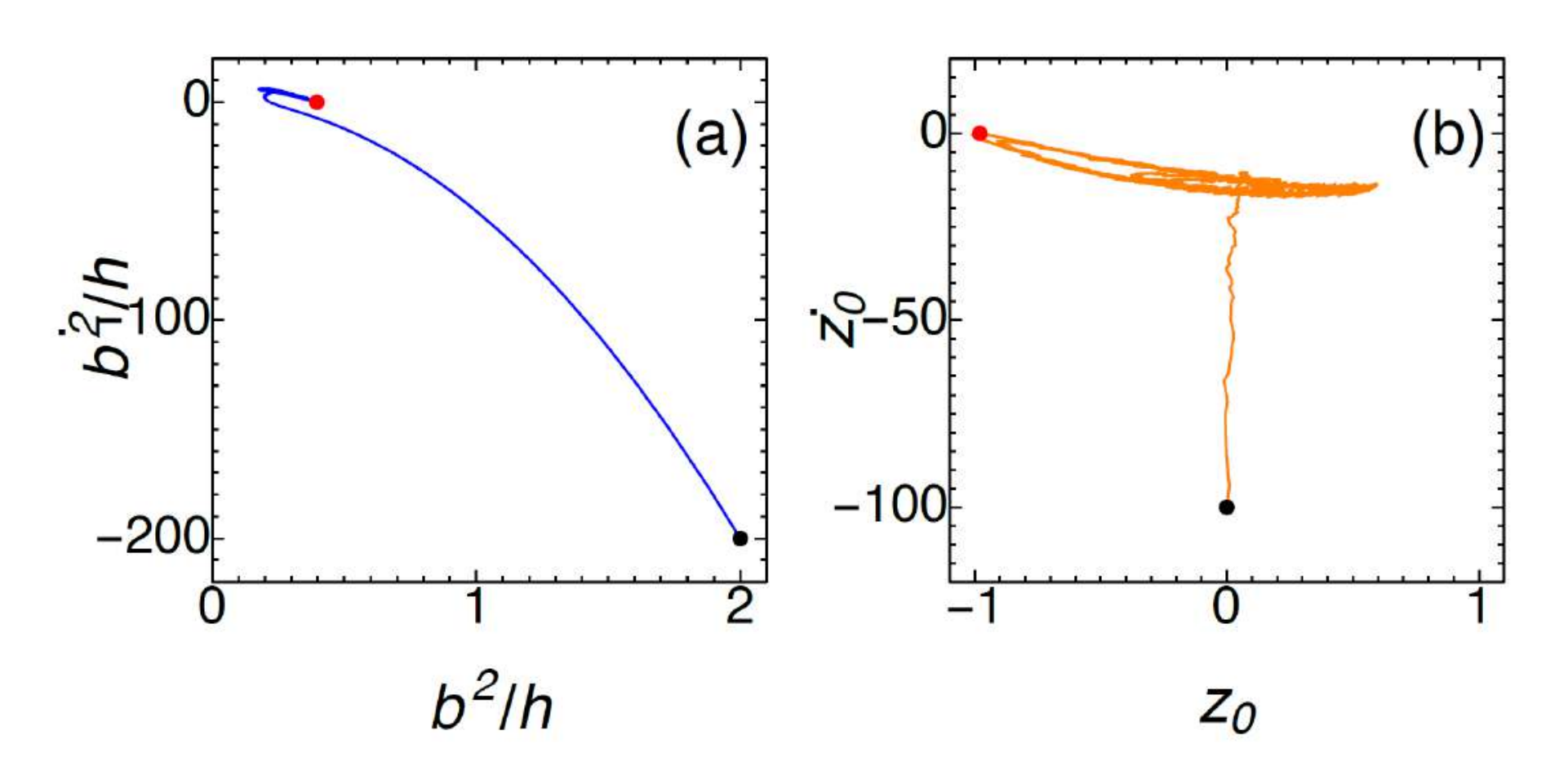}
  \caption{Full conditional dynamics of $(b^2,\dot{b^2})$ (a) and $(z_0,\dot{z_0})$ (b) of a single trajectory in the strong measurement regime ($\Gamma=100J$). The black point represents the starting point $(b^2=4 h, \, z_0=0)$ while the red one marks the stationary point.}
  \label{3-fig:strongWithJumps}
\end{figure}

If the measurement dominates the dynamics, i.~e. $\Gamma \gg J$,  the non-Hermitian dynamics dominates the evolution between two quantum jumps, In this case, the coordinates of the stationary point in this regime are $b^2(\infty)=4 h \sqrt{J/\Gamma} $ and $z_0 (\infty)=-1+J \omega^2/(2\Gamma)$, i. e. the width of the atomic distribution is extremely squeezed while the odd sites of the lattice tend to be empty.  Importantly, the evolution of the system is not oscillatory since the equations of motions around the stable point resemble an over-damped oscillator as the eigenvalues of the Jacobian matrix are $\lambda_{1,2}=\pm i\sqrt{J \Gamma}- \sqrt{J\Gamma}$ and $ \lambda_{3,4}=\pm 2i\sqrt{J \Gamma}- 2\sqrt{J \Gamma}$. In other words, the period  of an oscillation around the stable point and the damping time are approximately the same ($\Omega \Delta t_{d} \approx 1+ J \omega ^2/(2 \Gamma )$, see Figure~\ref{3-fig:strongNoJumps}). As we described in the previous paragraphs, the quantum jumps decrease the width of the atomic distribution even further and the full dynamics of $b(t)$ is not qualitatively different from the one determined by the differential equation (\ref{3-eq:b1}). In contrast, the evolution of the imbalance $z_0(t)$ is heavily affected by the photodetections, as illustrated in Figure~\ref{3-fig:strongWithJumps}. Specifically, we compare the dynamics due to the quantum jumps (\ref{3-eq:diff_exp1}) to the one due to the differential equation (\ref{3-eq:z01}): 
\begin{eqnarray}\label{3-diff_exp2}
z_0(t)=\frac{1}{2} \mathrm{e}^{-\sqrt{J \Gamma}t } \left[c(0) \sin (\sqrt{J \Gamma}t)+2 z_0(0) \cos (\sqrt{J \Gamma}t)\right].
\end{eqnarray}
Taking the difference between the two exponents we obtain 
\begin{eqnarray}
\Gamma \left(\frac{b^2}{2h}-\frac{1}{\sqrt{J \Gamma}}\right),
\end{eqnarray}
which is always positive, implying that the quantum jumps dominate the dynamics of the system taking $z_0$ away from its stationary point.

\subsection{Effect of the detector efficiency: measurement vs dissipation}

\begin{figure}[h]
\captionsetup{justification=justified}
\centering
\includegraphics[width=0.7\textwidth]{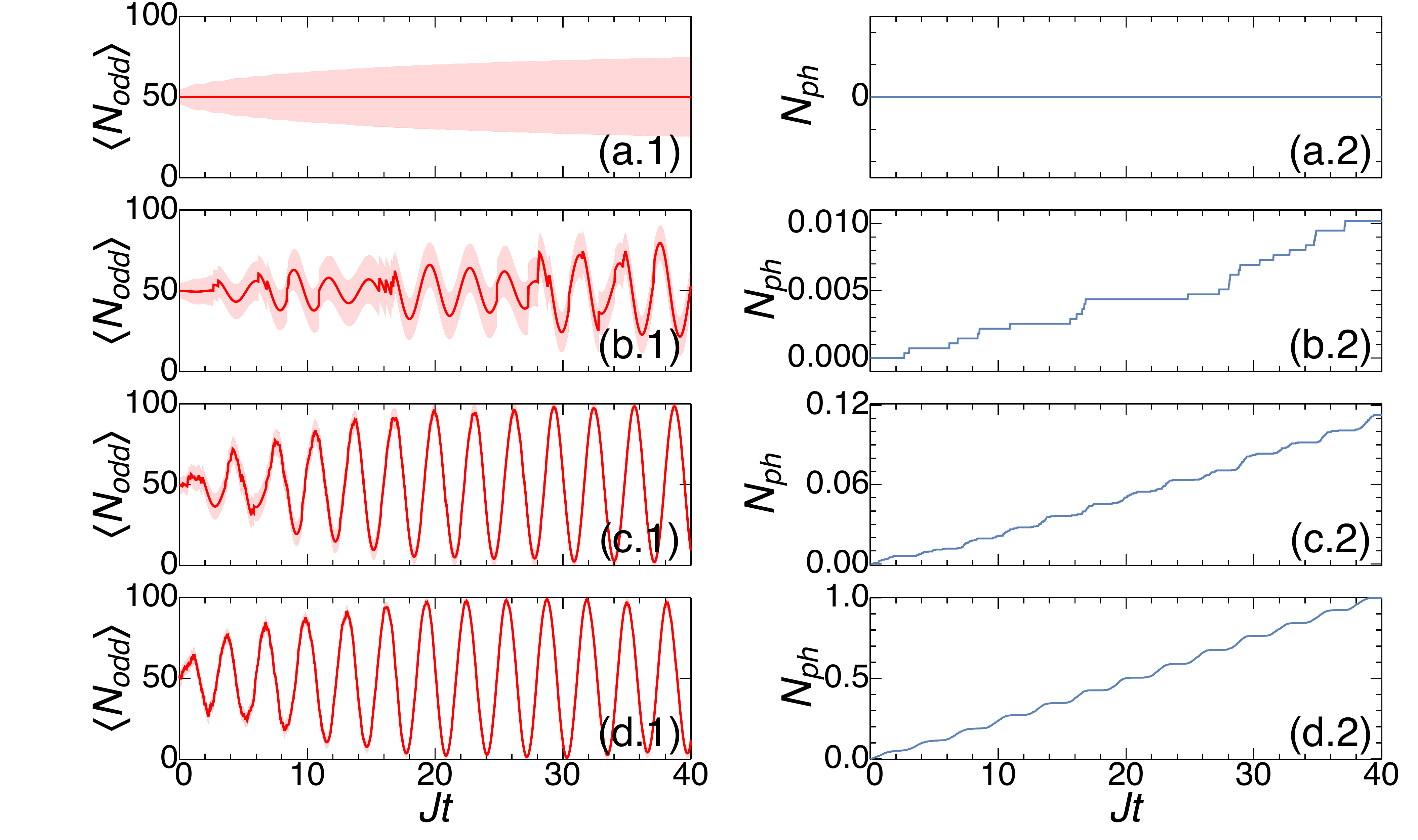}
\caption{Quantum Measurement vs dissipation. Conditional dynamics measuring $N_{\odd}$ for different detection efficiency obtained solving the SME for different efficiencies ($\eta=0,0.01,0.1,1$ for the panels (a), (b), (c), (d) respectively). Panels (1): atomic population of the odd sites. Panels (2): number of photons detected  $N_{ph}$ (normalized to the case $\eta=1$). The shaded area in the panels (1) represents the fluctuations $\sigma=\sqrt{\mathrm{Tr} \left( \hat{\rho}  \hat{N}_{\odd}^2\right) -\mathrm{Tr} \left( \hat{\rho}  \hat{N}_{\odd}\right)^2}$. Oscillations in the atomic population can be directly observed in the behavior of $N_{ph}(t)$. For zero efficiency (no detector), no oscillations develop, which corresponds to the dissipation. For finite efficiency the oscillations exist. ($N=100$, $\gamma/J=0.01$)}\label{3-fig:eff}
\end{figure}

The oscillatory dynamics we presented in the previous sections requires that all the photons leaving the optical cavity are successfully recorded by the detector, i. e., $\eta=1$, where $\eta$ is the detection efficiency. Nevertheless,  the effects we described in this chapter can be observed even if $\eta<1$ provided that enough photons are detected for each oscillation period so that it is possible to estimate the photocurrent. Figure~\ref{3-fig:eff} illustrates this by showing the conditional dynamics of the atomic system for different detection efficiencies when the measurement addresses the population of the odd lattice sites $\h{a}_1=C \h{N}_\odd$.

If the detector is not ideal, it is not possible to describe the conditional dynamics of the system by applying the quantum trajectory technique to the atomic wavefunction. Specifically, if some photons are ``missed'' by the detector, the quantum state of the system becomes mixed and needs to be described by the density matrix $\h{\rho}$ \cite{Wiseman}. Moreover, the evolution of this matrix follows the stochastic master equation (SME)
\begin{eqnarray}\label{3-sme}
\d \hat{\rho} (t)= \left\{   \d N \mathcal{G} [ \sqrt{\eta}\hat{c} ] - \d t \mathcal{H} [i \hat{H}_0 + \frac{\eta}{2} \hat{c}^\dagger \hat{c} ]  + \d t (1 - \eta)\mathcal{D}[\hat{c}] \right\}\hat{\rho}(t),
\end{eqnarray}
where  $\mathcal{G}, \mathcal{H}$, and  $\mathcal{D}$ are the superoperators 
\begin{gather}
\mathcal{G} [\hat{A} ]\hat{\rho}=\frac{\hat{A}\hat{\rho}\hat{A}^\dagger }{\text{Tr} \left[\hat{A}\hat{\rho}\hat{A}^\dagger \right]}-\hat{\rho},\\
\mathcal{H} [\hat{A} ]\hat{\rho}=\hat{A} \hat{\rho} + \hat{\rho} \hat{A}^\dagger - \text{Tr}\left[ \hat{A} \hat{\rho} + \hat{\rho} \hat{A}^\dagger \right],\\
\mathcal{D} [\hat{A} ]\hat{\rho}=\hat{A} \hat{\rho} \hat{A}^\dagger - \frac{1}{2}\left(  \hat{A}^\dagger  \hat{A} \hat{\rho} + \hat{\rho}  \hat{A}^\dagger  \hat{A}\right),
\end{gather}
and $\d N$ is the stochastic It\^o increment such that $E[\d N]=\eta \text{Tr}[\hat{c} \hat{\rho} \hat{c}^\dagger]\d t$. The physical quantity that is directly accessible in the experiments is $N_{ph}(t)$, i.e.,  the number of photons recorded by the detector up to time $t$. Importantly, this function is related to the jump operator and, in the limit where the timescale of the atomic dynamics is much slower than the typical interval between two photocounts, can be expressed as
\begin{eqnarray}\label{3-der}
\frac{\d N_{ph}}{\d t}=\eta \langle \hat{c}^\dagger \hat{c}\rangle (t),
\end{eqnarray}
where the symbol $\langle \hat{O} \rangle$ represents the expectation value of the operator $\hat{O}$ on a single realization of the SME. Equation (\ref{3-der}) allows us to estimate the minimum efficiency required for distinguishing the long-range oscillations induced by the measurement backaction. If the population of the odd sites of the optical lattices oscillates in time, the number of photons  recorded by the detector should show a growing ``staircase'' behavior with a characteristic time $2 \pi /J$ (see panel 2 of Figure~\ref{3-fig:eff} ). We can use this peculiar shape to identify the value of $\m{N_\odd}$: the tread corresponds to the times when $\m{N_\odd} \sim 0$ while the riser coincide with $\m{N_\odd} \sim N$. Therefore, if the detection efficiency allows to clearly resolve each step, the measurement backaction makes the atoms oscillate between odd and even sites retrieving the phenomena that we described in the previous sections. More quantitatively, defining $N_e(t)$ as the number of photons escaping the cavity, the value of $N_{ph}(t)$ follows a Bernoulli process with probability $\eta$ so that  $E[N_{ph}(t)]=\eta N_e(t)$ and $Var[N_{ph}(t)]=\eta (1-\eta) N_e(t)$. Importantly, the detection scheme we consider does not address single site properties since the measurement-induced spatial modes scatter light  collectively. For this reason, the photocurrent $ \eta \langle \hat{c}^\dagger \hat{c}\rangle$ scales as the square of the number of atoms loaded in the optical lattice. This property is crucial and highlights the fact that the oscillatory behavior of the atomic population is not a consequence of the detection of a \emph{single} photon but relies on many collective scattering events. Comparing the variance of $N_{ph}(t)$ to the number of detected photons in a single oscillation period, we estimate that the oscillatory dynamics is present if $\eta\gtrsim J /\gamma  N^2$, making the effects we described robust with respect to detection inefficiency. The dissipation corresponds to the zero detector efficiency, when as we see from the figure, the oscillations are completely absent.

\subsection{Stochastic differential equations and measurement}

In Section~\ref{3-sec:mf} we investigated the conditional evolution of the atomic system by analyzing the effect of  the non-Hermitian dynamics and the stochastic process described by the quantum jumps separately. However, it is possible to reach the same conclusions by treating these two effects in the same stochastic differential equation. Specifically, we model the evolution of the atomic wavefunction in terms of the stochastic Schr\"odinger equation (SSE)~\cite{Wiseman}
\begin{gather}
\d \!\ketz{\psi(t)}= \left[  \d  N(t) \left( \frac{\h{c}}{\sqrt{\m{\h{c}^\dagger \h{c}}(t)}} - \id \right) 
+ \d  t \left( \frac{ \m{\h{c}^\dagger \h{c}}(t)}{2} - \frac{\h{c}^\dagger \h{c}}{2} - i  \h{H}\right)\right] \ketz{\psi(t)}, \label{3-eq:SSE}
\end{gather}
where $\h{c}$ is the jump operator associated to the measurement, $\h{H}$ is the Hamiltonian generating the coherent dynamics of the system and $\d  N(t)$ is a stochastic increment that obeys the It\^o table
\begin{eqnarray}
&\d  N(t)^2=\d  N(t), \label{3-eq:Ito1}\\
&\d  N(t)\, \d  t =0.\label{3-eq:Ito2}
\end{eqnarray}
This SSE describes the atomic dynamics in a single experimental run, i. e. a single quantum trajectory. The stochastic term defines a point process which models the photocounts: if $\d  N(t)=1$ a photon is detected and the quantum jump operator is applied to wavefunction while if $\d  N(t)=0$ the system evolves deterministically. Importantly, the probability of detecting a photon in the (small) time interval $\delta t$ depends on the quantum state of the system and it is given by 
\begin{eqnarray}
p=\bok{\psi(t)}{\h{c}^\dagger \h{c}}{\psi(t)} \delta t.
\end{eqnarray}
In order to give a description of the measurement-induced oscillatory dynamics, we focus on the conditional evolution of the expectation values of few collective variables. From the SSE (\ref{3-eq:SSE}) and the  It\^o table (\ref{3-eq:Ito1})--(\ref{3-eq:Ito2}), we find a generalization of the Ehrenfest theorem for the conditional evolution of the observable  $\h{O}$
\begin{gather}
\d \m{\h{O}} (t) =\left( \frac{\m{\h{c}^\dagger \h{O} \h{c}}}{\m{\h{c}^\dagger \h{c}}} -\m{\h{O}} \right) \d  N(t) 
+\left(i \m{\left[ \h{H}, \h{O} \right]} - \frac{1}{2}\m{\left\{\h{c}^\dagger\h{c},\h{O} \right\}} + \m{\h{O}} \m{\h{c}^\dagger\h{c}}\right)  \d  t ,\label{3-eq:SSEobs}
\end{gather}
where the expectation values on the right hand side are computed at time $t$ and $[\cdot,\cdot]$ ($\{\cdot,\cdot\}$) is the (anti)commutator. Considering a probe that addresses the population of the odd sites, we can follow the evolution of the system by computing the expectation values of the number of atoms occupying the mode ($\h{N}_\odd$), the atomic current between odd and even sites ($\h{\Delta}$) and their fluctuations $\sigma_{AB}=\m{\h{A}\h{B} + \h{B}\h{A}}/2-\m{\h{A}}\m{\h{B}}$.  From equation (\ref{3-eq:SSEobs}) we find
\begin{gather}
\d \m{\h{N}_\odd}=\d  N \left[\frac{\m{\h{N}_\odd^3}}{\m{\h{N}_\odd^2}} - \m{\h{N}_\odd}\right] + \d  t \left[ -J \m{\h{\Delta}} - \gamma \left( \m{\h{N}_\odd^3} -\m{\h{N}_\odd} \m{\h{N}_\odd^2}\right)\right], \label{3-eq:SSE_dw1}\\
\d \m{\h{\Delta}}= \d  N \left[\frac{\m{\h{N}_\odd \h{\Delta} \h{N}_\odd}}{\m{\h{N}_\odd^2}} - \m{\h{\Delta}}\right]+ \d  t \left[ -2J \left( N - 2 \m{\h{N}_\odd} \right)  \right. \nonumber \\
\left.- \frac{ \gamma}{2} \left(\m{\h{N}_\odd ^2 \h{\Delta}} + \m{\h{\Delta} \h{N}_\odd^2} - \m{\h{\Delta}} \m{\h{N}_\odd } \right) \right], \label{3-eq:SSE_dw2}\\
\d \sigma^2_N=\d  N \left[\frac{\m{\h{N}_\odd^4}\m{\h{N}_\odd^2}-\m{\h{N}_\odd^3}^2}{\m{\h{N}_\odd^2}} -  \sigma^2_N \right]
 + \d  t \left[ -J \left( \m{\h{N}_\odd \h{\Delta}} + \m{\h{\Delta} \h{N}_\odd} - 2 \m{\h{\Delta}}\m{\h{N}_\odd}\right) \right. \nonumber \\
 \left. -\frac{ \gamma}{2} \left( 2\m{\h{N}_\odd^4} -4\m{\h{N}_\odd}\m{\h{N}_\odd^3} -2 \m{\h{N}_\odd^2}^2 +4 \m{\h{N}_\odd^2}\m{\h{N}_\odd}^2 \right)\right], \label{3-eq:SSE_dw3}\\
\d \sigma^2_\Delta=\d  N \left[\frac{\m{\h{N}_\odd \h{\Delta}^2 \h{N}_\odd}\m{\h{N}_\odd^2}-\m{\h{N}_\odd \h{\Delta} \h{N}_\odd}^2}{\m{\h{N}_\odd^2}} -  \sigma^2_\Delta \right]     
\nonumber \\
+ \d  t \left[ 4J \left( \m{\h{N}_\odd \h{\Delta}} + \m{\h{\Delta} \h{N}_\odd} - 2 \m{\h{\Delta}}\m{\h{N}_\odd}\right) \right.  \nonumber \\
\left. -\frac{ \gamma}{2} \left( \m{\h{N}_\odd^2 \h{\Delta}^2} + \m{\h{\Delta}^2 \h{N}_\odd^2}-2 \m{\h{\Delta}}(  \m{\h{N}_\odd^2 \h{\Delta}} + \m{\h{\Delta} \h{N}_\odd^2})\right.  
\left.  - 2 \m{\h{N}_\odd^2}(\m{\h{\Delta}^2}-2\m{\h{\Delta}}^2) \right)\right] \label{3-eq:SSE_dw4},
\end{gather}
\begin{gather}
\d \sigma_{\Delta N}=\d  N \left[\frac{\m{\h{N}_\odd \h{\Delta} \h{N}_\odd^2}\m{\h{N}_\odd^2}-\m{\h{N}_\odd \h{\Delta} \h{N}_\odd}\m{\h{N}_\odd^3}}{\m{\h{N}_\odd^2}} -  \sigma_{\Delta N} \right] + \nonumber \\ 
  +\d  t \left[ -J \left( \m{\h{\Delta}^2}-\m{\h{\Delta}}^2 +4 \m{\h{N}_\odd}^2 - 4\m{\h{N}_\odd^2} \right) \right.  \nonumber \\
\left. -\frac{ \gamma}{2} \left( \m{\h{N}_\odd^2 \h{\Delta} \h{N}_\odd} + \m{\h{\Delta}\h{N}_\odd^3}-2 \m{\h{\Delta}}\m{\h{N}_\odd^3} \right. \right.+ \nonumber \\
 \left.  \left.-\m{\h{N}_\odd}( \m{\h{N}_\odd^2 \h{\Delta}} + \m{\h{\Delta} \h{N}_\odd^2})- 2 \m{\h{N}_\odd^2}(\m{\h{\Delta}\h{N}_\odd}-2\m{\h{\Delta}}\m{\h{N}_\odd}) \right)\right], \label{3-eq:SSE_dw5}
\end{gather}
where the probability of a jump in a small time interval $\delta t$ is given by
\begin{eqnarray}
p= \delta t \gamma \m{\h{N}_\odd^2}.
\end{eqnarray}
The system (\ref{3-eq:SSE_dw1})--(\ref{3-eq:SSE_dw4}) is not closed since each equation depends on the expectation values of higher moments of $\h{N}_\odd$ and $\h{\Delta}$. However, we can give an approximate closed formulation of these equations by assuming that  $\m{\h{N}_\odd}$ and $\m{\h{\Delta}}$ are classical Gaussian variables so that all their moments can be expressed as a function of their mean and variance (for example $\m{\h{N}_\odd^3}\approx \m{\h{N}_\odd}^3+3  \m{\h{N}_\odd} \sigma^2_N $). Similarly to Section~\ref{3-sec:mf}, we focus on the large particle number limit $N\gg 1$ so that it is possible to neglect the variance of $\h{N}_\odd$ with respect to its squared value, i. e. $\sigma^2_N / \m{\h{N}_\odd}^2\sim 1/N \sim 0$. Taking into account these approximations, equations  (\ref{3-eq:SSE_dw1})--(\ref{3-eq:SSE_dw4}) simplify greatly and can be rewritten as
\begin{eqnarray}
&\d \m{\h{N}_\odd}=  \frac{2 \sigma^2_N}{\m{\h{N}_\odd}}\d  N -  \left( J \m{\h{\Delta}} + 2\gamma \m{\h{N}_\odd}\sigma^2_N\right) \d  t \label{3-eq:SSE_dw1_2},\\
&\d \m{\h{\Delta}}=  \frac{2 \sigma_{\Delta N}}{\m{\h{N}_\odd}} \d  N -2 \left[ J \left( N-2  \m{\h{N}_\odd}\right) +\gamma \m{\h{N}_\odd} \sigma_{\Delta N} \right] \d  t \label{3-eq:SSE_dw2_2},\\
&\d \sigma^2_N=-\frac{ 2\sigma^4_N}{\m{\h{N}_\odd}^2}\d  N-2 \left(J  \sigma_{\Delta N} + \gamma \sigma^4_N \right) \d  t \label{3-eq:SSE_dw3_2},\\
&\d \sigma^2_\Delta =- \frac{ 2\sigma^2_{\Delta N}}{\m{\h{N}_\odd}^2} \d  N  -2 \left(4 J  \sigma_{\Delta N} +\gamma \sigma^2_{\Delta N} \right) \d  t \label{3-eq:SSE_dw4_2},\\
&\d \sigma_{\Delta N}= \frac{2 \sigma_{\Delta N}\sigma_{N}^2}{\m{\h{N}_\odd}^2} \d  N +  \left[J  (4 \sigma^2_N - \sigma^2_{\Delta}) - 2 \gamma \sigma^2_N\sigma_{\Delta N} \right] \d  t \label{3-eq:SSE_dw5_2},
\end{eqnarray}
where the jump probability is given by $p= \delta t \gamma  \m{\h{N}_\odd}^2 $.  The deterministic terms in the equations for  $\m{\h{N}_\odd}$ and $\sigma_N^2$, i. e. the ones proportional to the time increment $\d t$, coincide with the differential equations we obtained in Section~\ref{3-sec:mf} using the mean field approximation. Specifically, we retrieve equations  (\ref{3-eq:b1}) and (\ref{3-eq:z01})  by setting $\sigma_N^2=2 N^2 b^2$ and $\m{\h{N}_\odd}=N(1+z_0)$. This confirms that the two different approaches we considered are consistent and lead to the same behavior for the collective variables addressed by the measurement. The main advantage of the system (\ref{3-eq:SSE_dw1_2})--(\ref{3-eq:SSE_dw5_2}) is that it allows us to describe the quantum jumps and the non-Hermitian dynamics in a single equation. We can use these expressions for gaining insight in the conditional evolution of the atomic system. In order to discuss the behavior of a ``typical'' quantum trajectory, we focus on the equations for the atomic imbalance $\m{\h{N}_\odd}$ and  the current $\m{\h{\Delta}}$ in the limit where the number of photons recorded by the detector can be approximated by a continuous function, i. e. the time interval between two photocounts is much smaller than the  timescale of the atomic dynamics. If this is the case, we can rewrite the It\^o increment as $\d  N=\m{\d  N}-\m{\d  N}+\d  N=\gamma \m{\h{N}_\odd}^2 \d  t + \sqrt{\gamma}\m{\h{N}_\odd} \d  W $ where $ \d  W$ is a Wiener increment representing the fluctuations in the photoncounts around the average value \cite{Wiseman,Ruostekoski2014,Ashida2015b}.  Substituting this expression in (\ref{3-eq:SSE_dw1_2})-(\ref{3-eq:SSE_dw1_3}) we find
\begin{eqnarray}
&\d \m{\h{N}_\odd}=  -J \m{\h{\Delta}} \d  t +    2 \sqrt{\gamma} \sigma^2_N \d  W  \label{3-eq:SSE_dw1_3},\\
&\d \m{\h{\Delta}}=  -2  J \left( N-2  \m{\h{N}_\odd}\right) \d  t  + 2 \sqrt{\gamma} \sigma_{\Delta N} \d  W,  \label{3-eq:SSE_dw2_3}\\
&\d  \sigma^2_N =  -2  \left( J \sigma_{\Delta N} + 2 \gamma \sigma_N^4 \right) \d  t - \frac{2 \sqrt{\gamma} \sigma_{N}^4}{\m{\h{N}_\odd}} \d  W,  \label{3-eq:SSE_dw3_3}\\
&\d  \sigma_{\Delta}^2 =  -4  \left( 2J \sigma_{\Delta N} +  \gamma \sigma_{\Delta N}^2 \right) \d  t - \frac{2 \sqrt{\gamma} \sigma_{\Delta N}^2}{\m{\h{N}_\odd}} \d  W,  \label{3-eq:SSE_dw4_3}\\
&\d  \sigma_{\Delta N} =   J \left( 4\sigma_{N}^2- \sigma_{\Delta}^2  \right) \d  t +\frac{2 \sqrt{\gamma} \sigma_{\Delta N}  \sigma_{N}^2}{\m{\h{N}_\odd}} \d  W.  \label{3-eq:SSE_dw5_3}
\end{eqnarray}
Neglecting the fluctuations in the photocounts, the equations for $ \m{\h{N}_\odd}$ and  $\m{\h{\Delta}}$ describe the evolution of a harmonic oscillator and confirm the emergence of the oscillatory behavior for the population of the odd sites of the lattice. Note that these oscillations are present even without measurement but here their behavior is fundamentally different: in the absence of continuous monitoring the amplitude of the oscillations is proportional to the atom imbalance of the initial state and its probability distribution tends to spread, i. e. the value of $\sigma_N^2$ increases in time. In contrast, here we observe that the uncertainty in the occupation of the spatial modes ($\sigma_N^2$) decreases in time (as suggested by equation ($\ref{3-eq:SSE_dw3_3}$)) and that full-exchange of atoms between the two spatial modes is possible even starting with a perfectly balanced state. 
An alternative formulation of equations (\ref{3-eq:SSE_dw1_3})-(\ref{3-eq:SSE_dw5_3}) can be obtained by rewriting them using the Stratonovich formalism:
\begin{eqnarray}
\frac{\d }{\d  t} \m{\h{N}_\odd}= - J \m{\h{\Delta}} - \gamma \frac{\d }{\d  t} \sigma_N^4 +  2 \sqrt{\gamma} \sigma^2_N \xi (t)\label{3-eq:strat1},\\
\frac{\d }{\d  t} \m{\h{\Delta}} = -2  J \left( N-2  \m{\h{N}_\odd}\right)  -\gamma \frac{\d }{\d  t} \sigma_{\Delta N}^2+ 2 \sqrt{\gamma} \sigma_{\Delta N} \xi (t)\label{3-eq:strat2},\\
\frac{\d }{\d  t} \sigma_N^2 =  \frac{-2  \left( J \sigma_{\Delta N} + 2 \gamma \sigma_N^4 \right) + \gamma \sigma_N^4 \frac{\d }{\d  t} \frac{1}{\m{\h{N}_\odd}^2}-\frac{2 \sqrt{\gamma} \sigma_{N}^8}{\m{\h{N}_\odd}} \xi(t)}{1-2 \gamma  \frac{\sigma_{N}^2}{\m{\h{N}_\odd}^2}}\label{3-eq:strat3},\\
\frac{\d }{\d  t} \sigma_\Delta^2=-4  \left( 2J \sigma_{\Delta N} +  \gamma \sigma_{\Delta N}^2 \right)  + \gamma \frac{\d }{\d  t} \frac{\sigma_{\Delta N}^4}{\m{\h{N}_\odd}^2}- \frac{2 \sqrt{\gamma} \sigma_{\Delta N}^2}{\m{\h{N}_\odd}} \xi(t)\label{3-eq:strat4},\\
\frac{\d }{\d  t} \sigma_{\Delta N} =   \frac{J \left( 4\sigma_{N}^2- \sigma_{\Delta}^2  \right) -  \gamma \sigma_{\Delta N} \frac{\d }{\d  t} \frac{ \sigma_{N}^4}{\m{\h{N}_\odd}^2}+ \frac{2 \sqrt{\gamma} \sigma_{\Delta N}  \sigma_{N}^2}{\m{\h{N}_\odd}} \xi(t)}{1+\gamma \frac{\sigma_{N}^4}{\m{\h{N}_\odd}^2 }},\label{3-eq:strat5}
\end{eqnarray}
where $ \xi(t)$ is a Wiener process. Combining Eq. (\ref{3-eq:strat1}) and (\ref{3-eq:strat2}), we find that the dynamics of the number of atoms in the odd sites can be described as  a forced harmonic oscillator:
\begin{eqnarray}
\frac{\d ^2}{\d  t^2} \m{\h{N}_\odd}=  2  J^2 \left( N-2  \m{\h{N}_\odd}\right) + F,
\end{eqnarray}
where the forcing term is given by
\begin{eqnarray}
F=\gamma \left( J \frac{\d }{\d  t} \sigma_{\Delta N}^2 - \frac{\d ^2}{\d  t^2} \sigma_N^4 \right) + 2 \sqrt{\gamma} \left [ \frac{\d }{\d  t} \left( \sigma_N^2 \xi(t) \right)- \sigma_{\Delta N} \xi(t)\right].
\end{eqnarray}
Therefore, the measurement introduces a quasi-periodic stochastic force $F$ that drives the system towards larger imbalance, increasing the amplitude of the oscillations of  $ \m{\h{N}_\odd}$.

\subsection{Extensions for multimode photonic systems}

Here we focused on multimode dynamics of ultracold atoms. However, it is reasonable to ask a question, whether the idea of combining the multimode unitary dynamics and quantum backaction of measurement can be extended to other systems.  Recently, significant effort has been made in the development of purely photonic systems with multiple path interferometers, which are one of the setups promising for applications in quantum technologies. A possible realization consists of multiple interconnected fibers, the so called photonic circuits or photonic chips \cite{Spring2013, Holleczek2015}. Indeed, quantum walks \cite{Elster2015} have been already discussed in the contexts of both ultracold atoms in optical lattices and single photons propagating and interfering in a multiple path interferometer. Both systems are the candidates for realizations of quantum simulations and quantum computation protocols.  

The tunneling of atoms in an optical lattice can be analogous to the propagation of photons in the waveguides and their transmission and reflection at beamsplitters (waveguide couplers). Already current technologies allow using single photons and photon pairs as input states for multiple waveguides \cite{Spring2013}.  The boson sampling is considered as a realization of essentially multimode quantum interference of bosons during their unitary evolution (i.e. the propagation through the photonic system).

The detection of photons can be considered in several ways. On the one hand, the non-destructive detection of photons is indeed very difficult to implement. Nevertheless some research is being carried out, which makes it reasonable to expect at least some progress in the future. First, the parametric down conversion produces pairs of entangled photons or beams. The detection of an idler beam represents a QND measurement of the signal beam~\cite{Wiseman}. The use of photon pairs is already consistent with current systems~\cite{Spring2013}. Second, a QND method of photodetection was realized using a cavity QED system~\cite{Reiserer2013}. It indeed remains a challenge to integrate various elements together. On the other hand, as several photons participate in the interference, even the standard detection of a small number of them, while being destructive, can be also considered as a non-fully projective measurement as the rest of photons continue to evolve after some photons are detected. For example, the photon subtraction technique was already shown to produce interesting nonclassical states of light and quantum correlations~\cite{Paternostro2011}. One can draw an analogy with ultracold atoms for the case of two BECs: when the small number of atoms is destructively detected after the matter-wave interference, the remaining atoms develop the phase coherence between two condensates due to the projection of quantum state~\cite{Dalibard}. Thus, while being an experimental challenge, the fully photonic realization of the competition between measurement backaction and unitary dynamics can lead to interesting developments in quantum technologies.

In summary of this section, we have shown that light scattering from ultracold gases in optical lattice can be used for partitioning the system into macroscopically occupied spatial modes with non-trivial overlap which preserve long-range coherence. We formulated an effective model for the dynamics of such modes at a single quantum trajectory, mapping each spatial mode to a single ``well'' and describing its properties in terms of collective variables.  The measurement backaction competes with the standard local dynamics and induces oscillatory dynamics on the atomic state. Depending on the spatial profile of the measurement operator, this competition can be exploited for creating multimode macroscopic superposition states which could have applications in metrology and quantum information. Importantly, these states are robust with respect to detection inefficiencies because of the global addressing of our measurement setup. We presented an analytical model that captures the emergence of large-scale collective oscillations with increasing amplitude for the case where only two spatial modes are present. Using the quantum trajectory formalism, we found that  the measurement backaction drives the system away from its stationary point and behaves as a quasi-periodic force acting on the atoms. Finally, we confirmed our finding by formulating an alternative description in terms of stochastic differential equations for the evolution of collective atomic observables. In the limit, where the time interval between two photocounts is much smaller than the  timescale of the atomic dynamics and the fluctuations in the photocount can be neglected, the atomic population of one of the modes evolves as a harmonic oscillator driven by a stochastic force.


\section{Quantum state reduction by the matter-phase-related measurements}

In this section, instead of considering coupling light to the on-site density,
  we consider the quantum backaction due to the measurement of
  matter-phase-related variables such as global phase coherence. We
  show how this unconventional approach opens up new opportunities to
  affect system evolution. We demonstrate how this can lead to a new
  class of final states different from those possible with dissipative
  state preparation or conventional projective measurements. These
  states are characterised by a combination of Hamiltonian and
  measurement properties thus extending the measurement postulate for
  the case of strong competition with the system's own evolution.

In Sec. 3.3.2, we have already demonstrated an example of the quantum wave-particle dualism, where our many-body system behaves like particles or waves depending on the varibale, which we chose for the measurement: either the density- or phase-related atomic variables. Here we will go further in theoretical details.

Light scatters due to its interaction with the dipole moment of the
atoms which for off-resonant light results in an effective coupling
with atomic density, not the matter-wave amplitude. Therefore, it is
challenging to couple light to the phase of the matter-field, as is
typical in quantum optics for optical fields. Most of the existing
work on measurement couples directly to atomic density operators. However, in this work (Sec. 1.8) we have already proved that it is possible to
couple to the the relative phase differences between sites in an
optical lattice by illuminating the bonds (i. e. the inter-site areas) between them. This is a multi-site generalisation
of previous double-well schemes \cite{cirac1996, Dalibard, RuostekoskiPRA1997, RuostekoskiPRA1998, rist2012}, although the physical
mechanism is fundametally different as it involves direct coupling to
the interference terms caused by atoms tunnelling rather than
combining light scattered from different sources.

 In this section, we go beyond any previous work by
studying this new feature of optical lattice cavity systems in the
context of measurement backaction. The quantum trajectory approach to
backaction induced dynamics has attracted
significant experimental interest in single atom cavity
\cite{hood1998} and single qubit circuit \cite{murch2013, roch2014}
QED systems. However, its study in the context of many-body dynamics
is much more recent. Here, it is the novel combination of measurement backaction as the
physical mechanism driving the dynamics and phase coherence as the
observable, which the optical fields couple to, that provides a
completely new opportunity to affect and manipulate the quantum state.

Here we present an example of quantum gas system. The general mathematical details of the generated states can be found in Ref. \cite{Kozlowski2017}, where we generalized our model and showed a novel type of a
projection due to measurement which occurs even when there is
significant competition with the Hamiltonian dynamics. This projection
is fundamentally different to dissipative steady states, standard
formalism eigenspace projections or the quantum Zeno effect
\cite{misra1977,Facchi2008, Raimond2010, Raimond2012, Signoles2014}
thus providing an extension of the measurement postulate to dynamical
systems subject to weak measurement. Such a measurement-based
preparation is unobtainable using the dissipative state engineering,
as the dissipation would completely destroy the coherence in this
case.

As in previous sections, the scattered light amplitude (i.e. its annihilation operator) is $a = C ( \hat{D} + \hat{B} )$. In general, it is easier for the light to couple to atom density that
is localized within the lattice sites rather than the density within the
bonds, i.e.~in between the lattice sites. This means that in most
cases $\D \gg \B$ and thus $a \approx \D$. However, it is possible to
arrange the light geometry in such a way that scattering from the
atomic density operators within a lattice site is suppressed leading
to a situation where light is only scattered from these bonds leading
to an effective coupling to phase-related observables, $a = C \B$
 (cf. Sec. 1.8). This does not mean that light actually scatters
from the matter phase. Light scatters due to its interaction with the
dipole moment of the atoms which for off-resonant light scattering is always proportional to the density
distribution. However, in an optical lattice, the interference of
matter waves between neighbouring sites leads to density modulations
which allows us to indirectly measure these phase observables. Here, we will
summarize the results of Sec. 1.8 and extend them to include the effects of measurement
backaction due to such coupling.

\subsection{Matter-phase QND measurements}

\begin{figure}[h]
\captionsetup{justification=justified}
  \centering
  \includegraphics[width=0.6\linewidth]{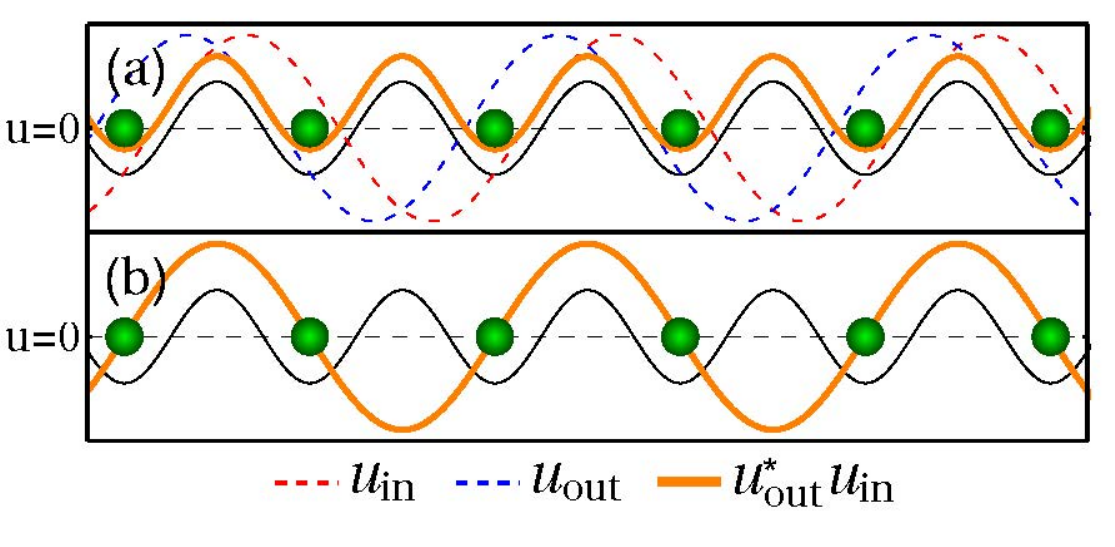}
  \caption{Light field arrangements which maximise coupling,
    $u_\mathrm{out}^*u_\mathrm{in}$, between lattice sites. The thin
    black line indicates the trapping potential (not to scale). (a)
    Arrangement for the uniform pattern $J_{m,m+1} = J_1$. (b)
    Arrangement for spatially varying pattern $J_{m,m+1}=(-1)^m J_2$;
    here $u_\mathrm{in}=1$ so it is not shown and $u_\mathrm{out}$ is
    real thus $u_\mathrm{out}^*u_\mathrm{in}=u_\mathrm{out}$.}
  \label{3-fig:setup}
\end{figure}

If we consider both incoming (probe) and outgoing (scattered) beams to be standing waves,
$u_\mathrm{in,out} = \cos(k^x_\mathrm{in,out} x +
\varphi_\mathrm{in,out})$ we can suppress the $\D$-operator
contribution by crossing the beams at angles such that $x$-components
of the wavevectors are $k^x_\mathrm{in,out} = \pi/d$, and the phase
shifts satisfy $\varphi_\mathrm{in} + \varphi_\mathrm{out} = \pi$ and
$\varphi_\mathrm{in} - \varphi_\mathrm{out} = \arccos[\mathcal{F}
[w^2(\b{r})](2 \pi / a) / \mathcal{F} [w^2(\b{r})] (0) ]/2$, where
$\mathcal{F}[f(\b{r})]$ denotes a Fourier transform of $f(\b{r})$. For clarity, this arrangement is illustrated in
Fig. \ref{3-fig:setup}(a). This ensures that $J_{m,m} = 0$ whilst
\begin{equation}
J_1 \equiv J_{m, m+1} = \mathcal{F} [w(\b{r} - \b{a}/2) w(\b{r} +
\b{a}/2)](2 \pi / a)/2 
\end{equation}
is a constant, and thus $\a = C \B_1$ ($\D = 0$, $\B = \B_1$) with
\begin{equation} 
  \B_1 = \sum_m^K J_1 \hat{p}_m = 2 J_1 \sum_k c^\dagger_k
  c_k \cos(ka),
\end{equation} 
where $\hat{p}_m = \bd_m b_{m + 1} + b_m \bd_{m + 1}$, and the second equality follows from converting to momentum space
via $b_m = \frac{1}{\sqrt{M}} \sum_k e^{-ikma} c_k$ and $c_k$
annihilates an atom with momentum $k$.

In order to correctly describe the dynamics of a single quantum
trajectory we introduce, as in previous sections, a non-Hermitian term to the Hamiltonian,
$-i \kappa a^\dag a$. As the jump operator itself, $a$ is linearly
proportional to the atom density, the new term introduces a quadratic
atom density term on top of the nonlocality due to the global nature
of the probing. Therefore, in order to focus on the competition
between tunnelling and measurement backaction we do not consider the
other (standard) nonlinearity due to the atomic interactions: $U =
0$. Therefore, $\B_1$ is proportional to the remaining part of the Bose--Hubbard Hamiltonian (describing tunneling)  and both
operators have the same eigenstates, i.e.~Fock states in the momentum
basis. We can thus rewrite as
\begin{equation}
  H = - \frac{J} {J_1} \B_1 - i \kappa |C|^2 \B_1^\dagger \B_1,
\end{equation}
which will naturally be diagonal in the $\B_1$ basis. Since it's
already diagonal we can easily solve its dynamics and show that the
probability distribution of finding the system in an eigenspace with
eigenvalue $B_1$ after $n$ photocounts at time $t$ is given by
\begin{equation} 
  p(B_1,n,t) = \frac{B_1^{2n}}{F(t)} \exp \left[ - 2
    \kappa |C|^2 B_1^2 t \right] p_0(B_1),
\end{equation} 
where $p_0(B_1)$ denotes the initial probability of observing $B_1$
\cite{MekhovPRA2009} and $F(t)$ is the normalisation
factor. This is exactly the form, which we derived earlier for the density operator measurements in Sec. 2.2, but now rewritten for the matter-amplitude variables $B_1$.

As in Secs. 2.2 and 2.3, this distribution has peaks at $B_1 = \pm \sqrt{n/2\kappa
  |C|^2 t}$ and an initially broad distribution will narrow down
around these two peaks with time and successive photocounts. The final
state is in a superposition, because we measure the photon number
$a^\dag \a$, and not the field amplitude. Therefore, the measurement is
insensitive to the phase of $a = C \B$ and we get a superposition of
$\pm B_1$. This means that the matter is still entangled with the
light as the two states scatter light with different phase which the
photocount detector cannot distinguish. However, this is easily
mitigated at the end of the experiment by switching off the probe beam
and allowing the cavity to empty out or by measuring the light phase
(quadrature) to isolate one of the components. Interestingly, this measurement will establish
phase coherence across the lattice, $\langle \bd_m b_n \rangle \ne 0$,
in contrast to density based measurements where the opposite is true,
Fock states with no coherences are favoured.

Unusually, we do not have to worry about the timing of the quantum
jumps, because the measurement operator commutes with the
Hamiltonian. This highlights an important feature of this measurement
-- it does not compete with atomic tunnelling, and represents a quantum
nondemolition (QND) measurement of the phase-related observable
\cite{BrunePRA1992}. This is in contrast to conventional density based
measurements which squeeze the atom number in the illuminated region
and thus are in direct competition with the atom dynamics (which
spreads the atoms), thus requiring strong couplings for a projection. Here a projection is achieved at any measurement
strength which allows for a weaker probe and thus effectively less
heating and a longer experimental lifetime.

\subsection{Beyond QND measurements of matter-phase variables: competition with tunneling}

It is also possible to achieve a more complex spatial pattern of
$J_{m, m+1}$ (cf. Sec. 1.8). This way the observable will no
longer commute with the Hamiltonian (and thus will no longer be QND),
but will still couple to the phase related operators. This can be done
by tuning the angles such that the wavevectors are
$k^x_\mathrm{in} = 0$ and $k^x_\mathrm{out} = \pi/d$ and the phase
shift of the outgoing beam is $\varphi_\mathrm{out} = \pm \pi/d$.
This yields
\begin{equation}
  (-1)^m J_2 \equiv J_{m,m+1} = - (-1)^m \mathcal{F} [w(\b{r} - \b{a}/2) w(\b{r} +
  \b{a}/2)](\pi / a) \cos (\varphi_\mathrm{in}),
\end{equation}
where $J_2$ is a constant. Now $a = C\B_2$ ($\D = 0$, $\B = \B_2$)
and the resulting coupling pattern is shown in
Fig. \ref{3-fig:setup}(b). The operator $\B_2$ is given by,
\begin{equation} 
  \B_2 = \sum_m^K (-1)^m J_2 \hat{p}_m = 2 i J_2 \sum_k c^\dagger_k
  c_{k - \pi/a} \sin(ka).
\end{equation} 
Note how the measurement operator now couples the momentum mode $k$
with the mode $k - \pi/a$. 

The measurement operator no longer commutes with the Hamiltonian so we
do not expect there to be a steady state as before. In order to
understand the measurement it will be easier to work in a basis in
which it is diagonal. We perform the transformation $\beta_k =
\frac{1}{\sqrt{2}} \left( c_k + i c_{k - \pi/a} \right)$,
$\tilde{\beta}_k = \frac{1}{\sqrt{2}} \left( c_k - i c_{k - \pi/a}
\right)$, which yields the following forms of the measurement operator
and the Hamiltonian: 
\begin{equation}
  \B_2 = 2 J_2 \sum_{\mathrm{RBZ}} \sin(ka) \left(
    \beta^\dagger_k \beta_k - \tilde{\beta}_k^\dagger \tilde{\beta}_k
  \right), 
\end{equation}
\begin{equation}
  \hat{H}_0 = 2 J \sum_{\mathrm{RBZ}} \cos(ka) \left(
    \beta_k^\dagger \tilde{\beta}_k + \tilde{\beta}^\dagger_k \beta_k
  \right),
\end{equation} 
where the summations are performed over the reduced Brilluoin Zone
(RBZ), $0 < k \le \pi/a$, to ensure the transformation is
canonical. We see that the measurement operator now consists of two
types of modes, $\beta_k$ and $\tilde{\beta_k}$, which are
superpositions of two momentum states, $k$ and $k - \pi/a$. Note how a
spatial pattern with a period of two sites leads to a basis with two
modes whilst a uniform pattern had only one mode, $c_k$.

\begin{figure}[h]
\captionsetup{justification=justified}
  \centering
  \includegraphics[width=0.5\linewidth]{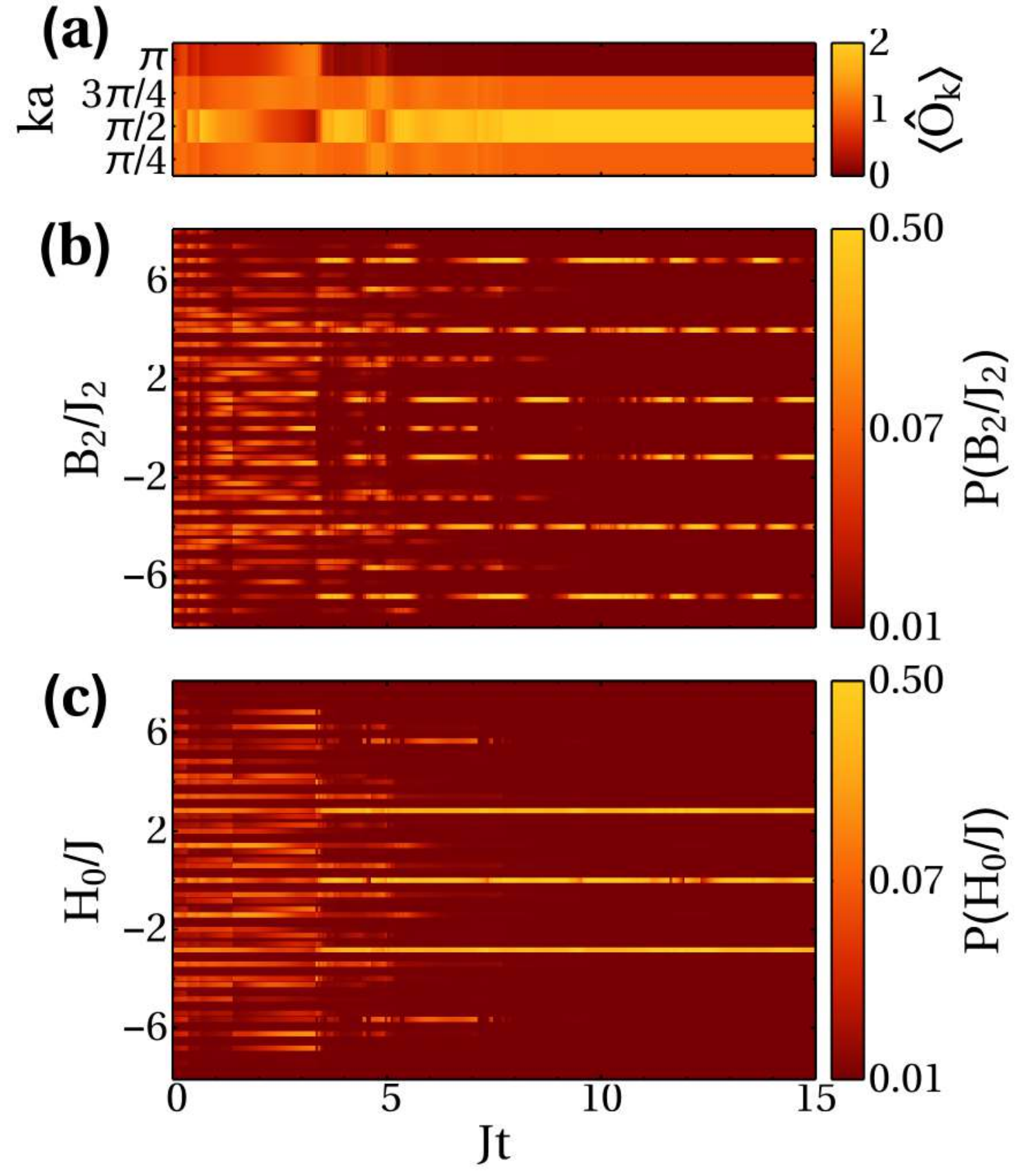}
  \caption{Subspace projections. Projection to a $\mathcal{P}_M$ space
    for four atoms on eight sites with periodic boundary
    conditions. The parameters used are $J=1$, $U=0$,
    $\kappa|C|^2=0.1$, and the initial state was
    $\ketz{0,0,1,1,1,1,0,0}$. (a) The $\langle \hat{O}_k \rangle =
    \langle \n_k + \n_{k - \pi/a} \rangle$ distribution becomes fully
    confined to its subspace at $Jt \approx 8$ indicating the system
    has been projected. (b) Populations of the $\B_2$ eigenspaces. (c)
    Population of the $\hat{H}_0$ eigenspaces. Once the projection is
    achieved at $Jt\approx8$ we can see from (b-c) that the system is
    not in an eigenspace of either $\B_2$ or $\hat{H}_0$, but it
    becomes confined to some subspace. The system has been projected
    onto a subspace, but it is neither that of the measurement
    operator or the Hamiltonian.}
  \label{3-fig:projections}
\end{figure}

Trajectory simulations confirm that there is no steady state.
However, unexpectedly, for each trajectory we observe that the
dynamics always ends up confined to some subspace as seen in
Fig. \ref{3-fig:projections} which is not the same for each
trajectory. In general, this subspace is not an eigenspace of the
measurement operator or the Hamiltonian. In
Fig. \ref{3-fig:projections}(b) it in fact clearly consists of multiple
measurement eigenspaces. This clearly distinguishes it from the
typical projection formalism. It is also not the quantum Zeno effect
which predicts that strong measurement can confine the evolution of a
system as this subspace must be an eigenspace of the measurement
operator \cite{misra1977, Facchi2008, Raimond2010, Raimond2012,
  Signoles2014}. Furthermore, the projection we see in
Fig. \ref{3-fig:projections} occurs for even weak measurement strengths
compared to the Hamiltonian's own evolution, a regime in which the
quantum Zeno effect does not happen. It is also possible to
dissipatively prepare quantum states in an eigenstate of a Hamiltonian
provided it is also a dark state of the jump operator, $a | \Psi
\rangle = 0$, \cite{Diehl2008}. However, this is also clearly not the
case here as the final state in Fig. \ref{3-fig:projections}(c) is not
only not confined to a single measurement operator eigenspace, it also
spans multiple Hamiltonian eigenspaces. Therefore, the dynamics
induced by $a = C\B_2$ projects the system into some subspace, but
since this does not happen via any of the mechanisms described above
it is not immediately obvious what this subspace is. 

A crucial point is that whilst single quantum trajectories might not
have a steady state, for dissipative systems the density matrix will
in general have a steady state which can undergo phase transitions as
the dissipative parameters are varied \cite{Kessler2012}. If we were
to average over many trajectories we would obtain such a steady state
for this system. However, we are concerned with measurement and not
dissipation. Whilst both are open systems, having knowledge of the
measurement outcome from the photodetector means we deal with pure
states that are the outcomes of individual measurements rather than an
ensemble average over all possible outcomes. This can reveal physical
effects which would be lost in a mixed state. The example in
Fig. \ref{3-fig:projections} shows how a single quantum trajectory can
become confined yet never approach any steady state -- measurement and
tunnelling still compete, albeit in a limited subspace. This subspace
will not in general be the same for each experimental trajectory, but
once the subspace is chosen, the system will remain there. This is
analogous to a QND measurement in which a system after the first
projection will remain in its chosen eigenstate, but this eigenstate
is not determined until the first projection takes place. However, if
we were to look at the dissipative steady state (by averaging
expectation values over many quantum trajectories), we would not see
these subspaces at all, because the mixed state is an average over all
possible outcomes, and thus an average over all possible subspaces
which on a single trajectory level are mutually exclusive. Here we consider only individual experimental runs, which are not
steady states themselves, but rather the individual pure state
components of the dissipative steady state that are obtained via the
weak measurement of $\B_2$. 

The mathematical details of such states can be found in Ref. \cite{Kozlowski2017}.

In summary of this section, we have investigated the measurement backaction resulting from
coupling light to an ultracold gas's phase-related observables. We
demonstrated how this can be used to prepare the Hamiltonian
eigenstates even if significant tunnelling is occuring as the
measurement can be engineered to not compete with the system's
dynamics. Furthermore, we have shown that when the observable of the
phase-related quantities does not commute with the Hamiltonian we
still project to a specific subspace of the system that is neither an
eigenspace of the Hamiltonian or the measurement operator. This is in
contrast to quantum Zeno dynamics \cite{misra1977, Facchi2008,
  Raimond2010, Raimond2012, Signoles2014} or dissipative state
preparation \cite{Diehl2008}. This projection is
essentially an extension of the measurement postulate to weak
measurement on dynamical systems where the competition between the two
processes is significant.


\section{Unconventional quantum Zeno dynamics and non-Hermitian evolution}

In Sec. 3.5 we demonstrated that a relatively strong, but not fully projective quantum measurement, can lead to a plethora of novel phenomena such as long-range correlated tunneling, entanglement, and correlations. In this section we provide further details and generalizations of these results.

  In this section, we show that the measurement leads to unconventional quantum Zeno
  dynamics with Raman-like transitions via virtual states outside the
  Zeno subspace. We extend this concept into the realm of
  non-Hermitian dynamics by showing that the stochastic competition
  between measurement and a system's own dynamics can be described by
  a non-Hermitian Hamiltonian. We obtain a solution for ultracold
  bosons in a lattice and show that a dark state of tunnelling is
  achieved as a steady state in which the observable's fluctuations
  are zero and tunnelling is suppressed by destructive matter-wave
  interference.

Frequent measurements can slow the evolution of a quantum system
leading to the quantum Zeno effect \cite{misra1977, Facchi2008} which
has been successfully observed in a variety of systems
\cite{Itano1990, Nagels1997, Kwiat1999, Balzer2000, Streed2006,
  Hosten2006, Bernu2008}. One can also devise measurements with
multi-dimensional projections which lead to quantum Zeno dynamics
where unitary evolution is uninhibited within this degenerate
subspace, i.e.~the Zeno subspace \cite{Facchi2008, Raimond2010,
  Raimond2012, Signoles2014}. In this section we go beyond conventional
quantum Zeno dynamics. By considering the case of measurement near,
but not in, the projective limit the system is still confined to a
Zeno subspace, but intermediate transitions are allowed via virtual
Raman-like processes (or two-photon processes well-known in optics). We show that this can be approximated by a
non-Hermitian Hamiltonian thus extending the notion of quantum Zeno
dynamics into the realm of non-Hermitian quantum mechanics joining the
two paradigms.

Non-Hermitian systems exhibit a variety of rich behaviour, such as
localisation \cite{Hatano1996, Refael2006}, $\mathcal{PT}$ symmetry
\cite{Bender1998,Giorgi2010,Zhang2013}, spatial order
\cite{Otterbach2014}, or novel phase transitions \cite{Lee2014a,
  Lee2014b}. Recent experimental results further motivate the study of
these novel phenomena \cite{Dembowski2001, Choi2010, Ruter2010,
  Barontini2013, Gao2015}. Non-Hermitian Hamiltonians commonly arise
in systems with decay or loss \cite{Dhar2015, Stannigel2014}, limited
by possibilities of controlling dissipation. Additionally, the
non-unitary time evolution is subject to discontinuous jumps applied
whenever decay events are detected requiring the postselection of
trajectories (so that the dynamics would be continuous and deterministic) \cite{Otterbach2014, Lee2014a, Lee2014b}. Here we
consider systems where the non-Hermitian term arises from measurement
and uncover a novel general mechanism that is independent of the
nature of the original Hamiltonian. It does not rely on losses or
postselection of exotic trajectories (but still gives continuous and deterministic dynamics without jumps), thus conceptually simplifying
experimental realizability of such intriguing effects. Furthermore, we
show that the physics can be much more complicated and lead to
dynamics beyond the conventional Hermitian and quantum Zeno dynamics
paradigms.

We will show that,
counter-intuitively, non-Hermitian dynamics causes two competing
processes, tunnelling and measurement, to cooperate to form a dark
state of the atomic dynamics with zero fluctuations in the observed
quantity without the need for an effective cavity potential which is
typically considered in self-organisation of atoms in a cavity.

  \subsection{Theoretical model: deterministic measurement-induced non-Hermitian dynamics without jumps}

\subsubsection{Suppression of coherences in the density matrix by strong
  measurment}

We consider a state described by the density matrix $\hat{\rho}$ whose
isolated behaviour is described by the Hamiltonian $H_0$ and when
measured the jump operator $\c$ is applied to the state at each
detection \cite{Wiseman}. The master equation describing
its time evolution when we ignore the measurement outcomes (i. e. the dissipation) is given by
\begin{equation}
  \dot{\hat{\rho}} = -i [ H_0 , \hat{\rho} ] + \c \hat{\rho} \cd - \frac{1}{2}(
  \cd\c \hat{\rho} + \hat{\rho} \cd\c ).
\end{equation}
We also define $\c = \lambda \opo$ and $H_0 = K \hx$. The exact
definition of $\lambda$ and $K$ is not so important as long as these
coefficients can be considered to be some measure of the relative size
of these operators. They would have to be determined on a case-by-case
basis, because the operators $\c$ and $H_0$ may be unbounded. If
these operators are bounded, one can simply define them such that
$||\opo|| \sim \mathcal{O}(1)$ and $||\hat{h}|| \sim \mathcal{O}(1)$. If
they are unbounded, one possible approach would be to identify the
relevant subspace in whose dynamics we are interested in and scale the
operators such that the eigenvalues of $\opo$ and $\hx$ in this subspace
are $\sim \mathcal{O}(1)$.

We will use projectors $P_m$ which have no effect on states within a
degenerate subspace of $\c$ ($\opo$) with eigenvalue $c_m$ ($o_m$), but
annihilate everything else. For convenience we will also use the
following definition $\hat{\rho}_{mn} = P_m \hat{\rho} P_n$ (these are
submatrices of the density matrix, which in general are not single
matrix elements). Therefore, we can write the master equation that
describes this open system as a set of equations
\begin{gather}
  \dot{\hat{\rho}}_{mn} =  -i K P_m \left[ \h \sum_r \hat{\rho}_{rn}\hx - \sum_r \hat{\rho}_{mr} \hx \right] P_n
 + \lambda^2 \left[ o_m o_n^* - \frac{1}{2} \left( |o_m|^2 + |o_n|^2 \right) \right] \hat{\rho}_{mn},\label{3-eq:master}
\end{gather}
where the first term describes coherent evolution whereas the second
term causes dissipation. 

Firstly, note that for the density submatrices for which $m = n$,
$\hat{\rho}_{mm}$, the dissipative term vanishes and they are thus
decoherence free subspaces and will form the Zeno
subspaces. Interestingly, any state that consists only of these
decoherence free subspaces, i.e.~
$\hat{\rho} = \sum_m \hat{\rho}_{mm}$, and that commutes with the
Hamiltonian, $[\hat{\rho}, \hat{H}_0] = 0$, will be a steady state.
This can be seen by substituting this ansatz into
Eq. \eqref{3-eq:master} which yields $\dot{\hat{\rho}}_{mn} = 0$ for all
$m$ and $n$. These states can be prepared dissipatively using known
techniques \cite{Diehl2008}, but it is not required that the state be
a dark state of the dissipative operator as is usually the case.

Secondly, we consider a large detection rate, $\lambda^2 \gg K$, for
which the coherences, i.e.~ the density submatrices $\hat{\rho}_{mn}$
for which $m \ne n$, will be heavily suppressed by
dissipation. Therefore, we can adiabatically eliminate these
cross-terms by setting $\dot{\hat{\rho}}_{mn} = 0$, to get
\begin{equation}
\label{3-eq:intermediate}
\hat{\rho}_{mn} = \frac{K}{\lambda^2} \frac{i P_m \left[ \hx \sum_r \hat{\rho}_{rn} - \sum_r \hat{\rho}_{mr} \hx \right] P_n } {o_m o_n^* - \frac{1}{2} \left( |o_m|^2 + |o_n|^2 \right)},
\end{equation}
which tells us that they are of order $K/\lambda^2 \ll 1$. One can
easily recover the projective Zeno limit by considering
$\lambda \rightarrow \infty$ when all the subspaces completely
decouple. However, it is crucial that we only consider
$\lambda^2 \gg K$, but not infinite. If the subspaces do not decouple
completely, then transitions within a single subspace can occur via
other subspaces in a manner similar to Raman transitions. In Raman
transitions population is transferred between two states via a third,
virtual, state that remains empty throughout the process. By avoiding
the infinitely projective Zeno limit we open the option for such
processes to happen in our system where transitions within a single
Zeno subspace occur via a second, different, Zeno subspace even though
the occupation of the intermediate states will remian negligible at
all times.

In general, a density matrix can have all of its $m = n$ submatrices,
$\hat{\rho}_{mm}$, be non-zero and non-negligible even when the
coherences are small. However, for a pure state this would not be
possible. To understand this, consider the state $| \Psi \rangle$ and
take it to span exactly two distinct subspaces $P_a$ and $P_b$
($a \ne b$). This wavefunction can also be written as
$| \Psi \rangle = P_a | \Psi \rangle + P_b | \Psi \rangle$. The
corresponding density matrix is thus given by
\begin{gather}
  \hat{\rho}_\Psi =  P_a | \Psi \rangle \langle \Psi | P_a + P_a | \Psi
                      \rangle \langle \Psi | P_b 
   + P_b | \Psi \rangle \langle \Psi | P_a +
       P_b | \Psi \rangle \langle \Psi | P_b.
\end{gather}
If the wavefunction has significant components in both subspaces then
in general the density matrix will not have negligible coherences,
$\hat{\rho}_{ab} = P_a | \Psi \rangle \langle \Psi | P_b$. Therefore,
a density matrix with small cross-terms between different Zeno
subspaces can only be composed of pure states that each lie
predominantly within a single subspace.

Therefore, in order for the coherences to be of order $K/\lambda^2$ we
would require the wavefunction components to satisfy
$P_a | \Psi \rangle \approx \mathcal{O}(1)$ and
$P_b | \Psi \rangle \approx \mathcal{O}(K/\lambda^2)$. This in turn
implies that the population of the states outside of the dominant
subspace (and thus the submatrix $\hat{\rho}_{bb}$) will be of order
$\langle \Psi | P_b^2 | \Psi \rangle \approx
\mathcal{O}(K^2/\lambda^4)$. Therefore, these pure states cannot exist
in a meaningful coherent superposition in this limit. This means that
a density matrix that spans multiple Zeno subspaces has only classical
uncertainty about which subspace is currently occupied as opposed to
the uncertainty due to a quantum superposition.

 \subsubsection{Determining the Zeno subspace}

We now consider how to estimate, which Zeno
subspace our system is in. Since the density matrix cross-terms are small we know
\emph{a priori} that the individual wavefunctions comprising the
density matrix mixture will not be coherent superpositions of
different Zeno subspaces. Therefore, each individual experiment will
at any time be predominantly in a single Zeno subspace with small
cross-terms and negligible occupations in the other subspaces. With no
measurement record our density matrix would be a mixture of all these
possibilities. However, we can try and determine the Zeno subspace
around which the state evolves in a single experiment from the number
of detections, $m$, in time $t$.

The detection distribution on time-scales shorter than dissipation (so
we can approximate as if we were in a fully Zeno regime) can be
obtained by integrating over the detection times \cite{MekhovPRA2009}
to get
\begin{equation}
  P(m,t) = \sum_n \frac{[|c_n|^2 t]^m} {m!} e^{-|c_n|^2 t} \mathrm{Tr} (\rho_{nn}).
\end{equation}
For a state that is predominantly in one Zeno subspace, the
distribution will be approximately Poissonian (up to
$\mathcal{O}(K^2 / \lambda^4$)). Therefore, in a single experiment we will measure
$m = |c_0|^2t \pm \sqrt{|c_0|^2t}$ detections. (Note, we have assumed
$|c_0|^2 t$ is large enough to approximate the distribution as
normal. This is not necessary, we simply use it here to not have to
worry about the asymmetry in the deviation around the mean value). The
uncertainty does not come from the fact that $\lambda$ is not
infinite. The jumps are random events with a Poisson
distribution. Therefore, even in the full projective limit we will not
observe the same detection trajectory in each experiment even though
the system evolves in exactly the same way and remains in a perfectly
pure state.

If the basis of $\c$ is continuous (e.g. free particle position or
momentum) then the deviation around the mean will be our upper bound
on the deviation of the system from a pure state evolving around a
single Zeno subspace. However, continuous systems are beyond the scope
of this work and we will confine ourselves to discrete systems. Though
it is important to remember that continuous systems can be treated
this way, but the error estimate (and thus the mixedness of the state)
will be different.

For a discrete system it is easier to exclude all possibilities except
for one. The error in our estimate of $|c_0|^2$ in a single experiment
decreases as $1/\sqrt{t}$ and thus it can take a long time to
confidently determine $|c_0|^2$ to a sufficient precision this
way. However, since we know that it can only take one of the possible
values from the set $\{|c_n|^2 \}$ it is much easier to exclude all
the other values.

In an experiment we can use Bayes' theorem to infer the state of our
system as follows
\begin{equation}
	p(c_n = c_0 | m) = \frac{ p(m | c_n = c_0) p(c_n = c_0) }{ p(m) },
\end{equation}
where $p(x)$ denotes the probability of the discrete event $x$ and
$p(x|y)$ the conditional probability of $x$ given $y$. We know that
$p(m | c_n = c_0)$ is simply given by a Poisson distribution with mean
$|c_0|^2 t$. $p(m)$ is just a normalising factor and $p(c_n = c_0)$ is
our \emph{a priori} knowledge of the state. Therefore, one can get the
probability of being in the right Zeno subspace from
\begin{align}
  p(c_n & = c_0 | m) = \frac{ p_0(c_n = c_0) \frac{ \left( |c_0|^2 t \right)^{2m} } {m!} e^{-|c_0|^2 t}} {\sum_n p_0(c_n) \frac{ \left( |c_n|^2 t \right)^{2m} } {m!} e^{-|c_n|^2 t}} \nonumber \\
	& = p_0(c_n = c_0) \left[ \sum_n p_0(c_n) \left( \frac{ |c_n|^2 } { |c_0|^2 } \right)^{2m} e^{\left( |c_0|^2 - |c_n|^2 \right) t} \right]^{-1},
\end{align}
where $p_0$ denotes probabilities at $t = 0$. In a real experiment one
could prepare the initial state to be close to the Zeno subspace of
interest and thus it would be easier to deduce the state. Furthermore,
in the middle of an experiment if we have already established the Zeno
subspace this will be reflected in these \emph{a priori} probabilities
again making it easier to infer the correct subspace. However, we will
consider the worst case scenario which might be useful if we don't
know the initial state or if the Zeno subspace changes during the
experiment, a uniform $p_0(c_n)$.

This probability is a rather complicated function as $m$ is a
stochastic quantity that also increases with $t$. We want it to be as
close to 1 as possible. In order to devise an appropriate condition
for this we note that in the first line all terms in the denominator
are Poisson distributions of $m$. Therefore, if the mean values
$|c_n|^2 t$ are sufficiently spaced out, only one of the terms in the
sum will be significant for a given $m$ and if this happens to be the
one that corresponds to $c_0$ we get a probability close to
unity. Therefore, we set the condition such that it is highly unlikely
that our measured $m$ could be produced by two different distributions
\begin{align}
  \sqrt{|c_0|^2 t} \ll ||c_0|^2 - |c_n|^2|| t, \forall n \ne 0 \\
  \sqrt{|c_n|^2 t} \ll ||c_0|^2 - |c_n|^2|| t, \forall n \ne 0
\end{align}
The left-hand side is the standard deviation of $m$ if the system was in subspace
$P_0$ or $P_n$. The right-hand side is the difference in the mean detections
between the subspace $n$ and the one we are interested in. The
condition becomes more strict if the subspaces become less
distinguishable as it becomes harder to confidently determine the
correct state. Once again, using $\c = \lambda \hat{o}$ where
$\hat{o} \sim \mathcal{O}(1)$ we get
\begin{equation}
  t \gg \frac{1}{\lambda^2} \frac{|o_{0,n}|^2} {(|o_0|^2 - |o_n|^2|)^2}.
\end{equation}
Since detections happen on average at an average rate of order
$\lambda^2$ we only need to wait for a few detections to satisfy this
condition. Therefore, we see that even in the worst case scenario of
complete ignorance of the state of the system we can very easily
determine the correct subspace. Once it is established for the first
time, the \emph{a priori} information can be updated and it will
become even easier to monitor the system.

However, it is important to note that physically once the quantum
jumps deviate too much from the mean value the system is more likely
to change the Zeno subspace (due to measurement backaction) and the
detection rate will visibly change. Therefore, if we observe a
consistent detection rate it is extremely unlikely that it can be
produced by two different Zeno subspaces so in fact it is even easier
to determine the correct state, but the above estimate serves as a
good lower bound on the necessary detection time.

\subsubsection{Quantum measurement vs. dissipation}

This is where quantum measurement deviates from dissipation. If we
have access to a measurement record we can infer which Zeno subspace
is occupied, because we know that only one of them can be occupied at
any time. The time needed to determine the correct state is 
\begin{equation}
  t \gg \frac{1}{\lambda^2} \frac{|o_n|^2}{\left( |o_m|^2 - |o_n|^2
    \right)^2 } \quad \forall m,n \quad m \ne n,
\end{equation}
which is faster than the system's internal dynamics as long as the
eigenvalues are distinguishable enough. Thanks to measurement we can
make another approximation. If we observe a number of detections
consistent with the subspace $P_m = P_0$ we can set
$\hat{\rho}_{mn} \approx 0$ for all cases when both $m \ne 0$ and
$n \ne 0$ leaving our density matrix in the form
\begin{equation}
  \label{3-eq:approxrho}
  \hat{\rho} = \hat{\rho}_{00} + \sum_{r\ne0} (\hat{\rho}_{0r} +
  \hat{\rho}_{r0}).
\end{equation}
We can do this, because the other states are inconsistent with the
measurement record. We know from the previous subsection that the system
must lie predominantly in only one of the Zeno subspaces and when that
is the case, $\hat{\rho}_{0r} \approx \mathcal{O}(K/\lambda^2)$ and
for $m \ne 0$ and $n \ne 0$ we have
$\hat{\rho}_{mn} \approx \mathcal{O}(K^2/\lambda^4)$. Therefore, this
amounts to keeping first order terms in $K/\lambda^2$ in our
approximation.

This is a crucial step as all $\hat{\rho}_{mm}$ matrices are
decoherence free subspaces and thus they can all coexist in a mixed
state decreasing the purity of the system without
measurement. Physically, this means we exclude trajectories in which
the Zeno subspace has changed (measurement isn't fully projective). By
substituting Eq. \eqref{3-eq:intermediate} into Eq. \eqref{3-eq:master} we
see that this happens at a rate of $K^2 / \lambda^2$. However, since
the two measurement outcomes cannot coexist any transition between
them happens in discrete transitions (which we know about from the
change in the detection rate as each Zeno subspace will correspond to
a different rate) and not as continuous coherent evolution. Therefore,
we can postselect in a manner similar to Refs. \cite{Otterbach2014,
  Lee2014a, Lee2014b}, but our requirements are significantly more
relaxed -- we do not require a specific single trajectory, only that it
remains within a Zeno subspace. Furthermore, upon reaching a steady
state, these transitions become impossible as the coherences
vanish. 

This approximation is analogous to optical Raman transitions
where the population of the intermediate state is neglected. Here, we can
make a similar approximation and neglect all but one Zeno subspace
thanks to the additional knowledge we gain from knowing the
measurement outcomes.

\subsubsection{The non-Hermitian Hamiltonian}

Rewriting the master equation using $\c = c_0 + \delta \c$, where
$c_0$ is the eigenvalue corresponding to the eigenspace defined by the
projector $P_0$ which we used to obtain the density matrix in
Eq. \eqref{3-eq:approxrho}, we get
\begin{equation}
\label{3-eq:finalrho}
\dot{\hat{\rho}} = -i \left( H_\text{eff} \hat{\rho} - \hat{\rho} H_\text{eff}^\dagger \right) + \delta \c \hat{\rho} \delta \cd,
\end{equation}
\begin{equation}
\label{3-eq:Ham}
H_\text{eff} = H_0 + i \left( c_0^*\c - \frac{|c_0|^2}{2} - \frac{\cd\c}{2} \right).
\end{equation}
The first term in Eq. \eqref{3-eq:finalrho} describes coherent evolution
due to the non-Hermitian Hamiltonian $H_\text{eff}$ and the second
term is decoherence due to our ignorance of measurement outcomes. When
we substitute our approximation of the density matrix
$\hat{\rho} = \hat{\rho}_{00} + \sum_{r\ne0} (\hat{\rho}_{0r} +
\hat{\rho}_{r0})$ into Eq. \eqref{3-eq:finalrho}, the last term
vanishes, $\delta \c \hat{\rho} \delta \cd = 0$. This happens, because
$\delta \c P_0 \hat{\rho} = \hat{\rho} P_0 \delta \c^\dagger = 0$. The
projector annihilates all states except for those with eigenvalue
$c_0$ and so the operator $\delta \c = \c - c_0$ will always evaluate
to $\delta \c = c_0 - c_0 = 0$. Recall that we defined
$\hat{\rho}_{mn} = P_m \hat{\rho} P_n$ which means that every term in
our approximate density matrix contains the projector $P_0$. However,
it is important to note that this argument does not apply to other
second order terms in the master equation, because some terms only
have the projector $P_0$ applied from one side,
e.g.~$\hat{\rho}_{0m}$. The term $\delta \c \hat{\rho} \delta \cd$
applies the fluctuation operator from both sides so it does not matter
in this case, but it becomes relevant for terms such as
$\delta \cd \delta \c \hat{\rho}$.

It is important to note that this term does not automatically vanish,
but when the explicit form of our approximate density matrix is
inserted, it is in fact zero. Therefore, we can omit this term using
the information we gained from measurement, but keep other second
order terms, such as $\delta \cd \delta \c \rho$ in the Hamiltonian
which are the origin of other second-order dynamics. This could not be
the case in a dissipative system.

Ultimately we find that a system under continuous measurement for
which $\lambda^2 \gg K$ in the Zeno subspace $P_0$ is described by the
deterministic non-Hermitian Hamiltonian $H_\text{eff}$ in
Eq. \eqref{3-eq:Ham} and thus obeys the following Schr\"{o}dinger
equation
\begin{equation}
 i \frac{\text{d} | \Psi \rangle}{\text{d}t} = \left[H_0 + i \left(
      c_0^*\c - \frac{|c_0|^2}{2} - \frac{\cd\c}{2} \right) \right] |
  \Psi \rangle.
\end{equation}
Of the three terms in the parentheses the first two represent the
effects of quantum jumps due to detections (which one can think of as
`reference frame' shifts between different degenerate eigenspaces) and
the last term is the non-Hermitian decay due to information gain from
no detections. It is important to emphasize that even though we
obtained a deterministic equation, we have not neglected the
stochastic nature of the detection events. The detection trajectory
seen in an experiment will have fluctuations around the mean
determined by the Zeno subspace, but there simply are many possible
measurement records with the same outcome. This is just like the fully
projective Zeno limit where the system remains perfectly pure in one
of the possible projections, but the detections remain randomly
distributed in time.

One might then be concerned that purity is preserved since we might be
averaging over many trajectories within this Zeno subspace. We have
neglected the small terms $\hat{\rho}_{m,n}$ ($m,n \ne 0$) which are
$\mathcal{O}(K^2/\lambda^4)$ and thus they are not correctly accounted
for by the approximation. This means that we have an
$\mathcal{O}(K^2/\lambda^4)$ error in our density matrix and the
purity given by
\begin{equation}
  \text{Tr}(\hat{\rho}^2) = \text{Tr}(\hat{\rho}^2_{00} + \sum_{m \ne
    0} \hat{\rho}_{0m}\hat{\rho}_{m0}) + \text{Tr}(\sum_{m,n\ne0}
  \hat{\rho}_{mn} \hat{\rho}_{nm}),
\end{equation}
where the second term contains the terms not accounted for by our
approximation and thus introduces an $\mathcal{O}(K^4/\lambda^8)$
error. Therefore, this discrepancy is negligible in our
approximation. The pure state predicted by $H_\text{eff}$ is only
an approximation, albeit a good one, and the real state will be mixed
to a small extent. Whilst perfect purity within the Zeno subspace
$\hat{\rho}_{00}$ is expected due to the measurement's strong
decoupling effect, the nearly perfect purity when transitions outside
the Zeno subspace are included is a non-trivial result. Similarly, in
Raman transitions the population of the neglected intermediate state is
also non-zero, but negligible. Furthermore, this equation does not
actually require the adiabatic elimination used in
Eq. \eqref{3-eq:intermediate} (note that we only used it to convince
ourselves that the coherences are small) and such situations may be
considered provided all approximations remain valid. In a similar way
the limit of linear optics is derived from the physics of a two-level
nonlinear medium, when the population of the upper state is neglected
and the adiabatic elimination of coherences is not required.
  
 We will now consider particular physical examples of this general theory.

  \subsection{Non-Hermitian dynamics in ultracold gases}

\subsubsection{Theoretical model}

We now focus on the setup of ultracold bosons in a lattice inside a cavity which
selects and enhances light scattered at a particular angle. One of its key advantages is the flexibility of
engineering the measurement $\c$ and the possibility of coupling to
both density and inter-site interference. The
global nature of such measurement is also in contrast to spontaneous
emission \cite{PichlerDaley2010, Sarkar2014}, local \cite{RempeScience2008,
  Hartmann2012, Vidanovic2014, Bernier2014, Daley2014} and
fixed-range addressing \cite{LesanovskyPRL2012, LesanovskyPRB2014} which are
typically considered in dissipative systems. Furthermore, it provides
the opportunity to extend quantum measurement and quantum Zeno
dynamics beyond single atoms into strongly correlated many-body
systems.

As in previous sections, the isolated system Hamiltonian $H_0$ is described by the Bose--Hubbard Model (BHM). The quantum
jump operator is given by $\c = \sqrt{2 \kappa} a$, where $\kappa$ is
the cavity relaxation rate, $a = C \hat{D}$ is the annihilation
operator of a photon in the cavity mode, $C$ is the coefficient of
Rayleigh scattering into the cavity. Additionally, we have the
necessary condition to be in the quantum Zeno regime
$\gamma/ J \gg 1$, where $\gamma = \kappa |C|^2$.

We will now consider the simplest case of global multi-site
measurement of the form $\hat{D} = \hat{N}_K = \sum_i^K \n_i$, where
the sum is over $K$ illuminated sites. Physically, this can be
realised by collecting the light scattered into a diffraction maximum. The effective Hamiltonian becomes
\begin{equation}
\label{3-eq:nHH2}
\hat{H}_\mathrm{eff} = \hat{H}_0 - i \gamma \left(  \delta \hat{N}_K \right)^2,
\end{equation}
where $ \delta \hat{N}_K = \hat{N}_K - N^0_K$ and $N^0_K$ is a
subspace eigenvalue. It is now obvious that continuous measurement
squeezes the fluctuations in the measured quantity, as expected, and
that the only competing process is the system's own dynamics.

In this case, if we adiabatically eliminate the density matrix
cross-terms and substitute Eq. \eqref{3-eq:intermediate} into
Eq. \eqref{3-eq:master} for this system, we obtain an effective
Hamiltonian within the Zeno subspace defined by $N_K$
\begin{equation}
  H_\varphi = P_\varphi \left[ H_0 - i \frac{J^2}{\gamma} \sum_{\substack{\langle i \in \varphi, j \in \varphi^\prime \rangle \\ \langle k \in \varphi^\prime, l \in \varphi \rangle}} b^\dagger_i b_j b^\dagger_k b_l \right] P_\varphi,
\end{equation}
where $\varphi = \{N_K\}$ denotes the set of states with $N_K$ atoms
in the illuminated area, $\varphi^\prime = \{N_K \pm 1\}$ denotes the
set of intermediate states and $P_\varphi$ is the projector onto
$\varphi$. We focus on the case when the second term is not only
significant, but also leads to dynamics within $\varphi$ that are not
allowed by conventional quantum Zeno dynamics accounted for by the
first term. The second term represents second-order transitions via
other subspaces which act as intermediate states much like virtual
states in optical Raman transitions. This is in contrast to the
conventional understanding of the Zeno dynamics for infinitely
frequent projective measurements (corresponding to
$\gamma \rightarrow \infty$) where such processes are forbidden
\cite{Facchi2008}. Thus, it is the weak quantum measurement that
effectively couples the states.
  
  \subsubsection{Small system example}

To get clear physical insight, we initially consider three atoms in
three sites and choose our measurement operator such that
$\hat{D} = \n_2$, i.e.~only the middle site is subject to measurement,
and the Zeno subspace defined by $n_2 = 1$. Such an illumination
pattern can be achieved with global addressing by crossing two beams
and placing the nodes at the odd sites and the antinodes at even
sites. This means that $P_\varphi H_0 P_\varphi = 0$. However, the
first and third sites are connected via the second term. Diagonalising
the Hamiltonian reveals that out of its ten eigenvalues all but three
have a significant negative imaginary component of the order $\gamma$
which means that the corresponding eigenstates decay on a time scale
of a single quantum jump and thus quickly become negligible. The three
remaining eigenvectors are dominated by the linear superpositions of
the three Fock states $|2,1,0 \rangle$, $|1, 1, 1 \rangle$, and
$|0,1,2 \rangle$. Whilst it is not surprising that these components
are the only ones that remain as they are the only ones that actually
lie in the Zeno subspace $n_2 = 1$, it is impossible to solve the full
dynamics by just considering these Fock states alone as they are not
coupled to each other in $\hat{H}_0$. The components lying outside of
the Zeno subspace have to be included to allow intermediate steps to
occur via states that do not belong in this subspace, much like
virtual states in optical Raman transitions.

\begin{figure}[h]
\captionsetup{justification=justified}
  \centering
	\includegraphics[width=0.7\linewidth]{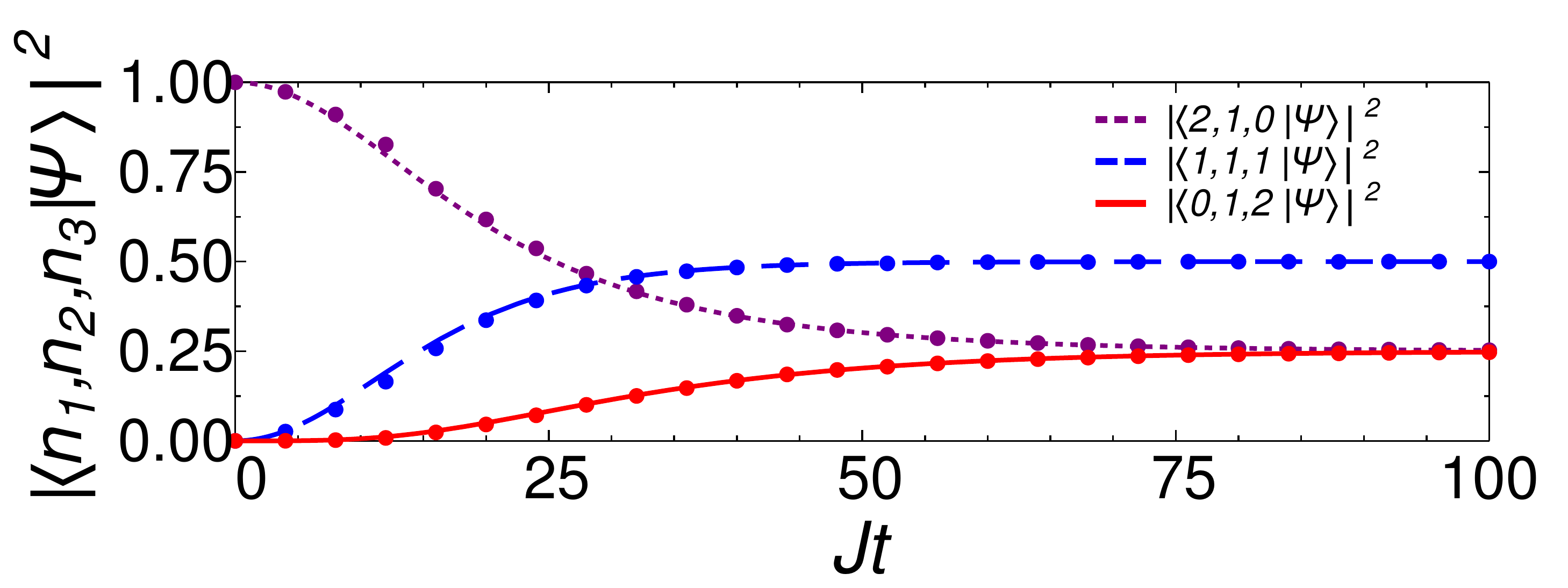}
	\caption{Populations of the Fock states in the
          Zeno subspace for $\gamma/J = 100$ and initial state
          $| 2,1,0 \rangle$. It is clear that quantum Zeno dynamics
          occurs via Raman-like processes even though none of these
          states are connected in $\hat{H}_0$. The dynamics occurs via
          virtual intermediate states outside the Zeno subspace. The
          system also tends to a steady state which minimises
          tunnelling effectively suppressing fluctuations. The lines
          are solutions to the non-Hermitian Hamiltonian, and the dots
          are points from a stochastic trajectory
          calculation.\label{3-fig:comp}}
\end{figure}

An approximate solution for $U=0$ can be written for the
$\{|2,1,0 \rangle, |1,1,1 \rangle, |0,1,2 \rangle\}$ subspace by
multiplying each eigenvector with its corresponding time evolution
\begin{equation}
  | \Psi(t) \rangle \propto \left( \begin{array}{c} 
  z_1 + \sqrt{2} z_2 e^{-6 J^2 t / \gamma} + z_3 e^{-12 J^2 t / \gamma} \\
  -\sqrt{2} \left(z_1 - z_3 e^{-12 J^2 t / \gamma} \right) \\ 
  z_1 - \sqrt{2} z_2 e^{-6 J^2 t / \gamma} + z_3 e^{-12 J^2 t /
                                     \gamma} \\ 
                                   \end{array} 
                                 \right), \nonumber
\end{equation}
where $z_i$ denote the overlap between the eigenvectors and the
initial state, $z_i = \langle v_i | \Psi (0) \rangle$, with
$| v_1 \rangle = (1, -\sqrt{2}, 1)/2$,
$| v_2 \rangle = (1, 0, -1)/\sqrt{2}$, and
$| v_3 \rangle = (1, \sqrt{2}, 1)/2$. The steady state as
$t \rightarrow \infty$ is given by
$| v_1 \rangle = (1, -\sqrt{2}, 1)/2$. This solution is illustrated in
Fig. \ref{3-fig:comp} which clearly demonstrates dynamics beyond the
canonical understanding of quantum Zeno dynamics as tunnelling occurs
between states coupled via a different Zeno subspace.

\subsubsection{Steady state of non-Hermitian dynamics}

A distinctive difference between BHM ground states and the final
steady state,
$[|2,1,0 \rangle - \sqrt{2} |1,1,1\rangle + |0,1,2\rangle]$, is that
its components are not in phase. Squeezing due to measurement
naturally competes with inter-site tunnelling which tends to spread
the atoms. However, from Eq. \eqref{3-eq:nHH2} we see the final state
will always be the eigenvector with the smallest fluctuations as it
will have an eigenvalue with the largest imaginary component. This
naturally corresponds to the state where tunnelling between Zeno
subspaces (here between every site) is minimised by destructive
matter-wave interference, i.e.~the tunnelling dark state defined by
$\hat{T} |\Psi \rangle = 0$, where
$\hat{T} = \sum_{\langle i, j \rangle} \bd_i b_j$. This is simply the
physical interpretation of the steady states we predicted for
\mbox{Eq. \eqref{3-eq:master}}. Crucially, this state can only be
reached if the dynamics is not fully suppressed by measurement and
thus, counter-intuitively, the atomic dynamics cooperate with
measurement to suppress itself by destructive interference. Therefore,
this effect is beyond the scope of traditional quantum Zeno dynamics
and presents a new perspective on the competition between a system's
short-range dynamics and global measurement backaction.

We now consider a one-dimensional lattice with $M$ sites so we extend
the measurement to $\hat{D} = \N_\text{even}$ where every even site is
illuminated (obtained by crossing two beams such that the nodes
coincide with odd sites and antinodes with even sites). The wavefunction in a Zeno
subspace must be an eigenstate of $\c$ and we combine this with the
requirement for it to be in the dark state of the tunnelling operator
(eigenstate of $H_0$ for $U = 0$) to derive the steady state. These
two conditions in momentum space are
\begin{equation}
  \hat{T} | \Psi \rangle = \sum_{\text{RBZ}} \left[ \bd_k b_k - \bd_{q} b_{q} \right] \cos(ka) |\Psi \rangle = 0, \nonumber
\end{equation}
\begin{equation}
  \Delta \N |\Psi \rangle = \sum_{\text{RBZ}} \left[ \bd_k b_{-q} + \bd_{-q} b_k \right] | \Psi \rangle= \Delta N |\Psi \rangle, \nonumber
\end{equation}
where $b_k = \frac{1}{\sqrt{M}} \sum_j e^{i k j a} b_j$,
$\Delta \hat{N} = \hat{D} - N/2$, $q = \pi/a - k$, $a$ is the lattice
spacing, $N$ the total atom number, and we perform summations over the
reduced Brillouin zone (RBZ), $-\pi/2a < k \le \pi/2a$, as the
symmetries of the system are clearer this way. Now we define
\begin{equation}
\hat{\alpha}_k^\dagger = \bd_k \bd_q - \bd_{-k} \bd_{-q},
\end{equation}
\begin{equation}
\hat{\beta}_\varphi^\dagger = \bd_{\pi/2a} + \varphi \bd_{-\pi/2a},
\end{equation}
where $\varphi = \Delta N / | \Delta N |$, which create the smallest
possible states that satisfy the two equations for $\Delta N = 0$ and
$\Delta N \ne 0$ respectively. Therefore, by noting that
$\left[ \hat{T}, \hat{\alpha}_k^\dagger \right] = \left[ \Delta \N,
  \hat{\alpha}_k^\dagger \right] = \left[ \hat{T},
  \hat{\beta}_\varphi^\dagger \right] = 0$ and
$\left[ \Delta \N, \hat{\beta}_\varphi^\dagger \right] = \varphi
\hat{\beta}_\varphi^\dagger$ we can now write the equation for the
$N$-particle steady state
\begin{equation}
\label{3-eq:ss}
| \Psi \rangle \propto \left[ \prod_{i=1}^{(N - |\Delta N|)/2} \left( \sum_{k = 0}^{\pi/2a} \phi_{i,k} \hat{\alpha}_k^\dagger \right) \right] \left( \hat{\beta}_\varphi^\dagger \right)^{| \Delta N |} | 0 \rangle, \nonumber
\end{equation}
where $\phi_{i,k}$ are coefficients that depend on the trajectory
taken to reach this state and $|0 \rangle$ is the vacuum state defined
by $b_k |0 \rangle = 0$. Since this is a dark state (an eigenstate of
$H_0$) of the atomic dynamics, this state will remain stationary even
with measurement switched-off. Interestingly, this state is very
different from the ground states of the BHM, it is even orthogonal to
the superfluid state, and thus it cannot be obtained by cooling or
projecting from an initial ground state. The combination of tunnelling
with measurement is necessary.

\begin{figure}[h]
\captionsetup{justification=justified}
  \centering
	\includegraphics[width=0.7\linewidth]{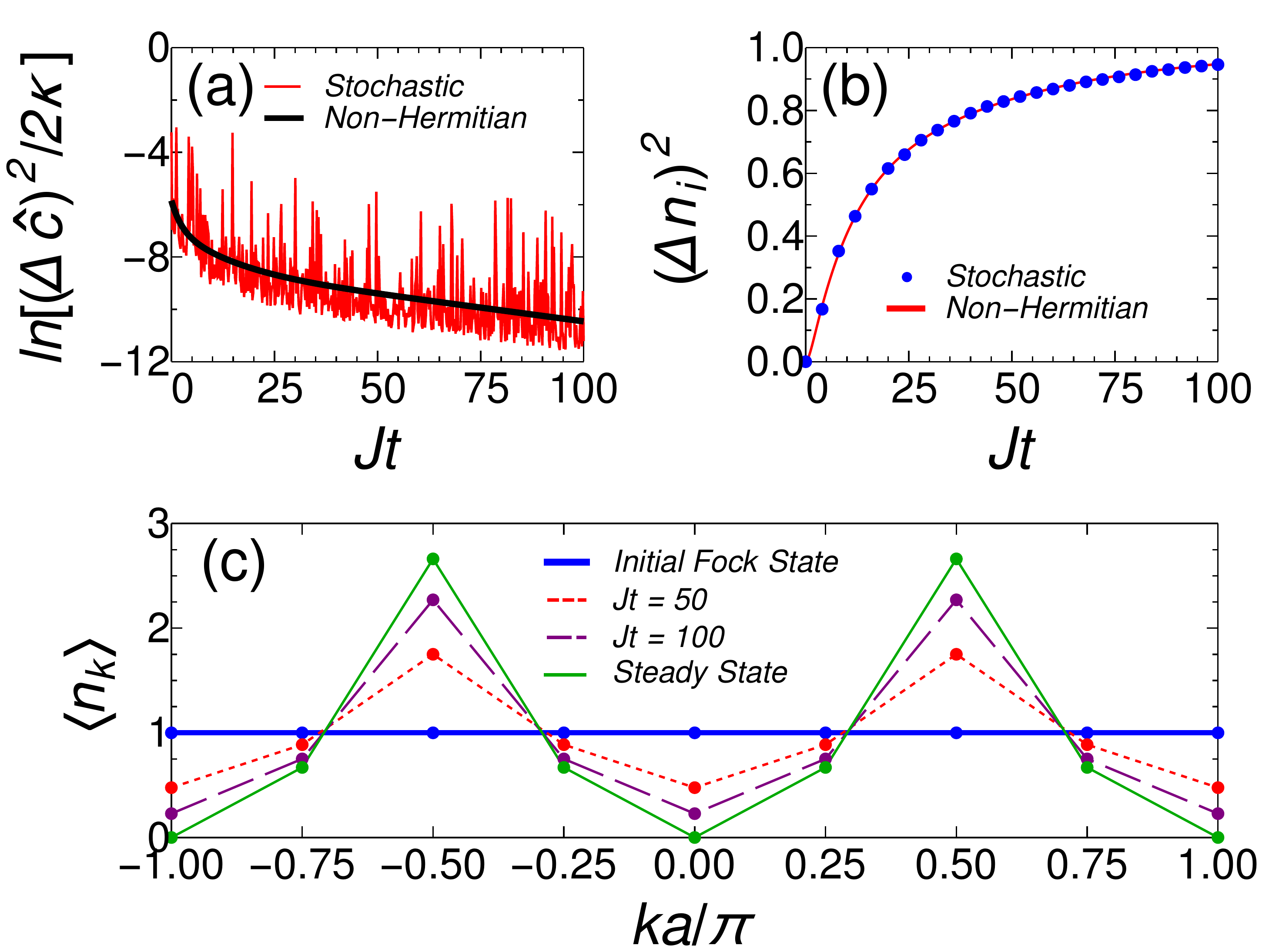}
	\caption{A trajectory simulation for eight
          atoms in eight sites, initially in
          $|1,1,1,1,1,1,1,1 \rangle$, with periodic boundary
          conditions and $\gamma/J = 100$. (a) The fluctuations in
          $\c$ where the stochastic nature of the process is clearly
          visible on a single trajectory level. However, the general
          trend is captured by the non-Hermitian Hamiltonian. (b) The
          local density variance. Whilst the fluctuations in the
          global measurement operator decrease, the fluctuations in
          local density increase due to tunnelling via states outside
          the Zeno subspace. (c) The momentum distribution. The
          initial Fock state has a flat distribution which with time
          approaches the steady state distribution of two identical
          and symmetric distributions centred at $k = \pi/2a$ and
          $k = -\pi/2a$.\label{3-fig:steady}}
\end{figure}

In order to prepare the steady state one has to run the experiment and
wait until the photocount rate remains constant for a sufficiently
long time. Such a trajectory is illustrated in Fig. \ref{3-fig:steady}
and compared to a deterministic trajectory calculated using the
non-Hermitian Hamiltonian. It is easy to see from
Fig. \ref{3-fig:steady}(a) how the stochastic fluctuations around the
mean value of the observable have no effect on the general behaviour
of the system in the strong measurement regime. By discarding these
fluctuations we no longer describe a pure state, but we showed how
this only leads to a negligible error. Fig. \ref{3-fig:steady}(b) shows
the local density variance in the lattice. Not only does it grow
showing evidence of tunnelling between illuminated and non-illuminated
sites, but it grows to significant values. This is in contrast to
conventional quantum Zeno dynamics where no tunnelling would be
allowed at all. Finally, Fig. \ref{3-fig:steady}(c) shows the momentum
distribution of the trajectory. We can clearly see that it deviates
significantly from the initial flat distribution of the Fock
state. Furthermore, the steady state does not have any atoms in the
$k=0$ state and thus is orthogonal to the superfluid state as
discussed.

To obtain a state with a specific value of $\Delta N$ postselection
may be necessary, but otherwise it is not needed.  The process can be
optimised by feedback control since the state is monitored at all
times \cite{Ivanov2014}. Furthermore, the form of the measurement
operator is very flexible and it can easily be engineered by the
geometry of the optical setup  which
can be used to design a state with desired properties.

 In summary of this section, we have presented a new perspective on quantum Zeno
dynamics when the measurement isn't fully projective. By using the
fact that the system is strongly confined to a specific measurement
eigenspace we have derived an effective non-Hermitian Hamiltonian. In
contrast to previous works, it is independent of the underlying system
and there is no need to postselect for a particular exotic trajectory
\cite{Lee2014a, Lee2014b}. Using the BHM as an example we have shown
that whilst the system remains in its Zeno subspace, it will exhibit
Raman-like transitions within this subspace which would be forbidden in
the canonical fully projective limit. Finally, we have shown that the
system will always tend towards the eigenstate of the Hamiltonian with
the best squeezing in the measured quantity and the atomic dynamics,
which normally tend to spread the distribution, cooperates with
measurement to produce a state in which tunnelling is suppressed by
destructive matter-wave interference. A dark state of the tunnelling
operator will have zero fluctuations and we provided an expression for
the steady state which is significantly different from the ground
state of the Hamiltonian. This is in contrast to previous works on
dissipative state preparation where the steady state had to be a dark
state of the measurement operator instead \cite{Diehl2008}.


\section{Measurement-induced antiferromagnetic order and density modulations in Fermi gases}

In this section, we show that quantum backaction of weak measurement can be used for tailoring long-range correlations of ultracold fermions, realizing quantum states with spatial modulations of the density and magnetization, thus overcoming usual requirement for a strong interatomic interactions. We propose detection schemes for implementing antiferromagnetic states and density waves. We demonstrate that such long-range correlations cannot be realized with local addressing, and they are a consequence of the competition between global but spatially structured backaction of weak quantum measurement and unitary dynamics of fermions.

The study of quantum gases trapped in optical lattice potentials is a truly multidisciplinary field~\cite{Lewenstein}. The experimental realization of toy Hamiltonians like the Hubbard model opened the opportunity of studying intriguing many-body effects in Fermi systems such as high temperature superconductivity and quantum magnetism. The latter one is particularity challenging to observe in ultracold gases  because of the extreme cooling it requires in order to create quantum states with very low entropy which exhibit antiferromagnetic (AFM) correlations. Recent experiments succeeded in realizing these states and investigated the effect of lattice geometry and dimensionality on the magnetic correlations of the ground state of the Hubbard model~\cite{Greif2015,Hart2014}. In these setups, the presence of AFM ordering is revealed by averaging the results of time-of-flight images over many different experimental runs. Moreover, classical light beams are used for manipulating, controlling and cooling the atoms. In this work, we show that the backaction arising from global spatially structured quantum measurement allows to engineer and detect quantum states presenting AFM correlations in a single experimental realization and in real-time, even in the absence of interactions between atoms with opposite spin. 

The dynamical AFM and density-wave states, presented in this section, resemble the bosonic oscillations described in Secs. 3.3 and 3.6.

\subsection{Theoretical model}

As the mojority of results of this chapter described bosons, we start by summarizing the main features of the fermionic model. The atomic dynamics is described by the usual Hubbard Hamiltonian
\begin{equation}\label{3-HHubbard}
H_0 =-\hbar J \sum_{\sigma=\up,\down} \sum_{\langle i,j\rangle} \hat{f}_{j,\sigma}^\dagger \hat{f}_{i,\sigma} +\hbar U \sum_i \hat{n}_{i,\uparrow}\hat{n}_{i,\downarrow},
\end{equation}
where $J$ is the tunneling amplitude, $U$ is the interaction energy between atoms with opposite spin, and $\hat{f}^\dagger_{j,\sigma}$ ($\hat{f}_{j,\sigma}$) creates (annihilates) an atom with spin $\sigma$ at the site $j$. 

The polarization of the probe beam defines which linear combination of the  spin-$\uparrow$ and spin-$\downarrow$ are addressed by the measurement scheme. For example, if the probe laser is circularly polarized ($L$ or $R$), the measurement process is sensitive only to one of the two spin species ($\a=C_L \hat{D}_\uparrow$ or $a=C_R \hat{D}_\downarrow$). Furthermore, considering the case of linearly polarized probe, the photons escaping the optical cavity carry information about the atomic density $\hat{\rho}_i=\hat{n}_{i\up}+\hat{n}_{i\down}$ and the magnetization $\hat{m}_i=\hat{n}_{i\up}-\hat{n}_{i\down}$ so that the annihilation operators for the cavity field are proportional to
$\hat{D}_x=\sum_{i=1}J_{ii} \hat{\rho}_{i}$ ($x$-polarized light) and $\hat{D}_y=\sum_{i=1}J_{ii} \hat{m}_{i}$ ($y$-polarized light). 

The state of the system is a result of the deterministic evolution given by the non-Hermitian Hamiltonian $H_\mathrm{eff}=H_0- i \hbar \hat{c}^\dagger \hat{c} /2$  and the stochastic quantum jumps when the operator $\hat{c}=\sqrt{2 \kappa} a$ is applied to the atomic state. Note that the non-Hermitian term in $H_\mathrm{eff}$ is characterized by the energy scale $\gamma=\kappa |C|^2$ which competes with the usual tunneling amplitude $J$ and on-site interaction~$U$, leading to new many-body dynamics not described by the Hubbard Hamiltonian and novel effects beyond the quantum Zeno limit, which we already considered.

In this section, we focus mainly on two different measurement schemes which partition the lattice in two spatial modes. The first one addresses the difference in occupation between odd and even lattice sites ($\hat{D}_{\sigma}=\hat{N}_\mathrm{\sigma,odd}-\hat{N}_\mathrm{\sigma,even}$). The second scheme we consider probes the number of atoms at the odd sites ($\hat{D}_{\sigma}=\hat{N}_\mathrm{\sigma,odd}$).

\subsection{Measurement-induced antiferromagnetic ordering}

We first focus on non-interacting fermions with two spin components at half filling ($N_{\up}=N_{\down}=L/2$, in this section we set $L$ as the number of lattice sites, so that not to confuse it with the magnetisation) and we detect the light scattered in the diffraction minimum. In this case, the ground state of the system is the Fermi Sea $\ketz{FS}$ where only single particle states with $k<k_F$ are occupied ($k_F$ being the Fermi wavevector) and the density and magnetization are uniform across the lattice. We use this state as a reference point, assuming that the atomic system is initialized in its ground state before the measurement take place. 
The monitoring process perturbs this state and induces AFM correlations that can drive the atomic state to a superposition of the Neel states $| \up\down\up\down ...\rangle$ and $| \down\up\down\up ...\rangle$. In contrast to previous works \cite{ritsch2013,Kramer2014,Morigi2015,Piazza2015, Caballero2015,Caballero2015a,Keeling2014superradiance,Piazza2014superradiance, Chen2014superradiance}, this state emerges as a consequence of the competition between measurement backaction and atomic tunneling in a single quantum trajectory and does not rely on the cavity potential. Probing the atomic system with linearly polarized light along the $y$ axis, the annihilation operator describing the photons escaping the cavity is  $a=C\sum_{i=1} (-1)^{i} \hat{m}_{i}=C(\hat{M}_{\mathrm{even}}-\hat{M}_{\mathrm{odd}})\equiv C \hat{M}_s$ so the measurement is directly addressing the staggered magnetization of the atomic system.  Moreover, the photon number operator $a^{\dagger} a$ does not depend on the sign of $\hat{M}_s$ and therefore does not distinguish between states with opposite magnetization profiles. Consequently, the conditional dynamics preserves $\m{\hat{M}_s}$ and the local magnetization remains the same as in the ground state ($\m{m_{i}}=0$ for all $i$). The quantum jumps tend to suppress states that do not present AFM correlations since the application of $\h{c}$ on the atomic state completely vanishes its components with $\m{\h{M}_s^2}=0$. Therefore, the detection process modifies the probability distribution of $\m{\hat{M}_s}$, making it bimodal with two symmetric peaks around $\m{\hat{M}_s}=0$ that reflects the degeneracy of $\h{a}^{\dagger}_{1} \h{a}_1$. The evolution of such peaks depends on the ratio $\gamma / J$ which determines whether the dynamics is dominated  by the quantum jumps or by the usual tunneling processes. Specifically, in the strong measurement regime ($\gamma \gg J$) the detection process freezes $\m{\h{M}_s^2}$ to a specific value stochastically determined by a particular series of quantum jumps. However, if $\gamma \ll J$ the measurement cannot inhibit dynamics and tunneling of atoms across the optical lattice leads to an oscillatory behavior, that, in the case of bosons, can be described analytically as we showed in Secs. 3.3 and 3.6.

\begin{figure}[h]
\captionsetup{justification=justified}
\centering
\includegraphics[width=0.7\textwidth]{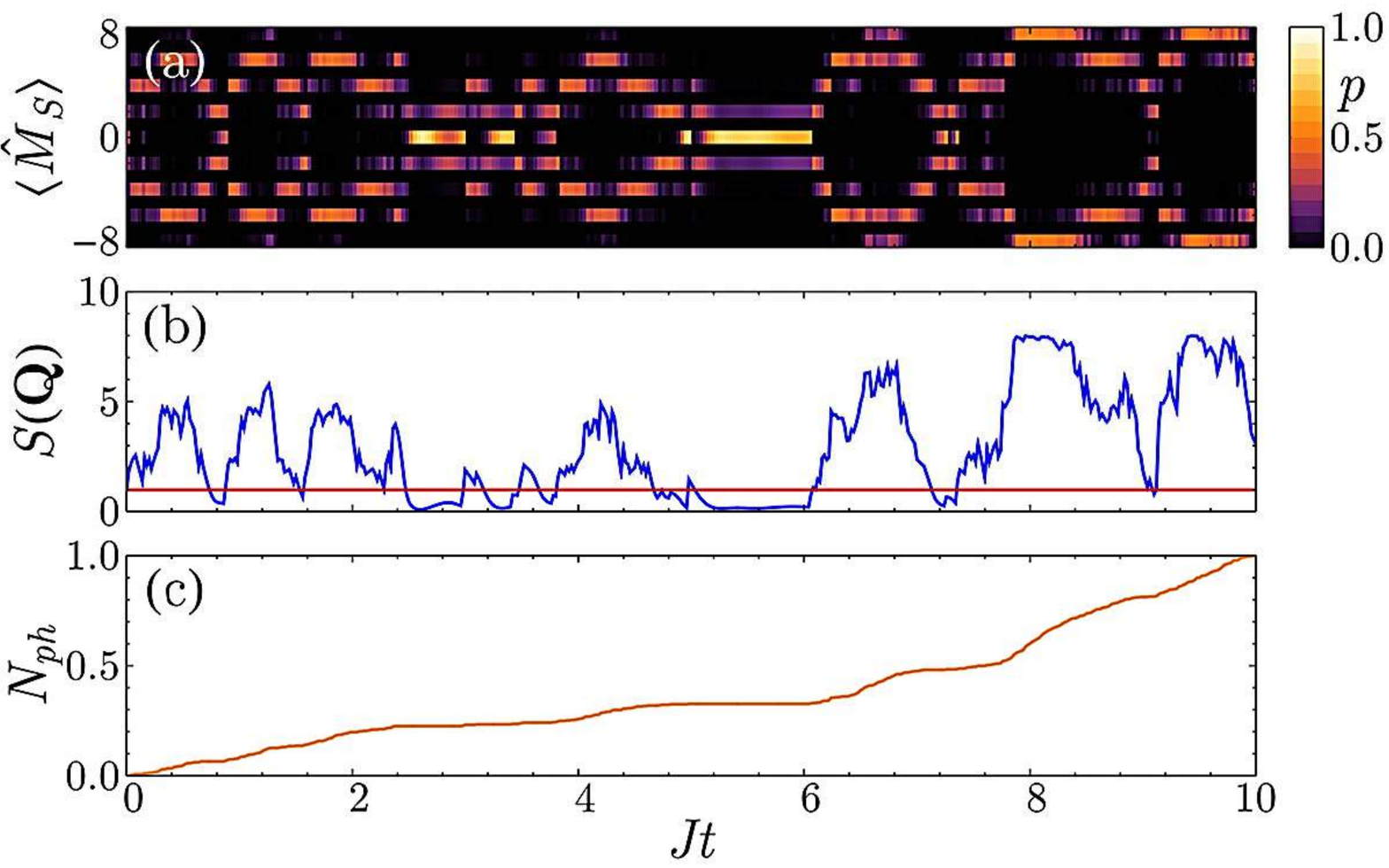}
\caption{Measurement-induced AFM order in a single experimental run (quantum trajectory). (a) The probability distribution of the staggered magnetization $\m{\hat{M}_s}$ presents two strong peaks as the state of the system is in a quantum superposition analogous to a Schr\"odinger cat state. (b) Comparison between the magnetic structure factor $S(\b{Q})$ for the ground state (red) and the conditional dynamics (blue), confirming the presence of AFM order. (c) Number of detected photons as a function of time (normalized to one). The derivative of this curve (photocount rate) is proportional to $S(\b{Q})$. $\gamma/J=1$, $U/J=0$, $N_\up=N_{\down}=4$, $L=8$.} \label{3-min}
\end{figure}

The presence of AFM correlations in the quantum state resulting from the conditional dynamics is revealed by computing the magnetic structure factor 
\begin{equation}
S(\b{q})=\frac{1}{L}\sum_{i,j} e^{i \b{q} \cdot (\b{r}_{i}-\b{r}_{j})} \left( \m{\h{m}_{i} \h{m}_{j}} - \m{\h{m}_{i}} \m{\h{m}_{j}}\right).
\end{equation}
Monitoring this quantity for each quantum trajectory, we find that the measurement induces a strong peak at $\b{q}=\b{Q}=\pi/d$ and creates an AFM state (Fig.~\ref{3-min}). Importantly, the value of $S(\b{Q})$ is directly accessible to the experiments since the probability for a photon to escape the optical cavity in a small time interval $\d t$ is proportional to $ \m{\h{c}^\dagger \h{c}} \d t$. This allows to observe the formation of AFM ordering in real-time by simply computing the photocount rate.  Note that the emergence of these intriguing quantum states is stochastic  and varies depending on the specific quantum trajectory, i. e., a single experimental set of photodetections. However, thanks to the photons escaping from the cavity,  it is possible to precisely determined when the AFM correlations are established and any subsequent  dynamics can be frozen by increasing the depth of the optical lattice. The resulting state can then be used for more advanced studies with applications to quantum simulations and quantum information. In contrast to condensed matter systems or usual cold atoms experiments where a strong repulsion between the atoms is necessary for establishing AFM order, our measurement scheme allows to obtain these states even for non-interacting fermions. Furthermore, the spatial period of the correlations imprinted by the measurement can be tuned changing the scattering angle and may lead to the realization of states with more complex spatial modulations of the magnetization. Therefore, global light scattering will help to simulate effects of long-range interactions in Fermi systems which are inaccessible in modern setups based on optical lattices with classical light.

\begin{figure}[h]
\captionsetup{justification=justified}
\centering
\includegraphics[width=0.7\textwidth]{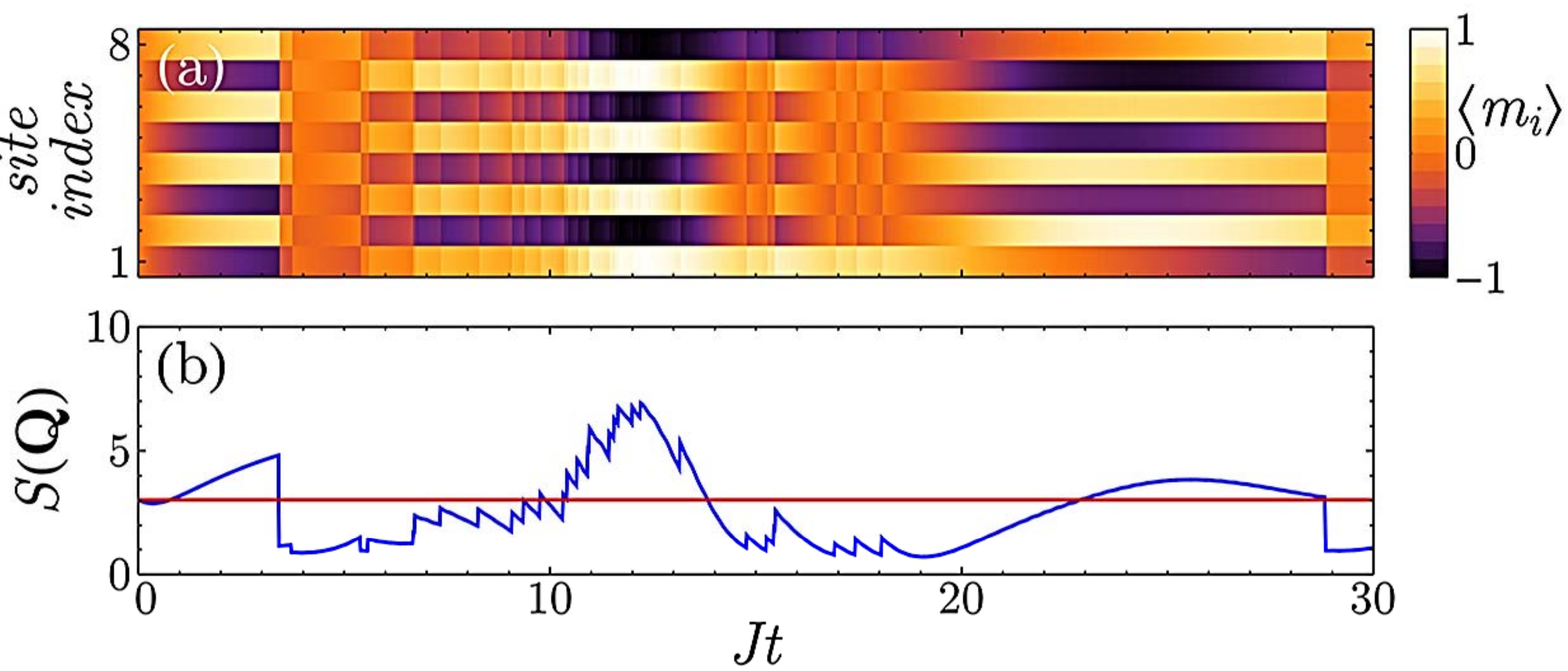}
\caption{Measurement-induced AFM order in a single experimental run (quantum trajectiry) for repulsive fermions. (a) The measurement process creates a modulation in the local magnetization, which is not present in the ground state. (b) Comparison between magnetic structure factor $S(\b{Q})$ for the ground state (red) and the conditional dynamics (blue), confirming the enhancement of AFM correlations. $\gamma/J=0.1$, $U/J=20$, $N_\up=N_{\down}=4$, $L=8$.} \label{3-minU}
\end{figure}

Considering the case of interacting fermions, we show that addressing only one of the two spin species affects the global density distribution and induces AFM correlations.  Specifically, we consider the measurement operator $\h{D}_{L}=\h{N}_{\mathrm{odd}\up}$, which probes only one spin species using circularly polarized light, and we focus on the strongly repulsive limit $U=20J$. The detection process is sensitive only to the spin-$\up$ density and therefore induces a periodic modulation of the spatial distribution of this species, favoring either odd or even lattice sites. However, since the two different spins are coupled by the repulsive interaction, spin-$\down$ atoms are also affected by the measurement and tend to occupy the lattice sites that are not occupied by the spin-$\up$ atoms, enhancing the AFM correlations of the ground state. Importantly, in this case the  AFM character of the atomic state is visible in the local magnetization~(Fig.~\ref{3-minU}) since the photon number operator can distinguish between states with opposite values of $\m{\hat{M}_s}$. The presence of a single peak in the probability distribution of $\m{\hat{M}_s}$ makes this scheme more robust to decoherence due photon losses than detecting the scattered light in the diffraction minimum.

\subsection{Measurement-induced density ordering}

\begin{figure}[h]
\captionsetup{justification=justified}
\centering
\includegraphics[width=0.6\textwidth]{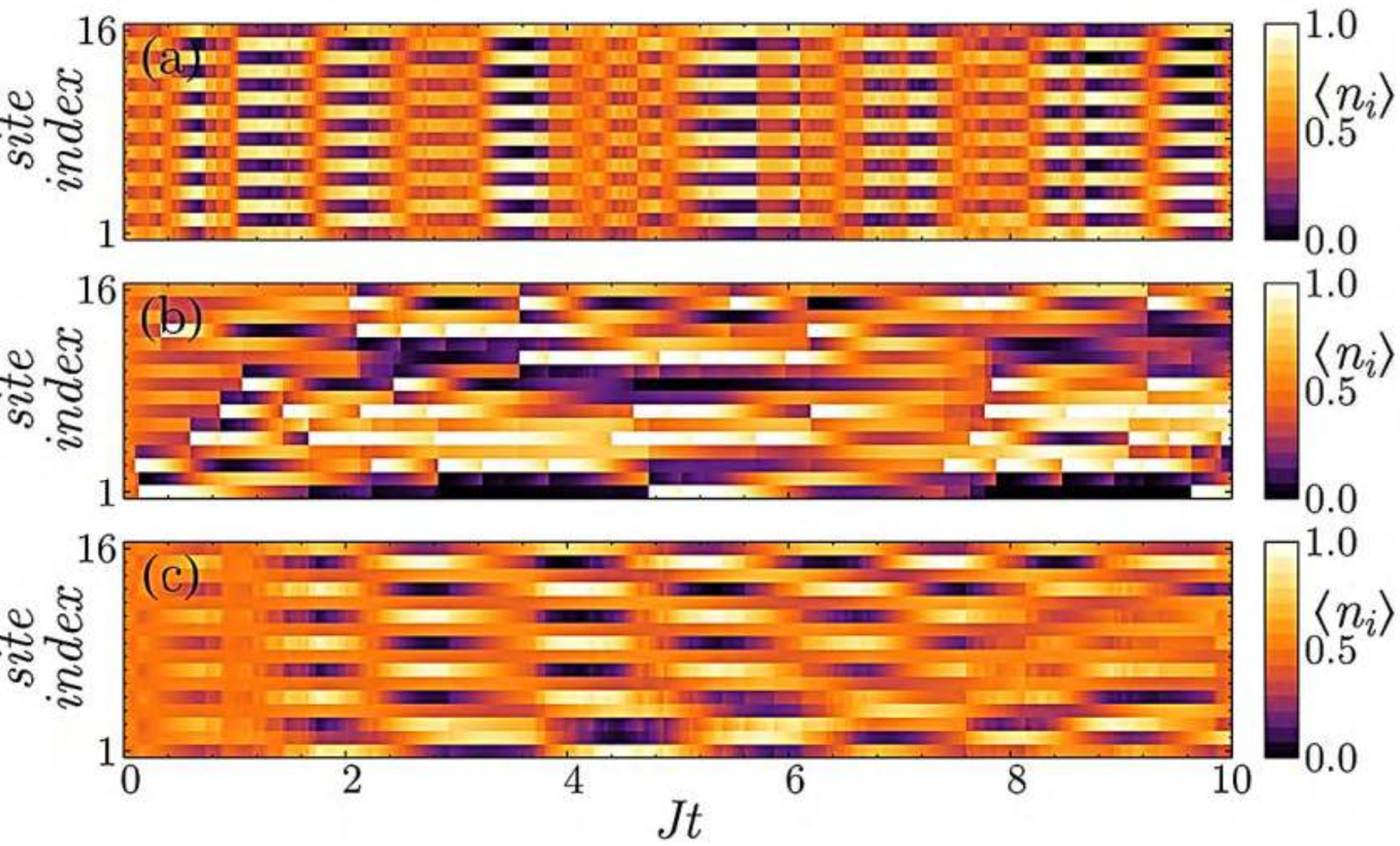}
\caption{Conditional evolution of the local density in a single experimental run probing the optical lattice with (a,c) global and (b) local addressing illuminating the odd lattice sites. (a)~The atomic population collectively oscillates between the two modes defined by the measurement establishing a spatial periodic modulation with period $2d$. (b)~The measurement process suppresses the local fluctuations independently for each lattice site. (c)~Changing the detection angle it is possible to tune the spatial period of the oscillations, establishing a density wave with period $3d$. $\gamma/J=1$, $U/J=0$, $N=N_\up=8$, $L=16$.} \label{3-frozen}
\end{figure}

We now turn to non-interacting polarized fermions, i.~e., all the atoms are in the same spin state and the atomic Hamiltonian only describes tunneling processes between neighboring lattice sites.  The effective dynamics emerging by measuring $\hat{c}\propto\h{N}_\mathrm{odd}$  modulates the atomic density across the lattice and depends on the ratio $\gamma/J$. If the measurement is weak, the atoms periodically oscillate between odd and even sites so that $\m{\h{N}_{\mathrm{odd}}-\h{N}_{\mathrm{even}}}\neq 0$, i. e. a state where a density wave with the period  of the lattice is established. Such configuration is usually a consequence of finite range interactions~\cite{Hirsch1984} and it has been observed in solid state systems~\cite{PhysRevB.16.801} and molecules in layer geometries~\cite{Block2012}.  Here, in contrast, the atoms do not interact but the cavity coupling with all the lattice sites mediates an effective interaction between them. Note that we have already proposed the global quantum nondemolition measurements for molecules in low dimensions in Chapter 1 \cite{MekhovLP2013}, which can link these fields even closer. Observing the photons leaving the cavity allows us to continuously monitor the state of the atoms, precisely determining when the density wave is established without the need of external feedback~\cite{Ivanov2014,Pedersen2014,Sherson2015}. 
If $\gamma \gg J$, the amplitude of the density wave remains constant on a timescale larger than $1/J$. Importantly, the fluctuations of the expectation value of the $\h{N}_{\mathrm{odd}}$ are strongly suppressed even if the on-site atomic density is not well-defined. This is a consequence of the global addressing of our measurement scheme: the jump operator does not distinguish between different configurations having the same $\h{N}_\mathrm{odd}$. This property is crucial for establishing correlations between distant lattice sites (Figure~\ref{3-frozen}). Local~\cite{Diehl2008} or fixed-range~\cite{LesanovskyPRL2012} addressing destroys coherence between different lattice sites and tends to project the atomic state to a single Fock state, failing to establish a density wave with a well-defined spatial period. Furthermore, the spatial period of the long-range correlations imprinted by the global measurement can be easily tuned changing the scattering angle. We illustrate this by considering the coefficients $J_{ii}=[1,\, 1/2,\, 0,\, 1/2,\, 1,\, 1/2,\, 0...]$ which can be obtained using standing waves crossing the lattice at an angle such that  $\b{k}_{\mathrm{1,0}}\cdot \b{r}=\pi/4$. With this scheme, the measurement partitions the lattice in $R=3$ non-overlapping spatial modes and leads to the emergence of density modulations with period~$3d$~(Figure~\ref{3-frozen}(c)).

\begin{figure}[h]
\captionsetup{justification=justified}
\centering
\includegraphics[clip, trim=0.3cm 5cm 0cm 6cm, width=0.7\textwidth]{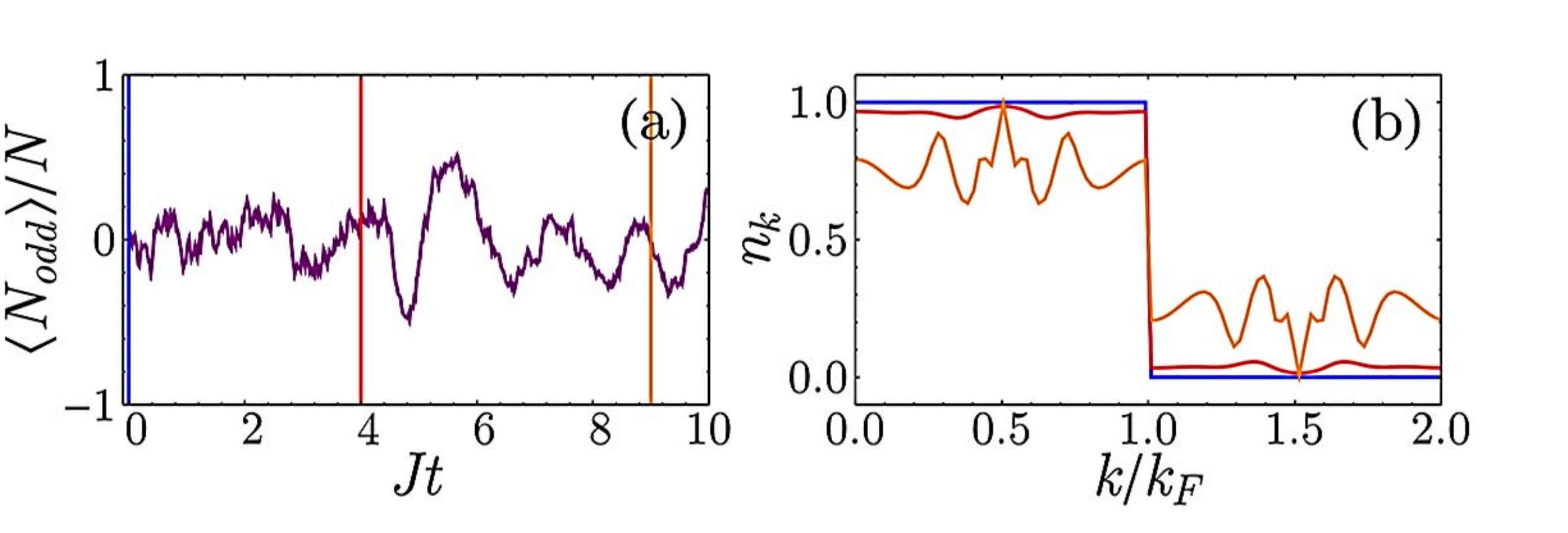}
\caption{Solution of the mean-field equations for a single experimental run. (a) The measurement imprints a density modulation on the atomic state as the number of atoms occupying the odd sites oscillates. (b) Occupation of the single particle momentum states for different times showing that the creation of particle-hole excitations is symmetric around $k_F$.  Different colors represent different times with reference to panel~(a). $\gamma/J=0.05$, $U/J=0$, $N=50$, $L=100$.} \label{3-MF}
\end{figure}

The measurement backaction changes the structure of the atomic state taking the system away from its ground state. The specific form of the excitations on top of the Fermi Sea depends on the spatial structure of the jump operator $\h{c}$ and, changing the spatial profile of $J_{ii}$, allows us to select which momentum states are affected by the measurement process. Defining $A_\b{k}$ to be the Fourier transform of $A_i=\sqrt{2 \kappa} C J_{ii}$, one has 
\begin{equation}
\h{c}=\sum_{\b{k},\b{p}\in BZ} A_\b{p} \h{f}^\dagger_{\b{k}} \h{f}_{\b{k}+\b{p}},
\end{equation}
where BZ indicates the first Brillouin zone. If $A_\b{k}$ presents a narrow peak around $\b{k}=0$, i.~e. the measurement probes the number of atoms rather homogeneously in an extended region of the lattice ($J_{ii}\sim \mathrm{const}$), the detection process creates particles and holes only on the Fermi surface, which is the typical scenario in conventional condensed matter systems.  In contrast to this, the setup we propose allows us to probe states that are deep in the Fermi Sea: if $\h{c}\propto \h{N}_\mathrm{odd}$ one has $A_\b{k}\propto \delta(\b{k})+\delta(\b{k}+\b{Q})$ and the resulting jump operator is
\begin{equation}
\h{c}\propto \sum_{\b{k}\in BZ} \h{f}^\dagger_{\b{k}} \h{f}_{\b{k}}+\h{f}^\dagger_{\b{k}} \h{f}_{\b{k}+\b{Q}}.
\end{equation}
Applying this expression on the ground states leads to $\h{c} \ketz{FS} \propto N \ketz{FS}/2 + \ketz{\Phi}$ where
\begin{equation}
\ketz{\Phi}=\sum_{\b{k}:|\b{k}+\b{Q}|<k_F}\h{f}^{\dagger}_{\b{k}}\h{f}_{\b{k}+\b{Q}} \ketz{FS}
\end{equation}
and $\braketz{\Phi}{FS}=0$. Therefore, the detection process creates particle-hole excitations with momenta that are symmetric around the wavevector $\b{Q}/2$ and are not necessarily confined around $k_F$. This symmetry is reflected in the occupation of the single particle states (Figure~\ref{3-MF}(b)) and can be better understood defining $\h{\beta}_{\b{k}} = (\h{f}_{\b{k}} +\h{f}_{\b{k}+\b{Q}} )/\sqrt{2}$ and rewriting the jump operator as $\h{c}=\sum_{\b{k}\in RBZ} \h{\beta}_{\b{k}}^\dagger \h{\beta}_{\b{k}}$ where $RBZ$ is the reduced Brillouin zone. Therefore, the measurement tends to freeze the number of particle-hole excitations with wavevectors that are symmetric superposition around~$\b{Q}/2$.

The emergence of long-range entanglement makes the numerical solution of the conditional dynamics in a single quantum trajectory an extremely challenging problem which is difficult to solve efficiently even with methods such as Matrix Product States (MPS)~\cite{Schollwock}. While, in general, the system dynamics for small and large particle numbers can be indeed different \cite{MekhovLP2011}, here we confirm our findings on larger systems by formulating a mean field theory for the stochastic evolution of single particle  occupation number $n_{\b{k}}=\m{\h{f}^{\dagger}_{\b{k}} \h{f}_{\b{k}}}$ and the order parameter $\alpha_{\b{k}}=\m{\h{f}^{\dagger}_{\b{k}} \h{f}_{\b{k}+\b{Q}}}$. In general, the evolution of the observable $\h{O}$ conditioned on the outcome of the measurement follows a generalization of the Ehrenfest theorem.  Between two quantum jumps the dynamics is deterministic and follows 
\begin{equation}
\frac{d}{dt} \m{{\h{O}}}= -i \m{[\h{H}_{0},\h{O}]} - \m{\{\h{c}^{\dagger}\h{c},\h{O}\}}+ 2 \m{\h{O}}\m{\h{c}^{\dagger}\h{c}},
\end{equation}
where $[\cdot,\cdot]$ ($\{\cdot,\cdot\}$) is the (anti)commutator. The photocurrent  follows a stochastic process where the photocounts are determined by the norm of the atomic wavefunction $\ketz{\Psi}$ which is given by 
\begin{equation}
\der{t} {\braketz{\Psi}{\Psi}}= -2 \m{\h{c}^{\dagger}\h{c}}.
\end{equation}
When a photon is detected, the quantum jump operator is applied to the atomic state and the value of $\h{O}$ changes~as 
\begin{equation}
 \m{\h{O}} \rightarrow \frac{\m{\h{c}^{\dagger} \h{O} \h{c}}}{\m{\h{c}^{\dagger}\h{c}}}.
\end{equation}
We apply a mean field treatment to these equations decoupling the terms with more than two operators as a function of $n_{\b{k}}$ and $\alpha_{\b{k}}$ and we solved them numerically (Fig.~\ref{3-MF}), confirming the emergence of density modulations even in large systems.

In this section, we have shown that measurement backaction on ultracold Fermi gases can be used for realizing intriguing quantum states characterized by a periodic spatial modulation of the density and the magnetization. We have demonstrated that this spatial structure is a consequence of the global nature of the coupling between atoms and light. The competition between measurement backaction and usual dynamics determined by the tunneling enables the study of quantum magnetism without requiring extreme cooling or strong interactions between the atoms.  Our method enables the possibility to engineer otherwise low entropy states (i.e. with small number of defects). Importantly, the formation of magnetic states can be observed in real-time by measuring the photons leaving the optical cavity, without the need of destructive techniques such as time-of-flight imaging. This opens new possibility for studying the dynamics of strongly correlated materials and the effect of magnetic ordering on superconducting states \cite{Mitrano} as the methods we described could be used for imprinting magnetic correlations on states with superconducting properties.


\section{Concluding remarks of Chapter 3}

We proved that the quantum backaction of a global measurement can efficiently compete with standard local processes in strongly correlated systems. This introduces a physically novel source of competition in research on quantum many-body systems. The competition becomes efficient due to the ability to spatially structure the global measurement at a microscopic scale comparable to the lattice period, without the need for single site addressing. The extreme tunability of the setup we considered allows us to vary the spatial profile of the measurement operator, effectively tailoring the long-range entanglement and long-range correlations present in the system. The competition between the global backaction and usual atomic dynamics leads to the production of spatially multimode macroscopic superpositions which exhibit large-scale oscillatory dynamics and could be used for quantum information and metrology. 

Such dynamical states show spatial density-density correlations with nontrivial periods and long-range coherence, thus having supersolid properties, but as an essentially dynamical version. For fermions, we showed the possibility of measurement-induced break-up and protection of strongly interacting fermion pairs. The macroscopic oscillations correspond to the generation of antiferromagnetic order and density waves of fermions. The measurement of atom-number-related variables and matter-phase-related variables demonstrates the wave-particle dualism of quantum mechanics represented in the many-body context.

In the strong measurement regime, the usual nearest-neighbour tunnelling is suppressed but the atoms can still tunnel across the lattice because of correlated tunnelling. Such globally paired tunneling due to a fundamentally novel phenomenon can enrich physics of long-range correlated systems beyond relatively short-range interactions expected from standard dipole-dipole interactions \cite{LewensteinPSFNJP,LewensteinExtBHM}. These nonlocal high-order processes entangle regions of the optical lattice that are disconnected by the measurement. Using different detection schemes, we showed how to tailor density-density correlations between distant lattice sites. Quantum optical engineering of nonlocal coupling to environment, combined with quantum measurement, can allow the design of nontrivial system-bath interactions, enabling new links to quantum simulations and quantum thermodynamics and extend these directions to the field of non-Hermitian quantum mechanics, where quantum optical setups are particularly promising~\cite{Lee2014a}. Further applications can include the simulations of superexchange interactions and dynamical gauge fields \cite{PRA2016-4}. Importantly, both systems and baths, designed by our method, can be strongly correlated systems with internal long-range entanglement. In general, we merged the paradigms of non-Hermitian physics and quantum Zeno effect in the many-body context demonstrating the Raman-like second order processes in quantum Zeno subspaces. 

Our predictions can be tested using both macroscopic measurements~\cite{EsslingerNat2010,HemmerichScience2012,ZimmermannPRL2014,KetterlePRL2011} as well as novel methods based on single-site resolution~\cite{BlochNature2011,Weitenberg2011,GreinerNature2009,Vengalattore}. A pathway to realize this is to combine several recent experimental breakthroughs: a BEC was trapped in a cavity, but without a lattice~\cite{EsslingerNat2010,HemmerichScience2012,ZimmermannPRL2014}; detection of light scattered from truly ultracold atoms in optical lattices was performed, but without a cavity~\cite{Weitenberg2011,KetterlePRL2011}, and more recently an optical lattice has been obtained in a cavity \cite{EsslingerNature2016, Hemmerich2015}. Furthermore, the single-atom and multi-particle quantum Zeno effect was observed by light scattering~\cite{Vengalattore,Barontini2015}.

The most important result of this chapter is the demonstration of the novel type of competitions in many-body systems. In the next chapter we will show how some of the predicted dynamical states can be stabilized using the feedback control. More importantly, we will show how such competition leads to the novel highly controllable phase transitions with the tunable universality class.

\clearpage

%% file: Chapter4.tex
\chapter{Feedback-induced phase transitions and their quantum control}\label{chapt4}

\section{Introduction and plan of the chapter}

In this chapter we will introduce the feedback control to the system we have considered before. 
Feedback is a general idea of modifying system behavior depending on the measurement outcomes. It spreads from natural sciences, engineering, and artificial intelligence to contemporary classical
and rock music. We will extend this concept into the realm of quantum phase transitions. This will have fundamental consequences: the quantum measurements will be able not only to dynamically compete with other processes (as shown in the previous chapter), but to induce novel phase transitions beyond the dissipative ones. Importantly, the quantum properties of phase transitions, e. g. their universality class, can be now controlled by the classical feedback loop. Therefore, we shift the paradigm of feedback from the control of quantum states (as known e. g. in quantum metrology) to the control of phase transitions in quantum systems.

In our classical world a similar transition appears, when a microphone is placed close to a loudspeaker (or two telephones are put together) resulting in a strong and noisy signal, sometimes used by rock bands (the Beatles being the first one). The difference is that in our work not only the telephones or feedback guitars are represented by quantum systems, but the noise (i.e. the fluctuations) are quantum as well. Controlling such fluctuations is our main focus. 

In Sec. 4.2 \cite{Mazzucchi2016Opt} we address the system considered in detail in the previous chapter: ultracold bosons and fermions trapped in an optical lattice. We show how the optical feedback creates strong correlations in bosonic and fermionic systems. It balances two competing processes, originating from different fields: quantum backaction of weak optical measurement and many-body dynamics, resulting in stabilization of the states found in the previous chapter: density waves, antiferromagnetic, and NOON states. Here the stabilization will appear above some critical value of the feedback strength, which manifests the feedback-induced phase transition. While considering a many-body system, we limit our self to a simple Markovian feedback leaving the demonstration of the full power of feedback control to the next section.

In Sec. 4.3 \cite{IvanovSciRep2020, IvanovPRL2020} we will consider a more general system, which at the same time is simpler in its treatment: an ensemble of effective spins, and will demonstrate new very fundamental phenomena. We show that applying feedback and weak measurements to a quantum system induces phase transitions beyond the dissipative ones. Feedback enables controlling essentially quantum properties of the transition, i.e., its critical exponent \cite{IvanovPRL2020} and universality class \cite{IvanovSciRep2020}, as it is driven by the fundamental quantum fluctuations due to the measurement. Feedback provides the non-Markovianity and nonlinearity to the hybrid quantum-classical system. Our approach will enable creation of novel quantum simulators of quantum baths, simulating effects similar to spin-bath problems, and creating new Floquet time crystals with tunable long-range (long-memory) interactions. 

We conclude this chapter in Sec. 4.4 and make a link to the next chapter of this work.   

Recently, the first experiment, where our predictions presented in this chapter can be tested, has been reported in Ref. \cite{EsslingerFeedback}. 

{\it The results of this chapter are based on the papers}  \cite{Mazzucchi2016Opt, IvanovSciRep2020, IvanovPRL2020}\footnote{For a more recent work cf. Ivanov D. A., Ivanova T. Yu., Caballero--Benitez S. F., Mekhov I. B. Tuning the universality class of phase transitions by feedback:
Open quantum systems beyond dissipation // \href{http://dx.doi.org/10.1103/PhysRevA.104.033719}{Phys. Rev. A} --- 2021. --- Vol. 104. --- P. 033719.}.


\section{Feedback control for creating strong correlations in many-body systems}

In this section, we address phenomena predicted in Secs. 3.3 and 3.9: macroscopic oscillations of density waves and ferromagnetic order due to the measurement backaction. We show that adding a feedback loop one can change the oscillation frequency and stabilize these states at any given imbalance value (in particular, the NOON states with maximal particle imbalance can be obtained). As such stabilization appears above a critical value of the feedback strength, this section is the first step towards the feedback-induced phase transitions predicted in the next section.

\subsection{Theoretical model}

We consider the system described in Sec. 1.2 with the addition of external feedback loop (see Figure~4.1). The information from the photodetections is used for applying feedback to the system. We will show that the feedback can dynamically stabilize interesting quantum states that can be targeted and obtained deterministically. 

\begin{figure}[h]
\captionsetup{justification=justified}
\centering
\includegraphics[width=0.5\textwidth]{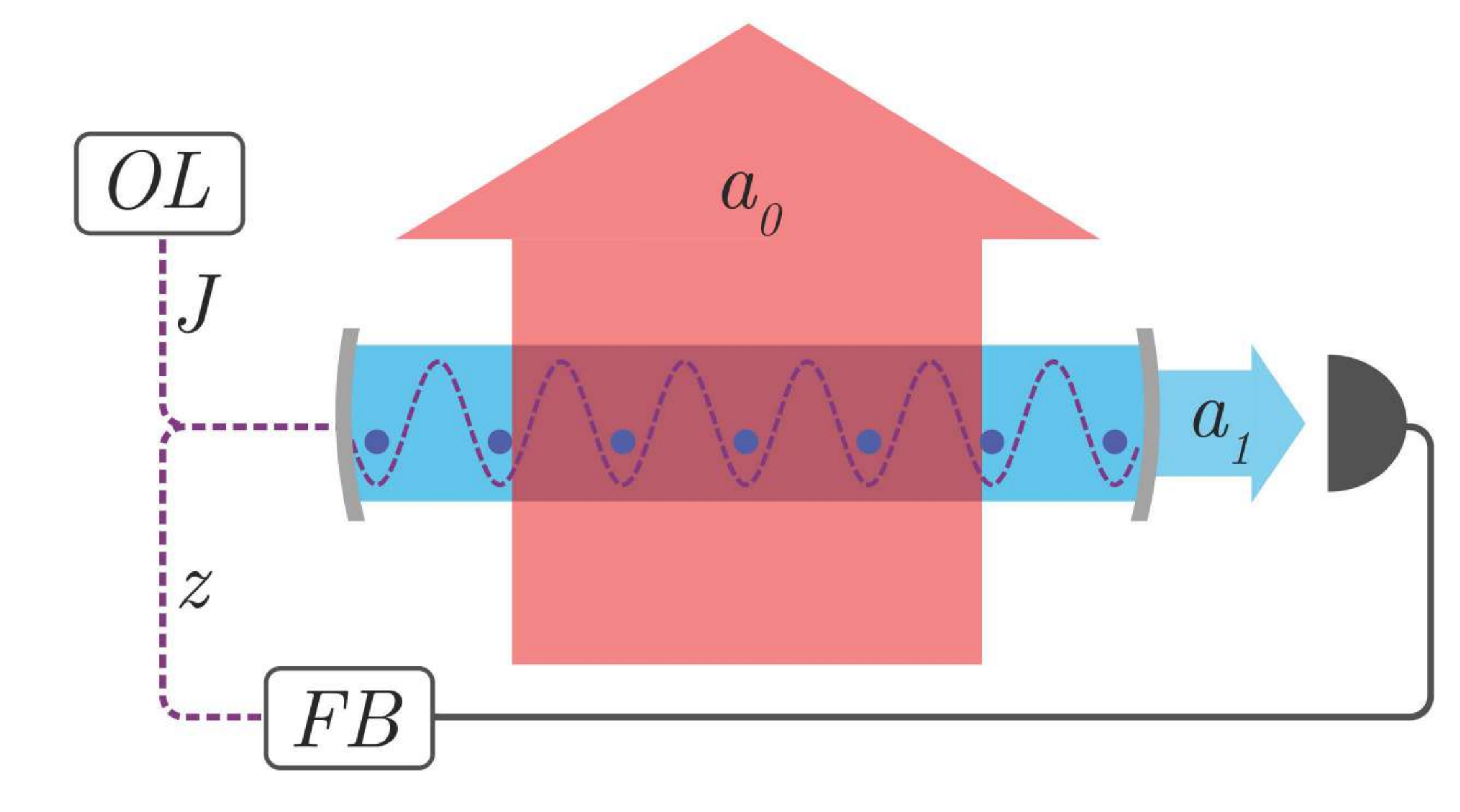}
\caption{Experimental setup. Ultracold atoms are loaded in an opical lattice (OL) inside an optical cavity and probed with a coherent light beam (mode $a_0$). The cavity enhances the light scattered orthogonally to $a_0$ and the photons escaping it are detected (mode $a_1$). Depending on the measurment outcome, a feedback (FB) loop with gain $z$ is applied for modulating the depth of the optical lattice.} \label{4-fig:setup}
\end{figure}

Similarly to the previous chapter, as in classical optics, the amplitude of the far off-resonant scattered light is proportional to the atomic density as $a_1=C \h{D}$, where 
\begin{align}\label{4-defC}
C=\frac{i U_{10} a_0}{i \Delta_p - \kappa}
\end{align}
is the Rayleigh scattering coefficient~\cite{Elliott2015,Mazzucchi2016PRA,Atoms}, $a_0$ is the amplitude of the classical coherent probe, $U_{10}=g_1 g_0/\Delta_a$, $g_l$ are the atom--light coupling constants, $\Delta_a$  and  $\Delta_p$ are respectively the atom--light and probe--cavity detunings, $\h{D}=\sum_j J_{jj} \h{n}_j$, $\h{n}_j=b^\dagger_j b_j$ is the density operator for the lattice site $j$ and $J_{jj}=\int w^2(\b{r}-\b{r}_j) u_{\mathrm{out}}^*(\b{r})  u_{\mathrm{in}}(\b{r})   \d \b{r}$. For fermions, different circular light polarizations couple to different spin states~\cite{Meineke2012, Sanner2012}: we exploit this property for probing linear combinations of the spin-$\uparrow$ and spin-$\downarrow$ atomic density changing the polarization of the probe beam. Focusing on the case of linearly polarized light ($a_{1x}$ and $a_{1y}$), we find that the measurement is sensitive to the local density  $a_{1x}=C \h{D}_x=C \sum_j J_{jj} \h{\rho}_j$ ($\h{\rho}_j=\h{n}_{j \uparrow} + \h{n}_{j \downarrow}$) or the local magnetization $a_{1y}=C \h{D}_y=C \sum_j J_{jj} \h{m}_j$ ($\h{m}_j=\h{n}_{j \uparrow} - \h{n}_{j \downarrow}$). 

In this section, we focus on two detection schemes. If the light modes are standing waves and their interference pattern has nodes at the odd sites of the lattice, one has $J_{jj}=0$ for odd and $J_{jj}=1$ for even sites. Therefore, the measurement probes the number of atoms occupying the even sites of the lattice and induces the formation of two spatial modes defined by lattice sites with different parity. The same mode structure can be obtained considering traveling or standing waves such that the wave vector of the cavity ($\b{k}_\mathrm{out}$) is along the lattice direction and the wavevector of the probe ($\b{k}_\mathrm{in}$) is orthogonal to it so that $(\b{k}_\mathrm{in}-\b{k}_\mathrm{out}) \cdot \b{r}_j=\pi j$. This configuration corresponds to detecting the photons that are scattered in the diffraction minimum at 90$^\circ$ and $J_{jj}=(-1)^j$, i. e., the jump operator is sensitive to the population difference between odd and even sites.

We focus on the outcome of a single experimental run and describe the conditional evolution of the atomic state using the quantum trajectories formalism. In general, the dynamics of a system subjected to continuous monitoring and feedback follows the master equation \cite{Wiseman}
\begin{equation}\label{4-eq:master}
\d \h{\rho} (t)= \left\{   \d N \left[ \mathrm{e}^{\mathcal{K}} \left (\mathcal{G} [ \hat{c} ]+1 \right) -1 \right] - \d t \mathcal{H} [i \hat{H}_0 + \frac{1}{2} \hat{c}^\dagger \hat{c} ] \right\}\h{\rho}(t),
\end{equation} 
where  $\h{c}=\sqrt{2 \kappa} a_1$ is the jump operator, $\d N$ is the stochastic It\^o increment such that $E[\d N]= \Trx[\hat{c} \h{\rho} \hat{c}^\dagger]\d t$,  $\mathcal{G}$ and  $\mathcal{H}$ are the superoperators
\begin{align}
\mathcal{G} [\hat{A} ]\h{\rho}=\frac{\hat{A}\rho\hat{A}^\dagger }{\Trx \left[\hat{A}\h{\rho}\hat{A}^\dagger \right]}-\h{\rho},\\
\mathcal{H} [\hat{A} ]\h{\rho}=\hat{A} \h{\rho}+ \h{\rho} \hat{A}^\dagger - \Trx\left[ \hat{A} \h{\rho} +\h{\rho} \hat{A}^\dagger \right],
\end{align}
$\h{H}_0$ describes the coherent (free) evolution of the system (in this section we consider non-interacting atoms) and the feedback loop acts on the master equation with the delay $\tau$ as $
\left[ \h{\rho}(t+\d t) \right]_{\mathrm{fb}}=  \exp \left[  \d N(t - \tau) \mathcal{K}  \right] \h{\rho}(t)$.

We consider the case where the feedback loop changes the depth of the atomic potential instantaneously ($\tau \rightarrow 0$), effectively modulating the value of the tunneling amplitude $J$ depending on the photocount rate. Within these assumptions, the superoperator $\mathcal{K}$ acts on the density matrix as 
\begin{align}
\mathcal{K} \h{\rho} = i [z\h{H}_0,\h{\rho}],
\end{align}
where the parameter $z$ describes the feedback strength. Assuming perfect detection efficiency and that the initial state of the system is in a pure state, we solve the master equation \eqref{4-eq:master} by simulating individual quantum trajectories. The evolution of the system is determined by the stochastic process described by the quantum jump operator $\h{d}=\sqrt{2 \kappa} \mathrm{e}^{i z \h{H}_0}\h{c}$ and the  non-Hermitian Hamiltonian $\h{H}_\mathrm{eff}= \h{H}_0 - i \hbar  \hat{c}^\dagger  \hat{c}/2$. In other words, $\h{H}_\mathrm{eff}$ generates the dynamics of the atomic system in between two consecutive photoemissions and,  when a photon escapes the optical cavity, the jump operator $\h{d}$ is applied to the atomic wavefunction. Note that $\h{d}$ describes both the effects of measurement backaction ($\h{c}$) and the feedback loop ($\mathrm{e}^{- i z \h{H}_0}$). Finally, we characterize the strength of the measurement process with the ratio $\gamma / J$ where $\gamma=\kappa |C|^2$: this quantity determines if the dynamics of the system in the absence of feedback is dominated by the tunneling or by the quantum jumps. 

\subsection{Stabilization of bosonic density waves}

\begin{figure}[h]
\captionsetup{justification=justified}
\centering
\includegraphics[width=0.99\textwidth]{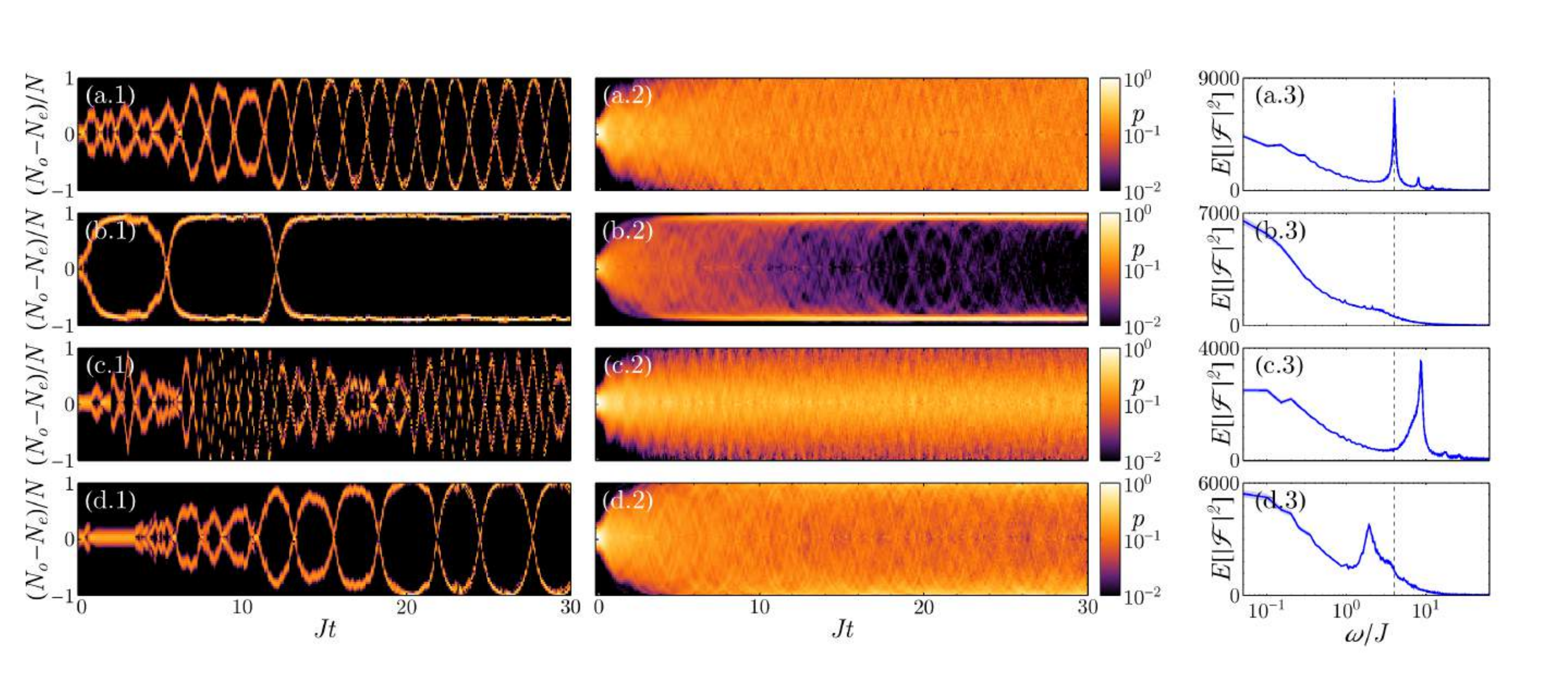}
\caption{Probability distribution of $\h{N}_\odd-\h{N}_\even$ (the curves have changing widths) for single quantum trajectories (Panels 1) and averages over 200 trajectories (Panels 2) for different values of the feedback gain $z$. Panel 3 shows the power spectrum of  $\m{\h{N}_\odd-\h{N}_\even}$ averaged over 200 trajectories. In the absence of feedback [Panel (a), $z=0$] the oscillations of the population of the odd sites are visible only in a single trajectory. For $z>z_c$ [Panel (b), $z=1.23 z_c$] the imbalance between odd and even sites is frozen for each quantum trajectory and $E \left[ |\mathcal{F}|^2\right]$ does not have a strong peak, indicating that $\h{N}_\odd-\h{N}_\even$ does not oscillate. For $z<z_c$ the  frequency of the oscillations can be tuned above [Panel (c)  $z=-4 z_c$] or below  [Panel (d) $z=0.8 z_c$] the frequency defined by the tunnelling amplitude $J$. Again, the oscillatory dynamics is visible only in a single quantum trajectory and the average probability distribution spreads quickly. $N=100$, $\gamma/J=0.02$, $J_{jj}=(-1)^j$, $z_c=0.0025$.}\label{4-fig:fig2}
\end{figure}

We first consider non-interacting bosons and demonstrate that the feedback process can be used for targeting specific quantum states even in the weak measurement regime ($\gamma \ll J$). Collecting the photons scattered in the diffraction minimum, the detection scheme probes the population imbalance between the odd  and even sites of the lattice ($\h{c}\propto \h{N}_{\odd}- \h{N}_{\even}$).  Importantly, since the intensity of the scattered light is $a_1^\dagger a_1$, the measurement is not sensitive to the sign of the imbalance and the atomic state remains in a superposition of states with opposite imbalances (a Schr\"odinger cat state). Moreover, as we demonstrated in Chapter 3, in the absence of feedback, the measurement backaction induces giant oscillations in the atomic population \cite{Mazzucchi2016PRA,Mazzucchi2016NJP} which resemble a dynamical supersolid state (Figure \ref{4-fig:fig2}a). Interestingly, these oscillations are visible only in a single experimental run (and not in average quantities), since their phase varies randomly between different quantum trajectories. Nevertheless, the oscillations are fully visible in the measured signal. We determine the frequency of such oscillations by computing the power spectrum of the photocurrent $\m{a_1^\dagger a_1} (t)$ for a single quantum trajectory: this quantity is directly accessible in the experiments and, being proportional  to $\m{(\h{N}_{\odd}- \h{N}_{\even})^2}$, allows to estimate the absolute value of the difference in population between odd and even sites. In order to characterize the behavior of all the trajectories, we calculate $\mathcal{F}=\int \m{(\h{N}_{\odd}- \h{N}_{\even})^2} e^{- i \omega t} \d t$  for each realization of the conditional evolution and we then average  $|\mathcal{F}|^2$ over several quantum trajectories ($E \left[ |\mathcal{F}|^2\right]$).  As expected, if the feedback is not present ($z=0$), $E \left[ |\mathcal{F}|^2\right]$ has a strong peak at $\omega= 4 J $ indicating that the oscillation frequency is determined by the tunneling amplitude.

\begin{figure}[h]
\captionsetup{justification=justified}
\centering
\includegraphics[width=.8\textwidth]{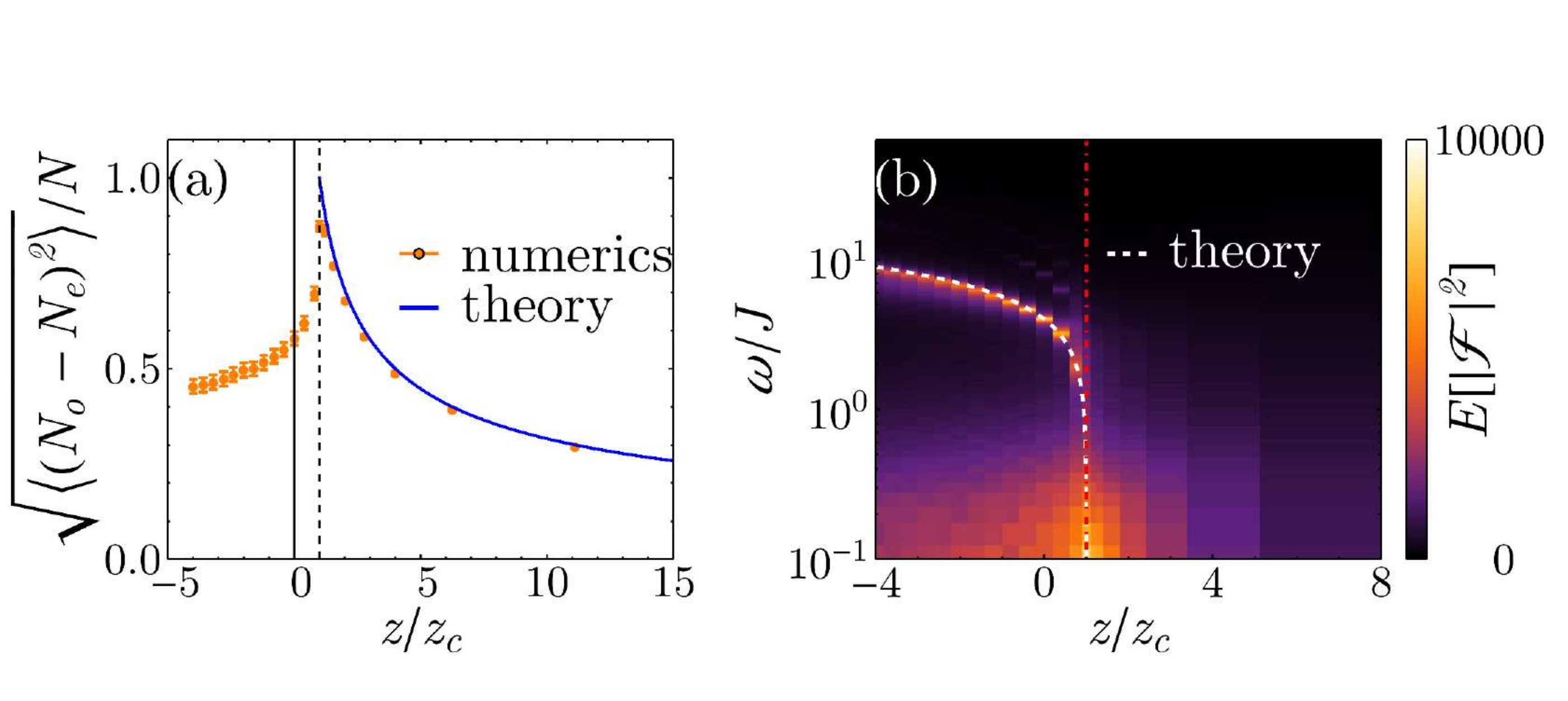}
\caption{Effects of measurement and feedback detecting the photons scattered in the diffraction minimum. Panel (a): Imbalance between odd and even sites as a function of the feedback strength. There is a very good agreement between the numerical results and the analytic expression derived in the text.  Panel (b): average power spectrum as a function of the feedback strength. The value of  $E \left[ |\mathcal{F}|^2\right]$ presents a strong peak for $z<z_c$, indicating that the trajectories are characterized by an oscillatory dynamics. The vertical dashed line marks $z=z_c$. $N=100$, $\gamma/J=0.02$, $J_{jj}=(-1)^j$, $z_c=0.0025$. } \label{4-fig:cfr1}
\end{figure}

Considering now the case when feedback is applied to the system, we find that there is a critical value for the parameter $z_c$ which defines two different dynamical regimes: if $z<z_c$ the expectation value $ \m{(\h{N}_{\odd}- \h{N}_{\even})^2}$ oscillates,  while if $z>z_c$ the imbalance reaches a steady state value that is deterministically defined by the parameter $z$ itself (Figure \ref{4-fig:fig2}). We explain this effect by looking at the effect of feedback on atoms: defining $\Delta t_n$ to be the time interval between the $(n-1)$-th and $n$-th quantum jump, the state of the system after $N_{ph}$ photocounts is
\begin{equation}
\ketz{\psi(t;N_{ph})}\propto\prod_{n=2}^{N_{ph}} \left[\mathrm{e}^{- i \h{H}_\mathrm{eff} \Delta t_n}\mathrm{e}^{i z \h{H}_0} \h{c} \right] \mathrm{e}^{- i \h{H}_\mathrm{eff} \Delta t_1} \ketz{\psi_0},
\end{equation} 
where $\ketz{\psi_0}$ is the initial state of the system. In the weak measurement regime ($\gamma \ll J$), we can focus on the terms depending linearly on $ \Delta t_n$ or $z$ and  neglect the commutators between $\h{H}_\mathrm{eff}$ and $\h{H}_0$ since it scales as $ \Delta t_n z$, thus
\begin{equation}
\mathrm{e}^{- i \h{H}_\mathrm{eff} \Delta t_n}\mathrm{e}^{i z \h{H}_0} \approx \mathrm{e}^{-i( \Delta t_n-z)\h{H}_0- \hbar \h{c}^\dagger \h{c} \Delta t_n /2}.
\end{equation} 
Therefore, the parameter $z$ defines an effective timescale which competes with the tunneling and the measurement processes. We find that it is possible to formulate a simple description of the dynamics of the atomic system by comparing the value of $z$ to the average time interval between two consecutive quantum jumps, i. e. $\overline{\Delta t}=1/(2 \gamma \m{\h{D}^\dagger \h{D}})$. Specifically, if $\Delta t_n \approx\overline{\Delta t}= z$, the feedback completely inhibits the tunneling  described by the Hamiltonian $\h{H}_0$ and the dynamics of the system is determined by the ``decay'' term in $\h{H}_\mathrm{eff}$. In this case, there are only two processes which contribute to the evolution of the system: the non-Hermitian dynamics (which tends to suppress the atom imbalance) and the quantum jumps (which drive the system towards large $\m{(\h{N}_{\odd}- \h{N}_{\even})^2}$). In the large time limit, these two effects balance each other and the system reaches a steady state where the probability distribution of $\h{N}_{\odd}- \h{N}_{\even}$ has two narrow peaks at opposite values. This regime is somehow analogous to the strong measurement regime (quantum Zeno effect)  and to the one described in Chapter 2, where the atomic dynamics was neglected. However, in these cases the system is confined to a quantum state that is an eigenvector of the jump operator whose eigenvalue is determined stochastically and ultimately depends on the initial state of the system. In contrast, here the introduction of a feedback loop allows us to deterministically select the final state of the system by tuning the value of $z$. Defining $\m{\h{D}^\dagger \h{D}}_{T}$ as the target imbalance one wants to obtain, the corresponding feedback gain realizing this specific configuration is $z=1/(2 \gamma \m{\h{D}^\dagger \h{D}}_{T})$ [see Figure~\ref{4-fig:cfr1}(a)]. Moreover, the target steady state is reached independently from the initial state of the system.

Since the value of the population imbalance between odd and even sites cannot exceed the total number of atoms, the maximum possible value for $\m{\h{D}^\dagger \h{D}}_{T}$ is $N^2$. This defines a critical $z$ under which the condition $\overline{\Delta t}= z$ cannot be fulfilled, $z_c=1/(2 \gamma N^2)$. As a consequence, for $z<z_c$ the state of the system does not reach a steady state and the measurement backaction establishes an oscillatory dynamics. Following the approach presented in Sec. 3.6, we find that carefully choosing the feedback gain it is possible to tune the frequency of the oscillations of $\m{(\h{N}_{\odd}- \h{N}_{\even})^2}$  according to $\omega=4 \sqrt{1-z/z_c}$ [see Figure~\ref{4-fig:cfr1}(b)]. Again, these oscillations are visible only analyzing single (but any) quantum trajectory and are not visible in the average probability distribution. The presence of two peaks in the probability distribution  $\m{\h{N}_{\odd}- \h{N}_{\even}}$ makes this measurment setup susceptible to decoherence due to photon losses~\cite{MekhovPRA2009}. However, this scheme can be made more robust by illuminating only the odd sites of the lattice so that $\h{c}\propto\h{N}_{\odd}$. In this case, the measurement operator probes the occupation of the odd sites and its probability distribution has only one strong peak.

\subsection{Stabilization of antiferromagnetic oscillations of fermions}

\begin{figure}[h!]
\captionsetup{justification=justified}
\centering
\includegraphics[width=.5\textwidth]{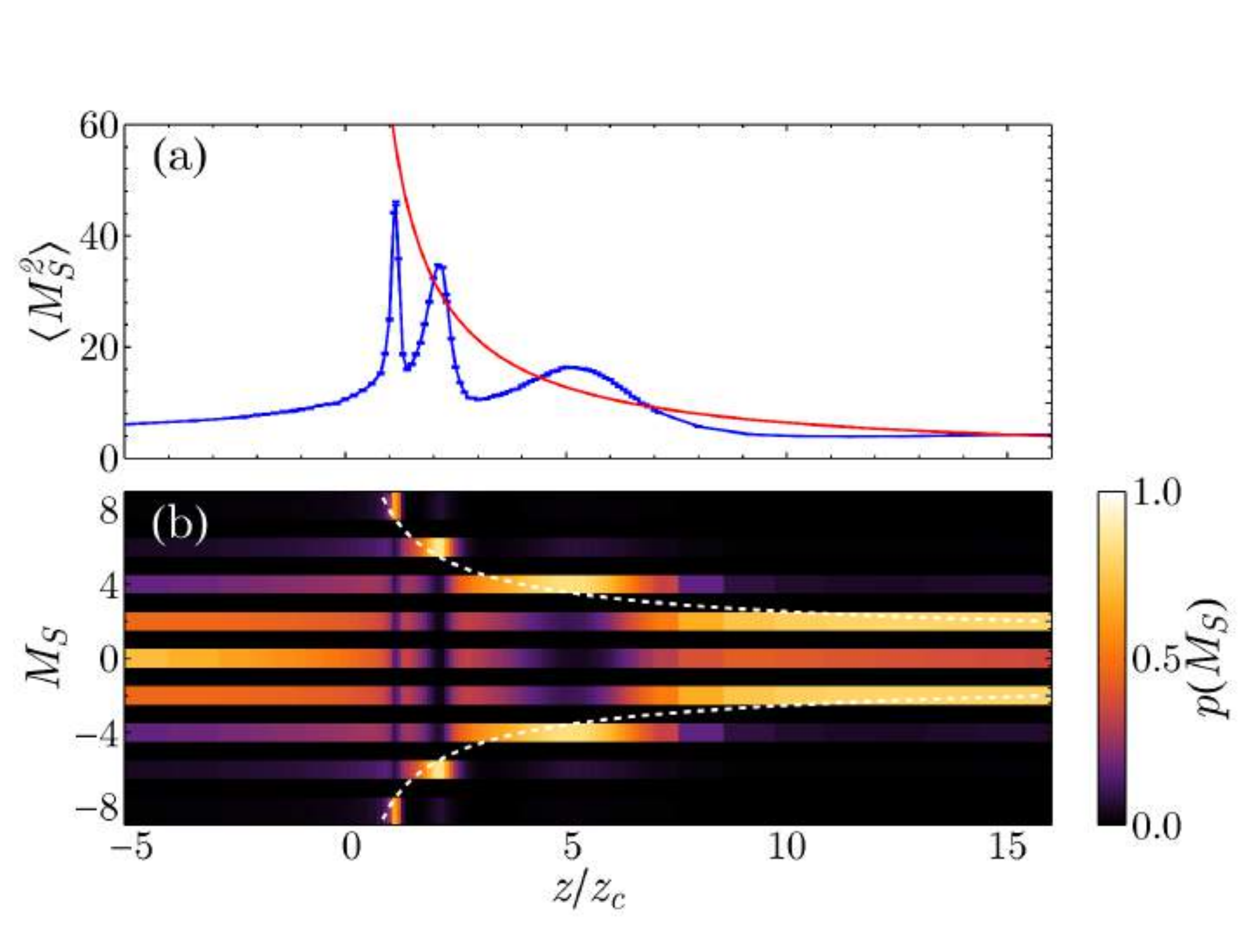}
\caption{Effects of measurement and feedback probing the staggered magnetization in a fermionic system. (a) Square of the staggered magnetization as a function of the feedback strength (blue line) compared to the analytic formula (red line) for a fermionic system. Note that the two curves do not have the same behavior because the analytic solution assumes that $\h{M}_S$ is a continuous variable while the numerical simulations are performed on a small system where $\h{M}_S$ assumes only discrete values. (b) Steady state value of the probability distribution of $\h{M}_S$ as a function of the feedback strength. The dashed line represents the theoretical prediction. The feedback loop stabilizes antiferromagnetic correlations. $N_\uparrow=N_\downarrow=4$, $\gamma/J=1$, $J_{jj}=(-1)^j$ , $z_c=1/128$.} \label{4-fig:fermi}
\end{figure}

We now turn to non-interacting fermions and we focus on the case where linearly polarized photons are detected so that the jump operator is sensitive to the staggered magnetization $\h{M}_S=\h{M}_\odd - \h{M}_\even$. If feedback is not present, the measurment backaction leads to quantum states characterized by antiferromagnetic ordering (cf. Sec. 3.9). However, these correlations follow an oscillatory dynamics and cannot be selected deterministically since they are a result of the competition between local tunneling processes and the stochastic quantum jumps. In analogy to the bosonic case, introducing a feedback loop allows us to obtain antiferromagnetic states with a predetermined staggered magnetization in each single quantum trajectory even in the absence of many-body interactions. Again, there is a critical value of the gain $z$ which sharply divides two regimes. If $z>z_c$ the system reaches a steady state such that $\m{\h{M}_S}=\sqrt{1/(2 \gamma z)}$ for each quantum trajectory. In contrast, if $z<z_c$ the value of  $\m{\h{M}}_S$ is not stationary and taking its expectation over many quantum trajectories we find that, on average, the atomic state does not present antiferromagnetic order. Figure~\ref{4-fig:fermi} illustrates this effect by showing the average over many realizations of the expectation value of the staggered magnetization and its probability distribution in the large time limit  as a function of the feedback gain. Note that the predicted value for $\h{M}_S$ agrees with the numerical results only qualitatively. This is because the analytic solution $\m{\h{M}_S}=\sqrt{1/(2 \gamma z)}$ treats the staggered magnetization as a continuous variable while  when performing a simulation on a small system only some discrete values of $\m{\h{M}_S}$ are possible. This effect is not surprising and it is rather analogous to our previous works, where the discreteness of the matter field leads to spectra with multiple peaks \cite{MekhovNaturePh2007} as in Sec. 1.10.

In this section, we demonstrated that the feedback control and optical measurement backaction can be used for engineering quantum states with long-range correlations which can be tuned by changing the spatial structure of the jump operator. We illustrated this by considering the case where the measurement induces two macroscopically occupied spatial modes and the feedback loop can stabilize interesting quantum phases such as supersolid-like states (i.e. the density waves with long-range matter-wave coherence) and states with antiferromagnetic correlations, depending on the value of the feedback gain $z$. This parameter determines the strength of the feedback loop and its net effect is to modulate the tunneling amplitude $J$ according to the measurment outcome.

The effects presented in this section can be generalized to other many-body (or simply multimode) physical systems such as optomechanical arrays \cite{Paternostro2011,Aspelmeyer2014}, superconducting qubits as used in circuit  QED \cite{Palacios2010,Paraoanu2011,Pirkkalainen2013,White2015}, ion strings, and even purely photonic systems (i.e. photonic chips or circuits) with multiple path interference, where, similarly to optical lattices, the quantum walks and boson sampling have been already discussed \cite{Spring2013,Nitsche2016,Brecht2015,Elster2015,Oren16}. Recently the ultra-strong light--matter coupling in a 2D electron gas in THz metamaterials has been obtained \cite{Zhang2016}, and developments have been made with respect to light induced  high-Tc superconductivity in real materials \cite{Cav1,Mitrano,Dieter1}. This further opens the possibility to engineer, what we propose here for atoms, in real solid state materials and hybrid devices for quantum technologies \cite{Kurizki2015} in the near future.


\section{Feedback control of the universality class of phase transitions}

In this section, we will consider a simpler but at the same time very general system: an ensemble of effective spins scattering light. In contrast, we will take the feedback, which is much more general than the simple Markovian (instantaneous) feedback considered in the previous section. As here we touch fundamental concepts of quantum and dissipative phase transitions and propose their generalizations, we prefer to start this section with a rather general introduction. In the end of this section we comment, how the system of ultracold atoms in optical lattices, which we considered so far, can be included in the model of this section. Nevertheless, the problem of tunable phase transitions in strongly interacting quantum gases in optical lattices still awaits its solution in the future.

The notion of quantum phase transitions (QPT) \cite{SachdevBook} plays a key role not only in physics of various systems (e.g. atomic and solid), but affects complementary disciplines as well, e.g., quantum information and technologies \cite{Osterloh}, machine learning \cite{Nieuwenburg} and complex networks \cite{Halu2013}. In contrast to thermal transitions, QPT is driven by quantum fluctuations existing even at zero temperature in closed systems. Studies of open systems advanced the latter case: the dissipation provides fluctuations via the system-bath coupling, and the dissipative phase transition (DPT) results in a nontrivial steady state \cite{Kessler2012,Daley}. 

Here we consider an open quantum system, which is nevertheless not a dissipative one, but is coupled to a classical measurement device. The notion of fundamental quantum measurement is broader than dissipation: the latter is its special case, where the measurement results are ignored in quantum evolution \cite{Wiseman}. We show that adding the measurement-based feedback can induce phase transitions. Moreover, this enables controlling quantum properties of the transition by tuning its critical exponent and thus the universality class. Such a feedback-induced phase transition (FPT) is driven by fundamentally quantum fluctuations of the measurement process, originating from the incapability of any classical device to capture the superpositions and entanglement of quantum world.

Feedback is a general idea of modifying system parameters depending on the measurement outcomes. It spreads from engineering to contemporary music (e.g. the Sampo device), including modeling  the Maxwell demon \cite{Murch2018,Masayuma2018,Koski2015} and reinforcement learning \cite{Petruccione}. Feedback control has been successfully extended to quantum domain \cite{Wiseman,HammererRMP, HaukePRA2013,HuardFB2016, Hacohen2016, HarocheBook,Mabuchi2004, Sherson2015,Sherson2016, Hush2013,Bouchoule2017,Hammerer2016PRA, Thomsen2002,St-K2012,Vuletic2007,Ivanova2014,Wallentowitz2004,Ivanova2016} resulting in quantum metrology aiming to stabilize nontrivial quantum states and squeeze (cool) their noise. The measurement backaction typically defines the limit of control, thus, playing an important but negative role \cite{Ivanova2005}. In our work, we shift the focus of feedback from quantum state control to phase transition control, where the measurement fluctuations drive transition thus playing an essentially positive role in the process as a whole.

Hybrid systems is an active field of quantum technologies, where various systems have been already coupled \cite{Kurizki2015}: atomic, photonic, superconducting, mechanical, etc. The goal is to use advantages of various components. In this sense, we address a hybrid quantum-classical system, where the quantum system can be a simple one providing the quantum coherence, while all other properties necessary for tunable phase transition are provided by the classical feedback loop: nonlinear interaction, non-Markovianity, and fluctuations. 

We will show that FPT leads to effects similar to particle-bath problems (e.g. spin-boson, Kondo, Caldeira-Leggett, quantum Browninan motion, dissipative Dicke models) describing very different physical systems from quantum magnets to cold atoms \cite{breuer2002,Leggett1987,LeHur2008, DomokosPRL2015, DomokosPRA2016,Scarlatella2016,Plenio2011}. While tuning quantum baths in a given system is a challenge, tuning the classical feedback is straightforward, which opens the way for simulating various systems in a single setup. This raises questions about quantum-classical mapping between Floquet time crystals \cite{Sacha2017,Eckardt2017} and long-range interacting spin chains  \cite{DavisPRL2019}. Our model is directly applicable to many-body systems, and as an example we consider ultracold atoms in a cavity. Such a setup of many-body cavity QED (cf. for reviews \cite{Mekhov2012,ritsch2013,Review2021}) was recently marked by experimental demonstrations of superradiant Dicke \cite{EsslingerNat2010}, lattice supersolid \cite{EsslingerNature2016, Hemmerich2015}, and other phase transitions \cite{LevPRL2018, Zimmermann2018}, as well as theory proposals \cite{Caballero2015,Caballero2015a, Rogers2014,Morigi2010,Niedenzu2013, Gopalakrishnan2009,Kollath2016, Diehl2013, DomokosPRL2015, DomokosPRA2016}. Nevertheless, effects we predict here require to go beyond the cavity-induced autonomous feedback \cite{St-K2017}.

\subsection{General model for an ensemble of spins} 

\begin{figure}[h]
\captionsetup{justification=justified}
\centering
\includegraphics[clip, trim=0cm 12.5cm 11.8cm 0cm, width=0.5\textwidth]{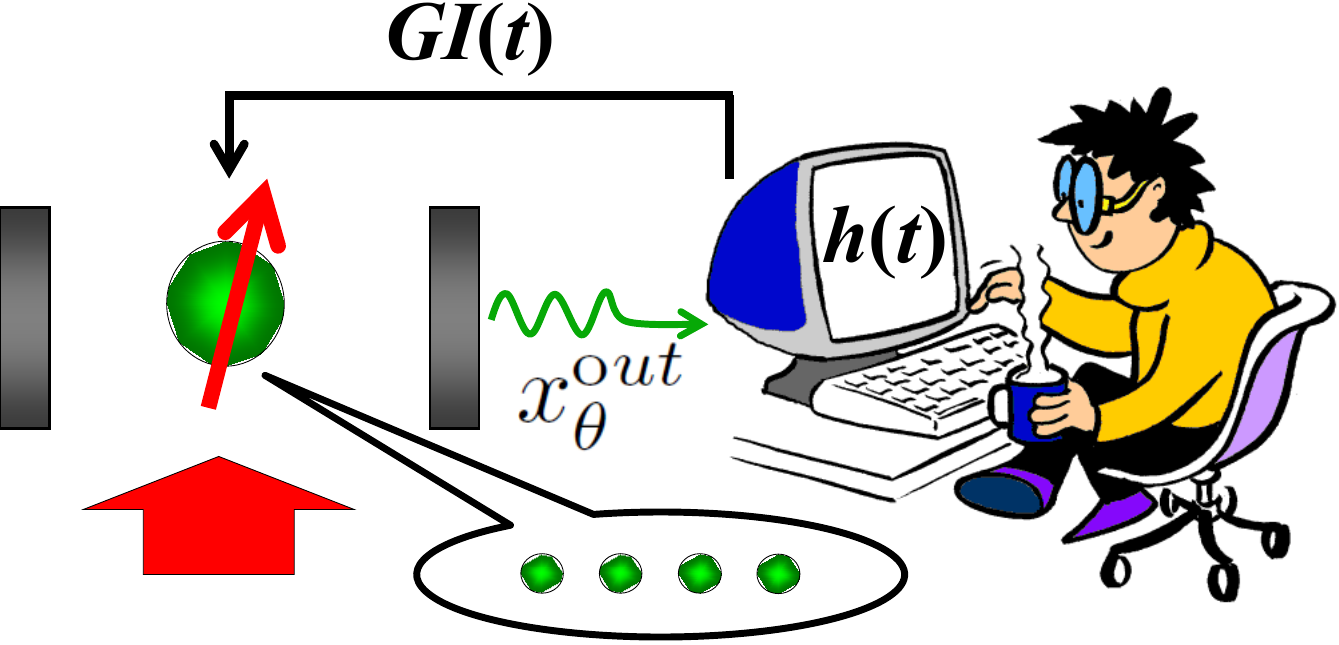}
\caption{\label{4-setup} Setup (details for a BEC system are given later in the text). Quantum dipoles (possibly, a many-body system) are illuminated by probe. Scattered light is measured and feedback acts on the system, providing non-Markovianity, nonlinearity, and noise, necessary for phase transition. Importantly, the feedback response $h(t)$ can be digitally tuned.} 
\end{figure}

Consider $N$ two-level systems (spins, atoms, qubits) coupled to a bosonic (light) mode, which may be cavity-enhanced (Fig. \ref{4-setup}). The Hamiltonian then reads
\begin{eqnarray}\label{4-Hamiltonian}
H=\delta a^\dag a +\omega_R S_z +\frac{2}{\sqrt{N}}S_x [g(a+a^\dag)+GI(t)],
\end{eqnarray}
which without the feedback term $GI(t)$ is the standard cavity QED Hamiltonian \cite{ScullyBook} describing the Dicke (or Rabi) model \cite{DomokosPRL2015, DomokosPRA2016} in the ultra-strong coupling regime \cite{HuardUltraS2018, Ustinov2017, Yoshihara2017, Forn2017} (without the rotating-wave approximation). Here $a$ is the annihilation operator of light mode of frequency $\delta$, $S_{x,y,z}$ are the collective operators of spins of frequency $\omega_R$, $g$ is the light--matter coupling constant. The Dicke model was first realized in Ref. \cite{EsslingerNat2010} using a Bose--Einstein condensate (BEC) in a cavity, and we relate our model to such experiments later in this section. Our approach can be readily applied to many-body settings as $S_x$ can represent various many-body variables (as we considered in the previous chapters of this work)  \cite{Elliott2015,Kozlowski2015PRA,Kozlowski2017}, not limited to the sum of all spins: e.g., fermion or spin (staggered) magnetization \cite{Mazzucchi2016PRA, Mazzucchi2016SciRep,LandiniPRL2018,LevPRL2018}  or combinations (e.g. atoms in odd and even sites) of strongly interacting atoms in arrays as in lattice experiments \cite{EsslingerNature2016, Hemmerich2015}.  

The feedback term $GI(t)$ has a form of the time-dependent operator-valued Rabi frequency rotating the spins ($G$ is the feedback coefficient and $I(t)$ is the control signal). We consider detecting the light quadrature $x_\theta^{\text out}(t)$ ($\theta$ is the local oscillator phase) and define $I(t)=\sqrt{2\kappa}\int_0^t h(t-z)\mathcal{F}[x_\theta^{\text out}(z)]dz$. Thus, the classical device continuously measures $x_\theta^{\text out}$, calculates the function $\mathcal{F}$, integrates it over time, and feeds the result back according to the term $GI(t)$. In a BEC system, the quasi-spin levels correspond to two motional states of atoms, and coupling of feedback to $S_x$ is achieved by modifying the trapping potential. Various forms of the feedback response $h(t)$ will play the central role in our work. The input-output relation \cite{walls2008quantum} gives $x_\theta^{\text out}=\sqrt{2\kappa}x_\theta-f_\theta/\sqrt{2\kappa}$, where the intracavity quadrature is $x_\theta=(a e^{-i\theta}+a^\dag e^{i\theta})/2$ and $\kappa$ is the cavity decay rate. The quadrature noise $f_\theta=(f_a e^{-i\theta}+f_a^\dag e^{i\theta})/2$ is defined via the Markovian noise operator $f_a$ [$\langle f_a(t+\tau)f_a(t)\rangle=2\kappa \delta(\tau)$] in the Heisenberg--Langevin equation:
\begin{eqnarray}\label{4-Heis}
\dot{a}=-i\delta a -i\frac{2g}{\sqrt{N}}S_x-\kappa a +f_a.
\end{eqnarray}

\subsection{Effective feedback-induced interactions} 

An illustration that feedback induces effective nonlinear interaction is used in quantum metrology \cite{Thomsen2002} for a simple cases such as $I(t)\sim x_\theta^{\text out}$. One sees this, if light can be adiabatically eliminated from Eq.~(\ref{4-Heis}), $a\sim S_x$. Then the effective Hamiltonian, giving correct Heisenberg equations for spins, contains the term $S_x^2$ leading to spin squeezing \cite{Thomsen2002} [cf. Eq. (\ref{4-Hamiltonian}) for $I(t)\sim x_\theta^{\text out}\sim S_x$]. Note that this is just an illustration and the derivation needs to account for noise as well. Nevertheless, we can proceed in a similar way and expect the interaction as $\int_0^t h(z)S_x(t)\mathcal{F}[S_x(t-z)]dz$. For the linear feedback, $\mathcal{F}[S_x]=S_x$, this term resembles the long-range spin-spin interaction in space: here we have a long-range (i.e. long-memory) ``interaction'' of spins with themselves in the past. The ``interaction length'' is determined by $h(t)$.

Such a time-space analogy was successfully used in spin-boson model \cite{Leggett1987,LeHur2008, VojtaPRL2005,VojtaPh2006}, describing spins in a bosonic bath of nontrivial spectral function: $\omega^s$ for small frequencies [$s=1$ for Ohmic, $s<1$ ($s>1$) for sub-(super-)Ohmic bath, see later in this section]. It was shown that a similar ``time-interaction'' term can be generated \cite{VojtaPRL2005,VojtaPh2006}. Moreover, an analogy with the spin chain and long-range interaction term in space $\sum_{i,j}S_iS_j/|r_i-r_j|^{s+1}$ was put forward and the break of the quantum-classical mapping was discussed \cite{VojtaPRL2005,VojtaPRL2012}. For $s=1$ a QPT of the Berezinskii--Kosterlitz--Thouless type was found \cite{LeHur2008}, while QPTs for the sub-Ohmic baths are still under active research \cite{Plenio2011,Plenio2018}.

In bath problems, such a long-memory interaction can be obtained only asymptotically \cite{VojtaPRL2005,VojtaPh2006}. Moreover, arbitrarily tuning the spectral properties of quantum baths in a given system is challenging (cf. \cite{Leppakangas2018} for quantum simulations of the spin-boson model and \cite{Nokkala2016,Nokkala2018} for complex network approach). In contrast, the feedback response $h(t)$ can be implemented and varied naturally, as signals are processed digitally, opening paths for simulating various problems in a single setup. The function
\begin{eqnarray}\label{4-h}
h(t)=h(0)\left(\frac{t_0}{t+t_0}\right)^{s+1}
\end{eqnarray}
will correspond to the spatial Ising-type interaction. The instantaneous feedback with $h(t)\sim\delta(t)$ will lead to ``short-range in time'' $S_x^2$ term, as in the Lipkin-Meshkov-Glick (LMG) model \cite{Parkins2008, Muniz2020} originating from nuclear physics. A sequence of amplitude-shaped time delays $h(t)\sim \sum_n \delta(t-nT)/n^{s+1}$ will enable studies of discrete time crystals \cite{Sacha2017,UedaTC2018,Demler2019,JakschNCom2019,JakschPRA2019} and Floquet engineering \cite{Eckardt2017} with long-range interaction $\sum_n S_x(t)S_x(t-nT)/n^{s+1}$, where the crystal period may be $T=2\pi/\omega_R$. This is in contrast to standard time crystals, where the parameter modulation is externally prescribed [e.g. periodic $g(t)$]. Here, the parameters are modulated depending on the system state (via $S_x$), i. e., self-consistently, as it happens in real materials e.g. with phonons. The “interaction in time” does not necessarily require the presence of standard atom-atom interaction in space. The global (all to all) interaction is given by constant $h(t)$. The Dicke model can be restored, even if light was adiabatically eliminated, by the exponentially decaying and oscillating $h(t)$ mimicking a cavity. 

Moreover, $h(t)$ can have minima, maxima, and even change its sign, which creates analogies with the molecular potentials (now in time, not in space) and rises intriguing questions about creation of time molecules and time-molecule crystals.

All such $h(t)$ can be realized separately or simultaneously to observe the competition between different interaction types. Our results do not rely on effective Hamiltonians \cite{Caballero2015a}, but this discussion motivates us to use in further simulations $h(t)$ given by Eq. (\ref{4-h}) that is unusual in feedback control.

\subsection{Feedback-induced phase transition} 

We show the existence of FPT with controllable critical exponent by linearizing (\ref{4-Hamiltonian}) and assuming the linear feedback: $\mathcal{F}[x_\theta^{\text out}]=x_\theta^{\text out}$. Using the bosonization by Holstein--Primakoff representation \cite{DomokosPRA2016}:  $S_z=b^\dag b - N/2$, $S_-=\sqrt{N-b^\dag b}b$, $S_+=b^\dag\sqrt{N-b^\dag b}$, $S_x=(S_++S_-)/2$, we get
\begin{eqnarray}\label{4-Hamiltonian-lin}
H=\delta a^\dag a +\omega_R b^\dag b +(b^\dag+b) [g(a+a^\dag)+GI(t)].
\end{eqnarray}
The bosonic operator $b$ reflects linearized spin ($S_x\approx\sqrt{N}X$), and the matter quadrature is $X=(b^\dag+b)/2$. 

As we have shown in Chapter 3, weak measurements constitute a source of competition with unitary dynamics, which is well seen in quantum trajectories formalism \cite{Daley,Ruostekoski2014, Pedersen2014, Molmer2016PRA, VasilyevPRL2018, VasilyevPRA2018, Sherson2018PRA}, underlining the distinction between measurements and dissipation. Thus they can affect phase transitions, including the many-body ones \cite{Mazzucchi2016PRA,UedaCrit2016, Bason2018}. Feedback was mainly considered for stabilizing interesting states \cite{Sherson2015,Mazzucchi2016Opt,Sherson2016, Hush2013,Bouchoule2017,Hammerer2016PRA}. Here, we focus on the QPT it induces. In this formalism, the operator feedback signal $I(t)$ in Eq. (\ref{4-Hamiltonian-lin}) takes stochastic values $I_c(t)$ conditioned on a specific set (trajectory) of measurement results $\langle x_\theta\rangle_c(t)$ \cite{Wiseman}: $I_c(t)=\sqrt{2\kappa}\int_0^t h(t-z)[\sqrt{2\kappa}\langle x_\theta\rangle_c(z)+\xi(z)]dz$, where $\xi(t)$ is white noise, $\langle \xi(t+\tau)\xi(t)\rangle=\delta(\tau)$. The evolution of conditional density matrix $\rho_c$ is then given by \cite{Wiseman}: $d\rho_c=-i[H,\rho_c]dt+\mathcal{D}[a]\rho_c dt+\mathcal{H}[a]\rho_c dW$, where $\mathcal{D}[a]\rho_c=2\kappa[a\rho_c a^\dag-(a^\dag a\rho_c + \rho_c a^\dag a)/2]$, $\mathcal{H}[a]\rho_c=\sqrt{2\kappa}[a e^{-i\theta}\rho_c+\rho_c a^\dag e^{i\theta}-\text{Tr}(a e^{-i\theta}\rho_c+\rho_c a^\dag e^{i\theta})\rho_c]$, $dW=\xi dt$. In general, averaging such stochastic master equation over trajectories does not necessarily lead to the master equation for unconditional density matrix $\rho$ used to describe DPTs. 

\begin{figure}[h]
\captionsetup{justification=justified}
\centering
\includegraphics[clip, trim=0.5cm 0.6cm 0.5cm 0.3cm, width=0.9\textwidth]{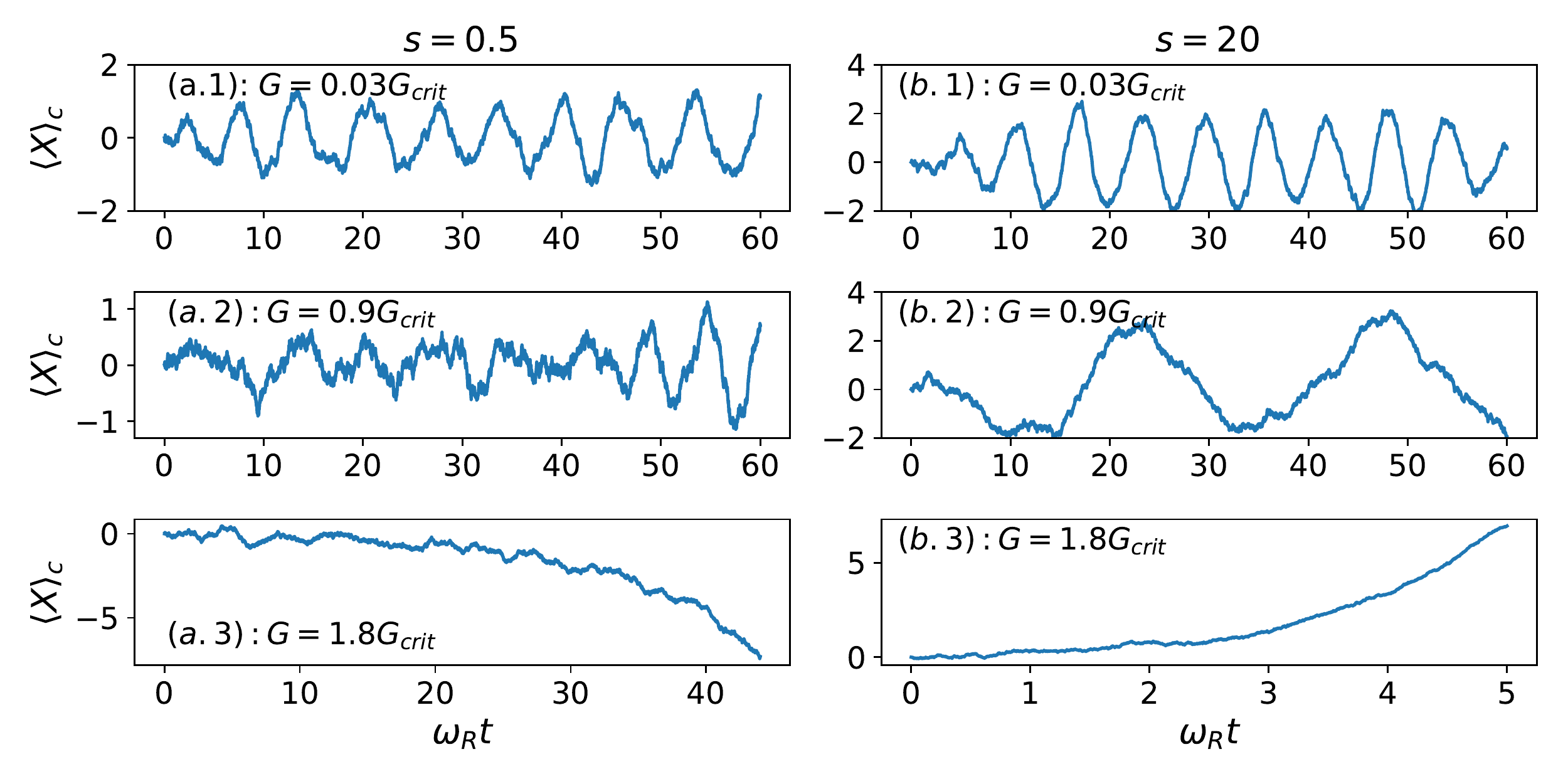}
\caption{\label{4-Fig2} Feedback-induced phase transition at a single trajectory. Conditional quadrature $\langle X\rangle_c$. For long-memory feedback [small $s=0.5$, panel (a)], approaching the transition at $G=G_\text{crit}$, the oscillatory trajectory becomes noisier and switches to slow growth. For fast feedback [large $s=20$, panel (b)], the oscillation frequency decreases (visualizing mode softening), and switches to fast growth. Even though the trajectories are stochastic, their frequencies and growth rates are the same for all experimental realizations. $g=\omega_R$, $\kappa=100\omega_R$, $\delta=\omega_R$, $\omega_Rt_0=1$. $h(0)=s$ gives the same $G_\text{crit}$ for all $h(t)$. Note the different scales on time axes. }
\end{figure}

Figure \ref{4-Fig2} compares trajectories for the spin quadrature $\langle X\rangle_c$ at various feedback constants $G$ and $s$ (\ref{4-h}). Crossing FPT critical point $G_\text{crit}$, the oscillatory solution changes to exponential growth. For large $s$ (nearly instant feedback), there is a frequency decrease before FPT and fast growth above it. For small $s$ (long memory), before FPT trajectories become noisier; the growth above it is slow. Note, that even though the trajectories are stochastic, their frequencies and growth rates are the same for all experimental realizations.

To get insight, we proceed with a minimal model necessary for FPT and adiabatically eliminate the light mode from Eq.~(\ref{4-Heis}): $a=(-2igX+f_a)/(\kappa+i\delta)$. This corresponds well to experiments \cite{EsslingerNat2010,EsslingerNature2016}, where $\kappa$ ($\sim$ MHz) exceeds other variables ($\sim$ kHz). The Heisenberg--Langevin equations for two matter quadratures then combine to a single equation describing matter dynamics:
\begin{gather}
\ddot{X}+\left(\omega^2_R  -\frac{4\omega_Rg^2\delta}{\kappa^2+\delta^2}\right)X -     \frac{4\omega_RGg\kappa}{\kappa^2+\delta^2}C_\theta \int_0^t{ h(t-z)X(z)dz}=F(t),\label{4-eqX}
\end{gather}
where $C_\theta=\delta \cos \theta+\kappa \sin \theta$. Here the frequency shift is due to spin-light interaction, the last term originates from the feedback. The steady state of Eq.~(\ref{4-eqX}) is $\langle X\rangle=0$, which looses stability, if the feedback strength $G>G_\text{crit}$. 

Note, that oscillations below $G_\text{crit}$ are only visible at quantum trajectories for conditional $\langle X\rangle_c$ (Fig. \ref{4-Fig2}). They are completely masked in the unconditional trivial solution $\langle X\rangle=0$. Thus, feedback can create macroscopic spin coherence $\langle X\rangle_c \ne 0$ at each single trajectory (experimental run) even below threshold. This is in contrast to dissipative systems, where the macroscopic coherence is attributed to $\langle X\rangle \ne 0$ above DPT threshold only.

The noise operator is $F(t)=-\omega_R[gf_a+G(\kappa-i\delta)e^{-i\theta}\int_0^t{h(t-z)f_a(z)dz/2}]/(\kappa+i\delta) +\text{H. c.}$ It has the following correlation function:
\begin{gather}
\langle F(t+\tau) F(t)\rangle=
\frac{\omega_R^2\kappa}{2(\kappa^2+\delta^2)}\{4g^2\delta(\tau)+     
 G^2(\kappa^2+\delta^2)\int_0^t h(z)h(z+\tau)dz +  \nonumber\\
2gG\left[(\kappa-i\delta)e^{-i\theta}h(\tau)+ 
(\kappa+i\delta)e^{i\theta}h(-\tau)\right]\}. \label{4-tau}
\end{gather}
We thus readily see how the feedback leads to the non-Markovian noise in spin dynamics, as the correlation function is not a delta-function.

Performing the Fourier transform of Eq.~(\ref{4-eqX}), one gets $D(\omega)\tilde{X}(\omega)=\tilde{F}(\omega)$, with the characteristic polynomial
\begin{eqnarray}\label{4-CharacterEq}
D(\omega)=\omega^2-\omega_R^2+\frac{4\omega_Rg^2\delta}{\kappa^2+\delta^2}+ 
\frac{4\omega_RGg\kappa}{\kappa^2+\delta^2}C_\theta H(\omega),
\end{eqnarray} 
where $\tilde{X}$, $\tilde{F}$, and $H(\omega)$ are transforms of $X$, $F$, and $h(t)$. The spectral noise correlation function is $\langle\tilde{F}(\omega)\tilde{F}(\omega')\rangle=S(\omega)\delta(\omega+\omega')$ with  
\begin{eqnarray}\label{4-SpNoise}
S(\omega)=\frac{\pi \omega_R^2\kappa}{\kappa^2+\delta^2}\left|2g+G(\kappa-i\delta)e^{-i\theta}H(\omega)\right|^2,
\end{eqnarray} 
whose frequency dependence again reflects the non-Markovian noise due to the feedback.

Even a simple feedback acting on spins leads to rich classical dynamics \cite{Kopylov2015}. Here we focus on the quantum case, but only for a simple type of phase transitions, where the eigenfrequency $\omega$ approaches zero \cite{Scarlatella2016} (``mode softening,'' visualized in quantum trajectories in Fig. \ref{4-Fig2}). From the equation $D(\omega)=0$ we find the FPT critical point for the feedback strength: 
\begin{eqnarray}\label{4-Gcrit}
G_\text{crit}H(0)=\frac{1}{4g\kappa C_\theta}[\omega_R(\kappa^2+\delta^2)-4g^2\delta],
\end{eqnarray}
where $H(0)=\int_0^\infty{h(t)dt}$. Without feedback ($G=0$) this gives very large $g_\text{crit}$ for LMG and Dicke transitions \cite{DomokosPRL2015, DomokosPRA2016}. Thus, feedback can enable and control these transitions, even if they are unobtainable because of large decoherence $\kappa$ or small light--matter coupling $g$.

\subsection{Tuning the quantum fluctuations, critical exponent, and universality class of phase transitions} 

We now turn to the quantum properties of FPT driven by the measurement-induced noise $F(t)$ (\ref{4-eqX}). While the mean-field solution is $\langle X\rangle=0$ below the critical point, $\langle X^2\rangle \ne 0$ exclusively due to the measurement fluctuations. From $D(\omega)\tilde{X}(\omega)=\tilde{F}(\omega)$ and noise correlations we get $\langle X(t+\tau)X(t)\rangle=\int_{-\infty}^\infty S(\omega)e^{i\omega\tau}/|D(\omega)|^2d\omega /(4\pi^2)$, giving  $\langle X^2\rangle$ for $\tau=0$.

\begin{figure}[h]
\captionsetup{justification=justified}
\centering
\includegraphics[clip, trim=0cm 0.3cm 0.35cm 0.37cm, width=0.9\textwidth]{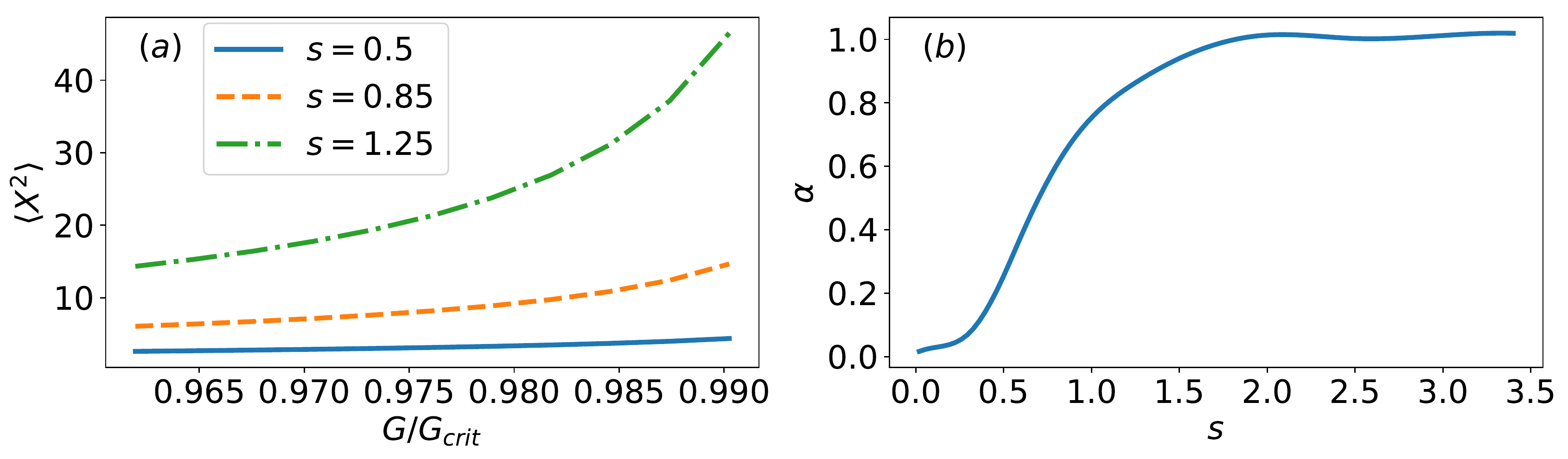}
\caption{\label{4-Fig3} Feedback control of the critical exponent of a phase transition (and of its universality class). (a) Growing fluctuations of unconditional matter quadrature $\langle X^2\rangle$ for various feedback exponents $s$. (b) Dependence of critical exponent $\alpha$ on feedback exponent $s$, proving opportunity for QPT control. $g=\omega_R$, $\kappa=100\omega_R$, $\delta=\omega_R$, $h(0)=s$, $\omega_Rt_0=1$.}
\end{figure}

To find the FPT critical exponent $\alpha$ we approximate the behavior near the transition point as $\langle X^2\rangle=A/|1-G/G_\text{crit}|^\alpha+B$, where $A,B=\text{const}$. Figure \ref{4-Fig3} demonstrates that the feedback can control the quantum phase transitions. Indeed, it does not only define the mean-field critical point (\ref{4-Gcrit}), but enables tuning the critical exponent as well. Varying the parameter $s$ of feedback response $h(t)$ (\ref{4-h}) allows one changing the critical exponent in a broad range and thus the universality class as well. This corresponds to varying the length of effective spin-spin interaction mentioned above. For $h(t)$  (\ref{4-h}), its spectrum is expressed via the exponential integral $H(\omega)=h(0)t_0e^{-i\omega t_0}E_{s+1}(i\omega t_0)$. At small frequencies its imaginary part behaves as $\omega^s$ for $s<1$, resembling the spectral function of sub-Ohmic baths. For large $s$, $\alpha$ approaches unity, as $h(t)$ becomes fast and feedback becomes nearly instant  such as interactions in open LMG and Dicke models, where $\alpha=1$ \cite{DomokosPRL2015, DomokosPRA2016,Oztop2012}. 

Note that a decaying cavity is well known to produce the autonomous exponential feedback \cite{St-K2017} $h(t)=\exp(-\kappa' t)$ [$H(\omega)=1/(i\omega+\kappa')$] crucial in many fields (e.g. lasers, cavity cooling, optomechanics, etc.) Such a simple $H(\omega)$ is nevertheless insufficient to tune the critical exponent and measurement-based feedback is necessary. 

The linearized model describes FPT near the critical point, but it does not give new steady state. The spin nonlinearity can balance the system (see below). However, the feedback with nonlinear $\mathcal{F}[x_\theta^{\text out}]$ can assure a new steady state even in a simple system of linear quantum dipoles (e.g. for far off-resonant scattering with negligible upper state population). It is thus the nonlinearity of the full hybrid quantum-classical system that is crucial.

\subsection{Reminder about quantum bath models}

In this section we give a concentrated reminder about the theory of bath models, which as we show can be simulated using the feedback. 

The Hamiltonian of the spin-boson model at zero temperature \cite{Leppakangas2018} extended for $N$ spins is 
\begin{eqnarray}\label{4-SB-Hamiltonian}
H=\sum_i\delta_i a_i^\dag a_i +\omega_R S_z +\frac{2}{\sqrt{N}}S_x \sum_i g_i(a_i^\dag+a_i).
\end{eqnarray}
It describes the interaction of spins with many (continuum) bosonic modes $a_i$ of frequencies $\delta_i$. The corresponding Heisenberg--Langevin equations are then given by
\begin{eqnarray} \label{4-HL-Res}
\dot{a_i}=-i\delta_i a_i -i\frac{2g_i}{\sqrt{N}}S_x, \nonumber\\
\dot{S_x}=-\omega_R S_y, \nonumber\\
\dot{S_y}=\omega_RS_x - \frac{2}{\sqrt{N}} S_z\sum_i g_i(a_i^\dag+a_i), \nonumber\\
\dot{S_z}=\frac{2}{\sqrt{N}}S_y\sum_ig_i(a_i^\dag+a_i).
\end{eqnarray}

Two equations for $S_x$ and $S_y$ can be joined: $\ddot{S_x} =-\omega_R^2S_x+2\omega_RS_z/\sqrt{N}\sum_i g_i(a_i^\dag+a_i)$. The modes $a_i$ can be formally found from the first Eq. (\ref{4-HL-Res}): $a_i(t) =a_i(0)e^{-i\delta_it}-2ig_i/\sqrt{N}\int_0^tS_x(\tau)e^{-i\delta_i(t-\tau)}d\tau$. This can be combined to $\sum_i g_i(a_i^\dag+a_i)=-4\int_0^\tau\beta(t-\tau)S_x(\tau)d\tau/\sqrt{N}-F_b(t)/\omega_R$, where
\begin{eqnarray}
\beta(t)=\sum_i g_i^2(\delta_i)\sin(\delta_i t)=\frac{1}{\pi}\int_0^{\infty}J(\omega)\sin\omega t d\omega, 
\end{eqnarray}
and the bath spectral function is
\begin{eqnarray}\label{4-J}
J(\omega)=\pi\sum_i g^2_i(\delta_i)\delta(\omega-\delta_i).
\end{eqnarray}
The noise operator $F_b(t)$ is determined by random initial values of the bosonic mode operators $a_i(0)$: $F_b(t)=-\omega_R\sum_i g_i [a_i(0)e^{-i\delta_it}+a_i^\dag(0)e^{i\delta_it}]$.

At the level of Heisenberg--Langevin equations, the linearized system can be obtained by assuming $S_z=-N/2$, $S_x=\sqrt{N}X$, and $S_y=\sqrt{N}Y$. Alternatively, it can be obtained from the Hamiltonian (\ref{4-SB-Hamiltonian}) using the Holstein--Primakoff representation as we explained before. The bosonized Hamiltonian then reads
\begin{eqnarray}
H=\sum_i\delta_i a_i^\dag a_i +\omega_R b^\dag b +(b^\dag+b) \sum_i g_i(a_i^\dag+a_i),
\end{eqnarray}
while the linearized equation for $S_x$ is reduced to the equation for the particle quadrature $X$:
\begin{eqnarray}\label{4-X-Res}
\ddot{X}+\omega_R^2 X- 4\omega_R\int_0^t \beta(t-z)X(z)dz = F_b(t).
\end{eqnarray}

Such Hamiltonian and operator equation correspond to the quadrature-quadrature coupling model and differ from the Caldeira-Leggett model only by the renormalized spin frequency $\omega_R$ \cite{breuer2002}. They describe the quantum Brownian motion as well \cite{breuer2002}. Equation (\ref{4-X-Res}) has a structure identical to Eq. (\ref{4-eqX}) obtained for the feedback model. The additional frequency shift in the case of feedback can be compensated either by modifying the spin frequency, or by adding the $\delta(t)$-term in the feedback response function $h(t)$. 

The bath spectral function $J(\omega)$ (\ref{4-J}) is usually approximated as $J(\omega)=\kappa_R(\omega/\omega_c)^sP_c(\omega)$, where $\omega_c$ is the cut-off frequency and $P_c(\omega)$ is the cut-off function. For $s=1$ the bath is called Ohmic, for $s<1$ it is sub-Ohmic, and for $s>1$ it is super-Ohmic. This corresponds to the bath response function asymptotically behaving as $1/t^{s+1}$ for large times \cite{LeHur2008}.  

Taking the Fourier transform of the differential equation (\ref{4-X-Res}) one gets
\begin{eqnarray}
\left(\omega^2-\omega_R^2+4\omega_RB(\omega)\right)X(\omega)=-\tilde{F_b}(\omega),
\end{eqnarray}
where $B(\omega)$ is the Fourier transform of $\beta(t)$. The spectral noise correlation function reads
\begin{eqnarray}
\langle\tilde{F_b}(\omega') \tilde{F_b}(\omega)\rangle=4\pi\omega_R^2J(\omega)\delta(\omega+\omega').
\end{eqnarray}

The time noise correlation function is
\begin{gather}
\langle F_b(t+\tau) F_b(t)\rangle=\frac{\omega_R^2}{\pi}\int_0^\infty J(\omega)e^{-i\omega\tau} d\omega = \nonumber \\ 
\frac{\omega_R^2}{\pi}\left[\int_0^\infty J(\omega)\cos\omega\tau d\omega - i\int_0^\infty J(\omega)\sin\omega\tau d\omega \right].
\end{gather}

A more standard way to write the differential equation is not via $\beta(t)$, but via $\gamma(t)$:
\begin{eqnarray}
\ddot{X}+\omega_R^2\left(1-\frac{2\gamma(0)}{\omega_R}\right) X+ 2\omega_R\int_0^t \gamma(t-z)\dot{X}(z)dz = F_b(t),
\end{eqnarray}
where $\dot{\gamma}(t)=-2\beta(t)$,
\begin{eqnarray}
\gamma(t)=\frac{2}{\pi}\int_0^{\infty}\frac{J(\omega)}{\omega}\cos\omega\tau d\omega. 
\end{eqnarray}
The frequency shift can be incorporated in the renormalized spin frequency. For the Ohmic bath, $\gamma(t)\sim \delta(t)$, and  the differential equation is reduced to that for a damped harmonic oscillator.

\subsection{Quantum bath simulators} 

Feedback control of QPTs enables simulating models similar to those for particle-bath interactions, e.g., spin-boson (SBM), Kondo, Caldeira-Leggett (CLM), quantum Brownian motion models. They were applied to various systems from quantum magnets to cold atoms with various spectral functions \cite{breuer2002, Leggett1987, LeHur2008,  DomokosPRL2015,  DomokosPRA2016, Scarlatella2016, Plenio2011}. Creating a quantum simulator, which is able to model various baths in a single device, is challenging, and proposals include, e.g., coupling numerous cavities or creating complex networks simulating multimode baths \cite{Leppakangas2018, Nokkala2016, Nokkala2018}. In contrast, the feedback approach is more flexible as tuning $h(t)$ of a single classical loop is feasible. E.g., for BEC (see below), the typical frequencies are in the kHz range, which is well below those of modern digital processors reaching GHz. Moreover, it can be readily extended for simulating broader class of quantum materials and qubits with nonlinear bath coupling \cite{ZhengPRB2018} and multiple baths \cite{VojtaPRL2012}.

The multi- (or large-) spin-boson models \cite{Anders2008, Winter2014, DomokosPRL2015,  DomokosPRA2016, Scarlatella2016} are based on Eq.~(\ref{4-Hamiltonian}) with sum over continuum of bosonic modes $a_i$ of frequencies $\delta_i$ distributed according to the spectral function $J(\omega)$. The feedback model reproduces exactly the form of bath dynamical equations for $S_{x,y,z}$ (for both linearized and nonlinear models discussed) if $\Im H(\omega)\sim J(\omega)-J(-\omega)$. The noise correlation function of linear CLM is $\langle\tilde{F}(\omega') \tilde{F}(\omega)\rangle=4\pi\omega_R^2J(\omega)\delta(\omega+\omega')$, whereas the feedback model contains $H(\omega)$ and additional light-noise term in Eq.~(\ref{4-SpNoise}). 

In bath models there is a delicate point of the frequency $\omega_R$ renormalization (``Lamb shift'') \cite{Leggett1987, DomokosPRL2015,  DomokosPRA2016, Scarlatella2016}. It may lead to divergences and necessity to repair the model \cite{Ford1988}. The feedback approach is flexible. The frequency shift in Eq.~(\ref{4-CharacterEq}) is determined by $GH(0)=G\int_0^\infty h(t)dt$ and can be tuned and even made zero, if $h(t)$ changes sign.

\subsection{Feedback-induced phase transition with nonlinear spins}

So far, we presented the properties of the phase transition using the linearized model with Hamiltonian (\ref{4-Hamiltonian-lin}). Such a linear model cannot give us the value of the stationary state above the critical point. Here we show the results of numerical simulations for a single nonlinear spin. The original Hamiltonian (\ref{4-Hamiltonian}) is 
\begin{eqnarray}\label{4-Hamiltonian-FB-2}
H=\delta a^\dag a +\omega_R S_z +\frac{2}{\sqrt{N}}S_x [g(a+a^\dag)+GI(t)].
\end{eqnarray}
The Heisenberg--Langevin equations are then given by
\begin{eqnarray} \label{4-HL-FB}
\dot{a}=-i\delta a -i\frac{2g}{\sqrt{N}}S_x-\kappa a +f_a, \nonumber\\
\dot{S_x}=-\omega_R S_y, \nonumber\\
\dot{S_y}=\omega_RS_x - \frac{2}{\sqrt{N}} [g(a+a^\dag)+GI(t)]S_z, \nonumber\\
\dot{S_z}=\frac{2}{\sqrt{N}}[g(a+a^\dag)+GI(t)]S_y.
\end{eqnarray}

\begin{figure}[h]
\captionsetup{justification=justified}
\centering
\includegraphics[clip, trim=0cm 0cm 0cm 0cm, width=0.5\textwidth]{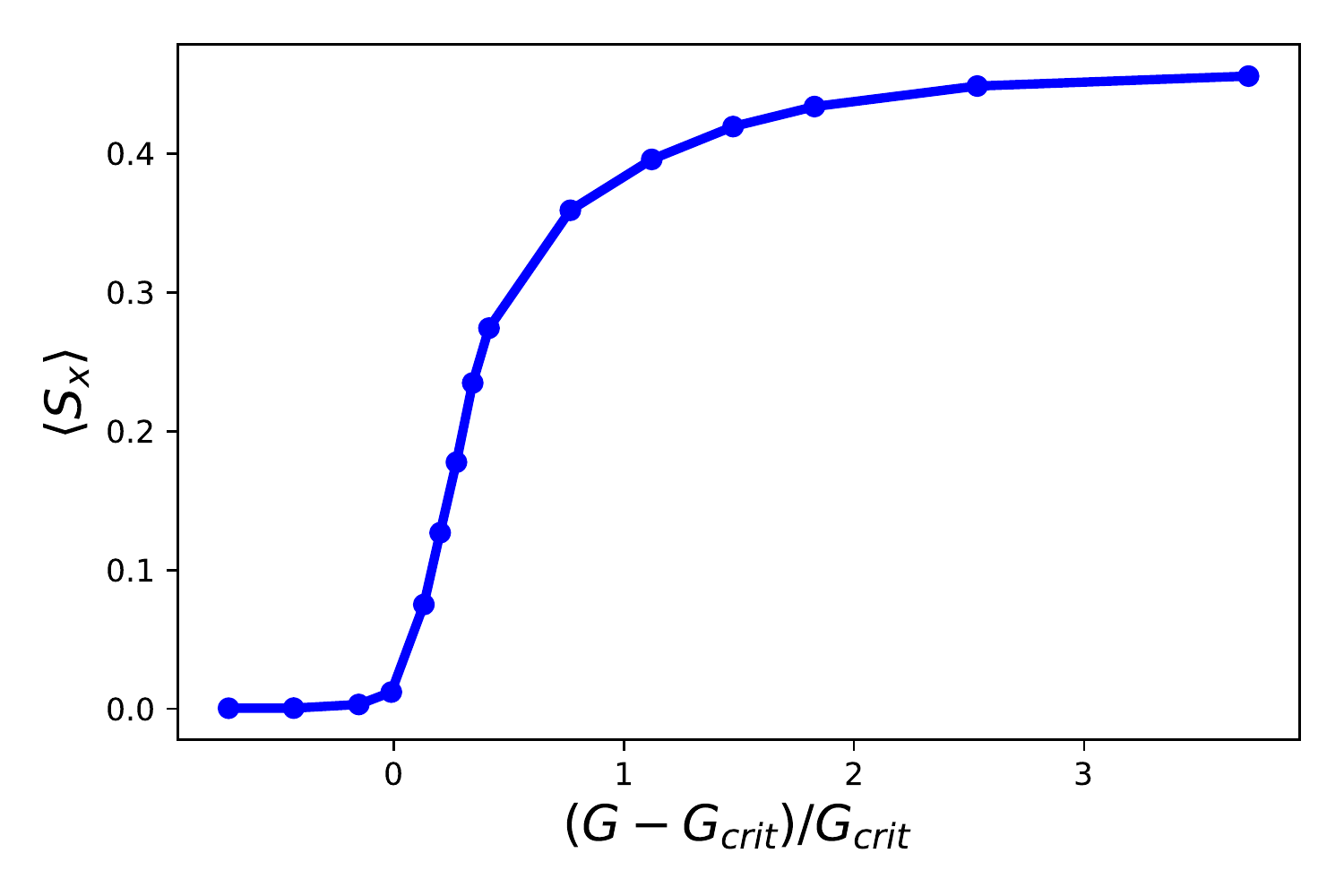}
\caption{\label{4-Suppl-Spin} Feedback-induced phase transition for a single spin. $\delta=\omega_R$, $g=0.1 \omega_R$,  $h(0)=s$,  $s=1$, $\kappa=10\omega_R$, $\omega_Rt_0=1$.} 
\end{figure}

Figure \ref{4-Suppl-Spin} shows the results of numerical simulations for the expectation value of the component $\langle S_x\rangle$ of a single spin ($N=1$). It shows the phase transition with $\langle S_x\rangle=0$ below $G_\text{crit}$ and  $\langle S_x\rangle \ne 0$ above the critical point. This is similar to the spin-boson model, where $\langle S_x\rangle \ne 0$ corresponds to the localization, while $\langle S_x\rangle = 0$ corresponds to the delocalized phase.

Such solutions can be obtained by calculating the conditional expectation values $\langle S_x(t)\rangle_c$ and then averaging them over multiple quantum trajectories, which gives the stationary unconditional expectation value $\langle S_x\rangle$. In general, the  measurement and feedback loop produce a single quantum trajectory. Averaging over many trajectories reproduces the results of quantum Heisenberg--Langevin equations (the equivalence between the quantum trajectory and quantum Heisenberg--Langevin approaches is demonstrated for feedback e.g. in Ref. \cite{Wiseman}). In turn, similar Heisenberg--Langevin approach describes the interaction of a particle with quantum baths as well.

The equations for $S_x$ and $S_y$ can be joined: $\ddot{S_x} =-\omega_R^2S_x+2\omega_RS_z/\sqrt{N}[g(a^\dag+a)+GI(t)]$. The light mode can be adiabatically eliminated from the first equation (\ref{4-HL-FB}) to give $g(a^\dag+a)+GI(t)=-[4g^2\delta S_x + 4g\kappa G C_\theta \int_0^t h(t-\tau)S_x(\tau)d\tau]/[\sqrt{N}(\kappa^2+\delta^2)]-F(t)/\omega_R$. These expressions have the same structure as those in the bath models. Various methods to treat the quantum Heisenberg--Langevin equations have been developed in quantum optics \cite{Davidovich,Andreev,ScullyBook}. The linearized equation for $S_x$ then reduces to Eq. (\ref{4-eqX}) for the particle quadrature $X$ ($S_x=\sqrt{N}X$):
\begin{eqnarray}
\ddot{X}+\left(\omega^2_R  -\frac{4\omega_Rg^2\delta}{\kappa^2+\delta^2}\right)X -     \frac{4\omega_RGg\kappa}{\kappa^2+\delta^2}C_\theta \int_0^t{ h(t-z)X(z)dz}=F(t),
\end{eqnarray}
which again has the same structure as the equation for $X$ in the bath model. The additional frequency shift in the case of feedback can be compensated either by modifying the spin frequency, or by adding the $\delta(t)$-term in the feedback response function $h(t)$. 

We would like to comment on the role of limited detector efficiency $\eta<1$. In the Heisenberg--Langevin approach, it can be taken into account by introducing an additional light noise corresponding to undetected photons leaked from the cavity, while the noise $f_a$ will still correspond to the detected photons (cf. Ref. \cite{Habibi}). The characteristic equation (\ref{4-CharacterEq}) will then take exactly the same form with $G$ replaced by $\sqrt{\eta}G$. Therefore, concerning the position of the critical point, the limited efficiency can be compensated by the increase of the feedback coefficient $G$. Indeed, when the detector efficiency approaches zero  ($\eta=0$) such that it can not be compensated in a real system, the role of feedback (detected photons) vanishes, and the feedback-induced phase transition reduces to a standard dissipative phase transition due to the undetected photons (i.e. the dissipation).

\subsection{Feedback control of quantum phase transitions in ultracold gases}

The model presented in this section can be realized using light scattering from the Bose--Einstein condensate (BEC). Two spin levels will correspond to two motional states of ultracold atoms (two matter waves with two different momenta). A very high degree of the light--matter interaction control has been achieved in several systems, where BEC was trapped in an optical cavity, and the Dicke and other supersolid-like phase transitions were obtained \cite{EsslingerNat2010,LevPRL2018, Zimmermann2018}. Experiments now include strongly correlated bosons in an optical lattice inside a cavity \cite{EsslingerNature2016, Hemmerich2015} and related works without a cavity \cite{Kettrle-selforg2017}. Here we propose one realization and underline that setups can be flexible and extendable for other configurations as well.

\begin{figure}[h]
\captionsetup{justification=justified}
\centering
\includegraphics[clip, trim=0cm 14cm 13cm 0cm, width=0.7\textwidth]{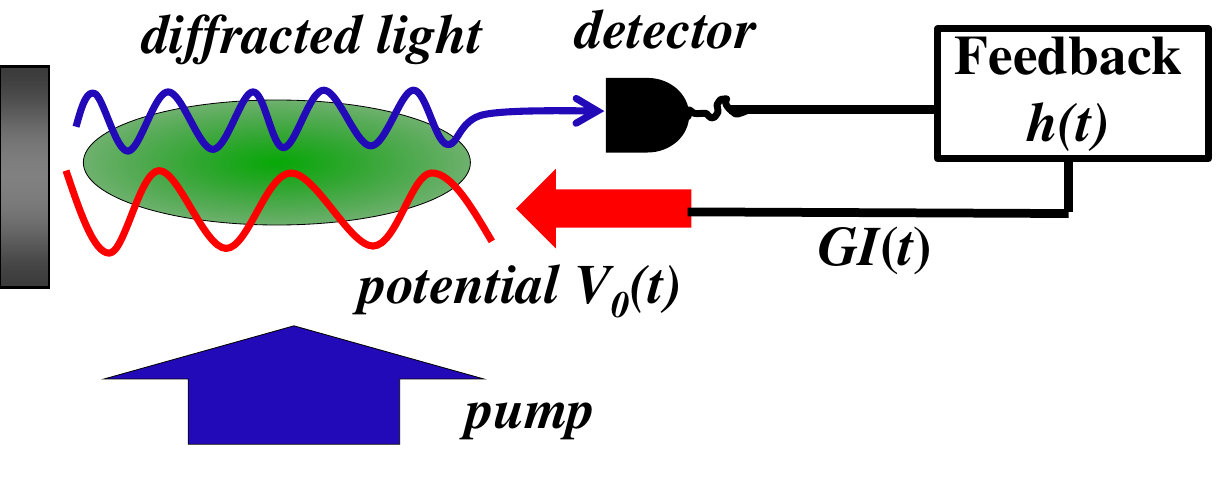}
\caption{\label{4-SupplSetup} Setup. A BEC is illuminated by the transverse pump, the scattered (diffracted) light is detected, feedback acts on the system via the change of the external periodic potential depth $V_0(t)$. Feedback provides the non-Markovianity, nonlinearity, and noise, necessary for the controllable  phase transition. Importantly, the feedback response $h(t)$ is tunable.} 
\end{figure}

We consider a BEC, elongated in the $x$ direction, illuminated by the pump (Fig. \ref{4-SupplSetup}).  (Note that in experiments one denotes as a pump the same laser beam, which we call the probe in this work). The quadrature of scattered light is measured and used for the feedback signal. The feedback is provided by the external trapping potential in the form of a standing wave, whose depth is varied according to the feedback signal: $V_0(t)\cos^2 k_0x$ ($k_0$ is the wave vector of laser beam creating the potential). The many-body Hamiltonian has a form (cf. review \cite{Mekhov2012}, we work in the units, where $\hbar=1$):
\begin{eqnarray}\label{4-Hamiltonian1}
H=\sum_l \omega_l a^\dag_l a_l +\int_0^L \Psi^\dag(x)H_{a1}\Psi(x)dx+\sum_l \zeta_l(a^\dag_l+a_l),
\end{eqnarray}
where $a_l$ are the annihilation operators of light modes of frequencies $\omega_l$ interacting with a BEC, $\zeta_l$ are the pumps of these modes (if they are shaped by cavities), $H_{a1}$ is the single-atom Hamiltonian, $\Psi(x)$ is the atom-field operator, and $L$ is the BEC length. For the far off-resonant interaction, $H_{a1}$ is determined by the interference terms between the light fields present \cite{Mekhov2012}:
\begin{eqnarray}\label{4-Hamiltonian-a1}
H_{a1}=\frac{p^2}{2m_a}+V_0(t)\cos^2k_0x+\frac{1}{\Delta_a}\sum_{l,m} g_l u^*_l(x)a^\dag_l  g_m u_m(x)a_m,
\end{eqnarray}
where the first term is atomic kinetic energy operator ($p^2=d^2/dx^2$, $m_a$ is the atom mass), $\Delta_a$ is the detuning between light modes and atomic transition frequency, $g_{l,m}$ is the light--matter coupling constants. $u_{l,m}$ are the geometrical mode functions of light waves, which can describe pumping and scattering at any angles to the BEC axis \cite{MekhovPRL2007,MekhovPRA2007}, which maybe convenient depending on the specific experimental realization. 

Here, to strongly simplify the consideration, we select the following geometry of light modes (cf. Fig. \ref{4-SupplSetup}). The pump is orthogonal to the BEC axis, thus its mode function is constant along $x$ (can be chosen as $u(x)=1$); its amplitude $a_\text{pump}$ is considered as a c-number. A single scattered light mode $a_1$ is non-negligible along $x$ direction, and its mode function is $u_1(x)=\cos k_1x$, where $k_1$ is the mode wave vector. There is no direct mode pumping, $\zeta_1=0$. The condition $k_0=k_1/2$ assures the maximal scattering of light into the mode $a_1$ (diffraction maximum). Thus, the pump diffracts into the mode $a_1$ from the atomic distribution. Thus, the wavelength of feedback field should be twice the mode wavelength. The matter-field operator can be decomposed in two modes \cite{DomokosPRL2015, DomokosPRA2016}:
\begin{eqnarray}\label{4-Psi}
\Psi(x)=\frac{1}{\sqrt{L}}c_0+\sqrt{\frac{2}{L}}c_1\cos k_1x,
\end{eqnarray}
where $c_{0,1}$ are the annihilation operators of the atomic waves with momenta $0$ and $k_1$ ($c^\dag_0c_0+c^\dag_1c_1=N$, $N$ is the atom number). Substituting Eqs. (\ref{4-Psi}) and (\ref{4-Hamiltonian-a1}) in Eq. (\ref{4-Hamiltonian1}) and neglecting in a standard way several terms \cite{DomokosPRL2010,EsslingerNat2010,DomokosPRL2015, DomokosPRA2016} (which however may appear to be important under specific conditions and thus enrich physics even further), we get the Hamiltonian (\ref{4-Hamiltonian})
\begin{eqnarray}\label{4-Hamiltonian-FB}
H=\delta a_1^\dag a_1 +\omega_R S_z +\frac{2}{\sqrt{N}}S_x [g(a_1+a_1^\dag)+GI(t)]
\end{eqnarray}
 with the following parameters: $\delta=\omega_1-\omega_\text{pump}+Ng_1^2/(2\Delta_a)$ (the detuning between mode and pump including the dispersive shift), $\omega_R=k_1^2/2m_a$ is the recoil frequency, $g=\Omega_\text{pump} g_1\sqrt{N/2}/\Delta_a$ with $\Omega_\text{pump}=g_\text{pump}a_\text{pump}$ being the pump Rabi frequency. The feedback signal is $GI(t)=\sqrt{N/8}V_0(t)$. The spin operators are $S_x=(c^\dag_1c_0+c_1c^\dag_0)/2$, $S_y=(c^\dag_1c_0-c_1c^\dag_0)/(2i)$, $S_z=(c^\dag_1c_1-c^\dag_0c_0)/2$. The main characteristic frequency of this system is the recoil frequency that for BEC experiments is $\omega_R=2\pi \cdot 4$kHz. This makes the feedback control feasible, as the modern digital procossing of the feedback signal can be much faster (up to the GHz values). Other experimental parameters such as $\delta$ and $g$ (depending on the pump amplitude) can be tuned in the broad range and, in particular, be close to kHz values. The cavity decay rate in Ref.  \cite{EsslingerNat2010} is $\kappa=2\pi\cdot 1.3$ MHz, which is much greater than kHz making the adiabatic elimination of the light mode to work very well.
 
There are other configurations relevant to our work. For example, for the BEC in a cavity setup \cite{EsslingerNat2010}, instead of creating the external potential $V_0$ with $k_0=k_1/2$, one can inject the feedback signal directly through the cavity mirror as the pump $\zeta_1(t)$. In this case, an additional laser with doubled wavelength is not necessary. While the Hamiltonian Eq. (\ref{4-Hamiltonian-FB}) will be somewhat different, after the adiabatic elimination of light mode, the equation for the matter quadrature operator $X$ (\ref{4-eqX}) will be the same with $\zeta_1(t)\sim I(t)$.

Our proposal can be realized using the lattice system, which we have presented in all chapters of this work.  In this case, instead of expanding $\Psi(x)$ in the momentum space, a more appropriate approach is to expand it in the coordinate space using localized Wannier functions. The spin operators will then be represented by sums of on-site atom operators $S_x\sim \sum_i J_{ii} n_i$, or the bond operators representing the matter-wave interference between neighboring sites $S_x\sim \sum_i J_{ij} b^\dag_i b_j$ \cite{Caballero2015,CaballeroNJP2016}. For example, the effective spin can correspond to the atom number difference between odd and even sites [for $J_{ii}=(-1)^i$] \cite{Mazzucchi2016PRA} (as in Chapter 3), it can represent the magnetization \cite{Mazzucchi2016PRA} or staggered magnetization \cite{Mazzucchi2016SciRep} of fermions, etc. (as we have seen in Chapter 3 as well). Such a strong flexibility in choosing the geometrical combination of many-body variables combined in the effective spin operator enables defining macroscopic modes of matter fields \cite{Elliott2015} and assures the competition between the long-range (but structured in space with a short period comparable to that of the lattice) light-induced interactions and short-range atom-atom interactions and tunneling on a lattice \cite{Mazzucchi2016PRA}. This will open the opportunities for the competition between the nontrivial feedback-induced interactions and many-body atomic interactions.

We believe that our proposal will extend the studies in the field of time crystals \cite{Sacha2017,UedaTC2018,Demler2019,JakschNCom2019,JakschPRA2019} and Floquet engineering \cite{Eckardt2017}. The time crystals is a recently proposed notion, where phenomena studied previously in space (e.g. spin chains, etc.) are now studied in the time dimension as well. Typically, the system is subject to the external periodic modulation of a parameter [e.g. periodic $g(t)$], which is considered as creating a “lattice in time”. Our approach makes possible introducing the effective “interaction in time.” This makes the modulation in the system not prescribed, but depending on the state of the system (via $S_x$). This resembles a true lattice in space with the interaction between particles. In other words, our model enables not only creating a “lattice in time,” but introducing the tunable “interaction in time” to such a lattice (without the necessity of having the standard particle-particle interaction in space).
 
Recently, the first experiment, where our predictions can be tested was reported in Ref. \cite{EsslingerFeedback}.

In summary of this section, we have shown that feedback does not only lead to phase transitions driven by the fundamental quantum measurement fluctuations, but controls its critical exponent and universality class. It induces effects similar to those of quantum bath problems, allowing their realization in a single setup, and enables studies of time crystals and Floquet engineering with long-range (long-memory) interactions. Experiments can be based on quantum many-body gases in a cavity \cite{EsslingerNat2010, EsslingerNature2016, Hemmerich2015, LevPRL2018, Zimmermann2018, Mazzucchi2016Opt}, and circuit QED, where ultra-strong coupling has been obtained \cite{Yoshihara2017, Forn2017} or effective spins can be considered \cite{Leppakangas2018, HuardUltraS2018,Ustinov2017}. Feedback methods can be extended by, e.g., measuring several outputs \cite{Hammerer2016PRA, Hacohen2016,HuardNCom2018} (enabling simulations of qubits and multi-bath SBMs \cite{VojtaPRL2012} with nonlinear couplings \cite{ZhengPRB2018}) or measuring various many-body atomic \cite{Elliott2015, Mazzucchi2016SciRep, Kozlowski2017, Mazzucchi2016Opt} or molecular \cite{MekhovLP2013} variables, which we proposed in this work. The first experiment, where our predictions can be tested, appeared in Ref. \cite{EsslingerFeedback}.


\section{Concluding remarks of Chapter 4}

Open dissipative systems have already shifted the paradigm of many-body physics and quantum information. In this chapter, we made the next step advancing these fields by exploiting the quantum nature of the measurement process in many-body systems.  The concept of fundamental measurement is broader than dissipation: the latter is its special case, where the measurement results are ignored. We demonstrated that measurements and feedback induce phase transitions beyond the dissipative ones. Moreover, we proved that tuning the universality class of the phase transitions is possible. Such transitions are driven by the fundamental measurement noise, originating from the incapability of any classical device to capture the superpositions and entanglement of the quantum world. Considering ultracold bosons and fermions, we demonstrated the phase transitions and stabilization of density wave and antiferromagnetic oscillations. In general, quantum measurements and feedback open a way to obtain novel phenomena untypical to both close unitary systems and open dissipative ones. Recently, the first experiment, where our predictions presented in this chapter can be tested, has been reported in Ref. \cite{EsslingerFeedback}. 

It will be intriguing to study, how more advanced methods than feedback control can influence quantum systems, for example, applying the digital methods of machine learning and artificial intelligence in real time.

So far, considering the quantum nature of light we focused on the backaction of light measurement on the state of a quantum gas. In the next chapter we will consider systems, where even the lattice trapping potential is quantum.

\clearpage

%% file: Chapter5.tex
\chapter{Quantum optical lattices}\label{chapt5}

\section{Introduction and plan of the chapter}

In previous chapters the atoms were assumed to be trapped in a prescribed potential formed by stationary laser light, which can be described by a c-number function. We have shown, how additional light probes can measure or modify the properties of the atomic quantum state due to the light--matter entanglement. In those examples, the quantum natures of both the atomic motion and probe light were important, whereas the potential was treated classically.

In this chapter, we address another regime, where the quantum nature and fluctuations  of the trapping light potential itself play a key role \cite{MekhovEPJD08,Caballero2015,CaballeroPRA2016}. Indeed, the potential is created by light, which is a quantum object. The goal is to consider the systems and phenomena, where the quantum nature of the potential cannot be neglected and the trapping potential is a quantum dynamical variable. It must be determined self-consistently with the solution for the quantum states of atoms trapped in that potential.

A natural implementation to study an ultracold quantum gas in the  quantized light is to load  atoms into a high-Q cavity. In this case, the light mode of the cavity (e.g., a single standing wave, in the simplest case) will form the trapping potential. The quantum properties of the light mode become more important, when the light intensity (the number of photons in a cavity) gets smaller and smaller. The relative value of quantum fluctuations then increases strongly, because the mean light amplitude can decrease down to zero. The striking property of the cavity configuration is that, although the light intensity decreases almost to zero, the depth of the optical potential in a cavity can be still kept very large. Indeed, the potential depth is given by the product of the photon number and the light--matter coupling coefficient. For example, the quantum potential formed by a single standing wave mode $a_1$ with the wave vector $k$ is given by $\hbar U_{11}a^\dag_1a_1\cos^2{kx}=\hbar (g^2_{1}/\Delta_a)a^\dag_1a_1\cos^2{kx}$ (see Eq. (1.3) and the disscustion after it), where $g_1$ is the light--matter coupling coefficient and $\Delta_a$ is the frequency detuning between the light and atomic resonance. In a cavity, the light-atom coupling coefficient can reach huge values. Thus, even the light field of a single photon can lead to a very deep optical potential inside a cavity. The trapping and cooling of atoms by a single photon inside a cavity has been already demonstrated experimentally \cite{Pinkse2000, HoodScience2000} (nevertheless, the atom in those works was not cold enough for its motion being quantized).

As we have shown in this work, the quantum states of atoms and light strongly depend on each other. For example, the presence of atoms shifts the cavity resonance and modifies the light amplitude because of light scattering. On the other hand, the atoms are trapped in that light field, which requires a self-consistent solution for the coupled light--matter quantum dynamics. Importantly, quantum mechanics allows the superpositions of several Fock states of photons forming the quantum potential. Therefore, one can consider a superposition of potentials of several depths. This rises an intriguing question about the possibility to have the superposition of several atomic phases, each of which is correlated to different Fock state of light (i.e. to the potentials of different depths). For example, one can imagine the potential provided by a very weаk cavity mode, where the Fock states of zero photons (the vacuum field), one photon and two photons are significant. Then, in principle, one can have a superposition, of the free particles (correlated with the zero photon number Fock state), the superfluid state (correlated with a potential provided by a single photon) and Mott insulator state (correlated with the two-photon Fock state). Of course, the stability of such an intriguing superposition and its robustness with respect to the decoherence should be carefully analyzed.

Another remarkable property of the quantum optical lattices in a cavity is the existence of the long-range interaction between the atoms beyond the standard Hubbard models. As demonstrated by the Hamiltonian and Heisenberg equations in Sec. 1.2, even if the tunneling is non-negligible only between the neighboring sites (as it is usual for the Bose--Hubbard model), even very distant atoms can still interact with each other via the common cavity light mode. As a consequence, the tunneling of very distant atoms gets correlated (sometimes, called co-tunneling). Mathematically, this is expressed by the coefficients in the generalized Bose--Hubbard model, which are not constants, but the operators depending on the whole set of light and atomic operators in the whole extended optical lattice. Thus, the cavity mediates the long-range interaction, whose properties can be tuned by the cavity parameters. Even for the simplest type of the neighboring interaction, the parameters of the Bose--Hubbard model influencing tunneling and atom-atom interaction can be controlled by the parameters such as probe-cavity detuning, probe intensity, probe-atom detuning, cavity linewidth, etc.

In Sec. 5.2  \cite{MekhovEPJD08, Mekhov2012} we start by giving an example of quantized atomic motion in a fully quantum optical lattice, where no classical potential is present. The fully quantum potential is difficult to treat even numerically, so we give only examples for the small atom numbers. We show, how the quantum fluctuations of light destroy the Mott insulator state and shift the Mott insulator -- superfluid (MI--SF) phase transition. The mean photon number in one of the examples is less than one, which provides very strong fluctuations of the quantum trapping potential. 

In all other sections, instead of straightforward numerical simulations we build an approximate mean-field theory, which enables us treating the macroscopic quantum gas and get more physical insight in the problem.

In Sec. 5.3  \cite{Caballero2015, CaballeroPRA2016} we clarify the three different types of optical lattices according to the role of light that creates them: classical lattices (used in almost all experiments and theories so far), dynamical, or semiclassical, lattices (demonstrated in several experiments with trapped BECs and fermions inside   optical cavities), and quantum lattices (not yet demonstrated, but are expected to be created in cavity systems). Our main achievement is predicting phenomena due to the quantum lattices and separating them from other simpler effect. As quantum effects will be observed in cavity setups, we study and predict new phenomena for dynamical lattices as well.

In Sec. 5.4  \cite{Caballero2015, CaballeroPRA2016} we introduce macroscopic spatial modes of atoms and explain their interaction. We bridge physics of systems with global collective and short-range interactions. This assures that novel quantum phases will posses properties of both types of systems.

In Sec. 5.5  \cite{CaballeroPRA2016} we present rather technical details of derivations of effective Hamiltonians. In the next sections we will focus more on the physical side of our results.

In Sec. 5.6  \cite{Caballero2015, Atoms, CaballeroNJP2016, CaballeroPRA2016} we present novel phases and modifications of known phases due to the quantum and dynamical natures of optical lattices. Here we consider light scattering in the orthogonal configuration of the probe and lattice. In this section we consider light scattering from atomic on-site densities. For quantum lattices, we predict the shift of the MI--SF phase transition induced by the quantum fluctuations of light.  For dynamical lattices, we predict particular properties of the lattice supersolid and density wave states, which were confirmed experimentally.

In Sec. 5.7 \cite{Caballero2015, CaballeroNJP2016, CaballeroPRA2016} we present results on the scattering from the matter-wave patterns (the so-called bonds), rather than from onsite densities considered in the previous section. We find novel phases of superfluid bond dimers and supersolid bond dimers. Moreover, we show that the paradigmatic short-range supersolid state can be obtained even if the light-induced interactions are long-range, which distinguishes this state from other lattice supersolids described in the previous section. This can be obtained due to the quantumness of an optical lattice.

In Sec. 5.8 \cite{Caballero2015, CaballeroPRA2016} we present results for scattering at the diffraction maximum from both densities and bonds. We demonstrate the shift of the MI--SF transition point and generation of a particular gapped quantum superposition state.

In Sec. 5.9 \cite{Caballero2015, CaballeroPRA2016} we consider many spatial atomic modes generated at the same time. We predict interplay between the supersolids and density waves of various periods.

In Sec. 5.10 \cite{Caballero2015, Caballero2015a} we briefly present our results on the generation of squeezed light from our system, which is possible, when the adiabatic elimination of light is not used. We show the full entangled light--matter state. We underline that the properties of novel quantum phases get imprinted on the quantum properties of scattered light, which can be measured without destroying the atomic system.

In Sec. 5.11 \cite{CaballeroPRA2016} we present our general view on the opportunities to build quantum simulators based on the collective light--matter interaction. We prove that if multiple probes and cavity modes are used, one can tune the effective interaction length between the atomic modes. This enables simulating systems with tunable long- and short-range interactions, which is extremely difficult to achieve in other physical systems.

In Sec. 5.12, we provide final conclusions of this chapter.

{\it The results of this chapter are based on the papers}  \cite{MekhovEPJD08, Mekhov2012, Caballero2015, Caballero2015a, Atoms, CaballeroNJP2016, CaballeroPRA2016}.


\section{Small atom number in a fully quantum potential}

In this section, we give some examples of numerical simulations for fully quantum potential (the classical trapping potential is zero, $V_\text{cl}=0$), but the small atom number. In the next section we will present a model for macroscopic atomic systems, but in addition to the quantum potential a classical potential will be present as well. 

\begin{figure}[h!]
\captionsetup{justification=justified}
\centering
\includegraphics[width=0.4\textwidth]{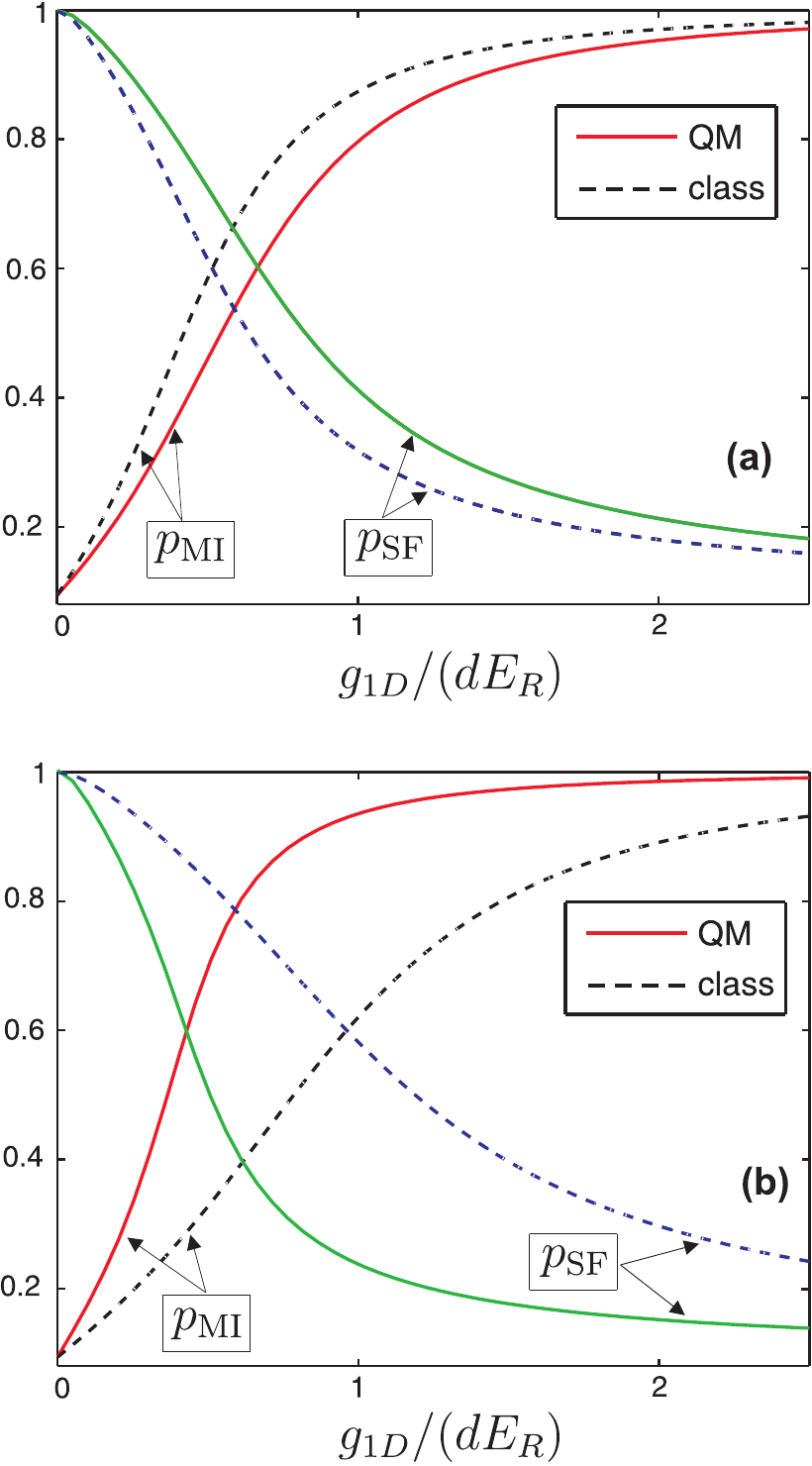}
\caption{\label{5-fig7JPB} Mott insulator to superfluid phase transition in a quantum optical lattice in a finite system. The influence of cavity on the transition is demonstrated by the comparison of the occupation probabilities $p_{\textrm{MI}}$ and $p_{\textrm{SF}}$ (to find the states $|\Psi_\text{MI}\rangle$ and $|\Psi_\text{SF}\rangle$) for a purely quantum field, i.e., $V_{cl}=0$, and a purely classical field, i.e., $\eta_1=0$, as a function of the dimensionless 1D on-site interaction strength $g_{1D}/(dE_R)$ ($d$ is the lattice period, $E_R$ is the recoil energy). Four atoms in four wells. We choose $\eta_1$ such that both potentials are of equivalent depth, $V=5.5E_R$, for zero on-site interaction ($g_{\textrm{1D}}=0$). The quantum (QM) and classical (class) cases are depicted with solid and dashed lines, respectively. $(U_{11},\kappa,\eta_1)=(-1,1/\sqrt{2},\sqrt{5.5})\omega_R$, where the recoil frequency is $\omega_R=E_R/\hbar$. The detuning between the probe and cavity frequencies (counted from the mean dispersive frequency shift) affects the position of the phase transition. In (a) this detuning is positive $\Delta_p-U_{11}N=\kappa$ and the transition point is shifted towards higher interaction strengths in comparison to that in a classical lattice. In (b) the detuning is negative $\Delta_p-U_{11}N=-\kappa$ and the transition point is shifted towards the smaller interaction strengths. (Joint work \cite{MekhovEPJD08}.)}
\end{figure}

In a joint paper \cite{MekhovEPJD08}, we analyzed a simple but fundamental example: the quantum phase transition between the MI and SF states in a fully quantum optical lattice in a finite system, significantly extending and correcting the preliminary model \cite{Maschler2005}. Figure \ref{5-fig7JPB} compares the phase transitions in quantum and classical optical lattices of the same mean potential depths. From the classical point of view, the lattices of the same depth assure the same physics. However, in quantum lattices, the potential depth is not the only important parameter. In particular, near the phase transition point the photon number fluctuations play a key role. More precisely, if the transition occurs for the potential depth given by the mean photon number $n_{\Phi}$, already the photon numbers $n_{\Phi}\pm 1$ are associated with different atomic phases. Thus the photon fluctuations drive the atomic fluctuations and hence the phase transition. Depending on the cavity parameters (e.g. the cavity frequency), the photon fluctuation can either suppress or enhance the atomic fluctuations and atomic hopping, therefore driving the system towards or outwards of the MI or SF states. In Fig. \ref{5-fig7JPB} one sees that the position of the phase transition in a cavity can be shifted towards either smaller or larger values of the atom-atom interaction strengths, depending on the detuning between the probe light and the cavity resonance. (The probe-cavity detuning counted from the mean dispersive shift of the cavity resonance is positive in Fig. \ref{5-fig7JPB}(a) and negative in Fig. \ref{5-fig7JPB}(b)). This demonstrates the tunability of the phase transition properties by the cavity parameters.

\begin{figure}[h!]
\captionsetup{justification=justified}
\centering
\includegraphics[width=0.4\textwidth]{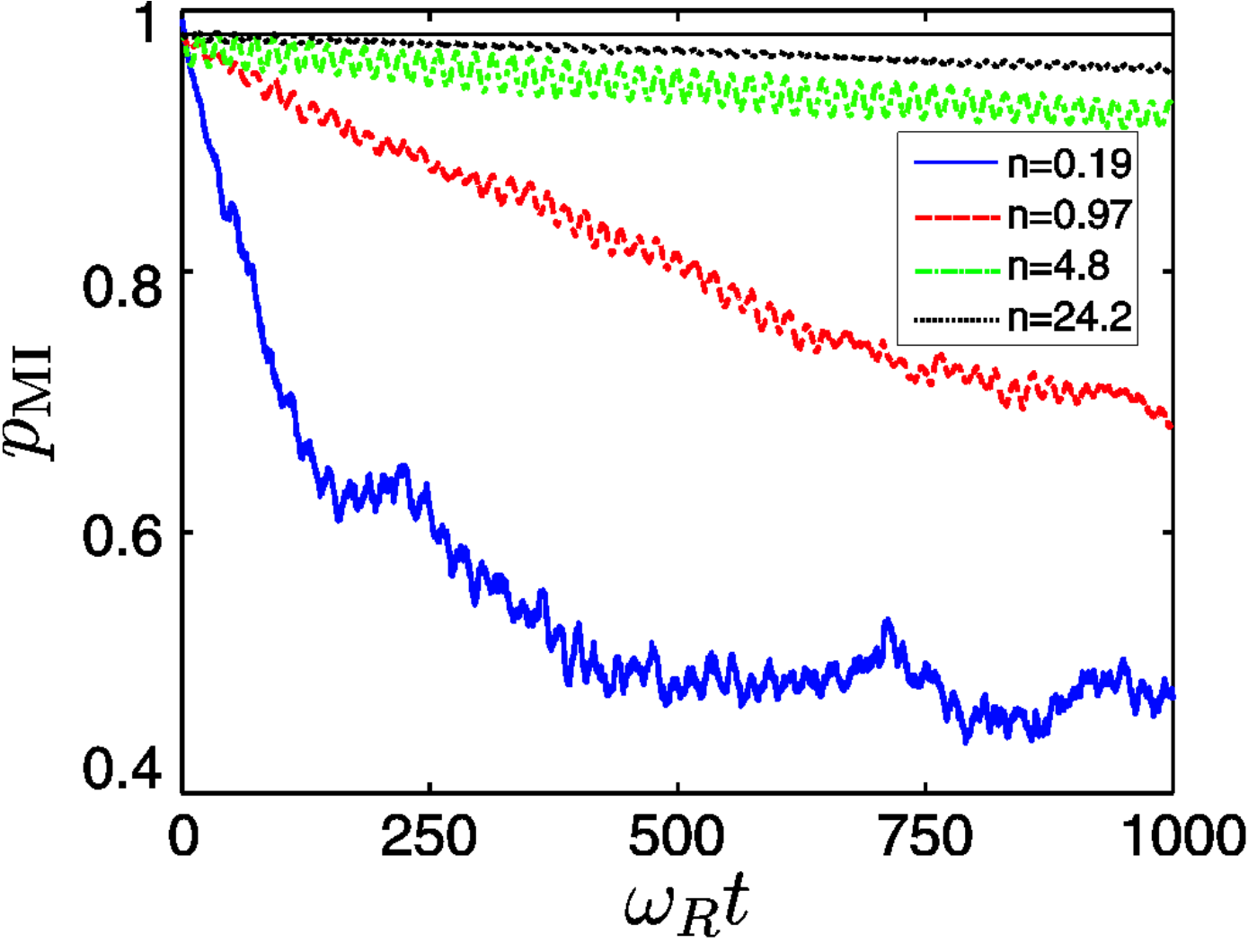}
\caption{\label{5-fig8JPB} Dynamics of the quantum state in a quantum potential. Probability $p_\text{MI}$ to find the MI state $|\Psi_\text{MI}\rangle$ for various photon numbers forming the quantum potential. In all cases the potential depth is the same ($8E_R$). Two atoms in two wells. Classically, the same potential depth would mean the same physics. However, in a quantum optical lattice, the evolution for the same depth, but different photon numbers is strikingly different. For the classical potential (constant line) the system stays in the MI state without any change. Decreasing the photon number in a cavity (the mean photon numbers are 24.2, 4.8, 0.97 and 0.19), the role of light fluctuations increases, which drives the atomic fluctuations destroying the MI state. The smaller the mean photon number, the stronger the system evolution deviates from the case of a classical optical potential. (Joint work \cite{MekhovEPJD08}.)}
\end{figure}

Figure \ref{5-fig8JPB} compares dynamics of the atomic states in classical and quantum lattices of various photon numbers. Importantly, the mean potential depths are chosen to be the same, so classically one should not expect any difference between all those curves. In contrast, one observes a striking difference in the evolutions of the probability to find the system in the Mott insulator state for different cavity photon numbers. First, for the classical potential there is no evolution and the system which started in the MI remains in it. For large photon numbers, the MI state gets destroyed with time, but does not significantly deviate from the classical solution (because the quantum fluctuations are much smaller than the mean photon numbers). However, for small photon numbers, the fluctuations are larger than the mean value and the system leaves the MI state completely. This is another example of how the photon fluctuations induce the atom hopping and atom fluctuations and radically change the quantum state of the system. Fluctuations towards lower photon number enhance tunneling and destroy the Mott phase.

Concerning the tunneling dynamics, the fundamental difference of the quantum lattices from their classical analogues is that the coupling of atoms to the leaking light mode opens a new dissipation channel for atomic dynamics. In quantum simulations with classical optical lattices, the main advantage of the ultracold atoms in contrast to the condensed matter systems is their almost complete isolation from the environment. Using the quantum optical lattices one can make a step further. The decoherence can be introduced in the atomic system in a very careful and controllable way: as we have shown in the previous chapters, the decoherence (via the measurement backaction) can be tailored using the optical geometry. 

After presenting numerical results on fully quantum lattices for small systems in this section, we will now turn to macroscopic atomic systems, where more analytical insight is possible.


\section{Classical, dynamical (semiclassical), and fully quantum optical lattices}

In this section, we clarify the physical context of three distinctive types of optical lattices (OL), depending on the degree of experimental complexity: 

(I) Classical optical lattices (COL), where the light potential is totally fixed or is changed externally (e.g. modulated). This is a case used in almost all quantum gas experiments to date and it is not a subject of this work. 

(II) Dynamical (or semiclassical) optical lattices (DOL), where the light scattering from atoms is strong enough to modify the trapping potential itself. Thus the self-consistent problem of a coupled dynamical system ``light potential -- quantum gas'' arises. This case was realized by several groups, where bosons  \cite{EsslingerNature2016,Hemmerich2015, Zimmermann2018, LevPRX2018,Naik2018} and more recently fermions \cite{Brantut2020} were trapped inside a cavity, which enhances the light--matter coupling. Nevertheless, the quantum effects of the trapping light have not been observed to date even in these advanced experiments: the light potential is dynamical, but its quantum properties are of no importance yet. In this work, we predict several novel quantum phases in dynamical OLs.

(III) Quantum optical lattices (QOL). The case of dynamical OL, but where the quantum properties of light affect the quantum many-body atomic state and vise versa. This regime, not yet observed experimentally, is the main target of our work. We demonstrate novel quantum phases of atoms solely due to the quantization of light potentials.

Modern experiments on ultracold atoms trapped in optical lattices enable to study quantum many-body phases with undeniable precision and target problems from several disciplines. Such optical potentials can be complicated in geometry, but are prescribed, i.e., they are created by external lasers and are not sensitive to atomic states. This limits the range of obtainable phases. 

Self-consistent light--matter states can be obtained, when scattered light modifies the trapping potential itself. This was achieved by trapping a Bose--Einstein condensate (BEC) ~\cite{EsslingerNat2010,HemmerichScience2012,ZimmermannPRL2014,LevPRL2019} and ultracold fermions \cite{Brantut2020} inside an optical cavity, which dramatically enhances the light--matter coupling, thus making the influence of reemission light comparable to that of external lasers. Such ``dynamical potentials" \cite{ritsch2013} enabled the structural Dicke phase transition and a state with supersolid properties~\cite{EsslingerNat2010}. A key effect observed so far is the dynamical dependence of light intensity (potential depth) on the atomic density. Although, the light becomes dynamical, its quantum properties are still not totally exploited even theoretically as works on atomic motion in quantum light were limited to few atoms~\cite{MekhovEPJD08,VukicsNJP2007,Kramer2014,RitschPRA2015}. Effects in dynamical OLs are analogous to semiclassical optics, where atomic excitations are quantum, while light is still classical. As the light and BEC are quantum objects, the quantum fluctuations of both were studied~\cite{DomokosPRA2014,Landig2015}, however, the fundamental reason of the structural Dicke phase transition can be traced back to the dynamical self-organization predicted~\cite{DomokosPRL2002} and observed~\cite{Vuletic2003} with thermal atoms and classical light. For single-mode cavities, dynamical light--matter coupling was shown to lead to several effects~\cite{Larson,Morigi2010,Hofstetter,Reza,Morigi2} yet to be observed with bosons and with non-interacting fermions~\cite{Keeling2014superradiance,Piazza2014superradiance,Chen2014superradiance}.  Multimode cavities extend the range of quantum phases further~\cite{Gopalakrishnan2009,Strack,Muller2012, Kramer2014, Kollar2015,LevPRX2018}. More recently optical lattices inside a cavity has been realized~\cite{EsslingerNature2016,Hemmerich2015}, where not only the Mott insulator and superfluid phases were obtained, but also the lattice supersolid and density waves. We present this type of states as well, and show that they are due to the dynamical rather than quantum character of light potentials. 

Importantly, in this work we show that, even in a single-mode cavity, the truly quantum potential leads to significant many-body effects beyond dynamical (semiclassical) ones. We demonstrate that the multimode spatial patterns of matter fields arise due to symmetry breaking resulting from the competition between imposed global light structure and standard local processes (tunneling and on-site interactions). We demonstrate that the efficient competition is  achieved due to the ability to structure the global interaction at a microscopic scale consistent with the lattice period. The physical origin of the atomic modes is the same as we discussed in Chapters 2 and 3. Nevertheless, here we will not focus on the measurement backaction, but rather on the interaction between such modes described by effective Hamiltonians. Such a competition in turn leads to novel many-body states, which are not limited to density-induced orders as in previous studies, but also represent long-range patterns of matter-field coherences (bonds~\cite{Lauchli2014}), leading, e.g., to far delocalized dimers, trimers, etc.  Importantly, we prove that our approach bridges models with global collective and short-range interactions, as new quantum phases possess properties of both, going beyond Dicke, Lipkin--Meshkov--Glick \cite{ParkinsPRL2008} and other simple spin-1/2 models. Nontrivial spatial patterns were obtained with classical atoms and light~\cite{LabeyrieNaturePhotonics2014}. Our work extends such efforts in the interdisciplinary field of optomechanics towards quantum multimode systems~\cite{Aspelmeyer2014}. The mechanisms we suggest, provide a general framework and a new set of tools, inaccessible in setups using classical OLs. This will strongly expand applications in quantum simulations. It will allow exploring fundamental issues concerning emergence of multimode generalizations of strongly correlated phases, such as gapped superfluids~\cite{Wen} and density waves~\cite{Gruner} as well as their interplay, giving rise to quantum solids~\cite{Pupillo}. The light-induced structure is similar to multi-component nonlinear sigma models ubiquitous in models of high-energy~\cite{Zohar,Stannigel2014}, condensed matter~\cite{Auerbach}, and relativistic~\cite{Witten} physics. Dimer phases can be used as building blocks for quantum spin-liquids simulations~\cite{Balents}.


\section{Model for interacting macroscopic modes of ultracold matter}

In this section, we consolidate the results of our general theoretical model presented in Sec.1.2 and make a starting point to build the mean-field theories of many-body phenomena in quantum and dynamical optical lattices. As we will be interested in novel quantum phases, we adopt a more condensed matter terminology and style of presentation than the one used in previous chapters using a more quantum optical language. In particular, we will be interested in obtaining various phase diagrams. To be consistent with other works on this subject, we now call the tunneling coefficient $J$ as $t_0$, which is a condensed matter tradition.

We consider the setup in Fig. 1.1 with atoms trapped in an OL inside a single-mode cavity. We will consider multiple cavity modes and multiple probes later in Sec. 5.11. The probe can be at any angle to the cavity axis (not necessarily at $90^\circ$). This can be described by the Hamiltonian $\HH=\HH^b+\HH^{a}+\HH^{ab}$, where $\HH^b$ is the regular Bose--Hubbard (BH) Hamiltonian. The light is described by  $\HH^{a}=\hbar\omega_c a_1^\dagger  a$ and the light-atom interaction is (cf. Sec. 1.2, where as usual for transverse probing we neglect the dispersive frequency shift and do not consider probing through the mirror $\eta_1$)
\begin{equation}
\HH^{ab}=U_{10}(a_0^* a_1\hat F^\dagger+a_0 a_1^\dagger\hat F)
\end{equation}
with
$
\hat F= \hat D+\hat B$ (we drop the subscripts of these three operators). $\hat D=\sum_{j}J_{j,j}\hat n_j$ is the diagonal (i.e. with a single site index $i$) coupling of light to on-site densities, $\hat B=\sum_{\langle i,j\rangle}J_{i,j}( \hat b^\dagger_i\hat b^{\phantom{\dagger}}_j+h.c.)$ is the off-diagonal (i.e. with two site indices $i$ and $j$) coupling to the inter-site densities reflecting the matter-field interference (also called bonds), as we seen it before. The sums go over illuminated sites $N_s$. $\HH^{ab}$  is a consequence of the quantum potential seen by atoms on top of BH model given by a classical OL with the hopping amplitude $t_0$ and on-site interaction $U$. 

The spatial structure of light gives a natural basis to define the atomic modes, as the coupling coefficients $J_{i,j}$ can periodically repeat in space. The symmetries broken in the system are inherited from such a periodicity: all atoms equally coupled to light belong to the same mode, while the ones coupled differently belong to different modes. We define operators corresponding to modes $\varphi$: $\hat F=\sum_\varphi\hat D_\varphi+\sum_{\varphi'}\hat B_{\varphi'}$, where
\begin{eqnarray}
\hat D_\varphi&=&J_{D,\varphi}\hat N_\varphi,\;\textrm{with}\;  \hat N_\varphi=\sum_{i\in\varphi}\hat n_i,
\\
\hat B_{\varphi'}&=&J_{B,\varphi'} \hat S_{\varphi'},\;\textrm{with}\; \hat S_{\varphi'}=\sum_{\langle i,j\rangle\in\varphi'}( \hat b^\dagger_i\hat b^{\phantom{\dagger}}_j+h.c.).
\end{eqnarray}
Thus, we replaced the representation of atomic operator $\hat F$ as a sum of microscopic on-site and inter-site contributions by the smaller sum of macroscopically occupied global modes with number density, $\hat N_\varphi$, and bond, $\hat S_\varphi$, operators. The physical origin of the atomic modes is the same as we discussed in Chapters 2 and 3. Nevertheless, here we will not focus on the measurement backaction, but rather on the interaction between such modes described by effective Hamiltonians.

The structures of density and bond modes can be nearly independent from each other, as we explained in Chapters 2 and 3. To be precise, for the homogeneous scattering in a diffraction maximum, $J_{i,j}=J_B$ and $J_{j,j}=J_D$, one spatial mode is formed. When light is scattered in the main diffraction minimum (at $90^{\circ}$ to the cavity axis), the pattern of light-induced modes alternates sign as in the staggered field, $J_{i,j}=J_{j,i}=(-1)^jJ_B$ and $J_{j,j}=(-1)^jJ_D$. This gives two spatial density modes (odd and even sites) and, as we will show, four bond modes. As we have shown in Sec. 1.8, the density and bond modes can be decoupled by choosing angles such that $J_D=0$ (by shifting the probe with respect to classical lattice thus concentrating light between the sites and assuring the zero overlap between Wannier and mode functions) or $J_B=0$~\cite{Kozlowski2015PRA}. Beyond this, additional modes get imprinted by the probe light at different angles such that each $R$-th site or bond scatters light with equal phases and amplitudes.  As we have shown previously in Secs. 2.9 and 3.2, this generates multimode structures of $R$ density modes and $2R$ bond modes. The prominent example of self-organization~\cite{DomokosPRL2002,Morigi2010,Hofstetter,Reza} is a special case of two density modes, while macroscopic effects related to the higher density modes and any bond modes have not been addressed before.

In general, the light and matter are entangled. In the steady state of light, it can be adiabatically eliminated and the full light--matter state can be then reconstructed as we will show in Sec. 5.10 \cite{Caballero2015a}. The effective atomic Hamiltonian is   
\begin{eqnarray}
 \HH^b_\mathrm{eff}=\HH^b+\frac{g_{\mathrm{eff}}}{2}(\hat F^\dagger \hat F+\hat F \hat F^\dagger),
 \label{5-effmodel}
\end{eqnarray}
where $g_{\mathrm{eff}}=\Delta_c|U_{10}a_0|^2/ (\Delta_c^2+\kappa^2)$. A key physical processes is that the ground state is reached [i.e. the energy (\ref{5-effmodel}) is minimized], when the system adapts (self-organize) in such a way that the light scattering term is maximized for $g_{\mathrm{eff}}<0$, and minimized for $g_{\mathrm{eff}}>0$. We will use such an energy minimization to calculate the phase diagrams in this chapter.

An important result is that the new terms beyond BH Hamiltonian (\ref{5-effmodel}) give the effective long-range light-induced interaction between the density and bond modes: 
\begin{gather}
\hat F^\dagger \hat F+\hat F \hat F^\dagger=
\sum_{\varphi,\varphi'}
[\gamma_{\varphi,\varphi'}^{D,D}
\hat N_\varphi^{\phantom{*}}
\hat N_{\varphi'}^{\phantom{*}}
+\gamma_{\varphi,\varphi'}^{B,B}
\hat S_\varphi^{\phantom{*}}\hat S_{\varphi'}^{\phantom{*}}
+\gamma_{\varphi,\varphi'}^{D,B}(
\hat N_\varphi^{\phantom{*}}\hat S_{\varphi'}^{\phantom{*}}
+
\hat S_{\varphi}^{\phantom{*}}\hat N_{\varphi'}^{\phantom{*}})],
\label{5-mdecomp}
\end{gather}
where $\gamma^{\nu,\eta}_{\varphi,\varphi'}=(J_{\nu,{\varphi}}^* J^{\phantom{*}}_{\eta,{\varphi'}}+c.c.)$. Thus, any symmetry broken by the light modes imprints the structure on the interaction of atomic modes. 

Fundamentally, Eq. (\ref{5-mdecomp}) displays the link between global interactions and the interaction resembling typical short-range one (usually appearing between the sites, while here between the modes). Thus, the resulting quantum phase will have properties of both collective and short-range systems. In this language, it is the term  $\hat N_\text{odd}^{\phantom{*}} \hat N_\text{even}^{\phantom{*}}$ that is responsible for supersolid properties of the self-organized state. (The standard supersolidity appears due to the short-term $\hat n_i^{\phantom{*}} \hat n_{i+1}^{\phantom{*}}$ interaction.) Our general approach enables to go far beyond typical global models (e.g. Dicke and Lipkin--Meshkov--Glick \cite{ParkinsPRL2008}) due to spatial structuring the global interaction thus assuring its effective competition with the short-range ones even in a single mode cavity. The bond interaction can be easily identified as pseudo-spin interaction via the Schwinger mapping \cite{Auerbach}. In addition, some components might be non-Abelian.

We decompose (\ref{5-mdecomp}) in mean-field contributions and fluctuations:
\begin{eqnarray}
\hat F^\dagger\hat F+\hat F\hat F^\dagger&=&\langle\hat F^\dagger\rangle\hat F^{\phantom{\dagger}}+\langle\hat F\rangle \hat F^\dagger+\delta\hat F^\dagger\hat F.
\end{eqnarray}
The last term $\delta\hat F^\dagger\hat F$ originates from the quantum light--matter correlations. It is the term that determines the quantum optical lattice (QOL), i. e. the essentially quantum nature of a quantum potential. Other terms originate from the dynamical but classical light, when the semiclassical approximation $a_1 \hat F^\dagger=\langle a_1\rangle \hat F^\dagger$ holds. 

In the following we will use the mean-field approach to this problem. We present all technical details in Sec. 5.5 and then will focus on the physical side of the results. Decorrelating operators at different sites, we obtain a mean-field theory that has nonlocal coupling between the matter modes and is local in fluctuations. For $\delta\hat D^\dagger\hat D$ these reduce to on-site number fluctuations. Importantly, this corresponds to the purely light-induced effective on-site interaction of atoms beyond the standard BH term. For $\delta\hat B^\dagger\hat B$, light--matter correlations include radically new terms beyond BH model: fluctuations of the order parameter and density coupling between neighboring sites, which appear due to two and four point quantum atomic correlations. 

In contrast to previous works, we will show non-negligible effects due to such terms, putting forward the quantumness of optical lattices. We will predict as well new effects due to the dynamical optical lattices. 

When the ground state of $\HH^b_\mathrm{eff}$ is achieved by maximizing scattering ($g_\mathrm{eff}<0$), a strong classical light emerges and small fluctuations can be neglected. In principle, even in the strong-light case, the light quantumness can play a role, because self-organized states can be in a superposition of several patterns and different light amplitudes are correlated to them. Nevertheless, in a realistic case with dissipation, the system quickly collapses to one of the semiclassical states~\cite{EsslingerPRL11}. We will show that quantum fluctuations play a key role in the opposite case, where scattering is minimized ($g_\mathrm{eff}>0$).  Here no classical light builds up and light fluctuations design the emergence of novel phases. To underline key phenomena, we will mainly consider cases with either density or bond modes. 

In summary of this section, we introduced a model, which bridges physics of global and short-range interacting systems. Therefore, the new quantum phases, which we will predict, will have properties of both types of such models.


\section{Technical details of deriving the mode decoupling and effective Hamiltonians}

In this section, we give rather technical details of calculations, and in the next sections will focus on the physical meaning of the results.

\subsection{Decomposition of the light-induced interaction}

The Hamiltonian for a single cavity and a single probe is
\begin{equation}
\HH_{\mathrm{eff}}=\HH^b+\frac{g_{\mathrm{eff}}}{2}\left(\hat F^\dagger\hat F^{\phantom{\dagger}}+\hat F^{\phantom{\dagger}}\hat F^\dagger\right).
\end{equation}
Next, we separate the light matter-correlations and dynamical terms in $\hat F^\dagger\hat F$  performing multimode on-site mean-field.

The $\hat D^\dagger\hat D$ (density coupling) terms can be expanded as
\begin{gather}
\hat D^\dagger\hat D+\hat D\hat D^\dagger \approx \sum_{\varphi,\varphi'}(J_{D,\varphi}^*J_{D,\varphi'}^{\phantom{*}}+c.c.)\langle\hat N_{\varphi'}^{\phantom{*}}\rangle(2\hat N_\varphi-\langle\hat N_\varphi^{\phantom{*}}\rangle)
+\delta \hat D^\dagger\hat D,
\label{5-DS1}
\\
\delta \hat D^\dagger\hat D=2\sum_\varphi |J_{D,\varphi}|^2\delta\hat N_\varphi^2,\label{5-DS2}
\\
\delta\hat N_\varphi^2=\sum_{i\in\varphi}(\hat n_i-\rho_i)^2,
\end{gather}
where $\langle\hat N_\varphi\rangle=\sum_{i\in\varphi}\rho_i$ is the mean number of atoms in the mode $\varphi$ and $\rho_i=\langle\hat n_i\rangle$ is the mean atom number at site $i$. The first term in Eq. (\ref{5-DS1}) is due to the dynamical properties of the light field, these terms exhibit non-local coupling between light-induced modes.  The terms in (\ref{5-DS2}) are the light--matter correlations and contain the effect due to quantum fluctuations, the QOL terms.

The $\hat B^\dagger\hat B$ (bond coupling) terms can be expanded as
\begin{gather}
\hat B^\dagger\hat B+\hat B\hat B^\dagger \approx \sum_{\varphi,\varphi'}(J_{B,\varphi}^*J_{B,\varphi'}^{\phantom{*}}+\mathrm{c.c.})\langle\hat S_{\varphi'}^{\phantom{*}}\rangle(2\hat S_\varphi-\langle\hat S_\varphi^{\phantom{*}}\rangle)
+\delta\hat B^\dagger\hat B,
\label{5-BS1}
\\
\delta\hat B^\dagger\hat B=2\sum_\varphi |J_{B,\varphi}|^2\delta \hat S_\varphi^2,
\label{5-BS2}
\end{gather}
\begin{eqnarray}
\delta \hat S_\varphi^2&=&
\sum_{\langle i,j \rangle\in\varphi}(
(\hat b_i^\dagger\hat b_j^{\phantom{\dagger}}+h.c.-\langle\hat b_i^\dagger\hat b_j^{\phantom{\dagger}}+ \mathrm{H.c.}\rangle)^2
\nonumber
\\
&+&
\sum_{\langle i,j,k \rangle\in\varphi}
\big(
\hat b^\dagger_i\hat b^\dagger_k(\hat b^{\phantom{\dagger}}_j)^2
+(\hat b^\dagger_j)^2\hat b^{\phantom{\dagger}}_i\hat b^{\phantom{\dagger}}_k
+
2\hat n_i^{\phantom{\dagger}}\hat b^\dagger_k\hat b^{\phantom{\dagger}}_j
+\hat b^\dagger_k\hat b^{\phantom{\dagger}}_j
\nonumber\\
&&\phantom{\sum_{\langle i,j,k}
}-(\hat b^\dagger_k\hat b^{\phantom{\dagger}}_j+\hat b^\dagger_i\hat b^{\phantom{\dagger}}_k+h.c.)\langle\hat b_i^\dagger\hat b_j^{\phantom{\dagger}}+ \mathrm{H.c.}\rangle\big),
\label{5-BS3}
\end{eqnarray}
where  $\langle i,j,k \rangle$ refers to $i$,$j$ nearest neighbor and $k$ is a nearest neighbor to the pair $\langle i,j\rangle$. The first term in (\ref{5-BS1}) is due to the dynamical properties of the light field and (\ref{5-BS2}) are due to the light--matter correlations. These are basically all the possible four-point correlations and tunneling processes between nearest neighbors, as higher order tunneling processes have much smaller amplitudes. The expectation value of the bond operators reduces to
 \begin{equation}
 \langle \hat S_{\varphi}\rangle=\sum_{\langle i, j\rangle\in\varphi}(\psi_i^\dagger\psi^{\phantom{\dagger}}_j+\psi_j^\dagger\psi_i^{\phantom{\dagger}})
 ,
 \end{equation}
where $\psi_i=\langle\hat b_i\rangle$  is the SF order parameter corresponding to the site $i$. The above is the sum of products of order parameters at nearest neighbor sites in the light-induced mode $\varphi$. 

 In fact for most purposes it is enough to consider
\begin{eqnarray}
\delta \hat S_\varphi^2&\approx&
\sum_{\langle i,j \rangle\in\varphi}
([\hat b_i^\dagger\hat b_j^{\phantom{\dagger}}+ \mathrm{H.c.}]-\langle\hat b_i^\dagger\hat b_j^{\phantom{\dagger}}+ \mathrm{H.c.}\rangle)^2
\\
&=&
\sum_{\langle i,j \rangle\in\varphi}\big[\hat b_i^{2\dagger}\hat b_j^{2\phantom{\dagger}}+\hat b_j^{2\dagger}\hat b_i^{2\phantom{\dagger}}+2\hat n_i\hat n_j+\hat n_i+\hat n_j
\nonumber\\
&-&2(\psi_i^*\psi_j^{\phantom{\dagger}}+\psi_j^*\psi_i^{\phantom{\dagger}})(\hat b_i^\dagger\hat b_j^{\phantom{\dagger}}+\hat b_j^\dagger\hat b_i^{\phantom{\dagger}})+(\psi_i^*\psi_j^{\phantom{\dagger}}+\psi_j^*\psi_i^{\phantom{\dagger}})^2\big],
\label{5-tp}
\end{eqnarray}
as these terms have a more significant effect in the effective Hamiltonian compared to the nearest-neighbor coupling to nearest neighbors (the $\langle i,j,k\rangle$ terms). These are the quantum fluctuations in the SF order parameters. The terms in the first line of Eq. (\ref{5-tp}) are due to two-particle hole excitations at adjacent sites. These will change the landscape of the supported quantum phases in the system, as they introduce  a mechanism to break translational invariance via a DW instability and the modification to quantum fluctuations. Importantly, the QOL can generate a DW instability. 

Note that for a homogeneous ideal superfluid state ($U=0$)  
\begin{equation}
\sum_{\varphi}\langle\delta \hat S_\varphi^2\rangle=\sum_{\varphi}\langle\hat S_\varphi\rangle=2zN_s|\psi|^2,
\end{equation}
where we have used the coordination number (the number of nearest neighbors)  defined as $z=2d$ for a $d$-dimensional square lattice. As we can expect, the fluctuations of the SF order parameter in this limit are { Poissonian}.

The terms that arise from the product of $\hat B$ and $\hat D$ (bond-density coupling) are
\begin{gather}
\hat B^\dagger\hat D+\hat D\hat B^\dagger+ \mathrm{H.c.} \approx 2\sum_{\varphi,\varphi'}(J_{B,\varphi}^*J_{D,\varphi'}^{\phantom{*}}+\mathrm{c.c.})
(\langle\hat S_{\varphi'}^{\phantom{*}}\rangle\hat N_\varphi
\nonumber\\
+\langle\hat N_\varphi^{\phantom{*}}\rangle\hat S_{\varphi'}^{\phantom{*}}
-\langle\hat S_{\varphi'}^{\phantom{*}}\rangle\langle\hat N_\varphi\rangle))
+(
\delta\hat B^\dagger\hat D+\delta\hat D\hat B^\dagger+ \mathrm{H.c.}),
\label{5-BDS1}
\end{gather}
\begin{gather}
\delta\hat B^\dagger\hat D+\delta\hat D\hat B^\dagger+ \mathrm{H.c.}=\sum_\varphi \big[(J_{B,\varphi'}^*J_{D,\varphi'}^{\phantom{*}}+\mathrm{c.c.})\delta \hat C_{\varphi'}
+ \mathrm{H.c.}\big],
\label{5-BDS2}
\end{gather}
\begin{equation}
\delta \hat C_{\varphi'}=
\sum_{\langle i,j \rangle\in\varphi'}
(\hat b_i^\dagger\hat b_j^{\phantom{\dagger}}+ \mathrm{H.c.}-
\langle\hat b_i^\dagger\hat b_j^{\phantom{\dagger}}+ \mathrm{H.c.}\rangle)(\hat n_i-\rho_i),
\label{5-CS1}
\end{equation}
where $\langle\delta \hat C_{\varphi'}\rangle$ 
is the sum of  the ``local'' covariances per mode given by the bond operator modes $\varphi'$. The additional terms $\delta\hat B^\dagger\hat B$, $\delta\hat D^\dagger\hat D$, $\delta\hat B^\dagger\hat D$, and $\delta\hat D^\dagger\hat B$ have a local character that alters the system at the quantum level, coupling the local densities to the local tunneling processes.  

Additional terms might be considered in the above expansions and their generalization is straightforward by removing the restriction over the sums beyond nearest neighbor.  It is evident from the decomposition and the expansion that the semiclassical terms, given by $\langle \hat F\rangle\hat F^\dagger$ and $\langle \hat F^\dagger\rangle\hat F$ have a non-local (global) character coupling all the illuminated sites and imprinting structure in the interaction.  Light scattering from the atoms can suppress or enhance quantum terms by properly choosing the detuning with respect to the cavity. In addition the light mode structure leads to a combination of novel phases of matter not supported without cavity light. When atoms scatter light maximally, the terms due to quantum fluctuations are strongly smeared out as their behavior scales with  $N_s$ compared with the factor of $N_s^2$ of semiclassical terms. This occurs when $g_\mathrm{eff}<0$ and the familiar scenario of self-organized states emerges. However, in the case when  $g_\mathrm{eff}>0$ quantum fluctuations become relevant as atoms scatter light minimally and the QOL becomes important. Thus, by suppressing self-organization one has access to the effects due to true quantum fluctuations otherwise not visible.

Now we will focus on the effective mean-field Hamiltonian components. We introduce on-site mean-field theory to represent the above terms defining superfluid order parameters per site such that $\langle\hat b_i\rangle=\psi_i$, and we consider for simplicity a square lattice in $d$ dimensions. Separating the short-range contributions due to quantum fluctuations ($\HH^F_{Q}$) and non-local contributions due to semiclassical terms ($\HH^F_{C}$) we obtain for $\hat F\hat F^\dagger+\hat F^\dagger\hat F=\HH^F_{Q}+\HH^F_{C}$,
\begin{gather}
\HH^F_Q=2\sum_i\left[|\gamma_{D,i}|^2(\hat n_i-\rho_i)^2
+
z(\gamma_{D,i}^*\gamma_{B,i}+\mathrm{c.c.})(\hat n_i\hat\beta_i+\hat\beta_i\hat n_i-2\langle\hat\beta_i\rangle\rho_i)
\right. \nonumber\\
+z|\gamma_{B,i}|^2(2\hat n_i\hat n_{-i}+\hat n_{i}+\hat n_{-i} -2\langle\hat\beta_i\rangle\hat\beta_i
\nonumber\\
\left.
+\langle\hat\beta_i\rangle^2+\langle\hat b^{\dagger 2}_i\rangle\hat b_{-i}^2+\langle\hat b^{\dagger 2}_{-i}\rangle\hat b_i^2+\langle\hat b^{2}_i\rangle\hat b_{-i}^{\dagger 2}+\langle\hat b^{2}_{-i}\rangle\hat b_i^{\dagger 2}-\langle\hat b^{\dagger 2}_i\rangle\langle\hat b^{2}_{-i}\rangle-\langle\hat b^{\dagger 2}_{-i}\rangle\langle\hat b^{2}_{i}\rangle)
\right],
\\
\HH^F_{C}=\sum_i\left[\langle\gamma_{D,i}^*(\hat D+\hat B)+\mathrm{H.c.}\rangle
(2\hat n_i-\rho_i) \right. \nonumber\\
\left.
+z\langle\gamma_{B,i}^*(\hat D+\hat B)+\mathrm{H.c.}\rangle
(2\hat\beta_i-\langle\hat\beta_i\rangle)\right],
\end{gather}
where $\hat \beta_i=\psi_{-i}^*\hat b^{\phantom{\dagger}}_{i}+\psi_{i}^*\hat b_{-i}^{\phantom{\dagger}}+\psi^{\phantom{*}}_{-i}\hat b_{i}^\dagger+\psi^{\phantom{*}}_{i}\hat b_{-i}^\dagger-(\psi_{-i}^*\psi_i^{\phantom{*}}+\mathrm{c.c.})$, $\langle\hat\beta_i\rangle=\psi_{-i}^*\psi_i+\mathrm{c.c.}$, with 
\begin{equation}
\langle\hat D\rangle=\sum_i\gamma_{D,i}\rho_i\quad \textrm{and}\quad\langle\hat B\rangle= z\sum_i\gamma_{B,i}\langle\hat\beta_i\rangle.
\end{equation}
We have used $\gamma_{D,i}=J_{i,i}$ and $\gamma_{B,i}=J_{i,nn(i)}=J_{i,-i}$ where $nn(i)$ is a nearest neighbor of  the site $i$. The sub-index $-i$ in operators means nearest neighbor of the site $i$. $\HH^F_{Q}$ are the quantum optical lattice contributions and $\HH^F_C$ are the dynamical contributions to the optical lattice.


\subsection{Effective Hamiltonians}

The representation of $\HH^F_Q$ and $\HH^F_C$ makes it clear that we can construct an effective mode representation in mean-field approximation for the full $\HH_\mathrm{eff}^b$ depending on the pattern of the $J$'s, such that
\begin{equation}
\HH^b_{\mathrm{eff}}\approx\HH^b+\frac{g_\mathrm{eff}}{2}(\HH^F_C+\HH^F_Q).
\end{equation}

The effective Hamiltonian considering only density coupling ($\hat D^\dagger\hat D+\hat D\hat D^\dagger$) is
\begin{gather}
\HH^b_{\mathrm{eff}}=
\sum_{\varphi}\Big[\sum_{i\in\varphi}\left(-t_0\hat\beta_i+\frac{U_\varphi}{2}\hat n_i(\hat n_i-1)
-2g_{\mathrm{eff}}|J_{D,\varphi}|^2\rho_i\hat n_i\right)
-\mu_\varphi\hat N_\varphi-g_{\mathrm{eff}}c_{D,\varphi}\Big],
\nonumber\\
\mu_\varphi=\mu-g_{\mathrm{eff}}\eta_{D,\varphi},
\\
U_\varphi=U+2g_{\mathrm{eff}}|J_{D,\varphi}|^2, \label{5-U}
\end{gather}
with $\eta_{D,\varphi}=\sum_{\varphi'}(J_{D,\varphi}^*J^{\phantom{*}}_{D,\varphi'}+\mathrm{c.c.})\langle \hat N_{\varphi'}\rangle-|J_{D,\varphi}|^2$ and $c_{D,\varphi}=\sum_{\varphi'}(J_{D,\varphi}^*J^{\phantom{*}}_{D,\varphi'}+\mathrm{c.c.})\langle\hat N_{\varphi'}\rangle \langle \hat N_{\varphi}\rangle/2-|J_{D,\varphi}|^2\sum_{i\in\varphi}\rho_i^2$. The many-body interaction $U_\varphi$ and the chemical potential $\mu_\varphi$ inherit the pattern induced by the quantum potential that depends on light-induced mode structure given by $\varphi$.  Thus, each mode component sees a different on-site interaction and chemical potential, while there is an additional dependency of the chemical potential on the density. The modification to the on-site interaction is the QOL effect, while the modification to the chemical potential is the DOL effect. As the $\hat \beta_i $ operator couples nearest neighbor sites, in principle one needs two on-site modes even for one light induced mode. However, this special case does not break translational symmetry and due to this $\psi_i=\psi_{-i}=\psi$. For more than one light induced mode the number of light induced modes will depend on the number of different values of $J_{D,\varphi}$. 

In the case of only off-diagonal bond scattering  ($\hat B^\dagger\hat B+\hat B\hat B^\dagger$), we have
\begin{gather}
\HH^b_{\mathrm{eff}}=\sum_{\varphi}\Big[-t_\varphi\hat S_\varphi+ g_{\mathrm{eff}}|J_{B,\varphi}|^2\delta \hat S_\varphi^2-g_{\mathrm{eff}}c_{B,\varphi}\Big]
+\sum_i\left(\frac{U}{2}\hat n_i(\hat n_i-1)-\mu\hat n_i\right),
\\
t_\varphi=t_0-g_{\mathrm{eff}}\eta_{B,\varphi},
\end{gather}
with 
$\eta_{B,\varphi}=\sum_{\varphi'}(J_{B,\varphi}^*J^{\phantom{*}}_{B,\varphi'}+\mathrm{c.c.})\langle \hat S_{\varphi'}\rangle$ and $c_{B,\varphi}=\sum_{\varphi'}(J_{B,\varphi}^*J^{\phantom{*}}_{B,\varphi'}+\mathrm{c.c.})\langle\hat S_{\varphi'}\rangle \langle \hat S_{\varphi}\rangle/2$. The effective tunneling amplitude $t_\varphi$ couples the SF components of all the light-induced modes $\varphi$, this is the DOL effect. The terms due to $\delta \hat S_\varphi^2$ induce non-trivial coupling between nearest neighbor sites and lead to the formation of a density wave instability with more than one light-induced mode, this is relevant whenever quantum fluctuations are not smeared out by the semiclassical contribution, this is the QOL effect.  

The full Hamiltonian including cross terms products of $\hat D$ and $\hat B$ can be written  as
\begin{gather}
\HH^b_{\mathrm{eff}}=
\sum_i\left[-z\left(t_{BD,i}-{g_\mathrm{eff}}(\gamma_{D,i}^*\gamma_{B,i}^{\phantom{*}}+\mathrm{c.c.})\hat n_i\right)\hat\beta_i-
\right.\nonumber\\
\left(\mu_{DB,i}- z g_\mathrm{eff}(\gamma_{B,i}^*\gamma_{D,i}^{\phantom{*}}+\mathrm{c.c.})\hat\beta_i\right)\hat n_i
\nonumber\\
+zg_{\mathrm{eff}}|\gamma_{B,i}^{\phantom{*}}|^2(2\hat n_i\hat n_{-i}+\hat n_{i}+\hat n_{-i}
+
\langle\hat b^{\dagger 2}_i\rangle\hat b_{-i}^2+\langle\hat b^{\dagger 2}_{-i}\rangle\hat b_i^2+\langle\hat b^{2}_i\rangle\hat b_{-i}^{\dagger 2}+\langle\hat b^{2}_{-i}\rangle\hat b_i^{\dagger 2})
\nonumber\\
+\left.\frac{U_{D,i}}{2}\hat n_i(\hat n_i-1)-c_{DB,i}
\right],
\label{5-fh}
\end{gather}
with effective nonlinear parameters
\begin{eqnarray}
t_{DB,i}&=&t_0-g_\mathrm{eff}\left(\langle\gamma_{B,i}^*(\hat D+\hat B)+\mathrm{H.c.}\rangle
-2z|\gamma_{B,i}^{\phantom{*}}|^2\langle\hat\beta_i\rangle\right),
\\
\mu_{DB_i}&=&\mu-g_\mathrm{eff}\left(\langle\gamma_{D,i}^*(\hat D+\hat B)+\mathrm{H.c.}\rangle
-2|\gamma_{D,i}^{\phantom{*}}|^2\rho_i-|\gamma_{D,i}|^2\right),
\\
U_{D,i}&=&U+2g_\mathrm{eff}|\gamma_{D,i}|^2,
\end{eqnarray}
\begin{gather}
c_{DB,i}=
\frac{zg_\mathrm{eff}\langle\hat\beta_i\rangle}{2}\langle\gamma_{B,i}^*(\hat D+\hat B)+\mathrm{H.c}\rangle
+\frac{g_\mathrm{eff}\rho_i}{2}
\left(
\langle\gamma_{D,i}^*(\hat D+\hat B)+\mathrm{H.c.}\rangle
-2|\gamma_{D,i}^{\phantom{*}}|^2\rho_i
\right)
\nonumber\\
+\frac{z g_\mathrm{eff}}{2}\left(2(\gamma_{D,i}^*\gamma_{B_i}^{\phantom{*}}+\mathrm{c.c.})\langle\hat\beta_i\rangle\rho_i-2|\gamma_{B,i}^{\phantom{*}}|^2(\langle\hat b^{\dagger 2}_i\rangle\langle\hat b^{2}_{-i}\rangle+\langle\hat b^{\dagger 2}_{-i}\rangle\langle\hat b^{2}_{i}\rangle-\langle\hat\beta_i\rangle^2)\right).
\end{gather}
The tunneling amplitudes $t_{DB,i}$ and the chemical potentials  $\mu_{DB,i}$ are renormalized by both semiclassical contributions and due to quantum fluctuations, both DOL and QOL contributions are relevant. The on-site interactions $U_{D,i}$ get modified by quantum fluctuations in the density (QOL effect), while the constants $c_{DB,i}$ are corrections to avoid over counting. The terms in  Eq. (\ref{5-fh}) contain the effect of fluctuations in the order parameter (QOL effect). These are an effective nearest neighbor interaction $\hat n_i\hat n_{-i}$, additional chemical potential shifts,  and all the two particle-hole excitations between nearest neighbors. 

The key aspect of our approach is to take advantage of the decomposition in the light induced mode basis to generate the effective Hamiltonians and analyze the competing emergent phases. The use of this basis will simplify greatly the estimation of the phase diagram of the system based on the effective model Hamiltonians, which are easy to interpret from their building blocks, while retaining enough relevant features to uncover the emergence of unconventional phases of quantum matter.

In the next sections we will present several geometrical configurations, where quantum phases appear either due to the quantum or dynamical natures of the optical lattice.


\section{Quantum and dynamical lattices: density scattering at $90^\circ$}

In this section, we will consider light scattering at $90^\circ$ due to the density-dependent operator $\hat{D}$: first quantum lattices with minimal scattering, and then dynamical lattices with maximal scattering. Only the second effect corresponds to the well-known self-organization, Dicke, or lattice supersolid phase transition. The first quantum effect is absolutely new.

Density-dependent, but classical light was previously shown to strongly modify the standard phase diagram of Mott insulator -- superfluid transition \cite{Morigi2010,Larson}, if plotted via the chemical potential $\mu$. Note, that if the phase diagrams are plotted via the density, rather than chemical potential, most of such modifications due to the classical light become invisible. Therefore, in this work we chose to plot the phase diagrams via the density (which is not typical to the condensed matter style) to focus more on the effects produced by the quantum light.

\subsection{Quantum lattices: shift of the Mott insulator -- superfluid transition point}

\begin{figure}[h!]
\captionsetup{justification=justified}
\begin{center}
\includegraphics[clip, trim=0cm 0cm 3.19cm 0cm, width=0.4\textwidth]{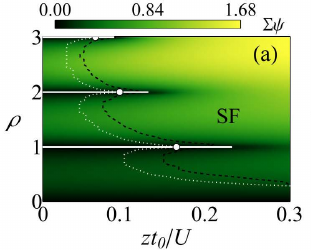}
\end{center}
\caption{Light scattering from on-site densities. 
Minimized scattering highlighting the quantumness of OL. Total order parameter $\Sigma\psi=|\psi_+|^2+|\psi_-|^2$, where $\psi_\pm$ are the SF order parameters of two modes, white lines correspond to MIs;  dashed lines are boundaries of SF with only two non-negligible Fock components. The system is homogeneous, $\rho_+=\rho_-$ and $\psi_+=\psi_-$. White points are the MI--SF  transition points without cavity light: the SF--MI can be significantly shifted by the quantum OL. 
Parameters: scattering at  $90^\circ$, $g_\mathrm{eff}= 10U/N_s$, the boundaries are for $g_\mathrm{eff}= 10U/N_s$ and $0$; $J_D=1.0$, $J_B=0$, $N_s=100$, $z=6$. }
\label{5-DminJD}
\end{figure}

For minimized scattering ($g_{\mathrm{eff}}>0$), the classical light cannot build up at all, and quantum fluctuations take the leading role (Fig. \ref{5-DminJD}).  As we showed in Sec. 5.5, at fixed density per site, the quantum light--matter fluctuations effectively renormalize the on-site interaction from $U$ to $U+ 2 g_{\mathrm{eff}}J_D^2$ in each component (this is well visible from Eq. (\ref{5-U})).  Thus, changing the light--matter coupling, one can shift the SF--MI transition point.  This occurs because the light-induced atomic fluctuations now enter the effective Hamiltonian, and these fluctuations need to be suppressed to minimize the energy. This favors MI state for $U$ smaller than that without cavity light, extending MI regions (Fig. \ref{5-DminJD}). Moreover, in a quantum OL, atoms can potentially enter MI even without atomic interaction. This provides absolute control on DW order formation.  In analogous fermionic systems DWs are relevant for the stability of superconducting phases~\cite{HiTc}. 

For incommensurate fillings, SF survives but with the smaller (suppressed) fluctuations as well: For convenience, we also plot the boundaries where the superfluid ground state is composed of mainly two lowest occupation Fock states depending on the filling factor (i.e. components with higher occupations are negligible). Note that with cavity light, the state becomes gapped with respect to adding more than one excitation, thus minimizing fluctuations. 
However there is no phase transition to this peculiar superfluid state. As this gapped state consists of only two Fock states, we denote it as the quantum superposition (QS) state and will describe it in more details in Sec. 5.8.

\subsection{Dynamical lattices: supersolids and density waves}

Scattering the probe at $90^\circ$ and maximizing light scattering ($g_{\mathrm{eff}}<0$), one breaks the translational symmetry.
Hence, the system can support density waves (DWs) and novel bond orders. The simultaneous occurrence of SF and DW orders is a supersolid (SS) phase~\cite{Pupillo}. SS  and DW have been predicted earlier due to classical maximized scattering~\cite{Hofstetter,Reza}. Their origin can be traced to the classical self-organization of thermal atoms in a cavity. 

Our finding is that for weak probe, DWs and SSs with only small density imbalance appear at half-integer filling, together with usual MI and SF. Moreover, we find that, in contrast, above the threshold (a quantum critical point)  $|g_{\mathrm{eff}}|N_s>U/2$, DWs and SSs with maximal imbalance are favored, while usual MI and SF are completely suppressed. 

Importantly, these our results about the existence of two kinds of DWs and SSs: with small and maximal imbalances, were confirmed experimentally in Ref. \cite{EsslingerNature2016}

In the case of diffraction minima we have the following pattern of the $J_{D,\varphi}$, in terms of the $\gamma$'s,
$
\gamma_D(i)=(-1)^{i+1}
$.
This generates a double sub-lattice structure, where it is convenient to define lattices $O$ and $E$ corresponding to the positive and negative case for the $\gamma$'s. The light induced effective structured interaction induces two modes for odd $(O)$ and even $(E)$ sites across the square classical optical lattice. We have then
\begin{eqnarray}
\hat D&=&J_D\sum_\nu(\hat n_{O,\nu}-\hat n_{E,\nu}),
\end{eqnarray}
and the effective Hamiltonian is
\begin{equation}
\HH^b_{\mathrm{eff}}=\HH^b+g_{\mathrm{eff}}|J_D|^2\left[\sum_\nu(\hat n_{O,\nu}-\hat n_{E,\nu})\right]^2,
\end{equation}
where the sum over $\nu$ goes over $N_s/2$ sites. The effective mean-field Hamiltonian following the general mean-field decoupling scheme is then (cf. Sec. 5.5)
 \begin{gather}
\HH^b_{\mathrm{eff}} \approx \HH^O_{\mathrm{eff}}+\HH^E_{\mathrm{eff}},
\\
\HH^\xi_{\mathrm{eff}} = \frac{N_s}{2}\Big[-z t_0\hat\beta-\mu_{\xi}\hat n_{\xi}
 +\frac{U_{\xi}}{2}\hat n_{\xi}(\hat n_{\xi}-1)
-g_{\mathrm{eff}}|J_D|^2\rho_{\xi}\hat n_{\xi}-g_{\mathrm{eff}}c_{D,\xi}\Big],
\\
\mu_{O/E}=\mu\pm 2g_{\mathrm{eff}}N_s|J_D|^2\Delta\rho,
\\
U_{O/E}=U+2g_{\mathrm{eff}}|J_D|^2,
\end{gather}
where $\xi=O/E$ with $c_{D,O/E}=\pm N_s|J_D|^2\Delta\rho\rho_{O/E}/2-|J_D|^2\rho_{O/E}^2/2$, $\hat\beta=\psi_{O}^*\hat b^{\phantom{\dagger}}_{E}+\psi_{E}^*\hat b_{O}^{\phantom{\dagger}}+\psi_{O}^{\phantom{*}}\hat b_{E}^\dagger+\psi_{E}^{\phantom{*}}\hat b_{O}^\dagger
-(\psi_{O}^*\psi^{\phantom{*}}_E+c.c.)$,
 and $\langle\hat\beta\rangle=(\psi^*_E\psi^{\phantom{*}}_O+\mathrm{c.c.})$. It is useful to define $\Delta\rho=(\rho_O-\rho_E)/2$ the emergent DW order parameter and the density $\rho=(\rho_O+\rho_E)/2$. As before, $\langle\hat n_{O/E}\rangle=\rho_{O/E}$ and $\langle\hat b_{O/E}\rangle=\psi_{O/E}$ are self-consistent constraints. We have assumed that these self-consistent parameters are homogeneous in each sub-lattice $O/E$. It is useful to regroup the $\hat\beta$ for the operators of each mode as
 \begin{equation}
 \hat\beta_{O/E}=2\big(\psi_{E/O}^*\hat b^{\phantom{\dagger}}_{O/E}+\psi_{E/O}^{\phantom{*}}\hat b_{O/E}^\dagger\big)
-(\psi_{O}^*\psi^{\phantom{*}}_E+\mathrm{c.c.}),
 \end{equation}
 so that then, the operator part of each sub-lattice Hamiltonian acts on its own sub-lattice Hilbert space, as $2\hat\beta=\hat\beta_O+\hat\beta_E$. Thus, the problem can be cast in terms of the global optimization problem to find the ground-state.

 The quantity $\Delta\rho$ measures the formation of density wave order  in the system in the stationary state.  Density wave order will be present in the system given that $\Delta\rho\neq 0$, this induces a checkerboard pattern in the density over the entire lattice. One can see from the above, that depending on the balance between the different couplings $\mu_{O/E}$ and the original parameters of the Bose--Hubbard model in the absence of quantum light $\HH^b(t_0,\mu,U)$, there exists the possibility for the system to be in different macroscopic phases in the steady state additional to the Mott-Insulator (MI) phases ($\Delta\rho=0$ and $|\psi_{O/E}|=0$) and the superfluid (SF) phases ($\psi_{O}=\psi_E\neq 0$). The system can be in a density wave (DW) insulating phase ($\Delta\rho\neq 0$ and $|\psi_{O/E}|=0$) or in a supersolid (SS) phase ($\Delta\rho\neq0$ and $|\psi_{O/E}|\neq0$).   Essentially, the  components of $\HH_{\mathrm{eff}}^b$ are just  Bose--Hubbard models for the sub-lattices $O/E$ coupled to each other via the chemical potentials $\mu_{O/E}$.  The long range effect of the cavity field is encoded in the dependency  between sub-lattices in the self-consistent parameters for the mean atom number per site and the superfluid order parameters. These terms due to the effective coupling between sub-lattices induce long-range order in our model, diagonal in $\rho_{O/E}$ and off-diagonal in $\psi_{O/E}$. Additionally, in our effective models, quantum fluctuations in each sub-lattice are modified by the light--matter interaction modifying the Hubbard $U$.

 \begin{figure}[h]
 \captionsetup{justification=justified}
\centering
\includegraphics[width=0.4\textwidth]{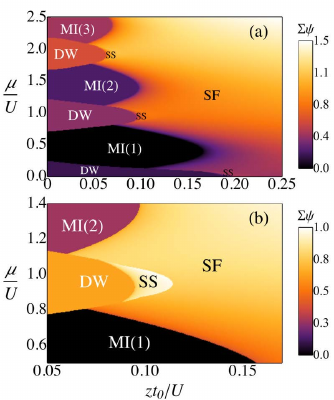}
\caption{(a) Phase diagram of two component density ordered states for $g_{\mathrm{eff}}<0$, when geometry of the probing is such that two spatial modes can occur  ($R=2$). The system is SF, without any spatial pattern as the interaction is decreased. The SF phase has total order parameter $\Sigma\psi=(|\psi_O|+|\psi_E|)/2\neq 0$ and $\psi_O=\psi_E$.  MI($n$) lobes appear for commensurate densities $n$, $\Sigma\psi=0$, $\rho=(\rho_O+\rho_E)/2=n$ with $\rho_O=\rho_E$. In between MI lobes, DW insulators form with $\Delta\rho=(\rho_O-\rho_E)/2\neq0$, maximal light scattering occurs then and $\Sigma\psi=0$. SS phases ($\Sigma\psi\neq 0$, $\Delta\rho\neq0$ ) are indicated near the boundary between DW and SF states. (b) Phase diagram closer to the DW-SS-SF transition at the tip of the DW insulators in between MI(1) and MI(2). The transition can have an intermediate SS state towards the homogeneous SF phase. Parameters are for (a) and (b): $J_D=1.0$, $J_B=0.0$, $g_{\mathrm{eff}}N_s/U=-0.5$, $N_s=100$, $z=6$ (3D). The color bar denotes $\Sigma\psi$ in the SF region. Quantum phases different from SF (labeled) are denoted by regions with different colors.}
\label{5-PD2mD}
\end{figure}

 \begin{figure}[h!]
  \captionsetup{justification=justified}
\centering
\includegraphics[width=0.4\textwidth]{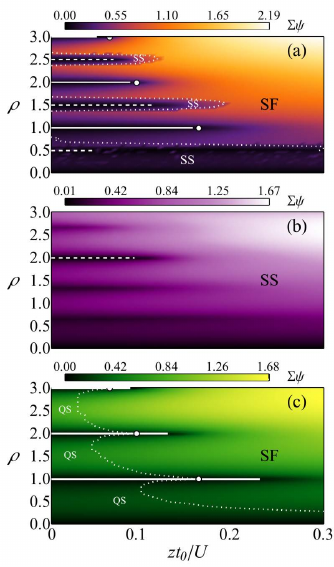}
\caption{ 
Phase diagrams at fixed density for two light induced modes ($R=2$). (a) The SF phase has total order parameter $\Sigma\psi=(|\psi_O|+|\psi_E|)/2\neq 0$ and $\psi_O=\psi_E$.  MI($n$) lobes appear for commensurate densities $n$, $\Sigma\psi=0$, $\rho=(\rho_O+\rho_E)/2=n$ with $\rho_O=\rho_E$, white lines. In between MI lobes, DW insulators form with $\Delta\rho=(\rho_O-\rho_E)/2\neq0$, maximal light scattering occurs then and $\Sigma\psi=0$, white dashed lines. SS phases ($\Sigma\psi\neq 0$, $\Delta\rho\neq0$ ) are indicated near the boundary between DW and SF states. (b) Large  $g_{\mathrm{eff}}<0$ regime. Only SS and DW exist (white dashed line). (c)  For convenience of comparison, we reproduce here again Fig. \ref{5-DminJD}. Large $g_{\mathrm{eff}}>0$ regime. MI insulators (white lines), regular SF and QS (two Fock component SF) exist. The two components in the system are the same. Dashed lines show the boundary below which QS occurs. White points denote the SF--MI transition point without cavity light. Parameters are for (a)  $g_{\mathrm{eff}}N_s=-0.5U$, (b) $g_{\mathrm{eff}}N_s=-1.25U$ (c) $g_{\mathrm{eff}}N_s=10U$;  $J_D=1.0$, $J_B=0.0$, $N_s=100$, $z=6$ (3D).  The color bar denotes $\Sigma\psi$ in the SF region. }
\label{5-FD2mDF}
\end{figure}

 The phase diagram of the system is shown in Fig. \ref{5-PD2mD} as a function of the chemical potential (being standard for the condensed matter presentation) and the effective tunneling amplitude $zt_0/U$. As a function of the density (being untypical for the condensed matter presentation) it is shown in Fig. \ref{5-FD2mDF} (a).  When light scatters maximally $g_{\mathrm{eff}}<0$ the modification due to the quantum fluctuations can be safely neglected as their contribution is strongly smeared out and we have  a dynamical OL (the contribution goes like $N_s$ vs. $N_s^2$ in contrast to the semiclassical contribution).  Depending on $g_{\mathrm{eff}}$ with respect to the on-site interaction $U$ there is the formation of DW  lobes in between the typical MI lobes in the system, at half integer fillings. In between the Mott regions as $U$ decreases at fixed $\mu/U$,  we find that SS phases can appear as  intermediate states from the DW towards the SF state as $U$ decreases. The size of the SS and DW phases is strongly influenced by the ratio $|g_{\mathrm{eff}}| N_s/U$.
This is similar to the case where nearest neighbors interaction is considered in an extended Bose--Hubbard model in addition to a soft-core (finite $U$) on-site interaction~\cite{Miyashita}, however here the coupling between the sub-lattices is via their difference in mean occupation (the DW order parameter). It is well known that the combination of soft-core bosons and nearest neighbors interactions stabilizes the SS phase against phase separation, we expect our system to  behave likewise. The number of photons scattered is  $\langle a_1^\dagger a_1 \rangle\propto\Delta\rho^2N_s^2$. Thus, when DW order occurs we expect a large signal in the detector as photons escape the cavity.

Interestingly, when the effective coupling exceeds the threshold, $|g_{\mathrm{eff}}||J_D|^2N_s\gg U$, the system can support only DW and SS phases. Moreovere, the density imbalance between them is maximal (all atoms go to either odd or even sites). MI and SF phases are not created at all. The phase diagram is presented in Fig. \ref{5-FD2mDF}(b). This occurs as the light induced interaction being effectively attractive for one of the modes in the system is equal or stronger than the repulsive on-site interaction. 

Such states with maximal imbalance for specific parameters, which we predicted, have been observed experimentally in Ref. \cite{EsslingerNature2016}. One can also compare this situation to the experiments with a BEC in a cavity without any lattice \cite{EsslingerNat2010}, where the supersolid state is formed with the maximal imbalance as well. Indeed, for a BEC without a lattice, the atomic interaction $U$ does not play very important role: the self-organization arises due to the competition between the light-induced interaction and tunneling. This is in coherence with our finding of the threshold value, which is easy to satisfy for small $U$.


\section{Quantum and dynamical lattices: scattering from matter-wave bonds at $90^\circ$}

In this section, we will consider light scattering at $90^\circ$ due to the matter-amplitude (or bond) dependent operator $\hat{B}$. We will predict novel phases neither discussed nor observed so far. First, we present the effect of dynamical OL and show the appearance of new phases of bonds: global superfluid dimers (SFD) and supersolid dimers (SSD). Second, we will demonstrate the effect of quantum OL, where the supersolid phase appears. Importantly, this supersolid arises due to the short-range interaction, which is surprisingly induced by the long-range light-mediated interaction. Thus, this supersolid phase is closer to the standard condensed matter expectations, where the short-range interaction is considered.

\subsection{Dynamical lattices: superfluid dimers and supersolid dimers of atomic bonds}

We find a novel phase transition, when light scattering from the bonds at $90^\circ$ ($J_{j,j+1}=(-1)^jJ_B$, $J_D=0$) is maximized ($g_\mathrm{eff}<0$).

The effective Hamiltonian of the system is then
\begin{gather}
\HH^b_{\mathrm{eff}}=\sum_{\xi=1}^4\HH^\xi_{\mathrm{eff}},
\\
\HH^\xi_{\mathrm{eff}}\approx\frac{N_s}{4}\big[\frac{z}{2} t_{\varphi_\xi}\hat\beta_\xi-\mu\hat n_\xi+g_\mathrm{eff}|J_B|^2\delta S_{\varphi_\xi}^2
+
U \hat n_\xi(\hat n_\xi-1) -g_{\mathrm{eff}}N_s|J_B|^2\tilde c_{B,{\varphi_\xi}}
\big]
\\
t_{\varphi_\xi}=t_0-g_{\mathrm{eff}}N_s|J_B|^2\tilde\eta_{B,{\varphi_\xi}},
\\
\hat\beta_\xi=
\big[
\psi_{\xi}^*\hat b^{\phantom{\dagger}}_{\xi+1}
+\psi_{\xi+1}^*\hat b_{\xi}^{\phantom{\dagger}}
+\psi_{\xi+1}^{\phantom{*}}\hat b_{\xi}^\dagger
+\psi_{\xi}^{\phantom{*}}\hat b_{\xi+1}^\dagger
-
(\psi_{\xi}^*\psi^{\phantom{*}}_{\xi+1}
+\mathrm{c.c.})
\big],
\end{gather}
\begin{eqnarray}
\tilde\eta_{B,\varphi_\xi}&=&\frac{z(-1)^{\xi+1}
}{8}\sum_{\xi'=1}^4(-1)^{\xi'+1}(\psi_{\xi'}^*\psi^{\phantom{*}}_{\xi'+1}+\mathrm{c.c.}),
\\
\tilde c_{B,{\varphi_\xi}}&=&\frac{z}{4}(\psi_{\xi}^*\psi^{\phantom{*}}_{\xi+1}+\mathrm{c.c.})\tilde{\eta}_{B,{\varphi_\xi}},
\end{eqnarray}
\begin{eqnarray}
\delta S_{\varphi_\xi}^2&=&\frac{z}{2}\big[\hat b_\xi^{2\dagger}\hat b_{\xi+1}^{2\phantom{\dagger}}+\hat b_{\xi+1}^{2\dagger}\hat b_{\xi}^{2\phantom{\dagger}}
+2\hat n_{\xi}\hat n_{\xi+1}+\hat n_{\xi}+\hat n_{\xi+1}
\nonumber\\
&+&\hat b_\xi^{2\dagger}\hat b_{\xi-1}^{2\phantom{\dagger}}+\hat b_{\xi-1}^{2\dagger}\hat b_{\xi}^{2\phantom{\dagger}}
+2\hat n_{\xi}\hat n_{\xi-1}+\hat n_{\xi}+\hat n_{\xi-1}
\nonumber\\
&-&2(\psi^*_\xi\psi^{\phantom{*}}_{\xi+1}+\mathrm{c.c.})\hat \beta_\xi+(\psi^*_{\xi}\psi^{\phantom{*}}_{\xi+1}+\mathrm{c.c.})^2
\nonumber\\
&-&2(\psi^*_\xi\psi^{\phantom{*}}_{\xi-1}+\mathrm{c.c.})\hat \beta_\xi+(\psi^*_{\xi}\psi^{\phantom{*}}_{\xi-1}+\mathrm{c.c.})^2
\big],
\end{eqnarray}
 where the component $\xi+4$ is the same as $\xi$, while $\langle\hat \beta_\xi\rangle=(\psi_{\xi}^*\psi^{\phantom{*}}_{\xi+1}
 +\mathrm{c.c.})$ and as usual $\langle\hat b_\xi\rangle=\psi_\xi$ defines the order parameters of the light induced modes.

\begin{figure}[h!]
   \captionsetup{justification=justified}
    \centering
    \includegraphics[width=0.9\textwidth]{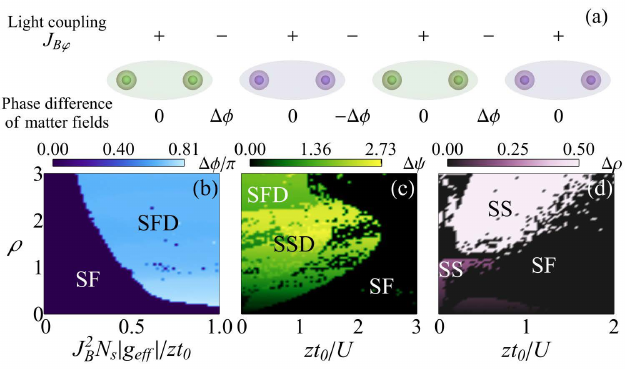}
    \caption{{Emergent dimer phases and quantum-light-induced supersolids due to scattering from bonds.} (a) Dimer structure for maximized scattering: matter-field coherence compensates for imposed light-field patterns. (b) Phase diagram for the phase difference $\Delta\phi \ne 0$ between superfluid dimers (SFD), when light scattering is maximized without on-site interaction (phase transition to $\Delta\phi = 0$ corresponds to standard SF). (c) Phase diagram for the difference in order parameters $\psi_{A/B}$ of the dimers $\Delta\psi=||\psi_A|^2-|\psi_B|^2|/2$ when light scattering is maximized; the on-site interaction exists and leads to supersolid dimers (SSD).      (d) Density wave order parameter $\Delta\rho=|\rho_A-\rho_B|/2$  for minimized scattering (i.e. the quantum optical lattice). The density components $\rho_{A/B}$ correspond to atomic populations of effective light-induced modes. 
     Regions with $\Delta\rho\neq0$ correspond to novel short-range supersolid phases;  $\Delta\rho=0$ corresponds to superfluid. 
    Parameters:  (b) $U=0$, (c) $g_\mathrm{eff}= -25U/N_s$, (d) $g_\mathrm{eff}=25U/N_s$;   $J_D=0$, $J_B=0.1$, $N_s=100$, and $z=6$.}
    \label{5-DminJB}
\end{figure}

Even in the absence of on-site interaction, a transition from normal SF to the superfluid dimer (SFD) state appears. SFD is a SF state in which the complex order parameter has alternating (zero and non-zero) phase difference between pairs of sites, and its amplitude is modulated as well. This occurs because of the competition between the kinetic energy BH terms, which promote a homogeneous SF, with the light-induced interaction that favors SF components with alternating phases across every other site [Fig. \ref{5-DminJB}(a)].  The phase of light interference flips from bond to bond [such that the phase difference at neighboring bonds is $\pi$, because we are in the diffraction minimum, Fig. \ref{5-DminJB}(a)]. In the Hamiltonian, this corresponds to alternating signs in front of the matter-field coherences between the neighboring sites (i.e. products of complex order parameters, in mean-field treatment). To minimize the energy (and maximize the light scattering) the quantum matter fields self-organize such that the matter-field phase difference between neighboring bonds flips as well to compensate for the imposed light-field phase flip. The dimer configurations have high degeneracy as the full many-body ground-state is composed of several equivalent arrangements of the phase pattern in space. The phase diagram is shown in~Fig. \ref{5-DminJB}(b). 

Moreover, in the presence of on-site interaction the system supports a transition to the supersolid dimer (SSD) state with modulated densities [Fig. \ref{5-DminJB}(c)]. The on-site interaction  suppresses atomic fluctuations, while as light scattering gets optimized, the density is unable to lock in a homogeneous density pattern leading to the additional density imbalance. Therefore, the phase modulation and density modulations coexist simultaneously while atoms retain mobility preventing the stabilization of an insulating phase.  

Note, that multimode bond structures can have very nontrivial spatial overlap and dimers (and their multimode generalizations as trimers, tetramers, etc., which can be obtained for tilted probe angle) extend over many sites demonstrating the interplay of the global and short-range properties. 

An analogy can be drawn concerning the properties of the bond dimer state with the well-known phase modulated state of superconducting fermions, the FFLO/LOFF state~\cite{FFLO1}. In the FFLO/LOFF state the superconducting order parameter varies in space periodically akin to the spatial phase variation seen in the dimer state. This can be traced back to the finite momentum transfer induced by light to the atoms via the bond coupling (addressing $\hat B$). This is similar to the finite momentum acquired by Fermi surface component mismatch in the fermionic system, forming finite momentum Cooper pairs which translates to the order parameter spatial variation. However, the dimer state can have in addition density modulation when interactions are present. Moreover these dimer states are akin to other condensed matter structures used in the study of strongly interacting quantum liquids~\cite{Balents} and could be used as building blocks for simulating them.

\subsection{Quantum lattices: short-range supersolids induced by long-range interactions}

We now switch to quantum optical lattices. Here we prove that there is a SS to SF transition that is solely driven by quantum correlations for minimized light scattering  ($g_{\mathrm{eff}}>0$), Fig. \ref{5-DminJB}(d). This occurs because the terms due to light--matter correlations in $\hat B^\dagger\hat B$ are not shadowed by semiclassical effects, as there is no classical light build up. Two-point tunneling correlations $\hat b^\dag_k \hat b_{k+1}\hat b_l \hat b^\dag_{l+1}$ introduce new terms in the BH model (for $k=l$), which couple densities at neighboring sites only: $\sum_{\langle i,j\rangle}\hat n_i\hat n_j$,  producing a DW instability even without strong light.   Density imbalance is energetically favoured and the atoms condense in a nearest-neighbour density pattern, while additional terms in $\HH^b_{\mathrm{eff}}$ favor atomic quantum fluctuations competing with the on-site interaction.

We have the two mode Hamiltonian:
\begin{gather}
\HH^b_{\mathrm{eff}}=\sum_{\xi=1}^2\HH^\xi_{\mathrm{eff}},
\\
\HH^\xi_{\mathrm{eff}}\approx\frac{N_s}{2}\big[z t_0\hat\beta_\xi-\mu\hat n_\xi+g_\mathrm{eff}|J_B|^2\delta S_{\varphi_\xi}^2
+
U \hat n_\xi(\hat n_\xi-1) 
\big],
\\
\hat\beta_\xi=
\big[
\psi_{\xi}^*\hat b^{\phantom{\dagger}}_{\xi+1}
+\psi_{\xi+1}^*\hat b_{\xi}^{\phantom{\dagger}}
+\psi_{\xi+1}^{\phantom{*}}\hat b_{\xi}^\dagger
+\psi_{\xi}^{\phantom{*}}\hat b_{\xi+1}^\dagger
-
(\psi_{\xi}^*\psi^{\phantom{*}}_{\xi+1}
+\mathrm{c.c.})
\big],
\\
\delta S_{\varphi_\xi}^2=z\big[\hat b_\xi^{2\dagger}\hat b_{\xi+1}^{2\phantom{\dagger}}+\hat b_{\xi+1}^{2\dagger}\hat b_{\xi}^{2\phantom{\dagger}}
+2\hat n_{\xi}\hat n_{\xi+1}+\hat n_{\xi}+\hat n_{\xi+1}
\nonumber\\
-2(\psi^*_\xi\psi^{\phantom{*}}_{\xi+1}+\mathrm{c.c.})\hat \beta_\xi+(\psi^*_{\xi}\psi^{\phantom{*}}_{\xi+1}+\mathrm{c.c.})^2
\big],
\end{gather}
where the component $\xi+2$ is the same as $\xi$. Here, the relevant contributions from the light matter coupling are the quantum fluctuations of the on-site coherences, the QOL. All the effect of the light--matter interaction reduces to the modification of quantum fluctuations of the $\hat B$ operator, which translate to the fluctuations in the order parameters.

Depending on the value of light--matter coupling strength $g_{\mathrm{eff}}$ compared to the other parameters of the system, we can have a SS phase. However, this SS phase is different from the density coupling case, discussed in the previous section, as it only depends on the density pattern between nearest neighbors, and does not have a global pattern as in all examples so far. Thus, the short-range processes induced by the long-range quantumness of light induce this transition. This is a closer analogy to the conventional scenario of supersolidity~\cite{RevSS, DasSarma, Pupillo,Ferlaino}, which is essentially a short-range effect due to the interaction $\hat n_i\hat n_{i+1}$.

Further details about the bond phases can be found in our papers \cite{CaballeroNJP2016,CaballeroPRA2016}, where we show the analogies of bond states with the valence bond states (VBS) and AKLT model. In addition, in these papers, we present the bond ordering obtained even by addressing the densities only (and not the bonds themselves). We present the phases arising due to the simultaneous interaction with both density $\hat{D}$ and bond $\hat{B}$ operators. We analyze the case of multiple cavity and probe modes as well. In general, the system we propose to study may help not only in the quantum simulations of existing materials, but in the design of novel solid state quantum materials never considered by the condensed matter community to date.


\section {Quantum lattices at diffraction maximum: shift of the Mott insulator -- superfluid phase transition}

In two previous sections we considered light scattering at the diffraction minimum. In this section we show the results about scattering into the diffraction maximum, where all atoms scatter identically to each other, and thus one atom mode is formed. 

\begin{figure}[h!]
\captionsetup{justification=justified}
\begin{center}
\includegraphics[width=0.7\textwidth]{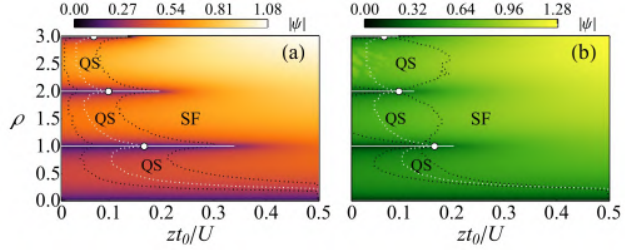}
\end{center}
\caption{Modifications of quantum phases due to quantum and semiclassical effects for homogenous scattering. (a) Phase diagram in terms of SF order parameter $\psi$ at fixed density for density-induced scattering. For minimal (maximal) scattering, MI boundaries (white lines) become extended (shortened) with respect to the transition point without cavity light (white point). This corresponds to suppression (enhancement) of quantum-light-induced atomic fluctuations. The behaviour of boundaries of the gapped QS state is similar (black dotted lines). The white dotted line is for gapless QS without cavity light. (b) Phase diagram for off-diagonal bond-induced scattering. The processes and lines are similar to (a), but arise due to renormalization of the tunnelling amplitude resulting from semiclassical light scattering.
(a) $g_\mathrm{eff}=25U/N_s$, boundaries are for $g_\mathrm{eff}=25U/N_s$, $0$, and $-12.5U/N_s$; $J_D=1.0$, $J_B=0$; (b)  $g_\mathrm{eff}=1.0U/N_s$, boundaries are for $g_\mathrm{eff}=1.0U/N_s$, $0$, and $-1.0U/N_s$; $J_D=0$, $J_B=0.05$; (a,b) $N_s=100$ and $z=6$ (3D).
\label{5-Dmax}}
\end{figure}

In the specific case of a single light-induced mode component in the diffraction maxima, we have: $\eta_D=2J_D^2(N_s\rho-1/2)$ and $\eta_B=2zJ_B^2N_s|\psi|^2$ with $\rho=\langle\hat n_i\rangle$ and $\psi=\langle\hat b_i\rangle$ for all sites and $z$ is the coordination number. This gives the effective tunneling amplitude $t_\varphi=t_0-2zg_{\mathrm{eff}}J_B^2N_s|\psi|^2$, the effective chemical potential $\mu_\varphi=\mu-g_{\mathrm{eff}}J_D^2(2(N_s-1)\rho-1)$  (where we have added all onsite density terms), and the effective interaction strength $U_\varphi=U+2g_{\mathrm{eff}}J_{D}^2$. At fixed density per site, it is the quantum light--matter correlations that effectively renormalize the on-site interaction from $U$ to $U+ 2 g_{\mathrm{eff}}J_D^2$ [see Fig. \ref{5-Dmax}(a)]. Thus, changing the light--matter coupling, one can shift the SF--MI transition point due to essentially the quantumness of light.

Choosing geometry, one can suppress the density scattering ($J_D=0$) and have all $J_B$'s equal. As no symmetry is broken, the bond self-organization does not emerge, but another semiclassical effect arises: tunneling is enhanced (suppressed) for maximum (minimum) scattering. This modifies the phase diagram (Fig. \ref{5-Dmax}(b)), because of nonlinear coupling of the SF order parameter $\psi=\langle\hat b_i\rangle$ to the tunneling amplitude $t_0$, which renormalizes to $t_0- 2 zg_{\mathrm{eff}} J_B^2N_s|\psi|^2$.

In the presence of cavity light, the gap opens in the SF state because the effective chemical potential depends on the density and renormalizes to $\mu-g_{\mathrm{eff}}J_D^2((2N_s-1)\rho-1)$. Without tunneling for $g_\mathrm{eff}>0$, the energy required to add a particle on top of the ground state is $\Delta E_{\mathrm{QS}}(\rho)=U\rho+g_\mathrm{eff}J_D^2(2N_s\rho+1)$ for incommensurate fillings between MI regions with fillings  $n$ and $n+1$, with  
 \begin{equation}
 \rho=\frac{U }{2g_\mathrm{eff}J_D^2N_s}\left(\frac{\mu}{U} -n\right)
 \end{equation}
  and the on-site number fluctuations $\Delta(\hat n_i)=(\rho-n)(1-\rho+n)$. For MI  regions at commensurate density $\rho=n$, the gap is $\Delta E_{\mathrm{MI}}(n)=U n+g_\mathrm{eff}J_D^2(2N_s n+1)$ and $\Delta(\hat n_i)=0$. { This means that only occupations of the lowest particle-hole excitations are allowed in between MI lobes. { Therefore in mean-field approximation this peculiar SF state is made of the quantum superposition of only two lowest Fock components that would satisfy the constraint on the density, instead of a coherent superposition of all possible fillings. Therefore, we use a term quantum superposition (QS) state for such a peculiar SF with a gap.}  The usual MI physics with renormalized  interaction strength occurs at commensurate fillings.} Similarly to SF--MI transition point, the boundaries of QS can be tuned (stabilized) due to the quantum correlations as $U$ is renormalized [Fig. \ref{5-Dmax}(a)]. The quantum superposition state can be understood as the SF state of minimal atomic fluctuations. { Note that a gap opens in the spectrum of excitations in the system, but because the particle filling is not commensurate with the lattice the system can't form an insulator and the best alternative energetically is the QS. As no symmetry has been broken in the system, the state is smoothly connected to the regular SF state and no phase transition occurs in the usual sense.}


\section{Multicomponent density orders for multiple atomic modes}

In this section, we will go beyond two atomic modes (at odd and even sites) and consider the density waves and supersolids of more complicated spatial structures.

\begin{figure}[h]
\captionsetup{justification=justified}
\begin{center}
\includegraphics[clip, trim=3.1cm 0cm 0cm 0cm, width=0.5\textwidth]{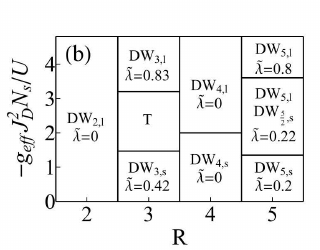}
\end{center}
\caption{Light scattering from on-site densities. 
Strongly interacting phase diagram for multiple number of modes $R$ (created for different probe angles) at half-filling. Quantum phases have a period of density-density correlations $q$ (in units of
lattice period), $\mathrm{DW}_{q,l}$ ($\mathrm{DW}_{q,s}$), $l$($s$) denotes large (small) density imbalance DW, $\tilde \lambda$ is the SF fraction~\cite{Mahan,Yang}. DW and SF order depend on the effective light--matter interaction strength $g_\mathrm{eff}$ and $R$. Horizontal lines denote phase boundaries between quantum phases.   
 Parameters:  scattering at various angles defining the mode number $R$, $J_D=1.0$, $J_B=0$, $t_0=0$, $N_s=100$.}
\label{5-DminJDx}
\end{figure}

Scattering at angles different from $90^\circ$ creates more than $R=2$ atomic modes ~\cite{Elliott2015}, and the light-imposed coefficients are $J_{j,j}=J_D \chi(j)$, where for traveling waves $\chi(j)=\exp\left({i 2\pi j}/{R} \right)$. Now, multiple terms $\hat N_\varphi^{\phantom{*}}
\hat N_{\varphi'}^{\phantom{*}}$ in Eq. (\ref{5-mdecomp}) become important. We present a phase diagram for various scattering angles (inducing  $R$ modes) for strong on-site interaction with maximal light scattering (Fig. \ref{5-DminJDx}). The $R$-mode induced pattern competes with on-site interaction modifying the density distribution. Hence, multiple DWs of period  $R$ (in units of lattice period) can coexist with SF forming multicomponent SSs. 

Interestingly, at half-filling for $R > 2$, odd, SS exists.  As $|g_\mathrm{eff}|$ changes from zero, checkerboard insulators form for even $R$ while different kinds of SS exist for odd $R$. The on-site interaction limits atomic fluctuations producing gapped SF components when SS exists. As $R$~increases, additional DWs with different periods and amplitudes emerge. These form unstable mixed state configurations ($T$) for $R=3$ and stable multicomponent SS for $R=5$. These occur in between regular SS phases with DWs of period $R$ and different amplitude. As the light induces atomic fluctuations, these generate competition between small and large amplitude DWs until the saturation in the SF component occurs. The system reaches a configuration, similar to the maximal imbalance DW state described above for scattering at $90^\circ$.

We will now consider a case of three modes in more details. One can choose to illuminate by a traveling wave, such that the projection of the difference between the cavity mode and probe wave vectors is $({\bf k}_{0}-{\bf k}_1){\bf r}_j=2\pi j/3$. This will induce $R=3$ spatial modes in the system, assuming that the lattice is sufficiently deep such that $\hat B\approx0$. We have then, that the light only couples to the density and the coupling is such that the mode operators $\hat D$ can be  written as
\begin{equation}
\hat D=J_D(\hat N_{1}+e^{\frac{i2\pi}{3}}\hat N_{2}+e^{\frac{i4\pi}{3}}\hat N_{3}),
\end{equation}
where each mode corresponds to a third of the lattice sites ($N_s/3$). The effective Hamiltonian of the system can be written as
\begin{eqnarray}
\HH^b_{\mathrm{eff}}&=&\sum_{\xi=1}^3\HH^\xi_{\mathrm{eff}},
\\
\HH^\xi_{\mathrm{eff}}&\approx&\frac{N_s}{3}\big[z t_0\hat\beta_\xi-\mu_\xi\hat n_\xi
+
U_\xi \hat n_\xi(\hat n_\xi-1)
\big],
\label{5-3mH}\\
\hat\beta_\xi&=&
\left[
(\psi_{\xi+1}^*+\psi_{\xi-1}^*)\hat b^{\phantom{\dagger}}_{\xi}
+(\psi_{\xi+1}^{\phantom{*}}+\psi_{\xi-1}^{\phantom{*}})\hat b_{\xi}^\dagger\phantom{\frac{1}{2}}\right.
\nonumber
\\
&-&\left.\frac{1}{2}
(\psi_{\xi}^*(\psi^{\phantom{*}}_{\xi+1}+\psi^{\phantom{*}}_{\xi-1})+\mathrm{c.c.})
\right],
\\
\mu_\xi&=&\mu-\frac{g_{\mathrm{eff}}N_s|J_D|^2}{3}\left(\rho_\xi-\frac{\rho_{\xi+1}+\rho_{\xi-1}}{2}\right),
\label{5-3mmu}
\\
U_\xi&=&U+2g_{\mathrm{eff}}|J_D|^2,
\end{eqnarray}
where the component $\xi+3$ is the same as $\xi$, similarly $\xi=0$ corresponds to $\xi=3$. From the above we can see that for minimal light scattering ($g_{\mathrm{eff}}>0$), the problem reduces to the Hamiltonian of the homogeneous system as density imbalance configurations are strongly suppressed and quantum fluctuations will shift the SF--MI transition as discussed previously. 

\begin{figure}[h!]
\captionsetup{justification=justified}
\centering
\includegraphics[width=0.4\textwidth]{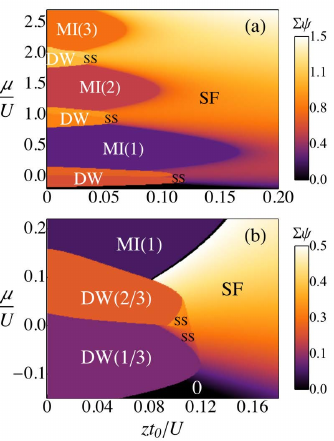}
\caption{(a) Phase diagram of multicomponent density ordered states, when geometry of the light probe is such that three spatial modes can occur  ($R=3$). The system is SF, without any spatial pattern as the interaction is decreased. The SF phase has total order parameter $\Sigma\psi=(|\psi_1|+|\psi_2|+|\psi_3|)/3\neq 0$ and $\psi_1=\psi_2=\psi_3=\psi$.  MI($n$) lobes appear for commensurate densities $n$, $\Sigma\psi=0$, $\rho=\rho_1=\rho_2=\rho_3$. In between MI lobes DW insulators form with $\langle\hat D\rangle\neq0$, maximal light scattering occurs then ($\Sigma\psi=0$). SS phases ($\Sigma\psi\neq 0$, $|\langle\hat D\rangle|^2\neq0$ ) are indicated near the boundary between DW and SF states. (b) Phase diagram closer to the DW-SS-SF transition at the tip of the DW insulators. The particular composition of the DW insulators is revealed, as depending on the chemical potential one has different values of DW phases. The transition can have an intermediate SS state toward the homogeneous SF phase. Parameters are for (a) and (b): $J_D=1.0$, $J_B=0.0$, $g_{\mathrm{eff}}N_s/U=-0.5$, $N_s=100$, $z=6$ (3D). The color bar denotes $\Sigma\psi$ in the SF region, other different quantum phases (labeled) are denoted by different colors.}
\label{5-PD3m}
\end{figure}

Nevertheless, we see that when there is maximal light scattering ($g_{\mathrm{eff}}<0$) a DW instability occurs.  The coupling between adjacent modes favours density imbalance, effectively one has for the ground state energy terms  of the form $\propto\rho_{\xi}\rho_{\xi-1}>0$ and $\propto\rho_{\xi}\rho_{\xi+1}>0$. These arise from the effective chemical potential $\mu_\xi$ in Eq. (\ref{5-3mH}), that in fact depends on the density of the three modes in the system, Eq. (\ref{5-3mmu}). In contrast to illuminating at $90^\circ$, here a self-organized state with three components occurs. The state is six fold degenerate, and the light amplitude gives the information about the DW order formation. The light amplitude is proportional to $\langle\hat D\rangle$, we have that when, $|\langle\hat D\rangle|^2=N_s^2|J_D|^2(\rho_1^2+\rho_2^2+\rho_3^2-\rho_1\rho_2-\rho_2\rho_3-\rho_1\rho_3)/9>0$ then the system maximizes light scattering and DW order is stablished. Depending on the competition between the light-induced interaction and the atomic on-site interaction we will have that the system will support DW insulators, MI insulators and even SS states, see Fig. \ref{5-PD3m}. The DW occur as one of the three components is strongly suppressed while the remaining two are uneven. These DW insulators appear in between MI lobes as the chemical potential is increased and have a critical value of $zt_0/U$ below the SF--MI transition.  As the system moves from the DW towards the SF state for small $U$ at fixed chemical potential, the system transitions via formation of  SS states, Fig. \ref{5-PD3m}.  The SS states we find, occur at the tip of the DW insulators, Fig. \ref{5-PD3m}(b).  It is interesting to note that the half-filled case will aways be in the SS state for large interaction as this state will be in the tip of the DW lobe, and it will shift to have a large (2/3) or small (1/3) DW while increasing $|g_{\mathrm{eff}}|$. There will be an intermediate region where the state will be better described as a mixed state of both configurations, all this consistent with the $t_0/U=0$ limit and exact diagonalization simulations~\cite{Caballero2015}. The density pattern that emerges in the system has a period of 3 in units of the lattice spacing.


\section{Squeezed light generation and the full light--mater entangled state }

In this section, we give an expression for the full entangled state of light and matter. The details of the derivation and its properties can be found in our paper \cite{Caballero2015a}. When the light is adiabatically eliminated (as we do in this chapter), the matter components are entangled to coherent states of light of different amplitudes (see for details \cite{Caballero2015}), as we have seen in Chapter 2. Nevertheless, when the light is not adiabatically eliminated and is treated as a dynamical variable (which we do in  Ref. \cite{Caballero2015a}), such light components become squeezed.

The general expression for the full light--matter state is the following:

\begin{equation}
|\Psi\rangle=\sum_{\varphi_q}\Gamma^b_{\varphi_q}(t)\beta_{\varphi_q}|\varphi_q\rangle_b|\alpha^{\phantom{\chi}}_{\varphi_q}+\alpha^\chi_{\varphi_q},\xi_{\varphi_q}\rangle_a,
\end{equation}
where the subscript $a$ ($b$) corresponds to the light (matter) part; $\Gamma^b(t)=\exp(-i\HH_{\mathrm{eff}}t)$, and $\Gamma^b_{\varphi_q}|\varphi_q\rangle_b=\hat\Gamma^b|\varphi_q\rangle_b$. The light components are squeezed coherent states  $|\alpha,\xi\rangle_a=D(\alpha)S(\alpha)|0\rangle_a$, with the squeezing operator $S(\xi)=\exp[(\xi^* a_1^2-\xi a_1^{\dagger 2})/2]$ and the displacement operator $D(\alpha)=\exp(\alpha a_1^\dagger-\alpha^* a_1)$.  The ground state of the effective Hamiltonian is $|\Psi\rangle_b=\sum_{\varphi_q}|\varphi_q\rangle_b$. The light amplitudes due to the projection of the matter structure are $\alpha_{\varphi_q}|\varphi_q\rangle_b=c\hat F|\varphi_q\rangle_b$, $\alpha^\chi_{\varphi_q}|\varphi_q\rangle_b=c\hat\eta|\varphi_q\rangle_b$, $c=U_{10}a_0/(\Delta_c+i\kappa)$ with 
$\hat \eta=\sum_{n=0}^{\infty}\tilde{U}^{n}\hat \chi_n$, $\tilde{U}=U/\Delta_c$, $\hat \chi_n$ are the linear combinations of the density mode and bond mode operators \cite{Caballero2015a}.
The weights due to the dynamical character of the light are 
$
\beta_{\varphi_q}=\exp(|c|^2\sum_{n=0}^{\infty}\tilde{U}^{2n}|\chi_{n,\varphi_q}|^2),
 $
 with $\chi_{n,\varphi_q}|\varphi_q\rangle_b=\hat\chi_n|\varphi_q\rangle_b$. 
In addition, the squeezing parameter amplitudes corresponding to the projection onto the matter sector are $\xi_{\varphi_q}|\varphi_q\rangle_b=\hat\xi|\varphi_q\rangle_b$ with
the squeezing amplitude operator $\hat\xi$, which is given by the linear superposition of atomic density mode and bond mode operators \cite{Caballero2015a}.  

Therefore, the structure of the strongly correlated matter gets imprinted in the quantum properties of light  via the squeezing parameter projections $\xi_{\varphi_q}|\varphi_q\rangle_b$. This generates a non-trivial superposition of squeezed coherent states entangled with the strongly correlated matter. 

Moreover, we showed that manipulating the matter, we can control the nonclassical features of light. This can be   accessible in an experiment via quantities such as the photon number and quadratures.  We showed how the quantum properties of light contain the information of matter-field coherences, density patterns of matter, and light--matter quantum correlations.


\section {Quantum simulators based on the global collective light--matter interaction: multiple light modes and tunable interaction length}

In this section, we present the opportunity to tune the effective interaction length between the atom modes. This leads us to a general view on the quantum simulations using the global light--matter interaction, which will include the features of the short-range systems. At the same time such quantum simulators will benefit from the collective enhancement of the light--matter interaction.

Ultracold gases loaded in an optical lattice is an ideal tool for studying the quantum degenerate regime of matter. Controlling the coupling between the atoms and light beams creating the optical lattice allows to realize simple models~\cite{Lewenstein} that were first formulated in different fields of physics from condensed matter to particle physics and  biological systems. These models would be useful for quantum simulation purposes and quantum information processing (QIP) applications. Specifically, one can realize effective Hamiltonians which contain short-range physical processes such as tunneling between neighbor lattice sites and on-site interactions. The implementation of long-range interactions that extends over many lattice sites is an extremely challenging task since it requires the use of more complex systems such as polar molecules~\cite{lahaye2009,Ferlaino} or Rydberg atoms~\cite{PohlLukin,Fleischhauer}. Moreover, spatial structure of the interaction itself is fixed by the physical system used (e.g. dipole-dipole interaction for molecules and Van der Waals interaction for Rydberg atoms) and cannot be changed. 

 In contrast to these examples, we show that by loading an optical lattice inside a cavity allows to engineer synthetic many-body interactions with an arbitrary spatial profile. These interactions are mediated by the light field and do not depend on fundamental processes, making them extremely tunable and suitable for realizing quantum simulations of many-body long-range Hamiltonians. In contrast to other proposals based on light-mediated interactions~\cite{Porras2006,Strack,IonsFR2012, Lesanovsky2013,PhotCrys2015}, we suggest a novel approach, where the shortening of the a priori infinitely long-range (global) light-induced interaction does not degrade the collective light--matter interaction, but rather benefits from it. In particular, in contrast to other proposals, where shortening of the interaction length requires increasing number of light modes, in our case, rather short-range interactions can be simulated with small number of light modes. Moreover, even a single mode cavity is enough to simulate some finite-range interactions. As a result, the quantum phase of matter posses properties of systems with both short-range and global collective interactions. The effective Hamiltonians can be an acceptable representation of an otherwise experimentally hard to achieve quantum degenerate system with finite range interactions.

We now generalize our model for many probe and cavity modes. The atomic system is probed with classical beams and the scattered light is selected and enhanced by the optical cavities (or by the modes of a single cavity). The  light from the probes has amplitudes  $\Omega_p$ (in units of the Rabi frequency). The probe-cavity detunning is $\Delta_{pc}=\omega_p-\omega_c$. The cavity modes couple with the atoms via the effective coupling strengths $g_{pc}= g_c \Omega_p/(2\Delta_{pa})$, with $g_c$ being the light--matter coupling coefficients of the cavities. The light--atom interaction Hamiltonian is then
\begin{equation}
\HH^{ab}=\sum_{c,p} \left(g_{pc}^*\hat a_c\hat F_{pc}^\dagger+g_{pc}\hat a_c^\dagger\hat F_{pc}\right)
\label{LMp}
\end{equation}
with
$
\hat F_{pc}= \hat D_{pc}+\hat B_{pc}$. $\hat D_{pc}=\sum_{j}J^{pc}_{jj}\hat n_j$ is the density coupling of light to the atoms, $\hat B_{pc}=\sum_{\langle i,j\rangle}J_{ij}^{pc}( \hat b^\dagger_i\hat b^{\phantom{\dagger}}_j+h.c.)$ is due to the inter-site densities reflecting matter-field interference (bonds). The sums go over illuminated sites $N_s$, and nearest neighbor pairs $\langle i,j\rangle$. As before, all atoms equally coupled to light belong to the same mode, while the ones coupled differently belong to different modes $\varphi$. Then we have for the atomic operators
\begin{equation}
\hat F_{pc}=\sum_{\varphi}J^{pc}_{D,\varphi}\hat N_\varphi+\sum_{\varphi'}J^{pc}_{B,\varphi'}\hat S_{\varphi'},
\end{equation}
where the light induced density  $\hat N_\varphi$ and  bond  $\hat S_{\varphi}$ mode operators are 
\begin{equation}
\hat N_\varphi=\sum_{i\in\varphi}\hat n_i ,\; \textrm{and}\; \hat S_{\varphi}=\sum_{\langle i,j\rangle\in\varphi}(\hat b_i^\dagger\hat b_j^{\phantom{\dagger}}+\hat b_j^\dagger\hat b_i^{\phantom{\dagger}}),
\end{equation}
with $J^{pc}_{D,\varphi}$ corresponding to the posible values of $J^{pc}_{ii}$ and $J^{pc}_{B,\varphi'}$ corresponding to $J^{pc}_{ij}$, where the pair $\langle i,j\rangle$ are the nearest neighbors.

The Hamiltonian Eq. (\ref{5-effmodel}) then generalizes to 
\begin{equation}
\HH_{\mathrm{eff}}=\HH^b+\sum_c\sum_{p,q}\left(\frac{g_{\mathrm{eff}}^{pqc}}{2}\hat F_{pc}^\dagger\hat F_{qc}^{\phantom{\dagger}}+\frac{(g_{\mathrm{eff}}^{pqc})^*}{2}\hat F_{pc}^{\phantom{\dagger}}\hat F_{qc}^\dagger\right)
\label{5-feff}
\end{equation}
with the effective coupling strengths
\begin{equation}
g_{\mathrm{eff}}^{pqc}=\frac{g^*_{pc}g_{qc}}{\Delta_{qc}+i\kappa_c}.
\end{equation}
 The sum over $c$ goes over the cavity modes (for a multimode cavity or several cavities) and $p$ and $q$ go over the probes.
 The new terms beyond BH Hamiltonian give the effective long-range light-induced interaction between density and bond modes that depend on geometry of the cavity modes and light probes injected into the system. We can rewrite the new terms as
\begin{gather}
\sum_c\sum_{p,q}\left(\frac{g_{\mathrm{eff}}^{pqc}}{2}\hat F_{pc}^\dagger\hat F_{qc}^{\phantom{\dagger}}+\frac{(g_{\mathrm{eff}}^{pqc})^*}{2}\hat F_{pc}^{\phantom{\dagger}}\hat F_{qc}^\dagger\right)= \nonumber\\
\sum_{\varphi,\varphi'}\sum_c\sum_{p,q}\Big[\tilde\gamma^{D,D}_{\varphi,\varphi'}(c,p,q)\hat N_\varphi^{\phantom{*}}
\hat N_{\varphi'}^{\phantom{*}}
+\tilde\gamma^{B,B}_{\varphi,\varphi'}(c,p,q)\hat S_\varphi^{\phantom{*}}\hat S_{\varphi'}^{\phantom{*}}+
\tilde\gamma^{D,B}_{\varphi,\varphi'}(c,p,q)(\hat N_\varphi^{\phantom{*}}\hat S_{\varphi'}^{\phantom{*}}
+\hat S_{\varphi'}^{\phantom{*}} \hat N_{\varphi}^{\phantom{*}})\Big]
\label{5-mdecompX}
\end{gather}
with
$
\tilde\gamma^{\mu,\nu}_{\varphi,\varphi'}(c,p,q)=[g_{\mathrm{eff}}^{pqc}(J_{\mu,{\varphi}}^{pc})^* J^{qc}_{\nu,{\varphi'}}+\mathrm{c.c.}]/2$, $\{\mu,\nu\}\in\{D,B\}
$. Cast in the above form, it is clear that the  terms are similar to  density-density interactions, exchange interactions, and the last two terms to a combination of the action of both multimode density and exchange. The different values of the effective interaction couplings  $\tilde\gamma$ between light-induced spatial mode operators $\hat N_\varphi$ and $\hat S_\varphi$ will determine the emergent phases in the system.
The structure constants together with the effective interaction couplings $g_{\mathrm{eff}}^{pqc}$ give an unprecedented ability to design an arbitrary interaction profile between the atoms. The new terms from Eq. (\ref{5-mdecompX}) provide the system with a plethora of new possibilities.   In principle  one can design arbitrary interactions patterns by choosing the geometry of the light: modifying angles, probe amplitudes, detunings and cavity parameters. Certainly, the potential of this to enrich cold atomic systems in terms of the simulation and design of interactions is vast. In what follows, we will discuss some simple examples of how the above Hamiltonian can be used to generate synthetic interactions for quantum simulation, i.e. the couplings between the atoms is not a consequence of fundamental physical processes but are mediated by the light field and depend on the geometrical arrangement of the probe beam and the optical cavity. 

\subsection{One probe and one cavity mode}

\begin{figure}[h]
\captionsetup{justification=justified}
\centering
\includegraphics[width=0.5\textwidth]{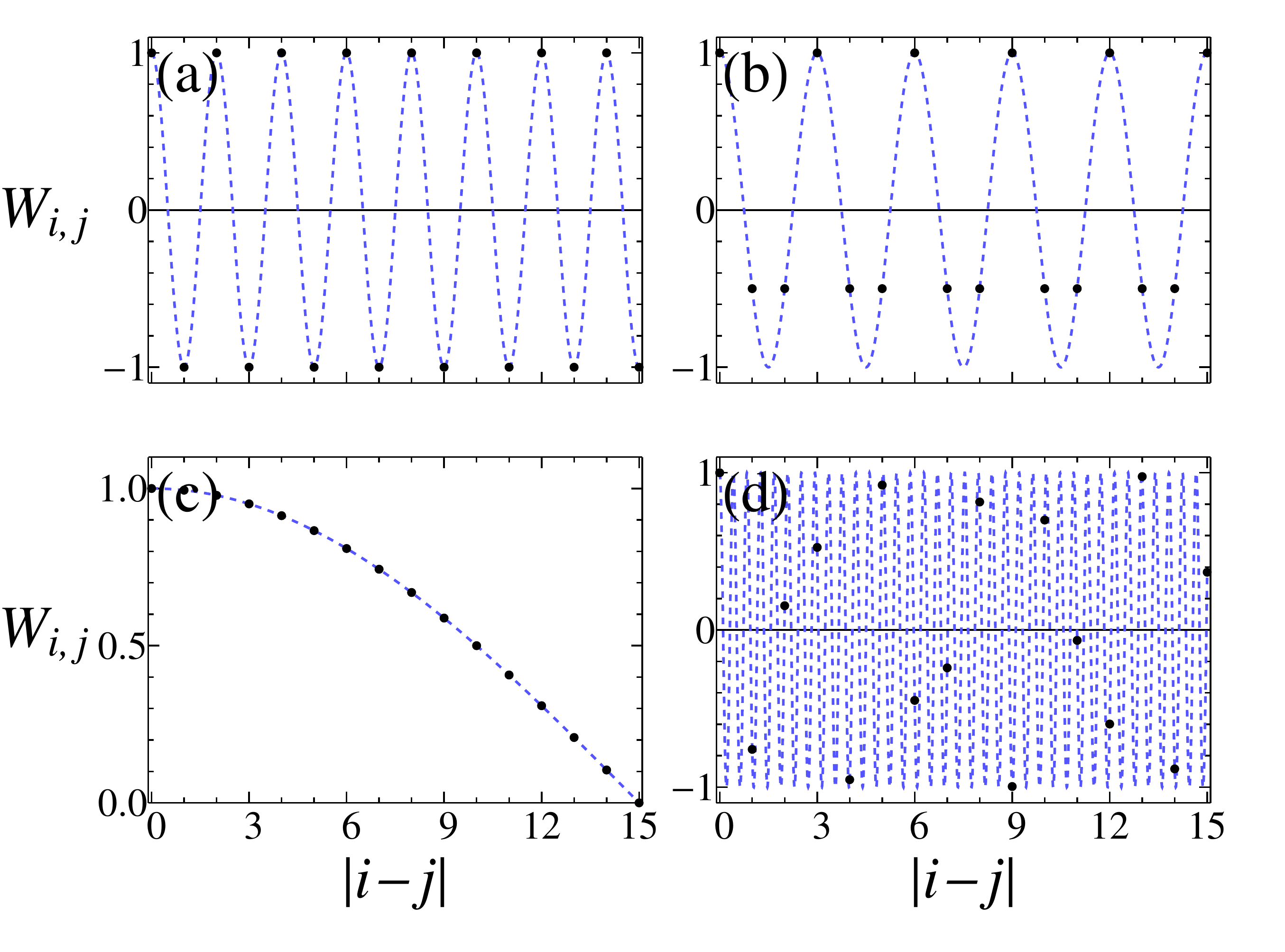}
\caption{Different interactions that can be implemented using a single cavity mode and a single probe for $L=16$ lattice sites and using traveling waves as mode functions. The panels illustrate the value of $W_{i,j}$ as a function of $|i-j|$ (normalized). Panels (a) and (b): interaction strength in presence of $R=2$ ($\theta_{c_1}=0$, $\theta_{p_1}=\pi/2$, $\lambda=2d$) and $3$ ($\theta_{c_1}=0$, $\cos \theta_{p_1}=1/3$, $\lambda=2d$) spatial modes. The value of $W_{i,j}$ depends solely on the mode distance. Panel (c): increasing the number of atomic modes leads to long-range interactions with a well-defined spatial profile ($\theta_{c_1}=0$, $\cos \theta_{p_1}=7/8$, $\lambda=2d$). Panel (d): the periodicity of $W_{i, j}$ is incommensurate to the lattice spacing, resulting in a ``disordered'' interaction. The dashed lines are a guide to the eye between modes choosen.}\label{5-fig:inter1}. 
\end{figure}

For a deep OL, the $\hat{B}_{cp}$ contribution to $\hat{F}_{cp}$ can be neglected. 
We find that the effective atom-atom interaction  can be re-written as
\begin{align}\label{5-eq:Heff1}
\HH_1^D&=g_\mathrm{eff}\sum_{i,j}W_{ij} \hat{n}_i \hat{n}_j
\end{align}
with $g_{\mathrm{eff}}=\Re[g_{\mathrm{eff}}^{111}]/2=\Delta_{11}|g_{11}|^2/(\Delta_{11}^2+\kappa_1^2)$, thus the effective interaction strength factors out. The specific dependence of the functions $J^{11}_{ii}$ defines the spatial profile of the interaction between the atoms via  the interaction matrix $W_{ij} = 2\Re \left[(J^{11}_{ii})^* J^{11}_{jj}\right]$. 
{
In order to illustrate this, we focus on a one-dimensional lattice and we consider traveling waves as mode functions for the light modes (i.~e. $u_c(\b{r})=e^{i \b{k}_c \cdot \b{r}}$ and $J^{11}_{jj}=e^{i (\b{k}_{c_1}-\b{k}_{p_1}) \cdot \b{r}_j}$) so that $W_{ij}$ becomes
\begin{align}\label{5-eq:IntMat1}
W_{ij}=\cos \left[(\b{k}_{c_1}-\b{k}_{p_1})\cdot (\b{r}_i-\b{r}_j)\right].
\end{align}
The cavity induces a periodic interaction between the atoms and, depending on the projection of $\b{k}_{c_1}-\b{k}_{p_1}$ along the lattice direction $\hat e_{\b{r}_i}$, its spatial period can be tuned to be commensurate or incommensurate to the lattice spacing. Specifically, if $J^{11}_{jj}=e^{i 2 \pi j /R} $, atoms separated by $R$ lattice sites scatter light with the same phase and intensity and the optical lattice is partitioned in $R$ macroscopically occupied regions (spatial modes) composed by non-adjacent lattice sites \cite{Elliott2015}. The specific geometric configuration of the light beams is determined by the angles $\theta_{c_1,p_1}$ between the wave vectors $\b{k}_{c_1,p_1}$ and $e_{\b{r}_i}$: in order to have $R$ spatial modes, one has to set
\begin{align}\label{5-geometry1}
\cos \theta_{p_1}=\cos \theta_{c_1} - \frac{\lambda}{d} \frac{1}{R},
\end{align}
where $\lambda$ is the wavelength of the light modes.} If this relation is fulfilled, $W_{ij}$ has periodicity $R$ lattice sites and, defining the operator $\hat{N}_j$ to be the population of the mode $j$, the interaction strength between two atoms belonging to the different modes $i$ and $j$ depends solely on their mode distance ($(i-j) \mathrm{mod}R$) and not their actual separation. In this case, we find that the mode-mode interaction Hamiltonian is given by
\begin{align}\label{5-interaction1}
\HH^D_1=2 g_\mathrm{eff} \sum_{i,j=1}^R \cos \left(\frac{2 \pi}{R}(i-j)\right) \hat{N}_i \hat{N}_j.
\end{align}
The case $R=2$ has been presented in detail in this work and realized experimentally \cite{EsslingerNature2016}, showing that new light-mediated interaction heavily affects the ground state of the Bose--Hubbard model inducing a new supersolid phase. Simply changing the angle between the cavity and the optical lattice, one can increase the number of spatial modes and therefore implement more complicated long-range interactions that cannot be obtained using molecules or Rydberg atoms [Fig.~\ref{5-fig:inter1}].

If the period of $W_{ij}$ is incommensurate to the lattice spacing, the spatial modes are not well-defined since each lattice sites scatters light with a different phase and amplitude. Therefore, the value of interaction matrix is no longer periodic and resembles disorder~\cite{Morigi2}. This behavior is analogous to the potential generated by the superposition of two incommensurate optical lattices~\cite{lewenstein2010} which has been used for generating controllable disordered potential, allowing the observation of Anderson localization and new quantum phases (Bose glass) in ultracold gases. Here, we move a step further and we create synthetic random interactions between the atoms, generalizing the Anderson model where disorder affects only the local potential and/or the tunneling amplitude.

\subsection{Multiple probes and one cavity mode}

The previous simple case, where only one cavity and one probe are present, allows to realize interactions that resemble a cosine profile. We now turn to a different monitoring scheme where the atoms are probed with $R$ different classical probes and scatter light to a single optical cavity.

In addition to the previous section, where the light mode of the cavity has contributions from all the sites of the optical lattice, all the probe beams concur to the value of the light field inside the cavity and the effective atom-atom interaction is 
\begin{align}\label{5-eq:Heff2}
\HH^D_1&=\sum_{p,q=1}^{R}\left(\frac{g^{pq1}_\mathrm{eff}}{2}\hat{D}_{p1}^\dagger \hat{D}_{q1} + \mathrm{H.c.}\right).
\end{align}

\begin{figure}[h]
\captionsetup{justification=justified}
\centering
\includegraphics[width=0.9\textwidth]{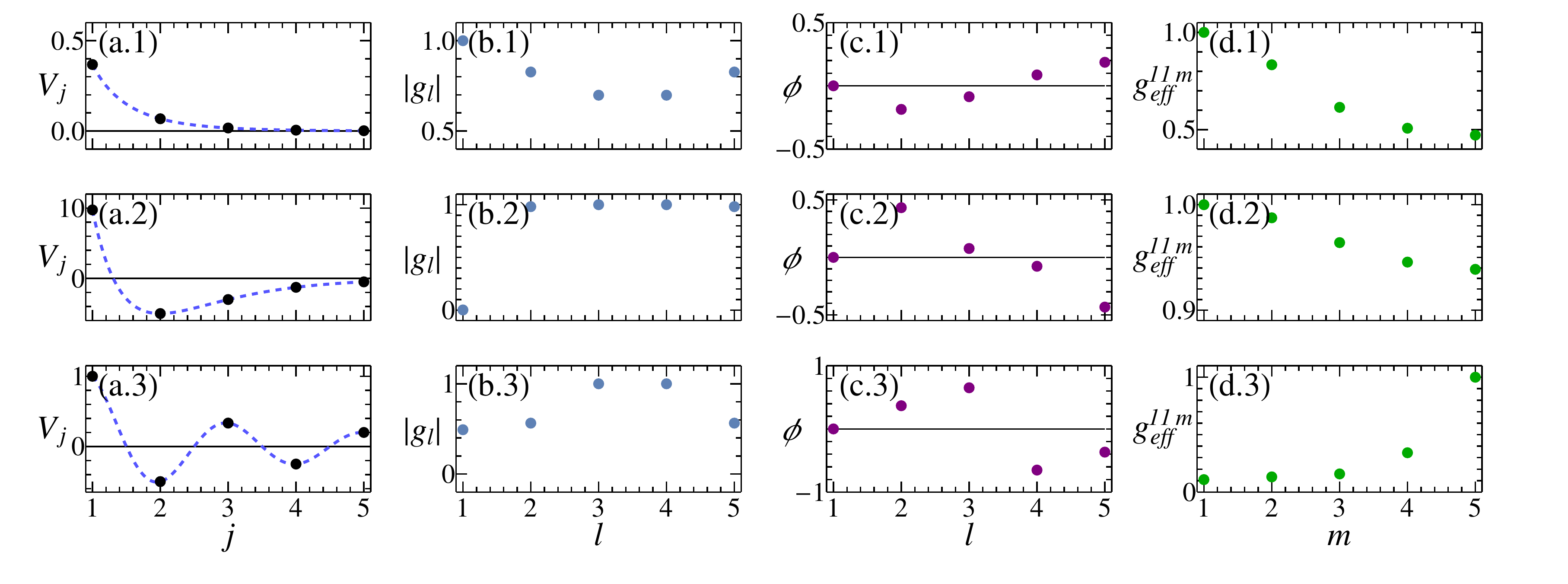}
\caption{Examples  of different synthetic interactions that can be implemented using traveling waves as mode functions. Panel (a) represents the effective synthetic interaction potential $V_j$. We show a Yukawa potential ($V_j=V_j = e^{-j}/j$) in panel (1), a Morse potential in panel (2)$ \left(V_j=5\left[ (1-e^{-(j-2)})^2-1 \right]\right)$ and a Bessel potential in panel (3) ($V_j=\pi y_0(\pi x)$), where $y_0$ is the Bessel function of the second kind.
Panels (b) and (c) show the parameters using a single cavity and five probes described by the effective intensity $|g_l|$ (normalized) and the phase $\phi=\mathrm{Arg}(g_l)$, the effective interaction is $W_{ij}\propto V_i^*V_j$ ($\theta_{c_1}=0$, $\cos \theta_{p_l}=1-2l/5$, $\lambda=2d$). Panel (d)  shows the effective couplings for five cavity modes and one probe corresponding to (a), the effective interaction is $W_{ij}\propto V_{|i-j|}$ ($\theta_{p_1}=0$, $\cos \theta_{c_l}=1-l/5$, $\lambda=2d$). The dashed lines are a guide to eye of the effective mode interaction.} \label{5-fig:potential}
\end{figure}

The spatial profile of the interaction described by (\ref{5-eq:Heff2}) can be tuned changing the light mode functions or/and the intensity and phase of the probe beams. As in the previous examples, we consider traveling waves as mode functions for the light and we note that 
\begin{gather}\HH^D_2=\gamma_{\mathrm{eff}}\sum_{l,m=1}^{R} \sum_{i,j}\left(g_{lc}^*g^{\phantom{*}}_{mc} e^{- i (\b{k}_{c_1}-\b{k}_{p_l}) \cdot \b{r}_i}
 e^{i (\b{k}_{c_1}-\b{k}_{p_m}) \cdot \b{r}_j}\hat{n}_i \hat{n}_j +  \mathrm{H.c.}\right)
 \label{5-eq:modefun1}
\end{gather}
with $\gamma_{\mathrm{eff}}=\Delta_{pc}/(\Delta_{pc}^2+\kappa_c^2)$, as the probes have the same wavelength $\lambda_{p_l}=\lambda_{p_m}$, such that all the detunings  are the same: $\Delta_{p_lc}=\Delta_{p_mc}=\Delta_{pc}$.  This is equivalent to the product of two Discrete Fourier Transforms (DFT) if $d(\b{k}_{c_1}-\b{k}_{p_l}) \cdot \hat e_{\b{r}_i}=2 \pi l/R$ for all the probe beams $l=1,2,...R$, $\hat e_{\b{r}_i}$ is the unit vector of $\b{r}_i$ and $d$ is the lattice spacing. {Importantly, these conditions fix only the directions of the probe beams ($\cos \theta_{p_l}=\cos \theta_{c_1} - \frac{\lambda}{d} \frac{l}{R}$) and not their intensities or phases, i. e., the coefficients $g_{p}$. }Furthermore, the probe beams divide the optical lattice in $R$ macroscopically occupied spatial modes, which interact according to the Hamiltonian
\begin{align}\label{5-interaction2}
\HH^D_{2}=\gamma_{\mathrm{eff}}\sum_{i,j=1}^R \left(V_i^* V_j \hat{N}_i \hat{N}_j + \mathrm{H.c.}\right),
\end{align}
where $V_j= \sum_{m=1}^{R} g_{mc}   e^{ i\frac{ 2 \pi m j}{R} }$ describes the strength of the interaction between the spatial modes defined by the light scattering. Tuning the probe intensities and their relative phase, the function $V_j$ can be modified to design to any spatial profile, as illustrated in Fig.~\ref{5-fig:potential}.  Importantly, Eq.~(\ref{5-interaction2}) implies that the coupling between the modes $i$ and $j$ is $W_{ij}\propto V_i^* V_j$ and therefore the interaction matrix $W_{ij}$ does not depend on the distance between the spatial modes. This is in contrast to the usual solid state physics scenarios where interactions do not depend on the specific position of two particles but only on their distance. The scheme we propose opens the possibility of studying new classes of interactions and effects not observable in conventional systems. It is worth mentioning that there is only one distance-dependent interaction function that can be realized with this setup: a cosine profile. Specifically, one has $W_{ij}=W_{|i-j|}$ only if $V_j$ is a pure phase (which can be obtained with one probe and one cavity).

{ In the above section, we assumed the light mode functions to be traveling waves which allowed us to give a simple description of the synthetic interactions in terms of DFT. Importantly, the definition of the spatial modes does not rely on this assumption but only on the fact that the coefficients $J_{jj}$ can have the same value on different lattice sites so that atoms in these positions scatter light with the same phase and intensity. For example, it is possible to realize the case $R=2$  by considering standing waves ($u_i(\b{r}) = \cos(\b{k}_i \cdot \b{r })$) crossed at such angles to the lattice that $\b{k}_0\cdot \b{r }$ is equal to $\b{k}_1\cdot \b{r }$ and shifted such that all even sites are positioned at the nodes, so $J_{ii}=1$ for $i$ odd while $J_{ii}=0$ for $i$ even or the $R=3$ case by imposing $\b{k}_{\mathrm{1,0}}\cdot \b{r}=\pi/4$ so that the coefficients $J_{ii}$ are $J_{ii}=[1,\, 1/2,\, 0,\, 1/2,\, 1,\, 1/2,\, 0...]$. If light mode functions are not traveling waves, the  general form of the interaction between the modes follows $V_j= \sum_{m=1}^{R} g_{mc} J_{jj}$ where $J_{jj}$ is not a simple phase factor. Therefore, in order to engineer a given long-range interaction it is not possible to use the simple DFT formalism for computing the coefficients $g_{mc}$ but one has to recur to numerical methods.
}

\subsection{Multiple cavity modes and one probe}

In order to obtain an interaction that depends solely on the distance between the atoms (or the modes), we turn to the case of one classical probe which scatter photons to $R$ light modes. This scheme can be realized using multiple cavities or a multimode cavity. 
Considering only the events when light is scattered by the atoms from the probe beam to one of the cavities and neglecting the photon scattering between different cavities, we find that the atom-atom interaction is
\begin{align}\label{5-eq:Heff3}
\HH^D_3&= \sum_{m=1}^{R} \left(\frac{g^{11m}_\mathrm{eff}}{2} \hat{D}_{1m}^\dagger \hat{D}_{1m} + \mathrm{H.c.}\right),
\end{align}
where $g^{11m}_{\mathrm{eff}}={\Delta_{1m}|g_{1m}|^2}/(\Delta_{1m}^2+\kappa_{1m}^2)$. Eq. (\ref{5-eq:Heff3}) is fundamentally different from (\ref{5-eq:Heff2}) since here only one sum is present and the interaction does not mix $\hat{D}$ operators with different indexes. This allows to engineer long-range interactions that depend on the distance between the lattice sites and are analogous to the usual two-body interactions studied in condensed matter systems. We illustrate this by considering traveling waves as mode functions for the light and (\ref{5-eq:Heff3}) becomes
\begin{align}\label{5-eq:interaction3}
\HH^D_{3}&=\sum_{m=1}^{R} \sum_{i,j}\left( g^{11m}_\mathrm{eff} e^{- i (\b{k}_{p_1}-\b{k}_{c_m}) \cdot (\b{r}_i-\b{r}_j)} \hat{n}_i \hat{n}_j + \mathrm{H.c.}\right).
\end{align}
Fixing the direction of the wave vectors of the cavity modes so that $d (\b{k}_{p_1}-\b{k}_{c_m}) \cdot\hat e_{\b{r}_i}=\pi l/R$ for all $l=1,2,...R$ (corresponding to the angles fulfilling $\cos \theta_{c_l}=\cos \theta_{p_1} - \frac{\lambda}{2d} \frac{l}{R}$), the light scattering process defines $R$ spatial modes and we can perform a DFT analysis. Here, instead we have a Discrete Cosine Transform.  Specifically, Eq.~(\ref{5-eq:interaction3}) reduces to 
\begin{equation}
\HH^D_3=2\sum_{i,j}V_{|i-j|} \hat{N}_i \hat{N}_{j},
\end{equation}
where $V_j= \sum_{m=1}^{R}g^{11m}_{\mathrm{eff}}\cos({\frac{ \pi m j}{R} })$. In contrast to the scheme we considered above, here the interaction between the mode $i$ and the mode $j$ depends solely on the distance between the modes ($i-j$) and can be shaped to any $V_j$ profile changing the detunings and the decay coefficients of the cavities. Therefore,   this scheme allows realizing  quantum simulators that are able to mimic long-range interactions with an arbitrary spatial profile, such that $W_{i,j}\propto V_{|i-j|}$.

{In summary of this section, we have seen that the global (infinitely long-range) light--matter interaction enables simulating systems with rather short-range and tunable interactions. Moreover, simulating short-range interaction requires just a small number of light modes. Indeed, the price for this is that we do not simulate an original problem of interacting atoms at sites, but replace it by an effective one, simulating the interaction between the global modes. The effective mode Hamiltonians can be an acceptable representation of an otherwise experimentally hard to achieve quantum degenerate system with finite range interaction. As we show in this work, such a global, but importantly spatially structured interaction, can still compete with intrinsic short-range processes leading to non-trivial phases. As a result, the quantum phases will have properties of systems due to both short-range and global processes, thus directly benefiting from the collective enhancement of the light--matter coupling. This will enable simulating systems with tunable long- and short-range interactions, which is extremely difficult to achieve in other physical systems.}

\section{Concluding remarks of Chapter 5}

In this chapter, we addressed the regime, where the quantumnes of trapping light potential plays a crucial role. We clearly differentiated the effects arising in classical optical lattices, dynamical (semiclassical) optical lattices, and quantum optical lattices. We have shown that quantum and dynamical optical lattices offer a new tool to engineer nonlocal many-body interactions with light-induced structures. These interactions can break symmetries by design and imprint a pattern that governs the origin of many-body phases. This effectively bridges physics of long-range and short-range interactions. The light and matter are entangled, forming non-trivial light--matter correlated states. We suggested how to generate not only multimode density and supersolid patterns, but nonlocal patterns of the matter-filed coherences (bonds) as well, in particular, the delocalized superfluid and supersolid dimers. We indicated the effects appearing solely due to the quantum fluctuations of light, where no strong light can build up. These are the phenomena appearing solely in the quantum optical lattices. Fermions in quantum optical lattices were considered in the work \cite{SantiagoFermi2017}\footnote{Spins in quantum lattices were considered in Ref. \cite{SantiagoPRL2022}. The reviews on dynamical and quantum lattices can be found, e. g., in Refs. \cite{Mekhov2012,ritsch2013,Review2021,LarsonBook}}.

We demonstrated that the ability to tune the effective interaction between the atomic modes opens a path for quantum simulations based on the collective light--matter interaction.

\clearpage

%% file: Conclusions.tex
\chapter*{Conclusions}							
\addcontentsline{toc}{chapter}{Conclusions}	

In this work we presented novel phenomena appearing at the crossroad of quantum optics and quantum atom optics, where the quantization of both the light and atomic motion are equally important. The presentation was organized according to the model complexity and approximations used: (I) Quantum nondemolition measurements of atomic properties by light. (II) Use of the fundamental notion of quantum measurement backaction for preparing novel states of many-body systems. (III) Introducing the quantum measurement as a novel source of competitions in many-body systems with a plethora of important consequences. (IV) Feedback control of phase transitions (tuning their universality class) in many-body systems. (V) Predicting novel effects due to the quantization of the light trapping potentials (quantum optical lattices). We applied our model to bosonic and fermionic atoms, as well as to dipolar molecules. 

The main system we considered in this work is a quantum gas in an optical lattice, nevertheless, we obtain several results, which can be applicable in a much broader research context. This includes the following findings. We extended the paradigm of feedback control from the state control to the control of quantum phase transitions, including tuning their universality class. We presented the quantum backaction of weak measurements as a novel source of competitions in many-body systems. We merged the paradigms of quantum Zeno dynamics and non-Hermitian physics. We introduced a novel type of quantum Zeno phenomena with Raman-like transitions well beyond the standard concept of Zeno dynamics. We proposed a concept of quantum simulators based on the collective light--matter interaction and, thus, global addressing of quantum particles. Our models for atoms in lattices can be applied to other arrays of quantum particles (qubits) resulting in new methods of quantum measurements and probing, quantum state preparation, generation of genuine multipartite mode entanglement in quantum arrays.  The other possible quantum particles include superconducting circuits (qubits), Rydberg atoms, ions, polaritons as well as other micro- and nanostructures. In general, quantum measurements and feedback open a way to obtain novel phenomena untypical to both close unitary systems and open dissipative ones in the context of many-body physics.

The main results of this work are the following.

1. We derived a model, which describes ultracold atoms (bosons and fermions) trapped in an optical lattice and interacting with one or several quantized modes of light. This model forms a basis for theoretical study of quantum optics of quantum gases.

2. We proved that under certain conditions the measurement of light represents nondestructive (up to the quantum nondemolition, QND, level) probe of many-body variables of an ultracold atomic system. This is in contrast to the absolute majority of modern methods, which are destructive. 

3. We found the relations between the measurable light properties and quantum statistical variables of a quantum gas such as fluctuations and multi-point spatial density correlations. Moreover, we proved that the distribution functions of various atomic variables can be directly mapped on the transmission spectrum of a high-Q cavity. In general, we proved that light measurements can distinguish between different many-body states of ultracold bosons and fermions, as well as few-body states of molecular complexes.  

4. We showed that light scattering is not only sensitive to the on-site atomic densities, but  also to the matter-field interference at its shortest possible distance in an optical lattice (the lattice period), which defines key properties such as tunneling, atom currents, and matter-wave phase gradients.

5. We showed that light scattering from atomic arrays constitutes a quantum measurement with a controllable form of the measurement backaction. We thus used the measurement as an active tool to prepare many-body atomic states such as number squeezed and macroscopic superposition states. Moreover, we demonstrated that the class of emerging many-body states can be chosen via the optical geometry and light frequencies.  

6. We proved that the backaction of quantum measurement constitutes a novel source of competitions in many-body systems, in addition to the standard tunneling and short-range atom interaction. As a general physical concept, new competitions can lead to new effects. We demonstrated a plethora of novel phenomena: the generation and macroscopic oscillations of matter modes, long-range correlated tunneling and genuinely multipartite mode entanglement, both protection and break-up of fermion pairs by the measurement, as well as the measurement-induced antiferromagnetic order. 

7. We predicted a new unconventional type of quantum Zeno dynamics due to the Raman-like transitions via the virtual states outside the Zeno subspace. We extended the notion of quantum Zeno dynamics into the realm of non-Hermitian quantum mechanics joining the two paradigms. 

8. We extended the concept of feedback control from the quantum state control (as known in quantum metrology) to the control of phase transitions in quantum systems. We showed that quantum weak measurements and feedback can induce phase transitions beyond the dissipative ones. Moreover, feedback allows controlling essentially quantum properties of phase transitions such as the critical exponents. Thus, we demonstrated tuning the universality class of a phase transition in a single setup\footnote{For a more recent work see Ref. \cite{IvanovPRA2021}}.  

9. We demonstrated that the quantum and dynamical natures of optical trapping potentials lead to new quantum phases of ultracold atoms unobtainable in comparable prescribed classical optical lattices. We demonstrated not only the density orders as lattice supersolid state and density waves, but the orders of the matter-wave amplitudes (bonds) such as superfluid and supersolid dimers. We formulated a concept of quantum simulators based on the collective light--matter interactions.

While several dynamical effects we predicted in this work have been already confirmed experimentally (the difference between small- and large-imbalanced lattice supersolids and density waves, the existence of feedback-induced phase transitions in an ultracold gas), the truly quantum effects are still awaiting their observation in systems, which go far beyond atomic systems only. It will be intriguing to study, how more advanced methods than feedback control can influence quantum systems, for example, applying the digital methods of machine learning and classical or quantum artificial intelligence in real time. This opens bright and long-term perspectives for the fields of quantum optics of quantum gases and open systems beyond dissipation, including both fundamental and applied areas of quantum science and quantum technologies.

\clearpage 

%% file: referencesIM.tex
\clearpage
\phantomsection
\addcontentsline{toc}{chapter}{\bibname}	

\bibliography{RefsAll}						